\documentclass[a4paper,11pt,openright,twoside]{book}

\usepackage[a4paper, total={6.8in, 8.9in}, bottom=1.5in]{geometry}
\usepackage{cite}
\usepackage[british]{babel}
\usepackage{kpfonts}
\usepackage[utf8]{inputenc}
\usepackage{layout}
\usepackage{color} 
\usepackage{braket}
\usepackage{multicol} 
\usepackage{epsfig}
\usepackage{cancel}
\usepackage{caption}
\usepackage{epsf}
\usepackage{fancyhdr}
\usepackage{feynmf} 
\usepackage{frontespizio}
\usepackage{rotating}
\usepackage{subfigure}
\usepackage{pstricks}
\usepackage{lettrine}
\usepackage{bbold}
\usepackage{calligra}
\usepackage{tikz}
\usepackage{subfigure}
\usepackage{mathrsfs}
\usepackage{curve2e}
\usepackage{setspace}
\usepackage{indentfirst}
\usepackage{emptypage}
\usepackage{relsize}
\usepackage{slashed}
\usepackage{mathrsfs}
\usepackage{stackengine}
\usepackage{calc}
\usepackage{hyperref}
\hypersetup{
	colorlinks,
	citecolor=green,
	filecolor=black,
	linkcolor=blue,
	urlcolor=black
}

\newcommand{\mO}{\mathcal{O}}

\let\a=\alpha   \let\b=\beta   \let\g=\gamma   \let\d=\delta
\let\e=\epsilon    \let\h=\eta     
    \let\k=\kappa  \let\l=\lambda  \let\m=\mu
\let\n=\nu      \let\x=\xi     \let\p=\pi      \let\r=\rho
\let\s=\sigma        
\let\c=\chi     \let\y=\psi    
\let\G=\Gamma  \let\D=\Delta   \let\L=\Lambda
     \let\P=\Pi      

\newcommand{\ro}{\rho}

\newcommand{\sdfrac}[2]{\mbox{\small$\displaystyle\frac{#1}{#2}$}}

\renewcommand{\Re}{\textrm{Re}}
\renewcommand{\Im}{\textrm{Im}}
\newcommand{\figref}[1]{Fig.~\ref{#1}}			
\newcommand{\tabref}[1]{Tab.~\ref{#1}}			
\newcommand{\secref}[1]{Section~\ref{#1}}		
\newcommand{\appref}[1]{Appendix~\ref{#1}}		
\newcommand{\chapref}[1]{Chapter~\ref{#1}}		
\newcommand*{\Scale}[2][4]{\scalebox{#1}{$#2$}}%
\newcommand{\app}[4]{F_{\!#1}\!
	\left(\left.\substack{\Scale[1]{ #2} \\[1.5ex] \Scale[1]{#3}}\right| #4 \right) }

\newcommand{\hpg}[5]{{}_{#1}\mbox{\rm F}_{\!#2}\!
	\left(\left.\substack{\Scale[1]{ #3} \\[1.5ex] \Scale[1]{#4}}\right| #5 \right) }

\newcommand{\hpgo}[2]{{}_{#1}\mbox{\rm F}_{\!#2}}

\newcommand{\equal}{&\!\!=\!\! &}

\newcommand{\Tr}{\text{Tr}}

\def\nbox#1#2{\vcenter{\hrule \hbox{\vrule height#2in
			\kern#1in \vrule} \hrule}}
\def\sq{\,\raise.5pt\hbox{$\nbox{.09}{.09}$}\,}
\def\sqb{\,\raise.5pt\hbox{$\overline{\nbox{.09}{.09}}$}\,}

\def\Box{\sq}
\numberwithin{equation}{section}

\usepackage{accents}

\renewcommand{\chaptermark}[1]{\markboth{#1}{}}

\usepackage{pifont}
\newcommand{\HRule}{\rule{\linewidth}{0.5mm}}

\begin{document}
\thispagestyle{empty}
\begin{titlepage}
	\begin{center}
		{\includegraphics[scale=0.12]{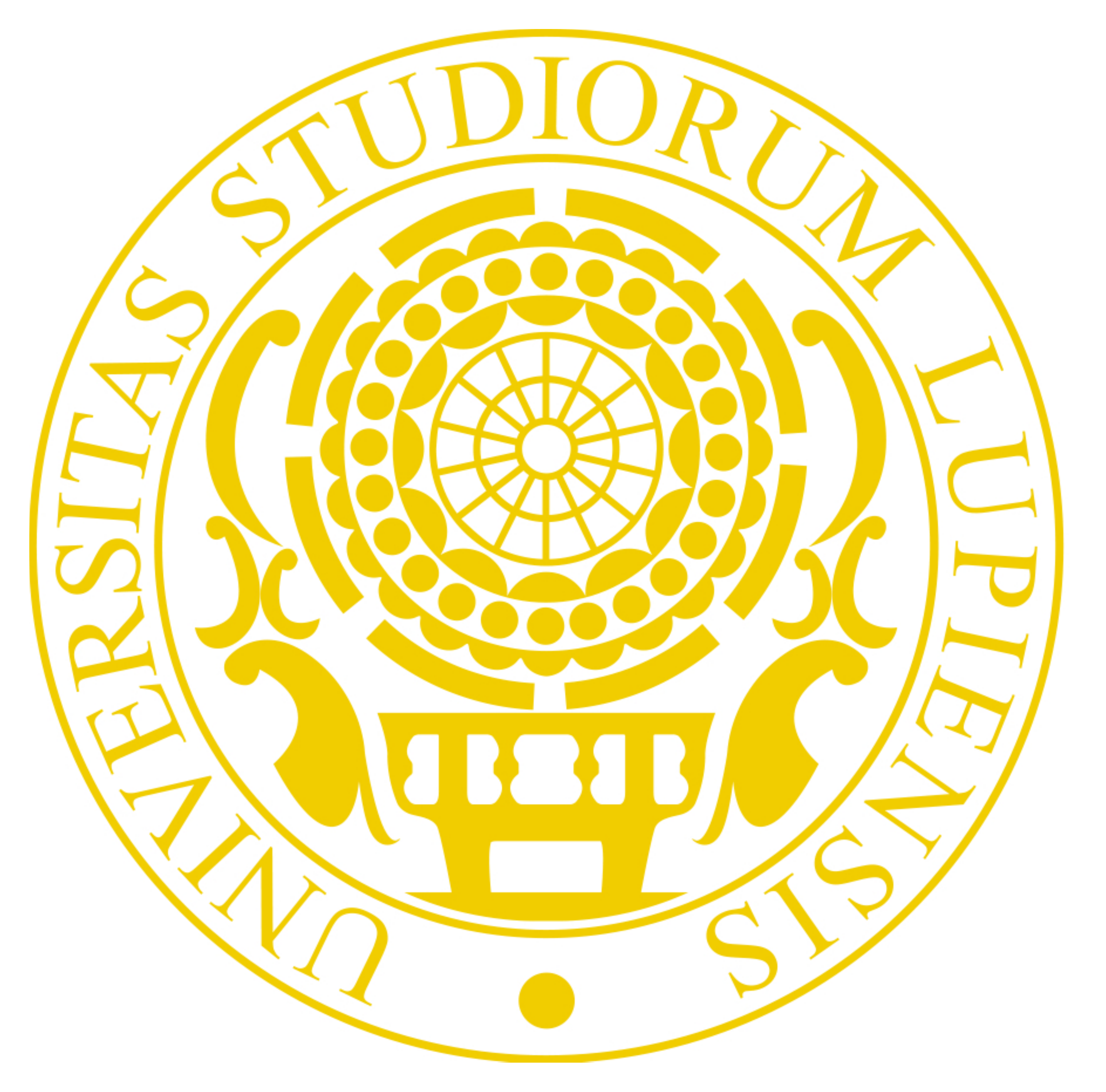}}\\[0.3cm]
		\scalebox{1.2}{\bf{Universit\`a del Salento}}\\[0.1cm]
		\HRule\\[0.3cm]
		\scalebox{1.5}{ Dipartimento di Matematica e Fisica ``Ennio De Giorgi''}	\\[2ex]
		\scalebox{1}{Corso di Dottorato di Ricerca in Fisica e Nanoscienze}\\[5cm]		
		{\Large { \bf{Conformal Symmetry in Momentum Space }}}\\[0.5cm]
		{\Large \bf { and Anomaly Actions in Gravity }}\\[4cm]
		\begin{minipage}{0.4\textwidth}
			\begin{flushleft}\large
				\textit{Supervisor}\\
				Prof.~Claudio Corian\`o
			\end{flushleft}
		\end{minipage}
		\begin{minipage}{0.4\textwidth}
			\begin{flushright} \large
				\textit{Candidate}\\
				Matteo Maria Maglio
			\end{flushright}
		\end{minipage}\\[3cm]
	\HRule\\
		Tesi di Dottorato in Fisica e Nanoscienze - XXXIII ciclo\\
		Anno Accademico 2019-2020
		\vfill
	\end{center}
\end{titlepage}
\cleardoublepage
\begin{flushright}
	\thispagestyle{empty}
	\textit{``If I had a world of my own, everything would be nonsense. \\
		Nothing would be what it is, because everything would be what it isn't. \\
		And contrary wise, what is, it wouldn't be. 
		\\And what it wouldn't be, it would. You see?''
		\\ Lewis Carroll}
\end{flushright}
\frontmatter
\newpage
\cleardoublepage
\tableofcontents
\chapter*{List of Publications}
\addcontentsline{toc}{chapter}{List of Publications}
\chaptermark{List of Publications}

\noindent The chapters of this thesis are based on the following research papers:
\begin{itemize}
	\item C. Corianò, M. M. Maglio, “\href{https://arxiv.org/abs/2005.06873}{Conformal Field Theory in Momentum Space and Anomaly Actions in Gravity: The Analysis of 3- and 4-Point Functions}”,     arXiv:2005.06873v2
	\item 	C. Corianò and M. M. Maglio, “\href{https://doi.org/10.1007/JHEP09(2019)107}{On Some Hypergeometric Solutions of the Conformal Ward Identities of Scalar 4-point Functions in Momentum Space}”, JHEP 09, 107 (2019)\\ doi:10.1007/JHEP09(2019)107 
	\item 	C. Corianò and M. M. Maglio, “\href{https://doi.org/10.1016/j.nuclphysb.2018.10.007}{The general 3-graviton vertex (TTT) of conformal field theories in momentum space in d = 4}”, Nucl. Phys. B 937, 56-134 (2018) \\ doi:10.1016/j.nuclphysb.2018.10.007 
	\item 	C. Corianò and M. M. Maglio, “\href{https://doi.org/10.1016/j.nuclphysb.2018.11.016}{Exact Correlators from Conformal Ward Identities in Momentum Space and the Perturbative TJJ Vertex}”, Nucl. Phys. B 938, 440-522 (2019)  \\
	doi:10.1016/j.nuclphysb.2018.11.016 
	\item C. Coriano, M. M. Maglio and E. Mottola, “\href{https://doi.org/10.1016/j.nuclphysb.2019.03.019}{TTT in CFT: Trace Identities and the Conformal Anomaly Effective Action}”, Nucl. Phys. B 942, 303-328 (2019) \\doi:10.1016/j.nuclphysb.2019.03.019  
	
\end{itemize}	
The following papers are related to the topics presented in this thesis but are not discussed in detail:
\begin{itemize}
	\item C. Corianò and M. M. Maglio, “\href{https://doi.org/10.3390/axioms9020054}{The Generalized Hypergeometric Structure of the Ward Identities of CFT's in Momentum Space in d > 2}”, Axioms 9, no.2, 54 (2020)\\ doi:10.3390/axioms9020054 
	\item 	C. Corianò, M. M. Maglio and D. Theofilopoulos, “\href{https://doi.org/10.1140/epjc/s10052-020-8089-1}{Four-Point Functions in Momentum Space: Conformal Ward Identities in the Scalar/Tensor case}”, Eur. Phys. J. C 80, no.6, 540 (2020) \\doi:10.1140/epjc/s10052-020-8089-1 
	\item M. N. Chernodub, C. Corianò and M. M. Maglio, “\href{https://doi.org/10.1016/j.physletb.2020.135236}{Anomalous Gravitational TTT Vertex, Temperature Inhomogeneity, and Pressure Anisotropy}”, Phys. Lett. B 802, 135236 (2020) \\  
	doi:10.1016/j.physletb.2020.135236 
	\item 	C. Corianò and M. M. Maglio, “\href{https://doi.org/10.1016/j.physletb.2018.04.003}{Renormalization, Conformal Ward Identities and the Origin of a Conformal Anomaly Pole}”, Phys. Lett. B 781, 283-289 (2018)  \\doi:10.1016/j.physletb.2018.04.003 
	\item 	C. Coriano, L. Delle Rose, M. M. Maglio and M. Serino, “\href{https://doi.org/10.1007/JHEP01(2015)091}{Electroweak Corrections to Photon Scattering, Polarization and Lensing in a Gravitational Background and the Near Horizon Limit}”, JHEP 01, 091 (2015)  doi:10.1007/JHEP01(2015)091  
\end{itemize}
\section*{Peer-reviewed conference proceedings }
\begin{itemize}
	\item C. Corianò, M. M. Maglio, A. Tatullo and D. Theofilopoulos, “\href{https://doi.org/10.22323/1.376.0080}{Dark Matter with Light and Ultralight Stückelberg Axions}”, PoS CORFU2019 (2020)  
	doi.org/10.22323/1.376.0080 
	\item C. Corianò, M. M. Maglio, A. Tatullo and D. Theofilopoulos, “\href{https://doi.org/10.22323/1.347.0072}{Exact Correlators from Conformal Ward Identities in Momentum Space and Perturbative Realizations}”, PoS CORFU2018 (2019), 072\\doi:10.22323/1.347.0072 
	\item C. Corianò and M. M. Maglio, “\href{https://doi.org/10.1051/epjconf/201819200047}{Conformal Ward Identities and the Coupling of QED and QCD to Gravity}”, EPJ Web Conf. 192 (2018), 00047 doi:10.1051/epjconf/201819200047 
\end{itemize}	
\chapter*{Introduction}
\addcontentsline{toc}{chapter}{Introduction}
\chaptermark{Introduction}

The study of conformal field theories (CFTs) has played a central role in theoretical/ mathematical physics for over half a century \cite{Kastrup:2008jn,DiFrancesco:1997nk}, with an impact on several different areas, from the theory of critical phenomena to string theory and, more recently, the AdS/CFT 
(AdS, Anti de Sitter)  correspondence. The latter allows establishing a link between gravitational and gauge forces in a specific supersymmetric setup.\\
In particular, such correspondence \cite{Maldacena:2003nj}, also known as gauge/gravity duality, has shown the importance of conformal symmetry in dimensions $d \geq 2$.
In $d=2$ the symmetry is infinite dimensional, and finite otherwise, for being defined by the generators of $SO(d,2)$.  \\
The relevance of the study of conformal correlation functions in $d>2$ 
is remarkable in areas as different as cosmology, condensed matter theory and particle phenomenology. For instance, recent experimental advances on Dirac and Weyl semimetals, have spurred a growing activity on the role played by CFT's and their anomalies (chiral, conformal) in the characterization of the fundamental properties of such materials \cite{Chernodub:2013kya,Ambrus:2019khr,Chernodub:2019tsx,Arjona:2019lxz,Gooth:2017mbd}.  \\
In our analysis, we will consider the case of $d > 2$, which is of utmost interest in physics. 
For quantum conformal invariant theories, the general idea is to develop a formalism that allows characterizing the structure of the corresponding correlation functions without resorting to a Lagrangian description. In this more general case, a CFT is essentially determined by a set of operators (primary fields) and by their descendants, which close an algebra via an operator product expansion. In principle, this allows to bootstrap correlators of the higher point from the lower point ones, by solving for the corresponding conformal blocks.    \\
Although in $d> 2$ the symmetry is less restrictive, correlation functions of CFT's up to 3-point functions can be determined in their tensorial structures and form factors only modulo few constants. These constraints take the form of conformal Ward identities (CWI's). 
We will show how it is possible to match general CFT results to ordinary Lagrangian formulations, in free field theory, limitedly to 3-point functions. The possibility of such direct match in momentum space, reproduces previous results of coordinate space \cite{Osborn:1993cr} but proceeds autonomously. This approach will allow investigating this class of theories in a framework that is quite close to studying the scattering amplitudes in ordinary perturbation theory. At the same time, as we are going to elaborate, it allows an entirely new perspective on the conformal anomaly's role in the dynamical breaking of such symmetry.  \\

\section*{The transition to momentum space}
\addcontentsline{toc}{section}{The transition to momentum space}

The solution of the CWI's in position (coordinate, configuration) space has been addressed long ago for 3-point functions\cite{Erdmenger:1996yc,Osborn:1993cr}. Most notably, Osborn and Petkou outlined a method to solve such identities in position space, indicating how the conformal anomaly could be included in a special and important correlators class. Their analysis covered correlators containing up to 3 insertions of stress energy tensors, beside conserved currents and scalar operators all of the generic scaling dimensions 
$(\Delta_i)$. \\
The idea of using conformal Ward identities (CWI's) to determine the structure of 3-point functions in momentum space was presented independently in \cite{Coriano:2013jba} and \cite{Bzowski:2013sza}, the second of which outlines a method that includes the tensor case. 
In \cite{Bzowski:2013sza}, Bzowski, McFadden and Skenderis (BMS)  have indicated a possible route to identify the solution of the conformal Ward identities (CWI's) of 3-point functions of tensor correlators, the most demanding one being the $TTT$, with three stress energy tensors $(T)$. The method builds the solution of the CWI's starting from the transverse-traceless components of such correlators and their related 2-point functions, such as the $TT$.
We will illustrate the method by working out the $TTO$ and TTT cases relatively straightforwardly, clarifying all the intermediate technical steps.  We will discuss all the simplifications that a matched perturbative analysis performed in \cite{Coriano:2018bsy,Coriano:2018bbe,Coriano:2018zdo} brings in the computation of the explicit expression of this and other similar vertices, which remains valid non perturbatively. Our interest in this analysis has grown out of previous studies in perturbative field theory, QCD and QED, of similar vertices (TJJ) \cite{Giannotti:2008cv,Armillis:2009pq,Armillis:2009im,Armillis:2010qk} where it has been shown that the breakings of conformal and chiral symmetries are associated to specific behaviour of a particular form factor of a tensorial 3-point function. It is therefore of interest to investigate if this phenomenon remains valid beyond perturbation theory. In a final section, we will show how a similar result also holds for the TTT correlator. \\
There are two main reasons why CFT in momentum space is essential for our understanding of the role of conformal symmetry in field theory. The first is that it allows establishing a direct connection with ordinary scattering amplitudes, in the analysis of which there has been significant progress up to very high perturbative orders. 
The second is the possibility of investigating the role of the conformal anomalies.  \\
Indeed, the operatorial expansion of two operators in coordinate space for any CFT, in general, does not address the issue of possible anomalies, since the operators are taken at different spacetime points. In this case, the CWI's have to be modified with the addition of ultralocal terms obtained by differentiating the anomaly functional. In other words, the corresponding anomalous CWI's are solved separately in their homogenous and inhomogeneous forms. At first one solves the equations when the coordinate points are kept separate. Then the contribution to their anomaly is added by hand as un ultralocal term, in the coincidence limit.\\
However, there is no physical understanding of how breaking a conformal symmetry by a quantum anomaly occurs if we stay in coordinate space. In this case, the exercise is purely formal, for a phenomenon - the breaking of an essential and possibly fundamental symmetry - which is crucial in so many physics areas.\\

\section*{Perturbative breakings in anomaly form factors}
\addcontentsline{toc}{section}{Perturbative breakings in anomaly form factors}

As just mentioned, perturbative analysis in momentum space \cite{Giannotti:2008cv,Armillis:2009pq,Armillis:2009im}, also in supersymmetric theories \cite{Coriano:2014gja}, have shown that the signature of such breaking is in the appearance of specific massless poles in correlators with a $T$ insertion. The structure that emerges from such virtual exchanges can be directly compared to the free field theory prediction. We can show that the explicit expression of the correlation functions is identical in the two cases. \\
Such a matching has recently been discussed in  \cite{Coriano:2018bsy,Coriano:2018bbe,Coriano:2018zdo}. We have shown how this test can be performed on a complex correlator such as the $TTT$. For the perturbative matching, in this case, one needs two sectors, a scalar and a fermion sector, linearly combined with arbitrary multiplicities of their particle content.  With a spin one running inside the loops, a third sector is necessary to account for its anomalies. \\  
This mapping of a general, non-perturbative result, to a perturbative one (a simple one-loop matching), allows proceeding with a drastic simplification of such correlators' expressions, providing the most straightforward realization of the form factors identified by the BMS decomposition. At this stage, one can determine the structure of the anomaly contributions in the several form factors present in a 3-point function.  \\
In the matched perturbative description, the conformal anomaly emerges from the renormalization of the longitudinal (or semilocal) terms of the correlator, which exhibits a specific pole structure. 
The distinction between such massless interactions, clearly related to renormalization, and other possible massless exchanges is that they are directly associated with the anomaly functional derivatives. \\
The pattern, in this case, generalizes what one obtains in the $TJJ$ case. 
We have shown in great detail in \cite{Coriano:2018zdo} that, in this correlator, the generation of a massless pole is a consequence of the process of regularization of one specific form factor - the anomaly form factor - as the spacetime dimensions $d$ tend to 4 $(d\to 4)$.    \\
It is then natural, from this perspective, to investigate whether a specific conformal anomaly action can account for these (nonlocal) anomaly contributions which are predicted both in free field theory and non perturbatively. For this reason, we have turned to a description of the non local Riegert action, which is expected to generate such terms directly. \\
Obviously, studies of 1-particle irreducible effective actions in gravitational backgrounds are common in the literature (see for instance \cite{Bastianelli:2004zp,Bastianelli:2007jv,Bastianelli:2012bz,Coriano:2012wp}), for instance in a $E/m$ expansion, in the presence of massive intermediate states.
More recent investigations of conformal correlators both for chiral and conformal anomalies are those of \cite{Bastianelli:2019zrq,Bonora:2014qla,Bonora:2017gzz}.  \\
For our goals, we will review the origin of the nonlocal Riegert action pedagogically, detailing the variational solution of the anomaly constraint by integration in field space, which takes to such expression.  
\section*{Massless exchanges as dynamical breaking of the conformal symmetry}
\addcontentsline{toc}{section}{Massless exchanges as dynamical breaking of the conformal symmetry}
Compared to local actions which introduce extra degrees of freedom - in the form of a dilaton or an axion - unified in a St\"uckelberg supermultiplet in the superconformal case - the nonlocal description appears to be dynamical. The pole is the result of a collinear (particle/antiparticle) exchange in the loop, captured by a local spectral density $(\sim\delta(s))$ in the dispersion variable ($s$).\\  
The emergence of composite intermediate massless states characterising the light-cone dynamics of the field operators in the theory predicts the nonlocal action, which is correctly reproduced by the matched perturbative theory. 
The equivalence between the perturbative and the non-perturbative realisations of specific correlation functions, for operators containing insertions of only T's and J's, and a careful analysis of the renormalisation in both approaches, shows that the appearance of massless exchanges in the two cases is not a spurious prediction of perturbation theory, but the signature of the anomaly. \\
Its relevance both in the context of condensed matter theory \cite{Chernodub:2019tsx,Arjona:2019lxz,Mottola:2019nui,Chernodub:2017jcp,Rinkel:2019kpo}, in the theory of Dirac and Weyl semimetals and in gravitational waves \cite{Mottola:2016mpl} is what makes it worthy of a close attention.
\section*{Local duality and dimensional transmutation}
\addcontentsline{toc}{section}{Local duality and dimensional transmutation}
The relevance of both local and nonlocal anomaly actions has been addressed repeatedly in the past, with results and predictions touching a wide range of phenomena, 
which are gradually being uncovered \cite{Chernodub:2019tsx,Mottola:2019nui,Mottola:2016mpl}. In \cite{Coriano:2019dyc} we have proposed that these actions parameterise the same anomaly phenomenon at two different ends (the UV and the IR) of a renormalisation group flow, with the possibility that a non-perturbative dynamics will connect the two. Massless excitations in the UV, captured by the anomaly diagrams, may turn into asymptotic degrees of freedom in the IR in the presence of nonpertubative interactions. Coherent interactions may be the cause of such behaviour in condensed matter theory, with the generation of a physical scale. The phenomenon of an "a  massless pole turning into a cut", with the cut associated to a mass scale, noticed in behaviour the spectral densities of anomaly form factors in the past \cite{Giannotti:2008cv,Coriano:2014gja}, is a generic phenomenon of anomalies, likely related to the emergence of a physical scale in a massless theory. \\
Anomalies come with specific superconvergent sum rules, with remarkable properties of their associated spectral densities. This phenomenon share similarity with local duality (quark/hadron duality) in QCD, due to the presence, also in this case, of a sum rule and a dispersion relation connecting the UV and the IR behaviour of specific cross-sections integrated over the energy.\\
In conformal theory, this is essentially linked to dimensional transmutation (DT), 
where a scale invariant theory develops a dynamical scale. \\
In QCD, DT is a non-perturbative effect, related to the breaking of classical scale invariance of the theory and the emergence of $\Lambda_{QCD}$, which is identified, in perturbation theory, from the singularity in momentum space of the running coupling. The perturbative predictions have little to say about confinement or chiral symmetry breaking in the theory, which characterises this phenomenon in a far more complex way than simple perturbative analysis.
In the case of quantum scale invariant theories, broken by anomalies, this phenomenon could be different. The symmetry holds at the quantum level, and could even be simpler.     
\section*{Overview of the thesis}
\addcontentsline{toc}{section}{Overview of the thesis}
The thesis is organised into different chapters. 
In the first chapter, we review the basic properties of conformal field theory in coordinate space. We study the implications of conformal invariance on the structure of $2$-, $3$- and $4$-point correlation functions by introducing the Conformal Ward Identities (CWI's). 

In the second chapter, we reformulate the conformal constraints in momentum space. We provide the solutions of the CWI's for the $2$-point functions involving scalar and tensor operators. We illustrate the hypergeometric character of the CWI's, and we discuss the general solution for the $3$-point function involving scalar primary fields. 

In the third chapter, we present the analysis of $4$-point functions in momentum space by investigating some scaling solutions of primary operators and showing that the corresponding CWI's hypergeometric character is preserved. The solutions that we present can be classified as being dual-conformal and conformal at the same time. We also discuss how to construct solutions of the CWI's in a specific kinematic approximation. 

In the fourth chapter, we present the general method of reconstructing any $3$-point function involving currents and multiple insertions of the stress energy tensor operator. We give a fully work-out example, the $TTO$ correlation function, and we present the general solution for the $TJJ$ and the $TTT$ correlators. 

In the fifth chapter, we discuss the connection between the general solutions of the CWI's and the perturbative realisations of the $TJJ$ correlator. In this way, we establish an essential link between perturbative and non-perturbative approaches to CFT, bringing significant simplifications of the general result. 

In the sixth chapter, we present a comparative study of the $3$-graviton vertex $TTT$ in momentum space. This analysis is an extension of the $TJJ$ presented in chapter five. In particular, we discuss the organisation of the result for the $TTT$, for its renormalised expression in $d=4$, focusing on the anomaly part of this correlator. Our analysis shows that such a contribution is not an artefact of a specific decomposition, but it is a general feature of CFT's and is related to renormalisation. 

In the seventh chapter, we show how to construct the part of the effective action related to the trace anomaly. We prove that the anomaly effective action, provided by Riegert, determines all the anomalous contributions to the $3$-point stress tensor correlations functions and in general higher point functions in $d=4$. 
\mainmatter
\chapter{Conformal symmetry in coordinate space}\label{chapter1}

\section{Conformal transformations}
\label{prima}

We present a brief review of the transformations which identify the conformal group in a flat $d-$dimensional Euclidean space $\mathbb{R}^d$ \cite{ DiFrancesco:1997nk,Fradkin:1996is}, showing it is locally isomorphic to $SO(1,d+1)$.
Conformal transformations may be defined as those transformations $x_\mu\to x'_\mu(x)$ that preserve the infinitesimal length up to a local factor $(\mu=1,2,\ldots d)$
\begin{equation}
	dx_\mu dx^\mu\to dx'_\mu dx'^\mu=\Omega(x)^{-2}dx_\mu dx^\mu.\label{lineint}
\end{equation}
In the infinitesimal form they are given by 
\begin{equation}
	x'_\mu(x)=x_\mu+a_\mu+\omega_{\mu\nu}x^\nu+\lambda x_\mu+b_\mu x^2-2b\cdot x\,x_\mu, \label{transf}
\end{equation}
with
\begin{equation}
	\Omega(x)=1-\sigma(x),\quad \sigma(x)=\lambda-2b\cdot x,\label{Om}
\end{equation}
and $b_\mu$ is a constant $d$-vector
The transformation in \eqref{transf} is composed of the parameters $a_\mu$ for the translations, $\omega_{\mu\nu}=-\omega_{\nu\mu}$ for boosts and rotations, $\lambda$ for the dilatations and $b_\mu$ for the special conformal transformations. The first three define the Poincar\`e subgroup, obtained for $\Omega(x)=1$, which leaves invariant the infinitesimal length. By considering the definition of the inversion
\begin{equation}
	x_\mu\to x'_\mu=\sdfrac{x_\mu}{x^2},
\end{equation}
the special conformal transformations can be realized as a translation preceded and followed by an inversion, leading to the finite transformation
\begin{equation}
	x^{\prime\,\mu}=\frac{x^{\mu}-b^{\mu}x^2}{1-2b\cdot x+b^2x^2}.
\end{equation}
Notice that an infinitesimal transformation
\begin{equation}
	x^\mu(x)\to x'^\mu(x)=x^\mu + v^\mu(x)
\end{equation}
is classified as an isometry if it leaves the metric $g_{\mu \nu}(x)$ invariant in form. If we denote with 
$g'_{\mu\nu}(x')$ the new metric in the coordinate system $x'$, then an isometry is such that
\begin{equation}
	g^\prime_{\mu\nu}(x')=g_{\mu\nu}(x').
	\label{met1}
\end{equation}
This condition can be inserted into the ordinary covariant transformation rule for $g_{\mu\nu}(x)$ to give 
\begin{equation}
	g^\prime_{\mu\nu}(x')=\frac{\partial x^\rho  }{\partial x'^\mu }\frac{\partial x^\sigma}{\partial x'^\nu} g_{\rho\sigma}(x)= g_{\mu\nu}(x'),
\end{equation}
from which one derives the Killing equation for the metric
\begin{equation}
	v^\alpha\partial_\alpha g_{\mu\nu} + g_{\mu\sigma} \partial_\nu v^\sigma + 
	g_{\sigma \nu} \partial_\mu v^\sigma =0.
\end{equation}
For a conformal transformation, according to \eqref{lineint}, the metric condition (\ref{met1}) is replaced by the condition 
\begin{equation}
	g'_{\mu\nu}(x')=\Omega^{-2} g_{\mu\nu}(x'),
\end{equation}
generating the conformal Killing equation (with $\Omega(x)= 1-\sigma(x)$)
\begin{equation}
	v^\alpha\partial_\alpha g_{\mu\nu} + g_{\mu\sigma} \partial_\nu v^\sigma + 
	g_{\sigma \nu} \partial_\mu v^\sigma=2 \sigma g_{\mu\nu},
\end{equation}
in accordance with \eqref{transf}. In the flat spacetime limit this becomes 
\begin{equation}
	\label{sigma}
	\partial_\mu v_\nu + 
	\partial_\nu v_\mu=2 \sigma\, \delta_{\mu\nu},\qquad  \sigma=\frac{1}{d} \partial \cdot v,
\end{equation}
that, except for $d=2$, it has the general solution
\begin{equation}
	v_\mu(x)=a_\mu+w_{\mu\nu}x^{\nu}+\lambda\,x_\mu+b_\mu\,x^2-2x_\mu\,b\cdot x,\qquad \sigma=\lambda-2b\cdot x. 
\end{equation}
For any such conformal transformation we may define a local orthogonal transformation by 
\begin{equation}
	\label{rot1}
	R^\mu_\alpha(x)=\Omega(x) \frac{\partial x'^\mu}{\partial x^\alpha}.
\end{equation}
We can first expand genericaly $R$ around the identity as 
\begin{equation}
	R=\mathbf{ 1  } + \left[\mathbf{\epsilon}\right] +\ldots
\end{equation}
with an antisymmetric matrix $\left[\epsilon\right]$, which can be re-expressed in terms of antisymmetric parameters 
($\tau_{\rho\sigma}$) and $1/2 \,d\, (d-1)$ generators $\Sigma_{\rho\sigma}$ of $SO(d)$ as 
\begin{eqnarray}
	\left[\epsilon\right]_{\mu\alpha}&=&\frac{1}{2} \tau_{\rho\sigma}\left(\Sigma_{\rho\sigma}\right)_{\mu\alpha}\nonumber \\
	\left(\Sigma_{\rho\sigma}\right)_{\mu\alpha}&=&\delta_{\rho\mu}\delta_{\sigma\alpha}-\delta_{\rho\alpha}\delta_{\sigma\mu}
\end{eqnarray}
from which, using also (\ref{rot1}) we derive a constraint between the parameters of the conformal transformation $(v)$ and the parameters $\tau_{\mu\alpha}$ of $R$
\begin{equation}
	R_{\mu\alpha}= \delta_{\mu\alpha} + \tau_{\mu\alpha}=\delta_{\mu\alpha} + \frac{1}{2}\partial_{[\alpha }v_{\mu]}\label{Rrepr}
\end{equation}
with $ \partial_{[\alpha }v_{\mu]}\equiv
\partial_{\alpha }v_{\mu}-\partial_{\mu }v_{\alpha}$.\\
\section{Conformal group}
Looking at the infinitesimal transformation \eqref{transf}, the generators of the conformal group are easily seen to be
\begin{equation}
	\begin{aligned}
		\text{(translation)}\quad&P_\mu=\partial_\mu,\\
		\text{(dilations)}\quad&D=x^\mu\partial_\mu,\\
		\text{(rotations)}\quad&L_{\mu\nu}=x_\nu\partial_\mu-x_\mu\partial_\nu,\\
		\text{(special conformal)}\quad&K_{\mu}=2x_\mu x^\nu\partial_\nu-x^2\partial_\mu.
	\end{aligned}
\end{equation}
These generators satisfy the following commutation rules that define the conformal algebra
\begin{equation}
	\begin{split}
		&[D,K_\mu]=K_\mu,\\
		&[D,P_\mu]=-P_\mu,\\
		&[P_\mu,K_\nu]=2\left(\d_{\mu\nu}\,D+L_{\mu\nu}\right),\\
		&[P_\mu,L_{\nu\rho}]=\delta_{\mu\rho}P_{\nu}-\delta_{\mu\nu}P_{\rho},\\
		&[K_\mu,K_\nu]=[P_\mu,P_\nu]=[D,D]=[D,L_{\mu\nu}]=0,\\
		&[K_\mu,L_{\rho\sigma}]=\delta_{\mu\ro}K_\sigma-\delta_{\mu\sigma}K_\rho,\\
		&[L_{\mu\nu},L_{\rho\sigma}]=\delta_{\mu\rho}L_{\nu\sigma}+\delta_{\nu\sigma}L_{\mu\rho}-\delta_{\nu\rho}L_{\mu\sigma}-\delta_{\mu\sigma}L_{\nu\rho}.
	\end{split}
	\label{spec}
\end{equation}
We define the generators $J_{ab}$, where $J_{ba}=-J_{ab}$ and $a,b\in {-1,0,1,\dots,d}$, as 
\begin{equation}
	\begin{aligned}
		&J_{\mu\nu}=L_{\mu\nu},&& J_{-1,\mu}=\frac{1}{2}\left(P_\mu-K_\mu\right),\\
		&J_{-1,0}=D,&& J_{0,\mu}=\frac{1}{2}\left(P_\mu+K_\mu\right),
	\end{aligned}
\end{equation}
such that the commutation rules \eqref{spec} are written in the compact and simpler form
\begin{equation}
	[J_{ab},J_{cd}]=\eta_{ac}\,J_{bd}+\eta_{bd}\,J_{ac}-\eta_{ad}\,J_{bc}-\eta_{bc}\,J_{ad}\label{LorAlg}
\end{equation}
where $\eta_{ab}=\text{diag}(-1,1,1,\dots,1)$. The commutation relations \eqref{LorAlg} define the algebra of the group $SO(1,d+1)$, and this shows the isomorphism between the conformal group in $d$-dimensional Euclidean space and the group $SO(1,d+1)$ with $1/2(d+2)(d+1)$ parameters.  
Notice from \eqref{spec} that rotations, dilations and translations form a subgroup of the full conformal group. This means that a theory invariant under translations, rotations and dilations can be not invariant under conformal transformations. 
\section{Representations of the Conformal Group}
We illustrate in this section how a general field transforms under conformal transformations, and in particular we are going to show the representations of the conformal group. 
We consider first the action of the conformal symmetry on a vector field.
Denoting with $\Delta_A$ the scaling dimensions of a vector field $A_\mu(x)'$,  its variation under a conformal transformation can be expressed via $R$ in \eqref{Rrepr} as
\begin{eqnarray}
	\label{trans1}
	A'^\mu(x')&=&\Omega^{\Delta_A} R^{\mu}_{\alpha} A^\alpha(x)\nonumber \\
	&=&\big(1-\sigma\,+\,\dots\big)^{\Delta_A}\bigg(\delta^\mu_\alpha+\frac{1}{2}\,\partial_{\alpha }v^{\mu}-\frac{1}{2}\,\partial^{\mu}v_{\alpha}\,+\,\dots\,\bigg)\,A^\alpha(x)
\end{eqnarray}
from which one can easily deduce that 
\begin{equation}
	\label{trans2}
	\delta A^\mu(x)\equiv A'^\mu(x)-A^\mu(x)=-(v\cdot \partial +\Delta_A \sigma)A^\mu(x) +\frac{1}{2} \partial^{[\alpha }v^{\mu]}A_\alpha(x), 
\end{equation}
which is defined to be the Lie derivative of $A^\mu$ in the $v$ direction
\begin{equation}
	L_v A^\mu(x) \equiv -\delta A^\mu(x).
\end{equation}
As an example, in the case of a generic rank-2 tensor field ($\phi^{I \, K}$) of scaling dimension $\Delta_\phi$, transforming according to a representation $D^I_J(R)$ of the rotation group $SO(d)$, (\ref{trans1}) takes the form 
\begin{equation}
	\phi'^{I\, K}(x')=\Omega^{\Delta_{\phi}} D^I_{I'}(R) D^K_{K'}(R) \phi^{I' \,K'}(x).
\end{equation}
In the case of the stress energy tensor ($D(R)=R$), with scaling (mass) dimension $\Delta_T$  $(\Delta_T=d)$ the analogue of (\ref{trans1}) is 
\begin{eqnarray}
	T'^{\mu\nu}(x')&=&\Omega^{\Delta_T} R^\mu_\alpha R^\nu_\beta T^{\alpha\beta}(x)\nonumber \\
	&=&\bigg(1- \Delta_T \sigma +\ldots\bigg)\bigg(\delta^\mu_\alpha+\frac{1}{2}\partial_{\alpha }v^{\mu}-\frac{1}{2}\partial^{\mu}v_{\alpha}+\ldots\bigg)
	\bigg(\delta^\nu_\beta+\frac{1}{2}\partial_{\beta }v^\nu-\frac{1}{2}\partial^{\nu }v_\beta+\ldots\bigg)\,T^{\a\b}(x).
\end{eqnarray}
Then one gets 
\begin{equation}
	\delta T^{\mu\nu}(x)=-\Delta_T\, \sigma\, T^{\mu\nu} -v\cdot \partial \,T^{\mu\nu}(x) +
	\frac{1}{2}\partial_{[\alpha }v_{\mu]}\,T^{\alpha\nu} +\frac{1}{2}\partial_{[\nu }v_{\alpha]}T^{\mu\alpha}.
\end{equation}
For a special conformal transformation (SCT) one chooses 
\begin{equation}
	v_{\mu}(x)=b_\mu x^2 -2 x_\mu b\cdot x
\end{equation}
with a generic parameter $b_\mu$ and  $\sigma=-2 b\cdot x$ (from \ref{sigma}) to obtain
\begin{equation}
	\delta T^{\mu\nu}(x)=-(b^\alpha x^2 -2 x^\alpha b\cdot x )\, \partial_\alpha  T^{\mu\nu}(x)   - \Delta_T \sigma T^{\mu\nu}(x)+
	2(b^\mu x_\alpha- b_\alpha x^\mu)T^{\alpha\nu} + 2 (b^\nu x_\alpha -b_\alpha x^\nu)\, T^{\mu\alpha}(x).
\end{equation}
It is sufficient to differentiate this expression respect to $b_\kappa$ in order to derive the form of the special conformal transformation $K^\kappa$ on $T$ in its finite form 
\begin{align}
	\mathcal{K}^\kappa T^{\mu\nu}(x)&\equiv\delta_\kappa T^{\mu\nu}(x) =\frac{\partial}{\partial b_\kappa} (\delta T^{\mu\nu})\notag\\
	&= -(x^2 \partial^\kappa - 2 x^\kappa x\cdot \partial) T^{\mu\nu}(x) + 2\Delta_T x^\kappa T^{\mu\nu}(x) +
	2(\delta_{\mu\kappa}x_\alpha -\delta_{\alpha}^{\kappa}x_\mu) T^{\alpha\nu}(x) + 2 (\delta^\kappa_\nu x_{\alpha} -\delta_\alpha^\kappa x_\nu )T^{\mu\alpha}. 
	\label{ith}
\end{align}
To summarize, having specified the elements of the conformal group, we can define a primary field $\mathcal{O}^i(x)$, where $i$ runs over the representation of the group which the field belongs to, through the transformation property under a conformal transformation $g$ belonging to the conformal group $SO(2,d)$ in the form
\begin{equation}\label{spintransf}
	\mathcal{O}^i(x)\xrightarrow{g}\mathcal{O}'^i(x')=\Omega(x)^\Delta\,D^i_j(g)\,\mathcal{O}^j(x),
\end{equation}
where $\Delta$ is the scaling dimension of the field and $D^i_j(g)$ denotes the representation of $O(d)$ acting on $\mathcal{O}'^i$. In the infinitesimal form we have
\begin{equation}
	\d_g\mathcal{O}^i(x)=-(L_g\mathcal{O})^i(x),\qquad \text{with}\quad L_g=v\cdot \partial+\Delta\sigma-\sdfrac{1}{2}\partial_{[\mu}v_{\nu]}\Sigma^{\mu\nu},
\end{equation}
where the vector $v_\mu$ is the infinitesimal coordinate variation $v_\mu=\d_gx_\m=x'_\mu(x)-x_\mu$ and $(\Sigma_{\mu\nu})^i_j$ are the generators of $O(d)$ in the representation of the field $\mathcal{O}^i$. The explicit form of the operators $L_g$ can be obtained from \eqref{transf} and \eqref{Om} and are given by
\begin{align}
	\text{translations}\qquad L_g&=a^\mu\partial_\mu,\\
	\text{rotations}\qquad L_g&=\sdfrac{\omega^{\mu\nu}}{2}[x_\nu\partial_\mu-x_\mu\partial_\nu]-\Sigma_{\mu\nu},\\
	\text{scale transformations}\qquad L_g&=\lambda\,[x\cdot \partial +\Delta],\\
	\text{special conformal transformations}\qquad L_g&=b^\mu[x^2\partial_\mu-2x_\mu\,x\cdot \partial-2\Delta\,x_\mu-2x_\nu\Sigma_\mu^{\ \nu}].
\end{align}

Let us now consider the subalgebra of the four-dimensional conformal algebra, corresponding to dilatations and Lorentz transformations. This allows us to label different representations of the conformal algebra with $(\D,j_L,j_R)$, where $\D$ is the scaling dimension and $j_L$, $j_R$ are Lorentz quantum numbers.  For any quantum field theory, unitarity implies that all the states in a representation must have a positive norm, imposing bounds on the unitarity representations. If one considers the compact subalgebra $\mathfrak{so}(2)\oplus\mathfrak{so}(4)$ of $\mathfrak{so}(4,2)$, its unitary representations are labelled with $(\D,j_L,j_R)$ and have to satisfy the constraints
\begin{equation}
	\begin{split}
		\D\ge 1+j_L\quad\text{for}\ \ j_R=0,&\qquad \D\ge 1+j_R\quad\text{for}\ \ j_L=0,\\
		\D\ge 2+j_L+j_R&\quad\text{for both}\ \ j_L,\,j_R\ne0.
	\end{split}
\end{equation}

In this context, for scalars we have $\D\ge1$, for vectors $\D\ge3$ and for symmetric traceless tensors $\D\ge4$. These bounds are saturated by a free scalar field $\phi$, a conserved current $J_\m$ and a conserved symmetric traceless tensor $T_{\m\n}$. In $d$ dimensions, the bound for fields of spin $s$ take the form
\begin{align}
	&\D\ge \sdfrac{d-2}{2},\hspace{2cm} s=0,\\
	&\D\ge \sdfrac{d-1}{2},\hspace{2cm} s=1/2,\\
	&\D\ge d+s-2,\hspace{1.53cm} s\ge1.
\end{align}
In a CFT fields transform under irreducible representations of the conformal algebra. In order to construct its irreduced representations for general dimensions, it is used the method of the {induced representations}. We briefly comment on this point.\\
First, we analyze the transformation properties of the field $\mathcal{O}$ at $x=0$. Then, with the help of the momentum vector $P^\m$, we may shift the argument of the field to an arbitrary point $x$, in order to obtain the general transformation rule. For rotations, we have postulated that 
\begin{equation}
	[L_{\m\n},\mathcal{O}(0)]=\mathcal{S}_{\m\n}\,\mathcal{O}(0),
\end{equation}
where $\mathcal{S}_{\m\n}$ is a finite-dimensional representation matrix of rotations, determining the spin of the field $\mathcal{O}(0)$. In addition, for the conformal algebra we postulate commutation relations with the dilatation operator $D$,
\begin{equation}
	[D,\mathcal{O}(0)]=\D_{\mathcal{O}}\,\mathcal{O}(0).
\end{equation}
This relation implies that $\mathcal{O}$ has the scaling dimension $\D_{\mathcal{O}}$, i.e. under dilations $x\mapsto x'=\l x$ (for a real $\lambda$) it transforms as 
\begin{equation}
	\mathcal{O}(x)\mapsto \mathcal{O}'(x')=\l^{-\D_\mathcal{O}}\,\mathcal{O}(x).
\end{equation}
In particular, a field $\mathcal{O}$, covariantly transforming under an irreducible representation of the conformal algebra, has a fixed scaling dimension and it is therefore an eigenstate of the dilatation operator $D$. In a conformal algebra it is sufficient to consider particular fields, the \textit{conformal primary fields}, which satisfy the commutation relation
\begin{equation}
	[K_\m,\mathcal{O}(0)]=0.\label{bound}
\end{equation}
By applying the commutation relations of $D$ with $P_\m$ and $K_\m$ to the eigenstates of $D$, one observes that $P_\m$ raises  while $K_\m$ lowers the scaling dimension  since
\begin{align}
	[D,[P_\m,\mathcal{O}(0)]]&=[P_\m,[D,\mathcal{O}(0)]]+[[D,P_\m],\mathcal{O}(0)]=(\D_\mathcal{O}+1)[P_\m,\mathcal{O}(0)],\\
	[D,[K_\m,\mathcal{O}(0)]]&=[K_\m,[D,\mathcal{O}(0)]]+[[D,K_\m],\mathcal{O}(0)]=(\D_\mathcal{O}-1)[K_\m,\mathcal{O}(0)].
\end{align}

As discussed previously, since in a unitary CFT there is a lower bound on the scaling dimensions of the fields, this implies that any conformal representation must contain operators of lowest dimension which, due to \eqref{bound}, are annihilated by $K_\m$ at $x^\n=0$. \\
In a given irreducible multiplet of the conformal algebra, conformal primary fields are those fields of lowest scaling dimension, determined by the relation \eqref{bound}. All the other fields, \emph{conformal descendants} of $\mathcal{O}$, are obtained by acting with $P_\m$ on such conformal primary fields. 
We can consider the operator $U(x)=~\exp\left(\hat P_\m\,x^\m\right)$ that, acting on $\mathcal{O}(0)$, gives
\begin{equation}
	U(x)\,\mathcal{O}(0)\,U^{-1}(x)=\mathcal{O}(x).
\end{equation}
Through this operator, we may deduce the commutation relations for a conformal primary field $\mathcal{O}(x)$, taking into account the relations of the conformal algebra. In order to show the procedure, we consider the case of $[P_\m,\mathcal{O}(x)]$. Expanding the operator $U(x)$ and using the Haussdorf formula
\begin{equation}
	U(x)\,\mathcal{O}(0)\,U^{-1}(x)=\sum_{n=0}^\infty\,\sdfrac{1}{n!}\,x^{\n_1}\dots x^{\n_n}[P_{\n_1},[\dots[P_{\n_n},\mathcal{O}(0)]\dots]],
\end{equation}
we obtain
\begin{align}
	[P_\m,U(x)\,\mathcal{O}(0)\,U^{-1}(x)]=\left[P_\m\ ,\ \sum_{n=0}^\infty\,\sdfrac{1}{n!}\,x^{\n_1}\dots x^{\n_n}[P_{\n_1},[\dots[P_{\n_n},\mathcal{O}(0)]\dots]]\right]=\partial_\m\mathcal{O}(x),
\end{align}
from which we deduce the commutation relations
\begin{equation}
	\begin{split}
		[P_\mu,\mathcal{O}(x)]&=\partial_\mu\mathcal{O}(x),\\
		[D,\mathcal{O}(x)]&=(\D_\mathcal{O}+x^\mu\partial_\mu)\mathcal{O}(x),\\
		[L_{\mu\nu},\mathcal{O}(x)]&=\left(\mathcal{S}_{\mu\nu}+x_\nu\partial_\mu-x_\mu\partial_\nu\right)\mathcal{O}(x),\\
		[K_\mu,\mathcal{O}(x)]&=\left[2x_\mu (x\cdot\partial)-x^2\partial_\mu+2\Delta_\mathcal{O} x_\mu-2x^\nu\mathcal{S}_{\mu\nu}\right]\mathcal{O}(x).
	\end{split}\label{transform}
\end{equation}
The correlation functions constructed with such primary fields transform covariantly and, in case of scalar conformal primaries, we have 
\begin{equation}
	\langle\mathcal{O}_1(x'_1)\dots \mathcal{O}_n(x'_n)\rangle=\left|\sdfrac{\partial x'_1}{\partial x_1}\right|^{-\D_1/d}\dots \left|\sdfrac{\partial x'_1}{\partial x_1}\right|^{-\D_n/d}\langle\mathcal{O}_1(x_1)\dots \mathcal{O}_n(x_n)\rangle.\label{invariance}
\end{equation}
In the case of fields of general spin there are extra terms depending on the transformation matrices defined in \eqref{spintransf}. Furthermore, in these theories we assume a unique vacuum state $\ket{0}$ that should be invariant under all global conformal transformations, and it is an eigenstate of the dilatation operator with eigenvalue zero. 
The theory can be coupled to a background metric $g_{\m\n}$, with specific properties under the action of the conformal group. This is intimately related to an important operator of any CFT which is the stress energy tensor. This operator, due to the constraints 
of conformal invariance, has the properties to be symmetric, traceless and conserved.\\
Once the theory is coupled to a curved background metric $g_{\mu\nu}$, it is natural to consider the action of the Weyl group for which the metric and the scalar primary fields transform as
\begin{equation}	
	\begin{split}
		g'_{\mu\nu}(x)&=\,e^{2\sigma(x)}\,g_{\mu\nu}(x),\\
		\mathcal{O}_i'(x)&=e^{-\sigma(x)\,\Delta_{i}}\,\mathcal{O}_i(x).
	\end{split}	\label{transformWeylx}
\end{equation}
The function $\sigma(x)$ allows to define a subgroup of the group of local Weyl transformation, induced by the conformal transformations, known as the conformal Weyl group. The form of the local function $\sigma(x)$ is written as
\begin{equation}
	\sigma(x)=\log\left(\frac{1}{1-2b\cdot x+b^2 x^2}\right),
\end{equation}
with $b$ any constant vector. For $b$ infinitesimal and using the expression of $\sigma(x)$, the transformation \eqref{transformWeylx} corresponds exactly to the conformal transformations given in \eqref{transf}. Every conformal field theory is Weyl invariant, and the scalar correlation function of primary fields given above transforms under the Weyl group as
\begin{equation}
	\label{distint}
	\langle\mathcal{O}_1(x_1)\dots \mathcal{O}_n(x_n)\rangle_{e^{2\s}g_{\m\n}}=e^{-\s(x_1)\D_1}\dots e^{-\s(x_n)\D_n}\langle\mathcal{O}_1(x_1)\dots \mathcal{O}_n(x_n)\rangle_{g_{\m\n}}.
\end{equation}
Notice the appearance in the equation above of $x$-dependent scalings $\sigma(x)$, respect to the global parameters of a conformal transformation, dependence which is typical of the Weyl symmetry. 
The relation between conformal and Weyl symmetry can be discussed in physical terms by the Weyl gauging procedure, which takes to Weyl gravity if a specific metric is taken to be a dynamical and not a compensator field. In this case, in each free-falling frame identified in a local region of spacetime, the local flat symmetry in the tangent space is controlled by the conformal group's generators rather than the Poincar\'e group.\\
It is worth noting that \eqref{distint} is valid at non-coincident point only, i.e. for $x_i\ne x_j$, $i\ne j$. This fact is due to the choice of the regularisation scheme that one has to consider in order to have a defined quantum field theory. The (dimensionful) regulator used in the scheme chosen, may break some of the symmetries of the theory that cannot be restored once the regulator is removed. For instance,  dimensional regularisation, which preserves Lorentz invariance, breaks Weyl invariance, and it is manifest in a local manner, i.e., it affects the correlation functions at coincident points only. The effects of the breaking of Weyl symmetry is manifest in the {trace} or {Weyl anomaly}.
In quantum field theory scale and conformal invariance are treated as equivalent, in the sense that all the realistic field theories which are scale invariant are also conformal invariant. Such enhancement of the symmetry (from dilatation invariance to conformal), whether it holds in general or not, has been discussed at length in the literature \cite{Nakayama:2013is}, with counterexamples which are quite unrealistic from the point of view of a local, covariant, relativistic field theory. In our case scale and conformal invariance will be taken as equivalent, as is indeed the case in ordinary Lagrangian field theories.  

\section{Correlation functions in CFT\label{secCFT}}
The presence of conformal symmetry in a quantum field theory is very powerful and restrictive. In fact, it imposes very strong constraints on the structure of the correlation functions. Up to $3$-point functions, the requirement of the conformal invariance fixes the structure of the correlator up to constants. 
Looking at the simplest case of $1$-point functions of a scalar operator $\mathcal{O}$, one finds from translational invariance they need to be constant. Then, scaling invariance requires that the constant has to vanish with the condition
\begin{equation}
	\braket{\mathcal{O}(x)}=0,\qquad \text{if}\ \Delta\ne0.
\end{equation}
The condition on the conformal dimensions of the operator $\Delta\ne0$ is satisfied looking at the unitarity bounds required for a unitarity CFT, for which all the operators, apart from the identity operator, have a strictly positive conformal dimension. Therefore, in a conformal theory $\braket{\mathcal{O}(x)}=0$, assuming that $\mathcal{O}$ is not proportional to the identity operator. 
Moving on to the next non trivial case, we consider $2$-point functions of two scalar operators $\mO_i$ and $\mO_j$ of conformal dimensions $\D_i$ and $\D_j$ and we construct their general form. From Poincar\`e invariance, these functions depend on the difference of coordinates and imposing scaling invariance one finds 
\begin{equation}
	\braket{\mO_i(x_i)\,\mO_j(x_j)}=\frac{C_{ij}}{|x_i-x_j|^{\Delta_i+\Delta_j}}.
\end{equation}
We have still to impose the invariance under the special conformal transformations. For such transformations we recall that
\begin{equation}
	\left|\sdfrac{\partial x'_i}{\partial x_i}\right|=\sdfrac{1}{(1-2b\cdot x_i+b^2 x_i^2)^d}.
\end{equation} 
Under this transformation the 2-point function transforms as
\begin{align}
	\braket{\mO'_1(x')\,\mO'_2(y')}&=\left|\sdfrac{\partial x'}{\partial x}\right|^{\D_1/d}\left|\sdfrac{\partial y'}{\partial y}\right|^{\D_2/d}\braket{\mO_1(x')\,\mO_2(y')}=\sdfrac{(\g_1\g_2)^{(\D_1+\D_2)/2}}{\g_1^{\D_1}\g_2^{\D_2}}\sdfrac{C_{12}}{|x_1-x_2|^{\D_1+\D_2}}
\end{align}
where $\g_i=(1-2b\cdot x_i+b^2 x_i^2)$. The invariance of the correlation function under special conformal transformations induces the constraint
\begin{equation}
	\sdfrac{(\g_1\g_2)^{(\D_1+\D_2)/2}}{\g_1^{\D_1}\g_2^{\D_2}}=1
\end{equation}
which is identically satisfied only if $\D_1=\D_2$. This means that two primary fields are correlated only if they have identical scaling dimensions
\begin{equation}
	\braket{\mO_1(x_1)\,\mO_2(x_2)}=\frac{C_{12}}{|x_1-x_2|^{2\D_1}}\d_{\D_1\D_2}.\label{2poin}
\end{equation}
This result can also be obtained analysing only the transformations of the correlation function under the operation of inversion $I^\mu$. Therefore, in order to analyse the implications of the action of the special conformal transformation, in many cases it is enough to consider the action of $I^\m(x)$. \\
A similar method can be applied to the 3-point functions. Poincar\'e invariance requires that
\begin{equation}
	\braket{\mO_1(x_1)\,\mO_2(x_2)\,\mO_3(x_3)}=f\big(\,|x_1-x_2|\, ,\, |x_1-x_3|\,,\,|x_2-x_3|\,\big),
\end{equation}
and the invariance under scale transformations 
\begin{equation}
	f\big(\,|x_1-x_2|\, ,\, |x_1-x_3|\,,\,|x_2-x_3|\,\big)=\l^{\D_1+\D_2+\D_3}f\big(\,\l|x_1-x_2|\, ,\, \l|x_1-x_3|'\,,\,\l|x_2-x_3|'\,\big)
\end{equation}
imposes another constraint that forces a generic 3-point function to take the following form
\begin{equation}
	\braket{\mO_1(x_1)\,\mO_2(x_2)\,\mO_3(x_3)}=\frac{C_{123}}{x_{12}^a\ x_{23}^b\ x_{13}^c}\label{3point}
\end{equation}
where $x_{ij}=|x_i-x_j|$. $C_{123}$ is a constant and the coefficients $a$, $b$, $c$ must verify the relation
\begin{equation}
	a+b+c=\D_1+\D_2+\D_3.
\end{equation}
The invariance of \eqref{3point} under special conformal transformations implies
\begin{equation}
	\sdfrac{C_{123}}{\g_1^{\D_1}\g_2^{\D_2}\g_3^{\D_3}}\sdfrac{(\g_1\g_2)^{a/2}(\g_2\g_3)^{b/2}(\g_1\g_3)^{c/2}}{x_{12}^a\ x_{23}^b\ x_{13}^c}=\frac{C_{123}}{x_{12}^a\ x_{23}^b\ x_{13}^c},
\end{equation}
that can be solved by the following set of equations
\begin{equation}
	\g_1^{a/2+c/2-\D_1}=1,\qquad\g_2^{a/2+b/2-\D_2}=1,\qquad\g_3^{b/2+c/2-\D_3}=1,
\end{equation}
and giving the constraints in terms of the conformal dimensions of the operators $\mO_j$ as
\begin{equation}
	a=\D_1+\D_2-\D_3,\qquad b=\D_2+\D_3-\D-1,\qquad c=\D_3+\D_1-\D_2.
\end{equation}
Therefore, the correlator of three primary fields is given by
\begin{equation}
	\braket{\mO_1(x_1)\,\mO_2(x_2)\,\mO_3(x_3)}=\frac{C_{123}}{x_{12}^{\D_1+\D_2-\D_3}\ x_{23}^{\D_2+\D_3-\D-1}\ x_{13}^{\D_3+\D_1-\D_2}}.
\end{equation}
$C_{123}$ and $\Delta_i$'s are usually referred to as "the conformal data". The strong constraints on $2$- and $3$- point function previously discussed are not enough to fix the correlation function of $4$-point functions and higher. In this case, indeed, one can construct scalar conformal invariants from the set of the external coordinates $\{x_j\}$.  Poincar\'e invariance implies that the variables appearing in the correlation function must be of the form $x_{ij}=|x_i-x_j|$, but in order for this to be scale invariant one has to construct ratios of such variables. We recall that, under the special conformal transformation, the distance separating two points $x_{i j}$ is mapped to
\begin{equation}
	|x'_i-x_j'|=\frac{|x_i-x_j|}{(1-2b\cdot x_i+b^2x_i^2)^{1/2}(1-2b\cdot x_j+b^2x_j^2)^{1/2}}.
\end{equation}
Therefore, the simplest objects that are conformally invariant are
\begin{equation}
	u_{ijkl}\equiv\frac{x_{ij}\,x_{kl}}{x_{ik\,}x_{jl}},\qquad v_{ijkl}=\frac{x_{ij}\,x_{kl}}{x_{il}\,x_{jk}},
\end{equation}
known as {anharmonic ratios} or {cross-ratios}. These ratios are well-defined as far as the points are distinct. Every function of these cross-ratios will be conformally invariant, and for this reason $4$- and higher-point functions are determined up to an arbitrary function of the cross-ratios, and are not uniquely fixed by conformal invariance. 
For instance, the 4-point function can be written as
\begin{equation}
	\braket{\mO_1(x_1)\,\mO_2(x_2)\,\mO_3(x_3)\,\mO_4(x_4)}=f\left(\frac{x_{12}x_{34}}{x_{13}x_{24}},\frac{x_{12}x_{34}}{x_{23}x_{14}}\right)\ \prod_{1\le i<j\le 4}^{4}\,x_{ij}^{\D_t/3-\D_i-\D_j}
\end{equation}
where $\D_t=\sum_{i=1}^4\D_i$ and $f$ is an undetermined function. In the case of 4-point functions there are two independent conformal ratios. In the case of $n$-point functions there are $n(n-3)/2$ independent conformal ratios. 

\section{The embedding space formalism}
The nonlinear action of the conformal group becomes simpler and more transparent in the formalism of the {embedding space} \cite{Simmons_Duffin_2014, Simmons-Duffin:2016gjk}. In fact, taking a Euclidean CFT in $d$-dimensions with the conformal group acting 
nonlinearly on $\mathbb{R}^d$, it is well known that it is locally isomorphic to $SO(1,d+1)$ and therefore it acts linearly on the space $\mathbb{R}^{1,d+1}$, called the {embedding space}. The procedure consists in embedding the target space $\mathbb{R}^d$ into $\mathbb{R}^{1,d+1}$.\\  
We introduce the light-cone coordinates in $\mathbb{R}^{1,d+1}$ as
\begin{equation}
	X^A=(X^+,X^-,X^a)\in\mathbb{R}^{1,d+1},
\end{equation}
where $a=1,\dots,d$ with the inner product
\begin{equation}
	X\cdot X=\eta_{AB}\,X^A\,X^B=-X^+\,X^-+X_a\,X^a.
\end{equation}
Using such coordinates, the condition $X^2=0$ defines an $SO(1,d+1)$ invariant subspace of dimension $d+1$, the null cone. Imposing the gauge condition $X^+=1$, that defines the {Poincar\`e section} of the embedding, we identify the projective null-cone with $\mathbb{R}^d$ and the null vectors take the form
\begin{equation}
	X=(1,x^2,x^\m)\label{poinc}.
\end{equation}
It is worth noting that the transformation of $X$ by $SO(1,d+1)$ is just a matrix multiplication, and the transformation
\begin{equation}
	X\to h\,X/(h\,X)^+, \qquad h\in SO(1,d+1)
\end{equation}
defines the nonlinear action of the conformal group on $\mathbb{R}^d$. Finally, taking two points $X_i$ and $X_j$, setting $X_{ij}=X_i-X_j$, on the Poincar\'e section where $X^2=0$, one obtains
\begin{equation}
	X_{ij}^2=-2X_i\cdot X_j=(x_i-x_j)^2=x_{ij}^2.\label{xxx}
\end{equation}
Now, we want to lift the fields defined on $\mathbb{R}^d$ to the embedding space. The first step is to consider a primary scalar field $\mO_d(x)$ in $\mathbb{R}^d$ with dimension $\D$, for which one can define a scalar on the entire null-cone, with the requirement
\begin{equation}
	\mO_{d+1,1}(\l X)=\l^{-\D}\mO_{1,d+1}(X),
\end{equation}
with the dimension of $\mO_d$ reflected in the degree of $\mO_{d+1}$.
For any such fields in $\mathbb{R}^{1,d+1}$ one can always perform a projection on $\mathbb{R}^d$ as
\begin{equation}
	\mO_{d}(x)=\left(X^+\right)^{\Delta}\mO_{1,d+1}(X(x)).
\end{equation}
The field $\mO_{d+1}(X)$ then transforms in a simple way under the conformal transformation $\mO_{d+1}(X)\to \mO_{d+1}(hX)$. 

Conformal invariance requires that the correlators containing $\mO_{d+1}(X)$ are invariant under $SO(d+1,1)$. 
In order to derive the structure of a correlator in the embedding space, one starts by 
writing down its most general form, covariant under  $SO(d+1,1)$, satisfying the appropriate scaling conditions. Then we substitute \eqref{poinc} and use \eqref{xxx} in order to obtain the expression in ordinary coordinate space. For example, the 2-point function of operators of dimension $\D$ is fixed by conformal invariance, homogeneity, and the null condition $X_i^2=0$, to take the form
\begin{equation}
	\braket{\mO(X_1)\,\mO(X_2)}=\frac{C_{12}}{(-2X_1\cdot X_2)^\D}.
\end{equation}
We notice that $-2X_1\cdot X_2$ is the only Lorentz invariant that we can construct out of two points, since on the projective null-cone the condition $X_j^2=0$ is satisfied. Now, using \eqref{xxx} we obtain
\begin{equation}
	\braket{\mO(x_1)\,\mO(x_2)}=\frac{C_{12}}{(X^2_{12})^\D}=\frac{C_{12}}{|x_1-x_2|^{2\D}}
\end{equation}
which is \eqref{2poin}. 
The next step is to extend the embedding formalism to tensor operators \cite{Ferrara:1973yt,Weinberg:2010fx,Costa:2011mg,SimmonsDuffin:2012uy}. We are going to illustrate the method for totally symmetric and traceless tensors, since such tensors transform with the simplest irreducible representation of $SO(1,d+1)$. 

Considering a tensor field of $SO(d+1,1)$, denoted as $\mathcal{O}_{A_1\dots A_n}(X)$, with the properties
\begin{itemize}
	\item defined on the null-cone $X^2=0$,
	\item traceless and symmetric,
	\item homogeneous of degree $-\Delta$ in $X$, i.e., $\mathcal{O}_{A_1\dots A_n}(\lambda\,X)=\lambda^{-\Delta} \mathcal{O}_{A_1\dots A_n}(X)$,
	\item transverse $X^{A_i}\mathcal{O}_{A_1\dots  A_{i}\dots A_n}(X)=0$ $j=1,\dots,n$,
\end{itemize}
it is clear that those are conditions rendering $\mathcal{O}_{A_1\dots A_n}(X)$ manifestly invariant under $SO(1,d+1)$. In order to find the corresponding tensor in $\mathbb{R}^d$, one has to restrict $\mathcal{O}_{A_1\dots A_n}(X)$ to the Poincar\'e section and project the indices as
\begin{equation}
	\mathcal{O}_{\mu_1\dots\mu_n}(x)=\frac{\partial\,X^{A_1}}{\partial x^\mu_1}\dots\frac{\partial\,X^{A_n}}{\partial x^\mu_n}\mathcal{O}_{A_1\dots A_n}(X).\label{transfemb1}
\end{equation}
For example, the most general form of the 2-point function of two operators with spin one and dimension $\D$ can be derived as
\begin{equation}
	\braket{\mO^A(X_1)\mO^B(X_2)}=\frac{C_{12}}{(X_1\cdot X_2)^\D}\left[\h^{AB}+\a\frac{X_2^A\,X_1^B}{X_1\cdot X_2}\right]\label{2pointemb},
\end{equation}
according to the rules given above. The transverse condition 
\begin{equation}
	X_1^A\braket{\mO^A(X_1)\mO^B(X_2)}=\frac{C_{12}}{(X_1\cdot X_2)^\D}\left[X_1^B+\a X_1^B\right]=0,
\end{equation}
implies that $\a=-1$. The projection on $\mathbb{R}^d$ using \eqref{transfemb1} leads to define the 2-point function in $\mathbb{R}^d$ as
\begin{equation}
	\braket{\mO^\m(x_1)\mO^\n(x_2)}=\frac{\partial X_1^A}{\partial x_1^\m}\frac{\partial X_2^B}{\partial x_2^\n}\braket{\mO^A(X_1)\mO^B(X_2)}.\label{transfemb}
\end{equation}
Finally, one needs to compute the terms
\begin{equation}
	\h_{AB}\frac{\partial X_1^A}{\partial x_1^\m}\frac{\partial X_2^B}{\partial x_2^\n},\qquad X_2^A\,X_1^B\frac{\partial X_1^A}{\partial x_1^\m}\frac{\partial X_2^B}{\partial x_2^\n},
\end{equation}
that are explicitly given as
\begin{align}
	\h_{AB}\sdfrac{\partial X_1^A}{\partial x_1^\m}\sdfrac{\partial X_2^B}{\partial x_2^\n}&=-\sdfrac{1}{2}\left(\sdfrac{\partial X_1^+}{\partial x_1^\m}\sdfrac{\partial X_2^-}{\partial x_2^\n}+\sdfrac{\partial X_1^+}{\partial x_1^\m}\sdfrac{\partial X_2^-}{\partial x_2^\n}\right)+\d_{ab}\sdfrac{\partial X_1^a}{\partial x_1^\m}\sdfrac{\partial X_2^b}{\partial x_2^\n}=\d_{\m\n}\\
	X_2^A\sdfrac{\partial X_1^A}{\partial x_1^\m}&=-\sdfrac{1}{2}\left(X_2^+\sdfrac{\partial X_1^-}{\partial x_1^\m}+X_2^-\sdfrac{\partial X_1^+}{\partial x_1^\m}\right)+\d_{ab}X_2^a\sdfrac{\partial X_1^b}{\partial x_1^\m}=x_2^\m-x_1^\m\\
	X_1^B\sdfrac{\partial X_2^B}{\partial x_2^\n}&=-\sdfrac{1}{2}\left(X_1^-\sdfrac{\partial X_2^+}{\partial x_2^\n}+X_1^+\sdfrac{\partial X_2^-}{\partial x_2^\n}\right)+\d_{ab}X_1^a\sdfrac{\partial X_2^b}{\partial x_2^\n}=x_1^\n-x_2^\n
\end{align}
which allow to define the following projective rules
\begin{equation}
	X_1^B\mapsto x_1^\n-x_2^\n,\qquad X_2^A\mapsto x_2^\m-x_1^\m,\qquad \h^{AB}\mapsto\d^{\m\n},\qquad X_1\cdot X_2\mapsto-1/2\,x_{12}^2.
\end{equation}
Therefore, from \eqref{transfemb} we infer the equation
\begin{equation}
	\braket{\mO^\m(x_1)\mO^\n(x_2)}=\frac{C_{12}\,I^{\m\n}(x_{12})}{x_{12}^{2\D}},\label{finres}
\end{equation}
where 
\begin{equation}
	I^{\m\n}(x)=\d^{\m\n}-\frac{2x^\m\,x^\n}{x^2}\label{Itensor}
\end{equation}
and $C_{12}$ is an undetermined constant. The $I^{\m\n}$ tensor appearing in \eqref{finres} plays an important role in the definition of the conformal structure. In fact, it is important in realizing the operation of inversion, and appears in all the correlation functions of tensor operators \cite{Osborn:1993cr, Erdmenger:1996yc}.\\
Using the same strategy, one can easily construct correlation functions of higher spin, and for a traceless spin-2 operator of dimension $\D$, we find
\begin{equation}
	\braket{\mO^{\m\n}(x_1)\mO^{\r\s}(x_2)}=\sdfrac{C_{12}}{x_{12}^{2\D}}\left[I^{\m\r}(x_{12})I^{\n\s}(x_{12})+I^{\m\s}(x_{12})I^{\n\r}(x_{12})-\sdfrac{2}{d}\d^{\m\n}\d^{\r\s}\right].\label{2poinembed}
\end{equation}
It is worth mentioning that expressions \eqref{finres} and \eqref{2poinembed} satisfy the requirement of being traceless and symmetric, with the possibility, in the case of conserved currents, of further specialization of this result. In fact, from \eqref{finres}, in the case of conserved currents $J^\mu$, the conservation Ward identities \eqref{Ward1} in \appref{transWard}
\begin{equation}
	\frac{\partial}{\partial x_1^{\mu_1}}\,\braket{J^{\mu_1}(x_1)J^{\mu_2}(x_2)}=0,
\end{equation}
imposes a strong condition on the conformal dimension. We can use in this expression the full form of the 2-point function given in \eqref{finres}, to derive the relation
\begin{align}
	0&=\sdfrac{\partial}{\partial x_1^\m}\left(\sdfrac{C_{12}\,I^{\m\n}(x_{12})}{x_{12}^{2\D}}\right)=C_{12}\left(-2\D\sdfrac{I^{\m\n}(x_{12})}{x_{12}^{2(\D+1)}}(x_{12})_\m+\sdfrac{1}{x_{12}^{2\D}}\sdfrac{\partial\,I^{\m\n}(x_{12})}{\partial x_1^\m}\right)\notag\\
	&=\sdfrac{C_{12}}{x_{12}^{2\D}}\left[-2\D\sdfrac{(x_{12})_\m}{x_{12}^2}\left(\d^{\m\n}-2\sdfrac{x_{12}^\m\,x_{12}^\n}{x_{12}^2}\right)-2(d-1)\sdfrac{x_{12}^\n}{x_{12}^2}\right]=-2(d-\D-1)\,C_{12}\,\sdfrac{x_{12}^\n}{x_{12}^{2(\D+1)}}.
\end{align}
This shows that a conformal primary operator of spin-1 is conserved if an only if its dimension is $\D=d-1$. A similar result holds for conserved energy momentum tensors $T^{\mu_i\nu_i}(x_i)$, by using the condition \eqref{transWardF} in \appref{transWard}
\begin{equation}
	\frac{\partial}{\partial x_1^{\mu_1}}\,\braket{T^{\mu_1\nu_1}(x_1)T^{\mu_2\nu_2}(x_2)}=0.
\end{equation}
In this case one has to use the relation 
\begin{align}
	0&=\sdfrac{\partial}{\partial x_1^\m}\left\{\sdfrac{C_{12}}{x_{12}^{2\D}}\left[I^{\m\r}(x_{12})I^{\n\s}(x_{12})+I^{\m\s}(x_{12})I^{\n\r}(x_{12})-\sdfrac{2}{d}\d^{\m\n}\d^{\r\s}\right]\right\}\notag\\
	&=-2\D\sdfrac{C_{12}}{x_{12}^{2\D+2}}(x_{12})_\m\left[I^{\m\r}(x_{12})I^{\n\s}(x_{12})+I^{\m\s}(x_{12})I^{\n\r}(x_{12})-\sdfrac{2}{d}\d^{\m\n}\d^{\r\s}\right]\notag\\
	&\hspace{2cm}+\sdfrac{C_{12}}{x_{12}^{2\D}}\sdfrac{\partial}{\partial x_1^\m}\left[I^{\m\r}(x_{12})I^{\n\s}(x_{12})+I^{\m\s}(x_{12})I^{\n\r}(x_{12})-\sdfrac{2}{d}\d^{\m\n}\d^{\r\s}\right].\label{3A}
\end{align}
Using the results
\begin{align}
	\sdfrac{\partial}{\partial x_1^\m}\,I^{\m\a}(x_{12})&=-2(d-1)\sdfrac{x_{12}^\a}{x_{12}^2},\\
	(x_{12})_{\m}\,I^{\m\a}(x_{12})&=-x_{12}^\a,\\
	\sdfrac{\partial}{\partial x_1^\m}\,I^{\n\a}(x_{12})&=-\sdfrac{2}{x_{12}^2}\left(\d^\n_\m\,x_{12}^\a+\d^\a_\m\,x_{12}^\n-2\sdfrac{(x_{12})_\m\,x_{12}^\n\,x_{12}^\a}{x_{12}^2}\right)
\end{align}
we can simplify \eqref{3A} in the form
\begin{align}
	0&=-2\D\sdfrac{C_{12}}{x_{12}^{2\D+2}}\left[-x_{12}^\r\,I^{\n\s}(x_{12})-x_{12}^\s\,I^{\n\r}(x_{12})-\sdfrac{2}{d}\d^{\r\s}x_{12}^\n\right]-2\sdfrac{C_{12}}{x_{12}^{2(\D+1)}}(d-1)\left[x_{12}^\r\,I^{\n\s}(x_{12})+x_{12}^\s\,I^{\n\r}(x_{12})
	\right]\notag\\
	&\hspace{0.1cm} -2\sdfrac{C_{12}}{x_{12}^{2(\D+1)}}\left[I^{\m\r}(x_{12})\left(\d^\n_\m\,x_{12}^\s+\d^\s_\m\,x_{12}^\n-2\sdfrac{(x_{12})_\m\,x_{12}^\n\,x_{12}^\s}{x_{12}^2}\right)+I^{\m\s}(x_{12})\left(\d^\n_\m\,x_{12}^\r+\d^\r_\m\,x_{12}^\n-2\sdfrac{(x_{12})_\m\,x_{12}^\n\,x_{12}^\r}{x_{12}^2}\right)\right]\notag\\
	&=-2\sdfrac{C_{12}}{x_{12}^{2(\D+1)}}\bigg\{(d-\D)\left[x_{12}^\r\,I^{\n\s}(x_{12})+x_{12}^\s\,I^{\n\r}(x_{12})\right]-\sdfrac{2\D}{d}\d^{\r\s}x_{12}^\n+2\,I^{\r\s}(x_{12})x_{12}^\n+4\sdfrac{x_{12}^\n\,x_{12}^\s\,x_{12}^\r}{x_{12}^2}\bigg\}\notag\\
	&=-2\sdfrac{C_{12}}{x_{12}^{2(\D+1)}}\bigg\{(d-\D)\left[x_{12}^\r\,I^{\n\s}(x_{12})+x_{12}^\s\,I^{\n\r}(x_{12})\right]-2\left(\sdfrac{\D}{d}-1\right)\d^{\r\s}x_{12}^\n\bigg\}.\label{A5}
\end{align}
In the curly brackets of \eqref{A5} there are independent tensor terms that do not add to zero. For this reason the only way to satisfy \eqref{A5} is when the coefficients of all the independent tensor structures are zero, implying that $\D=d$. We conclude that a primary operator of spin-2 is conserved if and only if its dimension is $\D=d$. 
These results ensure that the dimensions of $J^\m$ and $T^{\m\n}$ are protected in any CFT from any quantum correction. 
For the 3-point function one can proceed in a similar way, and observe that the problem can be greatly simplified by identifying all the possible tensor structures that can appear in a given correlation function \cite{Osborn:1993cr,Erdmenger:1996yc,Costa:2011mg}.

\section{Conformal Ward identities  in coordinate space}\label{CWIposition}
In this section we turn to a brief illustration of the derivation of the CWI's to be imposed on such correlators.
Considering a generating functional $Z[\phi_0^{(j)}]$, defined by
\begin{equation}
	Z\left[\phi_0^{(j)}\right]=\int\,D\Phi\,\exp\left(-S-\sum_{j}\int d^dx\,\phi_0^{(j)}(x)\,\mathcal{O}_j(x)\right),
\end{equation}	
one can define the correlation function by functional differentiation with respect to the sources as
\begin{align}
	\braket{\mathcal{O}_1(x_1)\dots \mathcal{O}_n(x_n)}=(-1)^n\left.\frac{\delta^n\,Z[\phi_0^{(j)}]}{\delta \phi_0^{(1)}(x_1)\dots\delta\phi_0^{(n)}(x_n)}\right|_{\phi_0^{(j)}=0}=\int\,D\Phi\,\mathcal{O}_1(x_1)\dots \mathcal{O}_n(x_n)\,e^{-S}.
\end{align}
Assuming that this correlation function is invariant under a symmetry transformation $g$ and the generating functional as well, since the measure is invariant under such a transformation, one obtains the Ward Identities for correlation functions of the form
\begin{equation}
	\sum_{i=1}^n\braket{\mathcal{O}_1(x_1)\dots\delta_g\mathcal{O}_i(x_i) \dots\mathcal{O}_n(x_n)}=0,\label{globalW}
\end{equation} 
where for a general infinitesimal conformal transformation defined by $g(x)$ the variation $\delta_g\mathcal{O}_i(x_i)$ is defined as
\begin{equation}
	\delta_g\mathcal{O}_i(x_i)=\left(-g\cdot\partial-\frac{\Delta}{d}\,\partial\cdot g-\frac{1}{2}\,\partial_{\,[\mu}g_{\,\nu]}\mathcal{S}_{\mu\nu}\right)\mathcal{O}_i(x_i),\label{transfSpecial}
\end{equation}
according to \eqref{transform}. In the case of translations the Ward identity we have the relation
\begin{equation}
	0=\sum_{j=1}^n\sdfrac{\partial}{\partial x_j^\m}\braket{\mO_1(x_1)\dots\mO_n(x_n)},
\end{equation}
implying that the correlation function depends on the differences $x_i-x_j$ only. Then the dilatation WI's can be easy constructed using the dilatation generator, for which we find
\begin{equation}
	\left[\sum_{j=1}^n\,\D_j+\sum_{j=1}^n\,x_j^\a\sdfrac{\partial}{\partial x_j^\a}\right]\braket{\mO_1(x_1)\dots\mO_n(x_n)}=0.
\end{equation}
where here we indicate with $\D_1,\dots,\D_n$ the dimensions of the conformal primaries operators $\mO_1,\dots,\mO_n$. Analogously to the previous cases, the special conformal WI's, corresponding to special conformal transformations, can be derived for the scalar case, and in particular we have
\begin{equation}
	\sum_{j=1}^n\left(2\D_j\,x_j^k+2x_j^\k\,x_j^\a\sdfrac{\partial}{\partial x_j\a}-x_j^2\sdfrac{\partial}{\partial x_{jk}}\right)\braket{\mO_1(x_1)\dots\mO_n(x_n)}=0,\label{special}
\end{equation}
where $k$ is a free Lorentz index. In the case of tensor operators one needs to add an additional term to the previous equation, the contribution $\mathcal{S}_{\mu\nu}$ in \eqref{transfSpecial}, the finite-dimensional representation of the rotations determining the spin for the field considered.
For example, if we assume that the tensor $\mO_j$ has $r_j$ Lorentz indices, i.e. $\mO_j=\mO_j^{\m_{j_1}\dots\m_{j_{r_j}}}$, for $j=1,2,\dots,n$, in this case we need to add the extra term
\begin{align}
	2\sum_{j=1}^n\sum_{h=1}^{r_j}\left[(x_j)_{\a_{j_h}}\d^{\k\m_{j_h}}-x_j^{\m_{j_h}}\d^\k_{\a_{j_h}}\right]\braket{\mO^{\m_{1_{\scalebox{0.5}{1}}}\m_{1_{\scalebox{0.5}{2}}}\dots\m_{1_{r_{\scalebox{0.4}{ 1}}}}}(x_1)\dots\  \mO^{\m_{j_{\scalebox{0.5}{1}}}\dots\a_{j_{\scalebox{0.5}{h}}}\dots\m_{j_{r_{\scalebox{0.4}{j}}}}}(x_j)\ \dots\ \mO^{\m_{n_{\scalebox{0.5}{1}}}\m_{n_{\scalebox{0.5}{2}}}\dots\m_{n_{r_{\scalebox{0.4}{n}}}}}(x_n)}\label{spinpart}
\end{align}
to the left-hand side of \eqref{special}. Finally, we can consider the WI associated with rotations in coordinate space, by taking for $\delta_g$ a Lorentz transformation in \eqref{transfSpecial}. As in the previous case, we first consider the $n$-point function of scalar operators, giving
\begin{equation}
	\sum_{j=1}^n\left(x_j^\n\sdfrac{\partial}{\partial x_{j\m}}-x_j^\m\sdfrac{\partial}{\partial x_{j\n}}\right)\braket{\mO_1(x_1)\dots\mO_n(x_n)}=0.\label{rot1x}
\end{equation}
For tensor operators one needs to add to the left-hand side of the previous equation the contribution
\begin{align}
	\sum_{j=1}^n\sum_{h=1}^{r_j}\left[\d_{\n\a_{j_h}}\d^{\m_{j_h}}_\m-\d_{\m\a_{j_h}}\d_\n^{\m_{j_h}}\right]\braket{\mO^{\m_{1_{\scalebox{0.5}{1}}}\m_{1_{\scalebox{0.5}{2}}}\dots\m_{1_{r_{\scalebox{0.4}{ 1}}}}}(x_1)\dots\  \mO^{\m_{j_{\scalebox{0.5}{1}}}\dots\a_{j_{\scalebox{0.5}{h}}}\dots\m_{j_{r_{\scalebox{0.4}{j}}}}}(x_j)\ \dots\ \mO^{\m_{n_{\scalebox{0.5}{1}}}\m_{n_{\scalebox{0.5}{2}}}\dots\m_{n_{r_{\scalebox{0.4}{n}}}}}(x_n)}.\label{rot2x}
\end{align}
There are simplifications in momentum space as a result of this symmetry as we are going to show in the next chapter.
\chapter{Conformal symmetry in momentum space}

\section{Introduction}
As already mentioned in the previous chapter, most of the current and past analysis in 
CFT has been focused around coordinate space. This is the natural domain where primary operators are introduced to discuss the fluctuations of physical systems around a certain critical point, as a function of the relative distances of the points of a given correlator. The operator algebra of a CFT  is endowed with the operator product expansion, which allows expressing the product of two operators at separate points in terms of an infinite series of local operators \cite{ Ferrara:1973yt,Dolan:2000ut, Poland:2018epd, Poland:2016chs}. A problem arises at coincident points, i.e. when two or more primary operators in the product approach the same spacetime point. For stress energy tensors, this limit introduces an anomaly in the operator algebra. The stress energy tensor is not traceless anymore, as expected in a CFT, and the conformal anomaly is the manifestation of the breaking of quantum symmetry.\\
As we move to momentum space, due to the Fourier transform acting on a correlation function's spacetime points, the integration also includes domains in which two or more coordinates coalesce. Therefore the anomaly contribution is naturally included in the expression of a correlator in momentum space \cite{Capper:1975ig, Deser:1976yx, Riegert:1984kt, Coriano:2017mux}. \\
One of the advantages of the momentum space approach to the determination of CFT correlators is to establish a link with the ordinary perturbative Feynman expansion. In particular, it allows to compare general results with explicit realizations of CFT's, where many methods are available. While the latter is directly connected with a specific Lagrangian realization, the analysis of the conformal Ward identities (CWI's) in momentum space, on the other hand, allows to investigate the operatorial content of a CFT in the most general way, whenever this is possible.
In even spacetime dimensions, the study of such constraints in momentum space finds essential applications in the context of the conformal anomaly action \cite{Giannotti:2008cv,Armillis:2009pq, Armillis:2010qk}, which has been investigated in the perturbative context in $d=4$.  \\
Besides the case of $d=4$, we also mention that in $d=3$ such correlators play an important role in the analysis of the gravitational perturbations and find wide applications in the investigation of non-gaussianities \cite{Maldacena:2011nz}, in holographic cosmology \cite{Bzowski:2011ab,Coriano:2012hd} and also in condensed matter \cite{Chernodub:2017jcp}.
More recently, a general formalism for the extension of such analysis to a De Sitter background, in the context of inflationary cosmology,  has been formulated \cite{Arkani-Hamed:2018kmz,Benincasa:2018ssx,Arkani-Hamed:2018bjr,Arkani-Hamed:2017fdk}.  
The general analysis of such correlators provides a complementary approach for those developed in the last two decades in the context of perturbative gauge theory amplitudes (see \cite{Henn:2014yza,Benincasa:2013faa} for an overview). The latter relies on specific Lagrangian realizations and supersymmetry.
\section{The dilatation equation}\label{dilMom}
We are now going, as a first step, to reformulate the conformal constraints (i.e. the conformal Ward identities) in  momentum space, proceeding from the scalar case and then moving to the tensor case.
We will be using some condensed notations in order to shorten the expressions 
of the transforms in momentum space. We will use the following conventions
\begin{eqnarray}
	\label{conv}
	&& \Phi(\underline{x})\equiv \braket{\mathcal{O}_1(x_1)\mathcal{O}_2(x_2)\ldots \mathcal{O}_n(x_n)} \qquad e^{i \underline{p x}}\equiv e^{i(p_1 x_1 + p_2 x_2 + \ldots p_n x_n)} \qquad \nonumber\\
	&& \underline{d p}\equiv dp_1 dp_2 \ldots d p_n \qquad 
	\Phi(\underline{p})\equiv \Phi(p_1,p_2,\ldots, p_n).\qquad 
\end{eqnarray}
with $\Phi$ denoting an n-point correlation functions of primary operators $\mathcal{O}_i$.
It will also be useful to introduce the total momentum $P=\sum_{j=1}^{n} p_j$.\\
The momentum constraint is enforced via a delta function $\delta(P)$ under integration. For instance, translational invariance of $\Phi(\underline{x})$ gives 
\begin{equation}
	\label{ft1}
	\Phi(\underline{x})=\int \underline{dp}\,\delta(P) \,e^{i\underline{p x}} \,\Phi(\underline{p}).
\end{equation}
In general, for an n-point function $\Phi(x_1,x_2,\ldots, x_n)=\braket{\mathcal{O}_1(x_1)\mathcal{O}_2(x_2)\ldots \mathcal{O}_n(x_n)} $, the condition of translational invariance  
\begin{equation}
	\braket{\mathcal{O}_1(x_1)\mathcal{O}_2(x_2)\ldots \mathcal{O}_n(x_n)}= \braket{\mathcal{O}_1(x_1+a)\mathcal{O}_2(x_2+a)\ldots \mathcal{O}_n(x_n+a)}
\end{equation}
generates the expression in momentum space of the form \eqref{ft1}, from which we can remove one of the momenta, conventionally the last one, $p_n$, which is replaced by 
$\overline{p}_n=-(p_1+p_2 +\ldots p_{n-1})$, giving
\begin{equation}
	\Phi(x_1,x_2,\ldots,x_n)=\int dp_1 dp_2... dp_{n-1}e^{i(p_1 x_1 + p_2 x_2 +...p_{n-1} x_{n-1} + 
		\overline{p}_n x_n)}\Phi(p_1,p_2,\ldots,\overline{p}_n).
\end{equation}
We start by considering the dilatation WI. \\
The condition of scale covariance for the fields $\mathcal{O}_i$ of scale dimensions $\Delta_i$ (in mass units)
\begin{equation}
	\label{scale1}
	\Phi(\lambda x_1,\lambda x_2,\ldots,\lambda x_n)=\lambda^{-\Delta} \Phi(x_1,x_2,\ldots, x_n), \qquad 
	\Delta=\Delta_1 +\Delta_2 +\ldots \Delta_n
\end{equation}
after setting $\lambda=1 +\epsilon$ and Taylor expanding up to $O(\epsilon)$ gives the 
scaling relation 
\begin{equation}
	\label{ft2}
	(D_n + \Delta)\Phi\equiv \sum_{j=1}^n \left(x_j^\alpha \frac{\partial}{\partial x_j^\alpha} +\Delta_j\right) \Phi(x_1,x_2,\ldots,x_n) =0.
\end{equation}
with 
\begin{equation}
	D_n=\sum_{j=1}^n x_j^\alpha\frac{\partial}{\partial x_j^\alpha}. 
\end{equation}
The expression of the dilatation equation in momentum space can be obtained either by a Fourier transform 
of (\ref{ft2}), or more simply, exploiting directly (\ref{scale1}). In the latter case, using the translational invariance of the correlator under the integral, by removing the $\delta$-function constraint, one obtains
\begin{align}
	\Phi(\lambda x_1,\lambda x_2,\ldots,\lambda x_n)&= \int d^d p_1 d^d p_2\ldots d^d p_{n-1} e^{i\lambda (p_1 x_1 + p_2 x_2 +...p_{n-1} x_{n-1} + 
		\overline{p}_n x_n)}\Phi(p_1,p_2,\ldots,\overline{p}_n)\notag \\
	&=\lambda^{-\Delta}  \int d^d p_1 d^d p_2\ldots d^d p_{n-1} e^{i(p_1 x_1 + p_2 x_2 +...p_{n-1} x_{n-1} + 
		\overline{p}_n x_n)}\Phi(p_1,p_2,\ldots,\overline{p}_n).
\end{align}
We perform the change of variables $ p_i=p'_i/\lambda$ on the right-hand-side (rhs) of the equation above (first line) with $d p_1...d p_{n-1}=(1/\lambda)^{d(n-1)} d^d p'_1 \ldots d^d p'_{n-1}$ to derive the relation 
\begin{equation}
	\frac{1}{{\lambda}^{d(n-1)}}\Phi\left(\frac{p_1}{\lambda},\frac{p_2}{\lambda},\ldots,\frac{\overline{p}_n}{\lambda}\right)=\lambda^{-\Delta} \Phi(p_1,p_2,\ldots,\overline{p}_n).
\end{equation}
Setting $\lambda=1/s$ this generates the condition
\begin{equation}
	s^{(n-1) d-\Delta}\Phi(s p_1,s p_2,\ldots, s \overline{p}_n)=\Phi(p_1,p_2,\ldots,\overline{p}_n)
\end{equation}
and with $s\sim 1 +\epsilon$, expanding at $O(\epsilon)$ we generate the equation 
\begin{equation}
	\label{sc1}
	\left[\sum_{j=1}^n \Delta_j  -(n-1) d -\sum_{j=1}^{n-1}p_j^\alpha \frac{\partial}{\partial p_j^\alpha}\right]
	\Phi(p_1,p_2,\ldots,\overline{p}_n)=0.
\end{equation}
There are some important comments to be made. The action of any differential 
operator which is separable on each of the coordinate $x_i$, once transformed to momentum space, violates the Leibnitz rule, if we want to differentiate only the independent momenta. This because of momentum conservation, which is a consequence of the translational invariance of the correlator. This point has been illustrated at length in \cite{Coriano:2018bbe}, to which we refer for further details. Notice that in \eqref{sc1} the sum runs over the first $n-1$ momenta. The equations, though, must be reduced to a scalar form and at that stage their hypergeometric structure \cite{Coriano:2013jba,Bzowski:2013sza} will appear clear in the following sections.

\section{Special Conformal WI's for scalar correlators}\label{specialMom}
We now turn to the analysis of the special conformal transformations in momentum space. 
Also in this case we discuss both the symmetric and the asymmetric forms of the equations, focusing our attention first on the scalar case. 
The Ward identity in the scalar case is given by
\begin{equation}
	\sum_{j=1}^{n} \left(- x_j^2\frac{\partial}{\partial x_j^\kappa}+ 2 x_j^\kappa x_j^\alpha \frac{\partial}
	{\partial x_j^\alpha} +2 \Delta_j x_j^\kappa\right)\Phi(x_1,x_2,\ldots,x_n) =0
\end{equation}
which in momentum space, using 
\begin{equation}
	x_j^\alpha\to -i \frac{\partial}{\partial p_j^\alpha} \qquad \frac{\partial}{\partial x_j^\kappa}\to i p_j^\kappa
\end{equation}
becomes 
\begin{equation}
	\sum_{j=1}^n \int \underline {d^d p}\left(p_j^\kappa \frac{\partial^2}{\partial p_j^\alpha \partial p_j^\kappa} -
	2 p_j^\alpha \frac{\partial^2}{\partial p_j^\alpha \partial p_j^\kappa}  -2 \Delta_j\frac{\partial}{\partial p_j^\kappa}\right) 
	e^{i\underline{p\cdot x}} \delta^d(P)\Phi(\underline{p})=0,
\end{equation}
where the action of the operator is only on the exponential. At this stage we integrate by parts, bringing the derivatives from the exponential to the correlator and on the Dirac $\delta$ function obtaining

\begin{equation}
	\int \underline{d^d p}e^{i \underline{p x}}  \,K_s^k\Phi(\underline{p})\delta^d(P) +\delta'_\textrm{term}  =0
\end{equation}
in the notations of \eqref{conv}, where we have introduced the differential operator acting on a scalar correlator in a symmetric form
\begin{equation}
	\,K_s^k=\sum_{j=1}^n\left(p_j^\kappa \frac{\partial^2}{\partial p_j^\alpha\partial p_j^\alpha} + 2(\Delta_j- d)\frac{\partial}{\partial p_j^\kappa}-2 p_j^\alpha\frac{\partial^2}{\partial p_j^\kappa \partial p_j^\alpha}\right).
\end{equation}
Using some distributional identities derived in \cite{Coriano:2018bbe}
\begin{align}
	\label{uno}
	\,K_s^\kappa \delta^d(P)&=\left(P^k\frac{\partial^2}{\partial P^\alpha \partial P_\alpha}  -2 P^\alpha \frac{\partial^2}{\partial P^\alpha\partial P^k} 
	+2 (\Delta- n\, d)\frac{\partial }{\partial P^k}\right)\delta^d(P)\notag\\
	&=2 d( d \,n -  d -\Delta)P^k\frac{\delta^d(P)}{P^2}\notag\\
	&=-2 ( d \,n -  d -\Delta)\frac{\partial}{\partial P_k} {\delta^d(P)},
	\qquad \qquad \Delta=\sum_{j=1}^n\Delta_j,
\end{align}
we obtain
\begin{align}
	\delta'_\textrm{term}&= \int {\underline{d^d p}}\,e^{i\underline{p\cdot x}}
	\left[ \frac{\partial}{\partial P^\alpha} \delta^d(P)\sum_{j=1}^n\left(p_j^\alpha\frac{\partial}{\partial p_j^\kappa}- p_j^\kappa\frac{\partial}{\partial p_j^\alpha}\right)\Phi(\underline{p}) \right.\notag\\
	&\hspace{5cm}\left..+ 2 \frac{\partial}{\partial P^\kappa} \delta^d(P)\left(\sum_{j=1}^n\left (\Delta_j -  p_j^\alpha 
	\frac{\partial}{\partial p_j^\alpha} \right)- (n-1) d \right)\Phi(\underline{p})\right].
\end{align}
Notice that such terms vanish \cite{Coriano:2018bbe} by using rotational invariance of the scalar correlator 
\begin{equation}
	\sum_{j=1}^n\left(p_j^\alpha\frac{\partial}{\partial p_j^\kappa}- p_j^\kappa\frac{\partial}{\partial p_j^\alpha}\right)\Phi(\underline{p})=0,
\end{equation}
as a consequence of the $SO(d)$ symmetry 
\begin{equation}
	\sum_{j=1}^n L_{\mu\nu}(x_j) \langle \mathcal{O}(x_1)\dots\mathcal{O}(x_n)\rangle =0, 
\end{equation}
with 
\begin{equation}
	L_{\mu\nu}(x)=i\left(x_\mu\partial_\nu - x_\nu\partial_\mu \right),
\end{equation}
and the symmetric scaling relation,
\begin{equation}
	\left(\sum_{j=1}^n \Delta_j - \sum_j^{n-1} p_j^\alpha 
	\frac{\partial}{\partial p_j^\alpha} - (n-1) d \right)\Phi(\underline{p})=0.
\end{equation}
Using \eqref{uno} and the vanishing of the $\delta'_\textrm{term}$ terms, the structure of the CWI on the correlator $\Phi(p)$ then takes the symmetric form 
\begin{equation}
	\label{sm1}
	\sum_{j=1}^n \int \underline {d^d p}e^{i\underline{p\cdot x}}\left(p_j^\kappa \frac{\partial^2}{\partial p_{j\alpha} \partial p_j^\alpha} -
	2 p_j^\alpha \frac{\partial^2}{\partial p_j^\alpha \partial p_j^\kappa}  + 2( \Delta_j-d)\frac{\partial}{\partial p_j^\kappa}\right) 
	\Phi(\underline{p})\delta^d(P)=0.
\end{equation}
This {\em symmetric} expression is the starting point in order to proceed with the elimination of one of the momenta, say $p_n$. 
Also in this case, one can proceed by following the same procedure used in the derivation of the dilatation identity, dropping the contribution coming from the dependent momentum $p_n$, thereby obtaining the final form of the equation

\begin{equation}
	\sum_{j=1}^{n-1}\left(p_j^\kappa \frac{\partial^2}{\partial p_j^\alpha\partial p_j^\alpha} + 2(\Delta_j- d)\frac{\partial}{\partial p_j^\kappa}-2 p_j^\alpha\frac{\partial^2}{\partial p_j^\kappa \partial p_j^\alpha}\right)\Phi(p_1,\ldots p_{n-1},\bar{p}_n)=0.\label{SpCWIs}
\end{equation}
Also in this case the differentiation respect to $p_n$ requires the chain rule. For a certain sequence of scalar single particle operators 
\begin{equation}
	\Phi(p_1,\ldots p_{n-1},\bar{p}_n)=\langle \mathcal{O}(p_1)\ldots \mathcal{O} (\bar{p}_n)\rangle 
\end{equation}
the Leibnitz rule is therefore violated. As we have already mentioned, the system of scalar equations obtained starting from the tensor one are, however, symmetric. For further details 
concerning this point and the arbitrariness in the choice of the independent momentum of the correlator we refer to  \cite{Coriano:2018bbe}. 

\section{Conformal constraints on 2-point functions\label{TwoPointSection}}
For the two-point functions the differential equations simplify considerably, being expressed in terms 
of just one independent momentum $p$, and take the form
\begin{equation}
	\label{ConformalEqMomTwoPoint}
	\left( - p_{\mu} \, \frac{\partial}{\partial p_{\mu}}  + \Delta_1 + \Delta_2 - d \right) G^{ij}(p) = 0 \,, 
\end{equation}
for the dilatation and 
\begin{equation}
	\left(  p_{\mu} \, \frac{\partial^2}{\partial p^{\nu} \partial p_{\nu}}  - 2 \, p_{\nu} \, \frac{\partial^2}{ \partial p^{\mu} 
		\partial p_{\nu} }    + 2 (\Delta_1 - d) \frac{\partial}{\partial p^{\mu}}\right) G^{ij}(p)  + 2 (\Sigma_{\mu\nu})^{i}_{k} \frac{\partial}{\partial 
		p_{\nu}}  G^{kj}(p)   = 0 \,,\label{ConformalEqMomTwoPoint2}
\end{equation}
for the special conformal Ward identities. We have defined $G^{ij}(p) \equiv \langle \mathcal O_1^i(p) \mathcal O_2^j(-p) \rangle$. The Eq.\eqref{ConformalEqMomTwoPoint} dictates the scaling behavior of the correlation function, while special conformal 
invariance allows a non zero result only for equal scale dimensions of the two operators $\Delta_1 = \Delta_2$, as we know from the 
corresponding analysis in coordinate space. \\
For the correlation function $G_S(p)$ of two scalar primary fields the invariance under the Poincar\'{e} group obviously 
implies that $G_S(p) \equiv G_S(p^2)$, so that the derivatives with respect to the momentum $p_\mu$ can be easily recast in terms of 
the variable $p^2$. \\
The invariance under scale transformations implies that $G_S(p^2)$ is a homogeneous function of degree 
$\alpha = \frac{1}{2}(\Delta_1 + \Delta_2 - d)$. 
At the same time, it is easy to show that \eqref{ConformalEqMomTwoPoint2} can be satisfied only if $\Delta_1 = \Delta_2$. 
Therefore conformal symmetry fixes the structure of the scalar two-point function up to an arbitrary overall constant $C$ as
\begin{equation}
	\label{TwoPointScalar}
	G_S(p^2) = \langle \mathcal O_1(p) \mathcal O_2(-p) \rangle = \delta_{\Delta_1 \Delta_2}  \, C\, (p^2)^{\Delta_1 - d/2} \, .
\end{equation}
If we redefine
\begin{equation}
	C=c_{S 12} \,  \frac{\pi^{d/2}}{4^{\Delta_1 - d/2}} \frac{\Gamma(d/2 - \Delta_1)}{\Gamma(\Delta_1)} 
\end{equation}
in terms of the new integration constant $c_{S 12}$, the two-point function reads as
\begin{equation}
	\label{TwoPointScalar2}
	G_S(p^2) =  \delta_{\Delta_1 \Delta_2}  \, c_{S 12} \,  \frac{\pi^{d/2}}{4^{\Delta_1 - d/2}} \frac{\Gamma(d/2 - \Delta_1)}{\Gamma(\Delta_1)} 
	(p^2)^{\Delta_1 - d/2} \,,
\end{equation}
and after a Fourier transform in coordinate space it takes the familiar form
\begin{equation}
	\langle \mathcal O_1(x_1) \mathcal O_2(x_2) \rangle \equiv \mathcal{F.T.}\left[ G_S(p^2) \right] =  \delta_{\Delta_1 \Delta_2} \,  c_{S 12} 
	\frac{1}{(x_{12}^2)^{\Delta_1}} \,,
\end{equation}
where $x_{12} = x_1 - x_2$. 
The ratio of the two Gamma functions relating the two integration constants $C$ and $c_{S 12}$ correctly reproduces the ultraviolet singular behavior of the correlation function and plays a role in the discussion of the origin of the scale anomaly.

Now we turn to the vector case where we define $G_V^{\alpha \beta}(p) \equiv \langle V_1^\alpha(p) V_2^\beta(-p) \rangle$. If the 
vector current is conserved, then the tensor structure of the two-point correlation function is entirely fixed by the 
condition $\partial^\mu V_\mu = 0$, as 
\begin{equation}
	\label{TwoPointVector0}
	G_V^{\alpha \beta}(p) =  \pi^{\alpha\beta}(p) \, f_V(p^2)\,, \qquad \qquad \mbox{with} \qquad
	\pi^{\alpha\beta}(p) = \delta^{\alpha \beta} -\frac{p^\alpha p^\beta}{p^2} 
\end{equation}
where $f_V$ is a function of the invariant square $p^2$ whose form, as in the scalar case, is determined by the conformal constraints. 
Following the same reasoning discussed previously we find that
\begin{equation}
	\label{TwoPointVector}
	G_V^{\alpha \beta}(p) = \delta_{\Delta_1 \Delta_2}  \, c_{V 12}\, 
	\frac{\pi^{d/2}}{4^{\Delta_1 - d/2}} \frac{\Gamma(d/2 - \Delta_1)}{\Gamma(\Delta_1)}\,
	\left( \eta^{\alpha \beta} -\frac{p^\alpha p^\beta}{p^2} \right)\
	(p^2)^{\Delta_1-d/2} \,,
\end{equation}
with $c_{V12}$ being an arbitrary constant. 
We recall that \eqref{ConformalEqMomTwoPoint2} gives consistent results for the two-point function in \eqref{TwoPointVector} only when the scale dimension $\Delta_1 = d - 1$. 

To complete this short excursus, we present the solution of the conformal constraints for the two-point function built out of two energy momentum tensor operators which are symmetric, conserved and traceless
\begin{equation}
	\label{EMTconditions}
	T_{\mu\nu} = T_{\nu\mu} \,, \qquad \qquad \partial^{\mu} T_{\mu\nu} = 0 \,, \qquad \qquad {T_{\mu}}^{\mu} = 0 \,.
\end{equation}
Exploiting the conditions defined in \eqref{EMTconditions} we can unambiguously define the tensor structure of the correlation 
function $G^{\alpha\beta\mu\nu}_T(p) = \Pi_{d}^{\alpha\beta\mu\nu}(p) \, f_T(p^2)$ with
\begin{equation}
	\label{TT}
	\Pi^{\alpha\beta\mu\nu}_{d}(p) = \frac{1}{2} \bigg[ \pi^{\alpha\mu}(p) \pi^{\beta\nu}(p) + \pi^{\alpha\nu}(p) \pi^{\beta\mu}(p) 
	\bigg] 
	- \frac{1}{d-1} \pi^{\alpha\beta}(p) \pi^{\mu\nu}(p) \,,
\end{equation}
and the scalar function $f_T(p^2)$ determined as usual, up to a multiplicative constant, by requiring the invariance under 
dilatations and special conformal transformations. We obtain
\begin{equation}
	\label{TwoPointEmt}
	G^{\alpha\beta\mu\nu}_T(p) = \delta_{\Delta_1 \Delta_2}  \, 
	c_{T 12}\,\frac{\pi^{d/2}}{4^{\Delta_1 - d/2}} \frac{\Gamma(d/2 - \Delta_1)}{\Gamma(\Delta_1)}\, 
	\Pi^{\alpha\beta\mu\nu}_{d}(p) \, (p^2)^{\Delta_1 - d/2} \,.
\end{equation}
As for the conserved vector currents, also for the energy momentum tensor the scaling dimension is fixed by \eqref{ConformalEqMomTwoPoint2}
and it is given by $\Delta_1 = d$. This particular value ensures that $\partial^\mu T_{\mu\nu}$ is also a quasi primary (vector) field. 

\section{More about 2-point functions}
\label{AppTwoPoint}
In this section we provide some details on the solutions of the conformal constraints for the two-point functions with conserved vector and tensor operators. \\
In the first case the tensor structure of the two-point function is uniquely fixed by the transversality condition $\partial^\mu V_\mu$ as
\begin{equation}
	G_V^{\alpha \beta}(p) = f(p^2) t^{\alpha \beta}(p)\,, \qquad \mbox{with} \quad t^{\alpha\beta}(p) = p^2 \delta^{\alpha\beta} - p^{\alpha} p^{\beta} \,.
\end{equation}
For the sake of simplicity, we have employed in the previous equation a slightly different notation with respect to \eqref{TwoPointVector0}, which, anyway, can be recovered with the identification $f(p^2) = f_V(p^2)/p^2$. \\
In order to exploit the invariance under scale and special conformal transformations it is useful to compute first and second order derivatives of the $t^{\alpha \beta}$ tensor structure. In particular we have
\begin{equation}
	\label{TDerivatives}
	\begin{split}
		t_1^{\alpha \beta, \mu}(p) &\equiv \frac{\partial }{\partial p_\mu} t^{\alpha \beta}(p) = 2 \, p^{\mu} \delta^{\alpha \beta} - p^{\alpha} \delta^{\mu \beta} - p^{\beta} \delta^{\mu \alpha} \,,  \\
		t_2^{\alpha \beta, \mu \nu}(p) &\equiv \frac{\partial^2 }{\partial p_\mu \, \partial p_\nu} t^{\alpha \beta}(p) = 2 \, \delta^{\mu \nu} \delta^{\alpha \beta} - \delta^{\nu \alpha} \delta^{\mu \beta} - \delta^{\nu \beta} \delta^{\mu \alpha} \,, 
	\end{split}
\end{equation}
with the properties
\begin{eqnarray}
	&& p_\mu t_1^{\alpha \beta, \mu}(p) = 2 \, t^{\alpha \beta}(p) \,, \qquad     t^{\alpha \beta}_{1\hspace{0.22cm},\alpha}(p) = - (d - 1) p^\beta \,, \notag \\
	&& p_\mu t_2^{\alpha \beta, \mu \nu}(p) = t_1^{\alpha \beta, \nu}(p) \,, \qquad    t^{\alpha \beta, \mu}_{2\hspace{0.45cm} \mu}(p) = 2(d-1) \delta^{\alpha \beta} \,. 
\end{eqnarray}
As we have already mentioned, the invariance under scale transformations implies that
\begin{equation}
	\label{FandLambda}
	f(p^2) = (p^2)^\lambda \qquad \qquad \mbox{with} \quad \lambda = \frac{\Delta_1 + \Delta_2 - d}{2} -1 \,,
\end{equation}
which can be easily derived from the first order differential equation in \eqref{ConformalEqMomTwoPoint} using \eqref{TDerivatives}. Having determined the structure of the scalar function $f(p^2)$, one can compute the derivatives appearing in \eqref{ConformalEqMomTwoPoint2}, namely the constraint following from invariance under the special conformal transformations
\begin{eqnarray}
	\label{GDerivatives}
	\frac{\partial}{\partial p_\mu} G_V^{\alpha \beta}(p) &=& (p^2)^{\lambda - 1} \bigg[ 2 \lambda \, p^\mu t^{\alpha \beta}(p) + p^2 \, t_1^{\alpha\beta, \mu}(p) \bigg] \,, \notag \\
	\frac{\partial^2}{\partial p_\mu \, \partial p_\nu} G_V^{\alpha \beta}(p) &=&(p^2)^{\lambda - 2} \bigg[ 4 \lambda (\lambda -1) p^{\mu} p^{\nu} t^{\alpha \beta}(p)  + 2 \lambda p^2 \delta^{\mu\nu}   t^{\alpha \beta}(p) + 2 \lambda  p^2 p^{\mu} t_1^{\alpha\beta,\nu}(p)  \notag \\
	&& \qquad +  \, 2 \lambda  p^2 p^{\nu} t_1^{\alpha\beta,\mu}(p) + (p^2)^2 t_2^{\alpha\beta, \mu\nu}(p) 
	\bigg] \,,
\end{eqnarray}
where we have used the definitions in \eqref{TDerivatives}.
Concerning the spin dependent part in \eqref{ConformalEqMomTwoPoint2}, we use the spin matrix for the vector field, which in our conventions is given by
\begin{equation}
	( \Sigma_{\mu\nu}^{(V)})^{\alpha}_{\beta} = \delta_{\mu}^{\alpha} \, \delta_{\nu \beta} - \delta_{\nu}^{\alpha} \, \delta_{\mu \beta} \,,
\end{equation}
and obtain
\begin{equation}
	\label{SigmaPart}
	2( \Sigma_{\mu\nu}^{(V)})^{\alpha}_{\rho} \frac{\partial}{\partial p_\nu} G_V^{\rho\beta}(p) = - (p^2)^{\lambda- 1} \bigg[ 2 \lambda \, p^\alpha {t_{\mu}}^{\beta}(p) + (d-1)p^2 p^\beta \delta_\mu^\alpha  + p^2 {t_{1 \, \mu}}^{\beta, \alpha}(p) \bigg] \,.
\end{equation}
Using the results derived in \eqref{GDerivatives} and \eqref{SigmaPart}, we have fully determined the special conformal constraint on the two-point vector function.
Then we can project \eqref{ConformalEqMomTwoPoint2} onto the three independent tensor structures, and set $\lambda$ to the value given in \eqref{FandLambda}, obtaining three equations for the scale dimensions $\Delta_i$ of the vector operators
\begin{equation}
	\begin{cases}
		(\Delta_1 - \Delta_2) (\Delta_1 + \Delta_2 - d) = 0 \,, \notag \\
		\Delta_1 - d +1 = 0 \,, \notag \\
		\Delta_2 - d +1 = 0 \,. \notag
	\end{cases}
	\\
\end{equation}
The previous system of equations can be consistently solved only for $\Delta_1 = \Delta_2 = d -1$, as expected. This completes our derivation of the vector two-point function which, up to an arbitrary multiplicative constant, can be written as in \eqref{TwoPointVector}. \\
The characterization of the two-point function with a symmetric, traceless and conserved rank-2 tensor follows the same lines of reasoning of the vector case. These conditions (see \eqref{EMTconditions}) fix completely the tensor structure of the two-point function as
\begin{equation}
	G_T^{\alpha\beta\mu\nu}(p) = g(p^2) \, T^{\alpha\beta\mu\nu}(p) 
\end{equation}
with
\begin{equation}
	T^{\alpha\beta\mu\nu}(p) =  \frac{1}{2} \bigg[ t^{\alpha\mu}(p) t^{\beta\nu}(p) + t^{\alpha\nu}(p) t^{\beta\mu}(p) 
	\bigg] 
	- \frac{1}{d-1} t^{\alpha\beta}(p) t^{\mu\nu}(p) \,.
\end{equation}
For consistency with the convention used in section \secref{TwoPointSection} we have set $g(p^2) \equiv f_T(p^2)/(p^2)^2$. \\
As in the previous case, we give the first and second order derivatives of the $T^{\alpha\beta\mu\nu}(p)$ tensor structure
\begin{eqnarray}
	\label{TTDerivatives}
	T_1^{\alpha\beta\mu\nu, \rho}(p) &\equiv& \frac{\partial}{\partial p_\rho} T^{\alpha\beta\mu\nu}(p) = \frac{1}{2} \bigg[ t_1^{\alpha\mu, \rho}(p) t^{\beta\nu}(p) 
	+ t^{\alpha\mu}(p) t_1^{\beta\nu, \rho}(p) + \left( \mu \leftrightarrow \nu \right) \bigg] \notag \\
	&& - \frac{1}{d-1} \bigg[ t_1^{\alpha\beta, \rho}(p) t^{\mu\nu}(p) + t^{\alpha\beta}(p) t_1^{\mu\nu, \rho}(p) \bigg] \,, \notag \\
	T_2^{\alpha\beta\mu\nu, \rho\sigma}(p) &\equiv& \frac{\partial}{\partial p_\rho \, \partial p_\sigma} T^{\alpha\beta\mu\nu}(p) = \frac{1}{2} \bigg[ 
	t_2^{\alpha\mu, \rho \sigma}(p) t^{\beta\nu}(p) + t_1^{\alpha\mu, \rho}(p) t_1^{\beta\nu, \sigma}(p) + t_1^{\alpha\mu, \sigma}(p) t_1^{\beta\nu, \rho}(p) \notag \\
	&& +  \, t^{\alpha\mu}(p) t_2^{\beta\nu, \rho \sigma}(p)+ \left( \mu \leftrightarrow \nu \right) \bigg]  \notag \\
	&& - \frac{1}{d-1} \bigg[  t_2^{\alpha\beta, \rho\sigma}(p) t^{\mu\nu}(p) +   t_1^{\alpha\beta, \rho}(p) t_1^{\mu\nu, \sigma}(p) + (\mu\nu) \leftrightarrow (\alpha\beta)  \bigg] \,,
\end{eqnarray}
together with some of their properties
\begin{eqnarray}
	p_\rho T_1^{\alpha\beta\mu\nu, \rho}(p) = 4 \, T^{\alpha\beta\mu\nu}(p) \,, \qquad
	p_\rho T_2^{\alpha\beta\mu\nu, \rho \sigma}(p) = 3 \, T_1^{\alpha\beta\mu\nu, \sigma}(p)  \,.
\end{eqnarray}
As we have already stressed, \eqref{ConformalEqMomTwoPoint} defines the scaling behavior of the two-point function, providing, therefore, that the functional form of $g(p^2)$ is given by
\begin{equation}
	g(p^2) = (p^2)^\lambda \qquad \mbox{with} \quad \lambda = \frac{\Delta_1 +\Delta_2 -d}{2} -2 \,.
\end{equation}
On the other hand, \eqref{ConformalEqMomTwoPoint2}, which represents the constraint from the special conformal transformations, fixes the scaling dimensions of the tensor operators. In this case the spin connection is given by
\begin{equation}
	(\Sigma_{\mu\nu}^{(T)})^{\alpha \beta}_{\rho \sigma } = \left( \delta_{\mu}^{\alpha} \, \delta_{\nu \rho} - \delta_{\nu}^{\alpha} \, \delta_{\mu \rho} \right) \delta_{\sigma}^{\beta}
	+ \left( \delta_{\mu}^{\beta} \, \delta_{\nu \sigma} - \delta_{\nu}^{\beta} \, \delta_{\mu \sigma} \right) \delta_{\rho}^{\alpha} \,.
\end{equation}
The algebra is straightforward but rather cumbersome due to the proliferation of indices. Here we give only the final result, which can be obtained by projecting \eqref{ConformalEqMomTwoPoint} and using of \eqref{TTDerivatives} in all the different independent tensor structures, giving
\begin{equation}
	\begin{cases}
		(\Delta_1 - \Delta_2) (\Delta_1 + \Delta_2 - d) = 0 \,, \notag \\
		\Delta_1 - d = 0 \,, \notag \\
		\Delta_2 - d = 0 \,, \notag
	\end{cases}
	\\
\end{equation}
which implies that $\Delta_1 = \Delta_2 = d$, as described in \eqref{TwoPointEmt}.

\section{Conformal Ward identities from the vector to the scalar form} 
We now come to illustrate the procedure for obtaining the conformal constraints in the form of partial differential equation with respect to scalar invariants.\\
We consider first the case of $3$-point function of scalar primary operators. Due to the translational invariance, the correlator can be expressed as a function of three invariants, i.e, the magnitudes of the momenta, defined as ${p}_i=\sqrt{p_i^2}$
\begin{equation}
	\braket{\mathcal{O}_1(p_1)\mathcal{O}_2(p_2)\mathcal{O}_3(\bar{p}_3)}=\Phi(p_1,p_2,p_3).
\end{equation} 
All the conformal WI's can be re-expressed in scalar form using the chain rules
\begin{equation}
	\label{chainr}
	\frac{\partial \Phi}{\partial p_i^\mu}=\frac{p_i^\mu}{  p_i}\frac{\partial\Phi}{\partial  p_i} 
	-\frac{\bar{p}_3^\mu}{  p_3}\frac{\partial\Phi}{\partial   p_3}, \quad i=1,2
\end{equation}
where $\bar{p}_3^\mu=-p_1^\mu-p_2^\mu$ and $p_3=\sqrt{(p_1+p_2)^2}$. By using this equation, the dilatation operator can be written as
\begin{equation}
	\sum_{i=1}^2\,p_i^\mu\,\frac{\partial }{{p_i}^\mu}\Phi(p_1,p_2,p_3)=\left(
	{p}_1\frac{ \partial }{\partial   p_1} +   p_2\frac{ \partial }{\partial   p_2} +   p_3\frac{ \partial }{\partial   p_3}\right)\Phi(p_1,p_2,p_3).
\end{equation}
Therefore, the scale equation becomes 
\begin{equation}
	\label{scale}
	\left(\sum_{i=1}^3\Delta_i -2 d - \sum_{i=1}^3    p_i \frac{ \partial}{\partial   p_i}\right)\Phi(p_1,p_2,\bar{p}_3)=0.
\end{equation}
It takes a straightforward but lengthy computation to show that the special conformal Ward identities in $d$ dimension take the form
\begin{align}
	0&={K}_{scalar}^{\kappa}\Phi(p_1,p_2,p_3)=\bigg(p_1^\kappa\,K_1+p_2^\kappa\,K_2+\bar{p}_3^\kappa\,K_3\bigg)\Phi(p_1,p_2,p_3)\notag\\
	&=p_1^\kappa\Big(K_1-K_3\Big)\Phi(p_1,p_2,p_3)+p_2^\kappa\Big(K_2-K_3\Big)\Phi(p_1,p_2,p_3),\label{SCWIs}
\end{align}
where we have used the conservation of the total momentum, with the $K_i$ operators defined as
\begin{equation}
	\label{Koper1}
	{ K}_i\equiv \frac{\partial^2}{\partial    p_i \partial    p_i} 
	+\frac{d + 1 - 2 \Delta_i}{   p_i}\frac{\partial}{\partial   p_i}, \quad i=1,2,3. 
\end{equation}
The equation \eqref{SCWIs} is satisfied if every coefficients of the independent four-momenta $p_1^\mu,\,p_2^\mu$ are equal to zero. This condition leads to the scalar form of the special conformal constraints 
\begin{equation}
	\frac{\partial^2\Phi}{\partial   p_i\partial   p_i}+
	\frac{1}{  p_i}\frac{\partial\Phi}{\partial  p_i}(d+1-2 \Delta_1)-
	\frac{\partial^2\Phi}{\partial   p_3\partial   p_3} -
	\frac{1}{  p_3}\frac{\partial\Phi}{\partial  p_3}(d +1 -2 \Delta_3)=0\qquad i=1,2,
	\label{3k1}
\end{equation}
and defining 
\begin{equation}
	\label{kij}
	K_{ij}\equiv {K}_i-{K}_j
\end{equation}
Eqs. \eqref{3k1} can be written in the coincise form 
\begin{equation}
	\label{3k2}
	K_{13}\,\Phi(p_1,p_2,p_3)=0 \qquad \textrm{and} \qquad K_{23}\,\Phi(p_1,p_2,p_3)=0.
\end{equation}

\section{Hypergeometric systems: the scalar case}\label{Sol3Point}

In this section, we illustrate the hypergeometric character of the CWI's, proven in the approach presented in \cite{Coriano:2013jba, Coriano:2018bsy, Coriano:2018bbe, Maglio:2019grh}. An independent analysis performed in \cite{Bzowski:2013sza} has connected the solutions of such equations to 3K (or triple-K) integrals. The analysis of \cite{Coriano:2013jba} proved that the fundamental basis for the most general solutions of such equations are given by Appell functions $F_4$.
We present the analysis of scalar $3$-point functions, and then we discuss the analogous of the $4$-point functions in the next chapter, elaborating on our extension, which is contained in \cite{Maglio:2019grh,Coriano:2020ccb}.
\subsection{Hypergeometric systems} 
We start with the following definitions of the Appell's hypergeometric functions $F_1(x,y)$, $F_2(x,y)$, $F_3(x,y)$, $F_4(x,y)$  
\begin{eqnarray} \label{appf1}
	\app 1{a;\;b_1,b_2}{c}{x,\,y} \equal \sum_{n=0}^{\infty} \sum_{m=0}^{\infty}
	\frac{(a)_{n+m}\,(b_1)_n\,(b_2)_m}{(c)_{n+m}\;n!\,m!}\,x^n\,y^m,\\ \label{appf2}
	\app 2{a;\;b_1,b_2}{c_1,c_2}{x,\,y} \equal \sum_{n=0}^{\infty} \sum_{m=0}^{\infty}
	\frac{(a)_{n+m}\,(b_1)_n\,(b_2)_m}{(c_1)_n\,(c_2)_m\;n!\,m!}\,x^n\,y^m,\\ \label{appf3}
	\app 3{\!a_1,a_2;\,b_1,b_2}{c}{x,\,y} \equal \sum_{n=0}^{\infty} \sum_{m=0}^{\infty}
	\frac{(a_1)_n(a_2)_m(b_1)_n(b_2)_m}{(c)_{n+m}\;n!\,m!}\,x^n\,y^m,\\ \label{appf4}
	F_4(a,b,c_1,c_2; x,y)\equiv\app 4{a;\;b}{c_1,c_2\,}{x,\,y} \equal \sum_{n=0}^{\infty} \sum_{m=0}^{\infty}
	\frac{(a)_{n+m}\,(b)_{n+m}}{(c_1)_n\,(c_2)_m\;n!\,m!}\,x^n\,y^m
\end{eqnarray}
that are bivariate generalizations of the Gauss hypergeometric series
\begin{equation} \label{gausshpg}
	\hpg21{A,\,B}{C}{z} = \sum_{n=0}^{\infty} 
	\frac{(A)_{n}\,(B)_n}{(C)_n\,n!}\,z^n,
\end{equation}
with the (Pochhammer) symbol $(\alpha)_{k}$ given by
\begin{equation}
	(\alpha)_{k}\equiv (\alpha,k)\equiv\frac{\Gamma(\alpha+k)}{\Gamma(\alpha)}=\alpha(\alpha+1)\dots(\alpha+k-1).\label{Pochh}
\end{equation}
An account of many of the properties of such functions and a discussion of the univariate cases, when the two variables coalesce, can be found in \cite{Vidunas1} and related works. They are solutions of equations generalizing Euler's hypergeometric equation 
\begin{equation} \label{eq:euler}
	z(1-z)\,\frac{d^2y(z)}{dz^2}+
	\big(C-(A+B+1)z\big)\frac{dy(z)}{dz}-A\,B\,y(z)=0,
\end{equation}
whose solution is denoted as $\hpgo21$, written in \eqref{gausshpg}.
This is classified as a Fuchsian equation with singularities at $z=0$, $z=1$ and $z=\infty$. 
When the two arguments $x,y$ of the Appell functions are algebraically related, they 
are referred to as univariate functions and satisfy Fuchsian ordinary
differential equations.\\
The proof that the CWIs of 3-point functions are hypergeometric systems of equations has been shown independently in \cite{Coriano:2013jba} and \cite{Bzowski:2013sza}. We recall that, in the case of Appell functions of type $F_4$ given in \eqref{appf4}, which are the relevant ones in all our discussion, such functions are solutions of the system of differential equations
\begin{equation}
	\label{F4diff.eq}
	\left\{\,\,
	\begin{aligned}
		&\\[-1.8ex]
		&\big[ x(1-x) \partial_{xx} - y^2 \partial_{yy}- 2 \, x \, y \partial_{xy}+  \left[ \gamma - (\alpha + \beta + 1) x \right] \partial_x- (\alpha + \beta + 1) y\,\partial_y  - \alpha \, \beta \big] F(x,y) = 0 \,, \notag \\[1ex]
		&\big[ y(1-y) \partial_{yy} - x^2 \partial_{xx} - 2 \, x \, y \partial_{xy} +  \left[ \gamma' - (\alpha + \beta + 1) y \right] \partial_y- (\alpha + \beta + 1) x \partial_x  - \alpha \, \beta \big] F(x,y) = 0 \,,\\[1.4ex] 
	\end{aligned}\right.
\end{equation}
as illustrated in \cite{Appell}, where $F(x,y)$ can be in the most general case a linear combinations of 4 independent functions $F_4$, hypergeometric of two variables $x$ and $y$. The univariate limits of the solutions are important, from the physical point of view, for the study of the behaviour of the corresponding correlation functions in special kinematics. An example has been discussed in \cite{Maglio:2019grh,Coriano:2020ccb,Coriano:2019nkw} in the case of 4-point functions. 

\subsection{The case of scalar 3-point functions}
\label{fuchs2}
For 3-point functions the momentum dependence of the correlator is parameterized uniquely by the three external invariant, $p_1^2, p_2^2$ and $p_3^3$, and we will denote with $p_i$ their magnitudes. The CWI's in momentum space, in this case, can be reduced to scalar equations by some manipulations, as discussed in \cite{Coriano:2013jba,Bzowski:2013sza,Coriano:2018bbe}.  
As we have presented in the previous sections, the scalar correlator 
$\Phi(p_1,p_2, p_3)$, in momentum space has to satisfy the two homogeneous conformal equations \eqref{3k2}, and the scaling equation \eqref{scale}. The hypergeometric character of the CWI's emerges in various ways. One may proceed from \eqref{SpCWIs} and introduce the change of variables
\begin{eqnarray}
	\frac{\partial}{\partial p_{1}^{\mu}}  &=   2 (p_{1\, \mu} + p_{2 \, \mu}) \frac{\partial}{\partial p_3^2} + \frac{2}{p_3^2}\left( 
	(1- x) p_{1 \, \mu}  - x  \,  p_{2 \, \mu} \right) \frac{\partial}{\partial x} - 2  (p_{1\, \mu} + p_{2 \, \mu}) \frac{y}{p_3^2} 
	\frac{\partial}{\partial y} \,, \notag \\
	\frac{\partial}{\partial p_{2}^{\mu}}  &= 2 (p_{1\, \mu} + p_{2 \, \mu}) \frac{\partial}{\partial p_3^2}   -   2  (p_{1\, \mu} + 
	p_{2 \, \mu}) \frac{x}{p_3^2} \frac{\partial}{\partial x}   + \frac{2}{p_3^2}\left( (1- y) p_{2 \, \mu}  - y  \,  p_{1 \, \mu} 
	\right) \frac{\partial}{\partial y}. \, 
\end{eqnarray}
with
\begin{equation}
	x=\frac{p_1^2}{p_3^2},\qquad y=\frac{p_2^2}{p_3^2}.\label{xy}
\end{equation}
Following the approach presented in \cite{Coriano:2013jba}, the ans\"atz for the solution can be taken of the form 
\begin{equation}
	\label{ans}
	\Phi(p_1,p_2,p_3)=(p_3^2)^{\,\Delta_t/2 -  d} x^{a}y^{b} F(x,y)
\end{equation}
with $\Delta_t=\Delta_1+\Delta_2+\Delta_3$. Here we are taking $p_3$ as "pivot" in the expansion, but we could have equivalently chosen as such any of the 3  momentum invariants $p_i^2$. \\
The solution $\Phi$ is required to be homogeneous of degree $\Delta_t-2 d$ under a scale transformation, according to \eqref{scale}, and in \eqref{ans} this is taken into account by the factor $p_3^{\Delta_t - 2 d}$. Inserting the ansatz in \eqref{3k2}, one derives the equation 
\begin{equation}
	K_{13}\Phi=4 (p_3^2)^{\Delta_t/2 -d -1} x^a y^b
	\left[x(1-x)\,\partial_{xx}  + (A x + \gamma)\,\partial_x -
	2 x y \partial_{xy}- y^2\,\partial{yy} + 
	D y\,\partial_y + \left(E +\frac{G}{x}\right)\right] F(x,y)=0
	\label{red}
\end{equation}
with
\begin{align}
	&A=D=\Delta_1 +\Delta_2 - 1 -2 a -2 b -\frac{3 d}{2}, &&\gamma(a)=2 a +\frac{d}{2} -\Delta_1 + 1,
	\notag\\
	&E=-\frac{1}{4}(2 a + 2 b +2 d -\Delta_1 -\Delta_2 -\Delta_3)(2 a +2 b + d -\Delta_1 -\Delta_2 +\Delta_3),&& G=\frac{a}{2}(d +2 a - 2 \Delta_1).
\end{align}
and
\begin{align}
	K_{23}\Phi &= 4( p_3^2)^{\Delta_t/2 - d -1} x^a y^b
	\left[  y(1-y)\,\partial {yy}   + (A' y + \gamma')\,\partial y -
	2 x y \partial_{xy}- x^2\,\partial_{xx}+ 
	D' x\,\partial_x + \left(E' +\frac{G'}{y}\right)\right]F(x,y)=0\label{red2}
\end{align}
with
\begin{align}
	&A'=D'= A,  &&\gamma'(b)=2 b +\frac{d}{2} -\Delta_2 + 1,\notag\\
	&E'= E,&&G'=\frac{b}{2}(d +2 b - 2 \Delta_2).
\end{align}
Notice that the elimination of the singularities $G/x=0$ and $G'/y=0$ in the equations guarantees, from the physical perspective,the analyticity of the solutions. This requirement is satisfied by considering the conditions on the Fuchsian exponents $a$ as
\begin{equation}
	\label{cond1}
	a=0\equiv a_0 \qquad \textrm{or} \qquad a=\Delta_1 -\frac{d}{2}\equiv a_1.
\end{equation}
From the equation $\textup{K}_{23}\Phi=0$ we obtain a similar condition for $b$, thereby fixing the two remaining (Fuchsian) indices
\begin{equation}
	\label{cond2}
	b=0\equiv b_0 \qquad \textrm{or} \qquad b=\Delta_2 -\frac{d}{2}\equiv b_1.
\end{equation}
The complete equivalence of the CWI's \eqref{3k2} with an hypergeometric system of equations is obtained by choosing such particular $(a,b)$ exponents in the non-scale invariant part of the ans\"atz. 
The four independent solutions of the CWI's then will all be characterized by the same 4 pairs of indices $(a_i,b_j)$ $(i,j=1,2)$.
Setting 
\begin{equation}
	\alpha(a,b)= a + b + \frac{d}{2} -\frac{\Delta_1 +\Delta_2 -\Delta_3}{2} \qquad \beta (a,b)=a +  b + d -\frac{\Delta_1 +\Delta_2 +\Delta_3}{2} \qquad 
	\label{alphas}
\end{equation}
the general solutions takes the form 
\begin{equation}
	\Phi(p_1,p_2,p_3)=p_3^{\Delta-2 d}\, \sum_{a,b} c(a,b,\vec{\Delta_t})\,x^a y^b \,F_4(\alpha(a,b), \beta(a,b); \gamma(a), \gamma'(b); x, y)
	\label{geneq} 
\end{equation}
where the sum runs over the four values $a_i, b_i$ $i=0,1$ with constants $c(a,b,\vec{\Delta_t})$ and $\vec{\Delta_t}=(\Delta_1,\Delta_2,\Delta_3)$. Defining 
\begin{align}
	&\alpha\equiv \alpha(a_0,b_0)=\frac{d}{2}-\frac{\Delta_1 + \Delta_2 -\Delta_3}{2},\, && \beta\equiv \beta(b_0)=d-\frac{\Delta_1 + \Delta_2 +\Delta_3}{2},  \notag\\
	&\gamma \equiv \gamma(a_0) =\frac{d}{2} +1 -\Delta_1,\, &&\gamma'\equiv \gamma(b_0) =\frac{d}{2} +1 -\Delta_2.
\end{align}
the 4 independent solutions can be re-expressed in terms of the parameters above as 
\begin{align}
	\label{F4def}
	S_1(\alpha, \beta; \gamma, \gamma'; x, y)\equiv F_4(\alpha, \beta; \gamma, \gamma'; x, y) = \sum_{n = 0}^{\infty}\sum_{m = 0}^{\infty} \frac{(\alpha)_{n+m} \, 
		(\beta)_{n+m}}{(\gamma)_n \, (\gamma')_m} \frac{x^n}{n!} \frac{y^m}{m!} 
\end{align}
with the definition of the Pochhammer symbol $(\l)_{k}$ given by
\begin{equation}
	(\l)_{k}=\frac{\Gamma(\l+k)}{\Gamma(\l)}=\l(\l+1)\dots(\l+k-1),\label{Pochh}
\end{equation}
and
\begin{align}
	\label{solutions}
	S_2(\alpha, \beta; \gamma, \gamma'; x, y) &= x^{1-\gamma} \, F_4(\alpha-\gamma+1, \beta-\gamma+1; 2-\gamma, \gamma'; x,y) \,, \notag\\
	S_3(\alpha, \beta; \gamma, \gamma'; x, y) &= y^{1-\gamma'} \, F_4(\alpha-\gamma'+1,\beta-\gamma'+1;\gamma,2-\gamma' ; x,y) \,, \notag\\
	S_4(\alpha, \beta; \gamma, \gamma'; x, y) &= x^{1-\gamma} \, y^{1-\gamma'} \, 
	F_4(\alpha-\gamma-\gamma'+2,\beta-\gamma-\gamma'+2;2-\gamma,2-\gamma' ; x,y) \, . 
\end{align}
for which the solution can be written in the final form
\begin{equation}
	\Phi(p_1,p_2,p_3)=(p_3^2)^{\Delta_t/2- d} \sum_{i=1}^4 \,c_i(\D_1,\D_2,\D_3)\,S_i (\alpha, \beta; \gamma, \gamma'; x, y)
\end{equation}
where $c_i$ are arbitrary coefficients which may depend on the scale dimensions $\D_i$ and on the spacetime dimension $d$. An equivalent version of the solution found above can be derived as in \cite{Bzowski:2013sza}, where it is written in terms of $K$ Bessel functions as
\begin{equation}
	\label{caz}
	\Phi(p_1,p_2,p_3)=\,C_{123}\, p_1^{\D_1-\frac{d}{2}}p_2^{\D_2-\frac{d}{2}}p_3^{\D_3-\frac{d}{2}}\int_0^\infty dx\,x^{\frac{d}{2}-1}\,K_{\D_1-\frac{d}{2}}(p_1\,x)\,K_{\D_2-\frac{d}{2}}(p_2\,x)\,K_{\D_3-\frac{d}{2}}(p_3\,x)
\end{equation}
where $C_{123}$ is an undetermined constant. This formalism will be used later in the analysis of the solution of the 4-point function.

\subsection{Symmetrizations}
Notice that in the scalar case, for ordinary correlators, one is allowed to require its complete symmetry under the exchange of the 3 external momenta and scaling dimensions, as discussed in \cite{Coriano:2013jba}. This reduces the four
constants of integration to just one overall. The 4 independent solutions are then all of the form $x^a y^b F_4$, with
$a$ and $b$ fixed by (\ref{cond1}) and (\ref{cond2}). Such values of the $(a,b)$ exponents in the part of the ans\"atz which is not scale invariant, are determined by the condition that the $1/x$ and $1/y$ contributions vanish in the PDE's, turning the CWI's into a hypergeometric system of two equations, whose structure is symmetric under the exhange of $x$ and $y$.\\
For tensor correlators such as the $TJJ$ or the $TTT$ an extensive use of the properties of the the hypergeometric operators $\textup{K}_{ij}$ allows to build the complete solutions for the form factors which parameterize each of these correlators \cite{Coriano:2018bsy,Coriano:2018bbe}. Imposing the symmetry conditions is, in general, rather cumbersome, and one has to rely on one of the few relations known for the Appell function $F_4$, specifically the inversion formula 
\begin{align}
	\label{transfF41}
	F_4(\alpha, \beta; \gamma, \gamma'; x, y) =& \frac{\Gamma(\gamma') \Gamma(\beta - \alpha)}{ \Gamma(\gamma' - \alpha) \Gamma(\beta)} (- y)^{- \alpha} \, F_4\left(\alpha, \alpha -\gamma' +1; \gamma, \alpha-\beta +1; \frac{x}{y}, \frac{1}{y}\right) \notag\\ 
	&+  \frac{\Gamma(\gamma') \Gamma(\alpha - \beta)}{ \Gamma(\gamma' - \beta) \Gamma(\alpha)} (- y)^{- \beta} \, F_4\left(\beta -\gamma' +1, \beta ; \gamma, \beta-\alpha +1; \frac{x}{y}, \frac{1}{y}\right) \,
\end{align}
which allows to reverse the ratios respect to the momentum chosen as pivot. The symmetrization, obviously, allows to reduce the number of constants. 
\subsection{Extracting the physical solution}
In order to clarify this subtle point, we illustrate the possible methods that can be followed in order to identify the unique physical solution of the hypergeometric equations. \\
Notice, as already mentioned above, that the four solutions \eqref{F4def} and \eqref{solutions} define the 
basis into which {\em any} solution can be expanded. Such basis allows to generate 
by linear combination any function which is symmetric in the external momenta, under the condition that the constants 
$c_i(\D_1,\D_2,\D_3)$ are appropriately chosen. This is exactly what \eqref{transfF41} allows to achieve. In fact, by using \eqref{transfF41}, the general symmetric solution can be identified - modulo a single overall constant - in the form 
{\cite{Coriano:2013jba}
	\begin{align}
		&\braket{\mathcal{O}(p_1)\,\mathcal{O}(p_2)\,\mathcal{O}(p_3)}=\big(p_3^2\big)^{-d+\frac{\Delta_t}{2}}\,C(\Delta_1,\Delta_2,\Delta_3,d)\notag\\
		&\Bigg\{\Gamma\left(\Delta_1-\frac{d}{2}\right)\Gamma\left(\Delta_2-\frac{d}{2}\right)\Gamma\left(d-\frac{\Delta_1+\Delta_2+\Delta_3}{2}\right)\Gamma\left(d-\frac{\Delta_1+\Delta_2-\Delta_3}{2}\right)\notag\\
		&\hspace{3cm}\times
		\,F_4\,\left(\frac{d}{2}-\frac{\Delta_1+\Delta_2-\Delta_3}{2},d-\frac{\Delta_t}{2},\frac{d}{2}-\Delta_1+1,\frac{d}{2}-\Delta_2+1;x,y\right)\notag\\[2ex]
		&\qquad+\,
		\Gamma\left(\frac{d}{2}-\Delta_1\right)\Gamma\left(\Delta_2-\frac{d}{2}\right)\Gamma\left(\frac{\Delta_1-\Delta_2+\Delta_3}{2}\right)\Gamma\left(\frac{d}{2}+\frac{\Delta_1-\Delta_2-\Delta_3}{2}\right)\notag\\
		&\hspace{3cm}\times x^{\Delta_1-\frac{d}{2}}\,F_4\,\left(\frac{\Delta_1-\Delta_2+\Delta_3}{2},\frac{d}{2}-\frac{\Delta_2+\Delta_3-\Delta_1}{2},\Delta_1-\frac{d}{2}+1,\frac{d}{2}-\Delta_2+1;x,y\right)\notag
	\end{align}
	\begin{align}
		&\qquad+\,
		\Gamma\left(\Delta_1-\frac{d}{2}\right)\Gamma\left(\frac{d}{2}-\Delta_2\right)\Gamma\left(\frac{-\Delta_1+\Delta_2+\Delta_3}{2}\right)\Gamma\left(\frac{d}{2}+\frac{-\Delta_1+\Delta_2-\Delta_3}{2}\right)\notag\\
		&\hspace{3cm}\times\,y^{\Delta_2-\frac{d}{2}}\,F_4\,\left(\frac{\Delta_2-\Delta_1+\Delta_3}{2},\frac{d}{2}-\frac{\Delta_1-\Delta_2+\Delta_3}{2},\frac{d}{2}-\Delta_1+1,\Delta_2-\frac{d}{2}+1;x,y\right)\notag\\[2ex]
		&\qquad+\,
		\Gamma\left(\frac{d}{2}-\Delta_1\right)\Gamma\left(\frac{d}{2}-\Delta_2\right)\Gamma\left(\frac{\Delta_1+\Delta_2-\Delta_3}{2}\right)\Gamma\left(-\frac{d}{2}+\frac{\Delta_1+\Delta_2+\Delta_3}{2}\right)\notag\\
		&\hspace{3cm}\times\,x^{\Delta_1-\frac{d}{2}}\,y^{\Delta_2-\frac{d}{2}}F_4\,\left(-\frac{d}{2}+\frac{\Delta_t}{2},\frac{\Delta_1+\Delta_2-\Delta_3}{2},\Delta_1-\frac{d}{2}+1,\Delta_2-\frac{d}{2}+1;x,y\right)\Bigg\}.\label{solfin}
	\end{align}
	One can verify that the symmetric solution above does not have any unphysical singularity in the physical region and it has the expected behaviour in the large momentum limit $p_3\gg p_1$, in agreement with the requirements discussed in \cite{Bzowski:2014qja}. In fact, one can check that the previous expression, in the limit $p_3\gg p_1$ (expressible also as $p_3^2,\,p_2^2\to\infty$ with $p_2^2/p_3^2\to1$ fixed), it behaves as
	\begin{align}
		\braket{\mathcal{O}(p_1)\,\mathcal{O}(p_2)\,\mathcal{O}(p_3)}\propto f(d,\Delta_i)\,p_3^{\Delta_1+\Delta_2+\Delta_3-2d}\left( 1 +O\left(\frac{p_1}{p_3}\right)\right)  \hspace{2cm}\text{if}\ \ \Delta_1>\frac{d}{2}&
	\end{align}
	and
	\begin{align}
		\braket{\mathcal{O}(p_1)\,\mathcal{O}(p_2)\,\mathcal{O}(p_3)}\propto g(d,\Delta_i)\,p_3^{\Delta_2+\Delta_3-\Delta_1-d}\,p_1^{2\Delta_1-d}\left(1 +O\left(\frac{p_1}{p_3}\right)\right)\hspace{2cm} \text{if}\ \ \Delta_1<\frac{d}{2}&,
	\end{align}
	with $f(d,\Delta_i)$ and $g(d,\Delta_i)$ depending only on the scaling and spacetime dimensions. Notice that this approach introduces the minimal set of independent 
	solutions.  The result above in \eqref{solfin} is in complete agreement with the direct computation performed by Davydychev \cite{Davydychev:1992xr} of the generalized master integrals, obtained by a Fourier transform of \eqref{corr} and the use of the Mellin-Barnes method. \\
	An alternative method consists in performing an explicit symmetrization of each of the four solutions and corresponding constants $c_j S_j$ $(j=1,2,3,4)$, obtained by permuting the $(p_i,\Delta_i)$ under the $\mathcal{S}_3$ permutation group. \\
	We remark that the method, in this case, introduces twenty-four functionally dependent contributions which, again, can be simplified by repeated use of \eqref{transfF41}. In this case, one discovers, after this simplification, that the symmetric solution so generated may manifest some unphysical singularities which disappear for a specific choice of the fundamental constants. A rather lengthy computation shows that the choice of such constants coincides with those presented in the solution \eqref{solfin}, originally given in \cite{Coriano:2013jba}, which involves the four basic solutions $S_j$ \eqref{F4def} and \eqref{solutions}.\\
	An alternative approach is based on the formalism of the 3K integrals developed in \cite{Bzowski:2013sza,Bzowski:2015yxv}, which for the 3-point function is automatically symmetric. In this case, the linear combination of the four solutions $S_i$ appearing in each 3K integral has been checked to be free of unphysical singularities in the region of convergence.  \\
	In the case of 4-point functions, the only method which appears manageable is the explicit symmetrization of the fundamental solutions accompanied by the requirement that the symmetric expression is free of unphysical singularities in the physical domain. We will be illustrating this point in the next chapter.
\chapter{Scalar 4-point function in momentum space}
\section{Introduction}
In this chapter, we are going to move to the analysis of 4-point functions in momentum space by investigating some scaling solutions of primary operators, showing that the hypergeometric character of the corresponding CWI's, already found in the case of 3-point functions, at least for such solutions, is preserved. Our analysis extends to  4-point functions  previous similar studies \cite{Coriano:2013jba,Bzowski:2013sza,Coriano:2018bsy,Coriano:2018bbe,Coriano:2018zdo,Bzowski:2015yxv,Bzowski:2018fql}, formulated for scalar correlators. \\
The solutions that we present, as we are going to elaborate, 
can be classified as being dual conformal (DC), as described in \cite{Drummond:2006rz,Drummond:2007aua,Drummond:2008vq}, {\em and} conformal (DCC) at the same time. They are constructed by requiring that they satisfy the first order differential conditions of dual conformal invariance, together with the second order ones coming from ordinary conformal symmetry. Both conditions are implemented and solved in momentum space.\\
The solutions that we identify can be written in two forms, either as generalized hypergeometrics, now functions of quartic ratios of momenta, or as integrals of 3 Bessel functions (3K integrals). It will be apparent, from our approach, that a central part in our analysis is played by the hypergeometric system of partial differential equations (PDE's) which emerge from CWI's once we select a specific ans\"atz for a given correlator in momentum space.

\subsection{Dual conformal ans\"atz (DCA) and conformal invariance in coordinate space}
We use specific (dual conformal) ans\"atze (DCA's) to reduce the system of CWI's to Appell's hypergeometric functions, by introducing specific factorization of the expression of the correlators in terms of a scaling factor and of a remaining scale-invariant function of some conformal ratios. The various DCA's allow us to build such exact solutions in momentum space, link 3- and 4- point functions, exemplifying well known previous results  \cite{Davydychev:1992xr, Usyukina:1992jd,Usyukina:1993ch,Broadhurst:1993ru,Eden:1998hh,Eden:2000mv} on ladder diagrams in perturbation theory, as mentioned above. These have provided the first examples of dual conformal symmetry in the planar limit for scalar ladders.\\
We solve the equations in two cases, for equal scalings $(\Delta_i=\Delta, i=1,...4)$ of the primary operators and for two separate scalings $(\Delta_1=\Delta_2=\Delta_x, \Delta_3=\Delta_4=\Delta_y)$. 
The choice of the ans\"atz in momentum space implies that the solutions that we are looking for are dual conformal, to begin with, and their Fourier transform to coordinate space is conformal as well. This last step is guaranteed if the ans\"atz satisfies the ordinary CWI's in coordinate space, which become second order PDE's in momentum space. \\
We show that for the solution of CWI's in momentum space that we derive, one can use the same formalism of the 3K integrals known for 3-point functions, though equivalent to their hypergeometric form.\\
By re-expressing the solutions generated by the ans\"atze as 3K integrals, the different ans\"atze are shown to determine a unique class of solutions, expressed just in terms of an overall constant and specific scaling dimensions. We will comment on the difference between such a result and those found in the computation of ladder diagrams in perturbation theories. Different dual conformal expressions - associated with a specific one, two-loop diagrams etc. - have, obviously, different analytic expressions.
\subsection{Approximate conformal solutions of primary operators and the Lauricella system}
Besides the search for exact solutions of the CWI's using the DCA in momentum space, in a second part of the chapter, we will focus our attention on some approximate solutions of the same CWI's (for primary operators) using a specific kinematic approximation. All our considerations apply to ordinary scattering amplitudes which are conformal in coordinate space, in particular to Feynman integrals of such type. We show that if we consider large $s$ and $t$ (Mandelstam) invariants in the correlators, with $-t/s$ fixed, suitable for a description of the same equations at a fixed angle, the CWI's simplify.\\
In this approximation, the equations will factorize the dependence on the external invariants $s, t$, from the remaining external mass invariants $p_i^2$. 
We show that the equations are fully compatible with asymptotic solutions which are logarithmic in  $-t/s$ in the Minkowski region, while the external mass invariants parameterize Lauricella functions, i.e. hypergeometric functions of 3 independent ratios. We show how such solutions and systems of equations can be equivalently described by the 3K integrals' natural generalization to 4K. We conjecture that this pattern may extend to even higher point functions when the external mass invariants are separated from the remaining invariants scalar products of 2 different momenta. It seems clear that such factorized ans\"atze capture the essential behaviour of these correlators in some particular kinematical limits, as it has been long known in the case of the Regge limit even at next-to-leading order in the gauge coupling, using conformal methods of t-channel unitarity \cite{Coriano:1995fj,Coriano:1996rj,Coriano:1994wk,Coriano:1995hx}. In all these cases the CFT constraints provide relatively simple predictions than the explicit  NLO computations performed in QCD, with new partial waves appearing at NLO in the conformal reconstruction of the evolution (BFKL) kernel at the same order. 
\subsection{Notational remarks} 
We will be denoting with $x_i$ the coordinate dependence of a correlator. We will reserve the symbols $y_i$ to denote the dual coordinates in momentum space of the same correlator, while the (incoming) four-momenta will be denoted as $p_i$. The variables $x$ and $y$ (without any lower positional index $i$) will be used to denote ratios in momentum space expressed in terms of the invariants built out of the momenta $p_i$. Instead, the two invariant ratios in coordinate space, defined below, will be denoted as $u(x_i)$ and $v(x_i)$. The same invariant ratios in the dual conformal coordinates will be denoted as $u(y_i)$ and $v(y_i)$. The generators of the dual conformal symmetry will carry a $y_i$ dependence, such as $D(y_i), K^\kappa(y_i)$ for dilatation and special conformal transformations. Their versions in momentum space will be denoted as $D(p_i), K^\kappa(p_i)$, where in all these cases $y_i\equiv(y_1\ldots y_4)$, $x_i\equiv(x_1,\ldots x_4)$ and $p_i\equiv (p_1,\ldots p_4)$. As will be hopefully clear in the following, $K^\kappa(y)$, the special conformal generator in momentum space but in the dual conformal coordinates is a first order differential operator while $K^\kappa(p_i)$ is second order. 
\section{Three- and four-point functions from conformal invariance for correlators of primaries}
In order to clarify the new features of 4-point functions respect to correlators of lower points, we start our discussion by reviewing the case of such correlators in coordinate space. For 3-point functions we summarize the approach used in the analysis of primary scalar 3-point functions directly in momentum space, discussed in previous studies \cite{Coriano:2013jba}.
We consider the simple case of a correlator of $n$ primary scalar fields $O_i(x_i)$, each of scaling dimension $\Delta_i$
\begin{equation}
	\label{defop}
	\Phi(x_1,x_2,\ldots,x_n)=\braket{O_1(x_1)O_2(x_2)\ldots O_n(x_n)}.
\end{equation}
Among these, 3- and 4-point functions (beside 2-point functions) in any CFT are significantly constrained in their general structure. Scalar 3-point functions of primary operators $\phi_i$ of scaling dimensions 
$\Delta_i$ $(i=1,2,3)$ are constrained to be of the form
\begin{equation}
	\label{corr}
	\langle \phi_1(x_1)\phi_2(x_2)\phi_3(x_3)\rangle =\frac{C_{123}}{ x_{12}^{\Delta_t - 2 \Delta_3}  x_{23}^{\Delta_t - 2 \Delta_1}x_{13}^{\Delta_t - 2 \Delta_2} },\qquad \Delta_t\equiv \sum_{i=1}^3 \Delta_i.
\end{equation}
$C_{123}$ is a constant which specifies the CFT (the "CFT data"). For 4-point functions the constraints determine the structure of the correlator in a less effective way. In that case one identifies the two cross ratios 
\begin{equation}
	\label{uv}
	u(x_i)=\frac{x_{12}^2 x_{34}^2}{x_{13}^2 x_{24}^2} \qquad v(x_i)=\frac{x_{23}^2 x_{41}^2}{x_{13}^2 x_{24}^2}
\end{equation}
and the general solution can be written in the form 
\begin{equation}
	\label{general}
	\langle \phi_1(x_1)\phi_2(x_2)\phi_3(x_3)\phi_4(x_4)\rangle= h(u(x_i),v(x_i))\, \frac{1}{\left(x_{12}^2\right)^\frac{\Delta_1 + \Delta_2}{2}\left(x_{3 4}^2\right)^\frac{\Delta_3 + \Delta_4}{2}}
\end{equation}
where $h(u(x_i),v(x_i))$ remains unspecified. We are going to show that the equations may constrain $h(u(x_i),v(x_i))$ to take a specific form in momentum space, if we look for a specific ans\"atz.

For scalar correlators the special CWI's are given by first order differerential equations 
\begin{equation}
	\label{SCWI0}
	K^\kappa(x_i) \Phi(x_1,x_2,\ldots,x_n) =0
\end{equation}
with 
\begin{equation}
	\label{transf1}
	K^\kappa(x_i) \equiv \sum_{j=1}^{n} \left(2 \Delta_j x_j^\kappa- x_j^2\frac{\partial}{\partial x_j^\kappa}+ 2 x_j^\kappa x_j^\alpha \frac{\partial}
	{\partial x_j^\alpha} \right)
\end{equation}
being the corresponding generator in coordinate space. The same operator, deprived of the scaling coefficients, will be denoted as $K_0^\kappa(x_i)$, i.e. 
\begin{equation}
	K_0^\kappa(x_i) \equiv \sum_{j=1}^{n} \left(2 x_j^\kappa x_j^\alpha \frac{\partial}
	{\partial x_j^\alpha} - x_j^2\frac{\partial}{\partial x_j^\kappa}\right).
\end{equation}
Conformal covariance and conformal invariance in coordinate space simply refer to the validity of \eqref{SCWI0} and of
\begin{equation}
	K_0^\kappa(x_i) \Phi(x_1,x_2,\ldots,x_n) =0
\end{equation}
respectively. Denoting with 
\begin{equation}
	\Phi(p_1,\ldots p_{n-1},\bar{p}_n)=\langle O_1(p_1)\ldots O_n(\bar{p}_n)\rangle 
\end{equation}
and 
\begin{equation}
	K^\kappa(p_i)\equiv\sum_{j=1}^{n-1}\left(2(\Delta_j- d)\frac{\partial}{\partial p_j^\kappa}+p_j^\kappa \frac{\partial^2}{\partial p_j^\alpha\partial p_j^\alpha} -2 p_j^\alpha\frac{\partial^2}{\partial p_j^\kappa \partial p_j^\alpha}\right)
\end{equation}
the Fourier transform of \eqref{defop} and of \eqref{transf1} respectively, the form of the second order differential equations is given by
\begin{equation}
	K^\kappa(p_i)\Phi(p_1,\ldots p_{n-1},\bar{p}_n)=0,\label{SCWI}
\end{equation}
where we have chosen $\bar{p}_n^\mu=-\sum_{i=1}^{n-1} p_i^\mu$ the n-th momentum, to be the linearly dependent one.
The action of the differential operators is realized on the shell of momentum conservation, where the $n$-th momentum, conventionally, will be taken as dependent from the previous ones. Coming to the dilatation WI's, in our conventions, a scale-covariant 
function in coordinate space 
\begin{equation}
	\phi(\lambda x_i)=\lambda^{-\Delta}\phi(x_i) 
\end{equation}
gives in momentum space 
\begin{equation}
	\phi(\lambda p_1\ldots \lambda \bar{p}_n)=\lambda^{-\Delta'}\phi(p_1\ldots \bar{p}_n),
\end{equation}
with 
\begin{equation}
	\Delta'\equiv \left(-\sum_{i=1}^n \Delta_i +(n-1) d\right)=-\Delta_t +(n-1) d.
\end{equation}
The corresponding equations are
\begin{equation}
	\label{scale12}
	D(x_i)
	\Phi(x_1,\ldots x_n)=0
\end{equation}
with 
\begin{equation}
	\label{scale11}
	D(x_i)\equiv\sum_{i=1}^n\left( x_i^\alpha \frac{\partial}{\partial x_i^\alpha} +\Delta_i\right)
\end{equation}
for scale covariant correlators, in the case of scale invariance turn into 
\begin{equation}
	D_0(x_i)
	\Phi(x_1,\ldots x_n)=0
\end{equation}
with $D_0(x_i)$ given by
\begin{equation}
	D_0(x_i)\equiv\sum_{i=1}^n\left( x_i^\alpha \frac{\partial}{\partial x_i^\alpha}\right). 
\end{equation}
In momentum space, the condition of scale covariance and invariance are respectively given by 
\begin{equation}
	D(p_i) \Phi(p_1\ldots \bar{p}_n)=0
\end{equation}
with
\begin{equation}
	D(p_i)\equiv\sum_{i=1}^{n-1}  p_i^\alpha \frac{\partial}{\partial p_i^\alpha} + \Delta'
\end{equation}
and 
\begin{equation}
	D_0(p_i) \Phi(p_1\ldots \bar{p}_n)=0
\end{equation}
with
\begin{equation}
	D_0(p_i)\equiv\sum_{i=1}^{n-1} p_i^\alpha \frac{\partial}{\partial p_i^\alpha} .
\end{equation}

Once we move to dual conformal coordinates in momentum space, denoted as $y_i$ below, it is important to keep clearly in mind the separation between actions of $K$ or $D$, such as those induced by their expressions in $x_i$ coordinates, from their second order in the $p_i$ variable. It is also common to refer to dual conformal symmetry to just an independent $SO(2,4)$ symmetry respect to the ordinary conformal symmetry of coordinate space (or of its Fourier image). 

\section{CWI's for scalar four-point functions}
In this section we discuss an extension of the method summarized in section \secref{Sol3Point} to 4-point functions. We follow a similar strategy, by choosing a specific set of variables to characterize the ans\"atz for the solution of the corresponding PDE's. In the case of 3-point functions it is quite clear that the special CWI's are two equations and one can explicitly show that they remain independent after we introduce the ans\"atz \eqref{ans}. In the class of solutions that we look for, with a specific ans\"atz, two of the three constraining equations are independent, while a third equation is automatically satisfied.\\
In the case of the four point function the correlator depends on six invariants that we will normalize as
$p_i=|\sqrt{p_{i}\,{\hspace{-0.09cm}}^2}|$, $i=1,\dots,4$, representing the magnitudes of the momenta, and $s=|\sqrt{(p_1+p_2)^2}|$, $t=|\sqrt{(p_2+p_3)^2}|$ the two Mandelstam invariants, redefined by a square root, and we write
\begin{equation}
	\braket{\mathcal{O}(p_1)\,\mathcal{O}(p_2)\,\mathcal{O}(p_3)\,\mathcal{O}(\bar{p}_4)}=\Phi(p_1,p_2,p_3,p_4,s,t).\label{invariant}
\end{equation}
This correlation function, to be conformally invariant, has to verify the dilatation Ward Identity
\begin{equation}
	\left[\sum_{i=1}^4\D_i-3d-\sum_{i = 1}^3p_i^{\mu}\frac{\partial}{\partial p_i^\mu}\right]\braket{\mathcal{O}(p_1)\,\mathcal{O}(p_2)\,\mathcal{O}(p_3)\,\mathcal{O}(\bar{p}_4)}=0,
\end{equation}
and the special conformal Ward Identities
\begin{equation}
	\sum_{i=1}^3\left[2(\D_i-d)\frac{\partial}{\partial p_{i\,\k}}-2p_i^\alpha\frac{\partial^2}{\partial p_i^\alpha\partial p_i^\kappa}+p_i^\kappa\frac{\partial^2}{\partial p_i^\alpha\partial p_{i\,\alpha}}\right]\braket{\mathcal{O}(p_1)\,\mathcal{O}(p_2)\,\mathcal{O}(p_3)\,\mathcal{O}(\bar{p}_4)}=0.
\end{equation}
One can split these equations in terms of the invariants of the four-point function written in \eqref{invariant}, by using the chain rules
\begin{align}
	\frac{\partial}{\partial p_{1\,\mu}}&=\frac{p_1^\mu}{p_1}\frac{\partial}{\partial p_1}-\frac{\bar{p}_4^\mu}{p_4}\frac{\partial}{\partial p_4}+\frac{p_1^\mu+p_2^\mu}{s}\frac{\partial}{\partial s},\\
	\frac{\partial}{\partial p_{2\,\mu}}&=\frac{p_2^\mu}{p_2}\frac{\partial}{\partial p_2}-\frac{\bar{p}_4^\mu}{p_4}\frac{\partial}{\partial p_4}+\frac{p_1^\mu+p_2^\mu}{s}\frac{\partial}{\partial s}+\frac{p_2^\mu+p_3^\mu}{t}\frac{\partial}{\partial t},\\
	\frac{\partial}{\partial p_{3\,\mu}}&=\frac{p_3^\mu}{p_3}\frac{\partial}{\partial p_3}-\frac{\bar{p}_4^\mu}{p_4}\frac{\partial}{\partial p_4}+\frac{p_2^\mu+p_3^\mu}{t}\frac{\partial}{\partial t},
\end{align}
where $\bar{p}_4^\mu=-p_1^\mu-p_2^\mu-p_3^\mu$. From this prescription the dilatation WI becomes
\begin{align}
	\bigg[(\D_t-3d)-\sum_{i=1}^4p_i\frac{\partial}{\partial p_i}-s\frac{\partial}{\partial s}-t\frac{\partial}{\partial t}\bigg]\Phi(p_1,p_2,p_3,p_4,s,t)=0,\label{Dilatation4}
\end{align}
with $\Delta_t=\sum_i \Delta_i$ is the total scaling,
and the special CWI's can be written as
\begin{equation}
	\sum_{i=1}^3\ p_i^\kappa\, C_i=0\label{primary},
\end{equation}
where the coefficients $C_i$ are differential equations of the second order with respect to the six invariants previously defined. Being $p_1^\k,\ p_2^\k,\ p_3^\k$, in \eqref{primary} independent, 
we derive three scalar second order differential equations for each of the three $C_i$, which must vanish independently.\\
At this stage the procedure to simplify the corresponding equations is similar to the one described in \cite{Coriano:2018bsy,Coriano:2018bbe}. A lengthy computation allows to rewrite the equations in the form
\begin{align}
	C_1&=\bigg\{\frac{\partial^2}{\partial p_1^2}+\frac{(d-2\D_1+1)}{p_1}\frac{\partial}{\partial p_1}-\frac{\partial^2}{\partial p_4^2}-\frac{(d-2\D_4+1)}{p_4}\frac{\partial}{\partial p_4}+\frac{(\D_3+\D_4-\D_1-\D_2)}{s}\frac{\partial}{\partial s}\notag\\[1.5ex]
	&\qquad+\frac{1}{s}\frac{\partial}{\partial s}\left(p_1\frac{\partial}{\partial p_1}+p_2\frac{\partial}{\partial p_2}-p_3\frac{\partial}{\partial p_3}-p_4\frac{\partial}{\partial p_4}\right)+\frac{(p_2^2-p_3^2)}{st}\frac{\partial^2}{\partial s\partial t}\bigg\}\,\Phi(p_1,p_2,p_3,p_4,s,t)=0\label{C1}
\end{align}
for $C_1$ and
\begin{align}
	C_2&=\bigg\{\frac{\partial^2}{\partial p_2^2}+\frac{(d-2\D_2+1)}{p_2}\frac{\partial}{\partial p_2}-\frac{\partial^2}{\partial p_4^2}-\frac{(d-2\D_4+1)}{p_4}\frac{\partial}{\partial p_4}+\frac{(\D_3+\D_4-\D_1-\D_2)}{s}\frac{\partial}{\partial s}\notag\\
	&\qquad+\frac{1}{s}\frac{\partial}{\partial s}\left(p_1\frac{\partial}{\partial p_1}+p_2\frac{\partial}{\partial p_2}-p_3\frac{\partial}{\partial p_3}-p_4\frac{\partial}{\partial p_4}\right)+\frac{(\D_1+\D_4-\D_2-\D_3)}{t}\frac{\partial}{\partial t}\notag\\
	&\qquad+\frac{1}{t}\frac{\partial}{\partial t}\left(p_2\frac{\partial}{\partial p_2}+p_3\frac{\partial}{\partial p_3}-p_1\frac{\partial}{\partial p_1}-p_4\frac{\partial}{\partial p_4}\right)+\frac{(p_2^2-p_4^2)}{st}\frac{\partial^2}{\partial s\partial t}\bigg\}\,\Phi(p_1,p_2,p_3,p_4,s,t)=0
	\label{C2}
\end{align}

\begin{align}
	C_3&=\bigg\{\frac{\partial^2}{\partial p_3^2}+\frac{(d-2\D_3+1)}{p_3}\frac{\partial}{\partial p_3}-\frac{\partial^2}{\partial p_4^2}-\frac{(d-2\D_4+1)}{p_4}\frac{\partial}{\partial p_4}+\frac{(\D_1+\D_4-\D_2-\D_3)}{t}\frac{\partial}{\partial t}\notag\\[1.5ex]
	&\qquad+\frac{1}{t}\frac{\partial}{\partial t}\left(p_2\frac{\partial}{\partial p_2}+p_3\frac{\partial}{\partial p_3}-p_1\frac{\partial}{\partial p_1}-p_4\frac{\partial}{\partial p_4}\right)+\frac{(p_2^2-p_1^2)}{st}\frac{\partial^2}{\partial s\partial t}\bigg\}\,\Phi(p_1,p_2,p_3,p_4,s,t)=0\label{C3}
\end{align}
for $C_2$ and $C_3$, in agreement with \cite{Arkani-Hamed:2018kmz}. One of the two equations that we will be solving will be  
$C_{13}\equiv C_1- C_3=0$ and it is convenient to present it explicitly

\begin{align}
	C_{13}&=\bigg\{\frac{\partial^2}{\partial p_1^2}+\frac{(d-2\D_1+1)}{p_1}\frac{\partial}{\partial p_1}-\frac{\partial^2}{\partial p_3^2}-\frac{(d-2\D_3+1)}{p_3}\frac{\partial}{\partial p_3}+\frac{(\D_3+\D_4-\D_1-\D_2)}{s}\frac{\partial}{\partial s}\notag\\
	&\qquad+\frac{1}{s}\frac{\partial}{\partial s}\left(p_1\frac{\partial}{\partial p_1}+p_2\frac{\partial}{\partial p_2}-p_3\frac{\partial}{\partial p_3}-p_4\frac{\partial}{\partial p_4}\right)+\frac{(\D_2+\D_3-\D_1-\D_4)}{t}\frac{\partial}{\partial t}\notag\\
	&\qquad+\frac{1}{t}\frac{\partial}{\partial t}\left(p_1\frac{\partial}{\partial p_1}+p_4\frac{\partial}{\partial p_4}-p_2\frac{\partial}{\partial p_2}-p_3\frac{\partial}{\partial p_3}\right)+\frac{(p_1^2-p_3^2)}{st}\frac{\partial^2}{\partial s\partial t}\bigg\}\,\Phi(p_1,p_2,p_3,p_4,s,t)=0.\label{Eq2}
\end{align}

\section{Dual Conformal/Conformal (DCC) examples}
\label{dccsection}
Before moving to a discussion of the DCA's and the character of the solutions that we are going to identify, we turn to some specific examples of perturbative 4-point functions which are both conformal and dual conformal at the same time (DCC). We recall that a dual conformal integral \cite{Drummond:2006rz,Drummond:2007aua,Drummond:2008vq} is a Feynman integral which, once rewritten in terms of some dual coordinates, under the action of $K^\kappa$, is modified by factors which depend only on the coordinates of the external points. The reformulation of the ordinary momentum integral in terms of such dual coordinates can be immediately worked out by drawing the associated dual diagram. 
\begin{figure}[t]
	\centering
	\raisebox{-0.5\height}{\includegraphics[scale=0.3]{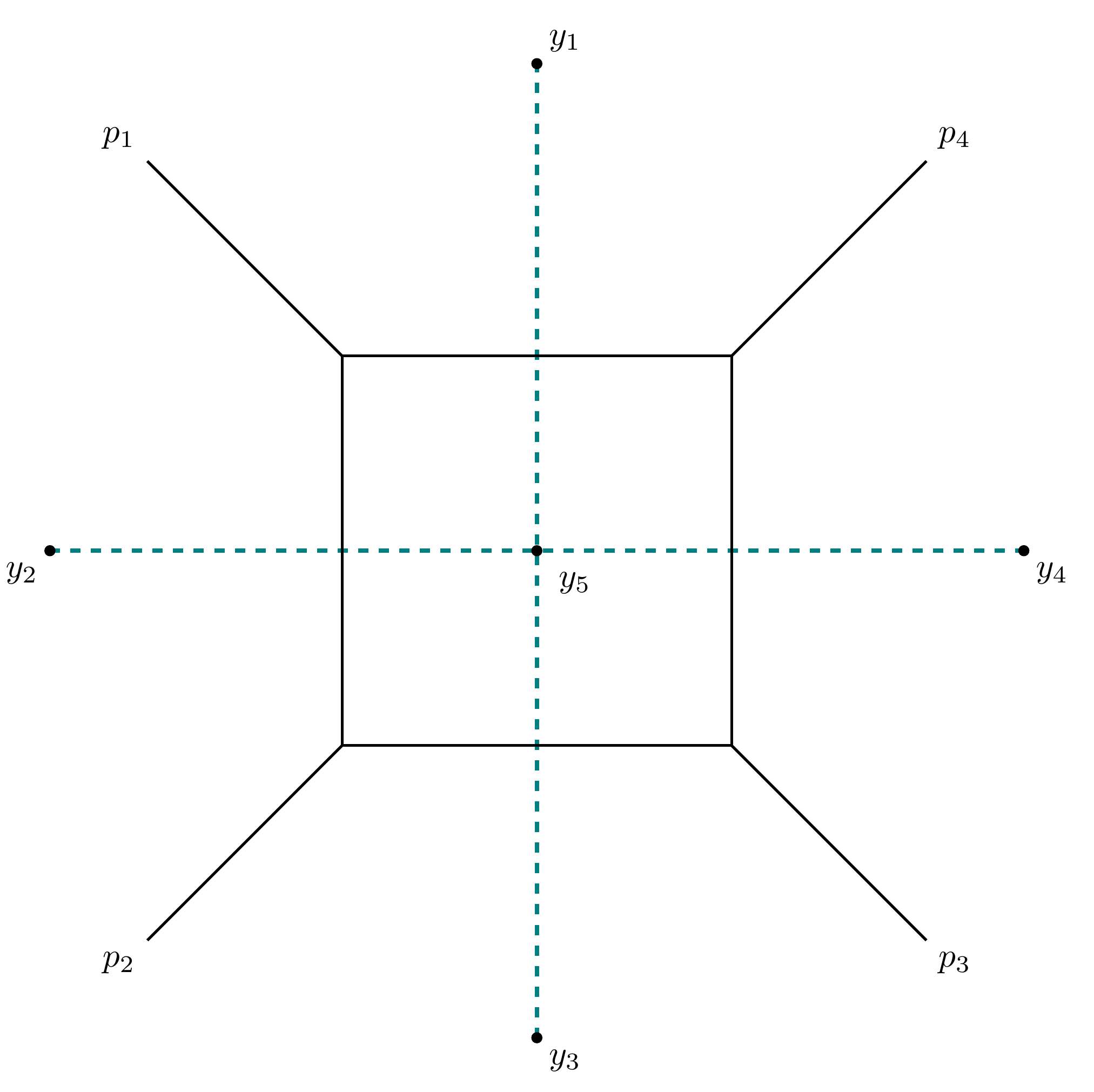}} \hspace{2cm}
	\raisebox{-0.5\height}{\includegraphics[scale=0.6]{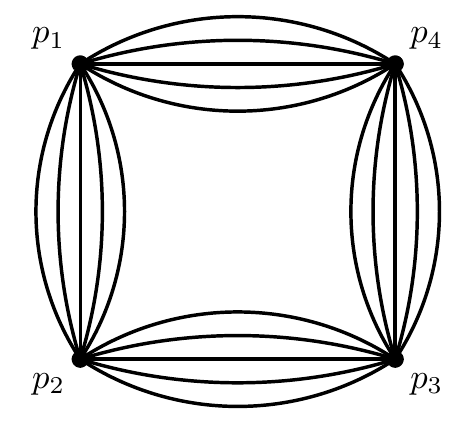}} 
	\caption{The box with its dual (left) and its higher scaling version (right). While the first is conformal for $d\ne2$ in ordinary conformal coordinates and for $d=4$ in dual coordinates, the right one is not conformal in coordinate and dual coordinate space at the same time. \label{Fig1}}
\end{figure}
We start from the ordinary box diagram (see Fig. \ref{Fig1})
\begin{equation}
	\Phi_{Box}(p_1,p_2,p_3,p_4)=\int \frac{d^d k}{k^2 (k + p_1)^2 (k + p_1 + p_2)^2 ( k + p_1 + p_2 + p_3)^2}
\end{equation}
and apply the redefinition in terms of momentum variables $y_i$
\begin{equation}
	k=y_{51}, \qquad p_1=y_{12}, \qquad p_2=y_{23},\qquad p_3=y_{34}
\end{equation}
with $y_{ij}=y_i-y_j$, thereby rewriting the integral in the form 
\begin{equation}
	\label{box1}
	\Phi_{Box}(y_1,y_2,y_3,y_4)=\int \frac{d^d y_5}{y_{15}^2 y_{25}^2 y_{35}^2 y_{45}^2}
\end{equation}
The action of $K^\kappa$ is realized in the form $\mathcal{I} \cdot \mathcal{T}\cdot \mathcal{I}$ (inversion, translation and inversion transformations) rather than as a differential action (by $K_0^\kappa$). We recall that under an inversion $(\mathcal{I})$ 
\begin{equation}
	\mathcal{I} (d^d y_5)=d^d y_5 {(y_5^2)}^{-d}  \qquad \mathcal{I}(y^2_{ij})=\frac{y_{ij}^2}{y_i^2y_j^2}
\end{equation}
and in order to have an expression which is invariant under special conformal transformation, it is necessary to include a pre-factor in $\Phi_{Box}$, in the form  
\begin{equation}
	s^2 t^2 \Phi_{Box}(p_1,p_2,p_3,p_4)= y_{13}^2 y_{24}^2  \Phi_{Box}(y_1,y_2,y_3,y_4)
\end{equation}
then its is easy to check that under the action of $\mathcal{I}$ the integrand 
\begin{equation}
	\mathcal{I} \left(\frac{d^d y_5 y_{13}^2 y_{2 4}^2}{y_{15}^2 y_{25}^2 y_{35}^2 y_{45}^2}\right)= 
	\left(\frac{d^d y_5   (y_5^2)^{4-d}  y_{13}^2 y_{2 4}^2}{y_{15}^2 y_{25}^2 y_{35}^2 y_{45}^2}\right)
\end{equation}
becomes invariant under the action of the special conformal transformation if $d=4$. Obviously, the invariance under the complete action $\mathcal{I T I}$ is ensured. It is easily checked that the integrand is also scale invariant. It is then clear that the expression of the box diagram can only be of the form 
\begin{equation}
	\label{ans1}
	\Phi_{Box}=\frac{1}{y_{13}^2 y_{24}^2}F\big(u(y_i),v(y_i)\big)
\end{equation}
with $u$ and $v$ given by 
\begin{equation}
	\label{uv2}
	u(y_i)=\frac{y_{12}^2 y_{34}^2}{y_{13}^2 y_{24}^2} \qquad v(y_i)=\frac{y_{23}^2 y_{41}^2}{y_{13}^2 y_{24}^2}
\end{equation}
For future purposes it will be convenient to define
\begin{equation}
	\label{quartic}
	x=\frac{p_1^2\,p_3^2}{s^2\,t^2},\qquad y=\frac{p_2^2\,p_4^2}{s^2\,t^2}
\end{equation}
being the two invariant ratios $u(y_i),v(y_i)$, now expressed directly in terms of the original momentum invariants. 
Notice that, by construction $u,v$ satisfy the first order equation in the $y$ variables
\begin{equation}
	\label{firstp}
	\begin{split}
		K_0^\kappa(y)\, u(y_i) &=\sum_{j=1}^{4} \left(- y_j^2\frac{\partial}{\partial y_j^\kappa}+ 2 y_j^\kappa y_j^\alpha \frac{\partial}
		{\partial y_j^\alpha} \right)  u(y_i) =0\\
		K_0^\kappa(y)\, v(y_i) &=\sum_{j=1}^{4} \left(- y_j^2\frac{\partial}{\partial y_j^\kappa}+ 2 y_j^\kappa y_j^\alpha \frac{\partial}
		{\partial y_j^\alpha} \right)  v(y_i) =0
	\end{split}
\end{equation}
while the action of $K_0^\kappa (p)$ on $x$ and $y$ will be nonzero.

Notice that while the two forms of the $K_0^\kappa$ operator $K_0^\kappa(x_i)$ (coordinate) and $K_0^\kappa(p_i)$ (momenta) are one the Fourier transform of the other, $x$ and $y$ in \eqref{quartic} are not the Fourier images of $u(x_i)$ and $v(x_i)$. \\
The box diagram is an example of a diagram which is dual conformal and conformal in $d=4$.  To show this point reconsider this diagram in coordinate space
\begin{equation}
	\Phi_{Box}(x_i)=\frac{1}{x_{12}^2 x_{23}^2 x_{34}^2 x_{41}^2},
\end{equation}
that we can rewrite in the form 
\begin{align}
	\label{dccc}
	\Phi_{Box}(x_i)&=\frac{1}{(x_{12}^2 x_{34}^2)^2}\left(\frac{x_{12}^2 x_{34}^2}{x_{23}^2 x_{41}^2}\right)=\frac{1}{(x_{12}^2 x_{34}^2)^2}\left( \frac{u(x_i)}{v(x_i)}\right)
\end{align}
which is the conformally covariant correlator generated by 4 scalar primary fields $(\phi_i)$ in $d=4$ with $\Delta_i=2$. Denoting with $\chi$ an ordinary scalar field of scaling dimension 1, and setting $\phi_i=\chi^2$
we would have 
\begin{align}
	\Phi_{Box}(x_i)&\equiv \langle \phi_1(x_1)\phi_2(x_2)\phi_3(x_3)\phi_4(x_4)\rangle = \langle \chi^2(x_1)\chi^2(x_2)\chi^2(x_3)\chi^2(x_4)\rangle \notag\\
	&= \frac{1}{(x_{12}^{2})^\Delta (x_{34}^2)^\Delta}h\big(u(x_i),v(x_i)\big)
\end{align}
with $\Delta=2$ and $h\big(u(x_i),v(x_i)\big)=u(x_i)/v(x_i)$. It is then obvious that the scalar box diagram satisfies the four constraints 
\begin{eqnarray}
	& K^\kappa(x_i) \Phi(x_i)=0\qquad  D(x_i)\Phi_{Box}(x_i)=0\\
	& K^\kappa(y_i) \Phi(y_i)=0\qquad  D(y_i)\Phi_{Box}(y_i)=0
\end{eqnarray}
in coordinates $x_i$ and dual (momentum) coordinates $y_i$ respectively as a system of first order PDE's. The system of equations can be all reported to momentum space in the form
\begin{align}
	& K^\kappa(p_i) \Phi_{Box}(p_i)=0\qquad  D(p_i)\Phi_{Box}(p_i)=0\\
	& K^\kappa(y_i) \Phi_{box}(y_i)=0\qquad  D(y_i)\Phi_{Box}(y_i)=0
\end{align}
as a system of second and first order constraints. 
We are going to discuss the solution of such constraints in detail, showing its unique hypergeometric structure.
\subsection{DCC solutions and the Feynman expansion: melonic contributions}
The case discussed above is a special one. In general, in fact, in perturbation theory,  it is possible to find solutions which are dual conformal or conformal, but not both, since some of the basic requirements are violated. \\
Consider the case of the perturbative melonic diagram shown in \figref{Fig1} where we have introduced a composite operator
\begin{equation}
	\phi(x_i)=\chi^{n+m}(x_i)  \qquad  n,m\in  \mathbb{N}
\end{equation}
in $d$ dimensions with $n+m=N\in\mathbb{N}$ fixed, which in free field theory generates the correlator 
\begin{equation}
	\langle \phi(x_1)\phi(x_2)\phi(x_3)\phi(x_4)\rangle = \frac{1}{x_{12}^{2 a(n)} x_{23}^{2 b(m)}x_{34}^{2 a(n) }x_{41}^{2 b(m)}}
\end{equation}
with
\begin{equation}
	a(n)=n \,\D,\quad b(m)=m \,\D,  \qquad \D= \frac{d-2}{2}
\end{equation}
which is conformally covariant since it can be re-expressed in the form 
\begin{equation}
	\langle \phi(x_1)\phi(x_2)\phi(x_3)\phi(x_4)\rangle = \frac{1}{\left(x_{12}^2 x_{34}^{2} \right)^{a(n) + b(n)}}\left( 
	\frac{u(x_i)}{v(x_i)}\right)^{b(n)}
\end{equation}
with the scaling dimension of $\phi$ given by $\left[\phi\right]= a(n) + b(m)$. In momentum space the corresponding integral is given by 
\begin{equation}
	\int \frac{d^d k}{(k^2)^{\nu_1} ((k+p_1)^2)^{\nu_2} ((k+p_1+p_2)^2)^{\nu_3} ((k+ p_1 + p_2+p_3)^2)^{\nu_4}}
\end{equation}
with $\nu_1=\nu_3=d/2-a(n)$ and $\nu_2=\nu_4=d/2-b(m)$. Mapping this expression to dual coordinate, invariance of the integrand under special conformal transformations requires that 
\begin{equation}
	m+ n=\frac{d}{d-2}
\end{equation}
which clearly shows that only $d=4$ allows to satisfy the dual conformal {\em and} conformal conditions, since $n+m$ has to be an integer. This brings us back to the ordinary box diagram. 

\subsection{DC symmetry and ladders}
We can slightly generalize the discussion presented above. It is convenient to introduce a more general notation, which can be used for the single, double etc. box diagrams, in order to set a distinction between correlators which are either dual conformal or conformal, or both. \\
The conformal behaviour of the box diagram in coordinate space $x_i$, for generic $d\ne2$ dimensions 
can be explicitly rewritten in the form
\begin{equation}
	\Phi_{Box}(x_i)=\frac{1}{(x^2_{13})^{d-2}(x^2_{24})^{d-2}}\,\phi^{(1)}\left(u(x_i),v(x_i)\right),\quad d\ne2
\end{equation}
where $\phi^{(1)}\left(u(x_i),v(x_i)\right)$ is the undetermined function of the conformal ratios in coordinate space. 
$\phi^{(1)}$ can be easily identified from \eqref{dccc} in $d=4$ in perturbation theory. Its expression in dual (momentum space) coordinates can be rewritten as 
\begin{equation}
	\Phi_{Box}(y_i)=\frac{1}{y^2_{13}\,y^2_{24}}\,\tilde{\phi}^{(1)}\left(u(y_i),v(y_i)\right),
\end{equation}
only in $d=4$. As elaborated above,  the box diagram can be both conformal and dual conformal invariant only in $d=4$. \\
Moving to the two-loop case, we consider the four-point ladder (planar) diagram (see Fig. \ref{dbbb}) and using the special conformal transformations, its expression takes the form
\begin{figure}[t]
	\centering
	\raisebox{-0.5\height}{\includegraphics[scale=0.4]{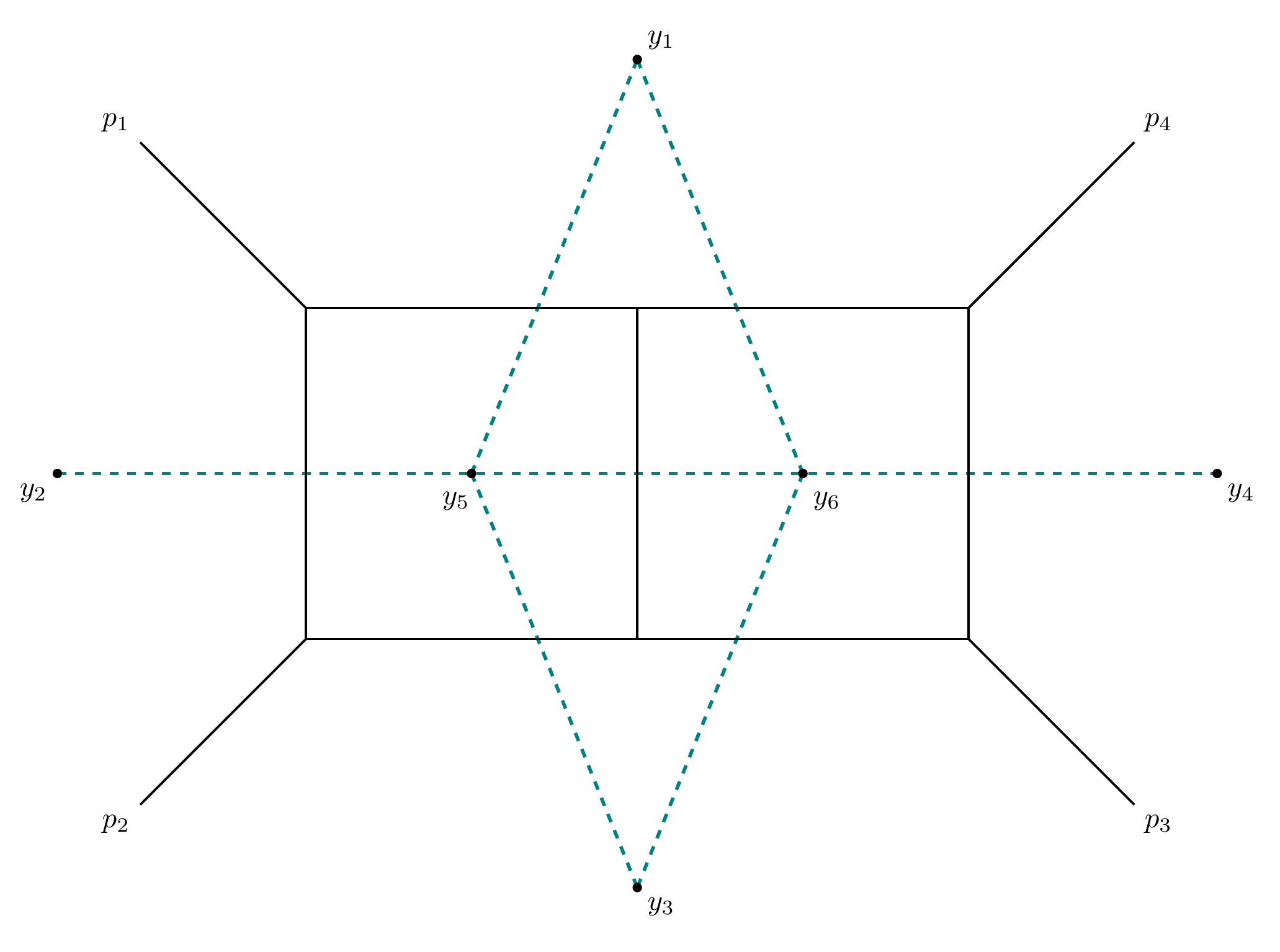}} 
	\caption{The two loop box digram with its dual. This diagram is not conformal in coordinate and dual coordinate space at the same time. }\label{dbbb}
\end{figure}
\begin{equation}
	\Phi_{2-Box}(x_i)=\frac{1}{(x^2_{13})^{4}(x^2_{24})^{4}}\,\phi^{(2)}\left(u(x_i),v(x_i)\right),
\end{equation}
valid for $d=6$, where also in this case $\phi^{(2)}\left(u(x_i),v(x_i)\right)$ is another function of the conformal ratios in coordinate space, different from the one obtained in the one-loop case. Moving to momentum space and then to dual coordinates, we find the dual conformal expression of the double box in this space as
\begin{equation}
	\Phi_{2-Box}(x_i)=\frac{1}{(y^2_{13})^{2}(y^2_{24})}\,\tilde{\phi}^{(2)}\left(u(y_i),v(y_i)\right),
\end{equation}
which holds for $d=4$.
It is obvious that the double box diagram can't be both conformal and dual conformal at the same time and does not provide a perturbative realization of the solution previously found using the CWI's.\\ 
Using the same argument one can prove that the 4-point n-loop ladder diagram in coordinate space is conformal covariant   only in $d=6$, taking the form
\begin{equation}
	\Phi_{n-Box}(x_i)=\frac{1}{(x^2_{13})^{4}(x^2_{24})^{4}}\,\phi^{(n)}\left(u(x_i),v(x_i)\right),
\end{equation}
valid for $n\ge 2$, where $\phi^{(n)}$ is a function of the conformal ratios. On the other hand, the same diagram in momentum space is dual conformal covariant  
\begin{equation}
	\Phi_{n-Box}(x_i)=\frac{1}{(y^2_{13})^{n}\,y^2_{24}}\,\tilde{\phi}^{(n)}\left(u(y_i),v(y_i)\right), 
\end{equation}
for $n\ge 2$ and only for $d=4$. This shows that the class of solutions that we have identified are only realized at one-loop level.

\subsection{The triangle diagram}
The triangle diagram, on the other hand, is truly special, if we follow the same reasonings as above. Given its general expression
\begin{equation}
	\label{j3}
	J(\nu_1,\nu_2,\nu_3) = \int \frac{d^d l}{(2 \pi)^d} \frac{1}{(l^2)^{\nu_3} ((l+p_1)^2)^{\nu_2} ((l-p_2)^2)^{\nu_1}}\, ,
\end{equation}
with generic indices for the Feynman propagators $(\nu_1,\nu_2,\nu_3)$, it is easy to verify that the condition of dual 
conformal invariance 
\begin{equation}
	\label{dci}
	d=\nu_1+\nu_2 +\nu_3
\end{equation}
allows to satisfy the DCC constraints in all dimensions. Such solutions are not obtained, in general, from free-field theories. We elaborate briefly on these points. A related discussion can be also found in \cite{Bzowski:2015pba}.\\
In fact, Eq. \eqref{j3} is the Fourier transform of a correlator of the form \eqref{corr}, for appropriate primary fields of scaling dimensions $\Delta_i$. Given some specific $\nu_i$, we can reverse-engineer three scalar primary fields of scalings $\Delta_i$ by the relations
\begin{equation}
	\Delta_1 = d - \nu_2 - \nu_3 \,, \qquad
	\Delta_2 = d - \nu_1 - \nu_3 \,, \qquad
	\Delta_3 = d - \nu_1 - \nu_2, \,
\end{equation}
in such a way that \eqref{corr} is respected. Equivalently, 
\begin{equation}
	\nu_1= \frac{1}{2} (d + \Delta_1 -\Delta_2-\Delta_3) \qquad
	\nu_2 = \frac{1}{2}(d - \Delta_1 + \Delta_2 -\Delta_3)\,, \qquad
	\nu_3 = \frac{1}{2}(d - \Delta_1 - \Delta_2 +\Delta_3).
\end{equation}

Using these relations, any conformal correlator of some scalar primaries of scaling $\Delta_i$'s, is 
bound to be of the form \eqref{corr}. The $\Delta_i's$  are trivially identified by the transform
\begin{align}
	& \int \frac{d^d p_1}{(2\pi)^d} \frac{d^d p_2}{(2\pi)^d} \frac{d^d p_3}{(2\pi)^d} \, (2\pi)^d \delta^{(d)}(p_1 + p_2 + p_3) \, 
	J(\nu_1,\nu_2,\nu_3) e^{- i p_1 \cdot x_1 - i p_2 \cdot x_2 - i p_3 \cdot x_3} \notag\\[1.5ex]
	&\hspace{1cm} = \frac{1}{4^{\nu_1+\nu_2+\nu_3} \pi^{3 d/2}}  \frac{\Gamma(d/2 - \nu_1) \Gamma(d/2 - \nu_2) \Gamma(d/2 - \nu_3)}{\Gamma(\nu_1) 
		\Gamma(\nu_2) \Gamma(\nu_3)} \Phi(x_1,x_2,x_3)  
	\label{oner}
\end{align}
with 
\begin{equation}
	\Phi(x_1,x_2,x_3)\equiv \frac{1}{(x_{12}^2)^{d/2- \nu_3} (x_{23}^2)^{d/2- \nu_1} (x_{31}^2)^{d/2- \nu_2}}\,
\end{equation}
being the expression of a scalar conformal 3-point function.
Therefore, the conformal constraints in coordinated space on $\Phi(x_1,x_2,x_3)$ are automatically satisfied, providing no new information, while in momentum space they amount to some significant differential conditions
\begin{equation}
	\label{cond12}
	\begin{split}
		K^\kappa(p_i)\,J(\nu_1,\nu_2,\nu_3) &=0 \\
		D(p_i)\,J(\nu_1,\nu_2,\nu_3)&=0
	\end{split}
\end{equation}
which need to be satisfied by the original integral $J$.\\
Eqs. \eqref{cond12} allow to obtain recursion relations among the class of master integrals associated to $J$  It can be also easily shown that the scale covariant condition, the second equation above, is equivalent to the integration by part rule used in the ordinary multiloop analysis of the master integrals \cite{Coriano:2013jba}.\\
We can follow a similar route with ordinary composite operators in free field theory, built out of scalar fields $\chi$ in $d$ dimensions, such as $\Phi=\chi^{2 n}$ with $\Delta_i=n(d-2)$. In this case the corresponding conformal 3-point function derived in free field theory is given by 
\begin{equation}
	\Phi(x_i)=\frac{1}{\left(x_{12}^2 x_{23}^2 x_{31}^2\right)^{n d'}}
\end{equation}
$(d'=(d-2)/2)$ which generates the master integral $J$ with $\nu_i=d/2 -nd'$. If we require the dual conformal condition 
\eqref{dci} to be valid, then this requires that $d= 6n/(3 n -1)$. For $d$ to be a physical dimension we require it to be an integer, and we are left only with the choice of $n=1$, which gives $d=3$. Therefore, the ordinary triangle diagram, if generated by a free CFT, is a DCC solution only in $d=3$.

\section{Factorized solutions of the CWI's  from DCA's}
Now we turn to discuss the solution of \eqref{C2} and of \eqref{Eq2}. As discussed above, we will consider possible solutions which are built around specific dual conformal ans\"atze, as illustrated in the previous sections. \\
The  equations involving $C_1$ and $C_3$ are both identically satisfied if the former equations \eqref{C2} and \eqref{Eq2} are. The number of independent equations, by using the ans\"atz that we are going to present below, will then reduce from 3 down to 2. We illustrate this procedure in some detail.

We choose  the ans\"atz
\begin{equation}
	\Phi(p_i,s,t)=\big(s^2t^2\big)^{n_s}\,F(x,y)\label{ansatz}
\end{equation}
where $n_s$ is a coefficient (scaling factor of the ans\"atz) that we will fix below by the dilatation WI, and the variables $x$ and $y$ are defined by the quartic ratios
\begin{equation}
	x=\frac{p_1^2\,p_3^2}{s^2\,t^2},\qquad y=\frac{p_2^2\,p_4^2}{s^2\,t^2}.
\end{equation}
We will comment in a later section on the significance of such a choice and on the way to set up the invariants in momentum space in general. We will re-express the equations in terms of these new variables which will replace $s$ and $t$.\\
By inserting the ans\"atz \eqref{ansatz} into the dilatation Ward Identities, and turning to the new variables $x$ and $y$, after some manipulations we obtain from \eqref{Dilatation4} the condition
\begin{align}
	\label{dil1}
	&\bigg[(\D_t-3d)-\sum_{i=1}^4p_i\frac{\partial}{\partial p_i}-s\frac{\partial}{\partial s}-t\frac{\partial}{\partial t}\bigg]\big(s^2t^2\big)^{c}\,F(x,y),\notag\\
	&=\big(s^2t^2\big)^{n_s}\big[(\D_t-3d)-4 n_s\big] \,F(x,y)=0
\end{align}
which determines $n_s=(\D_t-3d)/4$, giving
\begin{equation}
	\Phi(p_i,s,t)=\big(s^2t^2\big)^{(\D_t-3d)/4}\,F(x,y).\label{ansatz2}
\end{equation}
We will be using this specific form of the solution in two of the three equations ($C_2$ and $C_{13}$). The functional form of $F(x,y)$ will then be furtherly constrained. 

\subsection{Determining the solutions in the case of primaries with equal scalings}
In order to determine the conditions on $F(x,y)$ from \eqref{C2} and \eqref{Eq2},  we re-express these two equations in terms of $x$ and $y$ using several identities. In particular we will use the relations 
\begin{equation}
	\frac{\partial^2}{\partial s\partial t}F(x,y)=\frac{4}{st}\big[\big(
	x\,\partial_x +y\partial_y\big)F+\big(x^2\partial_{xx}+2xy\,\partial_{xy}+y^2\partial_{yy}\big)F\big],
	\label{onew}
\end{equation}
together with
\begin{align}
	\left(p_1\frac{\partial}{\partial p_1}+p_2\frac{\partial}{\partial p_2}-p_3\frac{\partial}{\partial p_3}-p_4\frac{\partial}{\partial p_4}\right)F(x,y)&=\left(2x\,\partial_x+2y\,\partial_y -2x\,\partial_x-2y\,\partial_y\right)F(x,y)=0,\\[1.5ex]
	\left(p_1\frac{\partial}{\partial p_1}+p_4\frac{\partial}{\partial p_4}-p_3\frac{\partial}{\partial p_3}-p_2\frac{\partial}{\partial p_2}\right)F(x,y)&=\left(2x\,\partial_x+2y\,\partial_y -2x\,\partial_x-2y\,\partial_y\right)F(x,y)=0.
	\label{twow}
\end{align}
Both relations can be worked out after some lengthy computations using the relations presented in \appref{Appendix0}. \\
We start investigating the solutions of these equations by assuming, as a first example, that the scaling dimensions of all the fields $\phi_i$ are  equal $\D_1=\D_2=\D_3=\D_4=\D$.\\
Using \eqref{onew} and \eqref{twow}, we write the first equation \eqref{C2} associated to $C_2$ in the new variable $x$ and $y$ as
\begin{align}
	&C_2= 4\big(p_2^2-p_4^2\big)(s^2)^{n_s-1}(t^2)^{n_s-1}\notag\\
	&\times\bigg[y(1-y)\partial_{yy} -2x\,y\,\partial_{xy}-x^2\partial_{xx}-(1-2n_s)x\,\partial_x+\left(1-\D+\frac{d}{2}-y(1-2n_s)\right)\,\partial_y-n_s^2\bigg] F(x,y)=0\label{first}
\end{align}
and the second one \eqref{Eq2} associated to $C_{13}$ as
\begin{align}
	&4\big(p_1^2-p_3^2\big)(s^2)^{n_s-1}(t^2)^{n_s-1}\notag\\
	&\times\bigg[x(1-x)\partial_{xx} -2x\,y\,\partial_{xy}-y^2\partial_{yy}-(1-2n_s)y\,\partial_y+\left(1-\D+\frac{d}{2}-x(1-2n_s)\right)\,\partial_x-n_s^2\bigg] F(x,y)=0\label{second}
\end{align}
where we recall that $n_s$ is the scaling under dilatations, now given by
\begin{equation}
	\label{scaled}
	n_s=\D-\frac{3d}{4}
\end{equation}
since $\D_t=4\, \D$. \\
By inspection, one easily verifies that \eqref{first} and \eqref{second} define a hypergeometric system of two equations whose solutions can be expressed as linear combinations of 4 Appell functions of two variables $F_4$, as in the case of 3-point functions discussed before. The general solution of such system is expressed as
\begin{align}
	\Phi(p_i,s,t) &=\big(s^2t^2\big)^{(\D_t-3d)/4}\, F(x,y) \notag\\
	F(x,y)&= \sum_{a,b} c(a,b,\vec\Delta_t) x^a y^b F_4\left(\a(a,b),\b(a,b),\gamma(a),\g'(b);x, y\right),\label{solution1}
\end{align}
with $\vec\Delta_t=\Delta(1,1,1,1)$ for being in the equal scaling case.  Notice that the solution is similar to that of the 3-point functions given by \eqref{geneq}, discussed before.\\
The general solution \eqref{solution1} has been written as a linear superposition of these with independent constants $c(a,b)$, labeled by the exponents $a,b$ 
\begin{align}
	&a=0,\,\D-\frac{d}{2},&&b=0,\,\D-\frac{d}{2},\label{FuchsianPoint}
\end{align}
which fix the dependence of the $F_4$  
\begin{align}
	\label{s2}
	&\a(a,b)=\frac{3}{4}d-\D+a+b,&&\b(a,b)=\frac{3}{4}d-\D+a+b,\notag\\
	&\gamma(a)=\frac{d}{2}-\D+1+2a,&&\g'(b)=\frac{d}{2}-\D+1+2b.
\end{align}\\
We are now going to show that the third CWI corresponding to $C_1$ is identically satisfied by choosing the solution identified in \eqref{solution1}. For this purpose we re-express the $C_1$ equation \eqref{C1} in terms of the $x$ and $y$ invariant ratios in the form 
\begin{equation}
	\begin{aligned}
		C_1&=4(p_2^2-p_3^2)(s^2)^{n_s-1}(t^2)^{n_s-1}\left[x^2 \partial_{xx} +2x\,y\,\partial_{xy}+y^2\,\partial_{yy}+(1-2n)x\,\partial_x+(1-2n)y\,\partial_y+n^2\right]\,F(x,y) \\[1.3ex]
		&+\frac{4\,(s^2)^{n_s}(t^2)^{n_s}}{p_1^2}\bigg[\,x^2\,\partial_{xx}-\frac{p_1^2}{p_4^2}\,y^2\,\partial_{yy}+\frac{(d-2\D+2)}{2}\,x\,\partial_x -\frac{(d-2\D+2)}{2}\frac{p_1^2}{p_4^2}\,y\,\partial_y \bigg]F(x,y) = 0.\label{C1mod}
	\end{aligned}
\end{equation}
One can observe that the first line of the previous expression is actually a linear combination of the \eqref{first} and \eqref{second}. After some lengthy algebra we can rewrite the equation coming from $C_1$ in the form
\begin{equation}
	(p_2^2+p_3^2)\bigg[2x\,\partial_{xx}-2y\,\partial_{yy}+(d-2\D+2)\partial_x-(d-2\D+2)\partial_y\bigg]F(x,y)=0.\label{C1new}
\end{equation}
In order to verify that the equation above is identically satisfied,  we use the following identities for the Appell hypergeometric function 
\begin{align}
	\partial_x\,F_4(a,b,c_1,c_2;x,y)&=\frac{a\,b}{c_1}\,F_4(a+1,b+1,c_1+1,c_2;x,y)\\
	\partial_y\,F_4(a,b,c_1,c_2;x,y)&=\frac{a\,b}{c_2}\,F_4(a+1,b+1,c_1,c_2+1;x,y)
\end{align}
\begin{equation}
	x\,\partial_x\, F_4(a,b,c_1,c_2;x,y)= (c_1-1)\big[F_4(a,b,c_1-1,c_2;x,y)-F_4(a,b,c_1,c_2;x,y)\big].
\end{equation} 
We can use these relations to derive the further relation 
\begin{align}
	x\,\partial_{xx}\, F_4(a,b,c_1,c_2;x,y)&=(c_1-1) \partial_x\,F_4(a,b,c_1-1,c_2;x,y)-c_1\,\partial_x\,F_4(a,b,c_1,c_2;x,y)\notag\\
	&=a\,b\big[F_4(a+1,b+1,c_1,c_2;x,y)-F_4(a+1,b+1,c_1+1,c_2;x,y)\big]
\end{align}
with an analogous expression obtained for the $y$ variable. Considering the general expression of $F(x,y)$ previously obtained in \eqref{solution1}, as $F(x,y)=x^a\,y^b\,F_4\left(\a(a,b),\b(a,b),\gamma(a),\g'(b);x, y\right)$ into \eqref{C1new} one indeed verifies that the equation
\begin{equation}
	\begin{aligned}
		0&=\bigg[2x\,\partial_{xx}-2y\,\partial_{yy}+(d-2\D+2)\partial_x-(d-2\D+2)\partial_y\bigg]x^a\,y^b\,F_4\left(\a(a,b),\b(a,b),\gamma(a),\g'(b);x, y\right)\\
		&=x^a\,y^b\bigg[2x\,\partial_{xx}-2y\,\partial_{yy}+(d-2\D+2+2a)\partial_x-(d-2\D+2+2b)\partial_y\bigg]\,F_4\left(\a(a,b),\b(a,b),\gamma(a),\g'(b);x, y\right)
	\end{aligned}
\end{equation}
is satisfied, if we choose $\a(a,b)$, $\b(a,b)$, $\gamma(a)$ and $\g'(b)$ as identified from \eqref{s2}.

Therefore one indeed verifies that equation $C_1$ vanishes on the chosen ans\"atz.

\subsection{Two independent operatorial scalings }
The solution obtained above in the equal scaling case can be extended to the more general case 
\begin{equation}
	\D_1=\D_3=\D_x,\qquad \D_2=\D_4=\D_y.
\end{equation}
In this case the CWI's  give the system of equations
\begin{equation}
	\left\{\,\,
	\begin{aligned}
		&\left[y(1-y)\partial_{yy} -2x\,y\,\partial_{xy}-x^2\partial_{xx}-(1-2n_s)x\,\partial_x+\left(1-\D_y+\frac{d}{2}-y(1-2n_s)\right)\,\partial_y-n_s^2\right] F(x,y)=0\\[2ex]
		&\bigg[x(1-x)\partial_{xx} -2x\,y\,\partial_{xy}-y^2\partial_{yy}-(1-2n_s)y\,\partial_y+\left(1-\D_x+\frac{d}{2}-x(1-2n_s)\right)\,\partial_x-n_s^2\bigg] F(x,y)=0
	\end{aligned}\right.
\end{equation}
where now $n_s$ is defined as 
\begin{equation}
	n_s=\frac{\D_x}{2}+\frac{\D_y}{2}-\frac{3}{4}d.
\end{equation}
whose solutions are expressed as
\begin{equation}
	\label{ress}
	\Phi(p_i,s,t)=\big(s^2t^2\big)^{(\D_t-3d)/4}\,\sum_{a,b} c(a,b,\vec\Delta_t) x^a y^bF_4\left(\a(a,b),\b(a,b),\gamma(a),\g'(b);x,y\right)
\end{equation}
with $\vec\Delta_t=(\Delta_x,\Delta_y,\Delta_x,\Delta_y)$, $\Delta_t=2 \Delta_x + 2 \Delta_y$ and the Fuchsian points are fixed by the conditions
\begin{align}
	&a=0,\,\D_x-\frac{d}{2}&&b=0,\,\D_y-\frac{d}{2}\notag\\
	&\a(a,b)=\frac{3}{4}d-\frac{\D_x}{2}-\frac{\D_y}{2}+a+b,&&\b(a,b)=\frac{3}{4}d-\frac{\D_x}{2}-\frac{\D_y}{2}+a+b,\notag\\
	&\gamma(a)=\frac{d}{2}-\D_x+1+2a,&&\g'(b)=\frac{d}{2}-\D_y+1+2b.
\end{align}
We pause for a moment to discuss the domain of convergence of such solutions. Such domain, for $F_4$, is bounded by the relation 
\begin{equation}
	\sqrt{x}+\sqrt{y}< 1, 
\end{equation}
which is satisfied in a significant kinematic region, and in particular at large energy and momentum transfers. Notice that the analytic continuation of \eqref{ress} in the physical region can be simply obtained by sending $t^2\to -t^2$ (with $t^2<0$) and leaving all the other invariants untouched. In this case we get
\begin{equation}
	\sqrt{p_1^2 p_3^2}  +\sqrt{p_2^2 p_4^2} < \sqrt{- s^2 t^2}.
\end{equation}
At large energy and momentum transfers the correlator exhibits a power-like behaviour of the form 

\begin{equation}
	\Phi(p_i,s,t)\sim \frac{1}{(- s^2 t^2)^{(3 d - \Delta_t)/4}}. 
\end{equation}
Given the connection between the function $F_4$ and the 3K integrals, we will reformulate this solution in terms of such integrals. They play a key role in the solution of the CWI's for tensor correlators, as discussed in \cite{Bzowski:2013sza} for 3-point functions.

\subsection{DCC solutions as 3K integrals}
The link between 3- and 4-point functions outlined in the previous section allows to re-express the solutions in terms of a class of parametric integrals of 3 Bessel functions, as done in the case of the  scalar and tensor correlators  \cite{Bzowski:2013sza}, with the due modifications.
We consider the case of the solutions characterized by $\D_1=\D_2=\D_3=\D_4=\D$ or $\D_1=\D_3=\D_x$ and\  $\D_2=\D_4=\D_y$. We will show that the solution can be written in terms of triple-K integrals which are connected to the Appell function $F_4$ by the relation 

\begin{align}
	& \int_0^\infty d x \: x^{\alpha - 1} K_\lambda(a x) K_\mu(b x) K_\nu(c x) =\frac{2^{\alpha - 4}}{c^\alpha} \bigg[ B(\lambda, \mu) + B(\lambda, -\mu) + B(-\lambda, \mu) + B(-\lambda, -\mu) \bigg], \label{3K}
\end{align}
where
\begin{align}
	B(\lambda, \mu) & = \left( \frac{a}{c} \right)^\lambda \left( \frac{b}{c} \right)^\mu \Gamma \left( \frac{\alpha + \lambda + \mu - \nu}{2} \right) \Gamma \left( \frac{\alpha + \lambda + \mu + \nu}{2} \right) \Gamma(-\lambda) \Gamma(-\mu) \times \notag\\
	& \qquad \times F_4 \left( \frac{\alpha + \lambda + \mu - \nu}{2}, \frac{\alpha + \lambda + \mu + \nu}{2}; \lambda + 1, \mu + 1; \frac{a^2}{c^2}, \frac{b^2}{c^2} \right), \label{3Kplus}
\end{align}
valid for
\begin{equation}
	\Re\, \alpha > | \Re\, \lambda | + | \Re \,\mu | + | \Re\,\nu |, \qquad \Re\,(a + b + c) > 0 \notag\\
\end{equation}
and the Bessel functions $K_\nu$ satisfy the equations 
\begin{align}
	\frac{\partial}{\partial p}\big[p^\b\,K_\b(p\,x)\big]&=-x\,p^\b\,K_{\b-1}(p x)\notag\\
	K_{\b+1}(x)&=K_{\b-1}(x)+\frac{2\b}{x}K_{\b}(x). \label{der}
\end{align}
In particular  the solution can be written as
\begin{equation}
	I_{\a\{\b_1,\b_2,\b_3\}}(p_1\,p_3; p_2\,p_4;s\,t)=\int_0^\infty\,dx\,x^\a\,(p_1\,p_3)^{\b_1}\,(p_2\,p_4)^{\b_2}\,(s\,t)^{\b_3}\,K_{\b_1}(p_1\,p_3\,x)\,K_{\b_2}(p_2\,p_4\,x)\,K_{\b_3}(s\,t\,x).\label{trekappa}
\end{equation}
Using  \eqref{der} one can derive several relations, 
such as 
\begin{align}
	\frac{\partial^2}{\partial p_1^2}I_{\a\{\b_1,\b_2,\b_3\}}&=-\,p_3^2\,I_{\a+1\{\b_1-1,\b_2,\b_3\}}+p_1^2\,p_3^4\,\,I_{\a+2\{\b_1-2,\b_2,\b_3\}}
\end{align}
which generate identities such as 
\begin{align}
	p_1^2\,p_3^2\,I_{\a+2\{\b_1-2,\b_2,\b_3\}}&=I_{\a+2\{\b_1,\b_2,\b_3\}}-2(\b_1-1)\,I_{\a+1\{\b_1-1,\b_2,\b_3\}}.
\end{align}
We refer to \appref{AppendixA} for more details and a complete list of identities for such integrals. Using these relations, the dilatation Ward identities \eqref{Dilatation4} take the form
\begin{equation}
	(\D_t-3d) I_{\a\{\b_1,\b_2,\b_3\}}+2p_1^2p_3^2\ I_{\a+1\{\b_1-1,\b_2,\b_3\}}+2p_2^2p_4^2\ I_{\a+1\{\b_1,\b_2-1,\b_3\}}+2s^2t^2\ I_{\a+1\{\b_1,\b_2,\b_3-1\}}=0
\end{equation}
where the arguments of the $I_{\a\{\b_1\b_2\b_3\}}$ function, written explicitly in \eqref{trekappa}, have been omitted for the sake of simplicity. The $I$ integrals satisfy the differential equations
\begin{align}
	\frac{1}{s}\frac{\partial}{\partial s}\left(p_1\frac{\partial}{\partial p_1}+p_2\frac{\partial}{\partial p_2}-p_3\frac{\partial}{\partial p_3}-p_4\frac{\partial}{\partial p_4}\right)I_{\a\{\b_1,\b_2,\b_3\}}&=0\\
	\frac{1}{t}\frac{\partial}{\partial t}\left(p_1\frac{\partial}{\partial p_1}+p_4\frac{\partial}{\partial p_4}-p_2\frac{\partial}{\partial p_2}-p_3\frac{\partial}{\partial p_3}\right)I_{\a\{\b_1,\b_2,\b_3\}}&=0
\end{align} 
which can be checked using the relations given in the same appendix, and we finally find
\begin{equation}
	(\D_t-3d+2\a+2-2\b_t) I_{\a\{\b_1,\b_2,\b_3\}}=0
\end{equation}
where $\b_t=\b_1+\b_2+\b_3$. In order to satisfy this equation the $\a$ parameter has to be equal to a particular value given by 
\begin{equation}
	\tilde{\a}\equiv \frac{3}{2}d+\b_t-1-\frac{\D_t}{2}.
\end{equation}
In the particular case $\D_i=\D$ the special conformal Ward identities are given by
\begin{equation}
	\left\{\begin{aligned}&\\[-1.2ex]
		&\bigg [\frac{\partial^2}{\partial p_1^2}+\frac{(d-2\D+1)}{p_1}\frac{\partial}{\partial p_1}-\frac{\partial^2}{\partial p_3^2}-\frac{(d-2\D+1)}{p_3}\frac{\partial}{\partial p_3}+\frac{(p_1^2-p_3^2)}{st}\frac{\partial^2}{\partial s\partial t}\bigg ]\,I_{\tilde\a\{\b_1,\b_2,\b_3\}}=0\\[1ex]
		&\bigg[\frac{\partial^2}{\partial p_2^2}+\frac{(d-2\D+1)}{p_2}\frac{\partial}{\partial p_2}-\frac{\partial^2}{\partial p_4^2}-\frac{(d-2\D+1)}{p_4}\frac{\partial}{\partial p_4}+\frac{(p_2^2-p_4^2)}{st}\frac{\partial^2}{\partial s\partial t}\bigg ]\,I_{\tilde\a\{\b_1,\b_2,\b_3\}}=0\\[1ex]
		&\bigg[\frac{\partial^2}{\partial p_3^2}+\frac{(d-2\D+1)}{p_3}\frac{\partial}{\partial p_3}-\frac{\partial^2}{\partial p_4^2}-\frac{(d-2\D+1)}{p_4}\frac{\partial}{\partial p_4}+\frac{(p_2^2-p_1^2)}{st}\frac{\partial^2}{\partial s\partial t}\bigg ]\,I_{\tilde\a\{\b_1,\b_2,\b_3\}}=0\\[-0.7ex]
		\label{neweq}
	\end{aligned}\right.
\end{equation}
and using the properties of Bessel functions they can be rewritten in a simpler form. The first equation, for instance, can be written as
\begin{equation}
	\label{oone}
	(p_1^2-p_3^2)\bigg( (d-2\D+2\b_1)\,I_{\tilde\a+1\{\b_1-1,\b_2,\b_3\}}-2\b_3\,I_{\tilde\a+1\{\b_1,\b_2,\b_3-1\}} \bigg)=0,
\end{equation}
which is identically satisfied if the conditions 
\begin{equation}
	\b_1=\D-\frac{d}{2},\qquad\b_3=0
\end{equation}
hold. In the same way we find that the second equation takes the form
\begin{equation}
	\label{otwo}
	(p_2^2-p_4^2)\bigg((d-2\D+2\b_2)\,I_{\tilde\a+1\{\b_1,\b_2-1,\b_3\}}-2\b_3\,I_{\tilde\a+1\{\b_1,\b_2,\b_3-1\}}\bigg)=0
\end{equation}
and it is satisfied if 
\begin{equation}
	\b_2=\D-\frac{d}{2},\qquad \b_3=0.
\end{equation}
One can check that the third equation 
\begin{equation}
	p_2^2(d-2\D+2\b_2)\,I_{\tilde\a+1\{\b_1,\b_2-1,\b_3\}}-p_1^2(d-2\D+2\b_1)\,I_{\tilde\a+1\{\b_1-1,\b_2,\b_3\}}-2(p_2^2-p_1^2)\b_3\,I_{\tilde\a+1\{\b_1,\b_2,\b_3-1\}}=0,
\end{equation}
generates the same conditions given by \eqref{oone} and \eqref{otwo}.
After some computations, finally the solution for the  4-point function, in this particular case, can be written as
\begin{equation}
	\braket{O(p_1)\,O(p_2)\,O(p_3)\,O(\bar{p}_4)}=\,\bar{\a} \,I_{\frac{d}{2}-1\left\{\D-\frac{d}{2},\D-\frac{d}{2},0\right\}}(p_1\, p_3;p_2\,p_4; s\,t),
\end{equation}
where $\bar{\a}$ is an undetermined constant. 

In the case $\D_1=\D_3=\D_x$ and $\D_2=\D_4=\D_y$, the special CWI's can be written as
\begin{equation}
	\left\{\begin{aligned}&\\[-1.9ex]
		&\bigg [\frac{\partial^2}{\partial p_1^2}+\frac{(d-2\D_x+1)}{p_1}\frac{\partial}{\partial p_1}-\frac{\partial^2}{\partial p_3^2}-\frac{(d-2\D_x+1)}{p_3}\frac{\partial}{\partial p_3}+\frac{(p_1^2-p_3^2)}{st}\frac{\partial^2}{\partial s\partial t}\bigg ]\,I_{\tilde\a\{\b_1,\b_2,\b_3\}}=0\\[1ex]
		&\bigg[\frac{\partial^2}{\partial p_2^2}+\frac{(d-2\D_y+1)}{p_2}\frac{\partial}{\partial p_2}-\frac{\partial^2}{\partial p_4^2}-\frac{(d-2\D_y+1)}{p_4}\frac{\partial}{\partial p_4}+\frac{(p_2^2-p_4^2)}{st}\frac{\partial^2}{\partial s\partial t}\bigg ]\,I_{\tilde\a\{\b_1,\b_2,\b_3\}}=0\\[1ex]
		&\bigg[\frac{\partial^2}{\partial p_3^2}+\frac{(d-2\D_x+1)}{p_3}\frac{\partial}{\partial p_3}-\frac{\partial^2}{\partial p_4^2}-\frac{(d-2\D_y+1)}{p_4}\frac{\partial}{\partial p_4}+\frac{(p_2^2-p_1^2)}{st}\frac{\partial^2}{\partial s\partial t}\bigg ]\,I_{\tilde\a\{\b_1,\b_2,\b_3\}}=0\\[1.3ex]
	\end{aligned}\right.
\end{equation}
whose solution is
\begin{align}
	\label{ssm1}
	\braket{O(p_1)\,O(p_2)\,O(p_3)\,O(\bar{p}_4)}&=\,\bar{\bar{\a}} \,I_{\frac{d}{2}-1\left\{\D_x-\frac{d}{2},\D_y-\frac{d}{2},0\right\}}(p_1\, p_3;p_2\,p_4; s\,t)
\end{align}
which takes a form similar to the one typical of the three-point function given in \eqref{caz}.\\
In order to identify the form of the unique solution we need to satisfy the symmetry constraints and the absence of unphysical singularities \cite{Bzowski:2013sza} in the domain of convergence. We will address the first issue below, while the second is discussed in \secref{convergence}, where we show that such singularities are not present. 

\subsection{Symmetric solutions as \texorpdfstring{$F_4$}{F4} hypergeometrics or 3K integrals. The equal scalings case}
The derivation of symmetric expressions of such correlators requires some effort, and can be obtained either by 
using the few known relations available for the Appell function $F_4$ or, alternatively (and more effectively), by resorting to the formalism of the 3K integrals. \\ 
A solution which is symmetric respect to all the permutation of the momenta $p_i$, expressed in terms of 3 of the four constants $c(a,b)$, after some manipulations, can be expressed in the form
\begin{align}
	\braket{O(p_1)O(p_2)O(p_3)O(p_4)}&=\notag\\
	&\hspace{-2cm}=\sum_{a,b}c(a,b)\Bigg[\,(s^2\,t^2)^{\D-\frac{3}{4}d}\left(\frac{p_1^2p_3^2}{s^2t^2}\right)^a\left(\frac{p_2^2p_4^2}{s^2t^2}\right)^bF_4\left(\a(a,b),\b(a,b),\gamma(a),\g'(b),\frac{p_1^2p_3^2}{s^2t^2},\frac{p_2^2p_4^2}{s^2t^2}\right)\notag\\
	&\hspace{-1.5cm}+\,(s^2\,u^2)^{\D-\frac{3}{4}d}\,\left(\frac{p_2^2p_3^2}{s^2u^2}\right)^{a}\left(\frac{p_1^2p_4^2}{s^2u^2}\right)^{b}F_4\left(\a(a,b),\b(a,b),\gamma(a),\g'(b),\frac{p_2^2p_3^2}{s^2u^2},\frac{p_1^2p_4^2}{s^2u^2}\right)\notag\\
	&\hspace{-1.5cm}+\,(t^2\,u^2)^{\D-\frac{3}{4}d}\,\left(\frac{p_1^2p_2^2}{t^2u^2}\right)^{a}\,\left(\frac{p_3^2p_4^2}{t^2u^2}\right)^{b}\,F_4\left(\a(a,b),\b(a,b),\gamma(a),\g'(b),\frac{p_1^2p_2^2}{t^2u^2},\frac{p_3^2p_4^2}{t^2u^2}\right)
	\Bigg]
	\label{fform1}
\end{align}
where the four coefficients $c(a,b)$'s given in \eqref{solution} are reduced to three by the constraint 
\begin{align}
	c\left(0,\D-\frac{d}{2}\right)=c\left(\D-\frac{d}{2},0\right).
\end{align}
Additional manipulations, in order to reduce even further the integration constants are hampered by absence of known 
relations for the Appell functions $F_4$. As already mentioned above, it is possible, though, to bypass the problem by turning to the 3K formalism. \eqref{fform1} can be further simplified using this formalism. 

\subsubsection{3K symmetrization in the equal scaling case}
In order to show this, \eqref{fform1} can be written in terms of a linear combination of 3K integrals as
\begin{align}
	\braket{\mathcal{O}(p_1)\mathcal{O}(p_2)\mathcal{O}(p_3)\mathcal{O}(p_4)}&= C_1\,I_{\frac{d}{2}-1\{\D-\frac{d}{2},\D-\frac{d}{2},0\}}(p_1\,p_3,p_2\,p_4,s\,t)\notag\\[1.2ex]
	&\hspace{-1.5cm}+C_2\, \,I_{\frac{d}{2}-1\{\D-\frac{d}{2},\D-\frac{d}{2},0\}}(p_2\,p_3,p_1\,p_4,s\,u)+C_3\, \,I_{\frac{d}{2}-1\{\D-\frac{d}{2},\D-\frac{d}{2},0\}}(p_1\,p_2,p_3\,p_4,t\,u)
\end{align}
by an explicit symmetrization of the momenta in the parametric integrals.  It is now much simpler to show that the symmetry under permutations forces the $C_i$ to take the same value, and the final symmetric result is given by
\begin{align}
	\braket{\mathcal{O}(p_1)\mathcal{O}(p_2)\mathcal{O}(p_3)\mathcal{O}(p_4)}&= C\bigg[\,I_{\frac{d}{2}-1\{\D-\frac{d}{2},\D-\frac{d}{2},0\}}(p_1\,p_3,p_2\,p_4,s\,t)\notag\\
	&\hspace{-1.5cm}+\, \,I_{\frac{d}{2}-1\{\D-\frac{d}{2},\D-\frac{d}{2},0\}}(p_2\,p_3,p_1\,p_4,s\,u)+\, \,I_{\frac{d}{2}-1\{\D-\frac{d}{2},\D-\frac{d}{2},0\}}(p_1\,p_2,p_3\,p_4,t\,u)\bigg],
\end{align}
written in terms of only one arbitrary overall constant $C$. We can use the relation between the 3K integrals and the $F_4$ written in \eqref{3K} and \eqref{3Kplus},  to re-express the final symmetric solution, originally given in Eq.\eqref{fform1}, in terms of a single constant in the form
\begin{align}
	\braket{\mathcal{O}(p_1)\mathcal{O}(p_2)\mathcal{O}(p_3)\mathcal{O}(p_4)}&=2^{\frac{d}{2}-4}\ \ C\,\sum_{\l,\m=0,\D-\frac{d}{2}}\x(\l,\m)\bigg[\big(s^2\,t^2\big)^{\D-\frac{3}{4}d}\left(\frac{p_1^2 p_3^2}{s^2 t^2}\right)^\l\left(\frac{p_2^2p_4^2}{s^2t^2}\right)^\m\nonumber\\
	&\hspace{-3cm}\times\,F_4\left(\frac{3}{4}d-\D+\l+\m,\frac{3}{4}d-\D+\l+\m,1-\D+\frac{d}{2}+\l,1-\D+\frac{d}{2}+\m,\frac{p_1^2 p_3^2}{s^2 t^2},\frac{p_2^2 p_4^2}{s^2 t^2}\right)\notag\\
	&+\big(s^2\,u^2\big)^{\D-\frac{3}{4}d}\left(\frac{p_2^2 p_3^2}{s^2 u^2}\right)^\l\left(\frac{p_1^2p_4^2}{s^2u^2}\right)^\m\notag\\
	&\hspace{-3cm}\times\,F_4\left(\frac{3}{4}d-\D+\l+\m,\frac{3}{4}d-\D+\l+\m,1-\D+\frac{d}{2}+\l,1-\D+\frac{d}{2}+\m,\frac{p_2^2 p_3^2}{s^2 u^2},\frac{p_1^2 p_4^2}{s^2 u^2}\right)\notag\\
	&+\big(t^2\,u^2\big)^{\D-\frac{3}{4}d}\left(\frac{p_1^2 p_2^2}{t^2 u^2}\right)^\l\left(\frac{p_3^2p_4^2}{t^2u^2}\right)^\m\notag\\
	&\hspace{-3cm}\times\,F_4\left(\frac{3}{4}d-\D+\l+\m,\frac{3}{4}d-\D+\l+\m,1-\D+\frac{d}{2}+\l,1-\D+\frac{d}{2}+\m,\frac{p_1^2 p_2^2}{t^2 u^2},\frac{p_3^2 p_4^2}{t^2 u^2}\right)\bigg].\label{finalSol}
\end{align}
where the coefficients $\x(\l,\m)$ are explicitly given by
\begin{equation}
	\begin{split}
		\x\left(0,0\right)&=\left[\Gamma\left(\frac{3}{4}d-\D\right)\right]^2\left[\Gamma\left(\D-\frac{d}{2}\right)\right]^2\\
		\x\left(0,\D-\frac{d}{2}\right)&=\x\left(\D-\frac{d}{2},0\right)=\left[\Gamma\left(\frac{d}{4}\right)\right]^2\Gamma\left(\D-\frac{d}{2}\right)\Gamma\left(\frac{d}{2}-\D\right)\\
		\x\left(\D-\frac{d}{2},\D-\frac{d}{2}\right)&=\left[\Gamma\left(\D-\frac{d}{4}\right)\right]^2\left[\Gamma\left(\frac{d}{2}-\D\right)\right]^2.
	\end{split}\label{xicoef}
\end{equation}
The solution found in \eqref{finalSol} is explicitly symmetric under all the possible permutations of the momenta and it is fixed up to one undetermined constant $C$. Eq. \eqref{finalSol} gives the final expression of the solution obtained from the first DCA \eqref{ansatz2}.

\section{Solutions from other DCA's }
The DCA from which we start is clearly not unique, since other types of factorized ans\"atze can be chosen in dual coordinate space. It is then resonable to ask whether the types of solutions that we have identified are truly unique, even if they are generated starting from a specific DCA. In order to answer such a question we turn to a different DCA and show that this is indeed the case. The intermediate steps of the derivation are rather involved, but one can obtain the same expression of the DCC solution obtained from \eqref{ansatz2}, given in \eqref{finalSol}, using some analytic continuations of the new solution generated by such a second ans\"atz. \\
For this purpose, we consider as a starting point a DCA of the form 
\begin{equation}
	\Phi=\left(p_1^2\,p_3^2\right)^{n_s}\,F\left(\frac{s^2t^2}{p_1^2p_3^2},\frac{p_2^2p_4^2}{p_1^2p_3^2}\right)
	\label{ansatz3}
\end{equation}
where all the scalings are taken to be equal $\D_i=\D$, $i=1,2,3,4$. Also in this case
the dilatation WI's fix the value of $n_s$ as in \eqref{scaled}, while the special WI's can be written as
\begin{equation}
	\left\{
	\begin{aligned}\\[-1.6ex]
		&\left[x(1-x)\partial_{xx}-2x y\partial_{xy}-y^2\partial_{yy}-y(d-\D+1)\partial_y+[1-(d-\D+1)x]\partial_x-\frac{d}{4}\left(\frac{3}{4}d-\D\right)\right] F(x,y)=0\\[1.5ex]
		&\left[x\partial_{xx}-y\partial_{yy}+\partial_x-\left(\frac{d}{2}-\D+1\right)\partial_y\right]F(x,y)=0\\[1.3ex]
	\end{aligned}\right.
\end{equation}
where we have defined $x=s^2t^2/(p_1^2p_3^2)$ and $y=p_2^2p_4^2/(p_1^2p_3^2)$. Subtracting the second equation from the first one we derive the system of equations
\begin{equation}
	\left\{
	\begin{aligned}\\[-1.6ex]
		&\left[x(1-x)\partial_{xx}-2x y\,\partial_{xy}-y^2\partial_{yy}-y(d-\D+1)\partial_y+[1-(d-\D+1)x]\partial_x-\frac{d}{4}\left(\frac{3}{4}d-\D\right)\right] F(x,y)=0\\[1.5ex]
		&\Bigg[y(1-y)\partial_{yy}-2x y\,\partial_{xy}-x^2\partial_{xx}-x(d-\D+1)\partial_x+\left[\left(\frac{d}{2}-\D+1\right)-(d-\D+1)x\right]\partial_x-\frac{d}{4}\left(\frac{3}{4}d-\D\right)\Bigg] F(x,y)=0\\[1.3ex]
	\end{aligned}\right. 
\end{equation}
which corresponds, once more, to a hypergeometric system of equations in two variables, corresponding to Appell's $F_4$. The general solution of such a system can be expressed as a linear combination of two $F_4$ functions as
\begin{align}
	\Phi&=\left(p_1^2\,p_3^2\right)^{\D-\frac{3}{4}d}\bigg[C_1 F_4\left(\frac{d}{4}\,,\,\frac{3}{4}d-\D\, ,\,1\,,\,\frac{d}{2}-\D+1\,;\frac{s^2t^2}{p_1^2p_3^2}\,,\,\frac{p_2^2p_4^2}{p_1^2p_3^2}\right)\notag\\
	&\hspace{2.5cm}+C_2\left(\frac{p_2^2p_4^2}{p_1^2p_3^2}\right)^{\D-\frac{d}{2}} F_4\left(\D-\frac{d}{4}\,,\,\frac{d}{4}\, ,\,1\,,\,1-\frac{d}{2}+\D\,;\frac{s^2t^2}{p_1^2p_3^2}\,,\,\frac{p_2^2p_4^2}{p_1^2p_3^2}\right)\bigg].\label{Solution1}
\end{align}
This solution corresponds to a very specific case, in which one of the 4 parameters of the general solution given by the 4 hypergeometric functions of type $F_4$ is fixed to $\g=1$. One can show that in this case the number of independent hypergeometric solutions is then reduced from 4 to 2.  However, at this stage, Eq. \eqref{Solution1} is symmetric only respect to the momentum exchanges $(p_1\leftrightarrow p_3)$ and $(p_2\leftrightarrow p_4)$. 
As a first step we can proceed by constructing the completely symmetric solution of the same system in the form
\begin{align}
	\Phi&=\left(p_1^2\,p_3^2\right)^{\D-\frac{3}{4}d}\bigg[C_1 F_4\left(\frac{d}{4}\,,\,\frac{3}{4}d-\D\, ,\,1\,,\,\frac{d}{2}-\D+1\,;\frac{s^2t^2}{p_1^2p_3^2}\,,\,\frac{p_2^2p_4^2}{p_1^2p_3^2}\right)\notag\\
	&\hspace{2.5cm}+C_2\left(\frac{p_2^2p_4^2}{p_1^2p_3^2}\right)^{\D-\frac{d}{2}} F_4\left(\D-\frac{d}{4}\,,\,\frac{d}{4}\, ,\,1\,,\,1-\frac{d}{2}+\D\,;\frac{s^2t^2}{p_1^2p_3^2}\,,\,\frac{p_2^2p_4^2}{p_1^2p_3^2}\right)\bigg]\notag\\
	&+\left(p_2^2\,p_3^2\right)^{\D-\frac{3}{4}d}\bigg[C_1 F_4\left(\frac{d}{4}\,,\,\frac{3}{4}d-\D\, ,\,1\,,\,\frac{d}{2}-\D+1\,;\frac{s^2u^2}{p_2^2p_3^2}\,,\,\frac{p_1^2p_4^2}{p_2^2p_3^2}\right)\notag\\
	&\hspace{2.5cm}+C_2\left(\frac{p_1^2p_4^2}{p_2^2p_3^2}\right)^{\D-\frac{d}{2}} F_4\left(\D-\frac{d}{4}\,,\,\frac{d}{4}\, ,\,1\,,\,1-\frac{d}{2}+\D\,;\frac{s^2u^2}{p_2^2p_3^2}\,,\,\frac{p_1^2p_4^2}{p_2^2p_3^2}\right)\bigg]\notag\\
\end{align}
\begin{align}
	&+\left(p_1^2\,p_2^2\right)^{\D-\frac{3}{4}d}\bigg[C_1 F_4\left(\frac{d}{4}\,,\,\frac{3}{4}d-\D\, ,\,1\,,\,\frac{d}{2}-\D+1\,;\frac{u^2t^2}{p_1^2p_2^2}\,,\,\frac{p_3^2p_4^2}{p_1^2p_2^2}\right)\notag\\
	&\hspace{2.5cm}+C_2\left(\frac{p_3^2p_4^2}{p_1^2p_2^2}\right)^{\D-\frac{d}{2}} F_4\left(\D-\frac{d}{4}\,,\,\frac{d}{4}\, ,\,1\,,\,1-\frac{d}{2}+\D\,;\frac{u^2t^2}{p_1^2p_2^2}\,,\,\frac{p_3^2p_4^2}{p_1^2p_2^2}\right)\bigg]\label{Phi1}
\end{align}
containing only the coefficients $C_1$ and $C_2$.  Considering the $(p_2\leftrightarrow p_4)$ exchange the solution will be given by
\begin{align}
	\Phi_{p_2\leftrightarrow p_4}&=\left(p_1^2\,p_3^2\right)^{\D-\frac{3}{4}d}\bigg[C_1 F_4\left(\frac{d}{4}\,,\,\frac{3}{4}d-\D\, ,\,1\,,\,\frac{d}{2}-\D+1\,;\frac{s^2t^2}{p_1^2p_3^2}\,,\,\frac{p_2^2p_4^2}{p_1^2p_3^2}\right)\notag\\
	&\hspace{2.5cm}+C_2\left(\frac{p_2^2p_4^2}{p_1^2p_3^2}\right)^{\D-\frac{d}{2}} F_4\left(\D-\frac{d}{4}\,,\,\frac{d}{4}\, ,\,1\,,\,1-\frac{d}{2}+\D\,;\frac{s^2t^2}{p_1^2p_3^2}\,,\,\frac{p_2^2p_4^2}{p_1^2p_3^2}\right)\bigg]\notag\\
	&+\left(p_4^2\,p_3^2\right)^{\D-\frac{3}{4}d}\bigg[C_1 F_4\left(\frac{d}{4}\,,\,\frac{3}{4}d-\D\, ,\,1\,,\,\frac{d}{2}-\D+1\,;\frac{t^2u^2}{p_4^2p_3^2}\,,\,\frac{p_1^2p_2^2}{p_4^2p_3^2}\right)\notag\\
	&\hspace{2.5cm}+C_2\left(\frac{p_1^2p_4^2}{p_2^2p_3^2}\right)^{\D-\frac{d}{2}} F_4\left(\D-\frac{d}{4}\,,\,\frac{d}{4}\, ,\,1\,,\,1-\frac{d}{2}+\D\,;\frac{t^2u^2}{p_4^2p_3^2}\,,\,\frac{p_1^2p_2^2}{p_4^2p_3^2}\right)\bigg]\notag\\
	&+\left(p_1^2\,p_4^2\right)^{\D-\frac{3}{4}d}\bigg[C_1 F_4\left(\frac{d}{4}\,,\,\frac{3}{4}d-\D\, ,\,1\,,\,\frac{d}{2}-\D+1\,;\frac{u^2s^2}{p_1^2p_4^2}\,,\,\frac{p_3^2p_2^2}{p_1^2p_4^2}\right)\notag\\
	&\hspace{2.5cm}+C_2\left(\frac{p_3^2p_2^2}{p_1^2p_4^2}\right)^{\D-\frac{d}{2}} F_4\left(\D-\frac{d}{4}\,,\,\frac{d}{4}\, ,\,1\,,\,1-\frac{d}{2}+\D\,;\frac{u^2s^2}{p_1^2p_4^2}\,,\,\frac{p_3^2p_2^2}{p_1^2p_4^2}\right)\bigg]
\end{align}
that can be rearranged in the form given in \eqref{Phi1} using \eqref{transfF41}.\\
After imposing the symmetry condition $\Phi_{p_2\leftrightarrow p_4}=\Phi$ under this particular permutation, we find a single degenerate condition on the ratios of $C_1$ and $C_2$ given by
\begin{equation}
	\frac{C_1}{C_2}=\left[\Gamma\left(\D-\frac{3}{4}d\right)\Gamma\left(1-\D+\frac{3}{4}d\right)\Gamma\left(1+\D-\frac{d}{2}\right)\right]\left[\Gamma\left(\D-\frac{d}{4}\right)\Gamma\left(1+\D-\frac{3}{4}d\right)\Gamma\left(1-\D+\frac{d}{2}\right)\right]^{-1}.
\end{equation}
This constraint fixes the solution up to one undetermined constant in the form
\begin{equation*}
	\begin{aligned}
		\Phi&=C_1\bigg\{\left(p_1^2\,p_3^2\right)^{\D-\frac{3}{4}d}\bigg[F_4\left(\frac{d}{4}\,,\,\frac{3}{4}d-\D\,,\,1\,,\,\frac{d}{2}-\D+1\,;\frac{s^2t^2}{p_1^2p_3^2}\,,\,\frac{p_2^2p_4^2}{p_1^2p_3^2}\right)\\[1.2ex]
		&\hspace{-0.3cm}+\frac{\Gamma\left(\D-\frac{d}{4}\right)\Gamma\left(1+\D-\frac{3}{4}d\right)\Gamma\left(1-\D+\frac{d}{2}\right)}{\Gamma\left(\D-\frac{3}{4}d\right)\Gamma\left(1-\D+\frac{3}{4}d\right)\Gamma\left(1+\D-\frac{d}{2}\right)}\left(\frac{p_2^2p_4^2}{p_1^2p_3^2}\right)^{\D-\frac{d}{2}} F_4\left(\D-\frac{d}{4}\,,\,\frac{d}{4}\, ,\,1\,,\,1-\frac{d}{2}+\D\,;\frac{s^2t^2}{p_1^2p_3^2}\,,\,\frac{p_2^2p_4^2}{p_1^2p_3^2}\right)\bigg]\notag\\[1.2ex]
		&+\left(p_2^2\,p_3^2\right)^{\D-\frac{3}{4}d}\bigg[F_4\left(\frac{d}{4}\,,\,\frac{3}{4}d-\D\, ,\,1\,,\,\frac{d}{2}-\D+1\,;\frac{s^2u^2}{p_2^2p_3^2}\,,\,\frac{p_1^2p_4^2}{p_2^2p_3^2}\right)\\[1.2ex]
		&\hspace{-0.3cm}+\frac{\Gamma\left(\D-\frac{d}{4}\right)\Gamma\left(1+\D-\frac{3}{4}d\right)\Gamma\left(1-\D+\frac{d}{2}\right)}{\Gamma\left(\D-\frac{3}{4}d\right)\Gamma\left(1-\D+\frac{3}{4}d\right)\Gamma\left(1+\D-\frac{d}{2}\right)} \left(\frac{p_1^2p_4^2}{p_2^2p_3^2}\right)^{\D-\frac{d}{2}} F_4\left(\D-\frac{d}{4}\,,\,\frac{d}{4}\, ,\,1\,,\,1-\frac{d}{2}+\D\,;\frac{s^2u^2}{p_2^2p_3^2}\,,\,\frac{p_1^2p_4^2}{p_2^2p_3^2}\right)\bigg]\\[1.2ex]
	\end{aligned}
\end{equation*}
\begin{equation}
	\begin{aligned}
		&+\left(p_1^2\,p_2^2\right)^{\D-\frac{3}{4}d}\bigg[F_4\left(\frac{d}{4}\,,\,\frac{3}{4}d-\D\, ,\,1\,,\,\frac{d}{2}-\D+1\,;\frac{u^2t^2}{p_1^2p_2^2}\,,\,\frac{p_3^2p_4^2}{p_1^2p_2^2}\right)\\[1.2ex]
		&\hspace{-0.3cm}+\frac{\Gamma\left(\D-\frac{d}{4}\right)\Gamma\left(1+\D-\frac{3}{4}d\right)\Gamma\left(1-\D+\frac{d}{2}\right)}{\Gamma\left(\D-\frac{3}{4}d\right)\Gamma\left(1-\D+\frac{3}{4}d\right)\Gamma\left(1+\D-\frac{d}{2}\right)}\left(\frac{p_3^2p_4^2}{p_1^2p_2^2}\right)^{\D-\frac{d}{2}} F_4\left(\D-\frac{d}{4}\,,\,\frac{d}{4}\, ,\,1\,,\,1-\frac{d}{2}+\D\,;\frac{u^2t^2}{p_1^2p_2^2}\,,\,\frac{p_3^2p_4^2},{p_1^2p_2^2}\right)\bigg]\bigg\}\label{finalsolution}
	\end{aligned}
\end{equation}
which can be shown to be symmetric under all the possible permutations of the momenta $(p_1,\,p_2,\,p_3,\,p_4)$.\\
We are now going to show the equivalence of such solution to \eqref{finalSol}, which is given by a 3K integral. We perform an analytic continuation of \eqref{finalsolution} using \eqref{transfF41} to obtain the intermediate expression
\begin{equation}
	\begin{aligned}
		\Phi&=\left(s^2\,t^2\right)^{\D-\frac{3}{4}d}C_1\Bigg[ \frac{\Gamma\left(\D-\frac{d}{2}\right)(-1)^{\D-\frac{3}{4}d}}{\Gamma\left(1+\D-\frac{3}{4}d\right)\Gamma\left(\frac{d}{4}\right)}\,F_4\left(\frac{3}{4}d-\D\,,\,\frac{3}{4}d-\D\, ,\,\frac{d}{2}-\D+1\,,\,\frac{d}{2}-\D+1\,;\,\frac{p_1^2p_3^2}{s^2t^2}\, ,\,\frac{p_2^2p_4^2}{s^2t^2}\right)\\[1.2ex]
		&\hspace{2.5cm}+\frac{\Gamma\left(\frac{d}{2}-\D\right)(-1)^{-\frac{d}{4}}}{\Gamma\left(1-\frac{d}{4}\right)\Gamma\left(\frac{3}{4}d-
			\D\right)}\left(\frac{p_1^2p_3^2}{s^2t^2}\right)^{\D-\frac{d}{2}} F_4\left(\frac{d}{4}\,,\,\frac{d}{4}\, ,\,1-\frac{d}{2}+\D\,,\,\,1+\frac{d}{2}-\D\,;\,\frac{p_1^2p_3^2}{s^2t^2},\frac{p_2^2p_4^2}{s^2t^2}\right)\\[1.2ex]
		&\hspace{2.5cm}+\frac{\Gamma\left(\frac{d}{2}-\D\right)(-1)^{-\frac{d}{4}}}{\Gamma\left(1-\frac{d}{4}\right)\Gamma\left(\frac{3}{4}d-
			\D\right)}\left(\frac{p_2^2p_4^2}{s^2t^2}\right)^{\D-\frac{d}{2}} F_4\left(\frac{d}{4}\,,\,\frac{d}{4}\, ,\,1+\frac{d}{2}-\D\,,\,\,1-\frac{d}{2}+\D\,;\,\frac{p_1^2p_3^2}{s^2t^2},\frac{p_2^2p_4^2}{s^2t^2}\right)\\[1.2ex]
		&\hspace{2.5cm}+\frac{\left[\Gamma\left(\frac{d}{2}-\D\right)\right]^2\Gamma\left(\D-\frac{d}{4}\right)(-1)^{\frac{d}{4}-\D}}{\Gamma\left(1+\frac{d}{4}-\D\right)\Gamma\left(\frac{d}{4}\right)\Gamma\left(\frac{3}{4}d-\D\right)\Gamma\left(\D-\frac{d}{2}\right)} \\
		&\hspace{2cm}\times \left(\frac{p_2^2p_4^2}{s^2t^2}\right)^{\D-\frac{d}{2}}\left(\frac{p_1^2p_3^2}{s^2t^2}\right)^{\D-\frac{d}{2}}F_4\left(\frac{3}{4}d-\D\,,\,\frac{3}{4}d-\D\, ,\,1+\frac{d}{2}-\D\,,\,\,1+\frac{d}{2}-\D\,;\,\frac{p_1^2p_3^2}{s^2t^2},\frac{p_2^2p_4^2}{s^2t^2}\right)\Bigg] \\[1.2ex]
		&\hspace{2cm} +[(p_1\leftrightarrow p_2)]+[(p_2\leftrightarrow p_3)].\label{final}
	\end{aligned}
\end{equation}
After some manipulations and using the properties of Gamma function
\begin{equation}
	\Gamma(a-b)=\frac{\Gamma(a)\Gamma(1-a)(-1)^b}{\Gamma(1-a+b)},\hspace{1cm}\frac{1}{\Gamma(a-b)}=\frac{\Gamma(1-a+b)(-1)^{-b}}{\Gamma(a)\Gamma(1-a)},
\end{equation}
we write \eqref{final} as
\begin{align}
	\Phi&=\left(s^2\,t^2\right)^{\D-\frac{3}{4}d}\frac{C_1}{\Gamma\left(\frac{d}{4}\right)\Gamma\left(\frac{d}{2}\right)\Gamma\left(1-\frac{d}{2}\right)\Gamma\left(\frac{3}{4}d-\D\right)\Gamma\left(\D-\frac{d}{2}\right)}\notag\\[1.2ex]
	&\hspace{-0.5cm}\times\Bigg\{ \left[\Gamma\left(\D-\frac{d}{2}\right)\right]^2\left[\Gamma\left(\frac{3}{4}d-\D\right)\right]^2\,F_4\left(\frac{3}{4}d-\D\,,\,\frac{3}{4}d-\D\, ,\,\frac{d}{2}-\D+1\,,\,\frac{d}{2}-\D+1\,;\,\frac{p_1^2p_3^2}{s^2t^2}\, ,\,\frac{p_2^2p_4^2}{s^2t^2}\right)\notag\\[1.2ex]
	&\hspace{-0.5cm}+\left[\Gamma\left(\frac{d}{4}\right)\right]^2\Gamma\left(\frac{d}{2}-\D\right)\Gamma\left(\D-\frac{d}{2}\right)\left(\frac{p_1^2p_3^2}{s^2t^2}\right)^{\D-\frac{d}{2}}  F_4\left(\frac{d}{4}\,,\,\frac{d}{4}\, ,\,1-\frac{d}{2}+\D\,,\,\,1+\frac{d}{2}-\D\,;\,\frac{p_1^2p_3^2}{s^2t^2},\frac{p_2^2p_4^2}{s^2t^2}\right)\notag
\end{align}
\begin{align}
	&\hspace{-0.5cm}+\left[\Gamma\left(\frac{d}{4}\right)\right]^2\Gamma\left(\frac{d}{2}-\D\right)\Gamma\left(\D-\frac{d}{2}\right)\left(\frac{p_2^2p_4^2}{s^2t^2}\right)^{\D-\frac{d}{2}} F_4\left(\frac{d}{4}\,,\,\frac{d}{4}\, ,\,1+\frac{d}{2}-\D\,,\,\,1-\frac{d}{2}+\D\,;\,\frac{p_1^2p_3^2}{s^2t^2},\frac{p_2^2p_4^2}{s^2t^2}\right)\notag\\[1.2ex]
	&\hspace{-0.5cm}+\left[\Gamma\left(\frac{d}{2}-\D\right)\right]^2\left[\Gamma\left(\D-\frac{d}{4}\right)\right]^2 \notag\\
	&\hspace{1cm}\times \left(\frac{p_2^2p_4^2}{s^2t^2}\right)^{\D-\frac{d}{2}}\left(\frac{p_1^2p_3^2}{s^2t^2}\right)^{\D-\frac{d}{2}}F_4\left(\frac{3}{4}d-\D\,,\,\frac{3}{4}d-\D\, ,\,1+\frac{d}{2}-\D\,,\,\,1+\frac{d}{2}-\D\,;\,\frac{p_1^2p_3^2}{s^2t^2},\frac{p_2^2p_4^2}{s^2t^2}\right)\Bigg\}\notag\\[1.2ex]
	&\hspace{2cm} +[(p_1\leftrightarrow p_2)]+[(p_2\leftrightarrow p_3)].
\end{align}
which takes the same form of solution \eqref{finalSol}. In fact this expression can be rewritten as
\begin{align}
	\Phi&=\frac{C_1}{\Gamma\left(\frac{d}{4}\right)\Gamma\left(\frac{d}{2}\right)\Gamma\left(1-\frac{d}{2}\right)\Gamma\left(\frac{3}{4}d-\D\right)\Gamma\left(\D-\frac{d}{2}\right)}\,\sum_{\l,\m=0,\D-\frac{d}{2}}\x(\l,\m)\bigg[\big(s^2\,t^2\big)^{\D-\frac{3}{4}d}\left(\frac{p_1^2 p_3^2}{s^2 t^2}\right)^\l\left(\frac{p_2^2p_4^2}{s^2t^2}\right)^\m\notag\\
	&\hspace{1cm}\times\,F_4\left(\frac{3}{4}d-\D+\l+\m,\frac{3}{4}d-\D+\l+\m,1-\D+\frac{d}{2}+\l,1-\D+\frac{d}{2}+\m,\frac{p_1^2 p_3^2}{s^2 t^2},\frac{p_2^2 p_4^2}{s^2 t^2}\right)\notag\\
	&+\big(s^2\,u^2\big)^{\D-\frac{3}{4}d}\left(\frac{p_2^2 p_3^2}{s^2 u^2}\right)^\l\left(\frac{p_1^2p_4^2}{s^2u^2}\right)^\m\notag\\
	&\hspace{1cm}\times\,F_4\left(\frac{3}{4}d-\D+\l+\m,\frac{3}{4}d-\D+\l+\m,1-\D+\frac{d}{2}+\l,1-\D+\frac{d}{2}+\m,\frac{p_2^2 p_3^2}{s^2 u^2},\frac{p_1^2 p_4^2}{s^2 u^2}\right)\notag\\
	&+\big(t^2\,u^2\big)^{\D-\frac{3}{4}d}\left(\frac{p_1^2 p_2^2}{t^2 u^2}\right)^\l\left(\frac{p_3^2p_4^2}{t^2u^2}\right)^\m\notag\\
	&\hspace{1cm}\times\,F_4\left(\frac{3}{4}d-\D+\l+\m,\frac{3}{4}d-\D+\l+\m,1-\D+\frac{d}{2}+\l,1-\D+\frac{d}{2}+\m,\frac{p_1^2 p_2^2}{t^2 u^2},\frac{p_3^2 p_4^2}{t^2 u^2}\right)\bigg]
	\label{finf}
\end{align}
with a different coefficient in front, but with the coefficients $\xi(\l,\m)$ being the same of \eqref{xicoef}, completing the proof. \\
Notice that we could have gone through the analytic proof of the equivalence, by using even a third DCA, for instance of the form 
\begin{equation}
	\Phi'=\left(p_2^2\,p_4^2\right)^{n'_s}\,F\left(\frac{s^2t^2}{p_2^2p_4^2},\frac{p_1^2p_3^2}{p_2^2p_4^2}\right)
\end{equation}
and following the same procedure described above, we would have obtained an hypergeometric system of equations with a solution of the form
\begin{align}
	\label{sol2}
	\Phi'&=\left(p_2^2\,p_4^2\right)^{\D-\frac{3}{4}d}\bigg[C_1 F_4\left(\frac{d}{4}\,,\,\frac{3}{4}d-\D\, ,\,1\,,\,\frac{d}{2}-\D+1\,;\,\frac{p_1^2p_3^2}{p_2^2p_4^2}\, ,\,\frac{s^2t^2}{p_2^2p_4^2}\right)\notag\\
	&\hspace{2.5cm}+C_2\left(\frac{p_1^2p_3^2}{p_2^2p_4^2}\right)^{\D-\frac{d}{2}} F_4\left(\D-\frac{d}{4}\,,\,\frac{d}{4}\, ,\,1\,,\,\,1-\frac{d}{2}+\D\,;\,\frac{p_1^2p_3^2}{p_2^2p_4^2},\frac{s^2t^2}{p_2^2p_4^2}\right)\bigg],
\end{align}
as in \eqref{Solution1}. It can be explicitly shown that also in this case, by repeating the steps illustrated above, from \eqref{sol2} one arrives to \eqref{finalSol}. \\
We have indeed shown that DCC solutions take a unique form, independently of the structure of the original DCA. If we combine the results of section \ref{dccsection} with those above, it is clear that the solutions that we have found represent DCC correlators for any spacetime dimensions, of which the box diagram and its melonic variants 
are the only perturbtive realization, limited to $d=4$. 

\section{Convergence of the 3K solution integral and absence of physical singularities}\label{convergence}
The absence of unphysical singularities in the domain of convergence of the solution found, given in \eqref{ssm1}, can be addressed as follows. \\
Considering the DCC solution, we have derived its explicit expression as
\begin{align}
	&I_{\frac{d}{2}-1\{\Delta-\frac{d}{2},\Delta-\frac{d}{2},0\}}(p_1p_3,p_2p_4,s,t)=\notag\\
	&\qquad=\,(p_1p_3)^{\Delta-\frac{d}{2}}(p_2p_4)^{\Delta-\frac{d}{2}}\int_{0}^\infty\,dx\,x^{\frac{d}{2}-1}\,K_{\Delta-\frac{d}{2}}(p_1p_3\,x)\,K_{\Delta-\frac{d}{2}}(p_2p_4\,x)\,K_{0}(st\,x).\label{Sol}
\end{align}
Notice that a possible singularity which could invalidate the convergence of \eqref{Sol} can be generated by the Bessel function $K_0(x)$ at small $x$, as evident from the expansions 
\begin{align}
	\label{questoK}
	&K_{\nu}(x)\simeq\,\sqrt{\frac{\pi}{2}}\,\frac{e^{-x}}{\sqrt{x}}+\dots&&\text{at large $x$},\\
	&K_{\nu}(x)\simeq\,x^\nu\,\frac{\Gamma(-\nu)}{2^{1+\nu}}+x^{-\nu}\,\frac{\Gamma(\nu)}{2^{1-\nu}}+\dots&&\text{at small $x$}.
\end{align}
The singularity in $K_0$ can be regulated using the replacement 
$K_0\to K_\epsilon$, with $\epsilon$ a small regulator parameter $(\epsilon >0)$. For this purpose we consider the regulated expression
\begin{align}
	&I_{\frac{d}{2}-1\{\Delta-\frac{d}{2},\Delta-\frac{d}{2},\epsilon\}}(p_1p_3,p_2p_4,s,t)=\notag\\
	&\quad=\,(p_1p_3)^{\Delta-\frac{d}{2}}(p_2p_4)^{\Delta-\frac{d}{2}}(st)^\epsilon\int_{0}^\infty\,dx\,x^{\frac{d}{2}-1}\,K_{\Delta-\frac{d}{2}}(p_1p_3\,x)\,K_{\Delta-\frac{d}{2}}(p_2p_4\,x)\,K_{\epsilon}(st\,x).\label{sol}
\end{align}
With this regularization, at large $x$ the integrand  of \eqref{Sol} can be written as
\begin{align}
	\,x^{\frac{d}{2}-1}\,K_{\Delta-\frac{d}{2}}(p_1p_3\,x)\,K_{\Delta-\frac{d}{2}}(p_2p_4\,x)\,K_{\epsilon}(st\,x)\simeq\,(\sqrt{p_1p_2}\sqrt{p_3p_4}\sqrt{st})^{-1}\left(\frac{\pi}{2}\right)^{\frac{3}{2}}\,x^{\frac{d-5}{2}}\,e^{-(p_1p_3+p_2p_4+st)x}
\end{align}
which is well-behaved in the asymptotic region in $x$ if the condition
\begin{align}
	p_1p_3+p_2p_4+st>0
	\label{thecond}
\end{align}
is satisfied.\\
Similarly, the same integrand at small $x$ gives 
\begin{align}
	&\,x^{\frac{d}{2}-1}\,K_{\Delta-\frac{d}{2}}(p_1p_3\,x)\,K_{\Delta-\frac{d}{2}}(p_2p_4\,x)\,K_{\epsilon}(st\,x)\notag\\
	&\simeq\,x^{\frac{d}{2}-1}\left((p_1p_3\,x)^{\Delta-\frac{d}{2}}\frac{1}{2^{1+\Delta-\frac{d}{2}}}\Gamma\left(\frac{d}{2}-\Delta\right)+(p_1p_3\,x)^{\frac{d}{2}-\Delta}\frac{1}{2^{1+\frac{d}{2}-\Delta}}\Gamma\left(\Delta-\frac{d}{2}\right)\right)\notag\\
	&\times\left((p_2p_4\,x)^{\Delta-\frac{d}{2}}\frac{1}{2^{1+\Delta-\frac{d}{2}}}\Gamma\left(\frac{d}{2}-\Delta\right)+(p_2p_4\,x)^{\frac{d}{2}-\Delta}\frac{1}{2^{1+\frac{d}{2}-\Delta}}\Gamma\left(\Delta-\frac{d}{2}\right)\right)\left((st\,x)^\epsilon\,\frac{\Gamma(-\epsilon)}{2^{1+\epsilon}}+(st\,x)^{-\epsilon}\,\frac{\Gamma(\epsilon)}{2^{1-\epsilon}}\right) \label{expSmallx}
\end{align} 
and expanding the last factor in the previous expression - for small values of  the regulator $\epsilon$ - this takes the form 
\begin{align}
	(st\,x)^\epsilon\,\frac{\Gamma(-\epsilon)}{2^{1+\epsilon}}+(st\,x)^{-\epsilon}\,\frac{\Gamma(\epsilon)}{2^{1-\epsilon}}\simeq -\log(st\,x)-\gamma+\log(2)+O(\epsilon),
\end{align}
By combining all the contributions,  \eqref{expSmallx} can be rewritten as
\begin{align}
	&\,x^{\frac{d}{2}-1}\,K_{\Delta-\frac{d}{2}}(p_1p_3\,x)\,K_{\Delta-\frac{d}{2}}(p_2p_4\,x)\,K_{\epsilon}(st\,x)\simeq\log(st\, x)\,x^{\frac{d}{2}-1\pm\left(\Delta-\frac{d}{2}\right)\pm\left(\Delta-\frac{d}{2}\right)} +O(\epsilon)
\end{align}
which converges if the condition
\begin{align}
	\frac{d}{2}-1\pm\left(\Delta-\frac{d}{2}\right)\pm\left(\Delta-\frac{d}{2}\right)>0
	\label{scling}
\end{align}
is satisfied which branches into four possible constraints. 
One can check that the bound \eqref{thecond} is satisfied in the physical region 
\begin{align} 
	p_1+p_2+p_3+p_4>0
\end{align}
since 
\begin{equation}
	p_1 p_3+p_2p_4+s t>p_1+p_2+p_3+p_4>0
\end{equation}
and the convergence of the 3K representation is guaranteed if \eqref{scling} is satisfied.
The condition \eqref{scling} can generate in the physical region some divergences which need an appropriate regularization, as pointed out in \cite{Bzowski:2015yxv, Bzowski:2015pba} in the case of 3-point functions. A similar analysis of the singularities in view of the previous constraints is underway and the regularization procedure will be presented in a separate work.

\section{CWI's at fixed angle  and the Lauricella hypergeometric functions}
From this section on we turn to an analysis of another class of solutions of the CWI's, approximate in their character, which also show the hypergeometric nature of the system of equations derived from the CWI's, if we investigate such equations in a special kinematical limit. 

The hypergeometric nature of the CWI's can be shown if we resort to some approximations. \\
The second class of solutions that we are going to discuss are obtained by assuming particular asymptotic values of the $s$ and $t$ invariants. In this case the solution is generated by inspecting the contribution coming from the operatorial term $D_{s t} $ defined below in Eq. \eqref{dst}, which vanishes if it acts on a function of the form $\log(t/s)$. Such solution, for dimensional reason, is unique, and can be included in a factorized ans\"atz in order to generate a solution of the full equations. As we are going to show, the choice of such ans\"atz takes to solutions in which the dependence on the external mass invariants $p_i^2$ and the $s,t$ invariants are completely factorized and describe asymptotic solutions of the equations for large $s$ and $t$ invariants. In this case the rapidity variable $y=\log(t/s)$ can be associated with the behaviour of the correlator at fixed angle (i.e. with $s/t$ fixed and $O(1)$). The remaining part of the solutions, in this case, are expressed as a system of generalized hypergeometric (Lauricella) functions. We will show that such solutions can be expressed in terms of 4-K integrals, that we will define. \\   
For this purpose is helpful to identify several contributions in the expressions of the $C_i's$ given above taking $C_{13}$ as an example. Beside the operator ${K}_i$ given by \eqref{Koper1}, we define in general the operators
\begin{align}
	&J_{ij}=p_i\frac{\partial}{\partial p_i}+p_j\frac{\partial}{\partial p_j}, \qquad \tilde{J}_{ik}=p_i\frac{\partial}{\partial p_i}-p_k\frac{\partial}{\partial p_k} ,
	\label{firstref}\\[1.2ex]
	&J_{ij,kl}=p_i\frac{\partial}{\partial p_i}+p_j\frac{\partial}{\partial p_j}-p_k\frac{\partial}{\partial p_k}-p_l\frac{\partial}{\partial p_l}=J_{ij}-J_{kl}
\end{align}
\begin{equation}
	h_{ij,kl}=\D_i+\D_j-\D_k-\D_l,\qquad
	D^{t}=\frac{1}{t}\frac{\partial}{\partial t},\qquad  D^{s}=\frac{1}{s}\frac{\partial}{\partial s}
\end{equation}
and
{\begin{equation}
		\label{dst}
		D^{s t}_{ij}\equiv\frac{(p_i^2-p_j^2)}{s t}\frac{\partial^2}{\partial s\partial t}=(p_i^2-p_j^2) D^s D^t.
	\end{equation}
	These notations turns necessary when discussing the contributions of the various operators appearing in the equations in a compact way, but we will also turn to their their original (extended) expressions in order to avoid using indices, whenever possible.
	
	For instance, $C_{13}$ will take the form 
	\begin{equation}
		C_{13}=\left( {K}_{13} + D^s J_{12,34} +h_{34,12}D^s + D^t J_{14,23} +h_{23,14}D^t + D_{13}^{s t}\right)\Phi=0,
	\end{equation}
	while $C_1$ will be given by 
	\begin{equation}
		C_1=\left({K}_{14} +D^s J_{12,34}+ h_{34,12}D^s +D_{23}^{st}\right)\Phi=0.
		\label{intform}
	\end{equation}
	Using the definitons above, each equation can be characterized in terms of the set of operators $(K, J, h D, D^{s t})$. 
	We recognize in ${K}_{ij}$ the typical operators appearing in 3-point functions, which emerge when every form factor is 
	expressed in terms of the three external mass invariant, with the $J_{ij,kj}$ vanishing when the scaling dimensions of the same invariants are suitably balanced. For instance, given a function of two variables $f(z_1,z_2)$, 
	we will have 
	\begin{equation}
		J_{ij,kl}\, f\left(\frac{p_i^2}{p_j^2},\frac{p_k^2}{p_l^2}\right)=0 \qquad J_{ij,kl}\, f\left(p_i^2 p_k^2,p_j^2 p_l^2\right)=0
	\end{equation}
	and similar equations obtained by suitably exchanging $i,j,k,l$. If all the external invariants are grouped into a single 
	variable, for a given function $g(z)$, similarly we will obtain, for instance,
	\begin{equation}
		J_{ij} \,g\left(\frac{p_i^2}{ p_j^2}\right)=0 \qquad \tilde{J}_{ij} \,g\left({p_i^2}{ p_j^2}\right)=0 \qquad  J_{ij,kl}\,g(p_i^2 p_j^2 p_k^2 p_l^2)=0 \qquad J_{ij,kl}\, g\left( \frac{p_i^2}{ p_j^2}\frac{ p_k^2}{ p_l^2}\right)=0.
		\label{lastref}
	\end{equation}

	Beside the exact solutions identified in the previous sections, the CWI's allows other classes of solutions which may be found using a limited set of assumptions on the $s,t$ dependence of the ans\"atz. Therefore, we will proceed with an analysis of the special CWI's, trying to find approximate solutions of Eqs. \eqref{C1}-\eqref{C3}. We will adopt the notations introduced in Eqs. \eqref{firstref}-\eqref{lastref} in order to refer to the various terms of the corresponding partial differential equations. For definiteness we will consider the case of Eq. \eqref{C1}, rewritten in the form \eqref{intform}. We will assume that $s$  and $t$ are both large invariants but we will keep their ratio fixed. In the Minkowski region this would correspond to investigating the contribution of such correlator for scatterings at fixed angle (i.e. $-t/s$ fixed).\\
	We notice that if look for a factorised solution of the form 
	\begin{equation}
		\Phi(p_1,p_2,p_3,p_4)\equiv\chi(s,t) \phi(p_i^2),
		\label{phi}
	\end{equation}
	where we separate the dependence on the the external mass invariants $p_i^2$ from the $s,t$, we can satisfy the dilatation WI \eqref{dil1}  in the form 
	
	\begin{align}
		\label{dil21}
		&\bigg[(\D_t-3d)-\sum_{i=1}^4p_i\frac{\partial}{\partial p_i}\bigg]\phi(p_i^2)=0\\
		&\left(s\frac{\partial}{\partial s}+t\frac{\partial}{\partial t}\right)\chi(s,t)=0
		\label{dil22}
	\end{align}
	with $\chi(s,t)\equiv \chi(s/t)$, i.e. an arbitrary function of the ratio of the two external invariants, describing energy and momentum transfers.\\
	At this stage we can proceed with a separation of the special CWI \eqref{intform} into the three equations
	\begin{align}
		&D_{23}^{st}\chi(s/t)=0  \label{x1}\\
		&(D^s J_{12,34}+ h_{34,12}D^s)\chi(s/t) \phi(p_i^2)=0 \label{x2}\\
		&{K}_{14}\phi=0 \label{x3}
	\end{align}
	of which we try to identify an asymptotic solution.\\
	Notice that a simple but exact solution of the first of the three equations above is logarithmic with $\chi(s/t)\sim \log(-t/s)$. It is also easy to check, by plugging this expression into the second equation, that 
	\begin{equation}
		\left( D^s J_{12,34}+ h_{34,12}D^s\right)\chi(s/t) \phi(p_i^2)\sim O(1/s^2,1/t^2) 
	\end{equation}
	and contributes insignificantly if the mass invariants $p_i^2$ stays bound. Indeed we will consider solutions of the ratios 
	$p_i^2/p_j^2$ where this occurs. For this reason the solution of the last equation \eqref{x3} has to satisfy also \eqref{dil21}. We are clearly choosing to assign all the scaling behaviour of the global solution \eqref{phi} on the external mass invariants. If we require that  $p_i^2\sim O(1) \ll s, t$ then we can independently search for exact solutions of \eqref{x3}. 
	\subsection{Factorized solutions as generalized hypergeometrics}
	We can generalize these considerations to all the three CWI's \eqref{C1}, \eqref{C2}, \eqref{C3}, generating the system of equations 
	\begin{equation}
		{K}_{14}\phi=0,\qquad {K}_{24}\phi=0,\qquad {K}_{34}\phi=0\label{CWILaur}
	\end{equation}
	where 
	\begin{align}
		{K}_i&=\frac{\partial^2}{\partial p_i^2}+\frac{(d-2\D_i+1)}{p_i}\frac{\partial}{\partial p_i},\qquad i=1,\dots,4\ ,\\
		{K}_{ij}&={K}_i-{K}_j\ .
	\end{align}
	
	An equivalent way to rearrange this operator is to use a change of variables from $(p_1^2,p_2^2,p_3^2,p_4^2)$ to $(x,y,z,p_4^2)$ where 
	\begin{equation}
		x=\sdfrac{p_1^2}{p_4^2},\quad y=\sdfrac{p_2^2}{p_4^2},\quad z=\sdfrac{p_3^2}{p_4^2}
	\end{equation}
	are the dimensionless rations x, y and z which must not to be confused with coordinate points in a three dimensional space. The ans\"atz for the solution can be taken of the form
	\begin{equation}
		\phi(p_1,p_2,p_3,p_4)=(p_4^2)^{n_s}\,x^a\,y^b\,z^c\,F(x,y,z),
	\end{equation}
	satisfying the dilatation Ward identity \eqref{dil21}
	
	with the condition
	\begin{equation}
		n_s=\frac{\D_t}{2}-\frac{3d}{2}
	\end{equation}
	With this ans\"atz the conformal Ward identities read as
	\begin{align}
		K_{14}\phi=&4p_4^{\D_t-3d-2}\,x^a\,y^b\,z^c\,\bigg[(1-x)x\sdfrac{\partial^2}{\partial x^2}-2x\,y\sdfrac{\partial^2}{\partial x\partial y}-y^2\sdfrac{\partial^2}{\partial y^2}-2x\,z\sdfrac{\partial^2}{\partial x\partial z}-z^2\sdfrac{\partial^2}{\partial z^2}-2y\,z\sdfrac{\partial^2}{\partial y\partial z}\notag\\
		&\hspace{2cm}+(Ax+\gamma)\sdfrac{\partial}{\partial x}+Ay\sdfrac{\partial}{\partial y}+Az\sdfrac{\partial }{\partial z}+\left(E+\sdfrac{G}{x}\right)\bigg]F(x,y,z)=0
	\end{align}
	with
	\begin{subequations}
		\begin{align}
			A&=\D_1+\D_2+\D_3-\sdfrac{5}{2}d-2(a+b+c)-1\\
			E&=-\sdfrac{1}{4}\big(3d-\D_t+2(a+b+c)\big)\big(2d+2\D_4-\D_t+2(a+b+c)\big)\\
			G&=\sdfrac{a}{2}\,\left(d-2\D_1+2a\right)\\
			\g&=\sdfrac{d}{2}-\D_1+2a+1
		\end{align}
	\end{subequations}
	Similar constraints are obtained from the equation $K_{34}\phi=0$ that can be written as
	\begin{align}
		{K}_{24}\phi=&4p_4^{\D_t-3d-2}\,x^a\,y^b\,z^c\,\bigg[-x^2\sdfrac{\partial^2}{\partial x^2}-2x\,y\sdfrac{\partial^2}{\partial x\partial y}+(1-y)y\sdfrac{\partial^2}{\partial y^2}-2x\,z\sdfrac{\partial^2}{\partial x\partial z}-z^2\sdfrac{\partial^2}{\partial z^2}-2y\,z\sdfrac{\partial^2}{\partial y\partial z}\notag\\
		&\hspace{2cm}+A'x\sdfrac{\partial}{\partial x}+(A'y+\g')\sdfrac{\partial}{\partial y}+A'z\sdfrac{\partial }{\partial z}+\left(E'+\sdfrac{G'}{x}\right)\bigg]F(x,y,z)=0
	\end{align}
	with
	\begin{subequations}
		\begin{align}
			A'&=\D_1+\D_2+\D_3-\sdfrac{5}{2}d-2(a+b+c)-1\\
			E'&=-\sdfrac{1}{4}\big(3d-\D_t+2(a+b+c)\big)\big(2d+2\D_4-\D_t+2(a+b+c)\big)\\
			G'&=\sdfrac{b}{2}\,\left(d-2\D_2+2b\right)\\
			\g'&=\sdfrac{d}{2}-\D_2+2b+1
		\end{align}
	\end{subequations}
	and finally, for the third condition coming from the conformal Ward identities
	\begin{align}
		{K}_{34}\phi=&4p_4^{\D_t-3d-2}\,x^a\,y^b\,z^c\,\bigg[-x^2\sdfrac{\partial^2}{\partial x^2}-2x\,y\sdfrac{\partial^2}{\partial x\partial y}-y^2\sdfrac{\partial^2}{\partial y^2}-2x\,z\sdfrac{\partial^2}{\partial x\partial z}+(1-z)z\sdfrac{\partial^2}{\partial z^2}-2y\,z\sdfrac{\partial^2}{\partial y\partial z}\notag\\
		&\hspace{2cm}+A''x\sdfrac{\partial}{\partial x}+A''y\sdfrac{\partial}{\partial y}+(A''z+\g'')\sdfrac{\partial }{\partial z}+\left(E''+\sdfrac{G''}{x}\right)\bigg]F(x,y,z)=0
	\end{align}
	with
	\begin{subequations}
		\begin{align}
			A''&=\D_1+\D_2+\D_3-\sdfrac{5}{2}d-2(a+b+c)-1\\
			E''&=-\sdfrac{1}{4}\big(3d-\D_t+2(a+b+c)\big)\big(2d+2\D_4-\D_t+2(a+b+c)\big)\\
			G''&=\sdfrac{c}{2}\,\left(d-2\D_3+2c\right)\\
			\g''&=\sdfrac{d}{2}-\D_3+2c+1
		\end{align}
	\end{subequations}
	
	It is worth noticing that in order to perform the reduction to the hypergeometric form of the equations, we need to set $G=0$, $G'=0$ and $G''=0$, which imply that the Fuchsian points $a,b,c$ have different values as
	\begin{subequations}
		\begin{align}
			a&=0,\,\D_1-\sdfrac{d}{2}\\
			b&=0,\,\D_2-\sdfrac{d}{2}\\
			c&=0,\,\D_3-\sdfrac{d}{2}.
		\end{align}
	\end{subequations}
	We find also that $E=E'=E''=-\a(a,b,c)\,\b(a,b,c)$ where
	\begin{align}
		\a(a,b,c)&=d+\D_4-\sdfrac{\D_t}{2}+a+b+c\notag\\
		\b(a,b,c)&=\sdfrac{3d}{2}-\sdfrac{\D_t}{2}+a+b+c
	\end{align}
	as well as $A=A'=A''=-(\a(a,b,c)+\b(a,b,c)+1)$, indeed
	\begin{align}
		A=A'=A''&=-(\a(a,b,c)+\b(a,b,c)+1)=\D_1+\D_2+\D_3-\sdfrac{5}{2}d-2(a+b+c)-1
	\end{align}
	and finally
	\begin{equation}
		\Gamma(a)=\frac{d}{2}-\Delta_1+2a+1\,,\qquad\g'(b)=\frac{d}{2}-\Delta_2+2b+1\,,\qquad\g''(c)=\frac{d}{2}-\Delta_3+2c+1.
	\end{equation}
	With this redefinition of the coefficients, the equations are then expressed in the form
	\begin{equation}
		\left\{
		\begin{matrix}
			&x_j(1-x_j)\sdfrac{\partial^2F}{\partial x_j^2}+\hspace{-1cm}\sum\limits_{\substack{\hspace{1.3cm}s\ne j\ \text{for}\ r=j}}\hspace{-1.1cm}x_r\hspace{0.2cm}\sum x_s\hspace{0.5ex}\sdfrac{\partial^2F}{\partial x_r\partial x_s}+\left[\g_j-(\a+\b+1)x_j\right]\sdfrac{\partial F}{\partial x_j}-(\a+\b+1)\sum\limits_{k\ne j}\,x_k\sdfrac{\partial F}{\partial x_k}-\a\,\b\,F=0\\[3ex]
			& (j=1,2,3)
		\end{matrix}\right.\label{systemLauricella}
	\end{equation}
	where for sake of simplicity we have re-defined $\g_1=\g$, $\ \g_2=\g'$ and $\g_3=\g''$ and $x_1=x$, $x_2=y$ and $x_3=z$. 
	This system of equations allows solutions in the form of the Lauricella hypergeometric function $F_C$ of three variables, defined by the series 
	\begin{equation}
		F_C(\a,\b,\g,\g',\g'',x,y,z)=\sum\limits_{m_1,m_2,m_3}^\infty\,\frac{(\a)_{m_1+m_2+m_3}(\b)_{m_1+m_2+m_3}}{(\g)_{m_1}(\g')_{m_2}(\g'')_{m_3}m_1!\,m_2!\,m_3!}x^{m_1}y^{m_2}z^{m_3}.
	\end{equation}
	where the Pochhammer symbol $(\l)_{k}$ with an arbitrary $\l$ and $k$ a positive integer not equal to zero, was previously defined in \eqref{Pochh}. The convergence region of this series is defined by the condition
	\begin{equation}
		\left|\sqrt{x}\right|+\left|\sqrt{y}\right|+\left|\sqrt{z}\right|<1.
	\end{equation}
	The function $F_C$ is the generalization of the Appell $F_4$ for the case of three variables.
	The system of equations \eqref{systemLauricella} admits 8 independent particular integrals (solutions) listed below
	\begin{align}
		&S_1(\a,\b,\g,\g',\g'',x,y,z)=F_C\big(\a,\b,\g,\g',\g'',x,y,z\big)\notag,\\
		&S_2(\a,\b,\g,\g',\g'',x,y,z)=x^{1-\g}\,F_C\big(\a-\g+1,\b-\g+1,2-\g,\g',\g'',x,y,z\big)\notag\,,\\
		&S_3(\a,\b,\g,\g',\g'',x,y,z)= y^{1-\g'}\,F_C\big(\a-\g'+1,\b-\g'+1,\g,2-\g',\g'',x,y,z\big)\notag\,,\\
		&S_4(\a,\b,\g,\g',\g'',x,y,z)=z^{1-\g''}\,F_C\big(\a-\g''+1,\b-\g''+1,\g,\g',2-\g'',x,y,z\big)\notag,\,\\
		&S_5(\a,\b,\g,\g',\g'',x,y,z)=x^{1-\g}y^{1-\g'}\,F_C\big(\a-\g-\g'+2,\b-\g-\g'+2,2-\g,2-\g',\g'',x,y,z\big)\,,\notag\\
		&S_6(\a,\b,\g,\g',\g'',x,y,z)=x^{1-\g}z^{1-\g''}\,F_C\big(\a-\g-\g''+2,\b-\g-\g''+2,2-\g,\g',2-\g'',x,y,z\big)\notag\,,\\
		&S_7(\a,\b,\g,\g',\g'',x,y,z)=y^{1-\g'}z^{1-\g''}\,F_C\big(\a-\g'-\g''+2,\b-\g'-\g''+2,\g,2-\g',2-\g'',x,y,z\big)\notag\,,\\
		&S_8(\a,\b,\g,\g',\g'',x,y,z)=x^{1-\g}y^{1-\g'}z^{1-\g''}\notag\\
		&\hspace{4cm}\times\,F_C\big(\a-\g-\g'-\g''+2,\b-\g-\g'-\g''+2,2-\g,2-\g',2-\g'',x,y,z\big)\,.\label{oneeq}
	\end{align} 
	where we have defined 
	\begin{align}
		\a&\equiv\a(0,0,0)=d+\D_4-\sdfrac{\D_t}{2}\notag\\	
		\b&\equiv\b(0,0,0)=\sdfrac{3d}{2}-\sdfrac{\D_t}{2}\notag\\
		\g&\equiv\gamma(0)=\frac{d}{2}-\Delta_1+1\notag\\
		\g'&\equiv\g'(0)=\frac{d}{2}-\Delta_2+1\notag\\
		\g''&\equiv\g''(0)=\frac{d}{2}-\Delta_3+1.
	\end{align}
	Finally the solution for $\phi$  can be written as
	\begin{equation}
		\phi(p_i^2)=p_4^{\D_t-3d}\sum_i C_i\ S_i(\a,\b,\g,\g',\g'',x,y,z)
	\end{equation}
	where $C_i$ are arbitrary constant and $S_i$, $i=1,\dots, 2^3$ are the independent solutions written above. \\
	To summarize, we have indeed shown that approximate solutions of the CWI's, describing the behaviour of the correlator at fixed angle 
	can be taken of the factorized form 
	\begin{equation}
		\Phi(p_1,p_2,p_3,p_4)\sim \log(-t/s) \phi(p_i^2).
	\end{equation}
	We should remark that other approximate solutions of similar form, containing higher powers of logarithms of $-t/s$ are also compatible with the asymptotic ans\"atz that we have presented here. Obviously, in such a case we would be requiring that the exact condition \eqref{x1} would be replaced by the new condition 
	\begin{equation}
		D_{23}^{st}\chi(s/t)=O(1/s^2,1/t^2)  
	\end{equation}
	which is asymptotically satisfied also by higher powers of $\log(-t/s)$. In general, under such weaker assumptions, approximate asymptotic solutions can be summarized in the more general form
	\begin{equation}
		\Phi(p_1,p_2,p_3,p_4)\sim f\left(\log(-t/s) \right)\phi(p_i^2).
	\end{equation}
	where $f$ can be take of the generic form 
	\begin{equation} 
		f\left(\log(-t/s)\right) = \sum_k c_k \log^k\left((-t/s)\right).
	\end{equation}
	In the next section we are going to show that for the $p_i^2$ dependence on the external mass invariants of the approximate solution, given by the Lauricella functions, their equivalence to 4-K integrals, generalizing previous results for 3-point functions. 
	\subsection{Lauricella's as 4-K integrals}
	It is interesting to show how the solutions found above can be reformulated in a way which resembles what found in the case of 3-point functions. As alredy mentioned, the 3K integrals provide an efficient alternative way to express the solutions for scalar 3-point functions in terms of Appell functions. We are now going to show that  hypergeometrics of 3-variables, which belong to the class of Lauricella functions, similarly, can be related to 4K integrals. 
	We write the solutions of such systems in the form
	\begin{align}
		I_{\a-1\{\n_1,\n_2,\n_3,\n_4\}}(a_1,a_2,a_3,a_4)&=\int_0^\infty\,dx\,x^{\a-1}\,\prod_{i=1}^4(a_i)^{\n_i}\,K_{\n_i}(a_i\,x)
		\label{4Kintegral}
	\end{align} 
	with the Bessel functions $I_\nu,J_\nu, K_\nu$ related by the identities
	\begin{align}
		I_\nu(x)&=i^{-\n}\,J_{\n}(i\,x)\\
		K_\nu(x)&=\frac{\pi}{2\sin(\pi\,\n)}\bigg[I_{-\n}(x)-I_\n(x)\bigg]=\frac{1}{2}\bigg[i^\nu\, \Gamma(\n)\Gamma(1-\n)\,J_{-\n}(i\,x)+i^{-\n}\,\Gamma(-\n)\Gamma(1+\n)\,J_\n(i\,x)\bigg]\label{Kscomp}
	\end{align}
	where we have used the properties of the Gamma functions
	\begin{equation}
		\frac{\pi}{\sin(\pi\n)}=\Gamma(\n)\,\Gamma(1-\n),\qquad-\frac{\pi}{\sin(\pi\n)}=\Gamma(-\n)\,\Gamma(1+\n).
	\end{equation}
	
	The structure of the CWI's \eqref{CWILaur} supports this formulation. The dilatation Ward identities in this case can be written as
	\begin{equation}
		\bigg[(\D_t-3d)-\sum_{i=1}^4p_i\frac{\partial}{\partial p_i}\bigg]I_{\a\{\b_1,\b_2,\b_3,\b_4\}}(p_1,p_2,p_3,p_4)=0
	\end{equation}
	and using the properties of 4K integrals in \appref{AppendixB} we derive the relation
	\begin{align}
		(\a-\b_t+1+\D_t-3d)I_{\a\{\b_1,\b_2,\b_3,\b_4\}}(p_1,p_2,p_3,p_4)=0
	\end{align}
	which is identically satisfied if the $\a$ exponent is equal to $\tilde{\a}$
	\begin{equation}
		\tilde{\a}=\b_t+3d-\D_t-1.
	\end{equation}
	The conformal Ward identities \eqref{CWILaur} can now be written as
	\begin{equation}
		\left\{
		\begin{aligned}
			{K}_{14}I_{\tilde{\a}\{\b_1,\b_2,\b_3,\b_4\}}&=0\\
			{K}_{24}I_{\tilde{\a}\{\b_1,\b_2,\b_3,\b_4\}}&=0\\
			{K}_{34}I_{\tilde{\a}\{\b_1,\b_2,\b_3,\b_4\}}&=0,
		\end{aligned}
		\right.
	\end{equation}
	generating the final relations
	\begin{equation}
		\left\{
		\begin{aligned}
			(d-2\D_4+2\b_4)I_{\tilde{\a}+1\{\b_1,\b_2,\b_3,\b_4-1\}}-(d-2\D_1+2\b_1)I_{\tilde{\a}+1\{\b_1-1,\b_2,\b_3,\b_4\}}&=0\\
			(d-2\D_4+2\b_4)I_{\tilde{\a}+1\{\b_1,\b_2,\b_3,\b_4-1\}}-(d-2\D_2+2\b_2)I_{\tilde{\a}+1\{\b_1,\b_2-1,\b_3,\b_4\}}&=0\\
			(d-2\D_4+2\b_4)I_{\tilde{\a}+1\{\b_1,\b_2,\b_3,\b_4-1\}}-(d-2\D_3+2\b_3)I_{\tilde{\a}+1\{\b_1,\b_2,\b_3-1,\b_4\}}&=0\\
		\end{aligned}
		\right.
	\end{equation}
	which are satisfied if
	\begin{align}
		\b_i=\D_i-\frac{d}{2},\qquad i=1,\dots,4
	\end{align}
	giving
	\begin{equation}
		\tilde{\a}=d-1.
	\end{equation}
	The final solution can be written as
	\begin{align}
		\phi(p_1,p_2,p_3,p_4)&=\bar{\bar{\a}}\, I_{d-1\left\{\D_1-\frac{d}{2},\D_2-\frac{d}{2},\D_3-\frac{d}{2},\D_4-\frac{d}{2}\right\}}(p_1,p_2,p_3,p_4)\notag\\
		&=\int_0^\infty\,dx\,x^{d-1}\,\prod_{i=1}^4(p_i)^{\D_i-\frac{d}{2}}\,K_{\D_i-\frac{d}{2}}(p_i\,x).
		,\label{4Kfin}
	\end{align}
	where $\bar{\bar{\a}}$ is a undetermined constant. \\
	Concerning the convergence of the approximate 4K solutions found in the fixed angle scattering limit at large $s$ and $t$, one can discuss the general conditions to be 
	imposed, by following a strategy quite similar to the one discussed in \secref{convergence}.
	
	The asymptotic limit at large and small x values, also in this case previously shown in 
	\eqref{questoK} et seq., gives the conditions
	\begin{align}
		&p_1+p_2+p_3+p_4>0
	\end{align}
	and
	\begin{align}
		&\frac{d}{2}-1\pm\left(\Delta_1-\frac{d}{2}\right)\pm\left(\Delta_2-\frac{d}{2}\right)\pm\left(\Delta_3-\frac{d}{2}\right)\pm\left(\Delta_4-\frac{d}{2}\right)>0,
	\end{align}
	respectively. Therefore the condition of convergence at large $x$ of the parametric representation of the 4K integral is verified within the physical region of the general scalar 4-point function. Also in this case, as in \secref{convergence}, a discussion of implications of such convergence constraints will be presented in a related work.

	\subsection{Connection with the Lauricella}
	The key identity necessary to obtain the relation between the Lauricella functions and the 4K integral takes the form
	\begin{align}
		\int_0^\infty dx\,x^{\a-1}\prod_{j=1}^3\,J_{\m_j}(a_j\,x)\,K_{\nu}(c\,x)&=2^{\a-2}\,c^{-\a-\l}\,\Gamma\left(\frac{\a+\l-\n}{2}\right)\Gamma\left(\frac{\a+\l+\n}{2}\right)\notag\\
		&\hspace{-3cm}\times\prod_{j=1}^3\,\frac{a_j^{\m_j}}{\Gamma(\m_j+1)}F_C\left(\frac{\a+\l-\n}{2},\frac{\a+\l+\n}{2},\m_1+1,\m_2+1,\m_3+1;-\frac{a_1^2}{c^2},-\frac{a_2^2}{c^2},-\frac{a_3^2}{c^2}\right)\notag\\
		&\centering\,\bigg[\l=\sum_{j=1}^3\,\m_j\,;\, \Re(\a+\l)>|\Re(\n)|,\,\Re(c)>\sum_{j=1}^{3}|\Im\,a_j|\bigg]\label{Prudnikov}.
	\end{align}
	One of the advantages of the use of the 4K integral expression of a solution is the simplified way by which the symmetry conditions can be imposed. In fact, by taking each of the 8 independent solutions identified in \eqref{oneeq}, and by rewriting them in the form of 4K integrals, we can impose the symmetry constraints far more easily.
	Then the general 4K integral in \eqref{4Kintegral}, using \eqref{Kscomp}, can be written as
	
	\begin{align}
		\Phi(p_i^2)&=2^{d-5}p_4^{\D_t-3d}\, C_{1234}\bigg\{\,\Gamma\left(\D_1-\frac{d}{2}\right)\Gamma\left(\D_2-\frac{d}{2}\right)\Gamma\left(\D_3-\frac{d}{2}\right)\,\Gamma\left(\frac{3d-\D_t}{2}\right)\Gamma\left(d+\D_4-\frac{\D_t}{2}\right)\notag\\
		&\hspace{1cm}\times F_C^{(3)}\left(\frac{3d}{2}-\frac{\D_t}{2}\,,\,d+\D_4-\frac{\D_t}{2},\frac{d}{2}-\D_1+1,\frac{d}{2}-\D_2+1,\frac{d}{2}-\D_3+1;\frac{p_1^2}{p_4^2},\frac{p_2^2}{p_4^2},\frac{p_3^2}{p_4^2}\right)\notag\\[1.5ex]
		&+\Gamma\left(\D_1-\frac{d}{2}\right)\Gamma\left(\D_2-\frac{d}{2}\right)\Gamma\left(\frac{d}{2}-\D_3\right)\,\Gamma\left(\frac{d}{2}-\frac{\D_t}{2}+\D_3+\D_4\right)\Gamma\left(d+\D_3-\frac{\D_t}{2}\right)\left(\frac{p_3^2}{p_4^2}\right)^{\D_3-\frac{d}{2}}\notag\\
		&\hspace{1cm}\times F_C^{(3)}\left(d-\frac{\D_t}{2}+\D_3\,,\,\frac{d}{2}-\frac{\D_t}{2}+\D_3+\D_4,\frac{d}{2}-\D_1+1,\frac{d}{2}-\D_2+1,1-\frac{d}{2}+\D_3;\frac{p_1^2}{p_4^2},\frac{p_2^2}{p_4^2},\frac{p_3^2}{p_4^2}\right)\notag\\[1.5ex]
		&+\Gamma\left(\D_1-\frac{d}{2}\right)\Gamma\left(\frac{d}{2}-\D_2\right)\Gamma\left(\D_3-\frac{d}{2}\right)\,\Gamma\left(d-\frac{\D_t}{2}+\D_2\right)\Gamma\left(\frac{d}{2}-\frac{\D_t}{2}+\D_2+\D_4\right)\left(\frac{p_2^2}{p_4^2}\right)^{\D_2-\frac{d}{2}}\notag\\
		&\hspace{1cm}\times F_C^{(3)}\left(d-\frac{\D_t}{2}+\D_2\,,\,\frac{d}{2}-\frac{\D_t}{2}+\D_2+\D_4,\frac{d}{2}-\D_1+1,1-\frac{d}{2}+\D_2,\frac{d}{2}-\D_3+1;\frac{p_1^2}{p_4^2},\frac{p_2^2}{p_4^2},\frac{p_3^2}{p_4^2}\right)\notag\\[1.5ex]
		&+\Gamma\left(\frac{d}{2}-\D_1\right)\Gamma\left(\D_2-\frac{d}{2}\right)\Gamma\left(\D_3-\frac{d}{2}\right)\,\Gamma\left(d-\frac{\D_t}{2}+\D_1\right)\Gamma\left(\frac{d}{2}-\frac{\D_t}{2}+\D_1+\D_4\right)\left(\frac{p_1^2}{p_4^2}\right)^{\D_1-\frac{d}{2}}\notag\\
		&\hspace{1.5cm}\times F_C^{(3)}\left(d-\frac{\D_t}{2}+\D_1\,,\,\frac{d}{2}-\frac{\D_t}{2}+\D_1+\D_4,1-\frac{d}{2}+\D_1,\frac{d}{2}-\D_2+1,\frac{d}{2}-\D_3+1;\frac{p_1^2}{p_4^2},\frac{p_2^2}{p_4^2},\frac{p_3^2}{p_4^2}\right)\notag\\[1.5ex]
		&+\Gamma\left(\D_1-\frac{d}{2}\right)\Gamma\left(\frac{d}{2}-\D_2\right)\Gamma\left(\frac{d}{2}-\D_3\right)\,\Gamma\left(\frac{\D_t}{2}-\D_1\right)\Gamma\left(\frac{d}{2}-\frac{\D_t}{2}+\D_2+\D_3\right)\left(\frac{p_2^2}{p_4^2}\right)^{\D_2-\frac{d}{2}}\left(\frac{p_3^2}{p_4^2}\right)^{\D_3-\frac{d}{2}}\notag\\
		&\hspace{1.5cm}\times F_C^{(3)}\left(\frac{d}{2}-\frac{\D_t}{2}+\D_2+\D_3\,,\,\frac{\D_t}{2}-\D_1,1+\frac{d}{2}-\D_1,1-\frac{d}{2}+\D_2,1-\frac{d}{2}+\D_3;\frac{p_1^2}{p_4^2},\frac{p_2^2}{p_4^2},\frac{p_3^2}{p_4^2}\right)\notag\\
		&+\Gamma\left(\frac{d}{2}-\D_1\right)\Gamma\left(\frac{d}{2}-\D_2\right)\Gamma\left(\D_3-\frac{d}{2}\right)\,\Gamma\left(\frac{\D_t}{2}-\D_3\right)\Gamma\left(\frac{d}{2}-\frac{\D_t}{2}+\D_1+\D_2\right)\left(\frac{p_1^2}{p_4^2}\right)^{\D_1-\frac{d}{2}}\left(\frac{p_2^2}{p_4^2}\right)^{\D_2-\frac{d}{2}}\notag\\[1.5ex]
		&\hspace{1.5cm}\times F_C^{(3)}\left(\frac{\D_t}{2}-\D_3\,,\,\frac{d}{2}-\frac{\D_t}{2}+\D_1+\D_2,1+\D_1-\frac{d}{2},1+\frac{d}{2}-\D_2,1-\frac{d}{2}+\D_3;\frac{p_1^2}{p_4^2},\frac{p_2^2}{p_4^2},\frac{p_3^2}{p_4^2}\right)\notag\\[1.5ex]
		&+\Gamma\left(\frac{d}{2}-\D_1\right)\Gamma\left(\D_2-\frac{d}{2}\right)\Gamma\left(\frac{d}{2}-\D_3\right)\,\Gamma\left(\frac{\D_t}{2}-\D_2\right)\Gamma\left(\frac{d}{2}-\frac{\D_t}{2}+\D_1+\D_3\right)\left(\frac{p_1^2}{p_4^2}\right)^{\D_1-\frac{d}{2}}\left(\frac{p_3^2}{p_4^2}\right)^{\D_3-\frac{d}{2}}\notag\\
		&\hspace{1.5cm}\times F_C^{(3)}\left(\frac{\D_t}{2}-\D_2\,,\,\frac{d}{2}-\frac{\D_t}{2}+\D_1+\D_3,1+\D_1-\frac{d}{2},1+\frac{d}{2}-\D_2,1-\frac{d}{2}+\D_3;\frac{p_1^2}{p_4^2},\frac{p_2^2}{p_4^2},\frac{p_3^2}{p_4^2}\right)\notag
	\end{align}
	\begin{align}
		&+\Gamma\left(\frac{d}{2}-\D_1\right)\Gamma\left(\frac{d}{2}-\D_2\right)\Gamma\left(\frac{d}{2}-\D_3\right)\,\Gamma\left(\frac{3d}{2}-\frac{\D_t}{2}\right)\Gamma\left(d-\frac{\D_t}{2}+\D_4\right)\left(\frac{p_1^2}{p_4^2}\right)^{\D_1-\frac{d}{2}}\left(\frac{p_2^2}{p_4^2}\right)^{\D_2-\frac{d}{2}}\left(\frac{p_3^2}{p_4^2}\right)^{\D_3-\frac{d}{2}}\notag\\
		&\hspace{0.5cm}\times F_C^{(3)}\left(\frac{3d}{2}-\frac{\D_t}{2}\,,\,d-\frac{\D_t}{2}+\D_4,1+\D_1-\frac{d}{2},1+\D_2-\frac{d}{2},1-\frac{d}{2}+\D_3;\frac{p_1^2}{p_4^2},\frac{p_2^2}{p_4^2},\frac{p_3^2}{p_4^2}\right)\bigg\}
	\end{align}
	where $C_{1234}$ is the only undetermined constant. 
	
	\section{Conclusions}
	In this chapter, we have investigated two classes of solutions of the CWI's of scalar primary correlators in momentum space. In the first class, we have identified solutions in the form of 4-point functions that are dual conformal and conformal simultaneously. Such solutions have been found using ans\"atze which reduces the equations to generalized hypergeometrics systems. The method extends previous analysis of 3-point functions for scalar and tensor correlators, limitedly to DCC solutions, which can be expressed in terms of 3K integrals, similarly to the case of ordinary 3-point functions. \\ 
	We have also discussed how one can construct solutions of the CWI's, by showing that at large $s$ and $t$, with a fixed $-t/s$, i.e. at fixed angle, the CWI's are approximated by a system of special hypergeometric equations, which can be solved by a specific factorized ansatz. In the ansatz, which is an exact solution of such a system, the dependence of the correlators on the external mass invariants is separated from the $s$ and $t$ invariants. We have shown that the solutions, in this case, take the form of Lauricella hypergeometric functions of 3 variables. The $s$ and $t$ dependence of the solutions is compatible with such correlators' structure at fixed angle in the asymptotic limit, due to the logarithmic $-t/s$ dependence, typical of such solutions.\\
	Finally, we have shown that the Lauricella solutions system is equivalent to some newly introduced 4K integrals. Would be very interesting to investigate whether this pattern can be extended to n-point functions, in the context of more realistic field theories such as QCD, for instance, following the analysis presented in \cite{Kidonakis:1998nf,Sterman:2002qn,Aybat:2006mz}.
\chapter{Reconstruction method for spinning correlators in momentum space}\label{tensstruc}
In this chapter, we present the general method of decomposition of $3$-point functions involving tensorial operators. This method, presented in \cite{ Bzowski:2014qja, Bzowski:2018fql,Bzowski:2017poo,Bzowski:2011ab}, is based on the reconstruction of the full $3$-point functions involving stress-energy tensors, currents, and scalar operators starting from the expressions of transverse and traceless part only. In order to show all the steps of the method, we will present a fully worked out example, the $\braket{T^{\m_1\n_1}T^{\m_2\n_2}\mO}$ correlation function. Furthermore, we will present the solutions for the $TJJ$ and $TTT$ correlators. These results will be significant in discussing the next chapters, where the matching between the general and perturbative realizations of these correlators is presented. 
This reconstruction method can be generalized to the 4-point function correlators as pointed out in \cite{Bzowski:2013sza,Coriano:2019nkw}. 

\section{The example of the $TTO$}
As an example consider a 3-point function of two transverse, traceless, symmetric rank-2 operators $T^{\m\n}$ and a scalar operator $\mO$.
By using the transverse and traceless projectors \eqref{Proj}, it is possible to write the most general form of the transverse and traceless part as
\begin{equation}
	\braket{t^{\m_1\n_1}(p_1)\,t^{\m_2\n_2}(p_2)\,\mO(p_3)}=\Pi^{\m_1\n_1}_{\a_1\b_1}(p_1)\Pi^{\m_2\n_2}_{\a_2\b_2}(p_2)\,X^{\a_1\b_1\a_2\b_2},\label{decomp}
\end{equation}
where $X^{\a_1\b_1\a_2\b_2}$ is a general tensor of rank-4 built from the metric and the momenta. By using the conservation of the total momentum and the properties of the projectors, one ends up with the general form of our 3-point function 
\begin{equation}
	\braket{t^{\m_1\n_1}(p_1)\,t^{\m_2\n_2}(p_2)\,\mO(p_3)}=\Pi^{\m_1\n_1}_{\a_1\b_1}(p_1)\Pi^{\m_2\n_2}_{\a_2\b_2}(p_2)\left[A_1\,p_2^{\a_1}p_2^{\b_1}p_3^{\a_2}p_3^{\b_2}+A_2\,p_2^{\a_1}p_3^{\a_2}\d^{\b_1\b_2}+A_3\,\d^{\a_1\a_2}\d^{\b_1\b_2}\right]\label{TTOdec}
\end{equation}
where we also use the symmetry properties of the projectors in $\m\leftrightarrow\n$, $\a\leftrightarrow\b$, and the coefficient $A_1$, $A_2$ and $A_3$ are the \emph{form factors}, scalar functions of momenta. By Lorentz invariance, these form factors are functions of the momentum magnitudes
\begin{equation}
	p_j=\sqrt{p_j^2},\quad j=1,2,3
\end{equation} 
and in \eqref{TTOdec} we suppressed the dependence of form factors on the momentum magnitudes, writing simply $A_j$ despite of $A_j(p_1,p_2,p_3)$. The symmetry condition under the exchange $(p_1,\mu_1\nu_1)\leftrightarrow(p_2,\mu_2\nu_2)$ reflects in the behaviour of the form factors under this permutation and in particular they have the following symmetric properties
\begin{align}
	A_i(p_1\leftrightarrow p_2)=A_i, \qquad i=1,2,3. \label{symA}
\end{align}

\subsection{The method of reconstruction}
In this section we review how the reconstruction method proposed in \cite{Bzowski:2013sza} works for the entire correlation functions and in particular for the $TTO$ case. We have introduced the transverse and traceless part of $TTO$, that can be defined, in terms of operators as
\begin{equation}
	t^{\m\n}(p)\equiv \Pi^{\m\n}_{\a\b}(p)\,T^{\a\b}(p),
\end{equation}
then we define the local part of $T^{\mu\nu}$ as the difference
\begin{equation}
	t^{\m\n}_{loc}\equiv\,T^{\m\n}-t^{\m\n}=\,\mathcal I^{\m\n}_{\a\b}\ T^{\a\b},
\end{equation}
where
\begin{equation}
	\mathcal{I}^{\m\n}_{\a\b}(p)=\sdfrac{p_\beta}{p^2}\left[2p^{(\m}\d^{\n)}_\a-\sdfrac{p_\a}{d-1}\left(\d^{\m\n}+(d-2)\sdfrac{p^\m p^\n}{p^2}\right)\right]+\sdfrac{\pi^{\m\n}(p)}{d-1}\,\d_{\a\b}.\label{itensor}
\end{equation}
This procedure can be done also for spin-$1$ conserved currents $J^\mu$ as illustrated in \cite{Bzowski:2013sza}. 
We now observe that in a CFT, all terms involving $t_{loc}$ can be computed by means of the transverse and trace Ward identities. One can therefore divide a 3-point function into two parts: the \emph{transverse-traceless} part, and the \emph{semi-local part} (indicated by subscript $loc$) expressible through the transverse Ward Identities. This translates, for the case of $TTO$, into considering the decomposition
\begin{equation}
	\braket{T^{\m_1\n_1}T^{\m_2\n_2}\mO}=\braket{t^{\m_1\n_1}t^{\m_2\n_2}\mO}+\braket{t_{loc}^{\m_1\n_1}T^{\m_2\n_2}\mO}+\braket{T^{\m_1\n_1}t_{loc}^{\m_2\n_2}\mO}-\braket{t_{loc}^{\m_1\n_1}t_{loc}^{\m_2\n_2}\mO}\label{reconstr}
\end{equation}
where the definition of the transverse and traceless part $\braket{t^{\mu_1\nu_1}t^{\mu_2\nu_2}\mathcal{O}}$ has been given in \eqref{TTOdec}. As previously mentioned, all terms on the right-hand side, apart from the first may be computed by means of Ward Identities, but more importantly, all these terms depend on 2-point functions only. Thus, the unknown information about the $3$-point function is encoded just in the transverse traceless part. 
In the following sections we will obtain more information about the form factors in the tensorial decomposition using the conformal Ward identities. In particular we will obtain constraints on the form factors in terms of differential equations that will be solved in terms of the triple-K integrals or hypergeometric functions.

\subsection{Transverse and trace Ward identities}\label{TraceTransverse}

We show in \appref{transWard} how to obtain the Ward identities starting from the general requirement that the generating functional is invariant under some symmetry transformations. The relevant Ward identities in the presence of external sources are
\begin{equation}
	\begin{split}
		&\nabla_\n\braket{T^{\m\n}}+\partial^\n\phi_0\braket{\mO}=0,\\
		&g_{\m\n}\braket{T^{\m\n}}+(d-\D)\phi_0\braket{\mO}=0,
	\end{split}\label{startpoint}
\end{equation}
that correspond to the transverse and trace Ward identities respectively. Recall that $\nabla_\n$ denotes the covariant derivative with respect to the background metric $g_{\m\n}$. Multiplying these relations by $\sqrt{-g}/2$, we can rewrite them in the form
\begin{align}
	&\partial_\m\left(\sdfrac{\d Z}{\d g_{\m\n}(x)}\right)+\G^\n_{\m\l}\sdfrac{\d Z}{\d g_{\m\l}(x)}+\sdfrac{1}{2}\partial^\n\phi_0(x)\sdfrac{\d Z}{\d \phi_0(x)}=0\label{transverse1}\\
	&g_{\m\n}\sdfrac{\d Z}{\d g_{\m\n}(x)}+\sdfrac{1}{2}(d-\D)\phi_0(x)\sdfrac{\d Z}{\d\phi_0}=0,\label{traceW}
\end{align}

where $\G$ is the usual Christoffel connection where we have introduced the  expectation value of the energy momentum tensor $(\langle T \rangle)$. In order to derive the transverse and trace Ward identities for the $TT\mO$ correlator, we vary twice \eqref{transverse1} and \eqref{traceW} with respect to the metric $g_{\m_2\n_2}(x_2)$ and the scalar field source $\phi_0(x_3)$. After the functional variations, if one switches off the sources, the result will be 
\begin{align}
	&\partial_\m\left.\left(\sdfrac{\d^3 Z}{\d g_{\m\n}(x)\d g_{\m_2\n_2}(x_2)\d\phi_0(x_3)}\right)\right|_{g_{\m\n}=\d_{\m\n}}+\sdfrac{1}{2}\big(\partial^\n\d(x-x_3)\big)\left.\left(\sdfrac{\d^2 Z}{\d \phi_0(x)\d g_{\m_2\n_2}(x_2)}\right)\right|_{g_{\m\n}=\d_{\m\n}}\notag\\
	&\hspace{2cm}+\left.\left[\big(\,\d^{\n(\n_2}\d^{\m_2)}_\l\partial_\m-\sdfrac{1}{2}\d_\m^{(\m_2}\d^{\n_2)}_\l\d^{\n\e}\partial_\e\big)\d(x-x_2)\right]\left(\sdfrac{\d^2 Z}{\d g_{\m\l}(x)\d\phi_0(x_3)}\right)\right|_{g_{\m\n}=\d_{\m\n}}=0
\end{align}
and restoring the correlation function through the definitions
\begin{align}
	\braket{T^{\m\n}(x)T^{\m_2\n_2}(x_2)\,\mO(x_3)}&=4\left.\left(\sdfrac{\d^3 Z}{\d g_{\m\n}(x)\d g_{\m_2\n_2}(x_2)\d\phi_0(x_3)}\right)\right|_{g_{\m\n}=\d_{\m\n}}\label{defTTO}\\
	\braket{T^{\m\n}(x)\,\mO(x_3)}&=2\left.\left(\sdfrac{\d^2 Z}{\d \phi_0(x_3)\d g_{\m\n}(x_2)}\right)\right|_{g_{\m\n}=\d_{\m\n}}
\end{align}
we obtain the transverse Ward identities in position space
\begin{align}
	\partial_\m\braket{T^{\m\n}(x)T^{\m_2\n_2}(x_2)\,\mO(x_3)}&=-2\,\d^{\n(\n_2}\d^{\m_2)}_\l\partial_\m\d(x-x_2)\braket{T^{\m\l}(x)\,\mO(x_3)}\notag\\
	&+\partial^\n\d(x-x_2)\braket{T^{\m_2\n_2}(x)\,\mO(x_3)}-\partial^\n\d(x-x_3)\braket{T^{\m_2\n_2}(x_2)\,\mO(x)}\label{transverse2}.
\end{align}

Using the same procedure we obtain the trace Ward identities in position space for the same correlation function
\begin{align}
	g_{\m\n}(x)\braket{T^{\m\n}(x)T^{\m_2\n_2}(x_2)\mO(x_3)}=(\D-d)\d(x-x_3)\braket{T^{\m_2\n_2}(x_2)\mO(x)}-2\d(x-x_2)\braket{T^{\m_2\n_2}(x)\mO(x_3)}.\label{trace2}
\end{align}

Multiplying \eqref{transverse2} and \eqref{trace2} with $\exp(ip_1x+ip_2x_2+ip_3x_3)$ and integrating over all $x$, $x_2$ and $x_3$, we obtain the Ward identities in momentum space
\begin{align}
	p_{1\m_1}\braket{T^{\m_1\n_1}(p_1)T^{\m_2\n_2}(p_2)\mO(p_3)}&=-2p_2^{\n_1}\braket{T^{\m_2\n_2}(p_1+p_2)\mO(p_3)}+2p_3^{\n_1}\braket{T^{\m_2\n_2}(p_2)\mO(p_1+p_3)}\notag\\
	&\hspace{-0.5cm}+2\d^{\n_1\n_2}\,p_{2\m_1}\braket{T^{\m_1\m_2}(p_1+p_2)\mO(p_3)}+2\d^{\n_1\m_2}\,p_{2\m_1}\braket{T^{\m_1\n_2}(p_1+p_2)\mO(p_3)}\\[2ex]
	g_{\m_1\n_1}\braket{T^{\m_1\n_1}(p_1)T^{\m_2\n_2}(p_2)\mO(p_3)}&=2(d-\D)\braket{T^{\m_2\n_2}(p_2)\mO(p_1+p_3)}+4\braket{T^{\m_2\n_2}(p_1+p_2)\mO(p_3)}
\end{align}
where $\D$ is the conformal dimension of the scalar operator $\mO$. 

Notice that any conformal two point function involving two operators of different spin or different scale dimensions is zero. For this reason the transverse and trace Ward identities can be expressed in the final form
\begin{equation}
	\begin{split}
		p_{1\m_1}\braket{T^{\m_1\n_1}(p_1)T^{\m_2\n_2}(p_2)\mO(p_3)}&=0\\
		g_{\m_1\n_1}\braket{T^{\m_1\n_1}(p_1)T^{\m_2\n_2}(p_2)\mO(p_3)}&=0.
	\end{split}\label{finWardTTO}
\end{equation}

In this particular case, the definition the 3-point function $\braket{TT\mO}$ in \eqref{defTTO} brings to a correlation function that is already transverse-traceless and does not contain longitudinal contributions. 

\subsection{Dilatation Ward identities}

It is simple to rewrite the dilatation Ward identities for the 3-point function $\braket{T^{\m_1\n_1}T^{\m_2\n_2}\mO}$ as
\begin{equation}
	\left[\sum_{j=1}^3\D_j-2d-\sum_{j=1}^{2}p_j^\a\sdfrac{\partial}{\partial p_j^\a}\right]\braket{T^{\m_1\n_1}(p_2)T^{\m_2\n_2}(p_2)\mO(p_3)}=0.\label{dilorig}
\end{equation}
We can consider the decomposition of the correlation function in terms of the transverse and traceless part and the semi-local part for which the previous equation can be read as
\begin{align}
	&\left[\sum_{j=1}^3\D_j-2d-\sum_{j=1}^{2}p_j^\a\sdfrac{\partial}{\partial p_j^\a}\right]\bigg\{\braket{t^{\m_1\n_1}(p_2)t^{\m_2\n_2}(p_2)\mO(p_3)}+\braket{t_{loc}^{\m_1\n_1}(p_2)t^{\m_2\n_2}(p_2)\mO(p_3)}\notag\\
	&\hspace{4cm}+\braket{t^{\m_1\n_1}(p_2)t_{loc}^{\m_2\n_2}(p_2)\mO(p_3)}+\braket{t_{loc}^{\m_1\n_1}(p_2)t_{loc}^{\m_2\n_2}(p_2)\mO(p_3)}\bigg\}=0.\label{dildec}
\end{align}

We are free to apply transverse-traceless projectors \eqref{Proj} in \appref{transWard} to \eqref{dildec}, in order to isolate equations for the form factors appearing in the decomposition of $\braket{t^{\m_1\n_1}t^{\m_2\n_2}\mO}$. Evaluating the action of the differential operator in \eqref{dildec} on the semi-local terms via the formulae in \appref{appendixB}, we find 
\begin{align}
	\Pi^{\r_1\s_1}_{\m_1\n_1}(p_1)\,\left(p_1^\l\sdfrac{\partial}{\partial p_1^\l}\,\mathcal{I}^{\m_1\n_1}_{\ \ \ \a_1}p_{1\,\b_1}\right)&=0\,\\
	\Pi^{\m\n}_{\a\b}(p)\,\mathcal{I}^{\a\b}_{\r\s}(p)&=0
\end{align}
where the $\mathcal{I}$ tensor is defined in \eqref{itensor} and is used to define the semi-local part of the correlation function. This expression implies that any correlation function with one and more than one insertion of $t_{loc}$ vanish when the dilatation operator and the projectors are applied. For this reason \eqref{dildec} can be easily written as
\begin{equation}
	\Pi^{\r_1\s_1}_{\m_1\n_1}(p_1)\,\Pi^{\r_2\s_2}_{\m_2\n_2}(p_2)\,\left[\sum_{j=1}^3\D_j-2d-\sum_{j=1}^{2}p_j^\a\sdfrac{\partial}{\partial p_j^\a}\right]\braket{t^{\m_1\n_1}(p_2)t^{\m_2\n_2}(p_2)\mO(p_3)}=0.\label{dilfin}
\end{equation}
In order to write down the equations of form factors we substitute in \eqref{dilfin} the decomposition of the 3-point function \eqref{TTOdec} 
\begin{align}
	&\Pi^{\r_1\s_1}_{\m_1\n_1}(p_1)\,\Pi^{\r_2\s_2}_{\m_2\n_2}(p_2)\,\left[\sum_{j=1}^3\D_j-2d-\sum_{j=1}^{2}p_j^\a\sdfrac{\partial}{\partial p_j^\a}\right]\notag\\
	&\hspace{2.5cm}\times\bigg\{\Pi^{\m_1\n_1}_{\a_1\b_1}(p_1)\Pi^{\m_2\n_2}_{\a_2\b_2}(p_2)\left[A_1\,p_2^{\a_1}p_2^{\b_1}p_3^{\a_2}p_3^{\b_2}+A_2\,p_2^{\a_1}p_3^{\a_2}\d^{\b_1\b_2}+A_3\,\d^{\a_1\a_2}\d^{\b_1\b_2}\right]\bigg\}=0
\end{align}
and using the relations in \appref{appendixB} we find
\begin{equation}
	\Pi^{\r_1\s_1}_{\a_1\b_1}(p_1)\Pi^{\r_2\s_2}_{\a_2\b_2}(p_2)\left[\sum_{j=1}^3\D_j-2d-\sum_{j=1}^{2}p_j^\a\sdfrac{\partial}{\partial p_j^\a}\right]\left[A_1\,p_2^{\a_1}p_2^{\b_1}p_3^{\a_2}p_3^{\b_2}+A_2\,p_2^{\a_1}p_3^{\a_2}\d^{\b_1\b_2}+A_3\,\d^{\a_1\a_2}\d^{\b_1\b_2}\right]=0.\label{dilfin2}
\end{equation}
We can act with the differential operator in \eqref{dilfin2} observing that there is no change in the independent tensor structures. Once that all the calculations are expressed, it is possible to obtain a set of equation for all the form factors. These equations result from the vanishing of the coefficient of the independent tensor structures in \eqref{dilfin2}. In these equation there will be terms like
\begin{equation}
	\sum_{j=1}^2\,p_j^\a \sdfrac{\partial}{\partial p_j^\a}\,A_n(p_1,p_2,p_3),\qquad n=1,2,3\label{terms}
\end{equation}
but the form factors are purely functions of the momenta magnitude by the Lorentz invariance. The action of momentum derivatives on form factors may be obtained using the chain rules,
\begin{equation}
	\begin{split}
		\sdfrac{\partial}{\partial p_{1\m}}&=\sdfrac{\partial p_1}{\partial p_{1\m}}\sdfrac{\partial}{\partial p_1}+\sdfrac{\partial p_2}{\partial p_{1\m}}\sdfrac{\partial}{\partial p_2}+\sdfrac{\partial p_3}{\partial p_{1\m}}\sdfrac{\partial}{\partial p_3}=\sdfrac{p_1^\m}{p_1}\sdfrac{\partial}{\partial p_1}+\sdfrac{p_1^\m+p_2^\m}{p_3}\sdfrac{\partial}{\partial p_3}\\
		\sdfrac{\partial}{\partial p_{2\m}}&=\sdfrac{p_2^\m}{p_2}\sdfrac{\partial}{\partial p_2}+\sdfrac{p_1^\m+p_2^\m}{p_3}\sdfrac{\partial}{\partial p_3}
	\end{split}\label{chainrule}
\end{equation}
noting that $p_3$ is fixed by the conservation of the total momentum $p^\m_3=-p^\m_1-p^\m_2$. Using these relations we may re-expressed \eqref{terms} purely in terms of the momentum magnitudes 
\begin{equation}
	\sum_{j=1}^2\,p_j^\a \sdfrac{\partial}{\partial p_j^\a}\,A_n(p_1,p_2,p_3)=\sum_{j=1}^3\,p_j \sdfrac{\partial}{\partial p_j}\,A_n(p_1,p_2,p_3),\qquad n=1,2,3.
\end{equation}

Therefore it is possible to rewrite the dilatation Ward identity \eqref{dilorig} for a three point function of three conformal primary operator of any tensor structure in terms of its form factors as
\begin{equation}
	\left[2d+N_n-\sum_{i=1}^3\D_i+\sum_{i=1}^{3}\,p_i\sdfrac{\partial}{\partial p_i}\right]\,A_n(p_1,p_2,p_3)=0,\label{dilfin3}
\end{equation}
{where $N_n$ is the number of momenta that $A_n$ multiply in the decomposition \eqref{TTOdec}, and as previously $\D_j$ denote the conformal dimensions of the operator in the 3-point function: in this case $\D_{1,2}=d$ for a stress-energy tensor and $\D_3$ depending on the particular scalar operator chosen. From \eqref{dilfin3} it is clear that the form factor $A_n$ has scaling degree
	\begin{equation}
		\deg(A_n)=\D_t-2d-N_n,\label{deg}
	\end{equation}
	where $\D_t=\D_1+\D_2+\D_3$, as also pointed out in \cite{Bzowski:2013sza}. 
	
	\subsection{Special conformal Ward identities}\label{specialconfward}
	
	We now extract scalar equations for the form factors in the same way of the previous section but use the special conformal Ward identities (SCWI's). Considering the SCWI's for the 3-point function $\braket{TT\mO}$  we obtain
	\begin{align}
		0&=\sum_{j=1}^{2}\left[2(\D_j-d)\sdfrac{\partial}{\partial p_j^\k}-2p_j^\a\sdfrac{\partial}{\partial p_j^\a}\sdfrac{\partial}{\partial p_j^\k}+(p_j)_\k\sdfrac{\partial}{\partial p_j^\a}\sdfrac{\partial}{\partial p_{j\a}}\right]\braket{\,T^{\m_1\n_1}(p_1)T^{\m_2\n_2}(p_2)\mO(p_3)}\notag\\
		&\hspace{0.5cm}+4\left(\d^{\k(\m_1}\sdfrac{\partial}{\partial p_1^{\a_1}}-\d^\k_{\a_1}\d^{\l(\m_1}\sdfrac{\partial}{\partial p_1^\l}\right)\braket{\,T^{\n_1)\a_1}(p_1)T^{\m_2\n_2}(p_2)\mO(p_3)}\notag\\
		&\hspace{0.5cm}+4\left(\d^{\k(\m_2}\sdfrac{\partial}{\partial p_2^{\a_2}}-\d^\k_{\a_2}\d^{\l(\m_2}\sdfrac{\partial}{\partial p_2^\l}\right)\braket{\,T^{\n_2)\a_2}(p_2)T^{\m_1\n_1}(p_1)\mO(p_3)}\equiv \mathcal{K}^\k\braket{\,T^{\m_1\n_1}(p_1)T^{\m_2\n_2}(p_2)\mO(p_3)}
	\end{align}
	where we have defined the $\mathcal K^\k$ operator for simplicity. As previously we can consider the decomposition of the 3-point function to obtain
	\begin{align}
		0&=\mathcal{K}^\k\braket{\,T^{\m_1\n_1}(p_1)T^{\m_2\n_2}(p_2)\mO(p_3)}\notag\\
		&=\mathcal{K}^\k\braket{\,t^{\m_1\n_1}t^{\m_2\n_2}\mO}+\mathcal{K}^\k\braket{\,t_{loc}^{\m_1\n_1}t^{\m_2\n_2}\mO}+\mathcal{K}^\k\braket{\,t^{\m_1\n_1}t_{loc}^{\m_2\n_2}\mO}+\mathcal{K}^\k\braket{\,t_{loc}^{\m_1\n_1}t_{loc}^{\m_2\n_2}\mO}.\label{specConf}
	\end{align}
	In order to isolate equations for the form factors appearing in the decomposition, we are free to apply transverse-traceless projectors. Using the properties in \appref{appendixB} and through a direct calculation we find that
	\begin{align}
		\Pi^{\r_1\s_1}_{\m_1\n_1}(p_1)\,\Pi^{\r_2\s_2}_{\m_2\n_2}(p_2)\,\mathcal{K}^\k\braket{\,t_{loc}^{\m_1\n_1}t^{\m_2\n_2}\mO}&=\sdfrac{4d}{p_1^2}\,\Pi^{\r_1\s_1\k}_{\hspace{0.7cm}\m_1}(p_1)\,\bigg(p_{1\n_1}\braket{T^{\m_1\n_1}(p_1)t^{\r_2\s_2}(p_2)\mO(p_3)}\bigg)\\
		\Pi^{\r_1\s_1}_{\m_1\n_1}(p_1)\,\Pi^{\r_2\s_2}_{\m_2\n_2}(p_2)\,\mathcal{K}^\k\braket{\,t^{\m_1\n_1}t_{loc}^{\m_2\n_2}\mO}&=\sdfrac{4d}{p_2^2}\,\Pi^{\r_2\s_2\k}_{\hspace{0.7cm}\m_2}(p_2)\,\bigg(p_{2\n_2}\braket{t^{\r_1\s_1}(p_1)T^{\m_2\n_2}(p_2)\mO(p_3)}\bigg)\\
		\Pi^{\r_1\s_1}_{\m_1\n_1}(p_1)\,\Pi^{\r_2\s_2}_{\m_2\n_2}(p_2)\,\mathcal{K}^\k\braket{\,t_{loc}^{\m_1\n_1}t_{loc}^{\m_2\n_2}\mO}&=0
	\end{align}
	and \eqref{specConf} takes the form
	\begin{align}
		0&=\Pi^{\r_1\s_1}_{\m_1\n_1}(p_1)\,\Pi^{\r_2\s_2}_{\m_2\n_2}(p_2)\,\bigg\{\mathcal K^\k\braket{\,t^{\m_1\n_1}(p_1)t^{\m_2\n_2}(p_2)\mO(p_3)}\notag\\
		&\hspace{1cm}+\sdfrac{4d}{p_1^2}\,\d^{\k\m_1}\,p_{1\r_1}\braket{T^{\n_1\r_1}(p_1)T^{\m_2\n_2}(p_2)\mO(p_3)}+\sdfrac{4d}{p_2^2}\,\d^{\k\m_2}\,p_{2\r_2}\braket{T^{\m_1\n_1}(p_1)T^{\m_2\r_2}(p_2)\mO(p_3)}\ \bigg\}.\label{specC}
	\end{align}
	The last two terms may be re-expressed in terms of 2-point functions via the transverse Ward identities, but by using \eqref{finWardTTO} these terms vanish. The remaining task is to rewrite the transverse and traceless component in terms of form factors and extract the conformal Ward identities for this particular scalar function of the magnitues of the momenta.\\
	By a direct calculation we find that the first term of \eqref{specC}, $\mathcal K^\k\braket{\,t\,t\,\mO}$, is transverse and traceless in the covariant indices with respect to the corresponding momenta, i.e.
	\begin{equation}
		\begin{split}
			\d_{\m_1\n_1}[\mathcal K^\k\braket{\,t^{\m_1\n_1}(p_1)t^{\m_2\n_2}(p_2)\mO(p_3)}]=0,&\qquad\d_{\m_2\n_2}[\mathcal K^\k\braket{\,t^{\m_1\n_1}(p_1)t^{\m_2\n_2}(p_2)\mO(p_3)}]=0\\
			p_{1\m_1}[\mathcal K^\k\braket{\,t^{\m_1\n_1}(p_1)t^{\m_2\n_2}(p_2)\mO(p_3)}]=0,&\qquad p_{\m_2}[\mathcal K^\k\braket{\,t^{\m_1\n_1}(p_1)t^{\m_2\n_2}(p_2)\mO(p_3)}]=0.
		\end{split}
	\end{equation}
	Using this result, and following the discussion in \secref{tensstruc}, we can write the most general form of $\mathcal K^\k\braket{\,t\,t\,\mO}$ as
	\begin{align}
		&\Pi^{\r_1\s_1}_{\m_1\n_1}(p_1)\,\Pi^{\r_2\s_2}_{\m_2\n_2}(p_2)\,\mathcal K^\k\braket{\,t^{\m_1\n_1}(p_1)t^{\m_2\n_2}(p_2)\mO(p_3)}\notag\\
		&=\Pi^{\r_1\s_1}_{\m_1\n_1}(p_1)\,\Pi^{\r_2\s_2}_{\m_2\n_2}(p_2)\,\bigg[\,p_1^\k\left(C_{11}\,p_2^{\m_1}p_2^{\n_1}p_3^{\m_2}p_3^{\n_2}+C_{12}\,p_2^{\m_1}p_3^{\m_2}\d^{\n_1\n_2}+C_{13}\,\d^{\m_1\m_2}\d^{\n_1\n_2}\right)\notag\\
		&\hspace{1.3cm}+p_2^\k\left(C_{21}\,p_2^{\m_1}p_2^{\n_1}p_3^{\m_2}p_3^{\n_2}+C_{22}\,p_2^{\m_1}p_3^{\m_2}\d^{\n_1\n_2}+C_{23}\,\d^{\m_1\m_2}\d^{\n_1\n_2}\right)\notag\\
		&\hspace{1.3cm}+\d^{\k\m_1}\left(C_{31}p_2^{\n_1}p_3^{\m_2}p_3^{\n_2}+C_{32}\d^{\m_2\n_1}p_3^{\n_2}\right)+\d^{\k\m_2}\left(C_{41}p_2^{\n_1}p_2^{\m_1}p_3^{\n_2}+C_{42}\d^{\m_1\n_2}p_2^{\n_1}\right)\bigg]\label{CWI}
	\end{align}
	{
		where now $C_{ij}$ are scalar differential equations involving the form factors $A_i$, $i=1,2,3$, written in terms of the momentum magnitudes $p_j$.
		The coefficients $C_{jk}$ in \eqref{CWI} are not all independent, indeed, $C_{1j}$ and $C_{2j}$, as well as $C_{3j}$ and $C_{4j}$, are pairwise equivalent, due to the symmetry under the exchange of the stress energy tensors in the correlator. The difference in the coefficients is manifest in the order of the corresponding differential equations. In all the cases of $3$-point function, the coefficients multiplying the momentum $p_i^\kappa$ ($\k$ is the special index related to the conformal operator $\mathcal{K}^\k$) are second order partial differential equations, and they are called \emph{primary} conformal Ward identities (CWI's). The coefficients of the structure $\delta^{\kappa\,\mu_i}$ are always first order partial differential equations, called \emph{secondary} CWI's in \cite{Bzowski:2013sza}. 
		We observe that the other terms in \eqref{specC}, differently from \eqref{CWI}, involve the $\delta^{\kappa\,\mu_i}$ structure, then they will contribute to the secondary CWI's. Furthermore, the primary CWI's are equivalent to the vanishing of the coefficients $C_{1j}$ and $C_{2j}$. 
	}%
	\subsection{Primary conformal Ward identities}
	
	In order to write the primary CWI's in a simpler form, we have to rearrange the expression of $C_{jk}$ using the dilatation Ward identities and some differential identities. We will show the explicit procedure for the first coefficient $C_{11}$ and the others coefficients follow the same prescription. In an explicit form the coefficient $C_{11}$ is expressed as
	\begin{equation}
		C_{11}= - \frac{2}{p_3} \left[ p_1 \frac{\partial^2}{\partial p_1 \partial p_3 } + p_2 \frac{\partial^2}{\partial p_2 \partial p_3 } \right] A_1 + \frac{d-1}{p_1} \frac{\partial}{\partial p_1} A_1 - \frac{\partial^2}{\partial p_1^2} A_1 + \frac{d-9}{p_3} \frac{\partial}{\partial p_3} A_1 - \frac{\partial^2}{\partial p_3^2} A_1\,.\label{C11}
	\end{equation}
	{
		Taking the dilatation Ward identities \eqref{dilfin3} for the $A_1$, and deriving it with respect to the magnitue of the momentum $p_3$ we obtain
		\begin{equation}
			\frac{\partial}{\partial p_3}\left(\sum_{i=1}^3\,p_i\,\frac{\partial}{\partial\,p_i}\, A_1\right) =\,\deg(A_1)\,\frac{\partial\,A_1}{\partial p_3}
		\end{equation}
		where the degree of the form factor is defined in \eqref{deg} and for $A_1$ is $\deg(A_1)=\Delta_t-2d-4=\Delta_3-4$. Then, after some simplification, we get
		\begin{equation}
			\left[  p_1 \frac{\partial^2}{\partial p_1 \partial p_3} +  p_2 \frac{\partial^2}{\partial p_3 \partial p_2}  +  p_3 \frac{\partial^2}{\partial p_3 \partial p_3} \right] A_1 =\,\bigg(\deg(A_1)-1\bigg)\,\frac{\partial\,A_1}{\partial p_3}\,.
			\label{eq:dilWI}
		\end{equation}
	}%
	By using \eqref{eq:dilWI}, we can re-expressed the first term in \eqref{C11} as
	\begin{equation}
		-\frac{2}{p_3}  \left[ p_1 \frac{\partial^2}{\partial p_1 \partial p_3} +  p_2 \frac{\partial^2}{\partial p_3 \partial p_2}  \right]  A_1 = \frac{\left( 2 - 2 \deg(A_1) \right)}{p_3} \frac{\partial}{\partial p_3} A_1 +2  \frac{\partial^2}{\partial p_3^2} A_1\,,
	\end{equation}
	and inserting this result into \eqref{C11} we simplify the form of the differential equation $C_{11}$ as
	\begin{equation}
		C_{11}= \left[ -\frac{\partial^2}{\partial p_1^2} + \frac{d-1}{p_1} \frac{\partial}{\partial p_1} \right] A_1 + \left[ \frac{\partial^2 }{\partial p_3^2}  + \frac{d+1-2\Delta_3}{p_3} \frac{\partial}{\partial p_3} \right] A_1 \,.  
		\label{eq:C11quasi}  
	\end{equation}
	In order to write the primary CWI's in a simple way, we define the following fundamental differential operators
	\begin{equation}
		\begin{split}
			\textup{K}_i &= \frac{\partial^2}{\partial p_i^2} + \frac{d+1-2\Delta_i}{p_i} \frac{\partial}{\partial p_i}  \qquad i=1,2,3    \\ 
			\textup{K}_{ij} &= \textup{K}_i - \textup{K}_j \,,
		\end{split}\label{Koper}
	\end{equation}
	where $\D_j$ is the conformal dimension of the j-th operator in the 3-point function under consideration. Through this definition the $C_{11}$ is re-expressed as
	\begin{equation}
		C_{11} = (\textup{K}_3 - \textup{K}_1 ) A_1 = \textup{K}_{31}A_1 \,.
	\end{equation}
	
	The procedure presented above allows to obtain a simple second-order differential equations and it can be applied in the same way to all the $C_{1j}$'s, $j=1,2,3$. 
	
	Usually, while performing the lengthy computations, one may encounter the term
	\begin{equation}
		-\frac{2}{p_3}  \left[ p_1 \frac{\partial^2}{\partial p_1 \partial p_3} +  p_2 \frac{\partial^2}{\partial p_3 \partial p_2}  \right]  A_n = \frac{\left( 2 - 2 \deg(A_n) \right)}{p_3} \frac{\partial}{\partial p_3} A_n +2  \frac{\partial^2}{\partial p_3^2} A_n \,,
	\end{equation}
	where also in this case $\textup{deg}(A_n)= \Delta_1 + \Delta_2+ \Delta_3 -2 d - N_n \,$.
	Here, $N_n$ represents the tensorial dimension of $A_n$, i.e. the number of momenta multiplying the form factors $A_n$ and the projectors $\Pi$. 
	
	In the $\braket{TT\mO}$ case, for instance, $\Delta_1=\Delta_2=d$, whereas $\Delta_3$ remains implicit because of the unknown nature of the generic scalar operator $\mathcal{O}(p_3)$ (e.g. if $\mathcal{O}=\phi^2$ one has $\Delta_3=d-2$). 
	In this case we have
	\begin{equation}
		-\frac{2}{p_3}  \left[ p_1 \frac{\partial^2}{\partial p_1 \partial p_3} +  p_2 \frac{\partial^2}{\partial p_3 \partial p_2}  \right]  A_n = \frac{\left( 2 - 2 (\Delta_3 - N_n)\right)}{p_3} \frac{\partial}{\partial p_3} A_n +2  \frac{\partial^2}{\partial p_3^2} A_n \,,
	\end{equation}
	where
	\begin{align} 
		N_1&=4    \qquad\textup{for} \qquad   A_1(p_1,p_2,p_3)\,\, p_2^{\alpha_1} p_2^{\beta_1} p_3^{\alpha_2} p_3^{\beta_2} \\
		N_2&=2 \qquad\textup{for} \qquad A_2(p_1,p_2,p_3)\,\, p_2^{\alpha_1} p_3^{\alpha_2} \delta^{\beta_1\beta_2}  \\
		N_3&=0    \qquad \textup{for}  \qquad  A_3(p_1,p_2,p_3)\,\, \delta^{\alpha_1\alpha_2} \delta^{\beta_1\beta_2}. 
	\end{align}
	In this way we can simplify all the primary coefficients $C_{i,j}$, $i=1,2$ and $j=1,2,3$.
	The primary CWI's are obtained, as previously discussed, when the coefficients $C_{1j}$ and $C_{2j}$ are equal to zero. One obtains
	\begin{equation}
		\begin{matrix}
			K_{31}\,A_1=0,&\qquad K_{13}A_2=8A_1,&\qquad K_{13}A_3=2A_2,\\[1.3ex]
			K_{23}A_1=0,&\qquad K_{23}A_2=8A_1,&\qquad K_{23}A_3=2A_2.
		\end{matrix}
	\end{equation} 
	Note that, from the definition \eqref{Koper}, we have
	\begin{equation}
		K_{ii}=0,\qquad K_{ji}=-K_{ij},\qquad K_{ij}+K_{jk}=K_{ik}
	\end{equation}
	for any $i,j,k\,\in\{1,2,3\}$. One can therefore subtract corresponding pairs of equations and obtain the following system of independent partial differential equations
	\begin{equation}
		\begin{matrix}
			K_{13}\,A_1=0,&\qquad K_{13}A_2=8A_1,&\qquad K_{13}A_3=2A_2,\\[1.3ex]
			K_{12}A_1=0,&\qquad K_{12}A_2=0,&\qquad K_{12}A_3=0.
		\end{matrix}\label{primaryCWI}
	\end{equation}

	\subsection{Secondary conformal Ward identities}\label{secondary2}
	As previously mentioned, the secondary conformal Ward identities are first-order partial differential equations. In order to write them compactly, we define the two differential operators
	\begin{align}
		\textup{L}_N&= p_1(p_1^2 + p_2^2 - p_3^2) \frac{\partial}{\partial p_1} + 2 p_1^2 p_2 \frac{\partial}{\partial p_2} + \left[ (2d - \Delta_1 - 2\Delta_2 +N)p_1^2 + (2\Delta_1-d)(p_3^2-p_2^2)  \right]\label{Ldef} \\
		\textup{R} &= p_1 \frac{\partial}{\partial p_1} - (2\Delta_1-d) \label{Rdef}\, 
	\end{align}
	as well as their symmetric versions
	\begin{align}
		&L'_N=L_N,\quad\text{with}\ p_1\leftrightarrow p_2\ \text{and}\ \D_1\leftrightarrow\D_2,\\
		&R'=R,\qquad\text{with}\ p_1\mapsto p_2\ \text{and}\ \D_1\mapsto\D_2.
	\end{align}
	In the $\braket{TT\mO}$ case one finds for the coefficients $C_{31}$ and $C_{32}$ the results
	\begin{align}
		C_{31}&= \left[ - \frac{1}{p_1} \left( p_1^2 + p_2^2 - p_3^2 \right) \frac{\partial}{\partial p_1} - 2 p_2 \frac{\partial}{\partial p_2}  + \frac{d}{p_1^2} (p_2^2-p_3^2) + (d-2)  \right]A_1 + \left[ \frac{d}{p_1^2} - \frac{1}{p_1} \frac{\partial}{\partial p_1}   \right] A_2   \nonumber \\&= -\frac{1}{p_1^2}\left( \textup{L}_2\, A_1 + \textup{R}\, A_2 \right) \\[1.5ex] 
		C_{32}&= \left[  \frac{1}{4 p_1} \left( p_1^2 + p_2^2 - p_3^2 \right) \frac{\partial}{\partial p_1} + \frac{1}{2} p_2 \frac{\partial}{\partial p_2}  + \frac{d}{4 p_1^2} (p_3^2-p_2^2) + \frac{(2-d)}{4}  \right]A_2 + \left[\frac{1}{p_1} \frac{\partial}{\partial p_1}   - \frac{d}{p_1^2}  \right] A_3   \nonumber \\&= \frac{1}{4 p_1^2}\left( \textup{L}_2\, A_2 + 4 \textup{R}\, A_3 \right)    
	\end{align}
	and for the last two coefficient $C_{41}$ and $C_{42}$
	\begin{align}
		C_{41}&= \left[ - \frac{1}{p_2} \left( p_1^2 + p_2^2 - p_3^2 \right) \frac{\partial}{\partial p_2} - 2 p_1 \frac{\partial}{\partial p_1}  + \frac{d}{p_1^2} (p_1^2-p_3^2) + (d-2)  \right]A_1 + \left[ \frac{d}{p_2^2} - \frac{1}{p_2} \frac{\partial}{\partial p_2}   \right] A_2   \nonumber \\&= -\frac{1}{p_2^2}\left( \textup{L'}_2\, A_1 + \textup{R'}\, A_2 \right) \\[1.5ex] 
		C_{42}&= \left[  \frac{1}{4 p_2} \left( p_1^2 + p_2^2 - p_3^2 \right) \frac{\partial}{\partial p_2} + \frac{1}{2} p_1 \frac{\partial}{\partial p_1}  + \frac{d}{4 p_2^2} (p_3^2-p_1^2) + \frac{(2-d)}{4}  \right]A_2 + \left[\frac{1}{p_2} \frac{\partial}{\partial p_2}   - \frac{d}{p_2^2}  \right] A_3   \nonumber \\&= \frac{1}{4 p_2^2}\left( \textup{L'}_2\, A_2 + 4 \textup{R'}\, A_3 \right) \,.
	\end{align}
	The secondary CWI's are equivalent to the vanishing of these coefficient because in \eqref{specC} the only term that is not zero is the transverse traceless part. This can be seen  by using the transverse and trace Ward identities \eqref{finWardTTO}, and one finds two independent secondary CWI's, namely
	\begin{equation}
		\begin{split}
			\textup{L}_2\, A_1 + \textup{R}\, A_2&=0\\
			\textup{L}_2\, A_2 + 4 \textup{R}\, A_3&=0.
		\end{split}\label{secondary}
	\end{equation}
	In fact, due to the symmetry properties of the form factor in \eqref{symA}, one realizes that the coefficients $C_{4i}$ are equivalent to the $C_{3i}$, with $i=1,2$.

\section{Solutions of the CWI's}\label{solution}

We have shown in \secref{Sol3Point} how to solve the conformal constraints for the scalar $3$-point functions. In particular, the solutions can be equivalently given either in terms of hypergeometric functions or of triple-K integrals, as discussed in \cite{Coriano:2013jba, Bzowski:2013sza} respectively. For tensorial $3$-point functions, the solutions can be found, equivalently, by both methods  \cite{Bzowski:2013sza,Coriano:2018bsy,Coriano:2018bbe,Bzowski:2015pba}. 
\subsection{Triple-K integrals}
In this section we use the formulation of the triple-K integrals, because they reflect naturally the symmetry properties of the correlation functions and also because of their analytical properties. 
We recall the definition of the general triple-K integral
\begin{equation}
	I_{\a\{\b_1\b_2\b_3\}}(p_1,p_2,p_3)=\int dx \,x^\a\prod_{j=1}^{3}\,p_j^{\b_j}\,K_{\b_j}(p_jx),\label{3Kint}
\end{equation}
where $K_\n$ is a modified Bessel function of the second kind defined as
\begin{equation}
	K_\n(x)=\sdfrac{\pi}{2}\sdfrac{I_{-\n}(x)-I_\n(x)}{\sin(\n\pi)},\ \n\notin\mathbb Z\qquad I_\n(x)=\left(\sdfrac{x}{2}\right)^{\n}\sum_{k=0}^\infty\sdfrac{1}{\G(k+1)\,\G(\n+1+k)}\,\left(\sdfrac{x}{2}\right)^{2k}.\label{Kdef}
\end{equation} 
with the property 
\begin{equation}
	K_n(x)=\lim_{\e\to0}K_{n+\e}(x),\quad n\in\mathbb Z.
\end{equation}
The triple-K integral in \eqref{3Kint} depends on four parameters: the power $\a$ of the integration variable $x$, and the three Bessel function indices $\b_j$. The argument of this integral are magnitudes of momenta $p_j$, $j=1,2,3$. One can notice the integral is invariant under the exchange $(p_j,\beta_j)\leftrightarrow(p_i,\beta_i)$, and we will see that this properties will reflect the symmetry properties of the correlation functions. We will use also the reduced version $J_{N\{k_1,k_2,k_3\}}$ of the triple-K integral defined as
\begin{equation}
	J_{N\{k_j\}}=I_{\frac{d}{2}-1+N\{\D_j-\frac{d}{2}+k_j\}},\label{defJintegr}
\end{equation}
that are mapped to the expression \eqref{3Kint} via the substitutions
\begin{equation}
	\a=\sdfrac{d}{2}-1+N,\qquad \b_j=\D_j-\sdfrac{d}{2}+k_j,\ j=1,2,3,
\end{equation}
where we have used the condensed notation $\{k_j\}=\{k_1\,k_2\,k_3\}$. These triple-K integrals may also be re-expressed using the Feynman parametrization as
\begin{equation}
	I_{\a\{\b_1\b_2\b_3\}}(p_1,p_2,p_3)=2^{\a-3}\G\left(\sdfrac{\a-\b_t+1}{2}\right)\G\left(\sdfrac{\a+\b_t+1}{2}\right)\int_{[0,1]^3}dX\,D^{\frac{1}{2}(\b_t-\a-1)}\prod_{j=1}^3\,x_j^{\frac{1}{2}(\a-1-\b_t)+\b_j}\label{Feyn}
\end{equation}
where $\b_t=\b_1+\b_2+\b_3$ and the integration extends over the unit interval $[0,1]$ for each of the $x_j$, $j=1,2,3$ with the standard measure $dX=dx_1\,dx_2\,dx_3\,\d(1-x_1-x_2-x_3)$ and with
\begin{equation}
	D=p_1^2\,x_2\,x_3+p_2^2\,x_1\,x_3+p_3^2\,x_1\,x_2
\end{equation}
being the usual denominator appearing in the Feynman representation. Furthermore, the expression of the triple-K integral is linked to the hypergeometric functions through \eqref{3K}. \\
In order to study the convergences of the triple-K integral we assume that all the parameters in \eqref{3Kint} are real. At large $x$, the Bessel functions have the asymptotic expansions
\begin{equation}
	I_\n(x)=\sdfrac{1}{\sqrt{2\pi}}\sdfrac{e^x}{\sqrt{x}}+\dots,\quad K_\n(x)=\sqrt{\sdfrac{\pi}{2}}\,\sdfrac{e^{-x}}{\sqrt{x}}+\dots,\ \n\in\mathbb R, 
\end{equation}
and one readily observes that  inserting the expansions above in \eqref{3Kint}, the integral converges at large $x$ for physical configurations of the momenta, with $p_1+p_2+p_3>0$. However, there may still be a divergence at $x=0$. Considering the asymptotic expansion of the Bessel functions at small $x$ as
\begin{equation}
	I_{\nu}(x)=\frac{x^\nu}{2^\nu\ \Gamma(\nu+1)}+\dots,\qquad K_\nu(x)= \frac{2^{\nu-1}\Gamma(\nu)}{x^\nu}+\dots
\end{equation}
in \eqref{3Kint}, one observes that the triple-K integral converges at small $x$ only if
\begin{equation}
	\a>\sum_{j=1}^{3}|\b_j|+1,\qquad p_1,p_2,p_3>0.\label{region}
\end{equation}
If $\a$ does not satisfy this inequality, the integrals must be defined by an analytic continuation. The quantity
\begin{equation}
	\d\equiv \sum_{j=1}^3|\b_j| -1-\a
\end{equation}
is the expected degree of divergence. When
\begin{equation}
	\a+1\pm\b_1\pm\b_2\pm\b_3=-2k\label{condition}
\end{equation}
for some non-negative integer $k$ and any choice of the $\pm$ sign, the analytic continuation of the triple-K integral generally has poles in the regularization parameter. From the representation \eqref{Feyn} these poles may either be present in the gamma function or in the Feynman pamareterization of the integral multiplying them, or both. As one can see from \eqref{Feyn} the Feynman integrals are finite if
\begin{equation}
	\sdfrac{1}{2}(\a-1-\b_t)+\b_j>1
\end{equation}
as $d\to4$ for all $j=1,2,3$, but generally diverge otherwise. 
More details about the convergence of these kind of integrals and on their regularization procedure, can be found in \cite{Bzowski:2014qja,Bzowski:2015yxv, Bzowski:2018fql, Bzowski:2015pba}.
\subsection{Dilatation Ward identities}

We will provide the solution to the dilatation Ward identities in terms of triple-K integrals. First notice that using the relations \eqref{identityBess} in \appref{transWard} we obtain
\begin{align}
	\sum_{j=1}^3p_j\sdfrac{\partial }{\partial p_j}\,I_{\a\{\b_k\}}&=-\sum_{j=1}^3\,p_j^2\,I_{\a+1\{\b_k-\d_{jk}\}}=-\sum_{j=1}^3\,p_j^2\,I_{\a+1\{\b_k+\d_{jk}\}}+2\sum_{j=1}^3\,\b_j\,I_{\a\{\b_k\}}\notag\\
	&=-\sum_{j=1}^3\,p_j^2\,I_{\a+1\{\b_k+\d_{jk}\}}+2\,\b_t\,I_{\a\{\b_k\}}\label{DCWIsol}
\end{align}
and by using
\begin{equation}
	\int_0^\infty dx\,x^{\a+1}\,\sdfrac{\partial}{\partial x}\left(\prod_{j=1}^{3}p_j^{\b_j}K_{\b_j}(p_j\,x)\right)=-\sum_{j=1}^3\,p_j^2\,I_{\a+1\{\b_k+\d_{jk}\}}+\,\b_t\,I_{\a\{\b_k\}}
\end{equation}
we can re-expressed \eqref{DCWIsol}, after an integration by parts as
\begin{equation}
	\sum_{j=1}^3p_j\sdfrac{\partial }{\partial p_j}\,I_{\a\{\b_k\}}=\int_0^\infty dx\,\sdfrac{\partial}{\partial x}\left(x^{\a+1}\,\prod_{j=1}^{3}p_j^{\b_j}K_{\b_j}(p_j\,x)\right)-(\a+1-\b_t)\,I_{\a\{\b_k\}}\label{DCWIsol2}
\end{equation}
The first term on the right-hand side leads to a boundary term at $x=0$. In the region of convergence \eqref{region}, all integrals in this expression are well-defined and the boundary term vanishes. Then the triple-K integral satisfies an equation analogous to the dilatation equation with scaling degree
\begin{equation}
	\deg\Big(I_{\a\{\b_k\}}\Big)=\b_t-\a-1,
\end{equation}
and for $J_{N\{k_j\}}$ we obtain
\begin{equation}
	\deg\Big({J_{N\{k_j\}}}\Big)=\D_t+k_t-2d-N.
\end{equation}
It is worth mentioning that these results are valid in the case $\a+1\pm\b_1\pm\b_2\pm\b_3\ne-2k$ for some non-negative $k$ and independent choice of signs.

From this analysis it is simple to relate the form factors to the triple-K integrals. In the $\braket{TT\mO}$ case, we know from \eqref{dilfin3} that the form factors have degrees $\deg(A_n)=\D_t-2d-N_n$. In general, if $A_n=\a_\n\,J_{N\{k_j\}}$, where $\a_N\in \mathbb R$ is a constant value, then we can determine the value of $N=N(A_n)$ just through a comparison of the degree, i.e.  
\begin{equation}
	\begin{matrix}
		\hspace{2.5ex}\deg\big(A_n\big)&\hspace{-4.5ex}=\D_t-2d-N_n\\[1.2ex]
		\deg\Big(\,J_{N\{k_j\}}\Big)&\hspace{-1.5ex}=\D_t+k_t-2d-N
	\end{matrix}\quad\implies\quad
	N(A_n)=N_n+k_t,
\end{equation}
where $k_t=k_1+k_2+k_3$. The dilatation Ward identities will fix the values of $N$ in the triple-K integral and then can write a first general expression for the form factors as
\begin{equation}
	\begin{split}
		A_1&=\a_1\,J_{4+k_t\{k_j\}},\\
		A_2&=\a_2\,J_{2+k_t\{k_j\}},\\
		A_3&=\a_3\,J_{k_t\{k_j\}}.
	\end{split}\label{result1}
\end{equation}
These relations will be more constrained by imposing the primary and secondary CWI's. 
\subsection{Primary CWI's}\label{primarysol}
We have analysed the basic properties of the triple-K integral. We now want to use the results in the previous section in order to write a solution of the primary CWI's. Using the relation \eqref{Fund} in \appref{AppendixJ} we show that for any $m,n=1,2,3$ 
\begin{equation}
	K_{nm}\,J_{N\{k_j\}}=-2k_n\ J_{N+1\{k_j-\d_{jn}\}}+2k_m\ J_{N+1\{k_j-\d_{jm}\}}\label{sol1}
\end{equation}
for $k_1,\,k_2,\,k_3,\,N\in\mathbb R$, and where $K_{nm}$ is the conformal scalar operator defined in \eqref{Koper} and $\delta_{jn}$ is the Kronecker delta. To make the above equation clearer, we consider the case with $K_{13}$ 
\begin{align}
	K_{13}\,J_{N\{k_j\}}&=-2k_1\ J_{N+1\{k_j-\d_{j1}\}}+2k_3\ J_{N+1\{k_j-\d_{j3}\}}\notag\\
	&=-2k_1\ J_{N+1\{k_1-1,k_2,k_3\}}+2k_3\ J_{N+1\{k_1,k_2,k_3-1\}},
\end{align}
from which one can write the two following equations
\begin{align}
	&K_{13}\,J_{N\{0,k_2,0\}}=0,\notag\\
\end{align}
for any $k_2$, then also $k_2=0$. We can generalize this result as
\begin{equation}
	K_{nm}\,J_{N\{0,0,0\}}=0,
\end{equation}  
for any $n,m=1,2,3$. Let us consider the primary CWIs for the $\braket{TT\mO}$ given in \eqref{primaryCWI} and we are going to solve these equations in terms of triple-K integrals. The approach that allows us to find the general solutions to these constraints, is to construct a set of primary CWIs homogeneous using \eqref{primaryCWI}, and applying on them recursively operators $K_{nm}$ as follows
\begin{equation}
	\begin{matrix}
		K_{12}A_1=0,&\qquad K_{13}A_1=0,&\qquad\\[1.2ex]
		K_{12}A_2=0,&\qquad K_{13}^2A_2=0,&\qquad K_{12}K_{13}A_2=0\\[1.2ex]
		K_{12}A_3=0,&\qquad K_{13}^3A_3=0,&\qquad K_{12}K^2_{13}A_3=0,&\qquad K_{12}K_{13}A_3=0.\\
	\end{matrix}\label{homog}
\end{equation}

Using the results \eqref{result1} we can obtain some constraints on the form factors by the homogeneous primary CWIs, and for $A_1$ we find 
\begin{equation}
	\begin{matrix}
		&K_{12}A_1=\a_1K_{12}\ J_{4+k_t\{k_j\}}=0\\[1.2ex]
		&K_{13}A_1=\a_1K_{13}\ J_{4+k_t\{k_j\}}=0
	\end{matrix}
	\,\quad \implies\quad k_1=k_2=k_3=0
\end{equation}
giving the solution $A_1=\a_1\,J_{4\{000\}}$. Observe that if we impose only one homogeneous equation, say $K_{13}A_1=0$, then the most general solution in terms of the triple-K integrals is $\a\,J_{N\{0,k_2,0\}}$ for any $\a,\,N,\,k_3\in\mathbb{R}$. For the second form factors $A_2$ we can resolve the first equation
\begin{equation}
	0=K_{12}A_2=\a_2K_{12}\ J_{2+k_t\{k_j\}}=-2\a k_1\,J_{3+k_t\{k_1-1,k_2,k_3\}}+ 2\a k_2\,J_{3+k_t\{k_1,k_2-1,k_3\}}\,\\[1.2ex]
\end{equation}
that give the constraint $k_1=k_2=0$. Using this information we can resolve the second homogeneous equation for $A_2$ as
\begin{equation}
	0=K_{13}^2A_2=\a_2K^2_{13}\ J_{2+k_3\{00 k_3}=4\a\,k_3(k_3-1)\,J_{4+k_3\{00,k_3-2\}}
\end{equation}
given two solutions $k_3=0$ and $k_3=1$. The last equation 
\begin{equation}
	K_{12}K_{13}A_2=\a_2K_{12}K_{13}\ J_{2+k_3\{0,0,k_3\}}=0
\end{equation}
is identically satisfied and does not give any further informations. Finally, we can write down the most general solution for the second form factor $A_2$ in terms of the triple-K integrals as
\begin{equation}
	A_2=\a_2\,J_{2\{000\}}+\a_{21}\,J_{3\{001\}}
\end{equation}
where we have introduced another constant $\a_{12}\in\mathbb R$, related to the solution $k_3=1,\,k_1=k_2=0$. In the same way, we can solve the equations for the $A_3$ form factor in the form
\begin{align}
	&K_{12}A_3=0\,\quad\implies\quad k_1=k_2=0,\\[1.4ex]
	&K_{13}^3A_3=8\a_3\,k_3(k_3-1)(k_3-2)\,J_{k_3+3\{00,k_3-3\}}=0,\quad\implies\quad k_3=0,\ k_3=1,\ k_3=2
\end{align}
with the other equations identically satisfied. To summarize, we can write down the most general solutions obtained in our approach in the form
\begin{align}
	A_1&=\a_1\,J_{4\{000\}}\\[1.1ex]
	A_2&=\a_2\,J_{2\{000\}}+\a_{21}\,J_{3\{001\}}\\[1.1ex]
	A_3&=\a_3\,J_{0\{000\}}+\a_{31}\,J_{1\{001\}}+\a_{32}\,J_{2\{002\}},
\end{align}
where all the $\a$ are numerical constants. Finally, the inhomogeneous parts of \eqref{primaryCWI} fix some of these constants. When the solution above is substituted into the primary CWI's \eqref{primaryCWI} 
\begin{align}
	K_{13}A_2&=8A_1\ \implies\ 2\a_{21}J_{4\{000\}}=8\a_1\,J_{4\{000\}}\\
	K_{13}A_3&=2A_2\ \implies \ 2\a_{31}\,J_{2\{000\}}+4\a_{32}\,J_{3\{001\}}=2\a_2\,J_{2\{000\}}+2\a_{21}\,J_{3\{001\}}
\end{align}
they imply that
\begin{equation}
	\a_{21}=4\a_1,\quad \a_{31}=\a_2,\quad\a_{32}=2\a_1.
\end{equation}
In conclusions, we have analyzed the primary CWI's for the $\braket{TT\mO}$ correlation function and finally we have found that the solutions can be expressed in the form 
\begin{equation}
	\begin{split}
		A_1&=\a_1\,J_{4\{000\}}\\[1.1ex]
		A_2&=\a_2\,J_{2\{000\}}+4\a_{1}\,J_{3\{001\}}\\[1.1ex]
		A_3&=\a_3\,J_{0\{000\}}+\a_{2}\,J_{1\{001\}}+2\a_{1}\,J_{2\{002\}},
	\end{split}\label{primarysolution}
\end{equation}
depending on three undetermined constants $\a_1,\,\a_2,\,\a_3\in\mathbb R$. The method to solve the primary CWI's is applicable to generic 3-point functions and we will see that the secondary CWI's - in this case for the $TTO$, but equivalently in all the other cases - will reduce the number of undetermined constants to just one.

\subsection{The analysis of the secondary CWI's}
We have shown that the primary CWI's fix the functional structure of the form factors and write the solutions as linear combinations of triple-K integrals with some undetermined constants. The role of the secondary CWI's, that are first order partial differential equations, is to fix the algebraic relation between the undetermined constants in order to give a final solution given as a linear combination of a minimal and independent number of constants.\\ 
We can now impose the secondary CWI's for the solutions obtained in \eqref{primarysolution}. If we substitute the full solutions to the primary CWI's into the secondary to extract more information about the constants, we may encounter some troubles in the purely algebraic computation. However, we can examine the relations in the zero-momentum limit and this procedure can simplify our algebraic equations.\\ 
In the zero-momentum limit $p_{3\m}\to0$, $p_{1\m}=-p_{2\m}=p_{\m}$, there are different expansion for the Bessel function. In particular, depending on the parameter $\nu$ one can expand as
\begin{align}
	\,K_{0}(p_3 x)&= \log\left(\frac{2}{p_3\,x}\right)-\gamma_E\\
	p_3^{\nu}\,K_{\nu}(p_3 x),&
	=\left[\sdfrac{2^{{\nu}-1}\G(\nu)}{x^{\nu}}+O(p_3^2)\right]+p_3^{2{\nu}}\left[2^{-{\nu}-1}\G(-{\nu})\,x^\nu+O(p_3^2)\right],\quad \nu\notin\mathbb{Z}\\
	p_3^{n}\,K_{n}(p_3 x),&
	=\left[\sdfrac{2^{{n}-1}\G(n)}{x^{n}}+O(p_3^2)\right]+p_3^{2{n}}\left[\frac{(-1)^{n+1}}{2^n\,\Gamma(n+1)}\,x^n\,\log\,p_3+f(\log(x))+O(p_3^2)\right],\quad n\in\mathbb{N}.
\end{align}
obtained from the expression of the modified Bessel functions K given in \cite{Prudnikov}, indeed for the case $\nu=n\in\mathbb{N}$ the Bessel function assumes the following form
\begin{align}
	K_{n}(x)=\lim_{\nu\to n}\,K_{\nu}(x)&=(-1)^{n+1}\,I_{n}(x)\,\log\frac{x}{2}+\frac{1}{2}\sum_{k=0}^{n-1}(-1)^k\frac{(n-k-1)!}{k!}\,\left(\frac{x}{2}\right)^{2k-n}\notag\\
	&\qquad +\frac{(-1)^n}{2}\sum_{k=0}^\infty\,\frac{\psi(k+n+1)+\psi(k+1)}{(n+k)!k!}\,\left(\frac{x}{2}\right)^{n+2k},
\end{align}
where $I_n(x)$ is the modified Bessel function of the first kind. For the further discussion, we will assume that $\beta_3>0$ that translates, in terms of scaling dimensions, in the condition $\D_3>\sdfrac{d}{2}$. The latter is always satisfied when the correlation function is constructed with conserved currents and stress-energy tensor. It is worth noting that the relation $K_{-\nu}(x)=K_{\nu}(x)$ is valid in any case. 
With the assumptions discussed above, we can compute the zero-momentum limit of the triple-K integrals as
\begin{equation}
	\lim_{p_3\to0}I_{\a\{\b_j\}}(p,p,p_3)=2^{\b_3-1}\G(\b_3)p^{\b_1+\b_2}\int_0^\infty\,dx\,x^{\,\a-\b_3}\,K_{\b_1}(px)\,K_{\b_2}(px)
\end{equation}
and using the relation in \cite{Prudnikov} 
\begin{align}
	\int_0^\infty\,dx\,x^{\,\a-1}\,K_{\n}(cx)\,K_{\m}(cx)&=\sdfrac{2^{\a-3}}{\G(\a)\,c^\a}\,\Gamma\left(\frac{\alpha+\mu+\nu}{2}\right)\Gamma\left(\frac{\alpha+\mu-\nu}{2}\right)\Gamma\left(\frac{\alpha-\mu+\nu}{2}\right)\Gamma\left(\frac{\alpha-\mu-\nu}{2}\right)\notag\\
	&\text{with}\quad \Re(c)>0,\,\Re(\alpha)>|\Re(\mu)|+|\Re(\nu)|,
\end{align}
we obtain the final result
\begin{equation}
	\lim_{p_3\to0}I_{\a\{\b_j\}}(p,p,p_3)=p^{\b_t-\a-1}\ \ell_{\a\{\b_j\}}\label{zeromom}
\end{equation}
where
\begin{equation}
	\ell_{\a\{\b_j\}}=\sdfrac{2^{\alpha-3}\Gamma(\beta_3)}{\G(\a-\beta_3+1)}\,\Gamma\left(\frac{\alpha+\beta_t+1}{2}-\beta_3\right)\Gamma\left(\frac{\alpha-\beta_t+1}{2}+\beta_1\right)\Gamma\left(\frac{\alpha-\beta_t+1}{2}+\beta_2\right)\Gamma\left(\frac{\alpha-\beta_t+1}{2}\right)\label{ldef}
\end{equation}
which is valid away from poles of the gamma function with the conditions $\a>\b_t-1$, $p>0$. 
\subsection{Solving the secondary equations}
As previously mentioned, the secondary CWI's allow to establish a relation among all the constants in the solutions of the primary CWI's. For this reason, we perform the zero momentum limit in order to extract these algebraic equations involving all the constants. In this section we illustrate this procedure for the $\braket{TTO}$. \\
First of all we have to explicit the secondary CWI's \eqref{secondary} using the expression of the form factors given by the primary \eqref{primarysolution}, and making use of the relations \eqref{1dJ}-\eqref{Fund} in \appref{transWard} we obtain
\begin{align}
	0&=L_2A_2+4RA_3=\a_2\bigg[-p_1^2(p_1^2+p_2^2-p_3^2)\,I_{\frac{d}{2}+2\,\left\{\frac{d}{2}-1,\,\frac{d}{2},\,\D_3-\frac{d}{2}\right\}}-2p_1^2p_2^2\,I_{\frac{d}{2}+2\,\left\{\frac{d}{2},\,\frac{d}{2}-1,\,\D_3-\frac{d}{2}\right\}}\notag\\
	&+d(p_3^2-p_2^2-p_1^2)\,I_{\frac{d}{2}+1\,\left\{\frac{d}{2},\,\frac{d}{2},\,\D_3-\frac{d}{2}\right\}}+2p_1^2\,I_{\frac{d}{2}+1\,\left\{\frac{d}{2},\,\frac{d}{2},\,\D_3-\frac{d}{2}\right\}}\bigg]
	+4\a_1\bigg[-p_1^2(p_1^2+p_2^2-p_3^2)\,I_{\frac{d}{2}+3\,\left\{\frac{d}{2}-1,\,\frac{d}{2},\,\D_3-\frac{d}{2}+1\right\}}\notag\\
	&-2p_1^2p_2^2\,I_{\frac{d}{2}+3\,\left\{\frac{d}{2},\,\frac{d}{2}-1,\,\D_3-\frac{d}{2}+1\right\}}+d(p_3^2-p_2^2-p_1^2)\,I_{\frac{d}{2}+2\,\left\{\frac{d}{2},\,\frac{d}{2},\,\D_3-\frac{d}{2}+1\right\}}+2p_1^2\,I_{\frac{d}{2}+2\,\left\{\frac{d}{2},\,\frac{d}{2},\,\D_3-\frac{d}{2}+1\right\}}\bigg]\notag\\
	&+4\a_3\,\bigg(-p_1^2\,I_{\frac{d}{2}\,\left\{\frac{d}{2}-1,\,\frac{d}{2},\,\D_3-\frac{d}{2}\right\}}-d\,I_{\frac{d}{2}-1\,\left\{\frac{d}{2},\,\frac{d}{2},\,\D_3-\frac{d}{2}\right\}}\bigg)+4\a_2\,\left(-p_1^2\,I_{\frac{d}{2}+1\,\left\{\frac{d}{2}-1,\,\frac{d}{2},\,\D_3-\frac{d}{2}+1\right\}}-d\,I_{\frac{d}{2}\,\left\{\frac{d}{2},\,\frac{d}{2},\,\D_3-\frac{d}{2}+1\right\}}\right)\notag\\
	&+8\a_1\,\left(-p_1^2\,I_{\frac{d}{2}+2\,\left\{\frac{d}{2}-1,\,\frac{d}{2},\,\D_3-\frac{d}{2}+2\right\}}-d\,I_{\frac{d}{2}+1\,\left\{\frac{d}{2},\,\frac{d}{2},\,\D_3-\frac{d}{2}+2\right\}}\right),\label{firstsecondary}
\end{align}
where we have used the definition of the reduced version of the triple-K integral 
\begin{equation}
	J_{N\{k_j\}}=I_{\frac{d}{2}-1+N\{\D_j-\frac{d}{2}+k_j\}}.
\end{equation}
Equally, one can write the other secondary CWI's will have the form
\begin{align}
	0&=L_2A_1+RA_2\notag\\
	&=\a_1\bigg[-p_1^2(p_1^2+p_2^2-p_3^2)\,I_{\frac{d}{2}+4\,\left\{\frac{d}{2}-1,\,\frac{d}{2},\,\D_3-\frac{d}{2}\right\}}-2p_1^2p_2^2\,I_{\frac{d}{2}+4\,\left\{\frac{d}{2},\,\frac{d}{2}-1,\,\D_3-\frac{d}{2}\right\}}+d(p_3^2-p_2^2-p_1^2)\,I_{\frac{d}{2}+3\,\left\{\frac{d}{2},\,\frac{d}{2},\,\D_3-\frac{d}{2}\right\}}\notag\\
	&\hspace{1.6cm}+2p_1^2\,I_{\frac{d}{2}+3\,\left\{\frac{d}{2},\,\frac{d}{2},\,\D_3-\frac{d}{2}\right\}}\bigg] +\a_2\left(-p_1^2\,I_{\frac{d}{2}+2\,\left\{\frac{d}{2}-1,\,\frac{d}{2},\,\D_3-\frac{d}{2}\right\}}-d\,I_{\frac{d}{2}+1\,\left\{\frac{d}{2},\,\frac{d}{2},\,\D_3-\frac{d}{2}\right\}}\right)\notag\\
	&\hspace{0.7cm}+4\a_1\,\left(-p_1^2\,I_{\frac{d}{2}+3\,\left\{\frac{d}{2}-1,\,\frac{d}{2},\,\D_3-\frac{d}{2}+1\right\}}-d\,I_{\frac{d}{2}+2\,\left\{\frac{d}{2},\,\frac{d}{2},\,\D_3-\frac{d}{2}+1\right\}}\right)\ .\label{secondsecondary}
\end{align}
We now consider the zero-momentum limit \eqref{zeromom} of these equations, obtaining the relation
\begin{align}
	0&=\lim_{p_3\to0}L_2A_2+4RA_3\notag\\[1.5ex]
	&=-2p^{\D_3}\bigg\{ 4\a_1	 \bigg[d\,\ell_{\frac{d}{2}+1\,\left\{\frac{d}{2},\,\frac{d}{2},\,\D_3-\frac{d}{2}+2\right\}}+\,\ell_{\frac{d}{2}+2\,\left\{\frac{d}{2}-1,\,\frac{d}{2},\,\D_3-\frac{d}{2}+2\right\}}+\,\ell_{\frac{d}{2}+3\,\left\{\frac{d}{2}-1,\,\frac{d}{2},\,\D_3-\frac{d}{2}+1\right\}}+\,\ell_{\frac{d}{2}+3\,\left\{\frac{d}{2},\,\frac{d}{2}-1,\,\D_3-\frac{d}{2}+1\right\}}\notag\\
	&+(d-1)\,\ell_{\frac{d}{2}+2\,\left\{\frac{d}{2},\,\frac{d}{2},\,\D_3-\frac{d}{2}+1\right\}}\bigg]\,+\,\a_2\bigg[\,2\,\ell_{\frac{d}{2}+1,\left\{\frac{d}{2}-1,\,\frac{d}{2}-1,\,\D_3-\frac{d}{2}+1\right\}}+\,\ell_{\frac{d}{2}+2\,\left\{\frac{d}{2}-1,\,\frac{d}{2},\,\D_3-\frac{d}{2}\right\}}+\,\ell_{\frac{d}{2}+2\,\left\{\frac{d}{2},\,\frac{d}{2}-1,\,\D_3-\frac{d}{2}\right\}}\notag\\
	&+(d-1)\,\ell_{\frac{d}{2}+1\,\left\{\frac{d}{2},\,\frac{d}{2},\,\D_3-\frac{d}{2}\right\}}+2d\,\ell_{\frac{d}{2}\,\left\{\frac{d}{2},\,\frac{d}{2},\,\D_3-\frac{d}{2}+1\right\}}\bigg]\,+\,2\a_3\,\bigg[d\,\ell_{\frac{d}{2}-1\,\left\{\frac{d}{2},\,\frac{d}{2},\,\D_3-\frac{d}{2}\right\}}+\,\ell_{\frac{d}{2}\,\left\{\frac{d}{2}-1,\,\frac{d}{2},\,\D_3-\frac{d}{2}\right\}}\bigg]\ \bigg\}\label{firstE}
\end{align}
\begin{align}
	0&=\lim_{p_3\to0}L_2A_1+RA_2\notag\\[1.5ex]
	&=p^{\D_3-2}\bigg\{\, -2\a_1\,\bigg[\,2d\,\ell_{\frac{d}{2}+2\,\left\{\frac{d}{2},\,\frac{d}{2},\,\D_3-\frac{d}{2}+1\right\}}+2\,\ell_{\frac{d}{2}+3\,\left\{\frac{d}{2}-1,\,\frac{d}{2},\,\D_3-\frac{d}{2}+1\right\}}+(d-1)\,\ell_{\frac{d}{2}+3\,\left\{\frac{d}{2},\,\frac{d}{2},\,\D_3-\frac{d}{2}\right\}}\notag\\
	&\qquad+\,\ell_{\frac{d}{2}+4\,\left\{\frac{d}{2}-1,\,\frac{d}{2},\,\D_3-\frac{d}{2}\right\}}+\,\ell_{\frac{d}{2}+4\,\left\{\frac{d}{2},\,\frac{d}{2}-1,\,\D_3-\frac{d}{2}\right\}}\bigg]\,-\,\a_2\,\bigg[\,d\,\ell_{\frac{d}{2}+1\,\left\{\frac{d}{2},\,\frac{d}{2},\,\D_3-\frac{d}{2}\right\}}+\,\ell_{\frac{d}{2}+2\,\left\{\frac{d}{2}-1,\,\frac{d}{2},\,\D_3-\frac{d}{2}\right\}}\bigg]\ \bigg\}.\label{secondE}
\end{align}
The two secondary CWI's in \eqref{firstE} and \eqref{secondE} scale homogeneously as $p^{\D_3}$ and $p^{\D_3-2}$ respectively. Thus, expanding the result, by using the definition \eqref{ldef}, we derive the relations
\begin{align}
	0&=\lim_{p_3\to0}L_2A_2+4RA_3=p^{\D_3}\,2^{\frac{d}{2}-4}\,\G\left(-\sdfrac{\D_3}{2}\right)\,\left[\G\left(\sdfrac{d-\D_3}{2}\right)\right]^2\,\G\left(d-\sdfrac{\D_3}{2}\right)\,\G\left(\D_3-\sdfrac{d}{2}\right)\sdfrac{1}{\G\left(d-\D_3\right)}\notag\\
	&\hspace{-0.4cm}\times\bigg\{-2\a_3(2d-\D_3)-2\a_1(2d-\D_3)(\D_3+2)(d-\D_3-2)(d-2\D_3)+\a_2\left(d-\sdfrac{\D_3}{2}\right)\big(d(\D_3+4)-\D_3(\D_3+8)\big)\bigg\}\\
	0&=\lim_{p_3\to0}L_2A_1+RA_2=p^{\D_3-2}\,2^{\frac{d}{2}-4}\,\G\left(1-\sdfrac{\D_3}{2}\right)\left[\G\left(\sdfrac{d-\D_3}{2}\right)\right]^2\G\left(d-\sdfrac{\D_3}{2}+2\right)\G\left(\D_3-\sdfrac{d}{2}\right)\sdfrac{(d-\D_3)^2}{\G(d-\D_3+2)}\notag\\
	&\times\bigg\{\ \a_1(\D_3+2)(d-\D_3-2)-\a_2\bigg\}
\end{align}
In order to satisfy the above equations, the only way is to set the coefficient in curly brackets to vanish, obtaining
\begin{align}
	\a_2&=\a_1(\D_3+2)(d-\D_3-2)\\
	\a_3&=\sdfrac{\a_1}{4}\D_3(\D_3+2)(d-\D_3-2)(d-\D_3).
\end{align}
From this analysis we observe that the final solution depends only on a single constant factor $\a_1$. 

\section{Regularisation and renormalization}\label{reg}
{ In this section we briefly discuss the procedure of regularisation of the triple-K integrals when some divergences occur. For more details about this procedure see \cite{Bzowski:2014qja,Bzowski:2015yxv, Bzowski:2017poo}.\\
	We will now focus on the special cases where the triple-K integral is singular, i.e., when the dimensions of operators satisfy one or more of the conditions
	\begin{equation}
		\a+1\pm\b_1\pm\b_2\pm\b_3=-2k\label{cond}
	\end{equation}
	for some non-negative integer $k$. In this region of the parameters, the general triple-K integral defined in \eqref{3Kint} diverges and we have to take in account a regularization procedure. In order to do that, we introduce the regulated parameters}
\begin{equation}
	\a\mapsto\bar{\a}=\a+u\e,\qquad \b_j\mapsto\bar{\b_j}=\b_j+v_j\e,\label{shift}
\end{equation}
where $u$ and $v_j$, $j=1,2,3$ are fixed but arbitrary numbers that specify the direction of the shift. In this way we regard the regulated triple-K integral 
\begin{equation}
	I_{\a\{\b_k\}}(p_1,p_2,p_3)\mapsto I_{\bar{\a}\{\bar{\b_k}\}}(p_1,p_2,p_3)
\end{equation}
as a function of the regulator $\e$ with all the momenta fixed. The divergences of the integrals manifest as a pole at $\e=0$. { It is worth mentioning that not all the choices of the shifts $u$, $v_j$ actually regulate the integral, but some useful choices exist. For instance, if we consider the scheme $u=v_1=v_2=v_3$, this is equivalent to perform a shift of the physical dimensions of the same quantity $2u\e$. In fact, the general mapping \eqref{shift} can be a viewed as shift of the spacetime and conformal dimensions as
	\begin{equation}
		d\mapsto d+2u\e,\qquad \D_j\mapsto\D_j+(u+v_j)\e,\ \ j=1,2,3
	\end{equation}
	where with the choice mentioned above we have an equal scaling of $d$ and $\Delta_j$. Another particular and useful choice is to consider $v_j=0$ for all $j=1,2,3$. This scheme preserves the $\b_j$ and hence the indices of the Bessel functions in the triple-K integral, and then the expansion to extract the pole structures in $\e$ will be easier to perform.\\ 
	We now consider the case of the $\braket{TT\mO}$ correlation function. The solutions for the form factors from the conformal Ward identities are written in terms of triple-K integrals as in  \eqref{primarysolution}. Looking at the first form factor $A_1$, given just in terms of $J_{4\{000\}}$, we observe that it can diverge, by using \eqref{cond}, if one or more of the conditions
	\begin{equation}
		\frac{d}{2}+4\pm\frac{d}{2}\pm\frac{d}{2}\pm\left(\D_3-\frac{d}{2}\right)=-2k,\qquad k\in\mathbb{Z}^+\label{diverge}
	\end{equation}
	are satisfied. One can use a special notation in order to refer to all the cases in which a singularity shows up. If  denote by $(\pm\pm\pm)$ the possible choices of the sign in \eqref{diverge},  the are six cases in total. However, in order to have a negative even number on the LHS of \eqref{diverge}, there are just two possible choices of signs, which are $(++-)$ and $(+++)$, with
	\begin{equation}
		\d_{+}=4-d+\D_3,\qquad\d_{-}=4-\D_3\label{delta+}
	\end{equation}
	where $\d$ is the left-hand side of \eqref{diverge} and the subscript $\pm$ indicate the sign choice  of the $\D_3-d/2$ term. We observe that in the case $\D_3=4+2k$ or $\D_3=d-4-2k$ the particular triple-K integral $J_{4\{000\}}$ has divergences and needs to be regulated. One needs to repeat this analysis for all the integrals appearing in the final form of all the form factors.\\
	To present a practical example, we consider the scalar operator in the $3$-point function $\braket{TT\mO}$ with conformal dimension $\D_3=1$ and in a $d=3$ dimensional CFT. Under these conditions, one can immediately see that the form factor $A_1$ does not contain divergences and indeed by using \eqref{delta+} we find $\delta_{\pm}\ne-2k$, with $k\in\mathbb{Z}^+$. The same does not happen for the other two form factors $A_2$ and $A_3$ in \eqref{primarysolution}.   
	Using the results in \cite{Bzowski:2013sza,Bzowski:2015yxv} we see that the only convergent integrals are $J_{4\{000\}}=I_{\frac{9}{2}\left\{\frac{3}{2},\frac{3}{2},-\frac{1}{2}\right\}}$ and $J_{3\{001\}}=I_{\frac{7}{2}\left\{\frac{3}{2},\frac{3}{2},\frac{1}{2}\right\}}$. The remaining integrals require a regularisation procedure. To do this, we choose the particular scheme $u=1$ and $v_j=0$, $j=1,2,3$ for which we can calculate all the shifted integrals of the form $J_{N+\e\{k_j\}}$ using \eqref{halfinteg} in \appref{transWard}, and we can expand the result around $\e=0$. For instance, if we take the integral
	\begin{equation}
		J_{2+\e\{000\}}=I_{\frac{5}{2}+\e\left\{\frac{3}{2},\frac{3}{2},-\frac{1}{2}\right\}}=\left(\frac{\pi}{2}\right)^{\frac{3}{2}}\frac{1}{\e\,p_3}+O(\e^0)
	\end{equation}
	since the regulator $\e$ in the triple-K integrals cannot be removed, we cannot easily take the zero-momentum limit to resolve the secondary CWI's. In order to obtain a finite expression of the form factors, one has to assume that the constants $\a_j$ in \eqref{primarysolution} depend on the regulator $\e$ as well. We make explicit the $\e$ dependence of the coefficient $\a_j$ by the definition
	\begin{equation}
		\a_j=\sum_{n=-\infty}^{\infty}\,\a_j^{(n)}\e^n,\qquad j=1,2,3\label{expans}.
	\end{equation}
	However, the constant $\a_1$ does not depend on the regulator because it multiplies an integral which is already finite, and for this reason we will choose $\a_1=\a^{(0)}_1$. In the next section we will present all the details of the analysis of the secondary CWI's for the case $\braket{TT\mO}$ when there are divergences. 
}

\subsection{Secondary CWI's with divergent triple-K integral} 
We analyse in details the method to extract information from the secondary CWI's when the relation \eqref{cond} holds and the triple-K integrals manifest a singular behaviour. 
Consider the solution to the primary CWI's for the $\braket{TT\mO}$ point function 
\begin{align}
	A_1&=\a_1\,J_{4\{000\}}\\
	A_2&=\a_2\,J_{2\{000\}}+4\a_{1}\,J_{3\{001\}}\\
	A_3&=\a_3\,J_{0\{000\}}+\a_{2}\,J_{1\{001\}}+2\a_{1}\,J_{2\{002\}}\label{TTOsol}
\end{align}
and the secondary CWI's
\begin{align}
	L_2A_1+RA_2=0,\qquad L_2A_2+4RA_3=0.
\end{align}
We consider the case previously mentioned with $\D_3=1$, $d=3$, recalling that the secondary CWI's are}
\begin{align}
0&=L_2A_1+RA_2\\
0&=L_2A_2+4RA_3.
\end{align}
{
These form factors contain the integrals $J_{2\{000\}}$, $J_{1\{001\}}$, $J_{0\{000\}}$, $J_{1\{-100\}}$ that are divergent and can be regulated by using the prescription described in the previous section. Then, expanding in power of $\epsilon$ we find
\begin{align}
	J_{2+\e\{002\}}&=I_{\frac{5}{2}+\e\left\{\frac{3}{2},\frac{3}{2},\frac{3}{2}\right\}}=-\left(\sdfrac{\pi}{2}\right)^\frac{3}{2}\sdfrac{p_1^3+2p_1^2(p_2+p_3)+2p_1(p_2^2+p_2\,p_3+p_3^2)+(p_2+p_3)(p_2^2+p_2p_3+p_3^2)}{(p_1+p_2+p_3)^2}+O(\e)\notag\\
	J_{2+\e\{000\}}&=I_{\frac{5}{2}+\e\left\{\frac{3}{2},\frac{3}{2},-\frac{1}{2}\right\}}=\left(\sdfrac{\pi}{2}\right)^\frac{3}{2}\sdfrac{1}{\e p_3}+O(\e^0)\notag\\
	J_{1+\e\{001\}}&=I_{\frac{3}{2}+\e\left\{\frac{3}{2},\frac{3}{2},\frac{1}{2}\right\}}=-\left(\sdfrac{\pi}{2}\right)^\frac{3}{2}\sdfrac{p_3}{\e}+O(\e^0)\notag\\
	J_{0+\e\{000\}}&=I_{\frac{1}{2}+\e\left\{\frac{3}{2},\frac{3}{2},-\frac{1}{2}\right\}}=-\left(\sdfrac{\pi}{2}\right)^\frac{3}{2}\sdfrac{p_1^2+p_2^2-p_3^2}{2\e p_3}+O(\e^0).\label{integrals}
\end{align}.
All the other integrals in the form factors are convergent and are explicitly given by
\begin{align}
	J_{4\{000\}}&=I_{\frac{9}{2}\left\{\frac{3}{2},\frac{3}{2},-\frac{1}{2}\right\}}=\left(\sdfrac{\pi}{2}\right)^\frac{3}{2}\sdfrac{3p_1^3+4p_1(3p_2+p_3)+(p_2+p_3)(3p_2+p_3)}{(p_1+p_2+p_3)^4}\label{J4000}\\
	J_{3\{001\}}&=I_{\frac{7}{2}\left\{\frac{3}{2},\frac{3}{2},\frac{1}{2}\right\}}=\left(\sdfrac{\pi}{2}\right)^\frac{3}{2}\sdfrac{(p_1+p_2+p_3)(p_1+2p_2+p_3)+p_1(p_1+3p_2+p_3)}{(p_1+p_2+p_3)^3}\label{J3001}.
\end{align}
Using the explicit expressions written above, the secondary CWI's turn into 
\begin{align}
	0&=L_2A_1+RA_2\notag\\
	&=-\a_1\left(\sdfrac{\pi}{2}\right)^\frac{3}{2}\sdfrac{9}{p_3}-\a_2\left(\sdfrac{\pi}{2}\right)^\frac{3}{2}\sdfrac{p_1^2(p_1+3p_2+p_3)}{p_3\,(p_1+p_2+p_3)^3}\notag\\
	&\quad-3\left(\sdfrac{\pi}{2}\right)^\frac{3}{2}\frac{\a_2}{p_3}\left(\sdfrac{(p_1+p_2)(p_1+p_2+p_3)+p_1p_2)}{(p_1+p_2+p_3)^2}-\log\left(p_1+p_2+p_3\right)+\frac{1}{\e}-\gamma_E\right)+O(\e)\label{F7}
\end{align}
\begin{align}
	0&=L_2A_2+4RA_3\notag\\[2ex]
	&=36\a_1\left(\sdfrac{\p}{2}\right)^\frac{3}{2}\,p_3+\frac{\alpha_2}{p_3}\left(\sdfrac{\p}{2}\right)^\frac{3}{2}\bigg[\frac{(15p_3^2-3p_2^2-p_1^2)}{\e}+f_1\Big(p_1,p_2,p_3,\g_E\Big)\bigg]\notag\\
	&\qquad+\alpha_3\left(\sdfrac{\p}{2}\right)^\frac{3}{2}\Bigg[\frac{2(p_1^2+3p_2^2-3p_3^2)}{p_3\,\e}+f_2\Big(p_1,p_2,p_3,\g_E\Big)\Bigg]+O(\e)\label{F8},
\end{align}
where $f_1$ and $f_2$ are two analytic functions of the momenta in the physical region $p_1+p_2+p_3>0$, with $\gamma_E$ denoting the Euler constant. Notice that  
the $f'$s will not be relevant in the discussion that will follow, since they will be multiplied by positive powers of the regulator $\epsilon$.
In order to take care of the divergences we make explicit the $\e$ dependence of the coefficients $\alpha_j$ through the definition \eqref{expans}. Referring to the solution \eqref{TTOsol} of the form factors, we observe that they must not depend on the regulator $\e$. Therefore we impose the conditions \cite{Bzowski:2013sza}

\begin{equation}
	\begin{split}
		&\a_1=\a_1^{(0)},\\
		&\a_2^{(n)}=0,\\
		&\a_3^{(n)}=0,
	\end{split}
	\qquad n=-1,-2,-3,\dots.
\end{equation}
Then, collecting all the terms in \eqref{F7} with respect to the powers of $\e$ and imposing that they are zero order by order we find the conditions
\begin{align}
	\a_2^{(0)}&=0,&&\text{$\e^{-1}$ order}\label{Cond1}\\
	\a_2^{(1)}&=-3\a_1^{(0)},&&\text{$\e^0$ order}\label{Cond2}\\
	&\dots\,,\label{Condn}
\end{align}
where we have defined $p_{123}=p_1+p_2+p_3$. 
The same procedure may be applied to the remaining secondary CWI's \eqref{F8} giving the conditions
\begin{align}
	\a_3^{(0)}&=0,&&\text{$\e^{-1}$ order}\label{Cond10}\\
	\a_3^{(1)}&=-\frac{3}{2}\a_1^{(0)},&&\text{$\e^{0}$ order}\label{Cond11}\\
	&\dots,
\end{align}
where we have used the results in \eqref{Cond1}-\eqref{Condn}. It is worth mentioning that the coefficients $\alpha_2^{(k)},\,\alpha_3^{(k)}$, $k=2,3,\dots$, do not contribute to the final regularized expression of the form factors when the regulator $\e$ is set to zero. Finally, we can give the regulated expression for the form factors in $d=3$ and $\Delta=1$ as
\begin{align}
	A_1&=\a_1^{(0)}\,J_{4\{000\}}\\
	A_2&=-3\a_2^{(0)}\,J^{Reg}_{2\{000\}}+4\a_{1}^{(0)}\,J_{3\{001\}}\\
	A_3&=-\sdfrac{3}{2}\a_1^{(0)}\,J^{Reg}_{0\{000\}}-3\a_{1}^{(0)}\,J^{Reg}_{1\{001\}}+2\a_{1}^{(0)}\,J^{Reg}_{2\{002\}}
\end{align}
where $J_{4\{000\}}$ and $J_{3\{000\}}$ are given by \eqref{J4000} and \eqref{J3001} respectively, and the other regularized integrals are explicitly written as
\begin{align}
	J^{Reg}_{2\{002\}}&=-\left(\sdfrac{\pi}{2}\right)^\frac{3}{2}\sdfrac{p_1^3+2p_1^2(p_2+p_3)+2p_1(p_2^2+p_2\,p_3+p_3^2)+(p_2+p_3)(p_2^2+p_2p_3+p_3^2)}{(p_1+p_2+p_3)^2},\\
	J^{Reg}_{2\{000\}}&=\left(\sdfrac{\pi}{2}\right)^\frac{3}{2}\sdfrac{1}{p_3},\\
	J^{Reg}_{0\{000\}}&=-\left(\sdfrac{\pi}{2}\right)^\frac{3}{2}\,p_3,\\
	J^{Reg}_{1\{001\}}&=-\left(\sdfrac{\pi}{2}\right)^\frac{3}{2}\sdfrac{p_1^2+p_2^2-p_3^2}{2p_3}.
\end{align}
Also in this case all the form factors are completely fixed modulo an overall constant $\alpha_1^{(0)}$. We notice that these results are in agreement with the relations between the constants obtained in the general case 
\begin{align}
	\a_2&=\a_1(\D_3+2)(d-\D_3-2)\\
	\a_3&=\sdfrac{\a_1}{4}\D_3(\D_3+2)(d-\D_3-2)(d-\D_3).
\end{align}
With the conditions $\Delta_3=1$ and $d=3+\e$, they are written as
\begin{equation}
	\a_2=-3\e\,\a_1^{(0)},\qquad\a_3=-\sdfrac{3}{2}\,\e\,\a_1^{(0)},
\end{equation}
which, order by order in $\e$, give the same constraints \eqref{Cond1}-\eqref{Cond2} and \eqref{Cond10}-\eqref{Cond11} discussed in this section. 
We have illustrated a method \cite{Bzowski:2013sza} for extracting the algebraic dependencies among the integration constants, using the primary and secondary CWI's. The analysis has been performed in the general case, where the regulator can be removed from all the triple-K integrals involved since we are merely avoiding unphysical singularities. In the following section, we will discuss the case in which the regulator cannot be removed, which takes to the ordinary regularisation of the integrals and the conformal anomaly generation.\\
As we have just discussed, the primary CWI's can be solved using triple-K integrals, and the solutions are substituted into the secondary CWI's. In general, the secondary CWI's lead to linear algebraic equations between the various constants appearing in the solutions of the primary conformal Ward identities. The precise form of the secondary CWI's depends on the information provided by the transverse Ward identities. In the $\braket{TT\mO}$, we have shown that the transverse Ward identity does not contribute to the secondary CWI's. In general, this is not the case in the $\braket{TTT}$. \\
Furthermore, we have also illustrated how to extract a set of algebraic equations for the constants, obtained from the secondary CWI's. This set of equations may be extracted by analysing the zero-momentum limit of the same equations when the regulator can be removed from all triple-K integrals. Finally, we have also shown how to regulate the integrals when the zero-momentum limit generates divergences. This procedure brought to another set of equations that, once solved, give relations among the constants. The final result of the 3-point function is unique up to one overall constant, which can be matched to a free field theory with a particular particle content. This approach is fully addressed in the $TJJ$ and $TTT$ correlators in the next sections. 

\section{The \texorpdfstring{$TJJ$}{} in momentum space}\label{TJJmom}
In this section, we apply the method presented in the previous sections to the $TJJ$ case. We will study the decomposition of the correlator in its transverse-traceless part and semi-local one. Due to the conservation Ward identities, the latter is given explicitly in terms of two point functions. Then we will study the differential equations related to the conformal invariance. These constraints for the form factors of the transverse traceless part will be explicitly solved, giving the correlator's general solutions, written in terms of hypergeometric functions or equivalently in terms of $3$K integrals. 

\subsection{Conservation Ward identities} 
To fix the form of the correlator we need to impose the transverse WI on the vector lines and the conservation WI for $T^{\mu\nu}$. 
In this section we briefly discuss their derivation and their explicit expressions.
We consider the functional 
\begin{equation}
	W[g,A] =\int {D\bar\psi}D\psi e^{-(S_0[g,\psi] + S_1[A,\psi ])}
	\label{GF}
\end{equation}
integrated over the fermions $\psi$, in the background of the metric $g_{\mu\nu}$ and of the gauge field $A_\mu^a$. In the case of a non abelian gauge theory the action is given by
\begin{align}
	S_0[g,A,\psi]&=-\frac{1}{4}\int d^4 x \sqrt{-g_x}F_{\mu\nu}^{{a}} F^{\mu\nu a} +\int d^4 x \sqrt{-g_x} i \bar{\psi}\gamma^\mu D_\mu \psi \notag\\
	S_1[g,A]&=\int  \sqrt{-g_x}\,J_\mu^{a} A^{\mu a}
\end{align} 
with
\begin{align}
	F_{\mu\nu}^{a} &=\nabla_\mu A_{\nu}^{a} - \nabla_{\nu} A_{\mu}^{ a} + g_c f^{a  b  c} A_\mu^{ b} A_\nu^c\notag\\ 
	&=\partial_\mu A_{\nu}^{a} - \partial_{\nu} A_{\mu}^{ a} + g_c f^{a  b  c} A_\mu^{ b} A_\nu^c, \notag\\
	\nabla_\mu A^{\nu a} &=\partial_\mu A^{\nu a} +\Gamma_{\mu\nu}^\lambda A^{\lambda a}
\end{align}
with $J^{\mu  a}=g_c\bar{\psi}\gamma^\mu T^{a}\psi$ denoting the fermionic current, with $T^a$ the generators of the theory and $\nabla_\mu$ denoting the covariant derivative in the curved background on a vector field. The local Lorentz and gauge covariant derivative $(D)$ on the fermions acts via the spin connection 
\begin{equation}
	D_\mu \psi=\left( \partial_\mu \psi  + A_\mu^a T^a +\frac{1}{4}\omega_{\mu}^{\underline a \underline b}\gamma_{\underline a \underline b}\right)\psi
\end{equation}
having denoted with $\underline{a}\underline{b}$ the local Lorentz indices.  A local Lorentz covariant derivative $(D)$ can be similarly defined for a vector field, say $V^{\underline a}$, via the Vielbein $e^{\underline a}_{\, \,\mu}$ and its inverse  $e^{\,\,\mu}_{\underline a}$
\begin{equation}
	D_\mu V^{\underline a}= \partial_\mu V^{\underline a} +\omega^{\underline a}_{\mu \underline b} V^{\underline b}
\end{equation}
with 
\begin{equation}
	\nabla_\mu V^\rho = e_{\underline a}^{\,\,\rho} D_\mu V^{\underline a}
\end{equation}
with the Christoffel and the spin connection related via the holonomic relation 
\begin{equation}
	\Gamma^\rho_{\mu\nu}=e^{\,\,\rho}_{\underline a}\left( \partial_\mu e^{\underline a}_{\,\,\nu} +\omega^{\underline a}_{\mu \underline b} e^{\underline b}_{\,\,\nu}\right).
\end{equation}
Diffeomorphism invariance of the generating functional (\ref{GF}) gives 
\begin{equation}
	\int d^d x \left(\frac{\delta W}{\delta g_{\mu\nu}}\delta g_{\mu\nu}(x) +  \frac{\delta W}{\delta A_{\mu}^a}\delta A_{\mu}^a(x)\right)=0
	\label{aver}
\end{equation}
where the variation of the metric and the gauge fields are the corresponding Lie derivatives, for a change of variables 
$x^\mu\to x^\mu + \epsilon^\mu(x)$
\begin{align}
	\delta A_\mu^a(x) &= -\nabla_\alpha A_\mu^a \epsilon^\alpha - A^a_\alpha \nabla_\mu \epsilon^\alpha\notag\\
	\delta g_{\mu\nu}&=-\nabla_\mu \epsilon_\nu  - \nabla_\nu \epsilon_\mu \label{lie}
\end{align}
while for a gauge transformation with a parameter $\theta^a(x)$
\begin{equation}
	\delta A_\mu^a=\underline{D}_\mu \theta^a \equiv\partial_\mu \theta^a + g_c f^{a b c} A_\mu^b \theta^c.
\end{equation}
Using \eqref{lie}, \eqref{aver} becomes 
\begin{align}
	0=&\,\left\langle\, \int d^4 x \left( \frac{\delta (S_0 + S_1)}{\delta g_{\mu\nu}} \delta g_{\mu\nu} +\frac{\delta S_1}{\delta A_\mu^a} \delta A_{\mu}^a\right)
	\right\rangle\notag\\
	=&\,\left\langle \int d^4 x \sqrt{-g_x}\left[\nabla_\mu T^{\mu\nu} +(\nabla_\mu A_\nu^a -\nabla_\nu A_\mu^a) J^{\mu a} +
	\nabla_\mu J^{\mu a} A_\nu^a\right]\epsilon^\nu(x)\right\rangle
	\label{interm}
\end{align}
while the condition of gauge invariance gives 
\begin{equation}
	\int d^d x \frac{\delta W}{\delta A_{\mu}^a}\delta A_{\mu}^a=\left\langle \int d^4 x \sqrt{-g_x}J^\mu_a\underline{D}_\mu \theta^a \right\rangle=0
\end{equation}
which, in turn, after an integration by parts, generates the gauge WI
\begin{equation}
	\langle \nabla_\mu J^{\mu a} \rangle = g_c f^{a b c} \langle J_\mu^b\rangle A^{\mu c}. 
\end{equation}
Inserting this relation into (\ref{interm}) we obtain the conservation WI
\begin{equation}
	\langle \nabla^\mu T_{\mu\nu}\rangle +F_{\mu\nu}^a \langle J^{\mu a}\rangle =0.
\end{equation}
In the abelian case, diffeomorphism and gauge invariance  then give the relations
\begin{equation}
	\begin{split}
		0&=\nabla_\n \braket{T^{\m \n}}+F^{\m\n}\braket{J_\n}\\
		0&=\nabla_\n\braket{J^\nu}
	\end{split}\label{transversetjj}
\end{equation}
with naive scale invariance gives the traceless condition
\begin{equation}
	g_{\mu\nu}\braket{T^{\m\n}}=0.\label{tracetjj}
\end{equation}
The functional differentiation of \eqref{transversetjj} and \eqref{tracetjj} allows to derive ordinary Ward identities for the various correlators. In the $TJJ$ case we obtain, after a Fourier transformation, the conservation equation 
\begin{align}
	\label{tr}
	p_{1\n_1}\braket{T^{\mu_1\nu_1}(p_1)\,J^{\m_2}(p_2)\,J^{\m_3}(p_3)}&=4\,\big[\d^{\m_1\m_2}p_{2\l}\braket{J^\l(p_1+p_2)\,J^{\m_3}(p_3)}-p_2^{\m_1}\braket{J^{\m_2}(p_1+p_2)\,J^{\m_3}(p_3)}\big]\notag\\
	&+4\,\big[\d^{\m_1\m_3}p_{3\l}\braket{J^\l(p_1+p_3)\,J^{\m_2}(p_2)}-p_3^{\m_1}\braket{J^{\m_3}(p_1+p_3)\,J^{\m_2}(p_2)}\big]
\end{align}
and vector current Ward identities
\begin{align}
	\label{x1}
	p_{2\m_2}\braket{T^{\mu_1\nu_1}(p_1)\,J^{\m_2}(p_2)\,J^{\m_3}(p_3)}&=0\\
	p_{3\m_3}\braket{T^{\mu_1\nu_1}(p_1)\,J^{\m_2}(p_2)\,J^{\m_3}(p_3)}&=0,
\end{align}
while the naive identity \eqref{tracetjj} gives the non-anomalous condition
\begin{equation}
	\label{x2}
	\d_{\m_1\n_1}\braket{T^{\mu_1\nu_1}(p_1)\,J^{\m_2}(p_2)\,J^{\m_3}(p_3)}=0,
\end{equation}
valid in the $d\ne4$ case. We recall that the 2-point function of two conserved vector currents $J_i$ $(i=2,3)$ \cite{Coriano:2013jba} in any conformal field theory in $d$ dimension is given by 
\begin{equation}
	\langle J_2^\alpha(p)J_3^\beta(-p) \rangle =\delta_{\Delta_2\, \Delta_3}\left(c_{123} \Gamma_J \right)\pi^{\alpha\beta}(p) (p^2)^{\Delta_2-d/2},
	\qquad \Gamma_J=\frac{\pi^{d/2}}{ 4^{\Delta_2 -d/2}}\frac{\Gamma(d/2-\Delta_2)}{\Gamma(\Delta_2)},\label{twoJJ}
\end{equation}
with $c_{123}$ an overall constant and $\Delta_2=d-1$. In our case $\D_2=\D_3=d-1$ and Eq. (\ref{tr}) then takes the form 
\begin{align}
	\label{2point}
	p_{1\mu_1}\braket{T^{\mu_1\nu_1}(p_1)\,J^{\m_2}(p_2)\,J^{\m_3}(p_3)}&=4 c_{123} \Gamma_J
	\left( \delta^{\nu_1\mu_2}\frac{p_{2\lambda}}{(p_3^2)^{d/2 -\Delta_2}} {\pi^{\lambda \mu_3}}(p_3) -
	\frac{p_2^{\nu_1}}{(p_3^2)^{d/2 -\Delta_2}}\pi^{\mu_2\mu_3}(p_3) \right.\notag\\
	& \left. + \delta^{\nu_1\mu_3}\frac{p_{3\lambda}}{(p_2^2)^{d/2-\Delta_2}}\pi^{\lambda \mu_2}(p_2) -\frac{p_3^{\nu_1}}{(p_2^2)^{d/2-\Delta_2}}\pi^{\mu_3\mu_2}(p_2)\right).
\end{align}
Explicit expressions of the secondary CWI's are determined using (\ref{x1}) and (\ref{x2}) and the explicit form (\ref{2point}).

\subsection{Conformal Ward Identities}\label{Skend}

Given the partial symmetry of the $TJJ$ correlator, one can choose as independent momenta either $p_1$ and $p_2$ or, more conveniently, $p_2$ and $p_3$, given the symmetry of the two $J$ currents.\\
With the first choice, outlined below, the current $J(p_3)$ is singlet under the (spin) Lorentz generators. With the second choice, the two currents are treated symmetrically and the stress energy tensor is treated as a singlet under the same generators. The derivation of the CWI's in this second case will be outlined in \secref{p1section}. The equations obtained in the two cases are obviously the same. \\
We discuss the conformal Ward identities for the $\braket{TJJ}$ correlation function in momentum space. The dilatation Ward identities take the form
\begin{equation}
	0=\left[\sum_{j=1}^3\Delta_j-(n-1)d-\sum_{j=1}^2\,p_j^\a\sdfrac{\partial}{\partial p_j^\alpha}\right]\braket{{T^{\mu_1\nu_1}(p_1)\,J^{\m_2}(p_2)\,J^{\mu_3}(\bar p_3)}}.\label{DilatationTJJ}
\end{equation}
To proceed towards the analysis of the constraints, it is essential to introduce the Lorentz covariant Ward identities
\begin{align}
	0&=\sum_{j=1}^{2}\left[p_j^\nu\sdfrac{\partial}{\partial p_{j\mu}}-p_j^{\mu}\sdfrac{\partial}{\partial p_{j\nu}}\right]\braket{{T^{\mu_1\nu_1}(p_1)\,J^{\m_2}(p_2)\,J^{\mu_3}(\bar p_3)}}\notag\\
	&\qquad+2\left(\delta^{\nu}_{\a_1}\d^{\mu(\mu_1}-\delta^{\mu}_{\alpha_1}\delta^{\nu(\mu_1}\right)\braket{{T^{\nu_1)\alpha_1}(p_1)\,J^{\m_2}(p_2)\,J^{\mu_3}(\bar p_3)}}\notag\\
	&\qquad+\left(\delta^{\nu}_{\a_2}\d^{\mu\mu_2}-\delta^{\mu}_{\alpha_2}\delta^{\nu\mu_2}\right)\braket{{T^{\mu_1\nu_1}(p_1)\,J^{\alpha_2}(p_2)\,J^{\mu_3}(\bar p_3)}}\notag\\
	&\qquad+\left(\delta^{\nu}_{\a_3}\d^{\mu\mu_3}-\delta^{\mu}_{\alpha_3}\delta^{\nu\mu_3}\right)\braket{{T^{\mu_1\nu_1}(p_1)\,J^{\mu_2}(p_2)\,J^{\alpha_3}(\bar p_3)}}\label{RotationTJJ},
\end{align}
where 
\begin{equation}
	\d^{\n(\m_1}\,\braket{{T^{\nu_1)\alpha_1}\,J^{\m_2}\,J^{\mu_3}}}\equiv\sdfrac{1}{2}\big(\d^{\n\m_1}\,\braket{{T^{\nu_1\alpha_1}\,J^{\m_2}\,J^{\mu_3}}}+\d^{\n\n_1}\,\braket{{T^{\mu_1\alpha_1}\,J^{\m_2}\,J^{\mu_3}}}\big),
\end{equation}
and finally the special conformal Ward identities
\begin{align}
	0&=\sum_{j=1}^{2}\left[2(\Delta_j-d)\sdfrac{\partial}{\partial p_j^\k}-2p_j^\a\sdfrac{\partial}{\partial p_j^\a}\sdfrac{\partial}{\partial p_j^\k}+(p_j)_\k\sdfrac{\partial}{\partial p_j^\a}\sdfrac{\partial}{\partial p_{j\a}}\right]\braket{{T^{\mu_1\nu_1}(p_1)\,J^{\m_2}(p_2)\,J^{\mu_3}(\bar p_3)}}\notag\\
	&\qquad+4\left(\d^{\k(\mu_1}\sdfrac{\partial}{\partial p_1^{\a_1}}-\delta^{\k}_{\alpha_1}\delta^{\l(\mu_1}\sdfrac{\partial}{\partial p_1^\l}\right)\braket{{T^{\nu_1)\alpha_1}(p_1)\,J^{\m_2}(p_2)\,J^{\mu_3}(\bar p_3)}}\notag\\
	&\qquad+2\left(\d^{\k\mu_2}\sdfrac{\partial}{\partial p_2^{\a_2}}-\delta^{\k}_{\alpha_2}\delta^{\l\mu_2}\sdfrac{\partial}{\partial p_2^\l}\right)\braket{T^{\mu_1\nu_1}(p_1)\,J^{\alpha_2}(p_2)\,J^{\mu_3}(\bar p_3)}.\label{SCWTJJ}
\end{align} 
which we will use in the next sections in order to determine the tensor structure of this correlator. 
\subsection{Tensorial decomposition of the correlator}
We can divide the 3-point function $TJJ$ into two parts: the \emph{transverse-traceless} part and the \emph{semi-local} part (indicated by subscript $loc$) expressible through the transverse and trace Ward Identities. These parts are obtained by using the projectors previously defined. 
We can then decompose the full 3-point function as follows
\begin{align}
	\braket{T^{\mu_1\nu_1}\,J^{\mu_2}\,J^{\mu_3}}&=\braket{t^{\mu_1\nu_1}\,j^{\mu_2}\,j^{\mu_3}}+\braket{T^{\mu_1\nu_1}\,J^{\mu_2}\,j_{loc}^{\mu_3}}+\braket{T^{\mu_1\nu_1}\,j_{loc}^{\mu_2}\,J^{\mu_3}}+\braket{t_{loc}^{\mu_1\nu_1}\,J^{\mu_2}\,J^{\mu_3}}\notag\\
	&\quad-\braket{T^{\mu_1\nu_1}\,j_{loc}^{\mu_2}\,j_{loc}^{\mu_3}}-\braket{t_{loc}^{\mu_1\nu_1}\,j_{loc}^{\mu_2}\,J^{\mu_3}}-\braket{t_{loc}^{\mu_1\nu_1}\,J^{\mu_2}\,j_{loc}^{\mu_3}}+\braket{t_{loc}^{\mu_1\nu_1}\,j_{loc}^{\mu_2}\,j_{loc}^{\mu_3}}.
\end{align}
All the terms on the right-hand side, apart from the first one, may be computed by means of transverse and trace Ward Identities. The exact form of the Ward identities depends on the exact definition of the operators involved, but more importantly, all these terms depend on 2-point function only. The main goal now is to write the general form of the transverse-traceless part of the correlator and to give the solution using the Conformal Ward identities. 

Using the projectors $\Pi$ and $\pi$ one can write the most general form of the transverse-traceless part as
\begin{equation}
	{\braket{t^{\m_1\n_1}(p_1)\,j^{\mu_2}(p_2)\,j^{\mu_3}(p_3)}}=\Pi^{\mu_1\nu_1}_{\alpha_1\beta_1}(p_1)\pi^{\mu_2}_{\alpha_2}(p_2)\pi^{\mu_3}_{\alpha_3}(p_3)\,\,X^{\alpha_1\beta_1\,\alpha_3\alpha_3},
\end{equation}
where $X^{\alpha_1\beta_1\,\alpha_3\alpha_3}$ is a general tensor of rank four built from the metric and momenta. We can enumerate all possible tensor that can appear in $X^{\alpha_1\beta_1\,\alpha_3\alpha_3}$ preserving the symmetry of the 
correlator, as illustrated in \cite{Bzowski:2013sza}
\begin{align}
	\langle t^{\mu_1\nu_1}(p_1)j^{\mu_2}(p_2)j^{\mu_3}(p_3)\rangle& =
	{\Pi_1}^{\mu_1\nu_1}_{\alpha_1\beta_1}{\pi_2}^{\mu_2}_{\alpha_2}{\pi_3}^{\mu_3}_{\alpha_3}
	\left( A_1\ p_2^{\alpha_1}p_2^{\beta_1}p_3^{\alpha_2}p_1^{\alpha_3} + 
	A_2\ \delta^{\alpha_2\alpha_3} p_2^{\alpha_1}p_2^{\beta_1} + 
	A_3\ \delta^{\alpha_1\alpha_2}p_2^{\beta_1}p_1^{\alpha_3}\right. \notag\\
	& \left. + 
	A_3(p_2\leftrightarrow p_3)\delta^{\alpha_1\alpha_3}p_2^{\beta_1}p_3^{\alpha_2}
	+ A_4\  \delta^{\alpha_1\alpha_3}\delta^{\alpha_2\beta_1}\right).\label{DecompTJJ}
\end{align}
where we have used the symmetry properties of the projectors, and the coefficients $A_i\  i=1,\dots,4$ are the form factors, functions of $p_1^2, p_2^2$ and $p_3^2$. This ansatz introduces a minimal set of form factors which will be later determined by the solutions of the CWI's. For future discussion, we will refer to this basis as to the $A$-basis. \\
We can now consider the dilatation Ward identities for the transverse-traceless part obtained by the decomposition of \eqref{DilatationTJJ}. We are then free to apply the projectors $\Pi$ and $\pi$ to this decomposition in order to obtain the final result
\begin{align}
	0&=\Pi^{\mu_1\nu_1}_{\alpha_1\beta_1}(p_1)\pi^{\mu_2}_{\alpha_2}(p_2)\pi^{\mu_3}_{\alpha_3}(p_3)\left[\sum_{j=1}^3\,\Delta_j-2d-\sum_{j=1}^{2}\,p_j^\alpha\sdfrac{\partial}{\partial p_j^\alpha}\right]\bigg[ A_1 p_2^{\alpha_1}p_2^{\beta_1}p_3^{\alpha_2}p_1^{\alpha_3} \notag\\[-1ex]
	&\qquad+ 
	A_2\delta^{\alpha_2\alpha_3} p_2^{\alpha_1}p_2^{\beta_1} + 
	A_3 \delta^{\alpha_1\alpha_2}p_2^{\beta_1}p_1^{\alpha_3} + 
	A_3(p_2\leftrightarrow p_3)\delta^{\alpha_1\alpha_3}p_2^{\beta_1}p_3^{\alpha_2}
	+ A_4 \delta^{\alpha_1\alpha_3}\delta^{\alpha_2\beta_1}\bigg].
\end{align}
It is possible to obtain from this projection a set of differential equations for all the form factors. These equations are expressed as
\begin{equation}
	\left[2d+N_n-\sum_{j=1}^{3}\Delta_j+\sum_{j=1}^2\,p_j^\alpha\sdfrac{\partial}{\partial p_j^\alpha}\right]\,A_n(p_1,p_2,p_3)=0,\label{DilatationFactor}
\end{equation}
where $N_n$ is the tensorial dimension of $A_n$, i.e. the number of momenta multiplying the form factor $A_n$ and the projectors $\Pi$ and $\pi$. \\
Turning to the special CWI's, ${\braket{TJJ}}$ in  \eqref{SCWTJJ}, we can write the same equation in the form 
\begin{equation}
	{K}^\k\,{\braket{T^{\mu_1\nu_1}(p_1)\,J^{\m_2}(p_2)\,J^{\mu_3}( p_3)}}=0,
\end{equation}
where $K^\k$ is the special conformal generator.
As before, we introduce the decomposition of the 3-point function to obtain
\begin{align}
	0&={K}^\kappa\bigg[ \braket{{t^{\mu_1\nu_1}\,j^{\mu_2}\,j^{\mu_3}}}+{\braket{t_{loc}^{\mu_1\nu_1}\,j^{\mu_2}\,j^{\mu_3}}}+{\braket{t^{\mu_1\nu_1}\,j_{loc}^{\mu_2}\,j^{\mu_3}}}+{\braket{t^{\mu_1\nu_1}\,j^{\mu_2}\,j_{loc}^{\mu_3}}}\notag\\
	&\hspace{2cm} +{\braket{t^{\mu_1\nu_1}_{loc}\,j_{loc}^{\mu_2}\,j^{\mu_3}}}+{\braket{t_{loc}^{\mu_1\nu_1}\,j_{loc}^{\mu_2}\,j^{\mu_3}}}+{\braket{t^{\mu_1\nu_1}\,j_{loc}^{\mu_2}\,j_{loc}^{\mu_3}}}+{\braket{t_{loc}^{\mu_1\nu_1}\,j_{loc}^{\mu_2}\,j_{loc}^{\mu_3}}}\bigg].
\end{align}
In order to isolate the equations for the form factors appearing in the decomposition, we are free to apply the projectors $\Pi$ and $\pi$ defined in \appref{appendixB}. Through a lengthy calculation we find
\begin{align}
	\Pi^{\rho_1\sigma_1}_{\mu_1\nu_1}(p_1)\pi^{\rho_2}_{\mu_2}(p_2)\pi^{\rho_3}_{\mu_3}(p_3)\ \,K^\k
	{\braket{t_{loc}^{\mu_1\nu_1}\,j^{\mu_2}\,j^{\mu_3}}}&=\Pi^{\rho_1\sigma_1}_{\mu_1\nu_1}\,\pi^{\rho_2}_{\mu_2}\,\pi^{\rho_3}_{\mu_3}\,\left[\sdfrac{4d}{p_1^2}\,\delta^{\kappa\mu_1}\,p_{1\alpha_1}\,{\braket{T^{\alpha_1\nu_1}J^{\mu_2}J^{\mu_3}}}\right]\notag\\
	\Pi^{\rho_1\sigma_1}_{\mu_1\nu_1}(p_1)\pi^{\rho_2}_{\mu_2}(p_2)\pi^{\rho_3}_{\mu_3}(p_3)\ \,K^\k
	\braket{t^{\mu_1\nu_1}\,j_{loc}^{\mu_2}\,j^{\mu_3}}&=\Pi^{\rho_1\sigma_1}_{\mu_1\nu_1}\,\pi^{\rho_2}_{\mu_2}\,\pi^{\rho_3}_{\mu_3}\,\left[\sdfrac{2(d-2)}{p_2^2}\delta^{\kappa\mu_2}\,p_{2\alpha_2}\,{\braket{T^{\alpha_1\nu_1}J^{\alpha_2}J^{\mu_3}}}\right]\notag\\
	\Pi^{\rho_1\sigma_1}_{\mu_1\nu_1}(p_1)\pi^{\rho_2}_{\mu_2}(p_2)\pi^{\rho_3}_{\mu_3}(p_3)\ \,K^\k
	\braket{t^{\mu_1\nu_1}\,j^{\mu_2}\,j_{loc}^{\mu_3}}&=\Pi^{\rho_1\sigma_1}_{\mu_1\nu_1}\,\pi^{\rho_2}_{\mu_2}\,\pi^{\rho_3}_{\mu_3}\,\left[\sdfrac{2(d-2)}{p_3^2}\,\delta^{\kappa\mu_3}p_{3\alpha_3}\,{\braket{T^{\alpha_1\nu_1}J^{\mu_2}J^{\alpha_3}}}\right]
	\label{sec}
\end{align}
and all the terms with at least two insertion of local terms are zero. T
We have verified, as expected, that the equations above remain invariant if we choose as independent momenta $p_2$ and $p_3$ while acting on $p_1$ indirectly 
by the derivative chain rule. More details on this analysis will be given in a section below. In this way we may rewrite \eqref{SCWTJJ} in the form
\begin{align}
	0&=\Pi^{\rho_1\sigma_1}_{\mu_1\nu_1}(p_1)\pi^{\rho_2}_{\mu_2}(p_2)\pi^{\rho_3}_{\mu_3}(p_3)\  \bigg(\,K^\k\,{\braket{T^{\mu_1\nu_1}(p_1)\,J^{\m_2}(p_2)\,J^{\mu_3}( p_3)}}\bigg)\notag\\
	&=\Pi^{\rho_1\sigma_1}_{\mu_1\nu_1}(p_1)\pi^{\rho_2}_{\mu_2}(p_2)\pi^{\rho_3}_{\mu_3}(p_3)\ \bigg\{\,K^\kappa\,{\braket{t^{\mu_1\nu_1}(p_1)\,j^{\mu_2}(p_2)\,j^{\mu_3}(p_3)}}+\sdfrac{4d}{p_1^2}\,\delta^{\kappa\mu_1}\,p_{1\alpha_1}\,{\braket{T^{\alpha_1\nu_1}(p_1)J^{\mu_2}(p_2)J^{\mu_3}(p_3)}}\notag\\
	&\hspace{2cm}+\sdfrac{2(d-2)}{p_2^2}\,\delta^{\kappa\mu_2}p_{2\alpha_2}\,\braket{{T^{\alpha_1\nu_1}J^{\alpha_2}J^{\mu_3}}}+\sdfrac{2(d-2)}{p_3^2}\,\delta^{\kappa\mu_3}p_{3\alpha_3}\,{\braket{T^{\alpha_1\nu_1}J^{\mu_2}J^{\alpha_3}}}\bigg\}.\label{StrucGenSWIS}
\end{align}
The equation above is an independent derivation of the corresponding BMS result, which is not offered in \cite{Bzowski:2013sza}. Notice that our derivation, which details the various contributions coming from the local terms in the $TJJ$, has been derived using heavily the Lorentz Ward identities.\\
The last three terms may be re-expressed in terms of 2-point functions via the transverse Ward identities. After other rather lengthy computations, we find that the first term in the previous expression, corresponding to the transverse traceless contributions, can be written in the form
\begin{align}
	&\Pi^{\rho_1\sigma_1}_{\mu_1\nu_1}(p_1)\pi^{\rho_2}_{\mu_2}(p_2)\pi^{\rho_3}_{\mu_3}(p_3)\ \bigg[\,K^\kappa\,\braket{{t^{\mu_1\nu_1}(p_1)\,j^{\mu_2}(p_2)\,j^{\mu_3}(p_3)}}\bigg]\notag\\
	&=\Pi^{\rho_1\sigma_1}_{\mu_1\nu_1}(p_1)\pi^{\rho_2}_{\mu_2}(p_2)\pi^{\rho_3}_{\mu_3}(p_3)\times\notag\\
	&\times\bigg[p_1^\kappa\left(C_{11}\,p_1^{\mu_3}p_2^{\mu_1}p_2^{\nu_1}p_3^{\mu_2}+C_{12}\,\delta^{\mu_2\mu_3}p_2^{\mu_1}p_2^{\nu_1}+C_{13}\delta^{\mu_1\mu_2}p_2^{\nu_1}p_1^{\mu_3}+C_{14}\delta^{\mu_1\mu_3}p_2^{\nu_1}p_3^{\mu_2}+C_{15}\delta^{\mu_1\mu_2}\delta^{\nu_1\mu_3}\right)\notag\\
	&\quad+p_2^\kappa\left(C_{21}\,p_1^{\mu_3}p_2^{\mu_1}p_2^{\nu_1}p_3^{\mu_2}+C_{22}\,\delta^{\mu_2\mu_3}p_2^{\mu_1}p_2^{\nu_1}+C_{23}\delta^{\mu_1\mu_2}p_2^{\nu_1}p_1^{\mu_3}+C_{24}\delta^{\mu_1\mu_3}p_2^{\nu_1}p_3^{\mu_2}+C_{25}\delta^{\mu_1\mu_2}\delta^{\nu_1\mu_3}\right)\notag\\	
	&\quad+\delta^{\mu_1\k}\left(C_{31}\,p_1^{\mu_3}p_2^{\nu_1}p_3^{\mu_2}+C_{32}\,\delta^{\mu_2\mu_3}p_2^{\nu_1}+C_{33}\,\delta^{\mu_2\nu_1}p_1^{\mu_3}+C_{34}\,\delta^{\mu_3\nu_1}p_3^{\mu_2}\right)\notag\\
	&\qquad+\delta^{\mu_2\k}\left(C_{41}\,p_1^{\mu_3}p_2^{\mu_1}p_2^{\nu_1}+C_{42}\,\delta^{\mu_1\mu_3}p_2^{\nu_1}\right)+\delta^{\mu_3\k}\left(C_{51}\,p_3^{\mu_2}p_2^{\nu_1}p_3^{\mu_2}+C_{52}\,\delta^{\mu_1\mu_2}p_2^{\nu_1}\right)\bigg]\label{StrucSWIS}
\end{align}
where now $C_{ij}$ are differential equations involving the form factors $A_1,\ A_2,\ A_3,\ A_4$ of the representation of the $\langle{tjj}\rangle$ in \eqref{DecompTJJ}. For any 3-point function, the resulting equations can be divided into two groups, the \emph{primary} and the \emph{secondary} conformal Ward identities. The primary are second-order differential equations and appear as the coefficients of transverse or transverse-traceless tensor containing $p_1^\kappa$ and $p_2^\kappa$, where $\kappa$ is the special index related to the conformal operator ${K}^\kappa$. The remaining equations, following from all other transverse or transverse-traceless terms, are then secondary conformal Ward identities and are first-order differential equations. 

\subsection{Primary CWI's}
\label{primsection}
From \eqref{StrucGenSWIS} and \eqref{StrucSWIS} one finds that the primary CWI's are equivalent to the vanishing of the coefficients $C_{1j}$ and $C_{2j}$ for $j=1,\dots, 5$.  The CWI's can be rewritten in terms of the operators defined in Eq.  (\ref{kij}) as
\begin{equation}
	\begin{split}
		0&=C_{11}=K_{13}A_1\\
		0&=C_{12}=K_{13}A_2+2A_1\\
		0&=C_{13}=K_{13}A_3-4A_1\\
		0&=C_{14}=K_{13}A_3(p_2\leftrightarrow p_3)\\
		0&=C_{15}=K_{13}A_4-2A_3(p_2\leftrightarrow p_3)
	\end{split}
	\hspace{1.5cm}
	\begin{split}
		0&=C_{21}=K_{23}A_1\\
		0&=C_{22}=K_{23}A_2\\
		0&=C_{23}=K_{23}A_3-4A_1\\
		0&=C_{24}=K_{23}A_3(p_2\leftrightarrow p_3)+4A_1\\
		0&=C_{25}=K_{23}A_4+2A_3-2A_3(p_2\leftrightarrow p_3)
	\end{split}\label{Primary}
\end{equation}

\subsection{Secondary CWI's}
\label{secsection}
The secondary conformal Ward identities are first-order partial differential equations and in principle involve the semi-local information contained in $j_{loc}^\m$ and $t^{\m\n}_{loc}$. In order to write them compactly, one defines the two differential operators
\begin{align}
	L_N&= p_1(p_1^2 + p_2^2 - p_3^2) \frac{\partial}{\partial p_1} + 2 p_1^2\, p_2 \frac{\partial}{\partial p_2} + \big[ (2d - \Delta_1 - 2\Delta_2 +N)p_1^2 + (2\Delta_1-d)(p_3^2-p_2^2)  \big] \label{Ldef2} \\
	R &= p_1 \frac{\partial}{\partial p_1} - (2\Delta_1-d) \label{Rdef2}\,. 
\end{align}
The reason for introducing such operators comes from (\ref{StrucSWIS}), once the action of 
$K^\kappa$ is made explicit. The separation between the two sets of constraints comes from the same equation, and in particular from the terms trilinear in the momenta within the square bracket. One needs also the symmetric versions of such operators
\begin{align}
	&L'_N=L_N,\quad\text{with}\ p_1\leftrightarrow p_2\ \text{and}\ \D_1\leftrightarrow\D_2,\\
	&R'=R,\qquad\text{with}\ p_1\mapsto p_2\ \text{and}\ \D_1\mapsto\D_2.
\end{align}
These operators depend on the conformal dimensions of the operators involved in the 3-point function under consideration, and additionally on a single parameter $N$ determined by the Ward identity in question. In the $\braket{TJJ}$ case one finds considering the structure of Eqs. \eqref{StrucGenSWIS} and \eqref{StrucSWIS} 
\begin{equation}
	\begin{split}
		C_{31}&=-\sdfrac{2}{p_1^2}\left[L_4 A_1+R A_3-R A_3(p_2\leftrightarrow p_3)\right]\\
		C_{32}&=-\sdfrac{2}{p_1^2}\left[L_2\,A_2-p_1^2(A_3-A_3(p_2\leftrightarrow p_3))\right]\\
		C_{33}&=-\sdfrac{1}{p_1^2}\left[L_4\,A_3-2R\,A_4\right]\\
		C_{34}&=-\sdfrac{1}{p_1^2}\left[L_4\,A_3(p_2\leftrightarrow p_3)+2R\,A_4-4p_1^2A_3(p_2\leftrightarrow p_3)\right]\\
	\end{split}
\end{equation}
\begin{equation}
	\begin{split}
		C_{41}&=\sdfrac{1}{p_2^2}\left[L'_3\,A_1-2R'A_2+2R'A_3\right]\\
		C_{42}&=\sdfrac{1}{p_2^2}\left[L'_1\,A_3(p_2\leftrightarrow p_3)+p_2^2(4A_2-2A_3)+2R'A_4\right]\\
		C_{51}&=\sdfrac{1}{p_3}\left[(L_4-L'_3)A_1-2(2d+R+R')A_2+2(2d+R+R')A_3(p_2\leftrightarrow p_3)\right]\\
		C_{52}&=\sdfrac{1}{p_3^2}\left[(L_2-L'_1)A_3-4p_3^2A_2+2p_3^2A_3(p_2\leftrightarrow p_3)+2(2d-2+R+R')A_4\right]
	\end{split}
\end{equation}
From (\ref{StrucGenSWIS}) and (\ref{StrucSWIS}) using (\ref{2point}) the secondary CWI's take the explicit form 
\begin{equation}
	\begin{split}
		&C_{31}=C_{41}=C_{42}=C_{51}=C_{52}=0, \qquad C_{32}=\frac{16\,d\, c_{123}\,\Gamma_J}{p_1^2}\left[ \frac{1}{(p_3^2)^{\sigma_0}} - \frac{1}{(p_2^2)^{\sigma_0}} \right],\\[2ex]
		& \hspace{1cm}C_{33}=\frac{16\,d\,c_{123}\,\Gamma_J}{p_1^2 (p_3^2)^{\sigma_0}}, \hspace{2.5cm} C_{34}=-\frac{16\,d\,c_{123}\,\Gamma_J }{p_1^2\,(p_2^2)^{\sigma_0}},
	\end{split}
\end{equation}
where in our $\sigma_0=d/2-\Delta_2$. 
Expressed in this form all the scalar equations for the $A_i$ are not apparently symmetric in the exchange of $p_2$ and $p_3$, and it may not be immediately evident that they can be recast in such a way that the symmetry is respected. 

\subsection{Solutions to the CWI's as 3K integrals}
Following the analysis in \secref{primarysol}, the primary CWI's \eqref{Primary} admit the solutions
\begin{equation}
	\label{SolTJJ}
	\begin{split}
		A_1&=\alpha_1\, J_{4\{000\}}\\
		A_2&=\alpha_1\, J_{3\{100\}}+\alpha_2\,J_{2\{000\}}\\
		A_3&=2\alpha_1\, J_{3\{001\}}+\alpha_3\,J_{2\{000\}}\\
		A_4&=2\alpha_1\,J_{2\{011\}}+\alpha_3\,\left(J_{1\{010\}}+ J_{1\{001\}}\right)+\alpha_4\, J_{0\{000\}}.
	\end{split}
\end{equation}
By applying the approach illustrated in \secref{reg} for the $\braket{TTO}$ to this case, the secondary CWI's \secref{secsection} give the conditions
\begin{align}
	\alpha_4^{(0)}&=-(d-2)\alpha_3^{(0)},\\
	\alpha_3^{(0)}&=\alpha_2^{(0)}=-d\alpha_1^{(0)}+\left(\frac{\pi}{2}\right)^{-\frac{3}{2}}\frac{2(-1)^n\left(c_{123} \Gamma_J \right)}{(d-2)!!},\\
	\alpha_4^{'(1)}&=\frac{(d-4)}{2}\alpha_2^{(0)}-\left(\frac{\pi}{2}\right)^{-\frac{3}{2}}\frac{2(-1)^n\left(c_{123} \Gamma_J \right)}{(d-4)!!},
\end{align}
for odd spacetime dimensions $d=2n+1$, $(n=1,2,\dots)$, and 
\begin{align}
	\alpha_4^{(0)}&=-(d-2)\alpha_3^{(0)},\\
	\alpha_3^{(0)}&=\alpha_2^{(0)}=-d\alpha_1^{(0)}+\frac{2^{3-n}(-1)^n\left(c_{123} \Gamma_J \right)}{(n-1)!!},\\
	\alpha_3^{(1)}&=\alpha_2^{(1)}=-d\alpha_1^{(1)}-2v\alpha_1^{(0)}+\frac{(-1)^n2^{3-n}}{(n-1)!}[\left(c_{123} \Gamma_J \right)(\gamma_E-\ln 2-H_{n-1})+\left(c_{123} \Gamma_J \right)^{(0)}],\\
	\alpha_4^{'(1)}&=-\frac{d(d-4)}{2}\alpha_1^{(0)}-\frac{2^{3-n}(-1)^n\,n\,\left(c_{123} \Gamma_J \right)}{(n-1)!!},
\end{align}
\begin{align}
	\alpha_4^{'(2)}&=-\frac{d(d-4)}{2}\alpha_1^{(1)}-2(d-2)v\alpha_1^{(0)}\notag\\
	&\hspace{2cm}+\frac{2^{3-n}(-1)^n\,v\,\left(c_{123} \Gamma_J \right)}{(n-1)!!}[-1+n(H_{n-1}-\gamma_E+\ln 2)]-\frac{(-1)^n2^{3-n}nv\left(c_{123} \Gamma_J \right)^{(0)}}{(n-1)!},
\end{align}
for even spacetime dimensions $d=2n$, $(n=2,3,\dots)$, as discussed in \cite{ Bzowski:2018fql,Bzowski:2017poo}. $H_n$ is the $n$-th harmonic number.
It is worth noticing that there are contributions from the two point function $JJ$ in \eqref{twoJJ} as expected. 
\subsection{Symmetric treatment of the \texorpdfstring{$J$}{} currents }
\label{p1section}
Let's now consider $p_1$ as dependent momentum, showing the equivalence of the CWI's with this second choice. As we have just mentioned above, this choice is the preferred one in the search for the solutions of the $TJJ$. In this case, the action of the spin (Lorentz) part of the transformation will leave the stress energy tensor 
as a singlet, acting implicitly on $p_1$ via the chain rule. As we are going to show, the resulting equations will be linear combinations of the original part.\\
The structure of the decomposition in \eqref{DecompTJJ} of the $\braket{TJJ}$ correlator is still valid but now the explicit form of the special conformal operator ${K}^\k$ has to be modified as
\begin{align}
	{K}^\k{\braket{T^{\m_1\n_1}\,J^{\m_2}\,J^{\m_3}}}&=\sum_{j=2}^{3}\left[2(\Delta_j-d)\sdfrac{\partial}{\partial p_j^\k}-2p_j^\a\sdfrac{\partial}{\partial p_j^\a}\sdfrac{\partial}{\partial p_j^\k}+(p_j)_\k\sdfrac{\partial}{\partial p_j^\a}\sdfrac{\partial}{\partial p_{j\a}}\right]{\braket{T^{\mu_1\nu_1}(\bar p_1)\,J^{\m_2}(p_2)\,J^{\mu_3}(p_3)}}\notag\\
	&\qquad+2\left(\d^{\k\mu_2}\sdfrac{\partial}{\partial p_2^{\a_2}}-\delta^{\k}_{\alpha_2}\delta^{\l\mu_2}\sdfrac{\partial}{\partial p_2^\l}\right){\braket{T^{\mu_1\nu_1}(\bar p_1)\,J^{\alpha_2}(p_2)\,J^{\mu_3}(p_3)}}\notag\\
	&\qquad+2\left(\d^{\k\mu_3}\sdfrac{\partial}{\partial p_3^{\a_3}}-\delta^{\k}_{\alpha_3}\delta^{\l\mu_3}\sdfrac{\partial}{\partial p_3^\l}\right){\braket{T^{\mu_1\nu_1}(\bar p_1)\,J^{\mu_2}(p_2)\,J^{\a_3}( p_3)}}
\end{align}
where $\bar p_1^\m=-p_2^\m-p_3^\m$. Considering the SCWI's for the 3-point function we can write
\[{K}^\k(p_2,p_3){\braket{T^{\m_1\n_1}(\bar p_1)\,J^{\m_2}(p_2)\,J^{\m_3}(p_3)}}=0,\]
in which we have stress the $p_2$ and $p_3$ dependence of the special conformal operator. Then one has to take the decomposition of the 3-point function as in \eqref{DecompTJJ} and using the relations \eqref{sec}, that are still valid in this case, one derives \eqref{StrucGenSWIS}, in which now the $K$ operator is defined in terms of $p_2$ and $p_3$ only. As in the previous case, one finds CWI's which are similar to those given in \eqref{StrucSWIS} 
\begin{align}
	&\Pi^{\rho_1\sigma_1}_{\mu_1\nu_1}(p_1)\pi^{\rho_2}_{\mu_2}(p_2)\pi^{\rho_3}_{\mu_3}(p_3)\ \bigg[{K}^\kappa\,{\braket{t^{\mu_1\nu_1}(p_1)\,j^{\mu_2}(p_2)\,j^{\mu_3}(p_3)}}\bigg]\notag\\
	&=\Pi^{\rho_1\sigma_1}_{\mu_1\nu_1}(p_1)\pi^{\rho_2}_{\mu_2}(p_2)\pi^{\rho_3}_{\mu_3}(p_3)\times\notag\\
	&\times\bigg[p_2^\kappa\left(\tilde{C}_{11}\,p_1^{\mu_3}p_2^{\mu_1}p_2^{\nu_1}p_3^{\mu_2}+\tilde{C}_{12}\,\delta^{\mu_2\mu_3}p_2^{\mu_1}p_2^{\nu_1}+\tilde{C}_{13}\delta^{\mu_1\mu_2}p_2^{\nu_1}p_1^{\mu_3}+\tilde{C}_{14}\delta^{\mu_1\mu_3}p_2^{\nu_1}p_3^{\mu_2}+\tilde{C}_{15}\delta^{\mu_1\mu_2}\delta^{\nu_1\mu_3}\right)\notag\\
	&\quad+p_3^\kappa\left(\tilde{C}_{21}\,p_1^{\mu_3}p_2^{\mu_1}p_2^{\nu_1}p_3^{\mu_2}+\tilde{C}_{22}\,\delta^{\mu_2\mu_3}p_2^{\mu_1}p_2^{\nu_1}+\tilde{C}_{23}\delta^{\mu_1\mu_2}p_2^{\nu_1}p_1^{\mu_3}+\tilde{C}_{24}\delta^{\mu_1\mu_3}p_2^{\nu_1}p_3^{\mu_2}+\tilde{C}_{25}\delta^{\mu_1\mu_2}\delta^{\nu_1\mu_3}\right)\notag\\	
	&\quad+\delta^{\mu_1\k}\left(\tilde{C}_{31}\,p_1^{\mu_3}p_2^{\nu_1}p_3^{\mu_2}+\tilde{C}_{32}\,\delta^{\mu_2\mu_3}p_2^{\nu_1}+\tilde{C}_{33}\,\delta^{\mu_2\nu_1}p_1^{\mu_3}+\tilde{C}_{34}\,\delta^{\mu_3\nu_1}p_3^{\mu_2}\right)\notag\\
	&\qquad+\delta^{\mu_2\k}\left(\tilde{C}_{41}\,p_1^{\mu_3}p_2^{\mu_1}p_2^{\nu_1}+\tilde{C}_{42}\,\delta^{\mu_1\mu_3}p_2^{\nu_1}\right)+\delta^{\mu_3\k}\left(\tilde{C}_{51}\,p_3^{\mu_2}p_2^{\nu_1}p_3^{\mu_2}+\tilde{C}_{52}\,\delta^{\mu_1\mu_2}p_2^{\nu_1}\right)\bigg].
\end{align}
In this case we obtain the primary WI's by imposing the vanishing of the coefficients $\tilde{C}_{ij}$, for $i=1,2$ and $j=1,\dots, 5$. In this way we get
\begin{equation}
	\label{listeq}
	\begin{split}
		0&=\tilde{C}_{11}=K_{21}A_1\\
		0&=\tilde{C}_{12}=K_{21}A_2-2A_1\\
		0&=\tilde{C}_{13}=K_{21}A_3\\
		0&=\tilde{C}_{14}=K_{21}A_3(p_2\leftrightarrow p_3)+4A_1\\
		0&=\tilde{C}_{15}=K_{21}A_4+2A_3
	\end{split}
	\hspace{1.5cm}
	\begin{split}
		0&=\tilde{C}_{21}=K_{31}A_1\\
		0&=\tilde{C}_{22}=K_{31}A_2-2A_1\\
		0&=\tilde{C}_{23}=K_{31}A_3 +4A_1\\
		0&=\tilde{C}_{24}=K_{31}A_3(p_2\leftrightarrow p_3)\\
		0&=\tilde{C}_{25}=K_{31}A_4+2A_3(p_2\leftrightarrow p_3)
	\end{split}
\end{equation}
and it is simple to verify that these equations are equivalent to those given in \eqref{Primary}. 
In the case of the secondary WI's we have to consider some further properties of the form factors. For instance the coefficient $\tilde{C}_{31}$ has the explicit form
\begin{align}
	\tilde{C}_{31}&=\sdfrac{2}{p_1^2}\bigg[p_2(p_1^2-p_2^2+p_3^2)\sdfrac{\partial}{\partial p_2}A_1-p_1^2p_3\sdfrac{\partial}{\partial p_3}A_1-p_2^2p_3\sdfrac{\partial}{\partial p_3}A_1+p_3^3\sdfrac{\partial}{\partial p_3}A_1-p_2\sdfrac{\partial}{\partial p_2}A_3-p_3\sdfrac{\partial}{\partial p_3}A_3\notag\\
	&\hspace{1cm}+p_2\sdfrac{\partial}{\partial p_2}A_3(p_2\leftrightarrow p_3)+p_3\sdfrac{\partial}{\partial p_3}A_3(p_2\leftrightarrow p_3)-6 (p_2^2-p_3^2)A_1-4(A_3-A_3(p_2\leftrightarrow p_3))
	\bigg]\label{C31new}
\end{align}
in which it is possible to substitute the derivative with respect to $p_3$ in terms of derivatives with respect to $p_2$ and $p_1$ using the dilatation Ward identities
\begin{equation}
	\sdfrac{\partial}{\partial p_3}\,A_n=\sdfrac{1}{p_3}\bigg[(d-2-N_n)A_n-\sum_{j=1}^2p_j\sdfrac{\partial}{\partial p_j}A_n\bigg]\label{rel}.
\end{equation}
Using the identity given above in \eqref{C31new}, one derives the relation
\begin{equation}
	\tilde{C}_{31}=\sdfrac{2}{p_1^2}\bigg[L_4\,A_1+R\,A_3-R\,A_3(p_2\leftrightarrow p_3)\bigg]
\end{equation}
with the identification of the differential operators $L$ and $R$ defined in \eqref{Ldef} and \eqref{Rdef}. In this way it is possible to show that all the coefficients related to the secondary Ward identities are the same of those obtained with $p_3$ as the dependent momentum. This argument proves that in spite of the choice of the dependent momentum, the scalar equations for the form factors related to the CWI's remain identical. 

\subsection{Form factors: the solution for $A_1$}

The solutions for the form factors $A_1-A_4$ can be derived using a similar, but modified approach discussed in \secref{fuchs2}, being the equations also inhomogenous. As previously we take as a pivot $p_1^2$, and assume a symmetry under the $(P_{23})$ exchange of $(p_2,\Delta_2)$ with $(p_3,\Delta_3)$ in the correlator. In the case of two photons $\Delta_2=\Delta_3=d-1$.\\
We start from $A_1$ by solving the two equations from \eqref{Primary}
\begin{equation}
	K_{21}A_1=0   \qquad K_{31}A_1=0.
\end{equation}
In this case we introduce the ansatz 
\begin{equation}
	A_1=p_1^{\Delta-2 d - 4}x^a y^b  F(x,y),
\end{equation}
and derive two hypergeometric equations, which are characterized by new values of the 4 defining parameters. 
We obtain 
\begin{equation}
	\label{A1tjj}
	A_1(p_1,p_2,p_3)=p_1^{\Delta-2 d - 4}\sum_{a,b} c^{(1)}(a,b,\vec{\Delta})\,x^a y^b \,F_4(\alpha(a,b) +2, \beta(a,b)+2; \gamma(a), \gamma'(b); x, y),
\end{equation}
with the expression of $\alpha(a,b),\beta(a,b), \gamma(a), \gamma'(b)$ as 
\begin{equation}
	\label{consA1}
	\begin{split}
		\alpha(a,b)&= a + b + \frac{d}{2} -\frac{1}{2}(\Delta_2 + \Delta_3 -\Delta_1),\\
		\beta(a,b)&= a + b + d -\frac{1}{2}(\Delta_1 + \Delta_2 +\Delta_3),
	\end{split}
\end{equation}
which are $P_{23}$ symmetric and 
\begin{equation} 
	\label{consA12}
	\begin{split}
		\gamma(a)& =2 a +\frac{d}{2} -\Delta_2 + 1,\\
		\gamma'(b)&=2 b +\frac{d}{2} -\Delta_3 + 1,
	\end{split}
\end{equation}
with $P_{23}\gamma(a)=\gamma'(b)$.
If we require that $\Delta_2=\Delta_3$, as in the $TJJ$ case, the symmetry constraints are easily implemented. 
Given that the 4 indices, if we choose $p_1$ as a pivot, are given by 
\begin{equation}
	a_0=0,\quad b_0=0,\quad a_1=\Delta_2- \frac{d}{2},\quad b_1=\Delta_3-\frac{d}{2} 
\end{equation}
clearly in this case $a=b$ and $\gamma(a)=\gamma(b)$. 
$F_4$ has the symmetry 
\begin{align}
	F_4(\a,\b; \g, \g' ; x,y)=F_4(\a,\b; \g', \g ; y,x),
\end{align}
and this reflects in the Bose symmetry of $A_1$ if we impose the constraint
\begin{equation}
	c^{(1)}(a_1,b_0)=c^{(1)}(a_0,b_1).
\end{equation}

\subsection{The solution for $A_2$}

The equations for $A_2$ are inhomogeneous. In this case the solution can be identified using some properties of the hypergeometric differential operators $K_i$, appropriately splitted. We recall that in this case they are 
\begin{align}
	K_{21}A_2 &= 2 A_1\label{inhomA2}\\
	K_{31}A_2& = 2A_1.\label{inhomA21}
\end{align}
We take an ansatz of the form 
\begin{equation}
	A_2(p_1,p_2,p_3)=p_1^{\Delta-2 d - 2}F(x,y)
\end{equation}
which provides the correct scaling dimensions for $A_2$. 
Observe that the action of $K_{21}$ and $K_{3}$ on $A_2$ can be rearranged as follows
\begin{align}
	K_{21} A_2&=4 x^a y^b p_1^{\Delta-2 d -4}\bigg( \bar{K}_{21}F(x,y) +\frac{\partial}{\partial x} F(x,y)\bigg)\\[1.5ex]
	K_{31} A_2&=4 x^a y^b p_1^{\Delta-2 d -4}\bigg( \bar{K}_{31}F(x,y) +\frac{\partial}{\partial y} F(x,y)\bigg)
\end{align}
where
\begin{align}
	\label{k1barA2}
	\bar{K}_{21}F(x,y)&=\bigg\{x(1-x) \frac{\partial^2}{\partial x^2} - y^2 \frac{\partial^2}{\partial y^2} - 2 \, x \, y \frac{\partial^2}{\partial x \partial y} +\big[  (\gamma(a)-1) - (\alpha(a,b) + \beta(a,b) + 3) x \big] \frac{\partial}{\partial x}\notag\\
	&\hspace{3cm}
	+ \frac{a (a-a_1)}{x} - (\alpha(a,b) + \beta(a,b) + 3) y \frac{\partial}{\partial y}  - (\alpha +1)(\beta +1) \bigg\} F(x,y),
\end{align}
and 
\begin{align}
	\label{k2barA2}
	\bar{K}_{31} A_2&=\bigg\{ y(1-y) \frac{\partial^2}{\partial y^2} - x^2 \frac{\partial^2}{\partial x^2} - 2 \, x \, y \frac{\partial^2}{\partial x \partial y} +  \big[ (\gamma'(b)-1)- (\alpha(a,b) + \beta(a,b) + 3) y \big]\frac{\partial}{\partial y}\notag\\
	&\hspace{2cm} +  \frac{b(b- b_1)}{y} - (\alpha(a,b) + \beta(a,b) + 3) x \frac{\partial}{\partial x}  -(\alpha(a,b) +1)(\beta(a,b) +1)\bigg\} F(x,y).
\end{align}
At this point observe that the hypergeometric function solution of the equation
\begin{equation}
	\label{ffirst1}
	\bar{K}_{21}F(x,y)=0
\end{equation}
can be taken of the form
\begin{equation}
	\label{rep1A2}
	\Phi_1^{(2)}(x,y)=p_1^{\Delta-2 d - 2}\sum_{a,b} c^{(2)}_1(a,b,\vec{\Delta})\,x^a y^b \,F_4(\alpha(a,b) +1, \beta(a,b)+1; \gamma(a)-1, \gamma'(b) ; x, y) 
\end{equation}
with $c^{(2)}_1$ a constant and the parameters $a,b$ fixed at the ordinary values $(a_i,b_j)$ as in the previous cases, in order to get rid of the $1/x$ and $1/y$ poles in the coefficients of the differential operators. The sequence of parameters in \eqref{rep1A2} will obviously solve the related equation 
\begin{equation}
	\label{secA2}
	{K}_{31}\Phi_1^{(2)}(x,y)=0. 
\end{equation}
\eqref{ffirst1} can be verified by observing that the sequence of  parameters 
$(\alpha(a,b)+1,\beta(a,b)+1 \gamma(a)-1)$ allows to define a solution of \eqref{k2barA2} set to zero, for an arbitrary $\gamma'(b)$, since this parameter does not play any role in the solution of the corresponding equation. 
The sequence $(\alpha(a,b)+1,\beta(a,b)+1, \gamma'(b))$, on the other hand, solves the homogeneous equations associated to $K_{31}$ (i.e. \eqref{secA2}) for any value of the third parameter of $F_4$, which in this case takes the value $\gamma(a)-1$.
A similar result holds for the mirror solution
\begin{equation} 
	\label{repA2}
	\Phi_2^{(2)}(x,y)= p_1^{\Delta-2 d - 2}\sum_{a,b} c_2^{(2)}(a,b,\vec{\Delta})\,x^a y^b \,F_4(\alpha(a,b) +1, \beta(a,b)+1; \gamma(a), \gamma'(b)-1 ; x, y) 
\end{equation}
which satisfies 
\begin{equation}
	\label{firstA2}
	\bar{K}_{31}\Phi_2^{(2)}(x,y)=0 \qquad  {K}_{21}\Phi_2^{(2)}(x,y)=0. 
\end{equation}
As previously remarked, the values of the exponents $a$ and $b$ remain the same for any equation involving either a $K_{i,j}$ or a $\bar{K}_{i j}$, as can be explicitly verified. This implies that the fundamental solutions of the conformal equations are essentially the 4 functions of the type $S_1,\ldots S_4$, for appropriate values of their parameters.  \\
At this point, to show that $F_1$ and $F_2$ is a solution of  \eqref{inhomA2} we use the property 
\begin{equation}
	\frac{\partial^{p+q} F_4(\alpha,\beta;\gamma_1,\gamma_2;x,y)}{\partial x^p\partial y^q} =\frac{(\alpha,p+q)(\beta,p+q)}{(\gamma_1,p)(\gamma_2,q)}
	F_4(\alpha + p + q,\beta + p + q; \gamma_1 + p ; \gamma_2 + q;x,y)
\end{equation}
which gives (for generic parameters $\alpha,\beta,\gamma_1,\gamma_2$)
\begin{align}
	& \frac{\partial F_4(\alpha,\beta;\gamma_1,\gamma_2;x,y)}{\partial x} =\frac{\alpha \beta}{\gamma_1}F_4(\alpha+1,\beta+1,\gamma_1+1,\gamma_2,x,y) \notag\\
	&  \frac{\partial F_4(\alpha,\beta;\gamma_1,\gamma_2;x,y)}{\partial y} =\frac{\alpha \beta}{\gamma_2}F_4(\alpha+1,\beta+1,\gamma_1,\gamma_2 +1,x,y).
\end{align} 
Obviously, such relations are valid whatever dependence the four parameters $\alpha,\beta,\gamma_1,\gamma_2$ may have on 
the Fuchsian exponents $(a_i,b_j)$. 
The actions of $K_{21}$ and $K_{31}$ on the the $\Phi_2^{(i)}$'s (i=1,2)  in \eqref{rep} are then given by 
\begin{align}
	&K_{21}\Phi_1^{(2)}(x,y) = 4p_1^{\Delta-2 d -4} \sum_{a,b} c^{(2)}_1(a,b,\vec{\Delta})\,x^a y^b \frac{\partial}{\partial x} \,F_4(\alpha(a.b) +1, \beta(a,b)+1; \gamma(a)-1, \gamma'(b) ; x, y)   \notag\\
	&= 4 p_1^{\Delta-2 d -4} \sum_{a,b} c^{(2)}_1(a,b,\vec{\Delta})\,x^a y^b \frac{(\alpha(a,b)+1)(\beta(a,b)+1)}{(\gamma(a)-1)}F_4(\alpha(a,b) +2, \beta(a,b)+2; \gamma(a), \gamma'(b); x, y) \notag\\
	&K_{31}\Phi_1^{(2)}(x,y)= 0\\[3ex]
	&K_{31}\Phi_2^{(2)}(x,y) = 4 p_1^{\Delta-2 d -4} \sum_{a,b} c^{(2)}_2(a,b,\vec{\Delta})\, x^a y^b \frac{\partial}{\partial y} \,F_4(\alpha(a,b) +1, \beta(a,b)+1; \gamma(a), \gamma'(b)-1 ; x, y) \notag\\
	&=4 p_1^{\Delta-2 d -4} \sum_{a,b} c^{(2)}_2(a,b,\vec{\Delta})\,  x^a y^b  \frac{(\alpha(a,b)+1)(\beta(a,b)+1)}{(\gamma'(b)-1)} F_4(\alpha(a,b) +2, \beta(a,b)+2; \gamma(a), \gamma'(b); x, y)
	\notag\\
	&K_{21} \Phi_2^{(2)}(x,y)=0,
\end{align}
where it is clear that the non-zero right-hand-side of both equations are proportional to the form factor $A_1$ given in \eqref{A1tjj}. Once this particular solution is determined,  \eqref{A1tjj}, by comparison, gives the conditions on $c_1^{(2)}$ and $c_1^{(2)}$ as
\begin{equation}
	\begin{split}
		c_1^{(2)}(a,b,\vec{\Delta})&=\frac{\gamma(a)-1}{2(\alpha(a,b)+1)(\beta(a,b)+1)}\ c^{(1)}(a,b,\vec \Delta)\,,\\
		c_2^{(2)}(a,b,\vec{\Delta})&=\frac{\gamma'(b)-1}{2(\alpha(a,b)+1)(\beta(a,b)+1)}\ c^{(1)}(a,b,\vec \Delta)\,.
	\end{split}
	\label{condc2A2}
\end{equation}
Therefore, the general solution for $A_2$ in the $TJJ$ case (in which $\g(a)=\g'(b)$ ) is given by superposing the solution of the homogeneous form of \eqref{A1tjj} and the particular one \eqref{rep1A2} and \eqref{repA2}, by choosing the constants appropriately using \eqref{condc2A2}. Its explicit form is written as
\begin{align}
	A_2&= p_1^{\Delta-2 d - 2}\sum_{a b} x^a y^b\Bigg[c^{(2)}(a,b,\vec{\Delta})\,F_4(\alpha(a,b)+1, \beta(a,b)+1; \gamma(a), \gamma'(b); x, y)\notag\\
	&\hspace{1cm}+ \frac{(\gamma(a)-1)\,c^{(1)}(a,b,\vec \Delta)}{2(\alpha(a,b)+1)(\beta(a,b)+1)}\bigg(F_4(\alpha(a,b) +1, \beta(a,b)+1; \gamma(a)-1, \gamma'(b); x, y)\notag\\
	&\hspace{7cm}+ F_4(\alpha(a,b) +1, \beta(a,b)+1; \gamma(a), \gamma'(b)-1; x, y)\bigg)\Bigg],
\end{align}
since $\g(a)=\g'(b)$. 
\subsection{The solution for $A_3$}
Using a similar strategy, the particular solution for the form factor $ A_3$ of the equations 
\begin{equation}
	K_{21} A_3 =0 \qquad K_{3 1} A_3=-4 A_1 \label{A3eq}
\end{equation}
can be found in the form 
\begin{equation}
	\Phi^{(3)}(x,y)= p_1^{\Delta-2 d -2} \sum_{a b} c_1^{(3)}(a,b,\vec\Delta) x^a y^b   F_4(\alpha(a,b)+1,\beta(a,b)+1; \gamma(a),\gamma'(b)-1;x,y). 
\end{equation}

Also in this case the inhomogeneous equation in \eqref{A3eq} fixes the integration constants to be those appearing in $A_1$ 
\begin{equation}
	c_1^{(3)}(a,b,\vec \Delta)=-\frac{\gamma'(b)-1}{(\alpha(a,b)+1)(\beta(a,b)+1)}\, c^{(1)}(a,b,\vec\Delta).
\end{equation}
Therefore the general solution of the equations \eqref{A3eq} can be written as
\begin{align}
	A_3=&p_1^{\Delta-2 d -2} \sum_{a b} x^a y^b\,\bigg[c^{(3)}(a,b,\vec\Delta)\,F_4(\a(a,b)+1,\b(a,b)+1;\g(a),\g'(b);x,y)\notag\\
	&\hspace{3cm}-\frac{(\gamma(a)-1)\,c^{(1)}(a,b,\vec\Delta) }{(\alpha(a,b)+1)(\beta(a,b)+1)}F_4(\alpha(a,b)+1,\beta(a,b)+1; \gamma(a),\gamma'(b)-1;x,y)\bigg]
\end{align} 
since $\g(a)=\g'(b)$ in the $TJJ$ case. 

\subsection{The $A_4$ solution}
The last pair of equations 
\begin{equation}
	K_{21} A_4 =-2 A_3 \qquad \qquad K_{31} A_4=-2 A_3 (p_2\leftrightarrow p_3)\label{eqA4tjj}
\end{equation}
admit three particular solutions
\begin{align}
	\Phi_1^{(4)}&= p_1^{\Delta-2d}\sum_{a b} x^a y^b\, c_1^{(4)}(a,b,\vec{\Delta})\,F_4(\alpha(a,b),\beta(a,b),\gamma(a)-1,\gamma'(b),x,y)\\
	\Phi_2^{(4)}&= p_1^{\Delta-2d}\sum_{a b} x^a y^b\, c_2^{(4)}(a,b,\vec{\Delta})\,F_4(\alpha(a,b),\beta(a,b),\gamma(a),\gamma'(b)-1,x,y)\\
	\Phi_3^{(4)}&= p_1^{\Delta-2d}\sum_{a b} x^a y^b\,c_3^{(4)}\,(a,b,\vec{\Delta})\,F_4(\a(a,b),\b(a,b),\g(a)-1,\g'(b)-1;x,y)
\end{align}
with the action of $K_{21}$ and $K_{31}$ on them as
\begin{align}
	K_{21}\Phi^{(4)}_1&=4p_1^{\Delta-2d-2}\sum_{a b} x^a y^b\, c_1^{(4)}(a,b,\vec{\Delta})\,\frac{\a(a,b)\b(a,b)}{(\g(a)-1)}\,F_4(\alpha(a,b)+1,\beta(a,b)+1,\gamma(a),\gamma'(b),x,y)\\
	K_{31}\Phi^{(4)}_1&=0\\[2.2ex]
	K_{21}\Phi^{(4)}_2&=0\\
	K_{31}\Phi^{(4)}_2&=4p_1^{\Delta-2d-2}\sum_{a b} x^a y^b\, c_2^{(4)}(a,b,\vec{\Delta})\,\frac{\a(a,b)\b(a,b)}{(\g'(b)-1)}\,F_4(\alpha(a,b)+1,\beta(a,b)+1,\gamma(a),\gamma'(b),x,y)\\[2.2ex]
	K_{21}\Phi^{(4)}_3&=4p_1^{\Delta-2d-2}\sum_{a b} x^a y^b\,c_3^{(4)}(a,b,\vec{\Delta})\,\frac{\a(a,b)\b(a,b)}{(\g(a)-1)}\,F_4(\a(a,b)+1,\b(a,b)+1,\g(a),\g'(b)-1;x,y)\\
	K_{31}\Phi^{(4)}_3&=4p_1^{\Delta-2d-2}\sum_{a b} x^a y^b\,c_3^{(4)}(a,b,\vec{\Delta})\,\frac{\a(a,b)\b(a,b)}{(\g'(b)-1)}\,F_4(\a(a,b)+1,\b(a,b)+1,\g(a)-1,\g'(b);x,y)
\end{align}
The inhomogeneous equations \eqref{eqA4tjj} fix the integration constants to be those appearing in $A_3$ and $A_3(p_2\leftrightarrow p_3)$ as
\begin{align}
	c_1^{(4)}&=-\frac{(\g(a)-1)}{2\,\a(a,b)\b(a,b)}\,c^{(3)}(a,b,\vec{\Delta})\\[1.2ex]
	c_2^{(4)}&=-\frac{(\g'(b)-1)}{2\,\a(a,b)\b(a,b)}\,c^{(3)}(a,b,\vec{\Delta})\\[1.2ex]
	c_3^{(4)}&=\frac{(\a(a,b)+1)(\b(a,b)+1)}{2\,\a(a,b)\b(a,b)}\,c^{(1)}(a,b,\vec{\Delta}).
\end{align}
Finally, using the properties $\g(a)=\g'(b)$, we give the general solution for the $A_4$ as
\begin{align}
	A_4&=p_1^{\D-2d}\,\sum_{ab}\,x^a\,y^b\bigg[c^{(4)}(a,b,\vec{\Delta})\,F_4(\a(a,b),\b(a,b),\g(a),\g'(b);x,y)\notag\\
	&\hspace{0.5cm}+\frac{(\a(a,b)+1)(\b(a,b)+1)}{2\,\a(a,b)\b(a,b)}\,c^{(1)}(a,b,\vec{\Delta})\,F_4(\a(a,b),\b(a,b),\g(a)-1,\g'(b)-1;x,y)\notag\\
	&\hspace{0.5cm}-\frac{(\g(a)-1)}{2\,\a(a,b)\b(a,b)}\,c^{(3)}(a,b,\vec{\Delta})\,\bigg(F_4(\alpha(a,b),\beta(a,b),\gamma(a)-1,\gamma'(b),x,y)\notag\\
	&\hspace{7.5cm}+\,F_4(\alpha(a,b),\beta(a,b),\gamma(a),\gamma'(b)-1,x,y)\bigg)
	\bigg]
\end{align}
Notice that, differently from this case, number of free constants can be significantly reduced in the case of a fully symmetric correlator, such as the $TTT$, where the  number of constants reduces to 4, as in \cite{Bzowski:2011ab}.   
\subsection{Summary}
\label{finfin}
To summarize, the solutions of the primary CWI' s in the $TJJ$ case are expressed as sums of 4 hypergeometric functions of universal indicial points 
\begin{equation}
	a_0 =0,\quad b_0=0,\quad a_1=\Delta_2- \frac{d}{2},\quad b_1=\Delta_3-\frac{d}{2}
\end{equation}
and parameters 
\begin{align}
	&\alpha(a,b)= a + b + \frac{d}{2} -\frac{1}{2}(\Delta_2 +\Delta_3 -\Delta_1)\,,  &&\beta (a,b)=a +  b + d -\frac{1}{2}(\Delta_1 +\Delta_2 +\Delta_3) \\
	&\gamma(a) =2 a +\frac{d}{2} -\Delta_2 + 1\,, &&\gamma'(b)=2 b +\frac{d}{2} -\Delta_3 + 1. \label{cons2}
\end{align}
where $\D_2=\D_3=d-1$ and $\D_1=d$. In particular they are given by
\begin{equation}
	\begin{split}
		A_1&=p_1^{\Delta-2 d - 4}\sum_{a,b} c^{(1)}(a,b,\vec{\Delta})\,x^a y^b \,F_4(\alpha(a,b) +2, \beta(a,b)+2; \gamma(a), \gamma'(b); x, y)
	\end{split}
\end{equation}
\begin{equation}
	\begin{split}
		A_2&= p_1^{\Delta-2 d - 2}\sum_{a b} x^a y^b\Bigg[c^{(2)}(a,b,\vec{\Delta})\,F_4(\alpha(a,b)+1, \beta(a,b)+1; \gamma(a), \gamma'(b); x, y)\\
		&\hspace{1cm}+ \frac{(\gamma(a)-1)\,c^{(1)}(a,b,\vec \Delta)}{2(\alpha(a,b)+1)(\beta(a,b)+1)}\bigg(F_4(\alpha(a,b) +1, \beta(a,b)+1; \gamma(a)-1, \gamma'(b); x, y)\\
		&\hspace{7cm}+ F_4(\alpha(a,b) +1, \beta(a,b)+1; \gamma(a), \gamma'(b)-1; x, y)\bigg)\Bigg]
	\end{split}
\end{equation}
\begin{align}
	A_3=&p_1^{\Delta-2 d -2} \sum_{a b} x^a y^b\,\bigg[c^{(3)}(a,b,\vec\Delta)\,F_4(\a(a,b)+1,\b(a,b)+1;\g(a),\g'(b);x,y)\notag\\
	&\hspace{3cm}-\frac{(\gamma(a)-1)\,c^{(1)}(a,b,\vec\Delta) }{(\alpha(a,b)+1)(\beta(a,b)+1)}F_4(\alpha(a,b)+1,\beta(a,b)+1; \gamma(a),\gamma'(b)-1;x,y)\bigg]
\end{align}
\begin{align}
	A_4&=p_1^{\D-2d}\,\sum_{ab}\,x^a\,y^b\bigg[c^{(4)}(a,b,\vec{\Delta})\,F_4(\a(a,b),\b(a,b),\g(a),\g'(b);x,y)\notag\\
	&\hspace{0.5cm}+\frac{(\a(a,b)+1)(\b(a,b)+1)}{2\,\a(a,b)\b(a,b)}\,c^{(1)}(a,b,\vec{\Delta})\,F_4(\a(a,b),\b(a,b),\g(a)-1,\g'(b)-1;x,y)\notag\\
	&\hspace{0.5cm}-\frac{(\g(a)-1)}{2\,\a(a,b)\b(a,b)}\,c^{(3)}(a,b,\vec{\Delta})\,\bigg(F_4(\alpha(a,b),\beta(a,b),\gamma(a)-1,\gamma'(b),x,y)\notag\\
	&\hspace{7.5cm}+\,F_4(\alpha(a,b),\beta(a,b),\gamma(a),\gamma'(b)-1,x,y)\bigg)
	\bigg]
\end{align}
in terms of the constants $c^{(i)}(a,b)$ given above. The method has the advantage of being generalizable to higher point functions, 
in the search of specific solutions of the corresponding correlation functions.

\section{The $TTT$ in momentum space}\label{secTTTgen}
In this section we are going to discuss the reconstruction method for the $TTT$ case, following the strategy presented in the previous sections. 
\subsection{Ward Identities}
We start with the analysis of the transverse and trace Ward identities. These relations follow from the analysis in \secref{TraceTransverse}. By taking two functional derivatives of \eqref{transverse1} and \eqref{traceW} and then in the limit $g_{\m\n}=\d_{\m\n}$ we obtain the canonical Ward identities for the $\braket{TTT}$ in position space
\begin{align}
	\partial_{\n_1}\braket{T^{\m_1\n_1}(p_1)T^{\m_2\n_2}(p_2)T^{\m_3\n_3}(p_3)}&=\big[\partial_{\,\n_1}\d(x_1-x_2)\big]\mathcal{B}^{\m_1\n_1\m_2\n_2}_{\hspace{1.1cm}\a\b}\braket{T^{\a\b}(x_1)T^{\m_3\n_3}(x_3)}\notag\\
	&+\big[\partial_{\n_1}\d(x_1-x_3)\big]\mathcal{B}^{\m_1\n_1\m_3\n_3}_{\hspace{1.1cm}\a\b}\braket{T^{\a\b}(x_1)T^{\m_2\n_2}(x_2)}\notag\\
	\d_{\m_1\n_1}\braket{T^{\m_1\n_1}(p_1)T^{\m_2\n_2}(p_2)T^{\m_3\n_3}(p_3)}&=-2\braket{T^{\m_2\n_2}(p_2+p_1)T^{\m_3\n_3}(x_3)}-2\braket{T^{\m_2\n_2}(p_2)T^{\m_3\n_3}(p_3+p_1)},\notag
\end{align}
where $\mathcal B$ is a constant tensors defined as
\begin{equation}
	\mathcal{B}^{\m_1\n_1\m_j\n_j}_{\hspace{1.2cm}\a\b}\equiv -2\, \d^{\m_1(\m_j}  \d{}^{\n_j)}{}_{(\a}\d^{\n_1}{}_{\b)} 
	+ \d^{\m_1\n_1}\,\d^{(\m_j}{}_{\a}  \d^{\n_j)}{}_{\b}.
\end{equation}

By Fourier transforming we obtain these canonical Ward identities in momentum space as
\begin{align}
	p_{1\n_1}\braket{T^{\m_1\n_1}(p_1)T^{\m_2\n_2}(p_2)T^{\m_3\n_3}(p_3)}&=p_{2\,\n_1}\mathcal{B}^{\m_1\n_1\m_2\n_2}_{\hspace{1.1cm}\a\b}\braket{T^{\a\b}(p_1+p_2)T^{\m_3\n_3}(p_3)}\notag\\
	&+p_{3\n_1}\mathcal{B}^{\m_1\n_1\m_3\n_3}_{\hspace{1.1cm}\a\b}\braket{T^{\a\b}(p_1+p_3)T^{\m_2\n_2}(p_2)}\label{transvttt}\\
	\d_{\m_1\n_1}\braket{T^{\m_1\n_1}(p_1)T^{\m_2\n_2}(p_2)T^{\m_3\n_3}(p_3)}&=-2\braket{T^{\m_2\n_2}(p_2+p_1)T^{\m_3\n_3}(x_3)}-2\braket{T^{\m_2\n_2}(p_2)T^{\m_3\n_3}(p_3+p_1)}\label{tracettt},
\end{align}
\subsection{Tensor decomposition}
We can now apply the general method presented above in order to reconstruct the $\braket{T^{\m_1\n_1}T^{\m_2\n_2}T^{\m_3\n_3}}$. For the decomposition of the transverse and traceless part of this correlation function we follow the analysis of \secref{tensstruc}. The decomposition of the $TTT$ can be written as
\begin{align}
	\braket{T^{\mu_1\nu_1}\,T^{\mu_2\n_2}\,T^{\mu_3\n_3}}&=\braket{t^{\mu_1\nu_1}\,t^{\mu_2\n_2}\,t^{\mu_3\n_3}}+\braket{T^{\mu_1\nu_1}\,T^{\mu_2\n_2}\,t_{loc}^{\mu_3\n_3}}+\braket{T^{\mu_1\nu_1}\,t_{loc}^{\mu_2\n_2}\,T^{\mu_3\n_3}}\notag\\
	&+\braket{t_{loc}^{\mu_1\nu_1}\,T^{\mu_2\n_2}\,T^{\mu_3\n_3}}-\braket{T^{\mu_1\nu_1}\,t_{loc}^{\mu_2\n_2}\,t_{loc}^{\mu_3\n_3}}-\braket{t_{loc}^{\mu_1\nu_1}\,t_{loc}^{\mu_2\n_2}\,T^{\mu_3\n_3}}\notag\\
	&-\braket{t_{loc}^{\mu_1\nu_1}\,T^{\mu_2\n_2}\,t_{loc}^{\mu_3\n_3}}+\braket{t_{loc}^{\mu_1\nu_1}\,t_{loc}^{\mu_2\n_2}\,t_{loc}^{\mu_3\n_3}},\label{reconstrTTT}
\end{align}
where the transverse traceless part consists of five form factors,
\begin{align}
	&\braket{t^{\m_1\n_1}(p_1)t^{\m_2\n_2}(p_2)t^{\m_3\n_3}(p_3)}
	= \Pi^{\m_1\n_1}_{\a_1\b_1}(p_1)
	\Pi^{\m_2\n_2}_{\a_2\b_2}(p_2)\Pi^{\m_3\n_3}_{\a_3\b_3}(p_3)\notag\\
	&\hspace{0.7cm}\times\Big[ A_1\,p_2^{\a_1} p_2^{\b_1} p_3^{\a_2} p_3^{\b_2} p_1^{\a_3} p_1^{\b_3}+ A_2\,\d^{\b_1\b_2} p_2^{\a_1} p_3^{\a_2} p_1^{\a_3} p_1^{\b_3} 
	+ A_2\,(p_1 \leftrightarrow p_3)\, \d^{\b_2\b_3}  p_3^{\a_2} p_1^{\a_3} p_2^{\a_1} p_2^{\b_1} \notag\\
	&\hspace{1.3cm}+ A_2\,(p_2\leftrightarrow p_3)\, \d^{\b_3\b_1} p_1^{\a_3} p_2^{\a_1}  p_3^{\a_2} p_3^{\b_2}+ A_3\,\d^{\a_1\a_2} \d^{\b_1\b_2}  p_1^{\a_3} p_1^{\b_3} + A_3(p_1\leftrightarrow p_3)\,\d^{\a_2\a_3} \d^{\b_2\b_3}  p_2^{\a_1} p_2^{\b_1} \notag\\
	&\hspace{2cm}
	+ A_3(p_2\leftrightarrow p_3)\,\d^{\a_3\a_1} \d^{\b_3\b_1}  p_3^{\a_2} p_3^{\b_2} + A_4\,\d^{\a_1\a_3} \d^{\a_2\b_3}  p_2^{\b_1} p_3^{\b_2} + A_4(p_1\leftrightarrow p_3)\, \d^{\a_2\a_1} \d^{\a_3\b_1}  p_3^{\b_2} p_1^{\b_3} \notag\\
	&\hspace{3.5cm}+ A_4(p_2\leftrightarrow p_3)\, \d^{\a_3\a_2} \d^{\a_1\b_2}  p_1^{\b_3} p_2^{\b_1} + A_5  \d^{\a_1\b_2}  \d^{\a_2\b_3}  \d^{\a_3\b_1}\Big],\label{tttdec}
\end{align}
and all the local parts depend on the two point functions $TT$ via the canonical Ward Identities. One notices that the form of the transverse traceless part in \eqref{tttdec} is manifestly invariant under the permutation group of the set $\{1,2,3\}$. 
In this case the special conformal Ward identities reads
\begin{align}
	0=&\mathcal{K}^\kappa\braket{T^{\mu_1\nu_1}(p_1)T^{\mu_2\nu_2}(p_2)T^{\mu_3\nu_3}(p_3)}\notag\\
	=&\sum_{j=1}^{2}\left(2(\Delta_j-d)\frac{\partial}{\partial p_{j\,\kappa}}-2p_j^\alpha\frac{\partial}{\partial p_j^\alpha}\frac{\partial}{\partial p_{j\,\kappa}}+(p_j)^\kappa\frac{\partial}{\partial p_j^\alpha}\frac{\partial}{\partial p_{j\,\alpha}}\right)\braket{T^{\mu_1\nu_1}(p_1)T^{\mu_2\nu_2}(p_2)T^{\mu_3\nu_3}(p_3)}\notag\\
	&\hspace{1cm}+4\left(\delta^{\kappa(\mu_1}\frac{\partial}{\partial p_1^{\alpha_1}}-\delta^\kappa_{\alpha_1}\d_\l^{(\m_1}\frac{\partial}{\partial p_{1\,\l}}\right)\braket{ T^{\nu_1)\alpha_1}(p_1)T^{\mu_2\nu_2}(p_2)T^{\mu_3\nu_3}(p_3)}\notag\\
	&\hspace{1cm}+4\left(\delta^{\kappa(\m_2}\frac{\partial}{\partial p_2^{\a_2}}-\delta^\kappa_{\a_2}\d^{(\m_2}_\l\frac{\partial}{\partial p_{2\,\l}}\right)\braket{ T^{\mu_2)\a_2}(p_2)T^{\mu_1\nu_1}(p_1)T^{\mu_3\nu_3}(p_3)}.
\end{align}

Considering the decomposition $T^{\m\n}=t^{\m\n}+t^{\m\n}{\hspace{-2ex}\vspace{-0.5ex}}_{\substack{\scalebox{0.5}{loc}}}$, we find that the action of the special conformal operator $\mathcal K^\k$ on the transverse and traceless part of the correlator is still transverse and traceless. For this reason we are free to apply the transverse-traceless projectors in order to isolate equations for the form factors appearing in the decomposition of $\braket{t^{\m_1\n_1}t^{\m_2\n_2}t^{\m_3\n_3}}$. Using the relations in \appref{appendixB} one finds
\begin{align}
	0=&\ \Pi^{\mu_1\nu_1}_{\alpha_1\beta_1}(p_1)\Pi^{\mu_2\nu_2}_{\alpha_2\beta_2}(p_2)\Pi^{\mu_3\nu_3}_{\alpha_3\beta_3}(p_3)\mathcal{K}^\kappa\braket{ t^{\alpha_1\beta_1}(p_1)t^{\alpha_2\beta_2}(p_2)t^{\alpha_3\beta_3}(p_3)}\notag\\
	&+\frac{4d}{p_1^2}\Pi^{\mu_1\nu_1}_{\alpha_1\beta_1}(p_1)\left[\delta^{\alpha_1\kappa}p_{1\gamma_1}\braket{ T^{\beta_1\gamma_1}(p_1)t^{\mu_2\nu_2}(p_2)t^{\mu_3\nu_3}(p_3)}\right]\notag\\
	&+\frac{4d}{p_2^2}\Pi^{\mu_2\nu_2}_{\alpha_2\beta_2}(p_2)\left[\delta^{\alpha_2\kappa}p_{2\gamma_2}\braket{t^{\mu_1\nu_1}(p_1)T^{\beta_2\gamma_2}(p_2)t^{\mu_3\nu_3}(p_3)}\right]\notag\\
	&+\frac{4d}{p_3^2}\Pi^{\mu_3\nu_3}_{\alpha_3\beta_3}(p_3)\left[\delta^{\alpha_3\kappa}p_{3\gamma_3}\braket{t^{\mu_1\nu_1}(p_1)t^{\mu_2\nu_2}(p_2)T^{\beta_3\gamma_3}(p_3)}\right]\notag\\
\end{align}
The last three terms are semi-local and may be re-expressed in terms of 2-point functions via \eqref{transvttt}. From the previous relation one can extract the CWI's as in \secref{specialconfward} and it is possible to write the most general form of the result as 
\begin{align}
	0&=\Pi^{\mu 1\nu 1}_{\alpha_1\beta_1}(p1) \Pi^{\mu 2\nu 2}_{\alpha_2\beta_2}(p2)\Pi^{\mu_3\nu_3}_{\alpha_3\beta_3}(p_3) \mathcal{K}^{\kappa} \braket{ \braket{t^{\alpha_1 \beta_1} (p_1) t^{\alpha_2 \beta_2} (p_2) t^{\alpha_3 \beta_3} (p_3)  }} \notag\\
	&= \Pi^{\mu 1\nu 1 }_{\alpha_1 \beta_1} (p_1) \Pi^{\mu 2\nu 2 }_{\alpha_2 \beta_2} (p_2)\Pi^{\mu_3\nu_3}_{\alpha_3\beta_3}(p_3) \times \Bigl\{ p_1^{\kappa} \Bigl( C_{1,1} \, p_2^{\alpha_1} p_2^{\beta_1} p_3^{\alpha_2} p_3^{\beta_2}p_1^{\alpha_3} p_1^{\beta_3} + C_{1,2}\, p_2^{\a_1} p_1^{\alpha_3} p_1^{\beta_3} p_3^{\alpha_2}\delta^{\b_1 \beta_2} \notag\\
	&\quad+ C_{1,3}\, p_2^{\beta_1} p_2^{\alpha_1} p_1^{\alpha_3} p_3^{\a_2}\delta^{\b_2 \beta_3}+ C_{1,4}\, p_2^{\a_1} p_3^{\alpha_2} p_1^{\alpha_3} p_3^{\beta_2}\delta^{\b_1 \beta_3}+ C_{1,5}\, p_1^{\beta_3} p_2^{\beta_1} \delta^{\alpha_1 \alpha_2}\delta^{\alpha_3 \beta_2}\notag\\
	&\quad+ C_{1,6}\, p_1^{\beta_3} p_1^{\alpha_3}\delta^{\alpha_1 \a_2}\delta^{\b_1 \beta_2}+ C_{1,7}\, p_1^{\beta_3} p_3^{\alpha_2}\delta^{\alpha_3 \a_1}\delta^{\b_1 \beta_2}+ C_{1,8}\, p_2^{\beta_1} p_2^{\alpha_1}\delta^{\alpha_2 \a_3}\delta^{\b_2 \beta_3}\notag\\
	&\quad+ C_{1,9}\, p_3^{\beta_2} p_3^{\alpha_2}\delta^{\alpha_1 \a_3}\delta^{\b_1 \beta_3}+ C_{1,10}\, \delta^{\alpha_1 \beta_2}\delta^{\alpha_2 \beta_3}\delta^{\alpha_3 \beta_1}\Bigr)+ \big(p_1^\kappa\leftrightarrow p_2^\kappa; \ C_{1,j}\leftrightarrow C_{2,j}\big) + \big( p_1^\kappa\leftrightarrow p_3^\kappa; \ C_{1,j}\leftrightarrow C_{3,j}\big)\notag\\
	& \quad\  + \delta^{\kappa \alpha_1} \Bigl( C_{3,1}\, p_{3}^{\alpha_2}p_{3}^{\beta_2} p_{1}^{\alpha_3}p_{1}^{\beta_3}p_{2}^{\beta_1} + C_{3,2}\, \delta^{\alpha_2 \beta_3} p_{3}^{\beta_2}  p_{1}^{\alpha_3} p_{2}^{\beta_1} + C_{3,3}\, \delta^{\alpha_2 \beta_1} p_{1}^{\alpha_3}  p_{3}^{\beta_2} p_{1}^{\beta_3}+ C_{3,4}\, \delta^{\alpha_3 \beta_1} p_{3}^{\alpha_2}  p_{3}^{\beta_2} p_{1}^{\beta_3}\notag\\
	&\qquad\qquad+ C_{3,5}\, \delta^{\alpha_2 \beta_1} \delta^{\alpha_3 \beta_2} p_{1}^{\beta_3}+C_{3,6}\, \delta^{\alpha_2 \beta_3} \delta^{\alpha_3 \beta_1} p_{3}^{\beta_2}+C_{3,7}\, \delta^{\alpha_2 \beta_3} \delta^{\alpha_3 \beta_2} p_{2}^{\beta_1}\Bigr)\notag\\
	&\qquad\qquad+[(\a_1,\b_1,p_2)\leftrightarrow (\a_2,\b_2,p_3)]+[(\a_1,\b_1,p_2)\leftrightarrow (\a_3,\b_3,p_1)]\Bigr\}\label{specialTTT}
\end{align}
where the coefficients $C_{j,k}$ are differential equations involving the form factors $A_1$, $A_2$, $A_3$, $A_4$, $A_5$. Each CWI's can the be presented in terms of the momentum magnitudes $p_j$. The primary CWI's appear as the coefficients of transverse ore transverse-traceless tensors containing $p_1^\k$, $p_2^\k$. The remaining equations, following from all other transverse or transverse-traceless terms, are then the secondary CWI's. Using the definitions of the operators $L_N$ and $R$ from \eqref{Ldef} and \eqref{Rdef} we can write the primary Ward identities
\begin{equation}\label{PrimaryTTT}
	\begin{aligned}[c]
		&\textup{K}_{13}A_1=0 \\[1.5ex]
		&\textup{K}_{13}A_2=8 A_1 \\[1.5ex]
		& \textup{K}_{13}A_3=2A_2 \\[1.5ex]
		& \textup{K}_{13}A_4=-4A_2(p_{2\leftrightarrow 3}) \\[1.5ex]
		& \textup{K}_{13}A_5=2\left[A_4-A_4(p_{3\leftrightarrow 1}) \right]\\[1.5ex]
	\end{aligned}
	\qquad
	\begin{aligned}[c]
		&\textup{K}_{32}  A_1=0 \\[1.5ex]
		&\textup{K}_{32}A_2=-8 A_1 \\[1.5ex]
		&\textup{K}_{32}A_3=-2A_2 \\[1.5ex]
		&\textup{K}_{32}A_4= 4A_2(p_{3\leftrightarrow 1})\\[1.5ex]
		&\textup{K}_{32}A_5=2A_4(p_{3\leftrightarrow 2})-2A_4 \\[1.5ex]
	\end{aligned}\qquad
	\begin{aligned}[c]
		&\textup{K}_{12}A_1=0 \\[1.5ex]
		&\textup{K}_{12}A_2=0\\[1.5ex]
		& \textup{K}_{12}A_3=0\\[1.5ex]
		& \textup{K}_{12}A_4=4\left[A_2(p_{1\leftrightarrow 3})-A_2(p_{2\leftrightarrow 3}) \right]\\[1.5ex]
		& \textup{K}_{12}A_5=2\left[A_4(p_{2\leftrightarrow 3})-A_4(p_{3\leftrightarrow 1}) \right]\\[1.5ex]
	\end{aligned}
\end{equation}
The solution follow from the analysis in \secref{primarysol},
\begin{subequations}
	\begin{align}
		A_1&=\a_1\,J_{6\{000\}}\\
		A_2&=4\a_1\,J_{5\{001\}}+\a_2\,J_{4\{000\}}\\
		A_3&=2\a_1\,J_{4\{002\}}+\a_2\,J_{3\{001\}}+\a_3\,J_{2\{000\}}\\
		A_4&=8\a_1\,J_{4\{110\}}-\a_2\,J_{3\{001\}}+\a_4\,J_{2\{000\}}\\
		A_5&=8\a_1\,J_{3\{111\}}+2\a_2\,\big(\,J_{2\{110\}}+J_{2\{101\}}+J_{2\{011\}}\big)+\a_5\,J_{0\{000\}},
	\end{align}\label{PrimSolTTT}
\end{subequations}
where $a_j$, $j=1,\dots,5$ are constants. If the integrals diverge, we should we need to regulate them. 
The independent secondary Ward identities are written in the form
\begin{align}
	C_{31}&= -\frac{2}{p_1^2}\Bigl( \textup{L}_6\, A_1 + \textup{R}\bigl[\, A_2-A_2(p_3\leftrightarrow p_2)\bigr]\Bigr)\notag\\
	&=-\frac{4d}{p_1^2}\text{coeff. of } p_3^{\mu_2}p_3^{\nu_2}p_1^{\mu_3}p_1^{\nu_3}p_2^{\mu_1} \text{ in } p_{1\nu_1}\braket{ T^{\mu_1\nu_1}(p_1)T^{\mu_2\nu_2}(p_2)T^{\mu_3\nu_3}(p_3)}\\
	C_{32}&= \frac{2}{p_1^2}\Bigl( \textup{L}_4\, A_2(p_3\leftrightarrow p_1) + \textup{R}\bigl[\, A_4(p_3\leftrightarrow p_2)-A_4\bigr]\Bigr)-4\left(A_2-A_2(p_3\leftrightarrow p_2)\right)\nonumber\\
	&=\frac{4d}{p_1^2}\text{coeff. of } \delta^{\mu_2\mu_3}p_3^{\nu_2}p_1^{\nu_3}p_2^{\mu_1} \text{ in } p_{1\nu_1}\braket{ T^{\mu_1\nu_1}(p_1)T^{\mu_2\nu_2}(p_2)T^{\mu_3\nu_3}(p_3)}\\
	C_{33}&= \frac{1}{p_1^2}\Bigl( \textup{L}_6\, A_2 + 2\textup{R}\bigl[\,2A_3-A_4(p_3\leftrightarrow p_1)\bigr]\Bigr)\notag\\
	&=\frac{4d}{p_1^2}\text{coeff. of } \delta^{\mu_2\mu_1}p_1^{\mu_3}p_1^{\nu_3}p_3^{\nu_2} \text{ in } p_{1\nu_1}\braket{ T^{\mu_1\nu_1}(p_1)T^{\mu_2\nu_2}(p_2)T^{\mu_3\nu_3}(p_3)}\\
	C_{35}&= -\frac{1}{p_1^2}\Bigl( \textup{L}_4\, A_4(p_3\leftrightarrow p_2)  - 2\textup{R}A_5\Bigr)+8A_3-2A_4(p_3\leftrightarrow p_1)\notag\\
	&=-\frac{4d}{p_1^2}\text{coeff. of } \delta^{\mu_2\mu_1}\delta^{\mu_3\nu_2}p_1^{\nu_3} \text{ in } p_{1\nu_1}\braket{ T^{\mu_1\nu_1}(p_1)T^{\mu_2\nu_2}(p_2)T^{\mu_3\nu_3}(p_3)}
\end{align}
\begin{align}
	C_{36}&= \frac{1}{p_1^2}\Bigl( \textup{L}_4\, A_4- 2\textup{R}A_5(p_3\leftrightarrow p_2)\Bigr)+8A_3(p_3\leftrightarrow p_2)-2A_4(p_3\leftrightarrow p_1)\notag\\
	&=\frac{4d}{p_1^2}\text{coeff. of } \delta^{\mu_3\mu_1}\delta^{\mu_2\nu_3}p_3^{\nu_2} \text{ in } p_{1\nu_1}\braket{ T^{\mu_1\nu_1}(p_1)T^{\mu_2\nu_2}(p_2)T^{\mu_3\nu_3}(p_3)}\\
	C_{37}&= -\frac{2}{p_1^2} \textup{L}_2\, A_3(p_3\leftrightarrow p_1)  -2A_4+2A_4(p_3\leftrightarrow p_2)\notag\\
	&=-\frac{4d}{p_1^2}\text{coeff. of } \delta^{\mu_2\mu_3}\delta^{\nu_3\nu_2}p_2^{\mu_1} \text{ in } p_{1\nu_1}\braket{ T^{\mu_1\nu_1}(p_1)T^{\mu_2\nu_2}(p_2)T^{\mu_3\nu_3}(p_3)}
\end{align}
The equations $C_{31}$ and $C_{32}$ are trivially satisfied in all the cases and do not impose any additional conditions on primary constants. In order to show this statement we can now render explicit the contributions on the right-hand sides of the secondary CWI's. We need the expression of the transverse WI's to do so. Starting from \eqref{transvttt} and writing explicit the tensor coefficient $\mathcal{B}$ one obtains 
\begin{align}
	&p_{1\n_1}\braket{T^{\m_1\n_1}(p_1)T^{\m_2\n_2}(p_2)T^{\m_3\n_3}(p_3)}=p_2^{\,\m_1}\braket{T^{\m_2\n_2}(p_1+p_2)T^{\m_3\n_3}(p_3)}+p_3^{\,\m_1}\braket{T^{\m_2\n_2}(p_2)T^{\m_3\n_3}(p_3+p_1)}\notag\\
	&\quad-2\d^{\m_1(\m_2}p_{2\a}\braket{T^{\n_2)\a}(p_1+p_2)T^{\m_3\n_3}(p_3)}-2\d^{\m_1(\m_3}p_{3\a}\braket{T^{\n_3)\a}(p_1+p_3)T^{\m_2\n_2}(p_2)}
\end{align}
and using the explicit form for the $TT$ 2-point function we find
\begin{align}
	&p_{1\n_1}\braket{T^{\m_1\n_1}(p_1)T^{\m_2\n_2}(p_2)T^{\m_3\n_3}(p_3)}=p_2^{\,\m_1}\big[\Pi^{\m_2\n_2\m_3\n_3}(p_3)\,A_{TT}(p_3)-\Pi^{\m_2\n_2\m_3\n_3}(p_2)\,A_{TT}(p_2)\big]\notag\\
	&\hspace{2cm}-2\big[\d^{\m_1(\m_2}p_{2\a}\Pi^{\n_2)\a\m_3\n_3}(p_3)\,A_{TT}(p_3)-\d^{\m_1(\m_3}p_{3\a}\Pi^{\n_3)\a\m_2\n_2}(p_2)\,A_{TT}(p_2)\big]\label{TTTtran}
\end{align}
where $A_{TT}(p_i)$ depends on the normalization of the coefficient of the 2-point function $TT$, which is defined as
\begin{equation}
	A_{TT}(p)=c_T\,\begin{cases}
		p^d&\text{if}\ d=3,5,7,\dots\\
		p^d(-\log(p^2)+\text{local})&\text{if}\ d=2,4,6,\dots\,.
	\end{cases}\label{ATT}
\end{equation}
In \eqref{specialTTT} the transverse Ward identities multiply three projection operators. Keeping in mind this information, we can drop from \eqref{TTTtran} those terms that vanish, if contracted with a projector. In this way we obtain
\begin{align}
	p_{1\n_1}\braket{T^{\m_1\n_1}(p_1)T^{\m_2\n_2}(p_2)T^{\m_3\n_3}(p_3)}&=2\,p_2^{\m_1}\d^{\n_2\m_3}\d^{\m_2\n_3}\big[A_{TT}(p_3)-A_{TT}(p_2)\big]\notag\\
	&+4\big[p_1^{\n_3}\d^{\m_1\m_2}\d^{\n_2\m_3}A_{TT}(p_3)+p_3^{\n_2}\d^{\m_1\m_3}\d^{\n_3\m_2}A_{TT}(p_2)\big].
\end{align}
From the previous expression we derive an explicit form of the secondary Ward identities as
\begin{align}
	&\textup{L}_6\, A_1 + \textup{R}\bigl[\, A_2-A_2(p_3\leftrightarrow p_2)\bigr]=0\label{Tsec1}\\[1.5ex] 
	&\textup{L}_4\, A_2(p_3\leftrightarrow p_1) + \textup{R}\bigl[\, A_4(p_3\leftrightarrow p_2)-A_4\bigr]+2p_1^2\left[A_2(p_3\leftrightarrow p_2)-A_2\right]=0\\[1.5 ex] 
	&\textup{L}_6\, A_2 + 2\textup{R}\bigl[\,2A_3-A_4(p_3\leftrightarrow p_1)\bigr]=0\\[1.5 ex] 
	&\textup{L}_4\, A_4(p_3\leftrightarrow p_2)  - 2\textup{R}A_5+2p_1^2\left[A_4(p_3\leftrightarrow p_1)-4A_3\right]=8d\,A_{TT}(p_3)\\[2 ex] 
	&\textup{L}_4\, A_4- 2\textup{R}A_5(p_3\leftrightarrow p_2)  +2p_1^2\left[A_4(p_3\leftrightarrow p_1)-4A_3(p_3\leftrightarrow p_2)  \right]=8d\,A_{TT}(p_2)\\[2 ex]  
	&\textup{L}_2\, A_3(p_3\leftrightarrow p_1) +p_1^2\left[A_4-A_4(p_3\leftrightarrow p_2)\right]=4d\,\big[A_{TT}(p_3)-A_{TT}(p_2)\big]\label{F14}
\end{align}
From the definition of $A_{TT}$ in \eqref{ATT} one observes that the secondary CWI's simplify in zero-momentum limit $p_3\to0$, $p_1=-p_2=p$. 

\subsection{Divergences}
We discuss the divergences of the solutions \eqref{PrimSolTTT} for the form factors that satisfy the primary CWI's \eqref{PrimaryTTT}. We have discussed the cases in which the triple-K integrals are divergent and in the $TTT$ we obtain 
\begin{equation}
	\left(\frac{d}{2}+N\right)\pm\left(\frac{d}{2}+k_1\right)\pm\left(\frac{d}{2}+k_2\right)\pm\left(\frac{d}{2}+k_3\right)=-2n,\qquad n\in\mathbb{N}\label{TTTdiv}
\end{equation}
where the parameters $N,k_1,k_2,k_3$ are those appearing in the solutions of the form factors written as reduced triple-K integrals $J_{N,\{k_1,k_2,k_3\}}$. 
As pointed out in \cite{Bzowski:2017poo}, the triple-K integrals appearing in \eqref{PrimSolTTT} satisfy the divergent condition \eqref{TTTdiv} in two different ways, represented as $(---)$ and $(--+)$ (and permutation thereof), where the signs $\pm$ are those in \eqref{TTTdiv}. The difference between these two types of singularities lies in the proper way of removing them, and in particular for the $(--+)$ there are no counterterms which allow to obtain a finite result. The only form factor in \eqref{PrimSolTTT} affected by this kind of divergences is $A_5$ and in particular through the integrals
\begin{equation}
	J_{2\{110\}}: (--+),\quad
	J_{2\{101\}}:(-+-),\quad
	J_{2\{011\}}:(+--),
\end{equation}
\begin{equation}
	J_{0\{000\}}:(+--),\ (+--),\ (+--)
\end{equation}
which manifest these types of divergences independently of the specific dimensions $d$. These divergences emerge as poles in $1/\e(u-v)$ or $1/(u-v)$, where $u$ and $v$ are the parameters that regularize the integrals. 
Furthermore, in the case of odd dimensions, from \eqref{divFormFactor}, the form factors do not have any physical \emph{ultralocal} singularities $(---)$ but only \emph{semilocal} ones of the type $(--+)$ that can be removed by a choice of the parameters $\a_i$. In particular, we observe that after an expansion in the zero momentum limit of the divergent part of $A_5$, we obtain
\begin{align}
	&2\a_2\,\big(\,J_{2\{110\}}+J_{2\{101\}}+J_{2\{011\}}\big)+\a_5\,J_{0\{000\}}\notag\\
	&=2^{\frac{d}{2}+u\e-3}\,p^{d-u\epsilon+3v\e}\Gamma\left(\frac{d}{2}+v\e\right)\Gamma\left(\frac{d}{2}+\frac{(u-v)\e}{2}\right)\Gamma\left(1+\frac{(u-v)\e}{2}\right)\notag\\
	&\Gamma\left(-\frac{d}{2}+\frac{(u-3v)\e}{2}\right)\Gamma\left(\frac{(u-v)\e}{2}\right)\Bigg\{\a_5\bigg[\e(u-v)+1\bigg]+\a_2\bigg[2d^2+\e\left(\frac{5d^2}{2}(u-v)+3du+5dv\right)
	\notag\\
	&+\e^2\big((3d+1)u^2+4(d+1)uv+(3-7d)v^2\big)+O(\e^3)\bigg]\Bigg\}.
\end{align}
Notice that the pole $1/((u-v)\e)$, coming from the gamma function $\Gamma[(u-v)\e/2]$, is present independently of the choice of the dimensions. In the case of odd dimensions $d=2n+1$, $1/((u-v)\e)$ is the only spurious pole that, once the expansion of the coefficients \eqref{expans} is taken in account, can be removed by the choice
\begin{equation}
	\begin{split}
		\a_5^{(0)}&=-2d^2\,\a_2^{(0)}\\
		\a_5^{(1)}&=-2d\left(4v\,\a_2^{(0)}+d\a_2^{(1)}\right)+(u-v)\a_5^{\prime\,(1)}
	\end{split}\label{redef1}
\end{equation}
where $\a_5^{\prime(1)}$ redefines the undetermined constant,  that will be determined by the secondary CWI's. With this choice, the limit $u\to v$ is well defined and does not produce any divergence, leaving $A_5$ with its physical divergences expressed only in terms of $1/\e$.

Conversely, in the case of even spacetime dimensions $d=2n$, other divergences arise from the gamma function $\Gamma(-d/2+(u-3v)\e/2)$ for which the redefinition of the constants will be
\begin{equation}
	\begin{split}
		\a_5^{(0)}&=-2d^2\,\a_2^{(0)}\\
		\a_5^{(1)}&=-2d\left(4v\,\a_2^{(0)}+d\a_2^{(1)}\right)+(u-v)\a_5^{\prime\,(1)}\\
		\a_5^{(2)}&=-8v^2\,\a_2^{(0)}-8v\,d\,\a_2^{(1)}-2d^2\a_2^{(2)}+(u-v)\a_5^{\prime(2)}
	\end{split}\label{redef2}
\end{equation}
allowing the cancellation of the semilocal divergences. 
In the next appendix we study the secondary CWI's in the two cases of even and odd spacetime dimensions, in order to deal with the two different types of divergences that arise.  
The other type of divergence $(---)$, as previously mentioned, can be removed by adding counterterms constructed out of the metric in a covariant way. In particular, by inspection of \eqref{TTTdiv}, we list the possible cases in which the form factors manifest $(---)$ divergences as
\begin{equation}
	\begin{split}
		&A_1\hspace{1.3cm}\text{diverges for} \,d=6+2n,\\
		&A_2\hspace{1.3cm}\text{diverges for} \,d=4+2n,\\
		&A_3,\,A_4 \hspace{0.6cm}\text{diverge for} \,d=2+2n,\\
		&A_5 \hspace{1.3cm}\text{diverges for} \,d=2n,
	\end{split}\qquad n\in\mathbb{N} \label{divFormFactor}
\end{equation}
which clearly depend on the dimensions $d$. 
}
\subsection{The secondary CWI's for the TTT correlator}\label{appendixH}
{ We have previously discussed redefinitions of the constants in order to remove the semilocal divergences. In this section, we discuss the solution of the secondary CWI's in two different cases, for odd and for even spacetime dimensions respectively. We will be using the zero momentum limit in the correlators, in which the triple-K integral behaves as
\begin{equation}
	\lim_{p_3\to 0}I_{\alpha\{\beta_1,\beta_2,\beta_3\}}(p_1,p_2,p_3)=p^{\beta_t-\alpha-1}\ell_{\alpha\{\beta_1,\beta_2,\beta_3\}}\label{limit}
\end{equation}
where $\beta_t=\beta_1+\beta_2+\beta_3$ and with \eqref{ldef}. It is worth nothing that the three independent differential equations 
\begin{align}
	&\textup{L}_6\, A_2 + 2\textup{R}\bigl[\,2A_3-A_4(p_3\leftrightarrow p_1)\bigr]=0\\[1.5 ex] 
	&\textup{L}_4\, A_4(p_3\leftrightarrow p_2)  - 2\textup{R}A_5+2p_1^2\left[A_4(p_3\leftrightarrow p_1)-4A_3\right]=8d\,A_{TT}(p_3)\\[2 ex]  
	&\textup{L}_2\, A_3(p_3\leftrightarrow p_1) +p_1^2\left[A_4-A_4(p_3\leftrightarrow p_2)\right]=4d\,\big[A_{TT}(p_3)-A_{TT}(p_2)\big],
\end{align}
reduce the number of undetermined constants from five to two. By applying the approach previously illustrated for the $\braket{TT\mO}$ to this case, and with the redefinitions \eqref{redef1} or \eqref{redef2}, the secondary CWI's give the condition
\begin{align}
	\a_3^{(0)}&=-d\left[2(d+2)\a_1^{(0)}+\a_2^{(0)}\right]-\left(\frac{\pi}{2}\right)^{-\frac{3}{2}}\frac{2d(-1)^n\,c_T}{(d-2)!!},\\
	\a_4^{(0)}&=2\a_3^{(0)}+(3d+2)\a_2^{(0)},\\
	\a_5^{\prime(1)}&=-d\left[2d(d+2)\a_1^{(0)}+\frac{1}{2}(d+6)\a_2^{(0)}\right]-\left(\frac{\pi}{2}\right)^{-\frac{3}{2}}\frac{2d(-1)^{n}c_T}{(d-2)!!},
\end{align}
for odd spacetime dimensions $d=2n+1$, $(n=1,2,3,\dots)$, and 
\begin{align}
	\a_3^{(0)}&=-d\left[2(d+2)\a_1^{(0)}+\a_2^{(0)}\right]-\frac{2^{3-n}(-1)^n\,c_T}{(n-1)!},\\
	\a_4^{(0)}&=2\a_3^{(0)}+(3d+2)\a_2^{(0)},\\
	\a_3^{(1)}&=-2u\big(4(d+1)\a_1^{(0)}+\a_2^{(0)}\big)-d\big(2(d+2)\a_1^{(1)}+\a_2^{(1)}\big)+\frac{(-1)^{n}2^{3-n}u}{(n-1)!}\bigg[c_t(H_{n-1}+\ln 2-\g_e)-c_T^{(0)}\bigg],\\
	\a_4^{(1)}&=2a_3^{(1)}+6u\a_2^{(0)}+(3d+2)\a_2^{(1)},\\
	\a_5^{\prime\,(1)}&=-d\left[2d(d+2)\a_1^{(0)}+\frac{1}{2}(d+6)\a_2^{(0)}\right]-\left(\frac{\pi}{2}\right)^{-\frac{3}{2}}\frac{2^{4-n}(-1)^{n}c_T}{(n-1)!!},\\
	\a_5^{\prime\,(2)}&=-2u\bigg(2d(3d+4)\a_1^{(0)}+(d+3)\a_2^{(0)}\bigg)-d\bigg(2d(d+2)\a_1^{(1)}+\frac{d+6}{2}\a_2^{(1)}\bigg)\notag\\,
	&\hspace{1cm}+\frac{(-1)^n2^{4-n}nu}{(n-1)!}\bigg[c_T\big(H_{n-1}-\frac{1}{n}+\ln 2-\g_e\big)-c_T^{(0)}
	\bigg]
\end{align}
for even spacetime dimensions $d=2n$, $(n=2,3,\dots)$, as dicussed in \cite{Bzowski:2017poo}. In this case we have also expanded the contributions proportional to the constant $c_T$ of the $2$-point function, because of the appearance of divergences in the two point function. $H_n=\sum_{k=1}^n1 /k$ is the n-th harmonic number.

As a final consideration, by looking at the solutions of the secondary CWI's, we notice that the three-point function $\braket{TTT}$  both for even and odd dimensions ($d>2$) depends only on two undetermined constants. This property is the 
cornerstone to prove the correspondence between the general approach and the perturbative realizations in the next chapters.
}

\subsection{Solutions of the primary CWI's by an operatorial method: the solution for $A_1$}
\label{fuchs}
The hypergeometric character of the CWI's was recognized independently in \cite{Coriano:2013jba} and in \cite{Bzowski:2011ab}. Here we will briefly overview the derivation of such equations in the case of the $TTT$, discussing a direct method of solutions that we have developed for the $TJJ$ in \cite{Coriano:2018bbe} and that we are going to generalize.\\
This method exploits the universality of the Fuchsian points of such equations, a property which holds for all the 3-point functions. It is a general characteristic of the CWI's associated with such correlators, as we have verified in several cases. 
The solutions of such equations take a form given by the Appell function's product $F_4$ times $x$ and $y$ as given in \eqref{xy}, raised at specific powers $a,b$ (indices), which are universal. An overall extra factor of the momentum ($p_3$) raised to a specific power is introduced in such a way to give the correct scaling behaviour of the solution for each form factor $A_i$. \\
For each system (i.e.~each form factor), we first solve the homogeneous equation, determining the general solution. We then add to this a particular solution of the inhomogeneous equation. The latter is obtained by a split of the differential operator $K_{ij}$, which can be performed in various ways. 
The split that we adopt in this case is different from the one used in \cite{Coriano:2018bbe}. \\
Next, we are going to extend the analysis presented for the $TJJ$  to the $TTT$ using an alternative splitting of the hypergeometric operators $K_{i j}$ in order to deal with the more complex structure of the global system of equations which should be satisfied by the form factors. \\
We start from $A_1$ by solving the two equations from \eqref{PrimaryTTT}
\begin{equation}
K_{13}A_1=0   \qquad K_{23}A_1=0.
\end{equation}
In this case we introduce the ansatz 
\begin{equation}
A_1=p_3^{\Delta-2 d - 6}x^a y^b  F(x,y)
\end{equation}
and derive two hypergeometric equations as previously, which are characterised by the indices 
$(a_i, b_j)$. We obtain 
\begin{equation}
\label{A1}
A_1(p_1,p_2,p_3)=p_3^{\Delta-2 d - 6}\sum_{a,b} c^{(1)}(a,b,\vec{\Delta})\,x^a y^b \,F_4(\alpha(a,b) +3, \beta(a,b)+3; \gamma(a), \gamma'(b); x, y) 
\end{equation}
with the expression of $\alpha(a,b),\beta(a,b), \gamma(a), \gamma'(b)$ as given before
\begin{align}
\label{cons1}
\alpha(a,b)&= a + b + \frac{d}{2} -\frac{1}{2}(\Delta_2 - \Delta_3 +\Delta_1) \notag\\
\beta(a,b)&= a + b + d -\frac{1}{2}(\Delta_1 + \Delta_2 +\Delta_3) 
\end{align}
and 
\begin{align} 
\label{cons2}
\gamma(a)& =2 a +\frac{d}{2} -\Delta_1 + 1 \notag\\
\gamma'(b)&=2 b +\frac{d}{2} -\Delta_2 + 1 .
\end{align}

Expressing the values of the scaling dimensions $\D_1=\D_2=\D_3=d$, then 
\begin{align}
A_1(p_1,p_2,p_3)=p_3^{d - 6}\sum_{a,b} c^{(1)}(a,b)\,x^a y^b \,F_4(\alpha(a,b) +3, \beta(a,b)+3; \gamma(a), \gamma'(b); x, y) 
\end{align}
where now
\begin{align}
&a=0,\frac{d}{2}, &&b=0,\frac{d}{2},\notag\\
&	\alpha(a,b)= a + b,&&\beta(a,b)= a + b -\frac{d}{2},\notag\\
&\gamma(a) =2 a -\frac{d}{2} + 1,&&\gamma'(b)=2 b -\frac{d}{2}+ 1.
\end{align}

One can implement the symmetry condition on the $A_1$ form factor which has to be completely symmetric in the exchange of $(p_1,p_2,p_3)$. The three conditions 
\begin{align}
A_1(p_1,p_3,p_2)&=A_1(p_1,p_2,p_3)\notag\\
A_1(p_3,p_2,p_1)&=A_1(p_1,p_2,p_3)\notag\\
A_1(p_2,p_1,p_3)&=A_1(p_1,p_2,p_3)
\end{align}
constrain the coefficient $c^{(1)}(a,b)$ and in particular we obtain
\begin{align}
c^{(1)}\left(\frac{d}{2},0\right)&=c^{(1)}\left(0,\frac{d}{2}\right)\notag\\
c^{(1)}(0,0)&=-\frac{(d-4) (d-2)}{(d+2) (d+4)}c^{(1)}\left(0,\frac{d}{2}\right)\notag\\
c^{(1)}\left(\frac{d}{2},\frac{d}{2}\right)&=\frac{\Gamma\left(-\frac{d}{2}\right)\Gamma\left(d+3\right)}{2\,\Gamma\left(\frac{d}{2}\right)} c^{(1)}\left(0,\frac{d}{2}\right)\label{CondA1}
\end{align}
generating a solution which depends only on one arbitrary constant that we identify as $C_1$ 
\begin{equation}
c^{(1)}\left(0,\frac{d}{2}\right)=C_1.\label{oneconst}
\end{equation}
\subsection{The solution for $A_2$ and the operatorial shifts}
The equation for $A_2$ is inhomogeneous, but the solution can be identified using some properties of the hypergeometric forms of such equations. We recall that in this case they are 
\begin{align}
K_{13}A_2 &= 8 A_1\label{inhom}\\\
K_{23}A_2 &= 8A_1.\label{inhom1}
\end{align}
The ansatz which is in agreement with the scaling behaviour of $A_2$  in this case is 
\begin{equation}
A_2(p_1,p_2,p_3)=p_3^{d - 4}\,x^a\,y^b\,F(x,y).
\end{equation}
At this stage we proceed with the splitting. We observe that the action of $K_{13}$ and $K_{23}$ on $A_2$ can be rearranged as follows
\begin{align}
K_{13} A_2&=4 x^a y^b p_3^{d -6}\bigg( \bar{K}_{13}F(x,y) +x\frac{\partial}{\partial x} F(x,y)+y\frac{\partial}{\partial y} F(x,y)+\bar{\b}F(x,y)\bigg)\label{rep}\\[1.5ex]
K_{23} A_2&=4 x^a y^b p_3^{d -6}\bigg( \bar{K}_{23}F(x,y)  +x\frac{\partial}{\partial x} F(x,y)+y\frac{\partial}{\partial y} F(x,y)+\bar{\b}F(x,y)\bigg)\label{rep2}
\end{align}
where
\begin{align}
\label{k1bar}
\bar{K}_{13}F(x,y)&=\bigg\{x(1-x) \frac{\partial^2}{\partial x^2} - y^2 \frac{\partial^2}{\partial y^2} - 2 \, x \, y \frac{\partial^2}{\partial x \partial y} +\big[  \gamma(a)- (\tilde\alpha(a,b) + \bar\beta(a,b) + 1) x \big] \frac{\partial}{\partial x} \notag\\
&\hspace{5cm}+ \frac{a (a-a_1)}{x} - (\tilde\alpha(a,b) + \bar\beta(a,b) + 1) y \frac{\partial}{\partial y}  - \tilde\alpha(a,b) \, \bar\beta (a,b)\bigg\} F(x,y)
\end{align}
and 
\begin{align}
\label{k2bar}
\bar{K}_{23} A_2&=\bigg\{ y(1-y) \frac{\partial^2}{\partial y^2} - x^2 \frac{\partial^2}{\partial x^2} - 2 \, x \, y \frac{\partial^2}{\partial x \partial y} +  \big[ \gamma'(b) - (\tilde\alpha(a,b) +\bar \beta(a,b) + 1) y \big] \frac{\partial}{\partial y}\notag\\
&\hspace{5cm}+  \frac{b(b- b_1)}{y} - (\tilde\alpha(a,b) + \bar\beta(a,b) + 1) x \frac{\partial}{\partial x}  - \tilde\alpha(a,b) \, \bar\beta(a,b) \bigg\} F(x,y)
\end{align}
with 
\begin{equation}
\tilde\alpha(a,b)=\alpha(a,b) + 3 \qquad \bar\beta(a,b)=\beta(a,b) +2
\end{equation}
At this point we notice that the hypergeometric function  that satisfy the system of equations
\begin{equation}
\left\{
\begin{split}
	\bar{K}_{23}F(x,y)&=0\\
	\bar{K}_{13}F(x,y)&=0\\
\end{split}
\right.
\end{equation}
can be taken of the form
\begin{equation}
\label{rep1}
\Phi_1^{(2)}(x,y)=\sum_{a,b} c^{(2)}_1(a,b)\,x^a y^b \,F_4(\alpha(a,b) +3, \beta(a,b)+2; \gamma(a), \gamma'(b) ; x, y)
\end{equation}
with $c^{(2)}_1$ constant depending on the parameters $a,b$ fixed at the ordinary values $(a_0,b_0), (a_1,b_0), (a_0,b_1)$ and $(a_1,b_1)$ as in the previous cases (\ref{cond1}) and (\ref{cond2}). 
The convention that we adopt on the indices appearing on the constants is as follows.\\
The superscript $(i)$ on the constant $c^{(i)}$, is the index of the corresponding form factors and it is used in the homogeneous solution of the corresponding set of equations. On the other hand, the subscript $j$, in the constant $c^{(i)}_j$ instead, specifies the particular (inhomogeneous) solution of the same system of equations for the form factor $A_i$.\\
For instance, if one considers the particular solution $\Phi_1^{(2)}$ in \eqref{rep1}, the constant $c_1^{(2)}$ is well defined using this convention. In fact it tells us that this solution is the first particular solution of the inhomogeneous set of equations for the $A_2$ form factors. It is worth mentioning that all these constants will be fixed, at the end, just in terms of the homogeneous ones that don't carry any subscript. \\
As previously remarked, the values of the exponents $a$ and $b$ remain the same for any equation involving either a $K_{i,j}$ or a $\bar{K}_{i j}$, as one can verify. \\
At this point, to show that $\Phi_1^{(2)}$  is a solution of  Eqs. (\ref{inhom}) we use the property 
\begin{equation}
\frac{\partial^{p+q} F_4(a,b;c_1,c_2;x,y)}{\partial x^p\partial y^q} =\frac{(a)_{p+q}(b)_{p+q}}{(c_1)_{p}(c_2)_{q}}
F_4(a + p + q,b + p + q; c_1 + p ; c_2 + q;x,y)
\end{equation}
using the Pochammer symbol previously defined, from which one derives the simpler relations 
\begin{align}
& \frac{\partial F_4(a,b;c_1,c_2;x,y)}{\partial x} =\frac{a b}{c_1}F_4(a+1,b+1,c_1+1,c_2,x,y) \notag\\
&  \frac{\partial F_4(a,b;c_1,c_2;x,y)}{\partial y} =\frac{a b}{c_2}F_4(a+1,b+1,c_1,c_2 +1,x,y).
\end{align} 
We will be also using the known relation on the shift of one parameter of $F_4$
\begin{equation}
\label{exs}
F_4(a,b,c_1-1,c_2;x,y)=F_4(a,b,c_1,c_2,x,y)+x\frac{\partial}{\partial\,x}\,F_4(a,b,c_1,c_2,x,y)
\end{equation}
that leads to the identity
\begin{equation}
x\,F_4(a+1,b+1,c_1,c_2;x,y)=\frac{(c_1-1)(c_1-2)}{a\,b}\bigg[F_4(a,b,c_1-2,c_2,x,y)-F_4(a,b,c_1-1,c_2,x,y)\bigg],
\end{equation}
and furthermore the symmetry relation 
\begin{align}
\label{transfF4}
F_4(\alpha, \beta; \gamma, \gamma'; x, y) &= \frac{\Gamma(\gamma') \Gamma(\beta - \alpha)}{ \Gamma(\gamma' - \alpha) \Gamma(\beta)} (- y)^{- \alpha} \, F_4\left(\alpha, \alpha -\gamma' +1; \gamma, \alpha-\beta +1; \frac{x}{y}, \frac{1}{y}\right) \notag\\
&\qquad+  \frac{\Gamma(\gamma') \Gamma(\alpha - \beta)}{ \Gamma(\gamma' - \beta) \Gamma(\alpha)} (- y)^{- \beta} \, F_4\left(\beta -\gamma' +1, \beta ; \gamma, \beta-\alpha +1; \frac{x}{y}, \frac{1}{y}\right) \,.
\end{align}
already used in \cite{Coriano:2013jba} in the analysis of a scalar case, in order to impose the symmetry under the exchange of two of the three momenta. 
All these relations can be used in order to consider the action of $K_{13}$ and $K_{23}$ on the the $\Phi_2^{(1)}$  in (\ref{rep}) and in \eqref{rep2}, obtaining 
\begin{align}
&K_{13}\Phi_1^{(2)}(x,y) = 4p_3^{d -6} \sum_{a,b} c^{(2)}_1(a,b)\,x^a y^b\bigg(x\frac{\partial}{\partial x} +y\frac{\partial}{\partial y} +(\b+2)\bigg)\,F_4(\alpha +3, \beta+2; \gamma, \gamma' ; x, y)   \notag\\
&= 4 p_3^{d -6} \sum_{a,b} c^{(1)}_2(a,b)\,x^a y^b\bigg[ (\alpha+3)\bigg(x\frac{(\beta+2)}{\gamma}F_4(\alpha +4, \beta+3; \gamma+1, \gamma'; x, y)\notag\\
&\hspace{1cm}+y\frac{(\beta+2)}{\gamma'}F_4(\alpha +4, \beta+3; \gamma, \gamma'+1; x, y)\bigg)+(\b+2)\,F_4(\alpha +3, \beta+2; \gamma, \gamma'; x, y) \bigg]
\end{align}
where, for simplicity, we have denoted with $\a=\a(a,b)$ and $\b=\b(a,b)$. At this point, using the following properties of  hypergeometric functions \cite{Appell}
\begin{align}
\frac{b}{c_1}\,x\,F_4(a+1,b+1,c_1+1,c_2,x,y)+\frac{b}{c_2}\,y\,F_4(a+1,b+1,c_1,c_2+1,x,y)&=F_4(a+1,b,c_1,c_2,x,y)\notag\\
-F_4(a,b,c_1,c_2,x,y)
aF_4(a+1,b,c_1,c_2,x,y)-bF_4(a,b+1,c_1,c_2,x,y)&=(a-b)F_4(a,b,c_1,c_2,x,y)
\end{align}
after some algebra, it is simple to verify that 
\begin{align}
K_{13}\Phi_1^{(2)}(x,y) &=4p_3^{ d -6} \sum_{a,b} c^{(2)}_1(a,b)\,x^a y^b\big(\b(a,b)+2\big)F_4(\a+3,\b+3,\g,\g',x,y)
\end{align}
and in the same way
\begin{align}
&K_{23}\Phi_1^{(2)}(x,y)= 4p_3^{ d -6} \sum_{a,b} c^{(2)}_1(a,b)\,x^a y^b\big(\b(a,b)+2\big)F_4(\a+3,\b+3,\g,\g',x,y).
\end{align}
The non-zero right-hand-side of both equations are proportional to the form factor $A_1$ given in (\ref{A1}). Once this particular solution is determined, \eqref{A1}, by comparison, gives the conditions on $c_1^{(2)}$ as
\begin{align}
c_1^{(2)}(a,b)&=\frac{2}{\big(\b(a,b)+2\big)}\,c^{(1)}(a,b)\label{condc2}.
\end{align}
Notice that the coefficient $c_1^{(2)}$ of the first particular solution of the inhomogeneous set of equations for $A_2$ is fixed in terms of the coefficient $c^{(1)}$ of homogeneous one of $A_1$.

Finally, we obtain the general solution for $A_2$ in the $TTT$ case (in which $\g=\g'$ ) by superposing the solution of the homogeneous 
system generated by \eqref{inhom} and \eqref{inhom1} and the inhomogeneous one \eqref{rep1}, with a condition on the constants given by \eqref{condc2}. Therefore, the general expression of the solution for $A_2$ is given by
\begin{align}
A_2&= p_3^{d - 4}\sum_{a b} x^a y^b\left[c^{(2)}(a,b)\,F_4(\alpha +2, \beta+2; \gamma, \gamma'; x, y)+ \frac{2\,c^{(1)}(a,b)}{\big(\b+2\big)}F_4(\alpha +3, \beta+2; \gamma, \gamma'; x, y)\right]\label{A2}
\end{align}
where also in this case we have used a short-hand notation $\a=\a(a,b)$, $\b=\b(a,b)$, $\g=\Gamma(a)$ and $\g'=\g'(b)$. \\
Let's discuss now the symmetry properties of the $A_2$ form factors. The latter in fact is symmetric under the exchange $p_1\leftrightarrow p_2$, and this condition has to be implemented in the form
\begin{equation}
A_2(p_2,p_1,p_3)=A_2(p_1,p_2,p_3).
\end{equation}
Using the properties of the hypergeometric functions previously written, such symmetry constraint relates  $c^{(1)}$ and $c^{(2)}$ for the 4 indices $a,b$ which label the homogeneous solutions in the form  
\begin{equation}
\label{CondA2TTT}
\begin{split}
	c^{(2)}\left(\frac{d}{2},0\right)&=c^{(2)}\left(0,\frac{d}{2}\right)\\
	c^{(1)}\left(\frac{d}{2},0\right)&=c^{(1)}\left(0,\frac{d}{2}\right)
\end{split}
\end{equation}
Notice that of the two equations above, the second is redundant since it is already present as a symmetry condition on $A_1$, as clear from \eqref{CondA1}. Only the first condition on $c^{(2)}$ is new. \\
We have already established that $A_1$ can be written in terms of only a single constant $C_1$, as evident from \eqref{CondA1} and \eqref{oneconst}. From the expression of $A_2$ in \eqref{A2} and using the property \eqref{CondA2TTT}, we can deduce that so far this form factors can be written in terms of four constants: $C_1$, $c^{(2)}\left(0,d/2\right)$, $c^{(2)}\left(d/2,d/2\right)$, $c^{(2)}\left(0,0\right)$. We will see that the symmetry condition on $A_5$ will put additional constraint on the coefficients of $A_2$, by allowing us to write this form factors in terms of only two independent constants. At the end, we will see that this iterative method will allow to identify a rather small set of independent  constants for each form factor and the entire solution.

\subsection{The solution for $A_3$}
Also in this case the system of two equations is  inhomogeneous 
\begin{equation}
\left\{
\begin{split}
	K_{13} A_3 &=2A_2\\ 
	K_{23} A_3&=2A_2 
\end{split}\right.
\label{eqA3}
\end{equation}
Using the same strategy of the previous section,  it is possible to find two particular solutions of such system using an operatorial split as above
\begin{align}
\Phi_1^{(3)}(x,y)&= p_3^{d -2} \sum_{a b} c_1^{(3)}(a,b)\, x^a y^b   F_4(\alpha+2,\beta+1; \gamma,\gamma';x,y)\\
\Phi_2^{(3)}(x,y)&= p_3^{ d -2} \sum_{a b} c_2^{(3)}(a,b) \,x^a y^b   F_4(\alpha+3,\beta+1; \gamma,\gamma';x,y)
\end{align}
and the action of $K_{13}$ and $K_{23}$ on them are respectively
\begin{align}
&\left\{
\begin{matrix}
	K_{13}\Phi^{(3)}_1=4p_3^{d-4}\sum_{a b} x^a y^b\, c_1^{(3)}(a,b)\,(\b+1)\,F_4(\alpha+2,\beta+2,\gamma,\gamma',x,y)\\[1.5ex]
	K_{23}\Phi^{(3)}_1=4p_3^{d-4}\sum_{a b} x^a y^b\, c_1^{(3)}(a,b)\,(\b+1)\,F_4(\alpha+2,\beta+2,\gamma,\gamma',x,y)
\end{matrix}\right.\\[2.2ex]
&\left\{\begin{matrix}
	K_{13}\Phi^{(3)}_2=8p_3^{d-4}\sum_{a b} x^a y^b\, c_2^{(3)}(a,b)\,(\b+1)\,F_4(\alpha+3,\beta+2,\gamma,\gamma',x,y)\\[1.5ex]
	K_{23}\Phi^{(3)}_2=8p_3^{d-4}\sum_{a b} x^a y^b\, c_2^{(3)}(a,b)\,(\b+1)\,F_4(\alpha+3,\beta+2,\gamma,\gamma',x,y).
\end{matrix}\right.
\end{align}
These equations have to be equal to the right hand side of \eqref{eqA3}, and this condition fixes the integration constants to be those appearing in $A_2$ as
\begin{align}
c_1^{(3)}(a,b)&=\frac{1}{2(\b+1)}\,c^{(2)}(a,b)\\
c_2^{(3)}(a,b)&=\frac{1}{2(\b+1)(\b+2)}\,c^{(1)}(a,b).
\end{align}
The general solution for $A_3$ can be obtained by adding to the particular solution above the general solution of the homogeneous system \eqref{eqA3}, for which
\begin{align}
A_3&=p_3^{d-2}\,\sum_{ab}\,x^a\,y^b\Big[c^{(3)}(a,b)\,F_4(\a+1,\b+1,\g,\g';x,y)\notag\\
&\hspace{1.5cm}+\frac{1}{2(\b+1)}\,c^{(2)}(a,b)\,F_4(\a+2,\b+1,\g,\g';x,y)\notag\\
&\hspace{1.5cm}+\frac{1}{2(\b+1)(\b+2)}\,c^{(1)}(a,b)F_4(\alpha+3,\beta+1,\gamma,\gamma',x,y)
\Big].
\end{align}
Imposing the symmetry condition on $A_3$ under the change $p_1\leftrightarrow p_2$
\begin{equation}
A_3(p_2,p_1,p_3)=A_3(p_1,p_2,p_3)
\end{equation}
we obtain additional constraints on the homogeneous coefficients $c^{(i)}(a,b)$, $i=1,2,3$ as
\begin{equation}
\label{CondA3TTT}
\begin{split}
	c^{(3)}\left(\frac{d}{2},0\right)=c^{(3)}\left(0,\frac{d}{2}\right)\\
	c^{(2)}\left(\frac{d}{2},0\right)=c^{(2)}\left(0,\frac{d}{2}\right)\\
	c^{(1)}\left(\frac{d}{2},0\right)=c^{(1)}\left(0,\frac{d}{2}\right).
\end{split}
\end{equation}
We observe that the last two conditions are already satisfied by the solutions of $A_1$ and $A_2$ and the new information follows from the first of the equations in \eqref{CondA3TTT}. At this stage the independent constants appearing in $A_3$ are seven, but this number will be reduced to three once that the symmetry condition on $A_5$ will be also taken into account. 
\subsection{The solution for $A_4$}\label{A4section}
The solutions by our method for the $A_4$ and $A_5$ form factors require a special treatment, due to exchanged momenta on the functional dependence of the form factors on the right hand side of the respective equations. This complication is not present in the case of the $TJJ$ \cite{Coriano:2018bbe}.
In particular, the primary WI's
\begin{equation}
\left\{\begin{split}
	K_{13} A_4 &=-4 A_2(p_2\leftrightarrow p_3)\\
	K_{23} A_4&=-4 A_2 (p_1\leftrightarrow p_3)
\end{split}
\right.\label{eqA4}
\end{equation}
involve the symmetrization $A_2(p_2\leftrightarrow p_3)$, that can be obtained from \eqref{A2} with the exchange $(p_2,\D_2)\leftrightarrow (p_3,\D_3)$ and the replacements
\begin{equation}
\begin{split}
	&x\to \tilde x=\sdfrac{x}{y},\qquad y\to \tilde y=\sdfrac{1}{y},\\
	&\a(a,b)\to \tilde{\a}(a,b)=a+b+\sdfrac{d}{2}-\sdfrac{\D_1-\D_2+\D_3}{2}=\a(a,b)-(\D_2-\D_3)\\
	&\b(a,b)\to \tilde\b(a,b)=a+b+d-\sdfrac{\D_1+\D_2+\D_3}{2}=\b(a,b)\\
	&\Gamma(a)\to \tilde\Gamma(a)=2a+\sdfrac{d}{2}-\D_1+1=\Gamma(a)\\
	&\g'(b)\to \tilde\g'(b)=2b+\sdfrac{d}{2}-\D_3+1=\g'(b)-(\D_2-\D_3)\\
\end{split}
\end{equation}
in the basic solution $F_4$. In the $TTT$ case, inserting the corresponding scaling dimensions, one has $\tilde\a(a,b)=\a(a,b)$ and $\tilde\g'(a,b)=\g'(a,b)$. \\
The two particular solutions of the inhomogeneous equations \eqref{eqA4} can be expressed in the form
\begin{equation}
\begin{split}
	\Phi_1^{(4)}&=p_3^{d -2} \sum_{a b} c_1^{(4)}(a,b)\, x^a y^b   F_4(\alpha+2,\beta+1; \gamma,\gamma';x,y)\\
	\Phi_2^{(4)}&=p_3^{d -2} \sum_{a b} c_2^{(4)}(a,b)\, x^a y^b   F_4(\alpha+1,\beta+1; \gamma-1,\gamma'-1;x,y)
\end{split}\label{solA4}
\end{equation}
where the action of $K_{13}, K_{23}$ on them gives
\begin{align}
&\left\{\begin{matrix}
	K_{13}\Phi_1^{(4)}=4p_3^{d-4}\sum_{a b} x^a y^b\, c_1^{(4)}(a,b)\,(\b+1)\,F_4(\alpha+2,\beta+2,\gamma,\gamma',x,y)\\[1.5ex]
	K_{23}\Phi_1^{(4)}=4p_3^{d-4}\sum_{a b} x^a y^b\, c_1^{(4)}(a,b)\,(\b+1)\,F_4(\alpha+2,\beta+2,\gamma,\gamma',x,y)
\end{matrix}\right.\\[2.2ex]
&\left\{\begin{matrix}
	K_{13}\Phi_2^{(4)}=4p_3^{d -4} \sum_{a b} c_2^{(4)}(a,b)\,\frac{(\a+1)(\b+1)}{(\g-1)} \,x^a y^b   F_4(\alpha+2,\beta+2; \gamma,\gamma'-1;x,y)\\[1.5ex]
	K_{23}\Phi_2^{(4)}=4p_3^{d -4} \sum_{a b} c_2^{(4)}(a,b)\,\frac{(\a+1)(\b+1)}{(\g'-1)} \,x^a y^b   F_4(\alpha+2,\beta+2; \gamma-1,\gamma';x,y).
\end{matrix}\right.
\end{align}
Writing the previous expressions explicitly for the four fundamental indices $a=0,\,d/2$ and $b=0,\,d/2$, one can compare the two solutions in order to extract information about the corresponding constants introduced in \eqref{solA4}.  
We relate the various constants using the intermediate steps worked out in \cite{Coriano:2018bbe, Coriano:2018bsy}, to which we refer for further details.\\ 
The general solution to \eqref{eqA4} is obtained by adding such particular solution to the general homogenous one, and can be written in the form
\begin{align}
A_4&=p_3^{d-2}\sum_{ab}x^ay^b\Bigg\{c^{(4)}(a,b)\,F_4(\a+1,\b+1,\g,\g',x,y)\notag\\
&+c^{(4)}_1(a,b)F_4(\a+2,\b+1,\g,\g',x,y)+\,c^{(4)}_2(a,b)F_4(\a+1,\b+1,\g-1,\g'-1,x,y)\notag\\
&+c^{(4)}_3(a,b)F_4(\a+1,\b+1,\g-1,\g',x,y)+\,c^{(4)}_4(a,b)F_4(\a+1,\b+1,\g,\g'-1,x,y)\Bigg\}
\end{align}
with the constants $c^{(4)}_i,\ i=1,2,3,4$ given in terms of $c^{(1)}(a,b)$ and $c^{(2)}(a,b)$ once we enforce the symmetry constraints, and with $\a=\a(a,b)$, $\b=\b(a,b)$, $\g=\Gamma(a)$ and $\g'=\g'(b)$, for simplicity.
The form factor $A_4$ is symmetric under the exchange $p_1\leftrightarrow p_2$
\begin{equation}
A_4(p_1,p_2,p_3)=A_4(p_2,p_1,p_3)
\end{equation}
and leads to the conditions \eqref{CondA1}, \eqref{CondA2TTT}, \eqref{CondA3TTT} and to
\begin{align}
&c^{(4)}\left(\frac{d}{2},0\right)=c^{(4)}\left(0,\frac{d}{2}\right),&&c_3^{(4)}\left(0,0\right)=c_4^{(4)}\left(0,0\right)\notag\\
&c_3^{(4)}\left(\frac{d}{2},\frac{d}{2}\right)=c_4^{(4)}\left(\frac{d}{2},\frac{d}{2}\right)&&c_3^{(4)}\left(\frac{d}{2},0\right)=c_4^{(4)}\left(0,\frac{d}{2}\right)\notag\\
&c_3^{(4)}\left(0,\frac{d}{2}\right)=c_4^{(4)}\left(\frac{d}{2},0\right)
\end{align}
\begin{align}
c^{(4)}_3\left(\frac{d}{2},0\right)&=\frac{d^2}{ (d+2) (d+4)} c^{(1)}\left(0,\frac{d}{2}\right)+\frac{d}{(d-2) } c^{(2)}(0,0)-\frac{d}{(d+2)} c^{(2)}\left(0,\frac{d}{2}\right)- c_3^{(4)}\left(0,\frac{d}{2}\right).
\end{align}

Using the relation given in \appref{fuchsTTT}, the general solution can be parameterized as
\begin{align}
&\begin{aligned}[c]
	c_2^{(4)}\left(0,0\right)&=-\frac{d^2}{(d+2)(d+4)}C_1\\
	c_2^{(4)}\left(0,\frac{d}{2}\right)&=-\frac{d^2}{(d+2)(d+4)}C_1\\
	c_2^{(4)}\left(\frac{d}{2},0\right)&=c_2^{(4)}\left(0,\frac{d}{2}\right)\\
	c_2^{(4)}\left(\frac{d}{2},\frac{d}{2}\right)&=\frac{d\sec\left(\frac{\pi\,d}{2}\right)\,\Gamma\left(1-\frac{d}{2}\right)^2}{(d+2)(d+4)\,\Gamma(-d)}\,C_1
\end{aligned}
\label{C41}
\end{align}

\subsection{The $A_5$ solution}
Also in the case of $A_5$ we have to repeat the approach presented in \secref{A4section}. In particular the primary Ward identities for $A_5$ is given by
\begin{equation}
\left\{
\begin{split}
	K_{13} A_5 &=2[A_4- A_4(p_1\leftrightarrow p_3)] \\
	K_{23} A_5&=2[A_4- A_4 (p_2\leftrightarrow p_3)]
\end{split}\right.
\label{eqA5}
\end{equation}
and this system of equations admit seven particular solutions. Combined with the homogeneous solution they give
\begin{align}
A_5&=p_3^{d}\sum_{ab}x^ay^b\Bigg\{\frac{1}{\b}\bigg[+c_1^{(5)}(a,b)\,F_4(\a+1,\b,\g-1,\g'-1,x,y)\notag\\
&\hspace{2cm}+c^{(5)}_2(a,b)F_4(\a+1,\b,\g,\g'-1,x,y)+\,c^{(5)}_3(a,b)F_4(\a+1,\b,\g,\g',x,y)\notag\\
&\hspace{-0.7cm}+c^{(5)}_4(a,b)F_4(\a+1,\b,\g-1,\g',x,y)+c^{(5)}_5(a,b)F_4(\a,\b,\g-1,\g',x,y)+c^{(5)}_6(a,b)F_4(\a,\b,\g,\g'-1,x,y)\bigg]\notag\\
&+\frac{1}{\a\,\b}\bigg[c^{(5)}_7(a,b)F_4(\a,\b,\g-1,\g'-1,x,y)+c^{(5)}(a,b)\,F_4(\a,\b,\g,\g',x,y)\bigg]\Bigg\}. 
\end{align}
In particular the coefficients $c^{(5)}_i$, $i=1,\dots,4$ are fixed by the use \eqref{eqA5}, and imposing the symmetry conditions on $A_5$
\begin{align}
A_5(p_3,p_2,p_1)&=A_5(p_1,p_2,p_3) \notag\\
A_5(p_2,p_1,p_3)&=A_5(p_1,p_2,p_3) \notag\\
A_5(p_1,p_3,p_2)&=A_5(p_1,p_2,p_3).
\end{align}
We have left to \appref{conditionTTT} more details on the identification of the independent constants which characterize this solution and the analogous solution for $A_4$. There are 5 constants overall for the system of primary WI's, in agreement with the result presented in  \cite{Bzowski:2013sza}, which reduce to 3 after imposing the constraints derived by secondary WI's. Such additional reduction can be performed as discussed in \cite{Bzowski:2013sza}. 
\subsection{Summary}
In this section we will briefly summarize the final solutions obtained for all the form factors. 
We obtain \\
\begin{equation}
A_1=p_3^{d - 6}\sum_{a,b} C_1\,f_1(a,b)\,x^a y^b \,F_4(\alpha(a,b) +3, \beta(a,b)+3; \gamma(a), \gamma'(b); x, y),
\end{equation}
\begin{align}
f_1\left(0,\frac{d}{2}\right)=f_1\left(\frac{d}{2},0\right)=1,\quad
f_1(0,0)=-\frac{(d-4) (d-2)}{(d+2) (d+4)},\quad
f_1\left(\frac{d}{2},\frac{d}{2}\right)=\frac{\Gamma\left(-\frac{d}{2}\right)\Gamma\left(d+3\right)}{2\,\Gamma\left(\frac{d}{2}\right)}
\label{CondA2}
\end{align}
where $f_1(a,b)$ takes four values for the four Fuchsian indices. In this case the function $f_1$ can be read from the expressions \eqref{CondA1};
\begin{align}
A_2&= p_3^{d - 4}\sum_{a b} x^a y^b\bigg[C_2\,f_2(a,b)\,F_4(\alpha +2, \beta+2; \gamma, \gamma'; x, y)+ \frac{2\,C_1}{\big(\b+2\big)}\,f_1(a,b)\,F_4(\alpha +3, \beta+2; \gamma, \gamma'; x, y)\bigg].
\end{align}
\begin{align} 
f_2\left(0,0\right)=\frac{d-2}{d+2},\quad f_2\left(\frac{d}{2},0\right)=f_2\left(0,\frac{d}{2}\right)=1,\quad
f_2\left(\frac{d}{2},\frac{d}{2}\right)2=\frac{\Gamma(-d/2)\Gamma(d+2)}{\Gamma(d/2)}
\end{align}
In the same way we write the explicit form of $A_3$ using the results in \appref{conditionTTT} as
\begin{align}
A_3&=p_3^{d-2}\,\sum_{ab}\,x^a\,y^b\Big[C_3\,f_3(a,b,d)\,F_4(\a+1,\b+1,\g,\g';x,y)+\frac{C_2}{2(\b+1)}\,f_2(a,b,d)\,F_4(\a+2,\b+1,\g,\g';x,y)\notag\\
&\hspace{5.3cm}+\frac{C_1}{2(\b+1)(\b+2)}\,f_1(a,b,d)F_4(\alpha+3,\beta+1,\gamma,\gamma',x,y)
\Big],
\end{align}
where 
\begin{align} 
f_3\left(\frac{d}{2},0\right)=f_3\left(0,\frac{d}{2}\right)=1,\quad f_3\left(0,0\right)=-1,\quad 
f_3\left(\frac{d}{2},\frac{d}{2}\right)=\frac{\Gamma(-d/2)\Gamma(d+1)}{\Gamma(d/2)}
\end{align} 
\begin{align}
A_4&=p_3^{d-2}\sum_{ab}x^ay^b\Bigg\{c^{(4)}(a,b)\,F_4(\a+1,\b+1,\g,\g',x,y)\notag\\
&\hspace{3cm}+c^{(4)}_1(a,b)F_4(\a+2,\b+1,\g,\g',x,y)+\,c^{(4)}_2(a,b)F_4(\a+1,\b+1,\g-1,\g'-1,x,y)\notag\\
&\hspace{3cm}+c^{(4)}_3(a,b)F_4(\a+1,\b+1,\g-1,\g',x,y)+\,c^{(4)}_4(a,b)F_4(\a+1,\b+1,\g,\g'-1,x,y)\Bigg\}
\end{align}
\begin{align}
&c^{(4)}\left(\frac{d}{2},0\right)=c^{(4)}\left(0,\frac{d}{2}\right)=C_4,\quad
c^{(4)}\left(0,0\right)=-C_4,\notag\\
&c^{(4)}\left(\frac{d}{2},\frac{d}{2}\right)=\frac{\Gamma\left(-\frac{d}{2}\right)\Gamma(d+1)}{\Gamma\left(\frac{d}{2}\right)}C_4-\frac{d^2\,\Gamma\left(-\frac{d}{2}\right)\Gamma(d+2)}{(d+1)(d+2)(d+4)\Gamma\left(\frac{d}{2}\right)}C_1
\end{align}
\begin{align}
c_1^{(4)}\left(\frac{d}{2},0\right)=c^{(4)}_1\left(0,\frac{d}{2}\right)=-C_2,\quad
c_1^{(4)}\left(0,0\right)=\frac{2}{(d+2)}\,C_2,\quad c_1^{(4)}\left(\frac{d}{2},\frac{d}{2}\right)=-\frac{2\,\Gamma\left(-\frac{d}{2}\right)\,\Gamma\left(d+2\right)}{(d+2)\,\Gamma\left(\frac{d}{2}\right)}C_2
\end{align}
\begin{align}
&c_2^{(4)}\left(0,0\right)=-\frac{d^2}{(d+2)(d+4)}C_1,\quad
c_2^{(4)}\left(0,\frac{d}{2}\right)=c_2^{(4)}\left(0,\frac{d}{2}\right)=-\frac{d^2}{(d+2)(d+4)}C_1,\notag\\
&c_2^{(4)}\left(\frac{d}{2},\frac{d}{2}\right)=-\frac{d^2\,\Gamma\left(d+2\right)\Gamma\left(-\frac{d}{2}\right)}{(d+1)(d+2)(d+4)\,\Gamma\left(\frac{d}{2}\right)}\,C_1,
\end{align}
\begin{align}
&c_3^{(4)}\left(0,0\right)=c_4^{(4)}\left(0,0\right)=0,&&	c_3^{(4)}\left(0,\frac{d}{2}\right)=c_4^{(4)}\left(\frac{d}{2},0\right)=\frac{d^2}{(d+2)(d+4)}C_1,\notag\\
&c_3^{(4)}\left(\frac{d}{2},0\right)=c_4^{(4)}\left(0,\frac{d}{2}\right)=0,&&
c_3^{(4)}\left(\frac{d}{2},\frac{d}{2}\right)=c_4^{(4)}\left(\frac{d}{2},\frac{d}{2}\right)=\frac{d^2\,\Gamma\left(d+2\right)\Gamma\left(-\frac{d}{2}\right)}{(d+1)(d+2)(d+4)\,\Gamma\left(\frac{d}{2}\right)}\,C_1.
\end{align}
Finally we give the form of the $A_5$ form factors as
\begin{align}
A_5&=p_3^{d}\sum_{ab}x^ay^b\Bigg\{\frac{1}{\b}\bigg[+c_1^{(5)}(a,b)\,F_4(\a+1,\b,\g-1,\g'-1,x,y)\notag\\
&\hspace{2cm}+c^{(5)}_2(a,b)F_4(\a+1,\b,\g,\g'-1,x,y)+\,c^{(5)}_3(a,b)F_4(\a+1,\b,\g,\g',x,y)\notag\\
&\hspace{-0.5cm}+\,c^{(5)}_4(a,b)F_4(\a+1,\b,\g-1,\g',x,y)+c^{(5)}_5(a,b)F_4(\a,\b,\g-1,\g',x,y)+\,c^{(5)}_6(a,b)F_4(\a,\b,\g,\g'-1,x,y)\bigg]\notag\\
&+\frac{1}{\a\,\b}\bigg[c^{(5)}_7(a,b)F_4(\a,\b,\g-1,\g'-1,x,y)+c^{(5)}(a,b)\,F_4(\a,\b,\g,\g',x,y)\bigg]\Bigg\}. 
\end{align}
where the coefficients are summarized in Appendix \ref{conditionTTT}. The global solution is fixed up to five independent constants. 

\section{Conclusions}

In this chapter, we have presented the general method to construct any 3-point functions involving tensorial operators, as first presented by \cite{Bzowski:2013sza}. Starting from the worked out example of the $TTO$ correlation function, we have moved our attention to constructing the general solution of the correlators $TJJ$ and $TTT$. We have shown that these correlators are completely fixed modulo some constants. We have presented two equivalent methods to solve the CWI's in terms of $3$K integrals and hypergeometric structure. As we have already pointed out, these two methods are in complete agreement with each other as shown by the final solution.\\
In the next chapters, we are going to show the complete matching between the general solution and the perturbative realization of those correlators. The undetermined constants are completely fixed when the field content of the theory is chosen.\\
\chapter{Perturbative results in CFT: TJJ case}\label{PerturbativeTJJ}

In this chapter, we turn to discuss the connection between the solutions of the CWI's presented in terms of $3K$ integrals and the perturbative $TJJ$ vertex \cite{Giannotti:2008cv,Armillis:2009pq}. The expressions for the form factors had been given in the F-basis of 13 form factors, which will be reviewed in the next section. We will have to recompute them to present them expressed in terms of the two basic fundamental master integrals $B_0$ and $C_0$ of the tensor reduction \cite{Coriano:2018bbe} rather than in their final form, given in \cite{Armillis:2009pq}.

\section{Introduction} 
The analysis of multi-point correlation functions in conformal field theory (CFT) is of utmost importance in high energy physics and in string theory, where exact results for lower (2- and 3-) point functions are combined with the operator product expansion (OPE) in order to characterize the structure of correlators of higher orders. This is the crucial motivation for a bootstrap program in $d=4$ spacetime dimensions. \\
The enlarged $SO(2,4)$ symmetry of CFT's - respect to Poincar\'{e} invariance - has been essential for establishing the form of some of their correlation functions. For 3-point functions, the solution of the conformal constraints in coordinate space allows to determine such correlators only up to few constants \cite{Osborn:1993cr, Erdmenger:1996yc}, which can then be fixed within a specific realization of a theory.  In the case of a Lagrangian realization of a given CFT, such constants are expressed in terms of its (massless) field content (number of scalars, vectors, fermions), according to relatively simple algebraic relations.  

Except for perturbative studies performed at Lagrangian level, such as in the case of the $N=4$ super Yang-Mills theory, which reaches considerably high orders in the gauge coupling expansion, most of these analyses are performed in coordinate space, with no reference to any specific Lagrangian. 

There are obvious reasons for this. The first is that the inclusion of the conformal constraints is more straightforward to obtain in coordinate space, compared to momentum space. The second is that the operator product expansion (OPE) in momentum space is challenging to perform, especially for correlators of higher orders ($\geq 3$ ), in the Minkowski region. 
However, some advantages are typical of a momentum space analysis, and these are related to the availability of dimensional regularization (DR), at least at perturbative level, and to the technology of master integrals which has allowed to compute large classes of multiloop amplitudes. \\ 
Another advantage has to do with the identification of the conformal anomaly \cite{Capper:1975ig}, which can be automatically extracted in DR (in $d$ spacetime dimensions), being proportional to the $1/(d-4)$ singularity of the corresponding correlators. Instead, in coordinate space, the anomaly contributions have to be added by hand by the inclusion of an inhomogeneous local term (i.e. by pinching all of its external coordinates), whose structure has to be inferred indirectly  \cite{Osborn:1993cr}. \\  
Finally, a crucial issue concerns the physical character of the anomaly, which does not find any simple particle interpretation in position space, while it is associated to the appearance of an anomaly pole in momentum space \cite{Giannotti:2008cv,2009PhLB..682..322A,Armillis:2009sm} in an uncontracted anomaly vertex. One finds that by a perturbative one-loop analysis of any anomalous correlator, the anomaly is always associated with such massless exchanges in the corresponding diagram. Therefore, it is possible to identify them as effective degrees of freedom induced by the anomaly, present in the 1PI (one-particle irreducible) effective action. The physical significance of such contributions has been stressed in several previous works \cite{Giannotti:2008cv,Armillis:2009pq, Armillis:2009sm} along the years. They have recently discussed in condensed matter theory in the context of topological insulators, and Weyl semimetals \cite{Chernodub:2017jcp,Rinkel:2016dxo}.\\
In this chapter, we will compare and extend the previous perturbative analysis of the $TJJ$ correlator with more recent ones based on the solution of the conformal Ward identities (CWI's) in momentum space \cite{Bzowski:2013sza,Bzowski:2015yxv,Bzowski:2017poo}. This correlator is the simplest one describing the coupling of gravity to ordinary matter in QED, and it has been investigated in perturbation theory from several directions \cite{Bastianelli:2012es,Bastianelli:2012bz,Bonora:2017gzz,Bonora:2014qla, Bastianelli:2016nuf}. 
\subsection{Direct Fourier transform and the reconstruction program}
In principle, one can move from coordinate space to position space in a CFT by a Fourier transform. This was the approach of \cite{Coriano:2012wp} for 3-point functions, which can be explicitly worked out by introducing a regulator ($\omega$) for the transform very much alike DR. The regulator serves as an intermediate step since some of the components of the correlators in position space are apparently non-transformable. It has been shown that $1/\omega$ poles generated by the transform {\em cancel} in all the correlators analyzed, giving a complete expression for these in momentum space. The result is expressed in terms of ordinary and logarithmic master integrals of Feynman type, for which, in the latter case, it is possible to derive recursion relations as for ordinary ones \cite{Coriano:2012wp}. The advantage of such an approach is of being straightforward and algorithmic. It may be essential and probably the only manageable way to re-express the bootstrap program of CFT's in momentum space beyond 3- and 4- point functions, from the original coordinate space analysis. 
Consistency with the analysis presented in \cite{Coriano:2012wp} implied rather directly that such logarithmic integrals had to be re-expressed in terms of ordinary Feynman integrals. It was shown in the same study that a free field theory entirely reproduced the TJJ correlator in coordinate space. Our analysis in momentum space is in complete agreement with this former result.

\subsection{Reconstruction}
An alternative method, discussed in the previous chapter, has been developed more recently,  based on the conformal Ward identities' direct solution in momentum space. The method has been proposed in \cite{Bzowski:2013sza} and \cite{Coriano:2013jba} for scalar 3-point functions and extensively generalized to tensor correlators in \cite{Bzowski:2013sza}.\\
Several issues related to the renormalization of the conformal Ward identities' solutions have been investigated in \cite{Bzowski:2015yxv, Bzowski:2017poo}, adopting the formalism of the 3K integrals (i.e. parametric integrals of 3 Bessel functions). 
Several analyses in momentum space, for specific applications, have been worked out \cite{Coriano:2012hd, Bzowski:2011ab}. However, the approach's generality is a significant feature of \cite{Bzowski:2013sza}, which reconstructs a tensor correlator starting from its transverse/traceless components and using the conservation/trace Ward identities (local terms). The latter are reconstructed from lower point functions.\\
The result is expressed in terms of two sets of primary and secondary conformal Ward identities (CWI' s), the first involving the form factors of the transverse/traceless contributions, which are parametrized on a symmetric basis, the second emerging from CWI's of lower point functions. For 3-point functions, the secondary CWI's involve conservation, trace and special WI's. In all the cases, the reconstructed solutions for 3-point functions can be given in terms of generalized hypergeometrics of type $F_4$, \cite{Coriano:2012wp}, also known as Appell's hypergeometric function of two variables ($F_4$), related to 3-K integrals \cite{Bzowski:2013sza}. 
\subsubsection{ The anomaly pole of the TJJ}
One of our analysis results will be to show how such contributions originate from the process of renormalization, taking as an example the case of the $TJJ$, filling in the intermediate steps of the discussion presented in \cite{Coriano:2018zdo}. We follow the general (BMS) approach introduced in \cite{Bzowski:2013sza} for the solution of the conformal constraints, which we detail in several of its parts, not offered in \cite{Bzowski:2013sza}. It has been compelling to proceed with an independent re-derivation of all the lengthy equations. The method can be directly generalized to higher point functions and implemented algorithmically, as we will show in a separate work.
When discussing the momentum space approach in CFT, there are several gaps in the literature, which are methodological and need to be addressed. These concern the correct form of the differential equations, the treatment of the derivatives of the 
Dirac, $\delta$'s induced by momentum conservation, violations of the Leibnitz rule for the special conformal transformations, or the Lorentz (spin) singlet operator's choice in the action of the conformal group on a specific correlator. These are points that we will address systematically.
We will illustrate how to merge the results of the BMS approach on the structure of the minimal set of (4) form factors 
(the $A-$basis), solutions of the CWI's for the $TJJ$ correlator, with a basis of 13 ones (the $F$-basis) defined in previous perturbative studies. We will show how to extract from the $F-$basis 4 combinations of the 13, and we will verify that they respect the scalar equations identified within the BMS approach. \\ 
This second basis is essential to prove that the WI's and the renormalization procedure for this correlator imply that the anomaly can be attributed to the appearance of an anomaly pole in a single tensor structure of nonzero trace. \\
In the next sections, we will show how the perturbative solutions for the $A_i$, which are given in an appendix, reproduce the exact BMS result in a simplified way. We use the cases of $d=3$ and $d=5$ to show the exact correspondence between the two. This correspondence is studied by fixing an appropriate normalization of the photon two-point functions. We show that the choice of different perturbative sectors (scalar, fermion) in both cases is sufficient to reproduce the entire nonperturbative result. This implies that only arbitrary constant in the nonperturbative solution, expressed in terms of the 3K integrals, has to simplify and be expressible in terms of simple integrals $B_0$ and $C_0$, the scalar 2- and 3-point functions. In our conclusions, we briefly comment on the possible origin of such simplifications. 
\section{Perturbative analysis in the Conformal case: QED and scalar QED}
\label{pchecks}
Here we are also going to introduce the diagrammatic expansion for the $TJJ$ in QED, since it will be needed when we are going to compare the general non-perturbative solution against the perturbative one in $d=3$ and $d=5$. \\
\begin{figure}[t]
	\centering
	\vspace{-1.3cm}
	\subfigure{\includegraphics[scale=0.18]{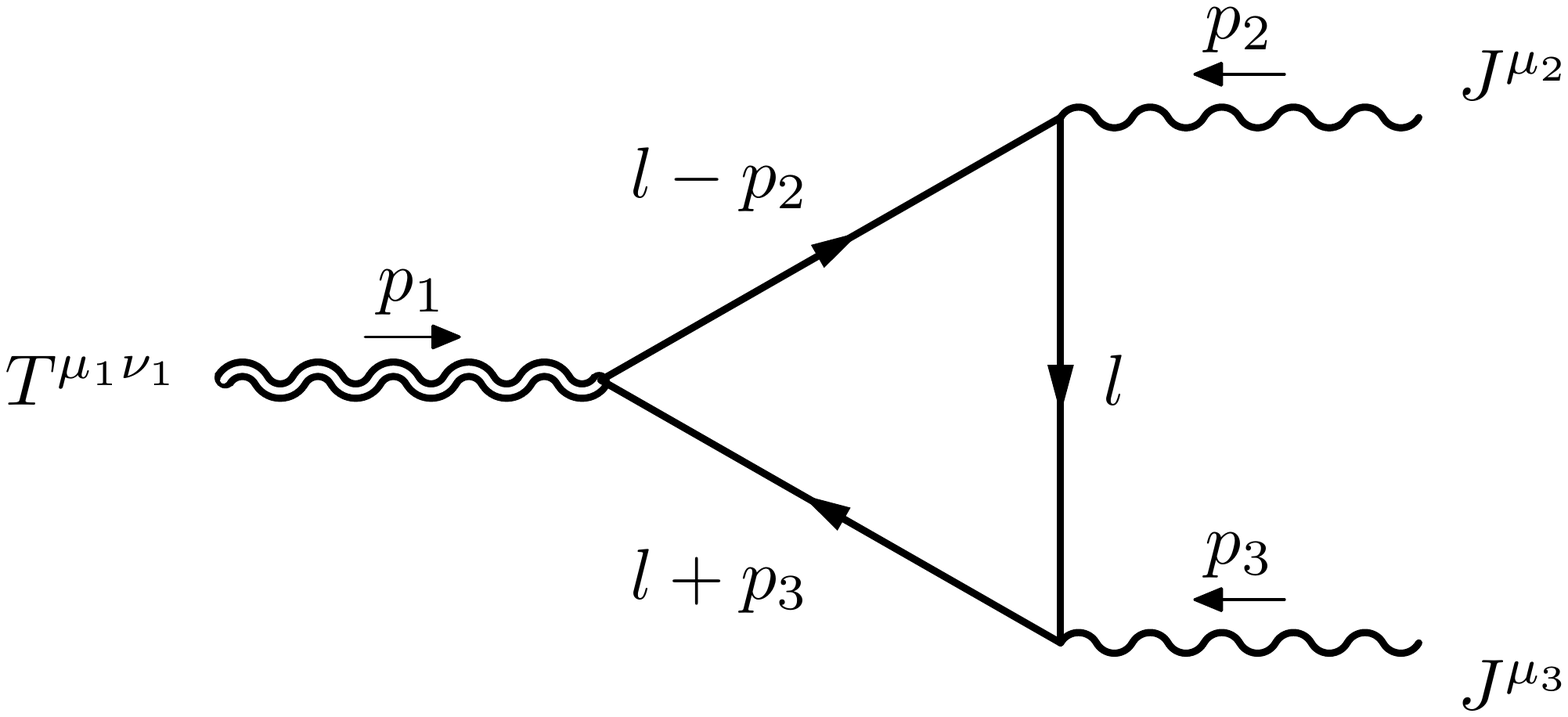}} \hspace{.3cm}
	\subfigure{\includegraphics[scale=0.18]{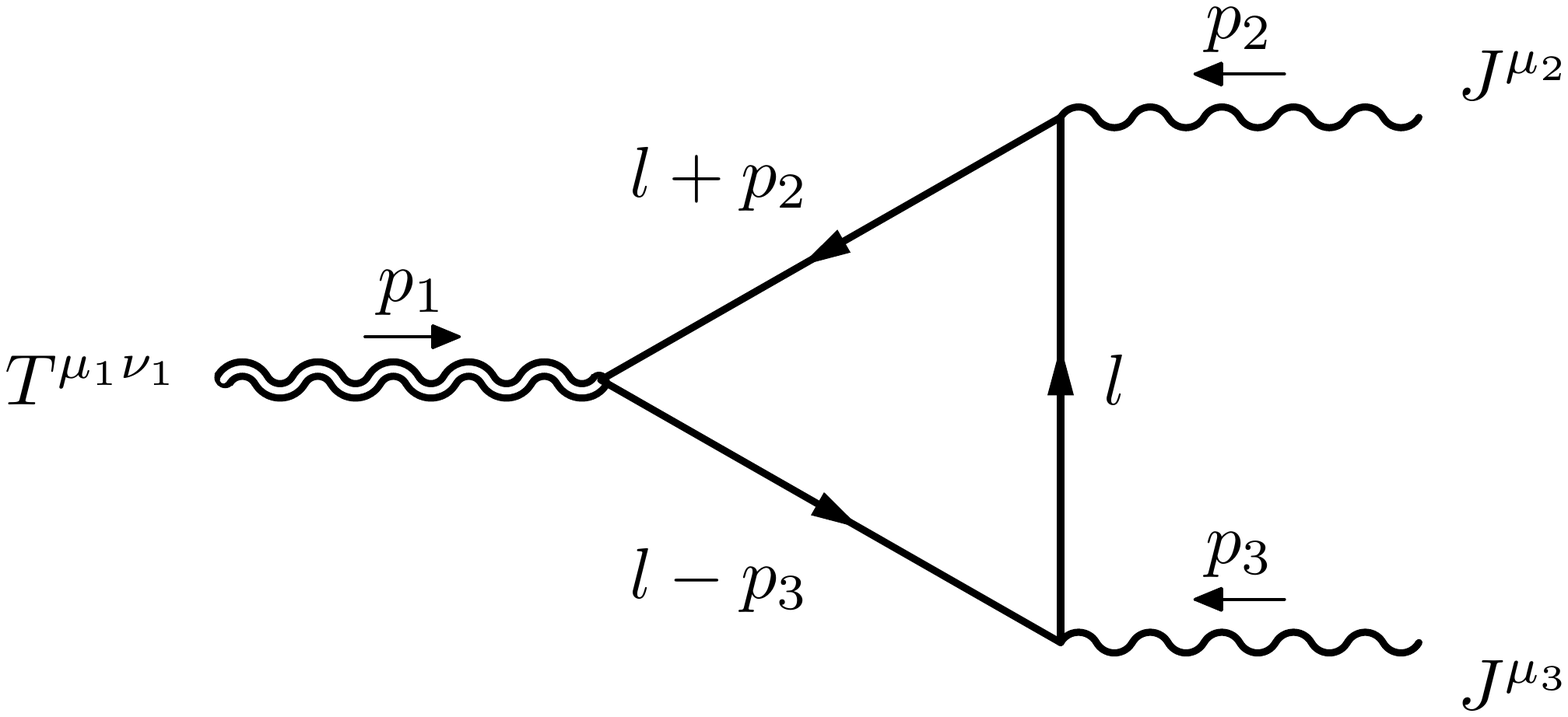}} \hspace{.3cm}
	\raisebox{.1\height}{\subfigure{\includegraphics[scale=0.14]{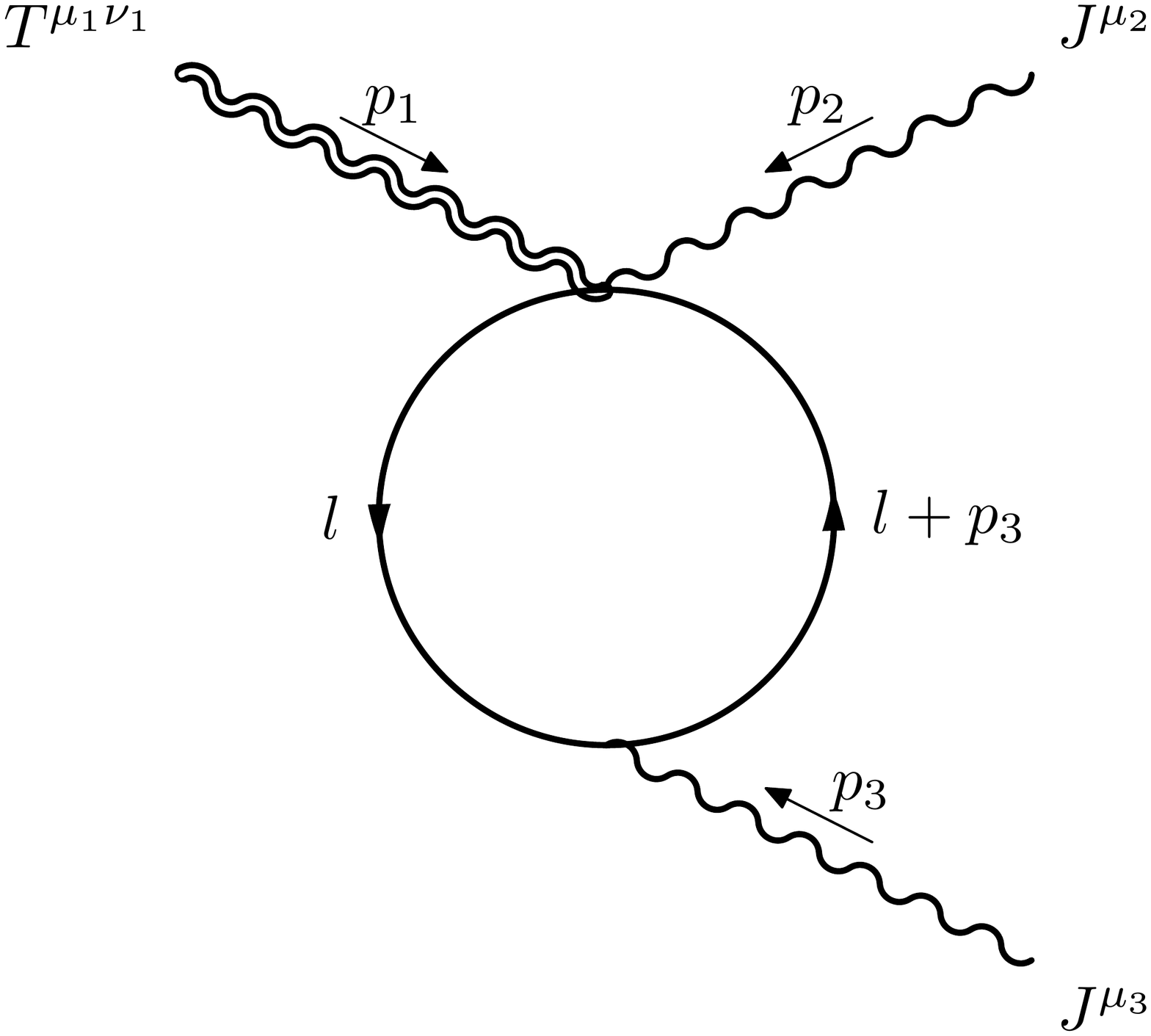}}}\hspace{.3cm}	\raisebox{.1\height}{\subfigure{\includegraphics[scale=0.14]{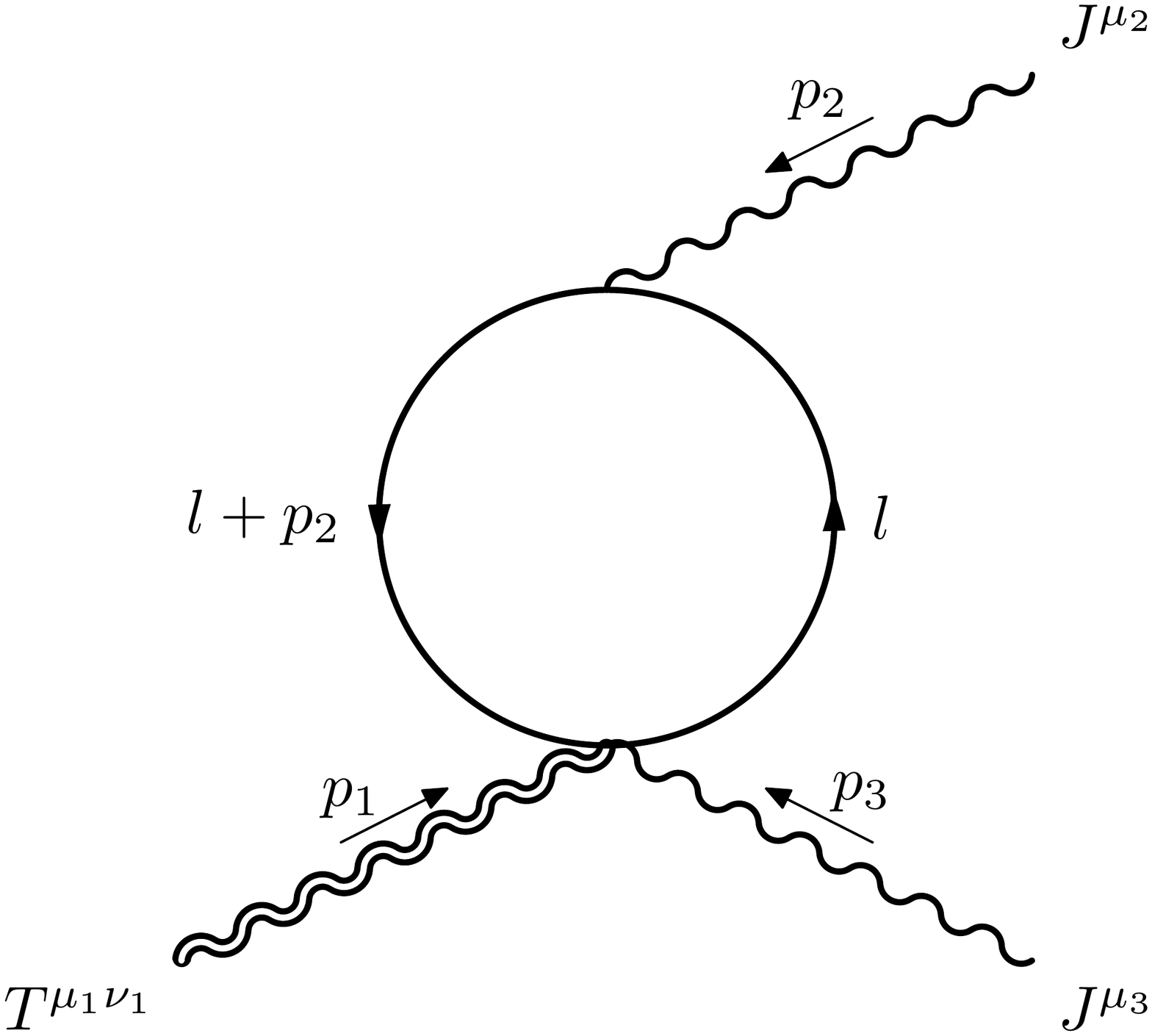}}}
	\vspace{-0.8cm}\caption{One-loop diagrams for the TJJ in QED. \label{Figura1}}
\end{figure}
The quantum actions for the fermion field is
\begin{align}
	S_{fermion}&=\sdfrac{i}{2}\int\, d^dx\ e\ e^{\m}_a\left[\bar{\psi}\g^a(D_\m\psi)-(D_\m\bar{\psi})\g^a\psi\right],
\end{align}
$e^\m_a$ is the vielbein and $e$ its determinant, with its covariant derivative $D_\m$ as
\begin{equation}
	D_\m=\partial_\m+ie\,A_\mu+\G_\m=\partial_\m+ie\,A_\mu+\sdfrac{1}{2}\Sigma^{ab}\,e^\s_a\nabla_\m\,e_{b\,\s}.
\end{equation}
The $\Sigma^{ab}$ are the generators of the Lorentz group in the case of a spin $1/2$-field. The gravitational field is expanded, as usual, in the form $g_{\mu\nu}=\eta_{\mu\nu} + h_{\mu\nu}$ around the flat background metric with fluctuations $h_{\mu\nu}$. As usual, the Latin anf Greek indices are related to the locally flat and curved backgrounds respectively.

In the one-loop approximation the contribution to the correlation functions are given by the diagrams in \figref{Figura1}, with vertices shown in \figref{Figura2} and explicitly written in \appref{Appendix1}. We calculate all the diagram contributions in momentum space for the fermion sector as
\begin{eqnarray}
	\Gamma^{\m_1\n_1\m_2\m_3}(p_2,p_3)&\equiv&\braket{T^{\m_1\n_1}(p_1)\,J^{\m_2}(p_2)\,J^{\m_3}(p_3)}_F\notag\\
	&&=2\,\bigg(\sum_{i=1}^2V_{F,i}^{\m_1\n_1\m_2\m_3}(p_1,p_2,p_3)+\sum_{i=1}^{2}W_{F,i}^{\m_1\n_1\m_2\m_3}(p_1,p_2,p_3)\bigg)
\end{eqnarray}
where the $V_{F,i}$ terms are related to the triangle topology contributions, while the $W_{F,i}$ terms denote the two bubble contributions in \figref{Figura1}. All these terms are explicitly given as
\begin{align}
	V_{F,1}^{\m_1\n_1\m_2\m_3}&=-\,\int\frac{d^d\ell}{(2\p)^d}\frac{\Tr\left[V^{\m_1\n_1}_{T\psi\bar\psi}(\ell-p_2,\ell+p_3)\left(\slashed{\ell}+\slashed{p}_3\right)V^{\m_2}_{J\psi\bar\psi}(\ell,\ell-p_2)\,\slashed{\ell}\,V^{\m_3}_{J\psi\bar\psi}(\ell,\ell+p_3)\,\left(\slashed{\ell}-\slashed{p}_2\right)\right]}{\ell^2\,(\ell-p_2)^2(\ell+p_3)^2}\\
	V_{F,2}^{\m_1\n_1\m_2\m_3}&=-\,\int\frac{d^d\ell}{(2\p)^d}\frac{\Tr\left[V^{\m_1\n_1}_{T\psi\bar\psi}(\ell-p_3,\ell+p_2)\left(\slashed{\ell}+\slashed{p}_2\right)V^{\m_2}_{J\psi\bar\psi}(\ell,\ell-p_3)\,\slashed{\ell}\,V^{\m_3}_{J\psi\bar\psi}(\ell,\ell+p_2)\,\left(\slashed{\ell}-\slashed{p}_3\right)	\right]}{\ell^2\,(\ell-p_3)^2(\ell+p_2)^2}\\
	W_{F,1}^{\m_1\n_1\m_2\m_3}&=-\,\int\frac{d^d\ell}{(2\p)^d}\frac{\Tr\left[V^{\m_1\n_1\m_2}_{TJ\psi\bar\psi}(\ell+p_3,\ell)\left(\slashed{\ell}+\slashed{p}_3\right)V^{\m_3}_{J\psi\bar\psi}(\ell,\ell+p_3)\,\slashed{\ell}\right]}{\ell^2\,(\ell+p_3)^2}\\[1.5ex]
	W_{F,2}^{\m_1\n_1\m_2\m_3}&=-\,\int\frac{d^d\ell}{(2\p)^d}\frac{\Tr\left[V^{\m_1\n_1\m_3}_{TJ\psi\bar\psi}(\ell+p_2,\ell)\left(\slashed{\ell}+\slashed{p}_2\right)V^{\m_2}_{J\psi\bar\psi}(\ell,\ell+p_2)\,\slashed{\ell}\right]}{\ell^2\,(\ell+p_2)^2}
\end{align}
\subsection{The $TJJ$ in scalar QED}
Now we turn to consider scalar QED. The action, in this case, can be written as
\begin{equation}
	S_{scalar}=\int\,d^dx\,\sqrt{-g}\,\left(\,\left|D_\m\,\phi\right|^2+\frac{(d-2)}{8(d-1)}\,R\,|\phi|^2\,\right)
\end{equation}
where $R$ is the scalar curvature and $\phi$ denotes a complex scalar. We have explicitly reported the coefficient of the term of improvement, and with $D_\m\phi=\partial_\m\phi+ie\,A_\m$ being the covariant derivative for the coupling to the gauge field $A_\mu$. 
At one-loop the contribution to the $TJJ$ is given by the diagram in \figref{Figura1}, with the obvious replacement of a fermion by a scalar in the internal loop corrections. In this case they are given by 
\begin{equation}
	\braket{T^{\m_1\n_1}(p_1)\,J^{\m_2}(p_2)\,J^{\m_3}(p_3)}_S=2\,\bigg(V_{S}^{\m_1\n_1\m_2\m_3}(p_1,p_2,p_3)+\sum_{i=1}^{3}W_{S,i}^{\m_1\n_1\m_2\m_3}(p_1,p_2,p_3)\bigg)
\end{equation}
where the $V_S$ terms are related to the triangle topology contribution and the $W_{S,i}$'s  are the three bubble contributions in \figref{Figura1}. All these are explicitly given as
\begin{align}
	V_{S}^{\m_1\n_1\m_2\m_3}(p_1,p_2,p_3)&=i^3\,\int\frac{d^d\ell}{(2\p)^d}\frac{V^{\m_1\n_1}_{T\phi\phi^*}(\ell-p_2,\ell+p_3)\,V^{\m_2}_{J\phi\phi^*}(\ell,\ell-p_2)\,V^{\m_3}_{J\phi\phi^*}(\ell,\ell+p_3)}{\ell^2\,(\ell-p_2)^2(\ell+p_3)^2}\\[1.5ex]
	W_{S,1}^{\m_1\n_1\m_2\m_3}(p_1,p_2,p_3)&=\frac{i^2}{2}\,\int\frac{d^d\ell}{(2\p)^d}\frac{V^{\m_1\n_1}_{T\phi\phi^*}(\ell+p_1,\ell)\,V^{\m_2\m_3}_{JJ\phi\phi^*}(\ell,\ell+p_1)}{\ell^2\,(\ell+p_1)^2}\\[1.5ex]
	W_{S,2}^{\m_1\n_1\m_2\m_3}(p_1,p_2,p_3)&=\frac{i^2}{2}\,\int\frac{d^d\ell}{(2\p)^d}\frac{V^{\m_1\n_1\m_2}_{TJ\phi\phi^*}(\ell+p_3,\ell)\,V^{\m_3}_{J\phi\phi^*}(\ell,\ell+p_3)}{\ell^2\,(\ell+p_3)^2}\\[1.5ex]
	W_{S,3}^{\m_1\n_1\m_2\m_3}(p_1,p_2,p_3)&=\frac{i^2}{2}\,\int\frac{d^d\ell}{(2\p)^d}\frac{V^{\m_1\n_1\m_3}_{TJ\phi\phi^*}(\ell+p_2,\ell)\,V^{\m_2}_{J\phi\phi^*}(\ell,\ell+p_2)}{\ell^2\,(\ell+p_2)^2}
\end{align}
\begin{figure}[t]
	\vspace{-1.3cm}
	\subfigure{\includegraphics[scale=0.16]{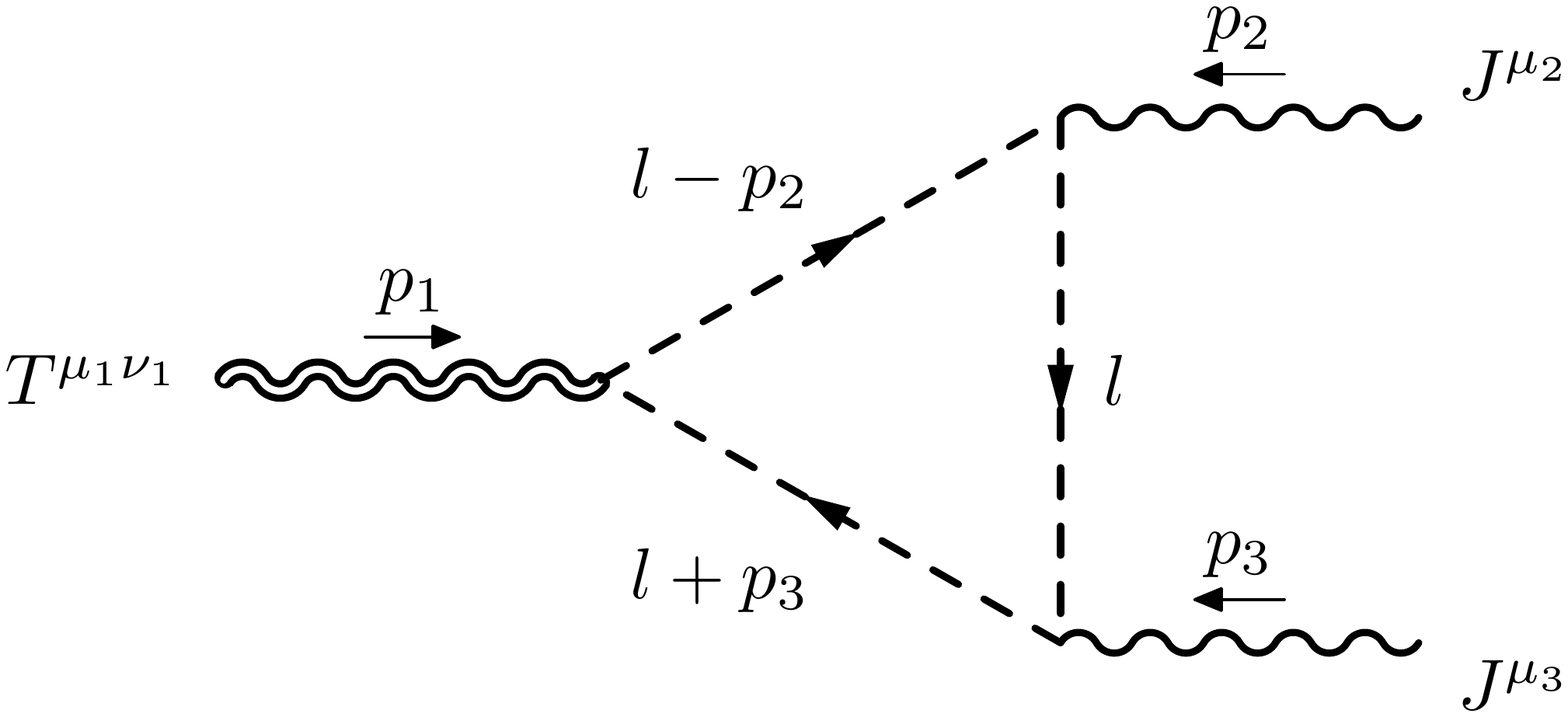}}\hspace{.3cm}
	\subfigure{\includegraphics[scale=0.16]{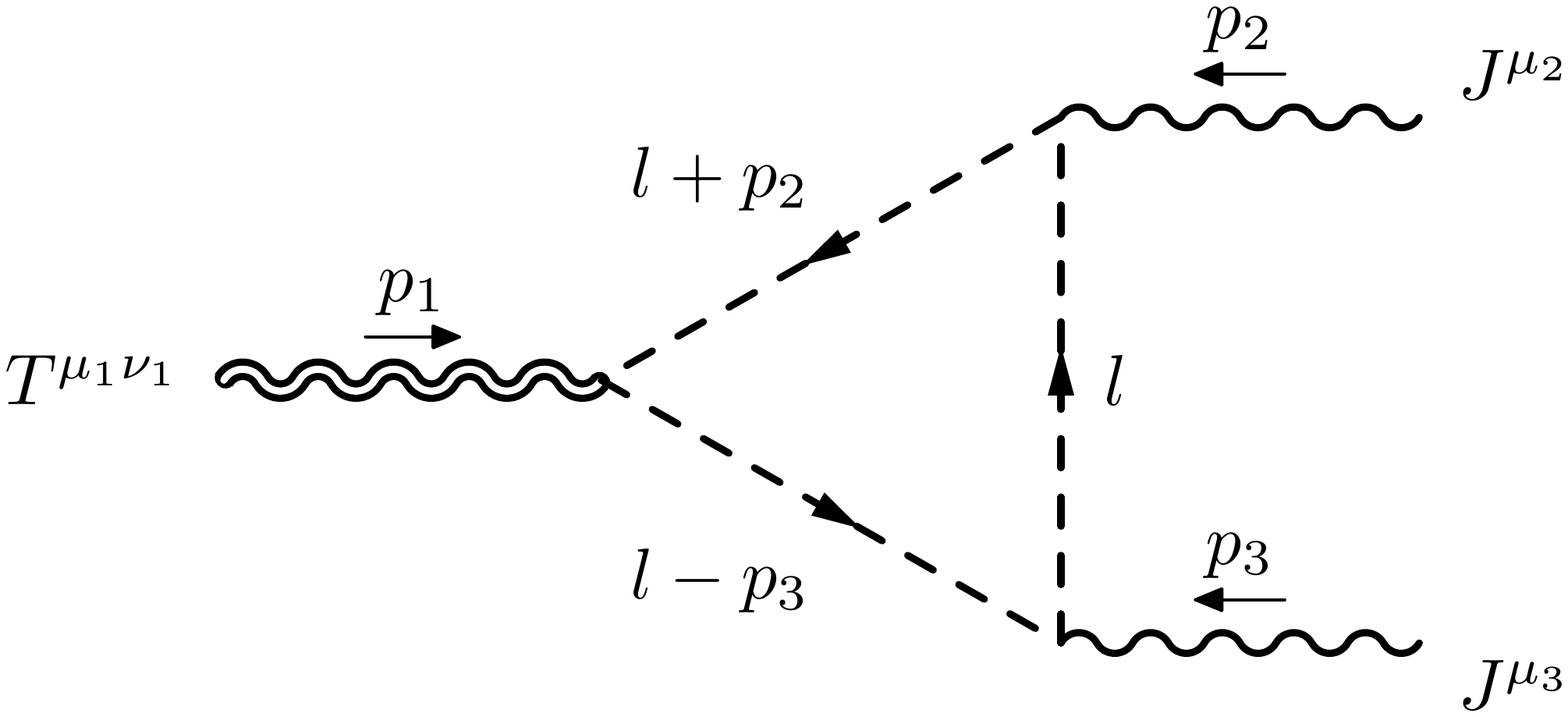}} \hspace{.3cm}
	\raisebox{0.06\height}{\subfigure{\includegraphics[scale=0.15]{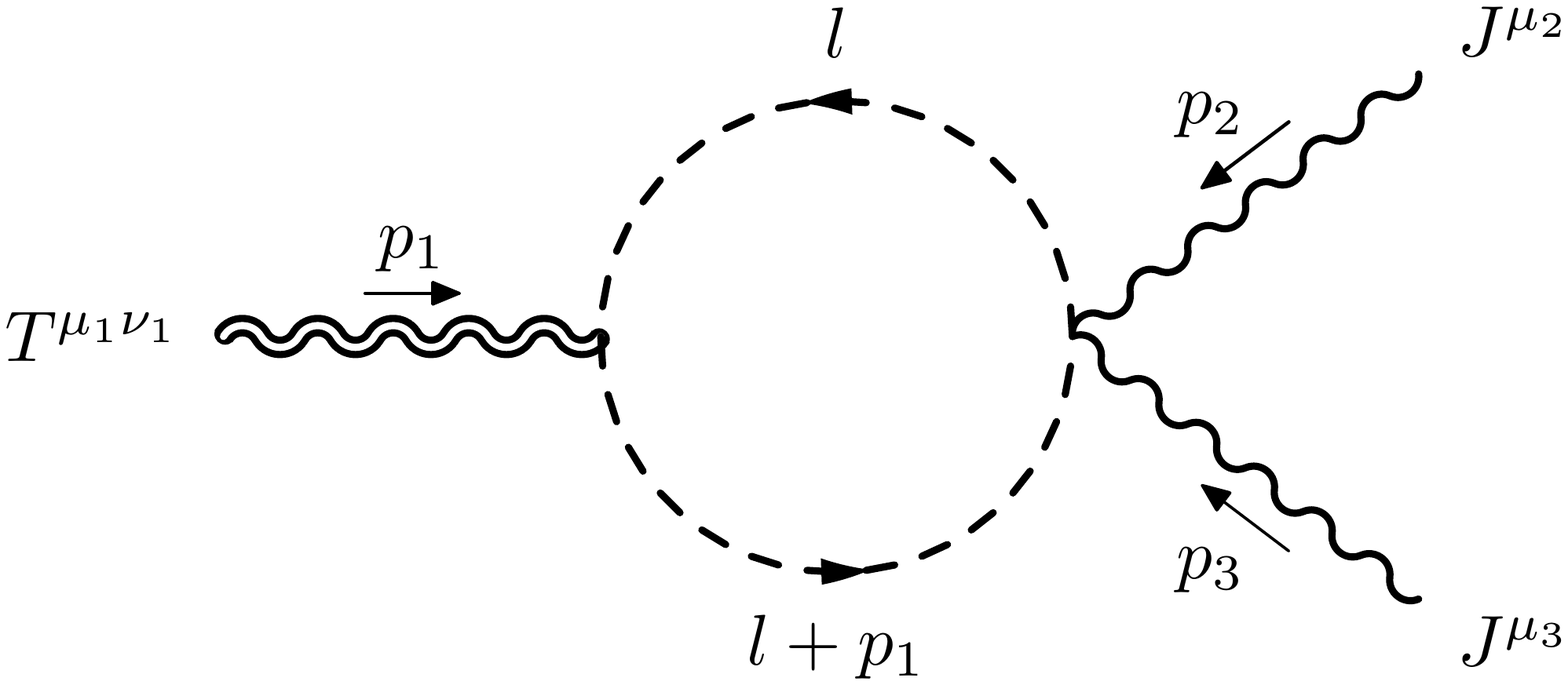}}}\hspace{.3cm}
	\raisebox{.11\height}{\subfigure{\includegraphics[scale=0.12]{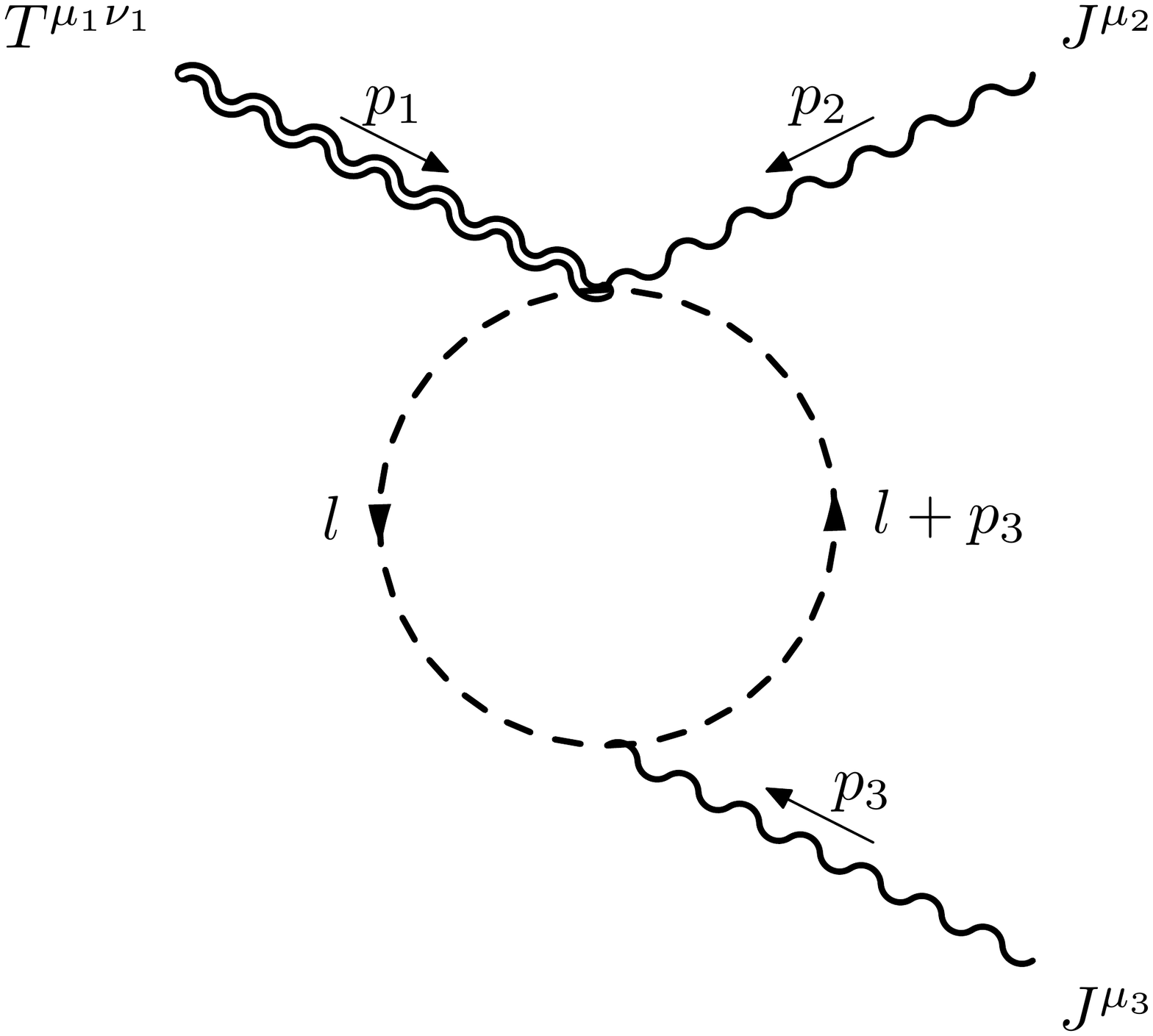}}}\hspace{.3cm}
	\raisebox{.11\height}{\subfigure{\includegraphics[scale=0.12]{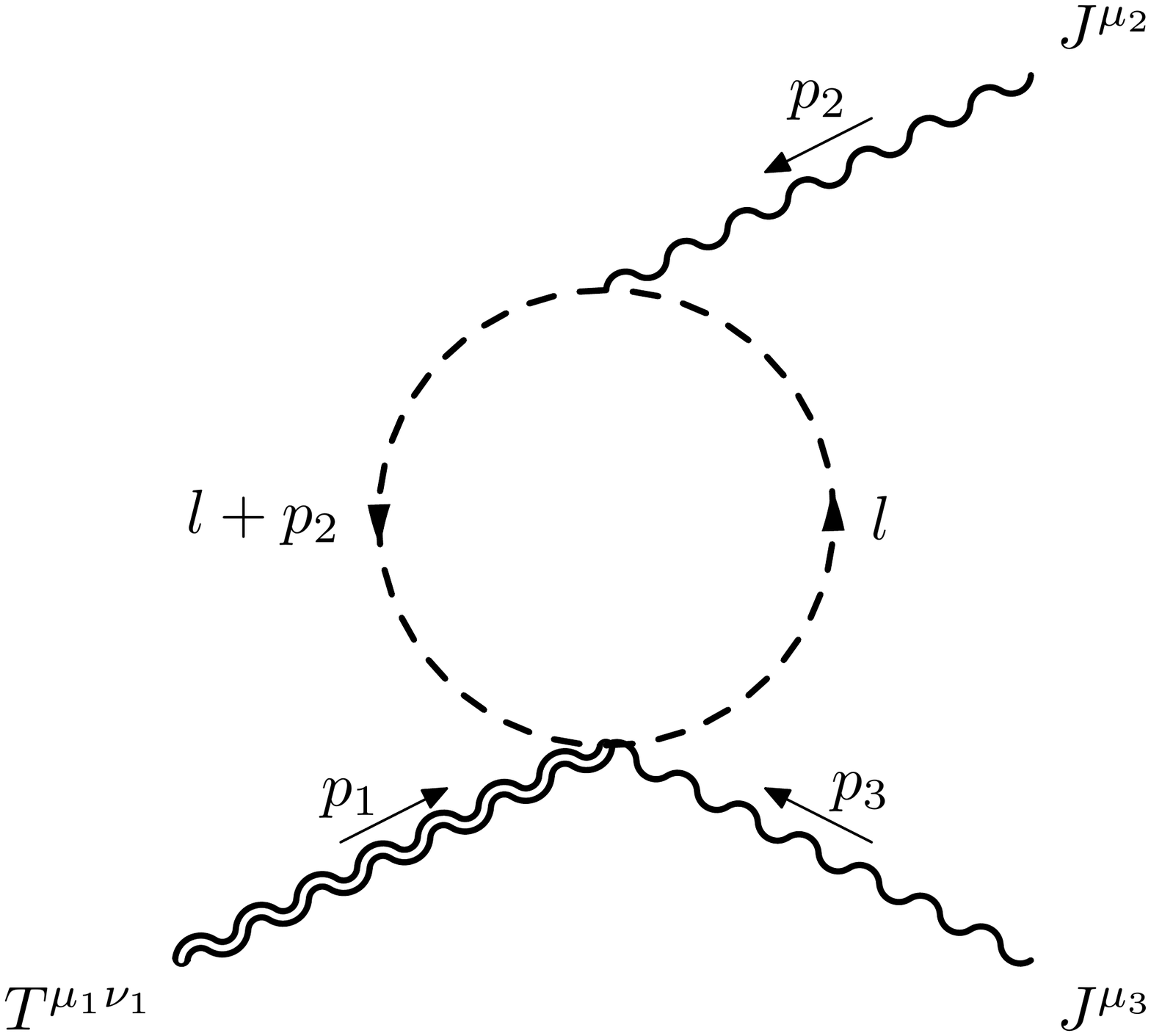}}}
	\vspace{-0.8cm}\caption{One-loop diagrams for the TJJ in the scalar QED. \label{Figura1bis}}
\end{figure}
where we have included the symmetry factors and the vertices are given in \appref{Appendix1}. 

\section{ The F-basis of the expansion for the $TJJ$ in QED}
The $TJJ$ can be expanded on the basis proposed in \cite{Giannotti:2008cv}, in terms of 13 independent tensors structures given in \tabref{genbasis}. In this scheme, the correlation function can be written as
\begin{equation}
	\Gamma^{\m_1\n_1\m_2\m_3}(p_2,p_3)=\sum_{i=1}^{13}\,F_i(p_1^2,p_2^2,p_3^2)\,t_i^{\m_1\n_1\m_2\m_3}(p_2,p_3),
\end{equation} 
where the invariant amplitudes $F_i$ are functions of the kinematic invariants $p_1^2=(p_2+p_3)^2$, $p_2^2$, $p_3^2$, and the $t_i^{\m_1\n_1\m_2\m_3}$ form the basis of independent tensor structures. 

This set of $13$ tensors is linearly independent in $d$ dimensions, for generic $p_1^2, p_2^2, p_3^2$ different from zero. Five of the $13$ are Bose symmetric, 
\begin{equation}
	t_i^{\mu_1\nu_1\mu_2\mu_3}(p_2,p_3) = t_i^{\mu_1\nu_1\mu_3\mu_2}(p_3,p_2)\,,\qquad i=1,2,7,8,13\,,
\end{equation}
while the remaining eight tensors are Bose symmetric pairwise
\begin{align}
	\label{pair}
	&t_3^{\mu_1\nu_1\mu_2\mu_3}(p_2,p_3) = t_5^{\mu_1\nu_1\mu_3\mu_2}(p_3,p_2)\,,\\
	&t_4^{\mu_1\nu_1\mu_2\mu_3}(p_2,p_3)= t_6^{\mu_1\nu_1\mu_3\mu_2}(p_3,p_2)\,,\\
	&t_9^{\mu_1\nu_1\mu_2\mu_3}(p_2,p_3) = t_{10}^{\mu_1\nu_1\mu_3\mu_2}(p_3,p_2)\,,\\
	&t_{11}^{\mu_1\nu_1\mu_2\mu_3}(p_2,p_3) = t_{12}^{\mu_1\nu_1\mu_3\mu_2}(p_3,p_2)\,.
\end{align}

\begin{table}[t]
	\begin{center}
		\begin{tabular}{|c|c|}
			\hline
			&\\[-1.7ex]
			i & $t_i^{\mu_1\nu_1\mu_2\mu_3}(p_2,p_3)$\\[1ex]
			\hline 
			&\\[-1.7ex]
			1 &
			$\left(k^2 \delta^{\mu_1\nu_1} - k^{\mu_1 } k^{\nu_1}\right) u^{\mu_2\mu_3}(p_2,p_3)$\\[1ex]
			\hline
			&\\[-1.7ex]
			2 &
			$\left(k^2\delta^{\mu_1\nu_1} - k^{\mu_1} k^{\nu_1}\right) w^{\mu_2\mu_3}(p_2,p_3) $ \\[1ex]
			\hline
			&\\[-1.7ex]
			3 & $\left(p_2^2 \delta^{\mu_1\nu_1} - 4 p_2^{\mu_1}  p_2^{\nu_1}\right)
			u^{\mu_2\mu_3}(p_2,p_3)$\\[1ex]
			\hline
			&\\[-1.7ex]
			4 & $\left(p_2^2 \delta^{\mu_1\nu_1} - 4 p_2^{\mu_1} p_2^{\nu_1}\right)
			w^{\mu_2\mu_3}(p_2,p_3)$\\[1ex]
			\hline
			&\\[-1.7ex]
			5 & $\left(p_3^2 \delta^{\mu_1\nu_1} - 4 p_3^{\mu_1} p_3^{\nu_1}\right)
			u^{\mu_2\mu_3}(p_2,p_3)$\\[1ex]
			\hline
			&\\[-1.7ex]
			6 & $\left(p_3^2 \delta^{\mu_1\nu_1} - 4 p_3^{\mu_1} p_3^{\nu_1}\right)
			w^{\mu_2\mu_3}(p_2,p_3) $\\[1ex]
			\hline
			&\\[-1.7ex]
			7 & $\left[p_2\cdot p_3\, \delta^{\mu_1\nu_1}
			-2 (p_3^{\mu_1} p_2^{\nu_1} + p_2^{\mu_1} p_3^{\nu_1})\right] u^{\mu_2\mu_3}(p_2,p_3)$ \\[1ex]
			\hline
			&\\[-1.7ex]
			8 & $\left[p_2\cdot p_3\, \delta^{\mu_1\nu_1}
			-2 (p_3^{\mu_1} p_2^{\nu_1} + p_2^{\mu_1} p_3^{\nu_1})\right] w^{\mu_2\mu_3}(p_2,p_3)$\\[1ex]
			\hline
			&\\[-1.7ex]
			9 & $\left(p_2\cdot p_3 \,p_2^{\mu_2}  - p_2^2 p_3^{\mu_2}\right)
			\big[p_2^{\mu_3} \left(p_3^{\mu_1} p_2^{\nu_1} + p_2^{\mu_1} p_3^{\nu_1} \right) - p_2\cdot p_3\,
			(\delta^{\mu_3\nu_1} p_2^{\mu_1} + \delta^{\mu_3\mu_1} p_2^{\nu_1})\big] $ \\[1ex]
			\hline
			&\\[-1.7ex]
			10 &$ \big(p_2\cdot p_3 \,p_3^{\mu_3} - p_3^2 p_2^{\mu_3}\big)\,
			\big[p_3^{\mu_2} \left(p_3^{\mu_1} p_2^{\nu_1} + p_2^{\mu_1} p_3^{\nu_1} \right) - p_2\cdot p_3\,
			(\delta^{\mu_2\nu_1} p_3^{\mu_1} + \delta^{\mu_2\mu_1} p_3^{\nu_1})\big] $ \\[1ex]
			\hline
			&\\[-1.7ex]
			11 & $\left(p_2\cdot p_3 \,p_2^{\mu_2} - p_2^2 p_3^{\mu_2}\right)
			\big[2\, p_3^{\mu_3} p_3^{\mu_1} p_3^{\nu_1} - p_3^2 (\delta^{\mu_3\nu_1} p_3^ {\mu_1}
			+ \delta^{\mu_3\mu_1} p_3^{\nu_1})\big] $ \\[1ex]
			\hline
			&\\[-1.7ex]
			12 & $\big(p_2\cdot p_3 \,p_3^{\mu_3} - p_3^2 p_2^{\mu_3}\big)\,
			\big[2 \, p_2^{\mu_2} p_2^{\mu_1} p_2^{\nu_1} - p_2^2 (\delta^{\mu_2\nu_1} p_2^ {\mu_1}
			+ \delta^{\mu_2\mu_1} p_2^{\nu_1})\big]$ \\[1ex]
			\hline
			&\\[-1.7ex]
			13 & $\big(p_2^{\mu_1} p_3^{\nu_1} + p_2^{\nu_1} p_3^{\mu_1}\big)\delta^{\mu_2\mu_3}
			+ p_2\cdot p_3\, \big(\delta^{\mu_2\nu_1} \delta^{\mu_3\mu_1}
			+ \delta^{\mu_2\mu_1} \delta^{\mu_3\nu_1}\big) - \delta^{\mu_1\nu_1} u^{\mu_2\mu_3} $\\
			& $-\big(\delta^{\mu_3\nu_1} p_2^{\mu_1}
			+ \delta^{\mu_3\mu_1} p_2^{\nu_1}\big)p_3^{\mu_2}
			- \big (\delta^{\mu_2\nu_1} p_3^{\mu_1}
			+ \delta^{\mu_2\mu_1} p_3^{\nu_1 }\big)p_2^{\mu_3} $ \\[1ex]
			\hline
		\end{tabular}
	\end{center}
	\caption{Basis of 13 fourth rank tensors satisfying the vector current conservation on the external lines with momenta $p_2$ and $p_3$. \label{genbasis}}
\end{table}
In the set are present two tensor structures
\begin{equation}
	\begin{split}
		&u^{\mu_2\mu_3}(p_2,p_3) \equiv (p_2\cdot p_3)\,\delta^{\mu_2\mu_3} - p_3^{\mu_2}p_2^{\mu_3}\,,\\
		&w^{\mu_2\mu_3}(p_2,p_3) \equiv p_2^2\, p_3^2\, \delta^{\mu_2\mu_3} + (p_2\cdot p_3)\, p_2^{\mu_2}p_3^{\mu_3}
		- p_3^2 p_2^{\mu_2}p_2^{\mu_3} - p_2^2 p_3^{\mu_2}p_3^{\mu_3}\,,
	\end{split}
	\label{uwdef}
\end{equation}
which appear in $t_1$ and $t_2$ respectively.
Each of them satisfies the Bose symmetry  requirement,
\begin{align}
	&u^{\mu_2\mu_3}(p_2,p_3) = u^{\mu_3\mu_2}(p_3,p_2)\,,\\
	&w^{\mu_2\mu_3}(p_2,p_3) = w^{\mu_3\mu_2}(p_3,p_2)\,,
\end{align}
and vector current conservation,
\begin{align}
	&p_{2\,\mu_2} u^{\mu_2\mu_3}(p_2,p_3) = 0 = p_{3\,\mu_3}u^{\mu_2\mu_3}(p_2,p_3)\,,\\
	&p_{2\,\mu_2} w^{\mu_2\mu_3}(p_2,p_3) = 0 = p_{3\,\mu_3}w^{\mu_2\mu_3}(p_2,p_3)\,,
\end{align}
obtained from the variation of gauge invariant quantities
$F_{\mu\nu}F^{\mu\nu}$ and $(\partial_{\mu} F^{\mu}_{\ \,\lambda})(\partial_{\nu}F^{\nu\lambda})$

\begin{align}
	&u^{\mu_2\mu_3}(p_2,p_3) = -\frac{1}{4}\int\,d^4x_2\,\int\,d^4x_3\ e^{\,i_2p\cdot x_2 + i p_3\cdot x_3}\ 
	\frac{\delta^2 \{F_{\mu\nu}F^{\mu\nu}(0)\}} {\delta A_{\mu_2}(x_2) A_{\mu_3}(x_3)} \,,
	\label{one}\\
	&w^{\mu_2\mu_3}(p_2,p_3) = \frac{1}{2} \int\,d^4x_2\,\int\,d^4x_3\ e^{ip_2\cdot x_2 + i p_3\cdot x_3}\
	\frac{\delta^2 \{\partial_{\mu} F^{\mu}_{\ \,\lambda}\partial_{\nu}F^{\nu\lambda}(0)\}} 
	{\delta A_{\mu_2}(x_2) A_{\mu_3}(x_3)}\,.\label{two}
\end{align}
All the $t_i$'s are transverse in their photon indices
\begin{equation}
	p_{2\,\mu_2} t_i^{\mu_1\nu_1\mu_2\mu_3}=0  \qquad p_{3\,\mu_3} t_i^{\mu_1\nu_1\mu_2\mu_3}=0.
\end{equation}
and $t_2\ldots t_{13}$ are traceless, $t_1$ and $t_2$ are tracefull. With this decomposition, the two vector Ward identities are automatically satisfied by all the amplitudes, as well as the Bose symmetry. \\ 
Coming to the conservation WI for the graviton line, this is automatically satisfied by the two tensor structures $t_1$ and $t_2$, which are completely transverse, while it has to be imposed on the second set ($t_3\ldots t_{13}$) giving three constraints  
\begin{align}
	- p_2^2 F_3 + (3 p_3^2 + 4 p_2\cdot p_3) F_5 + (2 p_2^2 + p_2\cdot p_3) F_7 - p_2^2 p_3^2 F_{10}
	- p_2^2 (p_2^2 + p_2\cdot p_3) F_9 + p_2^2 p_3^2 F_{11} &= 0\,, \label{WIcons1a}\\[2ex]
	p_2^2 F_4 - (3 p_3^2 + 4 p_2\cdot p_3) F_6 - (2 p_2^2 + p_2\cdot p_3) F_8 - p_2\cdot p_3 F_{10}
	+ (p_3^2 + 2 p_2\cdot p_3) F_{11} &= 0\,,\label{WIcons1b}\\[2ex]
	-p_2\cdot p_3 \,(p_2^2 + p_2\cdot p_3) F_9 - p_3^2 (p_3^2 + p_2\cdot p_3) F_{11} + F_{13} + \Pi(p_2^2)  &=0 \,,
	\label{cwi}
\end{align}
plus three symmetric additional ones, obtained by the exchange of the two photon momenta and the symmetries of the form factors corresponding to \eqref{pair}
\begin{equation}
	\begin{split}
		\label{pair1}
		F_5(p_2\leftrightarrow p_3)&=F_3\,,\\
		F_6(p_2\leftrightarrow p_3)&=F_4\,,\\
		F_{10}(p_2\leftrightarrow p_3)&=F_9\,,\\
		F_{12}(p_2\leftrightarrow p_3)&=F_{11}\,.
	\end{split}
\end{equation}
In other words, if we decided to identify from the $F_3\ldots F_{13}$ components a complete transverse traceless sector, using \eqref{pair1} and \eqref{cwi} we would identify only four components in this sector. Such four components, obviously, would be related to the transverse and traceless form factors $(A_1\ldots A_4)$ introduced in the parametrization presented in \cite{Bzowski:2013sza}. Their explicit expressions will be given below. An important aspect of the $F-$basis is that {\em only 1-form factor has to be renormalized} for dimensional reasons, the others being finite. Such form factor, $F_{13}$, plays an important role in the description of the behaviour of the trace parts 
of the same expansion, which involve $t_1$ and $t_2$ (i.e. $F_1$ and $F_2$).

\subsection{Dilatation Ward Identities in the \texorpdfstring{$F$}{}-basis}
The identification of the combination of form factors $F_i$ which span the transverse traceless sector of the correlator can proceed in several ways. In this and in the next section we will proceed by starting from its general expansion in the $F$-basis, and perform a transverse traceless projection, after acting on it with the dilatation and the special conformal transformations. 
This allows to gather the result of the action of the dilatation in terms of coefficients $D_i$ in the form 

\begin{subequations}
	\begin{align}
		0&=\Pi^{\m_1\n_1}_{\a_1\b_1}(p_1)\pi^{\m_2}_{\a_2}(p_2)\pi^{\m_3}_{\a_3}(p_3)\big\{D_1\,p_2^{\a_1}p_2^{\b_1}p_3^{\a_2}p_1^{\a_3}+ D_2\,\d^{\a_2\a_3} p_2^{\a_1}p_2^{\b_1} \notag\\
		&\hspace{4cm}+ D_3\d^{\a_1\a_2}p_2^{\b_1}+p_1^{\a_3}D_4\,\d^{\a_1\a_3}p_2^{\b_1}p_3^{\a_2}+D_5\,\d^{\a_1\a_3}\d^{\a_2\b_1}\big\}
	\end{align}
\end{subequations}
where $D_i$, $i=1,\dots,5$ are differential operator acting on the $F_i$ form factors. In order to verify the previous relation, the coefficients $D_i$ multiplying the independent tensor structures have to vanish, giving a set of differential equation for particular combination of the $F_i$'s. The first equation for $D_1$ will be of the form
\begin{align}
	&D_1=\left(\sum_{i=1}^3p_i\sdfrac{\partial}{\partial p_i}-(d-6)\right)\big[ 4(F_7-F_5-F_3)-2p_2^2F_9-2p_3^2F_{10}\big]=0,
\end{align}
and similarly for the other $D_i$'s, which correspond to
\begin{align}
	&D_2=\left(\sum_{i=1}^3p_i\sdfrac{\partial}{\partial p_i}-(d-4)\right)\big[2(p_1^2-p_2^2-p_3^2)(F_7-F_5-F_3)-4p_2^2p_3^2(F_6-F_8+F_4)-2F_{13}\big]=0\\
	&D_3=\left(\sum_{i=1}^3p_i\sdfrac{\partial}{\partial p_i}-(d-4)\right)\big[p_3^2(p_1^2-p_2^2-p_3^2)F_{10}-2p_2^2\,p_3^2 F_{12}-2F_{13}\big]=0\\
	&D_4=\left(\sum_{i=1}^3p_i\sdfrac{\partial}{\partial p_i}-(d-4)\right)\big[p_2^2(p_1^2-p_2^2-p_3^2)F_9-2p_2^2p_3^2F_{11}-2F_{13}\big]=0\\
	&D_5=\left(\sum_{i=1}^3p_i\sdfrac{\partial}{\partial p_i}-(d-2)\right)\big[(p_1^2-p_2^2-p_3^2)F_{13}\big]=0.
\end{align}
This allows us to identify specific combinations of the $F$'s which will span the transverse traceless sector of the $TJJ$.

\subsection{Special Conformal Ward identities in the \texorpdfstring{$F$}{}-basis}
A similar approach can be followed in the case of the primary and secondary CWI's.
Also in this case we project the special CWI's onto the transverse traceless sector, obtaining
\begin{align}
	0&=\Pi^{\rho_1\sigma_1}_{\mu_1\nu_1}(p_1)\pi^{\rho_2}_{\mu_2}(p_2)\pi^{\rho_3}_{\mu_3}(p_3)\ \bigg\{\,K^\kappa\,\braket{{t^{\mu_1\nu_1}(p_1)\,j^{\mu_2}(p_2)\,j^{\mu_3}(p_3)}}+\sdfrac{4d}{p_1^2}\,\delta^{\kappa\mu_1}\,p_{1\alpha_1}\,\braket{{T^{\alpha_1\nu_1}(p_1)J^{\mu_2}(p_2)J^{\mu_3}(p_3)}}\bigg\}\label{Constr}
\end{align}
where we have used the conservation Ward Identities
\begin{subequations}
	\begin{align}
		p_{2\m_2}\braket{{T^{\m_1\n_1}(p_1)\,J^{\m_2}(p_2)\,J^{\m_3}(p_3)}}&=0 \\
		p_{3\m_3}\braket{{T^{\m_1\n_1}(p_1)\,J^{\m_2}(p_2)\,J^{\m_3}(p_3)}}&=0.
	\end{align}
\end{subequations}

Also in this case one can express the first term in \eqref{Constr} in the form of \eqref{StrucSWIS} in order to isolate the primary and secondary WI's for the form factors. 
\subsection{Primary WI's}
A first set of primary conformal WI's is given by
\begin{subequations}
	\begin{align}
		0=&K_{13}\big[ 4(F_7-F_5-F_3)-2p_2^2F_9-2p_3^2F_{10}\big]\\[1ex]
		0=&K_{13}\big[2(p_1^2-p_2^2-p_3^2)(F_7-F_5-F_3)-4p_2^2p_3^2(F_6-F_8+F_4)-2F_{13}\big]+2\big[ 4(F_7-F_5-F_3)-2p_2^2F_9-2p_3^2F_{10}\big]\\[1ex]
		0=&K_{13}\big[p_3^2(p_1^2-p_2^2-p_3^2)F_{10}-2p_2^2\,p_3^2 F_{12}-2F_{13}\big]-4\big[ 4(F_7-F_5-F_3)-2p_2^2F_9-2p_3^2F_{10}\big]\\[1ex]
		0=&K_{13}\big[p_2^2(p_1^2-p_2^2-p_3^2)F_9-2p_2^2p_3^2F_{11}-2F_{13}]\\[1ex]
		0=&K_{13}\big[(p_1^2-p_2^2-p_3^2)F_{13}\big]-2\big[p_2^2(p_1^2-p_2^2-p_3^2)F_9-2p_2^2p_3^2F_{11}-2F_{13}\big]
	\end{align}
\end{subequations}
and a second set as
\begin{subequations}
	\begin{align}
		0=&K_{23}\big[ 4(F_7-F_5-F_3)-2p_2^2F_9-2p_3^2F_{10}\big]\\[1ex]
		0=&K_{23}\big[2(p_1^2-p_2^2-p_3^2)(F_7-F_5-F_3)-4p_2^2p_3^2(F_6-F_8+F_4)-2F_{13}\big]\\[1ex]
		0=&K_{23}\big[p_3^2(p_1^2-p_2^2-p_3^2)F_{10}-2p_2^2\,p_3^2 F_{12}-2F_{13}\big]-4\big[ 4(F_7-F_5-F_3)-2p_2^2F_9-2p_3^2F_{10}\big]\\[1ex]
		0=&K_{23}\big[p_2^2(p_1^2-p_2^2-p_3^2)F_9-2p_2^2p_3^2F_{11}-2F_{13}]+4\big[ 4(F_7-F_5-F_3)-2p_2^2F_9-2p_3^2F_{10}\big]\\[1ex]
		0=&K_{23}\big[(p_1^2-p_2^2-p_3^2)F_{13}\big]+2\big[p_3^2(p_1^2-p_2^2-p_3^2)F_{10}-2p_2^2\,p_3^2 F_{12}-2F_{13}\big]\notag\\
		&\hspace{6cm}-2\big[p_2^2(p_1^2-p_2^2-p_3^2)F_9-2p_2^2p_3^2F_{11}-2F_{13}\big].
	\end{align}
\end{subequations}
It is clear from the way in which we have organized the contributions in square brackets that they correspond to the same structures identified in the projections of the dilatation WI's. 
\subsection{Secondary WI's}
For completeness we list the secondary Ward Identities obtained in a similar way, which are given by
\begin{align}
	0&=L'_3\big[4(F_7-F_3-F_5)-2p_2^2F_9-2p_3^2F_{10}\big]+2R'\big[p_3^2(p_1^2-p_2^2-p_3^2)F_{10}-2p_2^2\,p_3^2 F_{12}-2F_{13}\big]\notag\\
	&\hspace{1cm}-2R'\big[2(p_1^2-p_2^2-p_3^2)(F_7-F_5-F_3)-4p_2^2p_3^2(F_6-F_8+F_4)-2F_{13}\big]\\[2.5ex]
	0&=L'_1\big[p_2^2(p_1^2-p_2^2-p_3^2)F_9-2p_2^2p_3^2F_{11}-2F_{13}\big]-2p_2^2\big[p_3^2(p_1^2-p_2^2-p_3^2)F_{10}-2p_2^2\,p_3^2 F_{12}-2F_{13}\big]\notag\\
	&\hspace{1cm}+4p_2^2\big[2(p_1^2-p_2^2-p_3^2)(F_7-F_5-F_3)-4p_2^2p_3^2(F_6-F_8+F_4)-2F_{13}\big]+2R'\big[(p_1^2-p_2^2-p_3^2)F_{13}\big]\notag\\[2.5ex]
\end{align}
\begin{align}
	0&=-\sdfrac{2}{p_1^2}\bigg\{L_4\big[4(F_7-F_3-F5)-2p_2^2F_9-2p_3^2F_{10}\big]+R\big[p_3^2(p_1^2-p_2^2-p_3^2)F_{10}-2p_2^2\,p_3^2 F_{12}-2F_{13}\big]\notag\\
	&\hspace{0.5cm}-R\big[p_2^2(p_1^2-p_2^2-p_3^2)F_9-2p_2^2p_3^2F_{11}-2F_{13}\big]\bigg\}+\sdfrac{4d}{p_1^2}\bigg[p_3^2(p_1\cdot p_3-p_1\cdot p_2-p_2\cdot p_3)F_{10}-4F_3p_1\cdot p_2\notag\\
	&\hspace{2cm}-p_2^2p_3^2(F_{11}-F_{12})+4F_5p_1\cdot p_3+2(p_1\cdot p_2-p_1\cdot p_3)F_7+p_2^2(p_1\cdot p_3-p_1\cdot p_2+p_2\cdot p_3)F_9\bigg]\\
	0&=-\sdfrac{2}{p_2^2}\Big\{\,L_2\big[2(p_1^2-p_2^2-p_3^2)(F_7-F_5-F_3)-4p_2^2p_3^2(F_6-F_8+F_4)-2F_{13}\big]\notag\\
	&\hspace{1cm}-p_1^2(p_1^2-p_2^2-p_3^2)(p_3^2F_{10}-p_2^2F_9)+2p_3^2p_2^2p_1^2(F_{12}-F_{11})\Big\}+\sdfrac{4d}{p_1^2}\big[(p_2^2-p_3^2)F_{13}\notag\\
	&\hspace{1cm}+(p_3^4+p_1^4-p_2^4-2p_1^2p_3^2)F_3+2p_2^2p_3^2(p_1^2+p_2^2-p_3^2)F_4+(2p_1^2p_2^2-p_1^4-p_2^4+p_3^4)F_5\notag\\
	&\hspace{1cm}+2p_2^2p_3^2(p_2^2-p_3^2-p_1^2)F_6+2p_2^2p_3^2(p_3^2-p_2^2)F_8+(p_2^4-p_3^4+p_1^2p_3^2-p_1^2p_2^2)F_7\big]\\[2.5ex]
	0&=-\sdfrac{1}{p_1^2}\Big\{L_4\big[p_3^2(p_1^2-p_2^2-p_3^2)F_{10}-2p_2^2\,p_3^2 F_{12}-2F_{13}\big]-2R\big[(p_1^2-p_2^2-p_3^2)F_{13}\big]\Big\}\notag\\
	&\hspace{1cm}+\sdfrac{4d}{p_1^2}\big[p_1\cdot p_3p_2\cdot p_3p_3^2F_{10}+p_\cdot p_2p_2^2p_3^2F_{12}-p_2^2F_{13}\big]\\[2.5ex]
	0&=-\sdfrac{1}{p_1^2}\Big\{L_4\big[p_2^2(p_1^2-p_2^2-p_3^2)F_9-2p_2^2p_3^2F_{11}-2F_{13}\big]+2R\big[(p_1^2-p_2^2-p_3^2)F_{13}\big]\notag\\
	&\hspace{0.5cm}-4p_1^2\big[p_2^2(p_1^2-p_2^2-p_3^2)F_9-2p_2^2p_3^2F_{11}-2F_{13}\big]\Big\}+\sdfrac{4d}{p_1^2}\big[p_2^2p_3^2\,p_1\cdot p_3 F_{11}+p_2^2F_{13}+p_1\cdot p_2 p_2^2 p_3\cdot p_2 F_9\big].
\end{align}
We are now going to use the results above in order to identify the link between the two transverse sections in the $F$-basis introduced by the perturbative expansion and the $A-$basis of the transverse traceless sector. Notice that the 13 form factors of the F-basis form a complete basis in $d$-dimensions, and have some nice properties, as we are going to emphasize below.

\subsection{Connection between the \texorpdfstring{$A-$}{} and the \texorpdfstring{$F-$}{} basis}
By a direct analysis of the previous primary and secondary constraint in the $F-$basis, using the equations given in Sections \ref{primsection} and \ref{secsection} for the 
$A_i$ form factors, we obtain the relations which define the mapping between the transverse traceless sectors in the two basis, which is given by
\begin{subequations}
	\begin{align}
		&A_1=4(F_7-F_3-F_5)-2p_2^2F_9-2p_3^2F_{10}\\
		&A_2=2(p_1^2-p_2^2-p_3^2)(F_7-F_5-F_3)-4p_2^2p_3^2(F_6-F_8+F_4)-2F_{13}\\
		&A_3=p_3^2(p_1^2-p_2^2-p_3^2)F_{10}-2p_2^2\,p_3^2 F_{12}-2F_{13}\\
		&A_3(p_2\leftrightarrow p_3)=p_2^2(p_1^2-p_2^2-p_3^2)F_9-2p_2^2p_3^2F_{11}-2F_{13}\\
		&A_4=(p_1^2-p_2^2-p_3^2)F_{13}.
	\end{align} \label{mapping}
\end{subequations}
It is worth noticing that the form factor $A_3$ and its corresponding $A_3(p_2\leftrightarrow p_3)$ are well-defined since 
\begin{equation}
	F_9(p_2\leftrightarrow p_3)=F_{10},\quad F_{11}(p_2\leftrightarrow p_3)=F_{12}.
\end{equation}

Going back to the full perturbative amplitude we can re-express the entire correlator as
\begin{equation}
	\braket{T^{\m_1\n_1}(p_1)\,J^{\m_2}(p_2)\,J^{\m_3}(p_3)}=\braket{t^{\m_1\n_1}(p_1)\,j^{\m_2}(p_2)\,j^{\m_3}(p_3)}+\braket{t^{\m_1\n_1}_{loc}(p_1)\,J^{\m_2}(p_2)\,J^{\m_3}(p_3)}
\end{equation}
where the semi-local term is expressed exactly as
\begin{align}
	&\braket{t^{\m_1\n_1}_{loc}(p_1)\,J^{\m_2}(p_2)\,J^{\m_3}(p_3)}=\notag\\
	&\qquad=\sdfrac{p_{1\b_1}}{p_1^2}\bigg[2p_1^{(\m_1}\d^{\n_1)}_{\a_1}-\sdfrac{p_{1\a_1}}{(d-1)}\left(\d^{\m_1\n_1}+(d-2)\sdfrac{p_1^{\m_1}p_1^{\n_1}}{p_1^2}\right)\bigg]\,\braket{T^{\a_1\b_1}(p_1)\,J^{\m_2}(p_2)\,J^{\m_3}(p_3)}
	\label{loc2}
\end{align}
and the transverse traceless part is reconstructed as 
\begin{align}
	&\braket{{t^{\m_1\n_1}(p_1)\,j^{\m_2}(p_2)\,j^{\m_3}(p_3)}}_{pert}=\Pi^{\m_1\n_1}_{\a_1\b_1}(p_1)\pi^{\m_2}_{\a_2}(p_2)\pi^{\m_3}_{\a_3}(p_3)\times\notag\\
	&\hspace{0.5cm}\times \bigg\{\ \big[4(F_7-F_3-F_5)-2p_2^2F_9-2p_3^2F_{10}\big]\,p_2^{\a_1}p_2^{\b_1}p_3^{\a_2}p_1^{\a_3}\notag\\
	&\hspace{1cm}+ \big[2(p_1^2-p_2^2-p_3^2)(F_7-F_5-F_3)-4p_2^2p_3^2(F_6-F_8+F_4)-2F_{13}\big]\ \d^{\a_2\a_3} p_2^{\a_1}p_2^{\b_1}\notag\\
	&\hspace{1cm} +\big[\,p_3^2(p_1^2-p_2^2-p_3^2)F_{10}-2p_2^2\,p_3^2 F_{12}-2F_{13}\big]\,\d^{\a_1\a_2}p_2^{\b_1}p_1^{\a_3}\notag\\
	&\hspace{1cm} +\big[p_2^2(p_1^2-p_2^2-p_3^2)F_9-2p_2^2p_3^2F_{11}-2F_{13}\big]\,\d^{\a_1\a_3}p_2^{\b_1}p_3^{\a_2}+ \big[(p_1^2-p_2^2-p_3^2)F_{13}\big]\d^{\a_1\a_3}\d^{\a_2\b_1}\,\bigg\}.
\end{align}
Notice that neither $F_1$ nor $F_2$ will be part of the local contributions since they are both 
completely traceless. Therefore in the $F$-basis, the contributions appearing in \eqref{loc2} will 
be combinations of $F_3\ldots F_{13}$ which are independent from the 4 combinations of the $F'$s identified by the mapping \eqref{mapping}.\\
Since we are still defining the correlator in $d-$dimensions, and it is conformal in this case, then its 
$d$-dimensional trace has to vanish. 
This condition brings in two additional constraints on the two form factors $F_1$ and $F_2$, which now enter into the analysis,  
\begin{align}
	F_1&=\frac{(d-4)}{p_1^2(d-1)}\big[F_{13}-p_2^2\,F_3-p_3^2\,F_5-p_2\cdot p_3\, F_7\big]\\
	F_2&=\sdfrac{(d-4)}{p_1^2(d-1)}\big[p_2^2\,F_4+p_3^2\,F_6+p_2\cdot p_3\,F_8\big],
	\label{confeq}
\end{align}
which will be important for the renormalization procedure and the identification of the anomaly term. \\
We remark that the independent analysis of $A_4$ \cite{Bzowski:2017poo}, which has essentially the same structure 
as $F_{13}$, as one can immediately realize from \eqref{mapping}, shows that in a general conformal field theory the singularity of $F_{13}$ can only be of order $1/(d-4)$ 
and not any higher. We are now going to test such general analysis to the specific case of QED at one-loop.
\subsection{\texorpdfstring{$F_{13}$}{} in QED}
The expressions of the thirteen $F_i$ form factors in $d-$ dimensions are shown in \cite{Coriano:2018bbe}. The renormalized results in $d=4$ have been given in \cite{Armillis:2009pq}. Their expressions are directly written in terms of the two master integrals   
\begin{align}
	B_0(p_1^2,0,0)&=\int \frac{d^d l}{i\pi^2}\frac{1}{l^2(l-p_1)^2}\equiv B_0(p_1^2)\\
	C_0 (p_1^2, p_2^2, p_3^2) &=
	\frac{1}{i \pi^2} \int d^d l \, \frac{1}{(l^2 ) \, ((l -p_2 )^2  ) \, ((l + p_3 )^2  )} 
\end{align}
and this is useful for the discussion of the action of the conformal generators on each of them. Introducing the variables $s=p_1^2$, $s_1=p_2^2$, $s_2=p_3^2$, $\s\equiv s^2-2s(s_1+s_2)+(s_1-s_2)^2$ and $\g\equiv s-s_1-s_2$, $F_{13}$ takes the form 
\begin{align}
	&F_{13, d}(s,s_1,s_2)=\frac{2 \pi ^2 e^2 s^2}{( d-2) ( d-1)  d \sigma ^2} \bigg\{-2 s_2 \big[( d^2-3 d+4) (s+s1)-4( d-1) s_1\big]\notag\\
	&+( d^2-3 d+4) \big[(s-s_1)^2+s_2^2\big]\bigg\}B_0(s)-\sdfrac{\pi ^2 e^2 s_1}{( d-2) ( d-1)  d \sigma ^2 \,\g} \bigg\{ d^3 \sigma ^2- d^2 \big[3 s^4-2 s^3 (7 s_1+9 s_2)\notag\\
	&+8 s^2 \left(3 s_1^2+3 s_1 s_2+4 s_2^2\right)-2 s (s_1-s_2)^2 (9 s_1+11 s_2)+5 (s_1-s_2)^4\big]+2  d \big[2 s^4-11 s^3 (s_1+s_2)\notag\\
	&+s^2 \left(21 s_1^2+24 s_1 s_2+19 s_2^2\right)+s \left(-17 s_1^3+5 s_1^2 s_2+25 s_1 s_2^2-13 s_2^3\right)+(s_1-s_2)^3 (5 s_1-3 s_2)\big]\notag\\
	&+8 s_1 \left(s^3-3 s^2 (s_1+s_2)+3 s \left(s_1^2-s_2^2\right)-(s_1-s_2)^3\right)\bigg\}B_0(s_1)-\sdfrac{\pi ^2 e^2 s_2}{( d-2) ( d-1)  d \sigma ^2\g} \bigg\{ d^3 \sigma ^2\notag\\
	&- d^2 \big[3 s^4-2 s^3 (9 s_1+7 s_2)+8 s^2 \left(4 s_1^2+3 s_1 s_2+3 s_2^2\right)-2 s (s_1-s_2)^2 (11 s_1+9 s_2)+5 (s_1-s_2)^4\big]\notag\\
	&+2  d \big[2 s^4-11 s^3 (s_1+s_2)+s^2 \left(19 s_1^2+24 s_1 s_2+21 s_2^2\right)+s \left(-13 s_1^3+25 s_1^2 s_2+5 s_1 s_2^2-17 s_2^3\right)\notag\\
	&+(s_1-s_2)^3 (3 s_1-5 s_2)\big]+8 s_2 \big[s^3-3 s^2 (s_1+s_2)-3 s \left(s_1^2-s_2^2\right)+(s_1-s_2)^3\big]\bigg\}B_0(s_2)\notag\\
	&+\sdfrac{4 \pi ^2 e^2 s^2 s_1 s_2 ( d \sigma +8 s_1 s_2)}{( d-2)  d \sigma ^2 \g}C_0(s,s_1,s_2).
\end{align}
which we will study in the $d\to 4$ limit. As discussed in \cite{Giannotti:2008cv, Armillis:2009pq}, the singularity of this form factor comes from the scalar form factor $\Pi(p^2)$ of the photon 2-point function $\braket{JJ}$. 
The singularity of this correlator will be at all orders of the form $1/\epsilon$ in a conformal theory and not higher. This is 
a crucial point in the proof which is clearly not satisfied in a non-conformal theory. In fact, the only available counterterm to regulate a conformal theory of this kind is given by 
\begin{equation}
	\frac{1}{\epsilon}\int d^4 x \sqrt{g} F_{\mu\nu}F^{\mu\nu}
\end{equation}
which renormalizes the 2-point function $\langle JJ \rangle$ and henceforth $F_{13}$. Explicit computations in QED at one-loop, where the theory is conformal, show that 
\begin{equation}
	\label{f13}
	F_{13}= G_0(p_1^2, p_2^2,p_3^3) -\frac{1}{2} \, [\Pi (p_2^2) +\Pi (p_3^2)]
\end{equation}
We just recall that the structure of the two-point function of two conserved vector currents of scaling dimensions $\Delta_1$ and $\Delta_2$ is given by \cite{Coriano:2013jba}
\begin{equation}
	\label{TwoPointVector2}
	G_V^{\alpha \beta}(p) = \delta_{\Delta_1 \Delta_2}  \, c_{V 12}\, 
	\frac{\pi^{d/2}}{4^{\Delta_1 - d/2}} \frac{\Gamma(d/2 - \Delta_1)}{\Gamma(\Delta_1)}\,
	\left( \delta^{\alpha \beta} -\frac{p^\alpha p^\beta}{p^2} \right)\
	(p^2)^{\Delta_1-d/2} \,,
\end{equation}
with $c_{V12}$ being an arbitrary constant. It will be nonvanishing only if the two currents share the same dimensions, and it is characterized just by a single pole (to all orders) ${1/\epsilon}$ in dimensional regularization. The divergence can be regulated with $d \to d - 2 \epsilon$, and expanding the product $\Gamma(d/2-\Delta)\,(p^2)^{\Delta - d/2}$ in \ref{TwoPointVector2} in a Laurent series around $d/2 - \Delta = -n$ (integer) one can extract the single pole in $1/\epsilon$ in the form \cite{Coriano:2013jba}
\begin{equation}
	\label{expansion}
	\Gamma\left(d/2-\Delta\right)\,(p^2)^{\Delta-d/2} = \frac{(-1)^n}{n!} \left( - \frac{1}{\epsilon} + \psi(n+1)  + O(\epsilon) \right) (p^2)^{n + \epsilon} \,,
\end{equation}
where $\psi(z)$ is the logarithmic derivative of the Gamma function, and $\epsilon$ takes into account the divergence of the two-point correlator for particular values of the scale dimension $\eta$ and of the space-time dimension $d$.
In the QED case, the renormalization involves only the master integrals $B_0(s_i)$, which gives ($d=4 -2 \epsilon$)
\begin{equation}
	\label{f13ren}
	F_{13}=\frac{2 \pi^2 e^2}{3 \epsilon} + F_{13}^{fin},
\end{equation}
implying that $F_1$ (from \eqref{confeq}) will be given by
\begin{equation}
	\label{limit}
	F_1=\left(\sdfrac{2}{3}\sdfrac{\epsilon}{s} \right)\left(\frac{2 \pi e^2}{3\epsilon} + F_{13}^{fin} \right),
\end{equation}
which in the $d\to 4$ limit gives 
\begin{equation}
	\lim_{d\to 4} F_1=-\frac{4}{9}\frac{\pi^2}{s},
\end{equation}
showing the appearance of an anomaly pole in the single form factor which is responsible for the trace anomaly.\\
It is quite obvious that the non-perturbative analysis of \cite{Bzowski:2017poo} in the $A-$basis and the perturbative ones in the $F-$basis are consistent. There is some additional  important information that we can extract in the latter basis if we go back to the two equations in \eqref{confeq}.

1. From the finiteness of all the form factors, except for $F_{13}$ which is regulated with a $1/\epsilon$ divergence, it is obvious that in the limit of $d\to 4$, as a result of \eqref{limit},
$F_1$ is nonvanishing and exhibits a $1/p_1^2\equiv 1/s$ behaviour. Therefore, the emergence of $F_1$ in $d=4$ as a form factor which accounts for the anomaly is a nice feature if this general analysis. It shows how to link an anomaly pole to the renormalization of a single form factor in the expansion of the correlator. 

2. At the same time, it is possible to check explicitly from its $d-$dimensional expression shown in \eqref{confeq} that the second form factor $F_2$ vanishes as $d\to 4$, proving that there will be one and only one tensor structure of nonzero trace.

3. Although the results above are fully confirmed by the previous perturbative analysis, 
they hold generically (non perturbatively) in the context of the conformal realizations of such correlator.\\ 
A natural question to ask is what happens to the tensor structures $t_2,\ldots t_{13}$ as we move from $d$ to 4 dimensions. The answer is quite immediate.  We contract such 
structures with the $d-$dimensional metric $g_{\mu\nu}(d)$ and perform the $d\to 4$ limit. One can easily check that $t_9,t_{10},t_{11}$ and $t_{12}$, remain traceless in any dimensions, 
while the remaining ones become traceless in this limit. For instance 
\begin{align}
	g_{\mu_1\nu_1}^{(d)}\,t_1^{\mu_1\nu_1\mu_2\mu_3}&=(d-4)\,p_1^2\,u^{\mu_2\mu_3}(p_2,p_3)\\
	g_{\mu_1\nu_1}^{(d)}\,t_2^{\mu_1\nu_1\mu_2\mu_3}&=(d-4)\,p_1^2\,w^{\mu_2\mu_3}(p_2,p_3),
\end{align}
and similarly for the others. Therefore, in the $d\to 4$ limit, the $F-$basis satisfies all the original constraints and the separation between traceless and trace-contributions which were described in \secref{pchecks}.
\begin{figure}[t]
	\centering
	\vspace{-1cm}
	\subfigure[]{\includegraphics[scale=0.9]{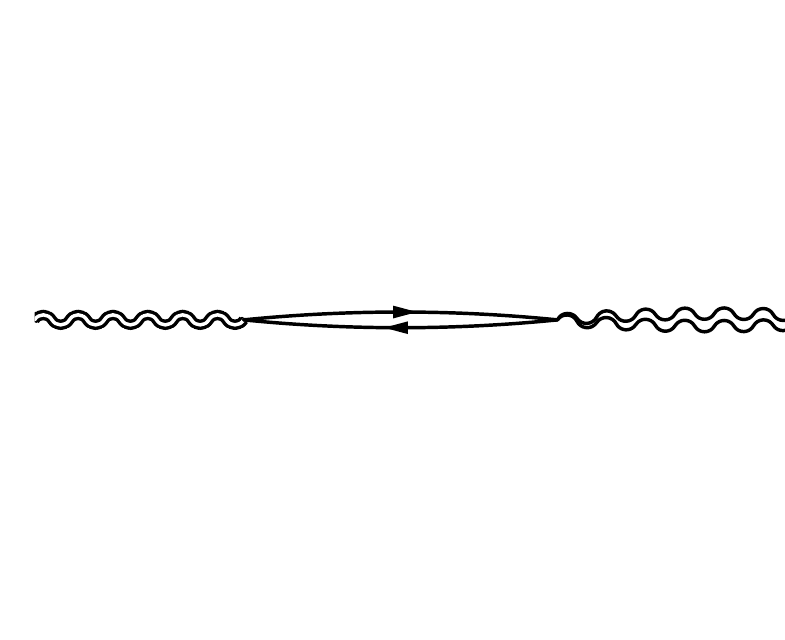}}\hspace{0.5cm}
	\subfigure[]{\includegraphics[scale=0.77]{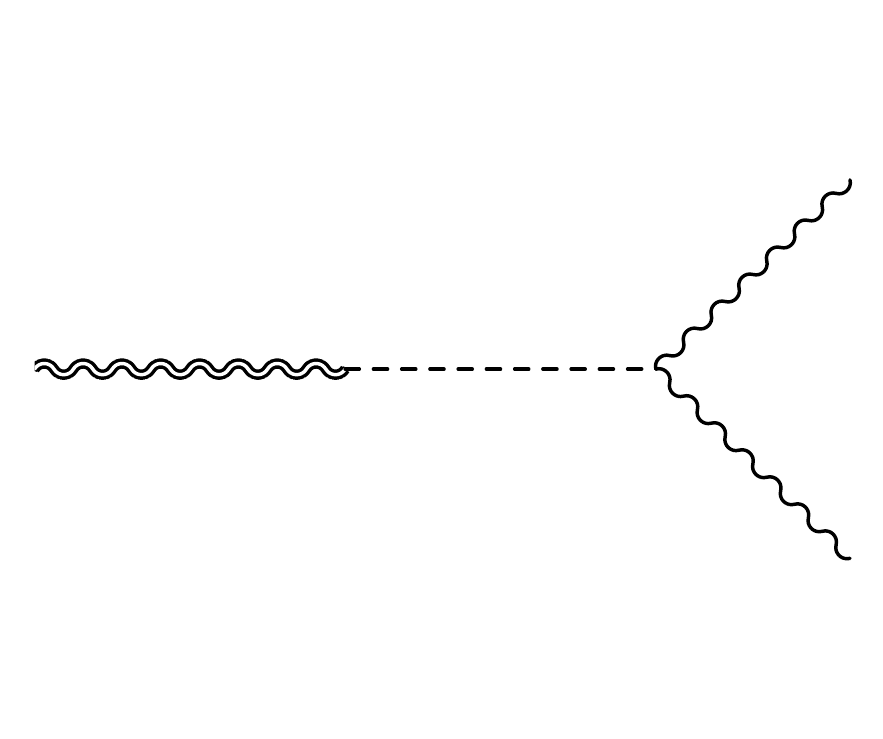}}
	\caption{\small Fig. (a): Dispersive description of the singularity of the spectral density $\rho(s)$ as a spacetime process. Fig. (b): The exchange of a pole as the origin of the conformal anomaly in the TJJ viewed in perturbation theory.}
	\label{collinear}
\end{figure}
\subsection{Implications}
We may summarize the result of this section by saying that the emergence of an anomaly pole in the $TJJ$ is not limited to perturbation theory but is a specific feature of the non-perturbative solution as well. The perturbative description offers a simple view of why this phenomenon takes place. In the dispersive representation of the unique form factor which is responsible for the appearance of the anomaly $(F_1)$, this phenomenon is related to the exchange of a collinear fermion/antifermion pair in the $s$ variable $(\rho(s)\sim \delta(s))$ (see Fig. \ref{collinear}). This configuration provides a contribution to the anomaly action of the form 
\begin{equation}
	\label{pole}
	\mathcal{S}_{pole}= - \frac{e^2}{ 36 \pi^2}\int d^4 x d^4 y \left(\square h(x) - \partial_\mu\partial_\nu h^{\mu\nu}(x)\right)  \square^{-1}_{x\, y} F_{\alpha\beta}(x)F^{\alpha\beta}(y)
\end{equation}
In view of the equivalence between the perturbative and the nonperturbative solution for the $A_i$ (and henceforth for the $F$- basis), which will be discussed in section 13, it is obvious that this phenomenon is lifted from its perturbative origin and acquires a general meaning. We refer to \cite{Giannotti:2008cv,Coriano:2014gja} for a general perturbative analysis 
of the spectral densities for such type of vertices.

There is no doubt that this is a one-loop phenomenon in QED which is obviously violated at higher orders, since the theory, in this case, ceases to be conformal. However, as we are going to show in the next section, the one-loop expression in QED reproduces the {\em entire} non perturbative conformal BMS solution, and this explains why our proof should be considered a definitive prove of the fact that, at least for this correlator, the exchange of effective massless interactions is the key signature of the conformal anomaly. 

\section{Conformal Ward identities and the perturbative master integrals}
We now turn to illustrate the action of the conformal generators on the $d-$dimensional expressions of the $F_i$, showing that they are indeed solutions of the corresponding CWI's. We will first elaborate on the action of the conformal generators on the simple master integrals $B_0$, $C_0$. Such an action will be reformulated in terms of the external invariant of each master integral, starting from the original definitions of the special conformal $K^\kappa$ and dilatation $D$ operators. \\
Understanding the way the conformal constraints work on perturbative realizations of conformal correlators is indeed important for various reasons. For instance, one can find, by a direct perturbative analysis simpler realizations of the general hypergeometric solutions of the CWI's for the $A_i$, while, at the same time, one can test the consistency of the general approach implemented in the solution of the conformal constraints which does not rely on an explicit Lagrangian but just on the data content of a given CFT. We are going to show that indeed such is the case and that the $d$-dimensional form factors given explicitly in the appendix of \cite{Coriano:2018bbe} satisfy all the conformal constraints corresponding to the dilatation and special conformal WI's.

\subsection{The conformal constraints on the external invariants}
The action of the special conformal transformations $K^\kappa$ and of the dilatations $D$ on them, will require that $K_{ij}$, $L_N$, $R$ in \eqref{kij}, \eqref{Ldef} and \eqref{Rdef} be expressed in terms of the three invariants $s, s_1$ and $s_2$.  Taking the four-momentum $p_1^\mu$ as dependent, we will be using the chain rules
\begin{align}
	\sdfrac{\partial}{\partial p_2^\m}&=-\sdfrac{p_{1\m}}{p_1}\sdfrac{\partial}{\partial p_1}+\sdfrac{p_{2\m}}{p_2}\sdfrac{\partial}{\partial p_2}\\
	\sdfrac{\partial}{\partial p_3^\m}&=-\sdfrac{p_{1\m}}{p_1}\sdfrac{\partial}{\partial p_1}+\sdfrac{p_{3\m}}{p_3}\sdfrac{\partial}{\partial p_3},
\end{align}
and by taking appropriate linear combinations of these relations we obtain the system of equations
\begin{equation}
	\begin{pmatrix}
		p_2^\m\sdfrac{\partial}{\partial p_2^\m}-p_2^\m\sdfrac{\partial}{\partial p_3^\m}\\[3ex]
		p_3^\m\sdfrac{\partial}{\partial p_2^\m}-p_3^\m\sdfrac{\partial}{\partial p_3^\m}
	\end{pmatrix}=
	\begin{pmatrix}
		&\\[-1ex]
		p_2& -\sdfrac{p_2\cdot p_3}{p_3}\\[3ex]
		\sdfrac{p_2\cdot p_3}{p_2}&-p_3\\[2ex]
	\end{pmatrix} \begin{pmatrix}
		\sdfrac{\partial}{\partial p_2}\\[3ex]
		\sdfrac{\partial}{\partial p_3}
	\end{pmatrix}\ .
\end{equation}

Solving the system above for the derivative of the magnitudes of the momenta we obtain the relations
\begin{subequations}
	\begin{align}
		\frac{\partial}{\partial p_2}&=\left(\frac{p_2\,p_3}{(p_2\cdot p_3)^2-p_2^2\,p_3^2}\right)\left[\frac{p_2\cdot p_3}{p_3}\left(p_3^\m\sdfrac{\partial}{\partial p_2^\m}-p_3^\m\sdfrac{\partial}{\partial p_3^\m}\right)-p_3\left(p_2^\m\sdfrac{\partial}{\partial p_2^\m}-p_2^\m\sdfrac{\partial}{\partial p_3^\m}\right)\right]\\[2ex]
		\frac{\partial}{\partial p_3}&=\left(\frac{p_2\,p_3}{(p_2\cdot p_3)^2-p_2^2\,p_3^2}\right)\left[p_2\left(p_3^\m\sdfrac{\partial}{\partial p_2^\m}-p_3^\m\sdfrac{\partial}{\partial p_3^\m}\right)-\frac{p_2\cdot p_3}{p_2}\left(p_2^\m\sdfrac{\partial}{\partial p_2^\m}-p_2^\m\sdfrac{\partial}{\partial p_3^\m}\right)\right].
	\end{align}\label{p2p3}
\end{subequations}
Afterwards, by taking other linear combinations we obtain
\begin{align}
	\left\{\begin{matrix}
		p_2^\m\sdfrac{\partial}{\partial p_2^\m}+p_3^\m\sdfrac{\partial}{\partial p_2^\m}=p_1\sdfrac{\partial}{\partial p_1}-\sdfrac{p_1\cdot p_2}{p_2}\sdfrac{\partial}{\partial p_2}\\[3ex]
		p_3^\m\sdfrac{\partial}{\partial p_3^\m}+p_2^\m\sdfrac{\partial}{\partial p_3^\m}=p_1\sdfrac{\partial}{\partial p_1}-\sdfrac{p_1\cdot p_3}{p_3}\sdfrac{\partial}{\partial p_3}
	\end{matrix}\right.\ .
\end{align}
which, combined together, give
\begin{align}
	p_1\frac{\partial}{\partial p_1}=&\left(\frac{p_2\,p_3}{(p_2\cdot p_3)^2-p_2^2\,p_3^2}\right)\bigg[\frac{p_2\cdot p_3}{p_2\,p_3}\big((p_2\cdot p_3)+p_2^2\big)\,p_3^\m\sdfrac{\partial}{\partial p_3^\m}\notag\\[2ex]
	&+\frac{p_2\cdot p_3}{p_2\,p_3}\big((p_2\cdot p_3)+p_3^2\big)\,p_2^\m\sdfrac{\partial}{\partial p_2^\m}-\frac{p_2}{p_3}\big((p_2\cdot p_3)+p_3^2\big)\,p_3^\m\sdfrac{\partial}{\partial p_2^\m}-\frac{p_3}{p2}\big((p_2\cdot p_3)+p_2^2\big)\,p_2^\m\sdfrac{\partial}{\partial p_3^\m}\bigg],\label{p1}
\end{align}

with the four-vector forms of the derivatives rearranged as
\begin{subequations}
	\begin{align}
		p_2^\m\frac{\partial}{\partial p_2^\m}\,B_0(s)&=\sdfrac{( d-4)}{2s}(s+s_1-s_2)\,B_0(s)\\ 
		p_3^\m\frac{\partial}{\partial p_2^\m}\,B_0(s)&=\sdfrac{( d-4)}{2s}(s-s_1+s_2)\,B_0(s)\\
		p_2^\m\frac{\partial}{\partial p_3^\m}\,B_0(s)&=\sdfrac{( d-4)}{2s}(s-s_1+s_2)\,B_0(s)\\
		p_3^\m\frac{\partial}{\partial p_2^\m}\,B_0(s)&=\sdfrac{( d-4)}{2s}(s+s_1-s_2)\,B_0(s).
	\end{align}
\end{subequations}
Inserting these expressions in \eqref{p2p3} and \eqref{p1} we obtain the relations
\begin{subequations}
	\begin{align}
		p_1\frac{\partial}{\partial p_1}B_0(s)=( d-4)\,B_0(s),\qquad\frac{\partial}{\partial p_2}B_0(s)=0,\qquad
		\frac{\partial}{\partial p_3}B_0(s)=0.
	\end{align}
\end{subequations}
The four-derivatives of $B_0(s_1)$ are computed in a similar way, obtaining
\begin{subequations}
	\begin{align}
		p_2^\m\frac{\partial}{\partial p_2^\m}\,B_0(s_1)&=( d-4)\,B_0(s_1)\\
		p_3^\m\frac{\partial}{\partial p_2^\m}\,B_0(s_1)&=\sdfrac{( d-4)}{2s_1}(s-s_1-s_2)\,B_0(s_1)\\
		p_2^\m\frac{\partial}{\partial p_3^\m}\,B_0(s_1)&=p_3^\m\frac{\partial}{\partial p_2^\m}\,B_0(s_1)=0,
	\end{align}
\end{subequations}
giving
\begin{subequations}
	\begin{align}
		p_1\frac{\partial}{\partial p_1}B_0(s_1)=0,\qquad\frac{\partial}{\partial p_2}B_0(s_1)=\sdfrac{( d-4)}{p_2}\,B_0(s_1),\qquad
		\frac{\partial}{\partial p_3}B_0(s_1)=0.
	\end{align}
\end{subequations}

Finally, for the scalar 2-point function $B_0(s_2)$ we obtain
\begin{subequations}
	\begin{align}
		p_2^\m\frac{\partial}{\partial p_2^\m}\,B_0(s_2)&=p_3^\m\frac{\partial}{\partial p_2^\m}\,B_0(s_2)=0\\
		p_2^\m\frac{\partial}{\partial p_3^\m}\,B_0(s_1)&=\sdfrac{( d-4)}{2s_2}(s-s_1-s_2)\,B_0(s_2)\\
		p_3^\m\frac{\partial}{\partial p_2^\m}\,B_0(s_1)&=( d-4)\,B_0(s_2),
	\end{align}
\end{subequations}
and
\begin{subequations}
	\begin{align}
		p_1\frac{\partial}{\partial p_1}B_0(s_2)=0,\qquad\frac{\partial}{\partial p_2}B_0(s_2,0,0)=0,\qquad
		\frac{\partial}{\partial p_3}B_0(s_2)=\sdfrac{( d-4)}{p_3}\,B_0(s_2).
	\end{align}
\end{subequations}

Finally, we consider the action of these operators on the scalar integral $C_0(s,s_1,s_2)$, obtaining the relations
\begin{subequations}
	\begin{align}
		p_2^\m\frac{\partial}{\partial p_2^\m}\,C_0(s,s_1,s_2)&=\sdfrac{( d-3)}{s}\big[B_0(s_1)-B_0(s_2)\big]+\sdfrac{( d-6)s+( d-4)(s_1-s_2)}{2s}C_0(s,s_1,s_2)\\[2ex]
		p_3^\m\frac{\partial}{\partial p_2^\m}\,C_0(s,s_1,s_2)&=\sdfrac{( d-3)}{s\,s_1}\big[(s_1-s)B_0(s_2)+sB_0(s)-s_1B_0(s_1)\big]\notag\\
		&\hspace{4cm}+\sdfrac{( d-4)(s-s_1)(s+s_1-s_2)}{2s\,s_1}C_0(s,s_1,s_2)\\[2ex]
		p_2^\m\frac{\partial}{\partial p_3^\m}\,C_0(s,s_1,s_2)&=\sdfrac{( d-3)}{s\,s_2}\big[(s_2-s)B_0(s_1)+sB_0(s)-s_2B_0(s_2)\big]\notag\\
		&\hspace{4cm}+\sdfrac{( d-4)(s-s_2)(s-s_1+s_2)}{2s\,s_2}C_0(s,s_1,s_2)\\[2ex]
		p_3^\m\frac{\partial}{\partial p_2^\m}\,C_0(s,s_1,s_2)&=\sdfrac{( d-3)}{s}\big[B_0(s_2)-B_0(s_1)\big]+\sdfrac{( d-6)s-( d-4)(s_1-s_2)}{2s}C_0(s,s_1,s_2).
	\end{align}
\end{subequations}
Using the relations given above, after some algebra, we obtain the expressions
\begin{subequations}
	\begin{align}
		\sdfrac{\partial}{\partial p_1}C_0&=\sdfrac{p_1}{\s\,s}\bigg\{ 2( d-3)\big[(s+s_1-s_2)\,B_0(s_1)+(s-s_1+s_2)\,B_0(s_2)-2s\,B_0(s)\big]\notag\\
		&\qquad+\big[( d-4)(s_1-s_2)^2-( d-2)s^2+2s(s_1+s_2)\big]\,C_0(s,s_1,s_2)\bigg\}\\[2ex]
		\sdfrac{\partial}{\partial p_2}C_0&=\sdfrac{p_2}{\s\,s_1}\bigg\{ 2( d-3)\big[(s+s_1-s_2)\,B_0(s)+(s_1+s_2-s)\,B_0(s_2)-2s_1\,B_0(s_1)\big]\notag\\
		&\quad+\big[2s_2\big(s_1-( d-4)s\big)+(s-s_1)\big(( d-4)s+( d-2)s_1\big)+( d-4)s_2^2\big]C_0(s,s_1,s_2)\bigg\}\\
		\sdfrac{\partial}{\partial p_3}C_0&=\sdfrac{p_3}{\s\,s_2}\bigg\{ 2( d-3)\big[(s-s_1+s_2)\,B_0(s)+(s_1+s_2-s)\,B_0(s_1)-2s_2\,B_0(s_2)\big]\notag\\
		&\qquad+\big[( d-4)(s-s_1)^2-( d-2)s_2^2+2s_2(s+s_1)\big]C_0(s,s_1,s_2)\bigg\}.
	\end{align}
\end{subequations}

At this point, one can write the operators $K$, $L_N$ and $R$ of the special CWI's in terms of the three invariants $s$, $s_1$ and $s_2$ as
\begin{equation}
	\frac{\partial}{\partial s}=\frac{1}{2p_1}\frac{\partial}{\partial p_1},\qquad\frac{\partial}{\partial s_1}=\frac{1}{2p_2}\frac{\partial}{\partial p_2},\qquad \frac{\partial}{\partial s_2}=\frac{1}{2p_3}\frac{\partial}{\partial p_3},
\end{equation}
and from the explicit form of the $K_i$ operator in \eqref{kij} we obtain
\begin{equation}
	K_i=\d_{i,j+1}\left[4s_j\sdfrac{\partial^2}{\partial s_j^2}+2( d+2-2\D_i)\,\sdfrac{\partial}{\partial s_j}\right],\quad i=1,2,3,
\end{equation}
where we have set $s_0\equiv s$.\\
In the same way, we find that the operators $L_N$ and $R$, from their defining expressions \eqref{Ldef} 
and \eqref{Rdef}, take the form

\begin{align}
	L_N&=2(s+s_1-s_2)\,s\frac{\partial}{\partial s}+4s\,s_1\,\frac{\partial }{\partial s_1}+ \big[(2d-\D_1-2\D_2+N)\,s+(2\D_1-d)(s_2-s_1)\big]\\
	R&=2\,s\,\frac{\partial}{\partial s}-(2\D_1-d),
\end{align}
while their symmetric versions are given by
\begin{align}
	L'_N&=L_N\ \ \text{with $s\leftrightarrow s_1$ and $\D_1\leftrightarrow\D_2$} \\
	R'&=R\ \ \text{with $s\mapsto s_1$ and $\D_1\mapsto\D_2$}.
\end{align}
We then obtain
\begin{subequations}
	\begin{align}
		K_{12}\ B_0(s)&=K_{13}\ B_0(s)=-\sdfrac{4( d-4)}{s}\,B_0(s,0,0),\qquad K_{23}\ B_0(s)=0\\
		K_{21}\ B_0(s_1)&=K_{23}\ B_0(s_1)=-\sdfrac{2( d-4)}{s_1}\,B_0(s_1),\qquad K_{13}\ B_0(s_1)=0\\
		K_{31}\ B_0(s_2)&=K_{32}\ B_0(s_2)=-\sdfrac{2( d-4)}{s_2}\,B_0(s_2),\qquad K_{12}\ B_0(s_2)=0,
	\end{align}
\end{subequations}
and 
\begin{subequations}
	\begin{align}
		&K_{12}\ C_0(s,s_1,s_2)\notag\\
		&\hspace{0.2cm}=\sdfrac{4( d-3)}{s\,s_1\,\s}\bigg[-\big(s^2+(s-2s_1)(s_1-s_2)\big)\,B_0(s_2)+s(s+5s_1-s_2)\,B_0(s)-2s_1\,(2s+s_1-s_2)\,B_0(s_1)\bigg]\notag\\
		&\hspace{0.5cm}+\sdfrac{1}{s\,s_1\,\s}\big[4( d-3)(3s-3s_1-s_2)\,s\,s_1+2( d-4)(s-2s_1)\s\big]\,C_0(s,s_1,s_2)
		\\[2ex]
		&K_{23}\ C_0(s,s_1,s_2)\notag\\
		&\hspace{0.2cm}=\sdfrac{4( d-3)}{s_1\,s_2\,\s}\bigg[(s_1-s_2)(s-s_1-s_2)\,B_0(s)+s_1(s_1-s+2s_2)\,B_0(s_1)+s_2(s-3s_1-s_2)\,B_0(s_2)\bigg]\notag\\
		&\hspace{0.5cm}+\sdfrac{1}{s_1\,s_2\,\s}\big[2( d-4)(s_1-s_2)[(s-s_1)^2-2s\,s_2+s_2^2]+4( d-2)(s_1-s2)\,s_1\,s_2\big]\,C_0(s,s_1,s_2)\\[2ex]
		&K_{13}\ C_0(s,s_1,s_2)\notag\\
		&\hspace{0.2cm}=\sdfrac{4( d-3)}{s\,s_2\,\s}\bigg[-\big(s^2+(s-2s_2)(s_2-s_1)\big)\,B_0(s_1)+s(s-s_1+5s_2)\,B_0(s)-2s_2(2s-s_1+s_2)\,B_0(s_2)\bigg]\notag\notag\\
		&\hspace{0.5cm}+\sdfrac{1}{s\,s_2\,\s}\big[4( d-3)(3s-3s_2-s_1)\,s\,s_2+2( d-4)(s-2s_2)\s\big]\,C_0(s,s_1,s_2),
	\end{align}
\end{subequations}
while the action of the operators $L_N$ and $R$ on $B_0(s)$ is given by
\begin{subequations}
	\begin{align}
		L_N\,B_0(s)&=\big[(N-2)s-4s_1+4s_2\big]\,B_0(s)\notag\\
		L'_N\,B_0(s)&=\big[(d+N-7)s_1+(d-2)(s_2-s)\big]\,B_0(s)\notag\\
		R\,B_0(s)&=-4B_0(s)\notag\\
		R'\,B_0(s)&=-(d-2)B_0(s),
	\end{align}
\end{subequations}
and on $B_0(s_1)$ 
\begin{subequations}
	\begin{align}
		L_N\,B_0(s_1)&=\big[(d+N-6)s+d(s_2-s_1)\big]\,B_0(s_1)\notag\\
		L'_N\,B_0(s_1)&=\big[(N-3)s_1-2s+2s_2\big]\,B_0(s_1)\notag\\
		R\,B_0(s_1)&=-d\,B_0(s_1)\notag\\
		R'\,B_0(s_1)&=-2\,B_0(s_1).
	\end{align}
\end{subequations}
Similarly, for $B_0(s_2)$ we obtain
\begin{subequations}
	\begin{align}
		L_N\,B_0(s_2)&=\big[(N-d+2)s+d(s_2-s_1)\big]\,B_0(s_2)\notag\\
		L'_N\,B_0(s_2)&=\big[(N-d+1)s_1+(d-2)(s_2-s)\big]\,B_0(s_2)\notag\\
		R\,B_0(s_2)&=-d\,B_0(s_2)\notag\\
		R'\,B_0(s_2)&=-(d-2)\,B_0(s_2),
	\end{align}
\end{subequations}
and for $C_0(s,s_1,s_2)$
\begin{subequations}
	\begin{align}
		L_N\,C_0(s,s_1,s_2)&=2(d-3)\big[B_0(s_1)-B_0(s_2)\big]+\big[(N-4)s-4(s_1-s_2)\big]C_0(s,s_1,s_2)\notag\\
		L'_N\,C_0(s,s_1,s_2)&=2(d-3)\big[B_0(s)-B_0(s_2)\big]+\big[(N-5)s_1-2(s-s_2)\big]C_0(s,s_1,s_2)\notag\\
		R\,C_0(s,s_1,s_2)&=\sdfrac{2(d-3)}{\s}\big[(s+s_1-s_2)\,B_0(s_1)+(s-s_1+s_2)\,B_0(s_2)-2s\,B_0(s)\big]\notag\notag\\
		&+\frac{2}{\s}\big[-(d-1)s^2+(d+1)s(s_1+s_2)-2(s_1-s_2)^2\big]C_0(s,s_1,s_2)\notag\\
		R'\,C_0(s,s_1,s_2)&=\sdfrac{2(d-3)}{\s}\big[(s+s_1-s_2)\,B_0(s)+(-s+s_1+s_2)\,B_0(s_2)-2s_1\,B_0(s_1)\big]\notag\notag\\
		&-\frac{2}{\s}\big[s(s_1-d\,s_1-2s_2)+(s_1-s_2)(d-2)s_1-s_2(s_1-s_2)+s^2\big]C_0(s,s_1,s_2) .
	\end{align}
\end{subequations}
Using the explicit expressions of the form factors $F_i$ presented in \cite{Coriano:2018bbe}, we have explicitly verified that they satisfy all the conformal constraints, once the action on the two master integrals is re-expressed according to the derivative rules that we have derived in this section in terms of the external invariants.

\section{Renormalization and Anomalous Ward Identities in QED }
In this final section we turn to the derivation of the anomalous CWI's for this correlator in QED. We recall that the singularity in the form factor $F_{13}$, which is the only one which is affected by the renormalization of the 
$F-$basis, can be removed by using the counterterm
\begin{equation}
	\int d^4x\,\sqrt{-g} F_{\mu\nu}F^{\mu\nu}.
\end{equation}
After the renormalization procedure in the F-basis, the renormalized expressions of the $A_i$ form factors are obtained using the mapping \eqref{mapping}. One can then derive renormalized anomalous (dilatation, primary and secondary) WI's. Using the explicit expressions of the $A_i$ given in \cite{Coriano:2018bbe}, and performing the renormalization, we obtain

\begin{align}
	\left(\sum_{i}^{3}p_i\sdfrac{\partial}{\partial p_i}+2\right)\,A_1&=0=-\m\sdfrac{\partial}{\partial \m}A_1\\
	\left(\sum_{i}^{3}p_i\sdfrac{\partial}{\partial p_i}\right)\,A_2^R&=\sdfrac{8\p^2\,e^2}{3}=-\m\sdfrac{\partial}{\partial \m}A_2^R\\
	\left(\sum_{i}^{3}p_i\sdfrac{\partial}{\partial p_i}\right)\,A_3^R(p_2\leftrightarrow p_3)&=\sdfrac{8\p^2\,e^2}{3}=-\m\sdfrac{\partial}{\partial \m}A_3^R\\
	\left(\sum_{i}^{3}p_i\sdfrac{\partial}{\partial p_i}-2\right)\,A_4^R&=-\sdfrac{4}{3}\p^2\,e^2(s-s_1-s_2)=-\m\sdfrac{\partial}{\partial \m}A_4^R.
\end{align}
for the dilatation WI's, while for the primary CWI's we obtain
\begin{equation}
	\begin{split}
		&K_{13}A_1=0\\
		&K_{13}A^R_2=-2A_1\\
		&K_{13}A^R_3=4A_1\\
		&K_{13}A^R_3(p_2\leftrightarrow p_3)=0\\
		&K_{13}A^R_4=2A^R_3(p_2\leftrightarrow p_3)-\sdfrac{16\,\pi^2e^2}{3}
	\end{split}
	\hspace{1.5cm}
	\begin{split}
		&K_{23}A_1=0\\
		&K_{23}A^R_2=0\\
		&K_{23}A^R_3=4A_1\\
		&K_{23}A^R_3(p_2\leftrightarrow p_3)=-4A_1\\
		&K_{23}A^R_4=-2A^R_3+2A^R_3(p_2\leftrightarrow p_3).
	\end{split}
\end{equation}
In all the equations $e$ is the renormalized charge and can be traded for $\beta(e)$ by the relation $\beta(e)/e=e^2/(12 \pi^2)$. 
For the secondary CWI's we obtain
\begin{equation}
	\begin{split}
		&L_4 A_1+R A_3^R-R A_3^R(p_2\leftrightarrow p_3)=0\\[1.5ex]
		&L_2\,A_2^R-s\,(A_3^R-A_3^R(p_2\leftrightarrow p_3))=\sdfrac{16}{9}\pi^2e^2\left[3s1\,B_0^R(s_1,0,0)-3s_2B_0^R(s_2,0,0)-s1+s2\right]+\sdfrac{24}{9}\pi^2e^2s\\[1.5ex]
		&L_4\,A^R_3-2R\,A^R_4=\sdfrac{32}{9}\pi^2\,e^2s_2\left[1-3B_0^R(s_2,0,0)\right]+\sdfrac{48}{9}\pi^2\,e^2\,s\\[1.5ex]
		&L_4\,A^R_3(p_2\leftrightarrow p_3)+2R\,A^R_4-4\,sA^R_3(p_2\leftrightarrow p_3)=\sdfrac{32}{9}\,\pi^2\,e^2s_1\left[3B_0^R(s_1,0,0)-1\right]\\[1.5ex]
		&L'_3\,A_1^R-2R'A_2^R+2R'A_3^R=0\\[1.5ex]
		&L'_1\,A_3^R(p_2\leftrightarrow p_3)+p_2^2(4A^R_2-2A^R_3)+2R'A^R_4=\sdfrac{16}{3}\pi^2\,e^2\,s_1,\\
	\end{split}
\end{equation}
where we have used the relations 
\begin{equation}
	\sdfrac{\partial}{\partial s_i}\,B_0^R(s_j,0,0)=-\sdfrac{\d_{ij}}{s_i}\  i=0,1,2
\end{equation}
where $s_0=s$ and 
\begin{align}
	\sdfrac{\partial}{\partial s}\,C_0&= \sdfrac{1}{s\,\s}\,\big[s(s_1+s_2-s)C_0+B_{0,R}(s_1)(s+s_1-s_2)+B_{0,R}(s_2)(s-s_1+s_2)-2s\,B_{0,R}(s)\big]\\
	\sdfrac{\partial}{\partial s_1}\,C_0&= \sdfrac{1}{s_1\,\s}\,\big[s_1(s+s_2-s_1)C_0+B_{0,R}(s)(s+s_1-s_2)+B_{0,R}(s_2)(s_1-s+s_2)-2s_1\,B_{0,R}(s_1)\big]\\
	\sdfrac{\partial}{\partial s_2}\,C_0&= \sdfrac{1}{s_2\,\s}\,\big[s_2(s+s_1-s_2)C_0+B_{0,R}(s_1)(s_2+s_1-s)+B_{0,R}(s)(s-s_1+s_2)-2s_2\,B_{0,R}(s_2)\big].
\end{align}
We have defined $\s=s^2-2s(s_1+s_2)+(s_1-s_2)^2$,  $B_{0,R}(s_i)\equiv B_{0,R}(s_i,0,0)=2-\log\left(-\sdfrac{s_i}{\mu^2}\right)$ and, for simplicity, $C_0\equiv C_0(s,s_1,s_2)$. \\
In $d=4$ the operator $K_i$ and $L_N$ take the forms
\begin{equation}
	K_1\equiv 4s\sdfrac{\partial^2}{\partial s^2}-4\sdfrac{\partial}{\partial s},\quad K_2\equiv 4s_1\sdfrac{\partial^2}{\partial s_1^2},\quad K_3\equiv 4s_2\sdfrac{\partial^2}{\partial s_2^2}
\end{equation}
\begin{align}
	L_N&\equiv 2s\,(s+s_1-s_2)\sdfrac{\partial}{\partial s}+4s\,s_1\sdfrac{\partial}{\partial s_1}+\left[(N-2)s+4(s_2-s_1)\right],\qquad R\equiv2s\,\sdfrac{\partial}{\partial s}-4\\
	L'_N&\equiv 2s_1\,(s+s_1-s_2)\sdfrac{\partial}{\partial s_1}+4s\,s_1\sdfrac{\partial}{\partial s}+\left[(N-3)s_1+2(s_2-s)\right],\qquad R'\equiv2s_1\,\sdfrac{\partial}{\partial s_1}-2.
\end{align}
\section{Comparing the non perturbative $3K$ solutions with the perturbative result in QED}
The perturbative analysis presented above provides a significant check of the consistency of the non perturbative approach for 3-point functions \cite{Bzowski:2013sza, Bzowski:2015yxv, Bzowski:2017poo}. However, at the same time, it carries a lot of insight about the connection between CFT's realized by free field theories and the structure of the corresponding nonperturbative solutions. In order to clarify this point we proceed 
with a direct comparison between the expressions of the $A_i$ given in \cite{Coriano:2018bbe} and the analogous results for the same correlator given in \cite{Bzowski:2013sza}. The $A_i$'s have been obtained by the recomputed $F_i$'s.

The final outcome of this analysis will be, by this direct check, that free field theories 
completely saturate the general nonperturbative solution, providing  drastic simplifications of the results presented in \cite{Bzowski:2013sza}. As already mentioned, the latter have been presented in the form of parametric integrals of 3 Bessel functions, which obviously amount to linear combinations of Appell functions, originally introduced in studies of the AdS/CFT correspondence. \\
To make our treatment self-contained we need to provide some basic description of the structure 
of the non perturbative solutions and then we will establish a direct link between these expressions and the simplified ones given in \cite{Coriano:2018bbe}, corresponding to the perturbative $A_i$'s. 
The BMS solutions take the form 
\begin{equation}
	\begin{split}
		A_1 &=\alpha_1 J_{4[000]}  \\
		A_2 &= \alpha_1 J_{3[100]} +\alpha_3 J_{2[000]}\\
		A_3 &=2 \alpha_1 J_{3[001]} +\alpha_3 J_{2[000]} \\
		A_4 &=  2 \alpha_1 J_{2[011]} +\alpha_3 J_{1[010]} +\alpha_4 J_{0[000]} 
		\label{nonper}
	\end{split}
\end{equation}
where the $J_{n[k_1 k_2 k_3]}$ are integrals corresponding to hypergeometric functions $F_4$ of 2 variables. In general  they are defined as 
\begin{equation}
	J_{n[k_1 k_2 k_3]}=I_{d/2 -1 + n[\beta_1\beta_2\beta_3]}
	\label{tworel}
\end{equation}
where 
\begin{equation}
	I_{\alpha[\sigma_1\sigma_2\sigma_3]}(p_1,p_2,p_3)=\int_0^\infty dx x^\alpha \prod_{j=1}^3 p_j^{\sigma_j}K_{\sigma_j}(p_j x)
\end{equation}
with $\beta_i=\Delta_i -d/2 + k_i$. In our case $\Delta_1=d$ and $\Delta_2=\Delta_3=d-1$. 
The parametric integral is expressed in terms of products of modified Bessel functions $K_\nu(x)$ of second kind.  The explicit expressions of such integrals have been worked out in \cite{Bzowski:2013sza}. All the $J$ integrals appearing in the solution correspond to master integrals of the form \cite{Coriano:2013jba}
\begin{equation}
	\label{davy}
	J(\nu_1,\nu_2,\nu_3) = \int \frac{d^d l}{(2 \pi)^d} \frac{1}{(l^2)^{\nu_3} ((l+p_1)^2)^{\nu_2} ((l-p_2)^2)^{\nu_1}}\, ,
\end{equation}
which can be directly connected to 3-point functions of scalar operators, of suitable scaling dimensions $\Delta_i$ by the relations 
\begin{align}
	& \int \frac{d^d p_1}{(2\pi)^d} \frac{d^d p_2}{(2\pi)^d} \frac{d^d p_3}{(2\pi)^d} \, (2\pi)^d \delta^{(d)}(p_1 + p_2 + p_3) \, 
	J(\nu_1,\nu_2,\nu_3) e^{- i p_1 \cdot x_1 - i p_2 \cdot x_2 - i p_3 \cdot x_3} \notag\\ 
	& = \frac{1}{4^{\nu_1+\nu_2+\nu_3} \pi^{3 d/2}}  \frac{\Gamma(d/2 - \nu_1) \Gamma(d/2 - \nu_2) \Gamma(d/2 - \nu_3)}{\Gamma(\nu_1) 
		\Gamma(\nu_2) \Gamma(\nu_3)}  \frac{1}{(x_{12}^2)^{d/2- \nu_3} (x_{23}^2)^{d/2- \nu_1} (x_{31}^2)^{d/2- \nu_2}}\,, \notag\\ 
	\label{oner}
\end{align}
with 
\begin{align}
	\label{etafromnu}
	\Delta_1 = d - \nu_2 - \nu_3 \,, \qquad
	\Delta_2 = d - \nu_1 - \nu_3 \,, \qquad
	\Delta_3 = d - \nu_1 - \nu_2 \, 
\end{align}
\begin{align}
	\label{etafromnu}
	\nu_1= \frac{1}{2} (d + \Delta_1 -\Delta_2-\Delta_3) \qquad
	\nu_2 = \frac{1}{2}(d - \Delta_1 + \Delta_2 -\Delta_3)\,, \qquad
	\nu_3 = \frac{1}{2}(d - \Delta_1 - \Delta_2 +\Delta_3)\,
\end{align}
An equivalent expression of the master integral $J(\nu_1,\nu_2,\nu_3)$ can be obtained combining the expressions above in the form
\begin{eqnarray}
	J(\nu_1,\nu_2,\nu_3)
	&=\frac{\pi^{-d/2} 2^{4 - 3 d/2} }{\Gamma(\nu_1)\Gamma(\nu_2)\Gamma(\nu_3)\Gamma(d-\nu_1-\nu_2-\nu_3)} p_1^{d/2-\nu_2-\nu_3} p_2^{d/2-\nu_1-\nu_3}p_3^{d/2-\nu_1-\nu_2}\notag\\
	& \times \int_0^\infty d x x^{d/2-1}K_{d-\nu_1-\nu_3} (p_1 x) K_{d-\nu_2-\nu_3} (p_2 x) K_{d-\nu_1-\nu_2} (p_3 x) 
\end{eqnarray}
An alternative expression for this integral can be found in \cite{Coriano:2013jba}. 
Notice that if we plug the scaling dimensions of $T^{\mu\nu}$ and of the current $J^\rho$, for instance in $J_{0[000]}$, we encounter divergences which need to be regulated using a rather complex scheme which has been discussed in \cite{Bzowski:2013sza}. This implies that the coefficients of the scalar 3-point functions contained in the $J$ integrals 
correspond to generalized master integrals with indices $\nu_i$ which are {\em real} numbers rather than just integers, as in the case of ordinary perturbation theory, for instance in QCD. This sets them apart from the standard integrals appearing in massless theories at higher orders. \\ 
Because of this, it is not obvious that the general 3K solution can be directly related to the simple master integrals which appear in the $A_i$ containing the two master integrals $B_0$ and $C_0$. In fact, all the relations presented in \cite{Bzowski:2013sza} which allow to connect various $J$ integrals are, therefore, unrenormalized and need to be expanded in a regulator in order to generate the final expressions for each form factor. This, unfortunately does not allow to recognize that such solutions can be drastically simplified. By the same token, the solutions that we have provided for the $A_i$ using the Fuchsian analysis of section (\ref{fuchs}) cannot be easily recognized for being related just to ordinary $B_0$ and $C_0$ master integrals. \\
We are now going to show that the same information provided by the $J$ integrals is entirely reproduced by the perturbative solution in a simplified way. This clearly implies that 
there are significant cancellations among the contributions coming from different $J$ integrals in the BMS solution, in whatever form they are written, which are far from being evident. The latter, obviously, remains essential in order to determine the minimal set of constants which characterize the general conformal solution and allow to establish a link between this and the perturbative result, but their significance probably stops here.   \\
In order to establish this link, we consider different field theories in various dimensions characterized by a generic 
number of degenerate massless fermions, say $n_\psi$, which, as we are going to show, will be taking the role of $\alpha_1$ in the final solution. Let's investigate this point.

As shown in \cite{Bzowski:2013sza}, the secondary CWI's allow to express the 4 constants $\alpha_i$ in terms of $\alpha_1$ and the normalization of the 2-point function $C_J$. These relations, in general, involve a regulator, except for specific dimensions. In $d=3$ and $d=5$, for instance, the correlator is finite and the relation between the perturbative and the non perturbative expressions of the $A_i$ is transparent and can be worked out analytically. \\ 
We start from the two point functions,  since the $C_J$ presented \cite{Bzowski:2013sza} and in free field theory (QED) must be matched. In \cite{Bzowski:2013sza} the normalization of the $JJ$ correlator is defined by the relation
\begin{equation}
	\braket{J^{\m}(p)\,J^{\n}(-p)}=C_J\,\pi^{\m\n}(p)\G\left(1-\frac{d}{2}+\frac{\epsilon}{2}\right)\,p^{d-2-\epsilon}\label{SkendCJ}
\end{equation}
while in our case, in conformal QED, we find in dimensional regularization
\begin{align}
	\braket{\ J^{\m}(p)\,J^{\n}(-p)\ }&=\frac{2\p^\frac{D}{2}\,e^2\,(d-2)}{d-1}\,p^2\,\p^{\m\n}(p)\,B_0(p^2,0,0)\notag\\
	&=\frac{2\p^\frac{d}{2}\,e^2\,(d-2)\,\p^{\m\n}(p)}{(d-1)\,\G\left(d-2\right)}\ \left[\G\left(\frac{d}{2}-1\right)\right]^2\G\left(2-\frac{D}{2}\right)p^{d-2}\label{QEDCJ},
\end{align}
having used the explicit expression of the two point scalar integral 
\begin{equation}
	{B}_0(p_1^2,0,0)=\sdfrac{1}{i\p^\frac{d}{2}}\int\,d^d \ell\ \frac{l}{\ell^2(\ell-p_1)^2}=\frac{\p^\frac{d}{2}\ \left[\G\left(\frac{d}{2}-1\right)\right]^2\G\left(2-\frac{d}{2}\right)}{\p^\frac{d}{2}\,\G\left(d-2\right)(p_1^2)^{2-\frac{d}{2}}}\label{B0ex1}
\end{equation}
In odd spacetime dimension $n$ ($d=n +\epsilon$), the limit $\epsilon\to0$ is finite and then the comparison between \eqref{SkendCJ} and \eqref{QEDCJ} gives the value for the normalization constant $C_J$ of the two point function $\braket{JJ}$ as
\begin{equation}
	C_J=\frac{2\p^\frac{d}{2}\,e^2\,(d-2)}{(d-1)\,\G\left(d-2\right)}\ \frac{\left[\G\left(\frac{d}{2}-1\right)\right]^2\G\left(2-\frac{d}{2}\right)}{\G\left(1-\frac{d}{2}\right)}\label{CJ}.
\end{equation}
It is simple to verify that the values of $C_J$ in the $d=3$ and $d=5$ are those given by \eqref{CJ}, in fact
\begin{align}
	C_J&\ \ \mathrel{\stackrel{\makebox[0pt]{\mbox{\normalfont\tiny $d=3$}}}{=}}\ \ \p^\frac{3}{2}e^2\, \left[\G\left(\frac{1}{2}\right)\right]^3\left[\G\left(-\frac{1}{2}\right)\right]^{-1}=-\frac{\pi^\frac{5}{2}\,e^2}{2}\\[2ex]
	C_J&\ \ \mathrel{\stackrel{\makebox[0pt]{\mbox{\normalfont\tiny $d=5$}}}{=}}\ \ \frac{3\,\p^\frac{5}{2}e^2}{2\,\G(3)}\, \left[\G\left(\frac{3}{2}\right)\right]^2\,\G\left(-\frac{1}{2}\right)\,\left[\G\left(-\frac{3}{2}\right)\right]^{-1}=-\frac{9\,\pi^\frac{7}{2}\,e^2}{32}
\end{align}
With this information, we can proceed further, showing that 
the expressions of the $A_i$'s presented in \eqref{mapping} agree with those determined in the nonperturbative solution \eqref{nonper}. 

\subsection{Explicit result in $d=3$ (QED and scalar QED)}\label{explicitd3}

We consider the fermion and the scalar contribution combined together, selecting an arbitrary number of massless fermions and scalars running inside the loops, with constants $n_F$ and $n_S$ related to this condition. \\
In the particular case of $d=3$  there are simplifications both in the exact and the perturbative solutions. The scalar integral ${B}_0$ is given by \eqref{B01}, for which in $d=3$ becomes
\begin{equation}
	{B}_0(s,0,0)=\frac{\p^{3/2}}{\,p_1}\label{B01}
\end{equation}
where $p_1=|p_1|=\sqrt{p_1^2}$. The scalar 3-point function ${C}_0$ can be simplified using the star-triangle rule for which
\begin{equation}
	\int\frac{d^{d}x}{[(x-x_1)^2]^{\a_1}\,[(x-x_2)^2]^{\a_2}\,[(x-x_3)^2]^{\a_3}}\ \  \mathrel{\stackrel{\makebox[0pt]{\mbox{\normalfont\tiny $\sum_i\a_i=d$}}}{=}}\ \  \frac{i\pi^{d/2}\n(\a_1)\n(\a_2)\n(\a_3)}{[(x_2-x_3)^2]^{\frac{d}{2}-\a_1}\,[(x_1-x_2)^2]^{\frac{d}{2}-\a_3}\,[(x_1-x_3)^2]^{\frac{d}{2}-\a_2}}\label{startriangle1}
\end{equation}
where
\begin{equation}
	\n(x)=\frac{\G\left(\frac{d}{2}-x\right)}{\G(x)},
\end{equation}
which holds only if the condition $\sum_i\a_i=d$ is satisfied. In the case $d=3$  \eqref{startriangle1} is proportional to the three point scalar integral, and in particular
\begin{align}
	{C}_0(p_1^2,p_2^2,p_3^2)&=\int\,\frac{d^d\ell}{i\p^\frac{d}{2}}\frac{1}{\ell^2(\ell-p_2)^2(\ell+p_3)^2}=\int\,\frac{d^dk}{i\p^\frac{d}{2}}\frac{1}{(k-p_1)^2(k+p_3)^2(k+p_3-p_2)^2}\notag\\[2ex]
	&=\frac{\left[\G\left(\frac{d}{2}-1\right)\right]^3}{\,(p_1^2)^{\frac{d}{2}-1}(p_2^2)^{\frac{d}{2}-1}(p_3^2)^{\frac{d}{2}-1}}\ \mathrel{\stackrel{\makebox[0pt]{\mbox{\normalfont\tiny $d=3$}}}{=}}\ \frac{\p^{3/2}}{\,p_1\,p_2\,p_3}.\label{C01}
\end{align}

The explicit expression of the form factors in $d=3$ using the perturbative approach to one loop order, can be found by substituting the expression of the scalar integral, using \eqref{B01} and \eqref{C01}, and then considering the limit $d\to3$ for all the form factors. The scalar and the fermion cases contribute equally, modulo an overall constant, giving 
\begin{align}
	A_{1,D=3}&=\left(\frac{\p^3\,e^2(8n_F+n_S)}{6}\right)\frac{2(4p_1+p_2+p_3)}{(p_1+p_2+p_3)^4}\\
	A_{2,D=3}&=\left(\frac{\p^3\,e^2(8n_F+n_S)}{6}\right)\frac{2\,p_1^2}{(p_1+p_2+p_3)^3}-\left(\frac{\p^3\,e^2\,(8n_F-n_S)}{2}\right)\frac{(2p_1+p_2+p_3)}{(p_1+p_2+p_3)^2}\\
	A_{3,D=3}&=\left(\frac{\pi ^3 e^2 (8 n_F+n_S)}{6} \right)\frac{(-2 p_1^2-3 p_1 p_2+3 p_1 p_3-p_2^2+p_3^2)}{(p_1+p_2+p_3)^3}-\left(\frac{\pi ^3 e^2 (8 n_F-n_S) }{2}\right)\frac{(2 p_1+p_2+p_3)}{ (p_1+p_2+p_3)^2}\\
	A_{4,D=3}&=\left(\frac{\pi ^3 e^2 (8 n_F+n_S)}{6}\right)\frac{ (2 p_1+p_2+p_3) (p_1^2-(p_2+p_3)^2+4 p_2 p_3)}{2(p_1+p_2+p_3)^2}\notag\\
	&\hspace{4cm}+\left(\frac{\pi ^3 e^2 (8 n_F-n_S) }{4}\right)\left(\frac{2p_1^2}{ (p_1+p_2+p_3)-p_2-p_3}\right)
\end{align}
which coincide with the solution given by BMS in the limit $d=3$, modulo the identification of the constants
\begin{equation}
	\a_1=\left(\frac{\p^3\,e^2(8n_F+n_S)}{6}\right),\qquad c_J=-\left(\frac{\p^\frac{5}{2}\,e^2\,(8n_F-n_S)}{8}\right).
\end{equation}
\subsection{Explicit result in $d=5$ (QED and scalar QED)}

The case $d=5$ is slightly more involved. In fact, while the result of ${B}_0$ is still the same, the explicit form of the ${C}_0$ needs some manipulations. We start from the star-triangle relation \eqref{startriangle1} 
\begin{align}
	\int\frac{d^{d}x}{[(x-x_1)^2]^{2}\,[(x-x_2)^2]\,[(x-x_3)^2]^{2}}&\ \mathrel{\stackrel{\makebox[0pt]{\mbox{\normalfont\tiny $d=5$}}}{=}}\ \frac{i\pi^{d/2}\n(2)\n(1)\n(2)}{[(x_2-x_3)^2]^{\frac{d}{2}-2}\,[(x_1-x_2)^2]^{\frac{d}{2}-2}\,[(x_1-x_3)^2]^{\frac{d}{2}-1}}\\[2ex]
	&=\frac{i\p^4}{2}\frac{1}{[(x_2-x_3)^2]^{1/2}\,[(x_1-x_2)^2]^{1/2}\,[(x_1-x_3)^2]^{3/2}}\label{eq}
\end{align}
and use an integration by parts to reduce the left-hand side. In particular, by setting $x_1\to p_1$, $x_2\to -p_3$, $x_3\to p_1-p_3$ we obtain
\begin{align}
	\int\frac{d^{d}x}{[(x-x_1)^2]^{2}\,[(x-x_2)^2]\,[(x-x_3)^2]^{2}}&=\int\frac{d^{d}x}{[(x-p_1)^2]^{2}\,[(x+p_3)^2]\,[(x-p_1+p_3)^2]^{2}}\notag\\[2ex]
	&=\int\frac{d^{d}\ell}{[(\ell)^2]^{2}\,(\ell-p_2)^2\,[(\ell+p_3)^2]^{2}}=\frac{i\p^4}{2\,p_1\,p_2\,p_3^3}
\end{align}
and 
\begin{align}
	\int\frac{d^{d}\ell}{[(\ell)^2]^{2}\,(\ell-p_2)^2\,[(\ell+p_3)^2]^{2}}&=-\frac{i\p^\frac{d}{2}(d-4)\big[(d-6)(s-s_1)^2-(d-4)s_2^2+2s_2(s+s_1)\big]}{4s\,s_1\,s_2^2}\ {C}_0(s,s_1,s_2)\notag\\[2ex]
	&\hspace{-5cm}-\frac{i\p^\frac{d}{2}\,(d-3)\big[(d-6)(s-s_1)+(d-4)s_2\big]}{2s\,s_1\,s_2^2}\,{B}_0(s,0,0)+\frac{i\p^\frac{d}{2}\,(d-3)\big[(d-6)(s-s_1)-(d-4)s_2\big]}{2s\,s_1\,s_2^2}\,{B}_0(s_1,0,0)\notag\\[2ex]
	&+\frac{i\p^\frac{d}{2}\,(d-3)\big[(d-6)(s+s_1)+(d-4)s_2\big]}{2s\,s_1\,s_2^2}\,{B}_0(s_2,0,0)
\end{align}
where $s=p_1^2$, $s_1=p_2^2$ and $s_2=p_3^2$ as previously. Inserting this expression into Eq. \eqref{eq} and solving for ${C}_0$, in $d=5$ one finds
\begin{equation}
	{C}_0(s,s_1,s_2)=\frac{\p^{3/2}}{p_1+p_2+p_3}.
\end{equation}
From \eqref{B0ex1} the $B_0$ is calculated in $d=5$ as
\begin{equation}
	{B}_0(s,0,0)=-\frac{\p^{3/2}}{4} p_1.
\end{equation}
Plugging these results into the general form of the $A_i$ given in appendix of \cite{Coriano:2018bbe} we find
\begin{align}
	A_{1,d=5}&=\frac{\pi ^4 e^2 (8 n_F+n_S)}{120 (p_1+p_2+p_3)^5} \Big[4 p_1^4+20 p_1^3 (p_2+p_3)+4 p_1^2 \left(7 \big(p_2+p_3\big)^2+6p_2 p_3\right)\notag\\
	&\hspace{2cm}+15 p_1 (p_2+p_3) \left(\big(p_2+p_3\big)^2+ p_2 p_3\right)+3 (p_2+p_3)^2 \left(\big(p_2+p_3\big)^2+ p_2 p_3\right)\Big]\\[2ex]
	A_{2,d=5}&=\frac{\pi ^4 e^2 p_1^2 (8 n_F+n_S) }{120 (p_1+p_2+p_3)^4}\Big[(p_1+p_2+p_3)^3+(p_1 p_2+p_1 p_3+p_2 p_3) (p_1+p_2+p_3)+3 p_1 p_2 p_3\Big]\notag\\
	&\quad+\frac{\pi ^4 e^2 (24 n_F-n_S)}{48 (p_1+p_2+p_3)^3} \Big[2 p_1^4+6 p_1^3 (p_2+p_3)+2 p_1^2 \left(5 \big(p_2+p_3\big)^2+p_2 p_3\right)\notag\\
	&\hspace{2cm}+9 p_1 \left(p_2^3+2 p_2^2 p_3+2 p_2 p_3^2+p_3^3\right)+3 (p_2+p_3)^2 \left(\big(p_2+p_3\big)^2-p_2 p_3\right)\Big]
\end{align}
\begin{align}
	A_{3,d=5}&=\frac{\pi ^4 e^2 (8 n_F+n_S)}{240 (p_1+p_2+p_3)^4} \Big[(-2 p_1^5-8 p_1^4 (p_2+p_3)-8 p_1^3 p_2 (2 p_2+3 p_3)\notag\\
	&+p_1^2 \left(-19 p_2^3-40 p_2^2 p_3+24 p_2 p_3^2+15 p_3^3\right)-3 (p_2-p_3) (p_2+p_3) \left(p_2^2+3 p_2 p_3+p_3^2\right) (4 p_1+p_2+p_3)\Big]\notag\\
	&+\frac{\pi ^4 e^2 (24 n_F-n_S)}{48 (p_1+p_2+p_3)^3} \Big[2 p_1^4+6 p_1^3 (p_2+p_3)+2 p_1^2 \left(5 p_2^2+6 p_2 p_3+5 p_3^2\right)\notag\\
	&\hspace{2cm}+9 p_1 \left(p_2^3+2 p_2^2 p_3+2 p_2 p_3^2+p_3^3\right)+3 (p_2+p_3)^2 \left(p_2^2+p_2 p_3+p_3^2\right)\Big]
	\\[2ex]
	A_{4,d=5}&=\frac{\pi ^4 e^2 (8 n_F+n_S) }{480 (p_1+p_2+p_3)^3}\Big[2 p_1^6+6 p_1^5 (p_2+p_3)+4 p_1^4 \left(2 (p_2+p_3)^2-p_2 p_3\right)\notag\\
	&\hspace{2cm}+p_1^2 \left((p_2+p_3)^2+p_2 p_3\right) \left(3 p_1 (p_2+p_3)-7 (p_2+p_3)^2+32 p_2 p_3\right)\notag\\
	&\hspace{3cm}-3 (p_2+p_3) \left((p_2+p_3)^2-4 p_2 p_3\right) \left((p_2+p_3)^2+p_2 p_3\right) (3 p_1+p_2+p_3)\Big]\notag\\
	&-\frac{\pi ^4 e^2 (24 n_F-n_S) }{96 (p_1+p_2+p_3)^2}\Big[2 p_1^5+4 p_1^4 (p_2+p_3)+4 p_1^3 \left(p_2^2+p_2 p_3+p_3^2\right)\notag\\
	&\quad-p_1^2 (p_2+p_3) \left(p_2^2-5 p_2 p_3+p_3^2\right)-6 p_1 \left(p_2^4+p_2^3 p_3+p_2 p_3^3+p_3^4\right)-3 (p_2+p_3)^3 \left(p_2^2-p_2 p_3+p_3^2\right)\Big]
\end{align}
in agreement with the solution given by BMS in the limit $d=5$, with the identifications 
\begin{equation}
	\a_1=\left(\frac{\pi ^4 e^2 (8 n_F+n_S)}{240} \right),\qquad c_J=-\left(\frac{3\pi ^\frac{7}{2} e^2 (24 n_F-n_S)}{128 }\right).
\end{equation}
This shows that, after fixing the normalization of the 2-point function, we are esentially left, both from the perturbative and the non perturbative side, with the same solution. 


\section{Comments} 
There are several apparent conclusions that we can draw from this comparative study of the $TJJ$ that we are going to summarise briefly.\\
The $A_i$ computed in perturbation theory satisfy the same anomalous conformal Ward identities as the non-perturbative ones, as shown by us in the previous sections for $d=4$.\\
They both satisfy homogeneous (non-anomalous) CWI's in general $(d)$ dimensions. \\
In $d=3$ and $d=5$, the two solutions match entirely, having consistently matched the $JJ$ 2-point function's normalisation in the two separate cases.\\
In $d=5$, where the corresponding field theory is nonrenormalisable, the perturbative computation still matches the non-perturbative one. This occurs because the theory's non-renormalizability does not play any role, matching the two theories a purely one-loop result (one loop saturation).\\ 
We conclude that, at least for the $TJJ$, free field theories in momentum space at one loop provide the same information derived from the non-perturbative solutions, and the two can be freely interchanged, being equivalent. The two analytical derivations are therefore expected to coincide for any dimension. As already mentioned, this implies that there should be significant cancellations among the contributions of the 3K integrals or those given by us in \secref{finfin} in such a way that they can be expressed in terms of the elementary master integrals $B_0$ and $C_0$.\\
The two expressions of $A_1$,  which is affected only by a single integration constant $\alpha_1$, in the two forms given in section \ref{TJJmom} and \eqref{nonper}, show quite directly that $J_{4 [000]}$ can be expressed in terms of the elementary master integrals present in the perturbative expansion. Direct checks for the other $J$ integrals appearing in the expressions of $A_2, A_3...$ and so on, are harder to perform, since each of these form factors depends on at least two $J$ integrals, as evident from Eq. (\ref{nonper}).\\
We should mention that this result {\em is not unexpected} since it had been shown in (\cite{Coriano:2012wp} section 3) that the solution for the $TJJ$ (but also for the TOO and $JJJ$) in coordinate space presented by Osborn and Petkou \cite{Osborn:1993cr} could be {\em completely reproduced } by the simple one-loop diagrams of scalar QED and QED using fermion and scalar sectors. These correlators were "completely integrable", in the words of \cite{Coriano:2012wp}, which meant that such expressions could proceed with a direct Fourier transform to momentum space without the need of introducing an extra regulator ($\omega$) for the transform. 

The perturbative ones mirror the non-perturbative solutions, showing that results obtained at one loop in the perturbative description are automatically inherited by the non-perturbative one, being identical in 3-point functions. The general solutions, both in coordinate and position space, can be reproduced entirely using a simple one-loop analysis. Since conformal symmetry for 3-point functions is "saturated" at one loop, this explains why radiative corrections in such correlators which preserve the conformal symmetry are necessarily bound to be proportional to the one-loop result, as in the case of the $AVV$ diagram, for instance.  

We are then entitled to come to immediate conclusions concerning the analysis presented in \cite{Armillis:2009pq} in regards to the emergence of anomaly poles in one-loop QED, which are not an artefact of perturbation theory, but are naturally part of the non-perturbative solution and are {\em protected by the conformal symmetry}. 
\section{Conclusions}
We have  presented a general discussion of the transition from position to momentum space in the analysis of tensor correlators, clarifying some aspects of 
the realisation of the conformal generators in this space. 
Then we have moved toward a direct analysis of the $TJJ$ vertex, detailing some of the involved intermediate steps, technically demanding, used in the BMS reconstruction of such correlator.  
In a recent work \cite{Coriano:2018zdo}, we have presented the physical motivations, derived from combined perturbative and non-perturbative studies of the $TJJ$ vertex, why anomaly poles in anomaly vertices should be considered the key signature of the conformal anomaly. The goal of this work has been to fill in the intermediate steps of our previous analysis. The presence of such massless degrees of freedom generated in the presence of anomalies shows that they are not an artefact of perturbation theory.  
We have studied a simple instantiation of 
such vertex in massless QED, and confronted it with the 
general approach for solving the conformal constraints in momentum space for scalars \cite{Armillis:2009pq,Bzowski:2013sza} and tensor correlators \cite{Bzowski:2013sza,Bzowski:2017poo}. 
Such solutions do not rely on any Lagrangian realisation and are, therefore, very general. In this way, it is possible to establish an essential connection between perturbative and non-perturbative approaches to analyse specific correlators, bringing significant simplifications of the general result.

Even though massless QED is not a conformal theory, the violation of conformality is associated to the $\beta$-function of the running coupling constant and, precisely for this reason, specific correlation functions containing one or more insertions of stress-energy tensors play, in this context, an important role, due to the trace anomaly. \\
There has been considerable interest in analysing the breaking of the conformal symmetry in realistic non-conformal theories such as QED and QCD, and in their manifestations through higher perturbative orders \cite{Kataev:1996ce,Broadhurst:1993ru}. In this respect, the analysis of exact solutions, at least for 3-point functions, may shed light on the manifestation of the conformal breaking, which is associated to a non-zero $\beta$ function and, according to our analysis and the analysis of \cite{Giannotti:2008cv,2009PhLB..682..322A,Armillis:2009pq}, is accompanied by the appearance of an anomaly pole in the effective action.\\
We have shown that the one-loop result in perturbation theory saturates the non-perturbative solutions. This implies that the principal master integrals necessary to expand the non-perturbative result should turn out to be just $B_0$ and $C_0$, in agreement with former analysis in coordinate space. These analyses share the goal of establishing the form of the anomaly action and address its nonlocality, given their role in describing conformal symmetry breaking. Therefore, it is not surprising that such massless interactions have received attention in investigating the role of the chiral and the conformal anomalies in topological insulators and Weyl semi-metals, as recently emphasised in \cite{Chernodub:2017jcp,Rinkel:2016dxo}. 
The study of such materials provides a direct application of the properties of the vertices that we have been interested in.
The parallel study of perturbative and non-perturbative methods stops at the level of 3-point functions since higher point functions need to be bootstrapped. However, the connection found in our work indicates that the possibility of studying a skeleton expansion of realistic field theories, with conformal vertices arrested to one-loop order,  
could provide a complementary way to investigate the bootstrap program for higher point functions is CFT's.\\
One obvious question, emerging from this analysis, is if this result can be generalised to more complex correlators, such as the TTT, where more integration constants are present and for which a coordinate space analysis of such correlator has not been related to the free field theory result. We address this issue in the next chapter.

\chapter{Perturbative results in CFT: TTT case}\label{PerturbativeTTT}
In this chapter we show the correspondence between the perturbative realizations of CFT correlators and the general solution obtained by solving the CWI's \cite{Coriano:2018bsy, Coriano:2018bbe,Coriano:2018zdo}. 
In particular, we consider the correlation function $\braket{TTT}$ in general $d$ dimensions and for particular choices of the spacetime dimensions, and we also study the behaviour of these correlators in $d = 4$, where a breaking of conformal invariance appears, made manifest by the trace anomaly.
\section{Canonical and trace Ward Identities}
The conformal constraints for the $TTT$ correspond to dilatation and special conformal transformations, beside the 
usual Lorentz symmetries. Generically
\begin{equation}
	\sum_{j=1}^3G_g(x_j)\braket{T^{\m_1\n_1}(x_1)\,T^{\mu_2\nu_2}(x_2)\,T^{\mu_3\nu_3}(x_3)}=0,
\end{equation}
where $G_g$ are the generators of the infinitesimal symmetry transformations. Among these, the conservation WI in flat space of the stress-energy tensor can be obtained by requiring the invariance of $\mathcal{W}[g]$ under diffeomorphisms of the background metric 
\begin{equation}
	\mathcal{W}[g]=\mathcal{W}[g']
\end{equation}
where $g'$ is the transformed metric under the general infinitesimal coordinate transformation $x^\m\to x'^\mu=x^\mu -\e^\mu$
\begin{equation}
	\d g_{\m \n }=\nabla_\m \epsilon_\n +\nabla_\n \e_\m.
\end{equation} It generates the relation
\begin{equation}
	\nabla_\n \braket{T^{\m \n}}=0
	\label{transverse}
\end{equation}
while naive scale invariance gives the traceless condition
\begin{equation}
	g_{\mu\nu}\braket{T^{\m\n}}=0.\label{trace}
\end{equation}
These have been the only constraints taken into account in previous perturbative studies of the $TJJ$  
\cite{Giannotti:2008cv,Armillis:2009pq,Armillis:2010qk} and $TTT$ \cite{Coriano:2012wp}. The functional differentiation of \eqref{transverse} and \eqref{trace} allows to derive ordinary Ward identities for the various correlators. 
For the three point function case these take the form 
\begin{align} \label{WI3PFcoordinate}
	\partial_\nu\langle T^{\mu\nu}(x_1)T^{\rho\sigma}(x_2)T^{\alpha\beta}(x_3) \rangle
	&=
	\bigg[\langle T^{\rho\sigma}(x_1)T^{\alpha\beta}(x_3)\rangle\partial^\mu\d(x_1,x_2) + 
	\langle T^{\alpha\beta}(x_1)T^{\rho\sigma}(x_2)\rangle\partial^\mu\d(x_1,x_3) \bigg]\notag\\ 
	&\quad-
	\bigg[\delta^{\mu\rho}\langle T^{\nu\sigma}(x_1)T^{\alpha\beta}(x_3)\rangle
	+     \delta^{\mu\sigma}\langle T^{\nu\rho}(x_1)T^{\alpha\beta}(x_3)\rangle\bigg]\partial_\nu\d(x_1,x_2)\notag\\
	&\quad-
	\bigg[\delta^{\mu\alpha}\langle T^{\nu\beta}(x_1)T^{\rho\sigma}(x_2)\rangle
	+ \delta^{\mu\beta}\langle T^{\nu\alpha}(x_1)T^{\rho\sigma}(x_2)\rangle\bigg]\partial_\nu\d(x_1,x_3)\, .
\end{align}
In order to move to momentum space we fix some conventions. The Fourier transform of the correlators is defined as  
\begin{align}
	\braket{T^{\mu_1\nu_1}(p_1)\,T^{\m_2\n_2}(p_2)\,T^{\m_3\n_3}(p_3)}=\int d^d x_1 d^d x_2 d^d x_3 e^{i\left( p_1\cdot x_1 + p_2\cdot x_2 + p_3\cdot x_3\right)}\braket{T^{\mu_1\nu_1}(x_1)\,T^{\m_2\n_2}(x_2)\,T^{\m_3\n_3}(x_3)}
\end{align} 
and similarly for the 2-point function. Translational invariance  introduces an overall $\delta(P)$  with 
$P$ being the sum of all the (incoming) momenta, with the generation of derivative terms $\delta'(P)$, after the action of the special conformal transformations on the integrand.
Such terms can be investigated rigorously using the theory of tempered distributions, formulated using a symmetric basis. The analysis has been presented in \cite{Coriano:2018bbe} for a Gaussian basis, to which we refer for more details. In our conventions, we have chosen $p_3$ as the dependent momentum $ {p_3}\to -p_1 - p_2$. Eq. \eqref{WI3PFcoordinate} becomes
\begin{align}
	p_{1\n_1}\braket{T^{\mu_1\nu_1}(p_1)\,T^{\m_2\n_2}(p_2)\,T^{\m_3\n_3}({p_3})}&=-p_2^{\m_1}\braket{T^{\m_2\n_2}(p_1+p_2)T^{\m_3\n_3}({p_3})}-{p_3^{\m_1}}\braket{T^{\m_2\n_2}(p_2)T^{\m_3\n_3}(p_1+{p_3})}\notag\\
	&\hspace{-1.5cm}+p_{2\a}\left[\d^{\m_1\n_2}\braket{T^{\mu_2\a}(p_1+p_2)T^{\m_3\n_3}({p_3})}+\d^{\m_1\m_2}\braket{T^{\nu_2\a}(p_1+p_2)T^{\m_3\n_3}({p_3})}\right]\notag\\
	&\hspace{-1.5cm}+{p_{3\a}}\left[\d^{\m_1\n_3}\braket{T^{\mu_3\a}(p_1+{p_3})T^{\m_2\n_2}(p_2)}+\d^{\m_1\m_3}\braket{T^{\nu_3\a}(p_1+{p_3})T^{\m_2\n_2}(p_2)}\right].
	\label{long}
\end{align}
In the next section, in order to clarify that differentiation in $p_3$ has to be performed with the chain rule, we will denote with 
$\bar{p}_3^\mu\equiv -p_1^\mu - p_2^\mu$, the dependent momentum, and the independent 4-momenta will be $p_1^\mu$ and $p_2^\mu$. 
Concerning the naive identity \eqref{trace}, it generates the non-anomalous condition
\begin{equation}
	g_{\m_1\n_1}\braket{T^{\mu_1\nu_1}(p_1)\,T^{\m_2\n_2}(p_2)\,T^{\m_3\n_3}(p_3)}=0
\end{equation}
valid in the $d\ne4$ case. \\
After renormalization this equation is modified by the contribution of the conformal 
anomaly, given by the general expression
\begin{align} \label{TraceAnomaly}
	g_{\mu\nu}(z)\langle T^{\mu\nu}(z) \rangle
	&=
	\sum_{I=F,S,G}n_I \bigg[\beta_a(I)\, C^2(z) + \beta_b(I)\, E(z)\bigg]  
	+ \frac{\kappa}{4}n_G F^{a\,\mu\nu}\,F^a_{\mu\nu} (z) \notag\\
	&\equiv \mathcal{A}(z,g) \, ,
\end{align}
by considering only the scheme independent terms with
\begin{equation}
	\begin{split}
		\b_a(S)&=-\frac{3\p^2}{720}\,,\hspace{1cm}\b_b(S)=\frac{\p^2}{720}\,,\\
		\b_a(F)&=-\frac{9\p^2}{360}\,,\hspace{1cm}\b_b(F)=\frac{11\p^2}{720}\,\\
		\b_a(G)&=-\frac{18\p^2}{360}\,,\hspace{1.0cm}\b_b(G)=\frac{31\p^2}{360}\,
	\end{split}
	\label{choiceparm}
\end{equation}
being the contributions to the $\beta$ functions coming from scalars $(S)$, fermions $(F)$ and vectors $(G)$. We have defined the 
two tensors  
\begin{align}
	C^2&=R_{abcd}R^{abcd}-\sdfrac{4}{d-2}R_{ab}R^{ab}+\sdfrac{2}{(d-2)(d-1)}R^2,\hspace{0.7cm}
	&E=R_{abcd}R^{abcd}-4R_{ab}R^{ab}+R^2
\end{align}
being the square of the Weyl conformal tensor and the Euler-Poincar\'e density respectively, while $R_{abcd}$ is the Riemann curvature tensor and $R_{ab}$ and $R$ are the Ricci tensor and the Ricci scalar, respectively. Then we get the anomalous WI
\begin{align}
	&g_{\mu_1\nu_1}\braket{ T^{\mu_1\nu_1}(p_1)T^{\mu_2\nu_2}(p_2)T^{\mu_3\nu_3}(p_3)}\notag\\
	&\hspace{2cm}=
	4 \, \mathcal A^{\mu_2\nu_2\mu_3\nu_3}(p_2,p_3)
	- 2 \, \braket{ T^{\mu_2\nu_2}(p_1+p_2)T^{\mu_3\nu_3}(p_3)} - 2 \, \braket{ T^{\mu_2\nu_2}(p_2)T^{\mu_3\nu_3}(p_1+p_3)}\notag\\
	&\hspace{2cm}=
	4 \, \bigg[ \beta_a\,\big[C^2\big]^{\mu_2\nu_2\mu_3\nu_3}(p_2,p_3)+ \beta_b\, \big[E\big]^{\mu_2\nu_2\mu_3\nu_3}(p_2,p_3) \bigg]\notag\\
	&\hspace{3cm}- 2 \, \braket{ T^{\mu_2\nu_2}(p_1+p_2)T^{\mu_3\nu_3}(p_3)} - 2 \, \braket{ T^{\mu_2\nu_2}(p_2)T^{\mu_3\nu_3}(p_1+p_3)}. \label{munu3PFanomaly}
\end{align}
We just remark that the solutions of all the conformal constraints, in this study, are obtained by working with the non-anomalous expressions of the corresponding CWI's, while the anomaly contributions, as in (\ref{munu3PFanomaly}), are obtained only after taking the $d\to 4$ limit of the general solution and the inclusion of the corresponding counterterms. 
\section{Lagrangian realizations and reconstruction}
In this section, we turn to the central aspect of our analysis, which will allow us to extend the results of the $TJJ$ correlator presented in \chapref{PerturbativeTJJ} to the $TTT$ \cite{Coriano:2018bbe}. We will be using the free field theory realizations of such a correlator to study the structure of the conformal Ward identities in momentum space and, in particular, the form of the anomalous Ward identities once the anomaly breaks the conformal symmetry. We will work in DR and adopt the $\overline{MS}$ renormalization scheme. Our analysis hinges on the correspondence between the exact result obtained by solving the CWI's and the perturbative one. \\
It is clear that the general solutions presented in the former chapters, though derived regardless of any perturbative picture, become entirely equivalent to the latter if, for a given spacetime dimension, we have a sufficient number of independent sectors in the Lagrangian realization that allow us to reproduce the general one. For instance, in $d=3,4$ such correspondence is exact since the number of constants in the solution coincides with the number of possible independent sectors in the free field theory.\\
In principle, one could proceed with an analysis of the general solutions - such as those presented in the previous chapters - as $d\to 4$, by going through a very involved process of extraction of the singularities from their general expressions in terms of $F_4$.\\
However, this can be avoided once the general results for the $A_i$'s are matched to the perturbative ones. As already mentioned, this brings in an important simplification of the final result for the $TTT$, which is expressed in terms of the simple $\log$ present in $B_0$ and the scalar 3-point function $C_0$. The latter is of type $F_4(a,b;c_1,c_2;x,y)$ in $d=4$, but takes a simpler form when the master integral $C_0$ is expressed as
\begin{align}
	&  C_0 ( p_1^2,p_2^2,p_3^2) = \frac{ 1}{p_3^2} \Phi (x,y),
\end{align}
with the function $\Phi (x, y)$ defined as
\cite{Usyukina:1993ch}
\begin{align}
	\Phi( x, y) &=& \frac{1}{\lambda} \biggl\{ 2 [Li_2(-\rho  x) + Li_2(- \rho y)]  +
	\ln \frac{y}{ x}\ln \frac{1+ \rho y }{1 + \rho x}+ \ln (\rho x) \ln (\rho  y) + \frac{\pi^2}{3} \biggr\},
	\label{Phi}
\end{align}
with
\begin{align}
	\lambda(x,y) = \sqrt {\Delta},
	\qquad  \qquad \Delta=(1-  x- y)^2 - 4  x  y,
	\label{lambda} \\
	\rho( x,y) = 2 (1-  x-  y+\lambda)^{-1},
	\qquad  \qquad x=\frac{p_1^2}{p_3^3} \, ,\qquad \qquad y= \frac {p_2^2}{p_3^2}\, .
\end{align}

This has the important implication that the study of the specific unitarity cuts in the diagrammatic expansions of the correlator \cite{Giannotti:2008cv, Coriano:2014gja} which are held responsible for the emergence of the anomaly, acquire a simple particle interpretation and are not an artifact of perturbation theory. Once the general correspondence between Lagrangian and non-Lagrangian solutions is established, we will concentrate on showing how renormalization is responsible for the emergence of specific anomaly poles 
in this correlators. We anticipate that the vertex will separate, after renormalization, into a traceless part and in an anomaly part, following the same pattern of the $TJJ$ \cite{Coriano:2018zdo}. From that point on, one can use just the Feynman expansion to perform complete further analysis of this vertex at one loop, with no loss of generality whatsoever. 
\subsection{Perturbative sectors}
In this section we define our conventions used for the perturbative sectors in analogy with the $TJJ$ case discussed in the previous chapter. The quantum actions for the scalar and fermion field are respectively
\begin{align}
	S_{scalar}&=\sdfrac{1}{2}\int\, d^dx\,\sqrt{-g}\left[g^{\m\n}\nabla_\m\phi\nabla_\n\phi-\c\, R\,\phi^2\right]\\
	S_{fermion}&=\sdfrac{i}{2}\int\, d^dx\,e\,e^{\m}_a\left[\bar{\psi}\g^a(D_\m\psi)-(D_\m\bar{\psi})\g^a\psi\right],
\end{align}
where $\c =(d-2)/(4d-4)$ for a conformally coupled scalar in $d$ dimensions, and $R$ is the Ricci scalar. $e_\m^a$ is the vielbein and $e$ its determinant,  with the covariant derivative $D_\m$ given by 
\begin{equation}
	D_\m=\partial_\m+\Gamma_\m=\partial_\m+\sdfrac{1}{2}\Sigma^{ab}\,e^\s_a\nabla_\m\,e_{b\,\s}.
\end{equation}
The $\Sigma^{ab}$ are the generators of the Lorentz group in the spin $1/2$ representation. The Latin indices are related to the flat space-time and the Greek indices to the curved space-time. 
For $d=4$ there is an additional conformal field theory described in terms of free abelian vector fields with the action
\begin{equation}
	S_{abelian}=S_{M}+S_{gf}+S_{gh}
\end{equation}
where the three contributions are the Maxwell action, the gauge fixing contribution and the ghost action
\begin{align}
	S_M&=-\sdfrac{1}{4}\int d^4x\,\sqrt{-g}\,F^{\m\n}F_{\m\n},\\
	S_{gf}&=-\sdfrac{1}{\xi}\int d^4x\,\sqrt{-g}\,(\nabla_\m A^\m)^2,\\
	S_{gh}&=\int d^4x\,\sqrt{-g}\,\,\partial^\m\bar c\,\partial_\m \,c.
\end{align}
The computation of the vertices of each theory can be done by taking (at most) two functional derivatives of the action with respect to the metric, since the vacuum expectation values of the third derivatives correspond to massless tadpoles, which are zero in DR. 
They are given in the \appref{Appendix1}.

\noindent Since we are interested in the most general Lagrangian realization of the $\braket{TTT}$ correlator in the conformal case, this can be obtained only by considering the scalar and fermion sectors in general $d$ dimensions.  

\subsection{Scalar sector}

\begin{figure}[t]
	\centering
	\vspace{-1.5cm}
	\subfigure{\includegraphics[scale=0.2]{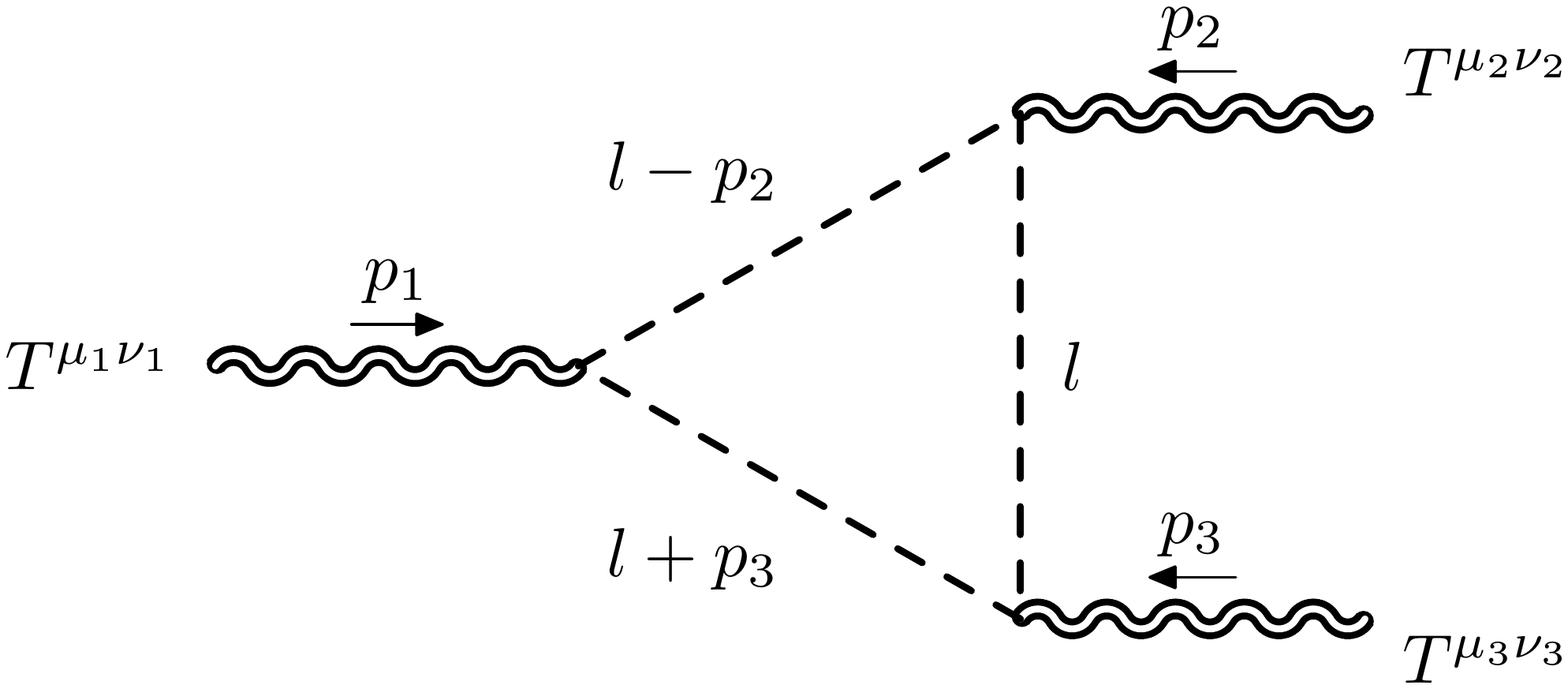}} \hspace{.3cm}
	\subfigure{\includegraphics[scale=0.2]{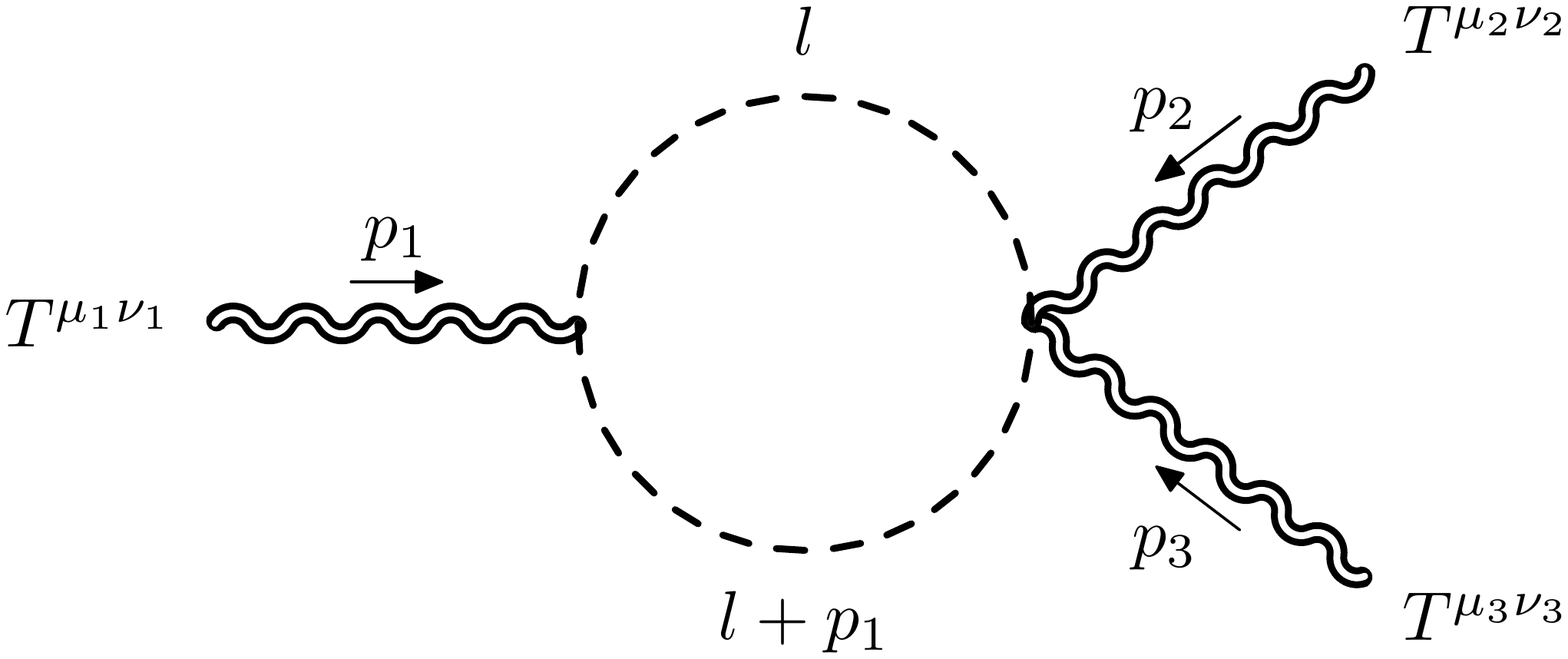}} \hspace{.3cm}
	\raisebox{.12\height}{\subfigure{\includegraphics[scale=0.16]{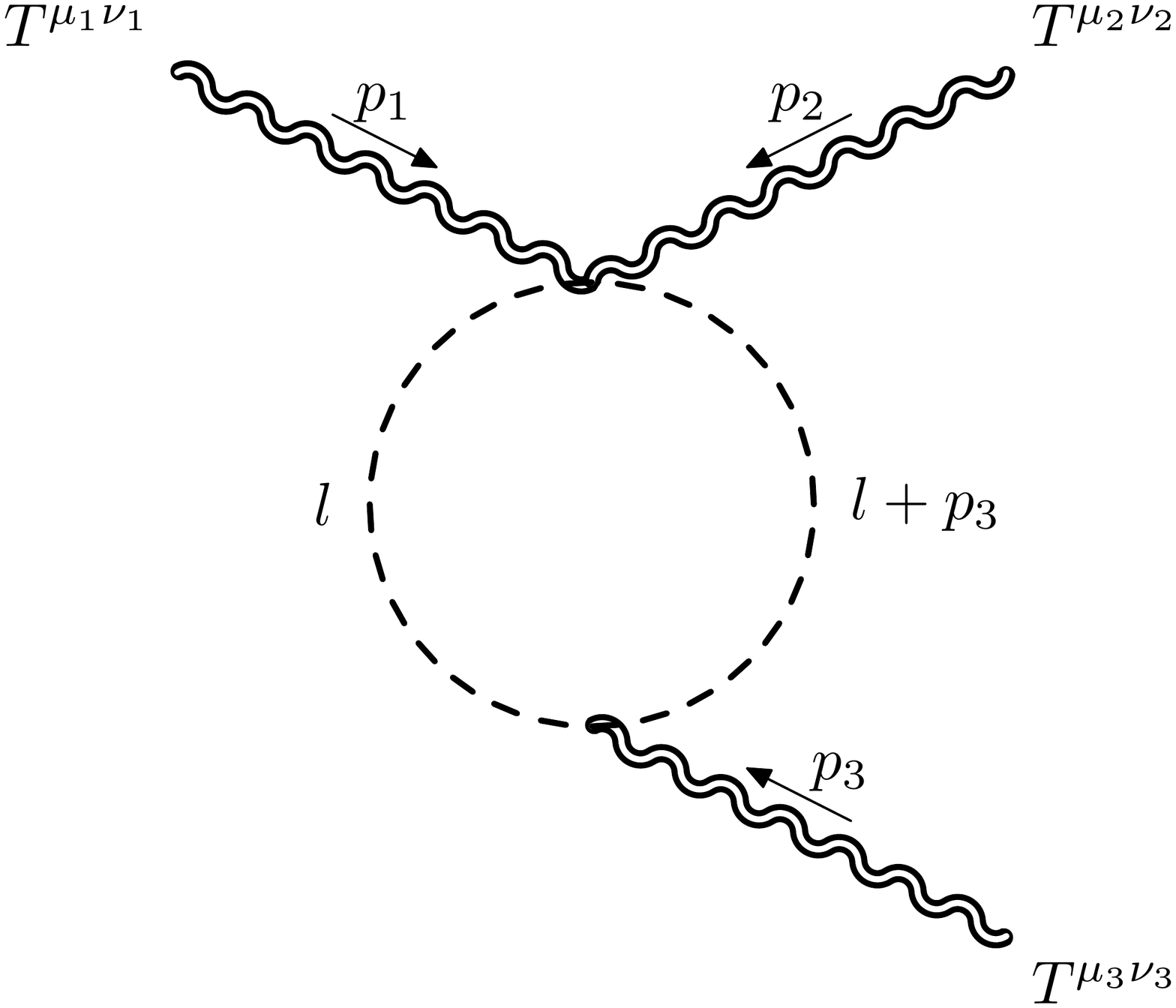}}\hspace{.3cm}}
	\raisebox{.12\height}{\subfigure{\includegraphics[scale=0.16]{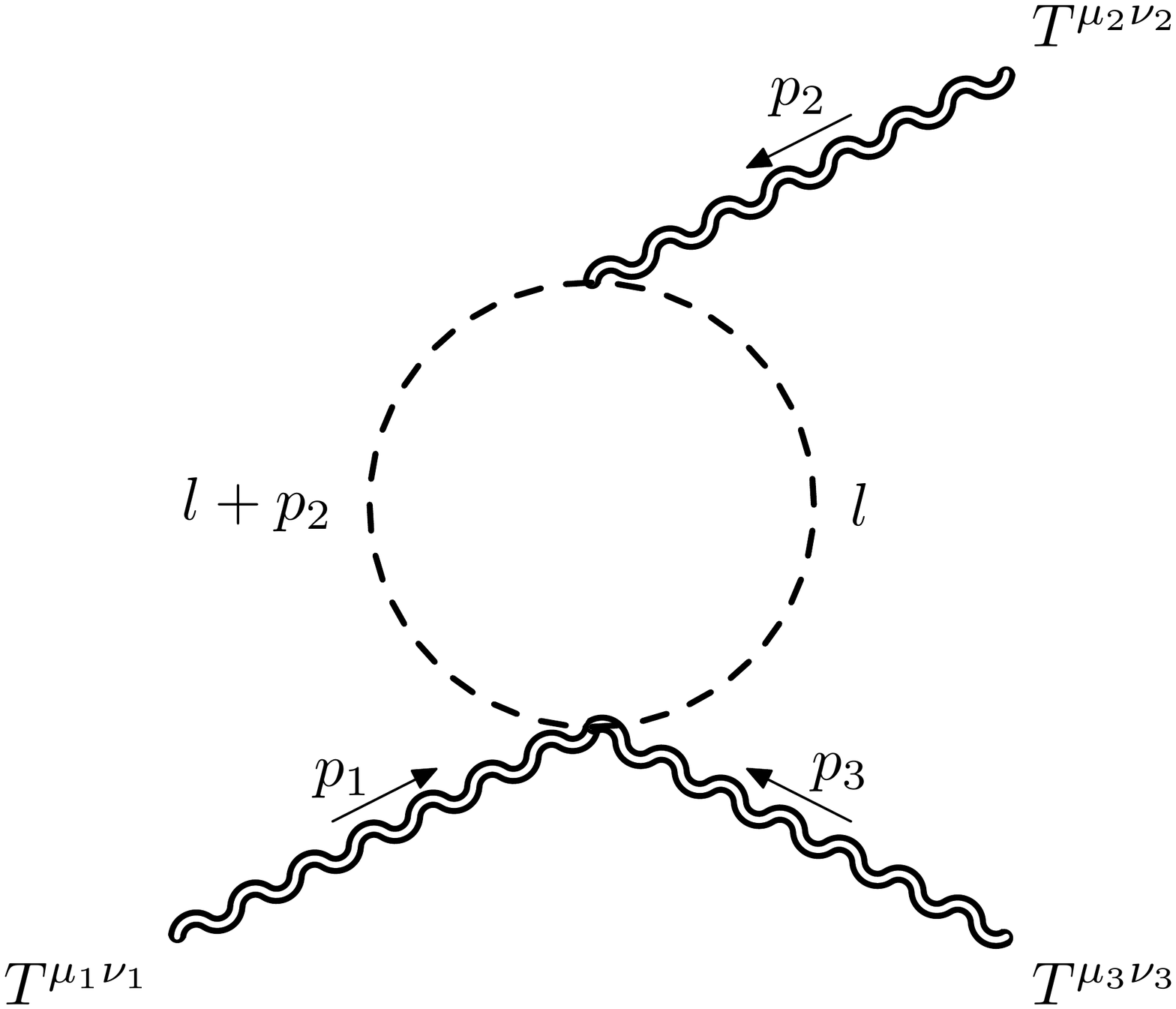}}}
	\vspace{-0.8cm}\caption{One-loop scalar diagrams for the three-graviton vertex.\label{Feynman1}}
\end{figure}

We start from the scalar sector. In the one-loop approximation the contributions to the correlation function are given by the diagrams in \figref{Feynman1}. Using the Feynman rules listed in Appendix \ref{Appendix1}, we calculate all the terms in the defining relation of the $TTT$ in momentum space, for the scalar sector, as
\begin{align}
	\braket{T^{\m_1\n_1}(p_1)T^{\m_2\n_2}(p_2)T^{\m_3\n_3}(p_3)}_S=\, -V_{S}^{\m_1\n_1\m_2\n_2\m_3\n_3}(p_1,p_2,p_3)+\sum_{i=1}^3W_{S,i}^{\m_1\n_1\m_2\n_2\m_3\n_3}(p_1,p_2,p_3)\label{ScalarPart}
\end{align}
where $V_S$ is related to the triangle diagrams in \figref{Feynman1} and $W_{S,i}$ terms are the three bubble contributions labelled by the index $i$, with $i=1,2,3$. These contribution are given by
\begin{align}
	V_{S}^{\m_1\n_1\m_2\n_2\m_3\n_3}(p_1,p_2,p_3)& =\int \sdfrac{d^d\ell}{(2\pi)^d}\sdfrac{V^{\m_1\n_1}_{T\phi\phi}(\ell-p_2,\ell+p_3)V^{\m_2\n_2}_{T\phi\phi}(\ell,\ell-p_2)V^{\m_3\n_3}_{T\phi\phi}(\ell,\ell+p_3)}{\ell^2(\ell-p_2)^2(\ell+p_3)^2}\notag\\
	W_{S,1}^{\m_1\n_1\m_2\n_2\m_3\n_3}(p_1,p_2,p_3)& =\sdfrac{1}{2}\int \sdfrac{d^d\ell}{(2\pi)^d}\sdfrac{V^{\m_1\n_1}_{T\phi\phi}(\ell,\ell+p_1)V^{\m_2\n_2\m_3\n_3}_{TT\phi\phi}(\ell,\ell+p_1)}{\ell^2(\ell+p_1)^2}\notag\\
	W_{S,2}^{\m_1\n_1\m_2\n_2\m_3\n_3}(p_1,p_2,p_3)& =\sdfrac{1}{2}\int \sdfrac{d^d\ell}{(2\pi)^d}\sdfrac{V^{\m_3\n_3}_{T\phi\phi}(\ell,\ell+p_3)V^{\m_1\n_1\m_2\n_2}_{TT\phi\phi}(\ell,\ell+p_3)}{\ell^2(\ell+p_3)^2}\notag\\
	W_{S,3}^{\m_1\n_1\m_2\n_2\m_3\n_3}(p_1,p_2,p_3)& =\sdfrac{1}{2}\int \sdfrac{d^d\ell}{(2\pi)^d}\sdfrac{V^{\m_2\n_2}_{T\phi\phi}(\ell,\ell+p_2)V^{\m_1\n_1\m_3\n_3}_{TT\phi\phi}(\ell,\ell+p_2)}{\ell^2(\ell+p_2)^2}
\end{align}
where we have included a symmetry factor $1/2$. The calculation of the integral can be simplified by acting with the projectors $\Pi$ on \eqref{ScalarPart} in order to write the form factors of the transverse and traceless part of the correlator, as in \eqref{tttdec}
\begin{align}
	\braket{t^{\m_1\n_1}(p_1)t^{\m_2\n_2}(p_2)t^{\m_3\n_3}(p_3)}_S&=\,\Pi^{\m_1\n_1}_{\a_1\b_1}(p_1)\Pi^{\m_2\n_2}_{\a_2\b_2}(p_2)\Pi^{\m_3\n_3}_{\a_3\b_3}(p_3)\notag\\
	&\times\bigg[ -V_{S}^{\a_1\b_1\a_2\b_2\a_3\b_3}(p_1,p_2,p_3)+\sum_{i=1}^3W_{S,i}^{\a_1\b_1\a_2\b_2\a_3\b_3}(p_1,p_2,p_3)\bigg]
\end{align}

\subsection{Fermion sector}
As in the scalar sector, also in this case we calculate in the one-loop approximation the contribution to the correlation function of the fermion sector by the diagrams in \figref{Feynman2}. These contributions can be written as 
\begin{align}
	\braket{T^{\m_1\n_1}(p_1)T^{\m_2\n_2}(p_2)T^{\m_3\n_3}(p_3)}_F=-\, \sum_{j=1}^2V_{F,j}^{\m_1\n_1\m_2\n_2\m_3\n_3}(p_1,p_2,p_3)+\sum_{j=1}^3W_{F,j}^{\m_1\n_1\m_2\n_2\m_3\n_3}(p_1,p_2,p_3)\label{FermionExpa}
\end{align}
using notations similar to the scalar case. In this case we take into account two possible orientations for the fermion in the loop. 
\begin{figure}[t]
	\centering
	\vspace{-2cm}
	\subfigure{\includegraphics[scale=0.2]{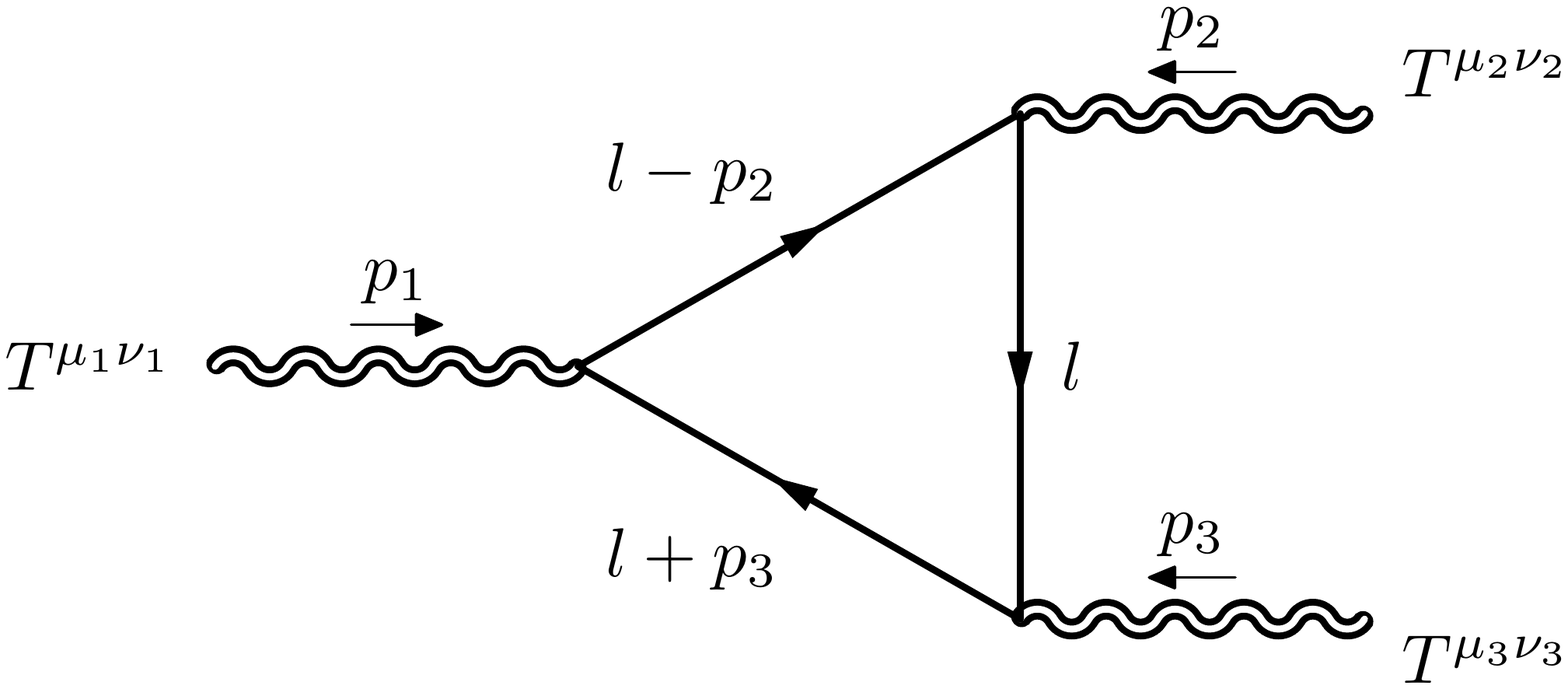}} \hspace{.3cm}
	\subfigure{\includegraphics[scale=0.2]{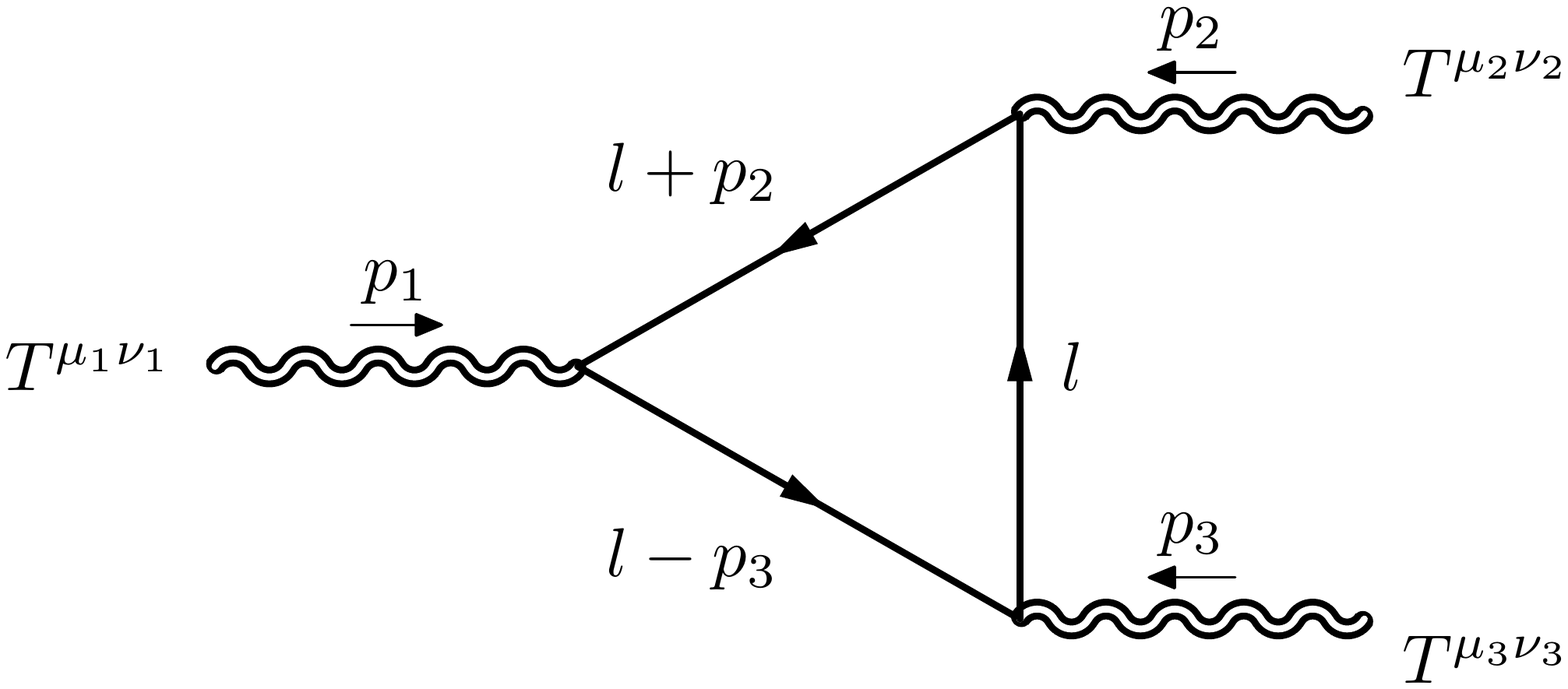}} \hspace{.3cm}
	\subfigure{\includegraphics[scale=0.2]{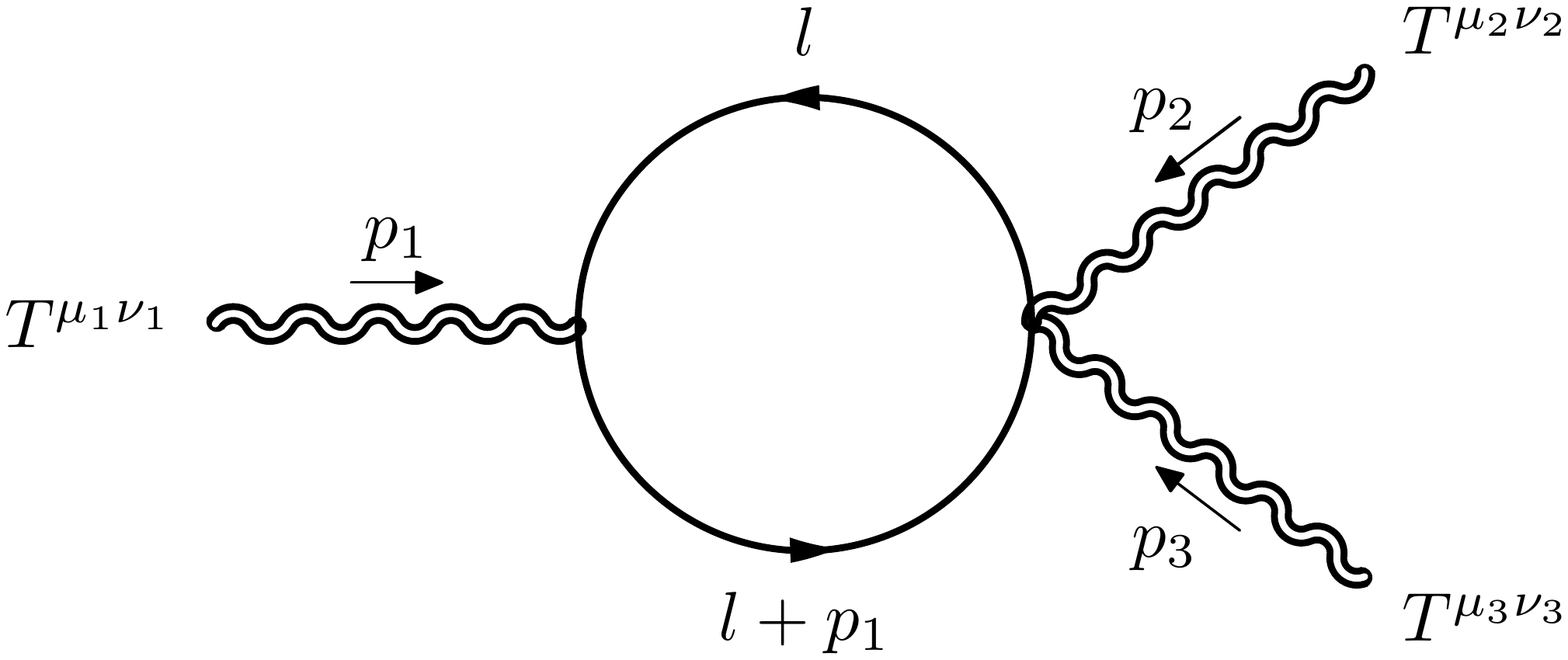}} \hspace{.3cm}\\
	\vspace{-2.5cm}
	\raisebox{.12\height}{\subfigure{\includegraphics[scale=0.16]{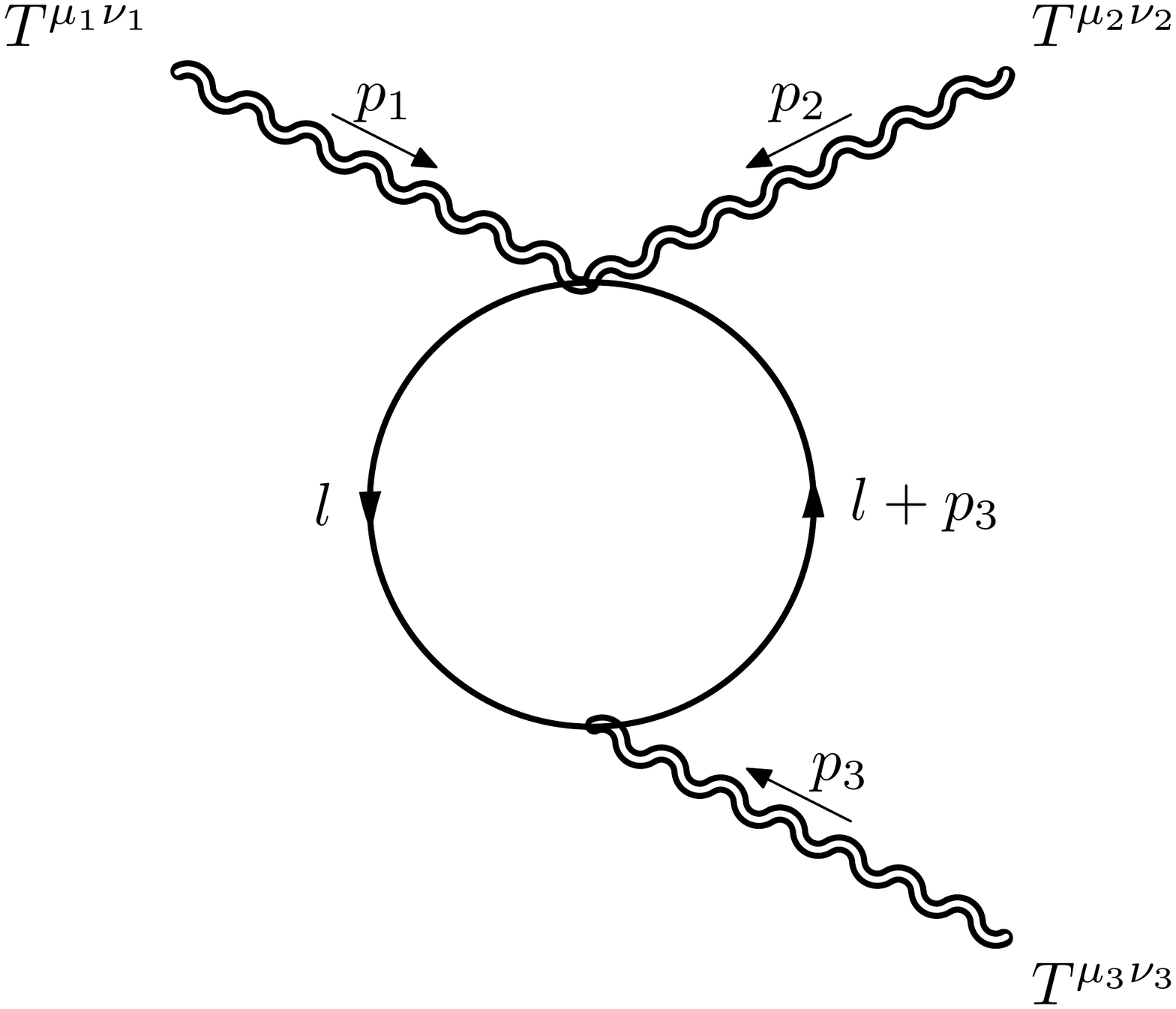}}\hspace{.3cm}}
	\raisebox{.12\height}{\subfigure{\includegraphics[scale=0.16]{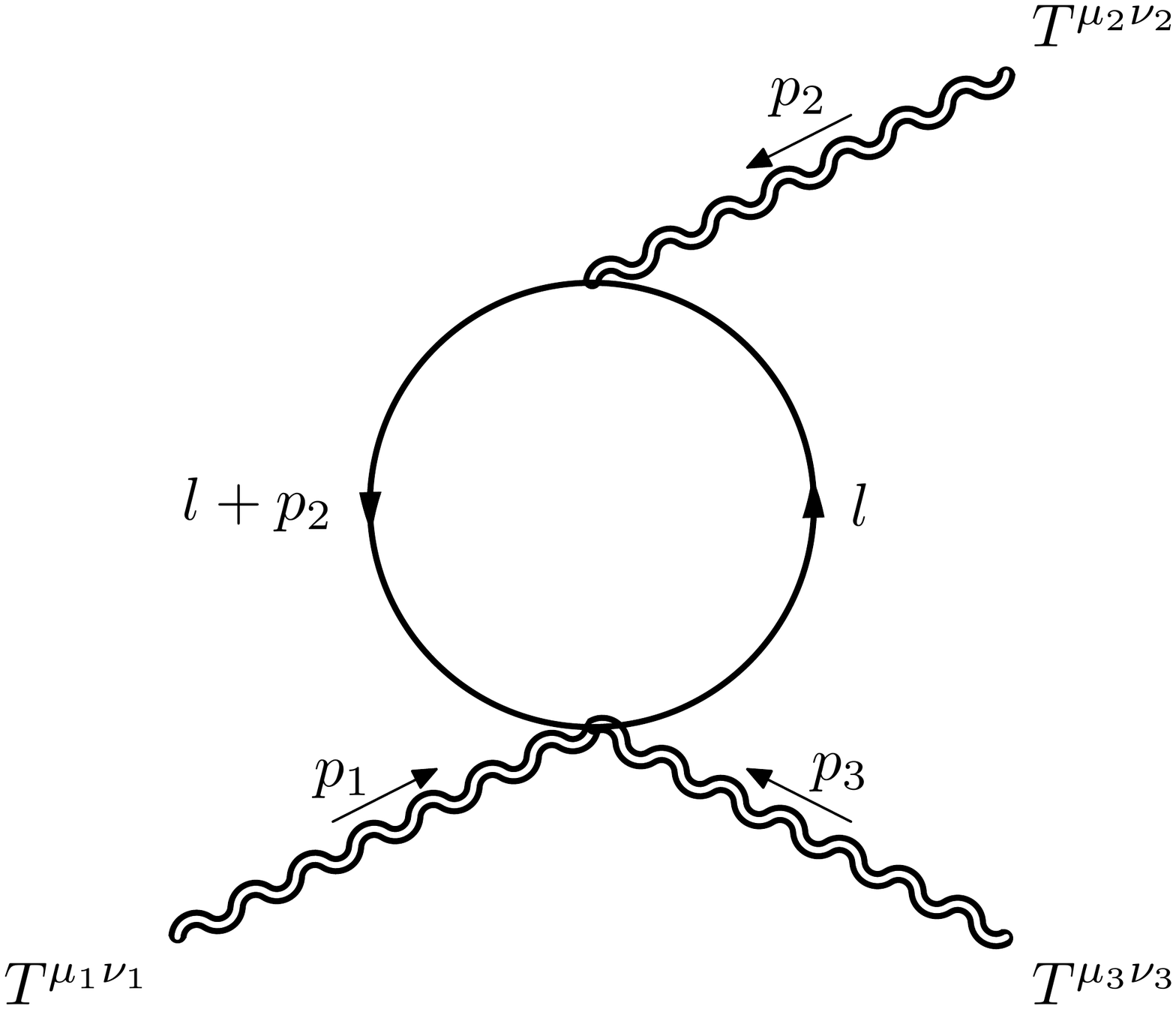}}}
	\vspace{-0.8cm}\caption{One-loop fermion diagrams for the three-graviton vertex.\label{Feynman2}}
\end{figure}
Explicitly the terms in \eqref{FermionExpa} are given by
\begin{align}
	V_{F,1}^{\m_1\n_1\m_2\n_2\m_3\n_3}(p_1,p_2,p_3)&=-\int \sdfrac{d^d\ell}{(2\pi)^d}\sdfrac{\Tr\big[V^{\m_1\n_1}_{T\bar\psi\psi}(\ell+p_3,\ell-p_2)(\slashed{\ell}+\slashed{p}_3)V^{\m_3\n_3}_{T\bar\psi\psi}(\ell,\ell+p_3)\slashed{\ell}V^{\m_2\n_2}_{T\bar\psi\psi}(\ell-p_2,\ell)(\slashed{\ell}-\slashed{p}_2)\big]}{\ell^2(\ell-p_2)^2(\ell+p_3)^2}\\[1ex]
	V_{F,2}^{\m_1\n_1\m_2\n_2\m_3\n_3}(p_1,p_2,p_3)&=V_{F,1}^{\m_1\n_1\m_3\n_3\m_2\n_2}(p_1,p_3,p_2)\\
	W_{F,1}^{\m_1\n_1\m_2\n_2\m_3\n_3}(p_1,p_2,p_3)&=-\int \sdfrac{d^d\ell}{(2\pi)^d}\sdfrac{\Tr\big[V^{\m_1\n_1}_{T\bar\psi\psi}(\ell,\ell+p_1)\ \slashed{\ell}\ V^{\m_2\n_2\m_3\n_3}_{TT\bar\psi\psi}(\ell+p_1,\ell)(\slashed{\ell}+p_1)\big]}{\ell^2(\ell+p_1)^2}\\[2ex]
	W_{F,2}^{\m_1\n_1\m_2\n_2\m_3\n_3}(p_1,p_2,p_3)&=W_{F,1}^{\m_3\n_3\m_1\n_1\m_2\n_2}(p_3,p_1,p_2)\\[1ex]
	W_{F,3}^{\m_1\n_1\m_2\n_2\m_3\n_3}(p_1,p_2,p_3)&=W_{F,1}^{\m_2\n_2\m_1\n_1\m_3\n_3}(p_2,p_1,p_3)
\end{align}
By a direct computation one can verify that the spin part of the two-gravitons and two-fermions vertex does not contribute to the correlation function. \\
Acting with the projectors transverse and traceless $\Pi$ we obtain the form factors $A_i$, $i=1,\dots,5$. For instance, in the fermion sector we obtain
\begin{align}
	\braket{t^{\m_1\n_1}(p_1)t^{\m_2\n_2}(p_2)t^{\m_3\n_3}(p_3)}_F&=\,\Pi^{\m_1\n_1}_{\a_1\b_1}(p_1)\Pi^{\m_2\n_2}_{\a_2\b_2}(p_2)\Pi^{\m_3\n_3}_{\a_3\b_3}(p_3)\notag\\
	&\times\bigg[ -\sum_{j=1}^2V_{F,j}^{\a_1\b_1\a_2\b_2\a_3\b_3}(p_1,p_2,p_3)+\sum_{j=1}^3W_{F,j}^{\a_1\b_1\a_2\b_2\a_3\b_3}(p_1,p_2,p_3)\bigg]
\end{align}
Also in this case the number of fermion families is kept arbitrary and we will multiply the result by a constant $n_F$ to account for it. It will be essential for matching this contribution to the general non-perturbative one. 

\section{Comparisons with the conformal solutions in $d=3$ and $d=5$}

\subsection{Normalization of the two point function}
\label{compare}
In order to investigate the correspondence between the conformal and the perturbative solutions 
we briefly recall the result for the $\braket{TT}$ correlator, that we will need in order to investigate the match between the general conformal solution and its perturbative realization. 
Here we start from a general analysis, based on the CFT solution for this correlator, with a specific application to the case of odd spacetime dimensions, where no renormalization is needed. We will come back to the same correlator in a following section, when we will address the issue of its renormalization in $d=4$.

The $TT$ is fixed by conformal invariance in coordinate space to take the form 
\begin{align}
	\label{2PFOsborn}
	\langle T^{\mu \nu}(x) \, T^{\alpha \beta} (y)\rangle =
	\frac{C_T}{(x-y)^{2d}} \, \mathcal{I}^{\mu\nu ,\alpha\beta}(x-y) \, ,
\end{align}
with 
\begin{align} \label{Inversion}
	\mathcal{I}^{\mu \nu,\alpha \beta} (s) =
	I^{\mu\rho}(x-y)I^{\nu\sigma}(x-y) {\epsilon_T}^{\rho\sigma,\alpha\beta} \, ,
\end{align}
where
\begin{align}
	I^{\mu\nu}(x)=
	\delta^{\mu\nu} - 2 \frac{x^\mu x^\nu}{x^2}
\end{align}
and 
\begin{align}\label{epsilon}
	{\epsilon_T}^{\mu\nu,\alpha\beta} =
	\frac{1}{2} \, (\delta^{\mu\alpha} \delta^{\nu \beta} + \delta^{\mu \beta} \delta^{\nu \alpha} \bigl)
	- \frac{1}{d} \, \delta^{\mu \nu} \delta^{\alpha \beta}.
\end{align}
It is not difficult to check that \eqref{2PFOsborn} can be cast in the form 
\begin{align}
	\langle T^{\mu \nu}(x) \, T^{\alpha \beta} (y)\rangle =
	\frac{C_T}{4 (d-2)^2 d (d+1)} \Delta^{(d) \mu\nu\alpha\beta}(\partial) \frac{1}{((x-y)^2)^{(d-2)}}
\end{align}
where 
\begin{align} \label{TransverseDeltaCoord}
	\hat{\Delta}^{(d)\,\mu\nu\alpha\beta}(\partial)
	&=
	\frac{1}{2}\left( \hat{\Theta}^{\mu\alpha} \hat{\Theta}^{\nu\beta} +\hat{ \Theta}^{\mu\beta}\hat{ \Theta}^{\nu\alpha} \right)
	- \frac{1}{d-1}\hat{ \Theta^{\mu\nu}} \hat{\Theta}^{\alpha\beta}\, ,
	\quad
	\text{with}
	\quad
	\hat{\Theta}^{\mu\nu} = \partial^\mu  \partial^\nu - \delta^{\mu\nu} \, \Box  \notag\\
	\partial_\mu \, \hat{\Delta}^{(d)\,\mu\nu\alpha\beta}(\partial)
	&=
	0 \, , \quad
	\delta_{\mu\nu} \, \hat{\Delta}^{(d)\,\mu\nu\alpha\beta}(\partial) = 0   \, 
\end{align}
for any function on which it acts.
Using the representation 
\begin{align} \label{fund}
	\frac{1}{(x^2)^\alpha}&=\equiv C(\alpha) \, \int d^d l \, \frac{e^{i l\cdot x}}{(l^2)^{d/2 - \alpha}} \nonumber \\
	C(\alpha) &=\frac{1}{4^{\alpha}\,\pi^{d/2}} \frac{\Gamma(d/2 - \alpha)}{\Gamma(\alpha)}
\end{align}
it can be re-expressed in the form 

\begin{align}
	\label{firstform}
	\braket{T^{\m\n}(p)T^{\alpha\beta}(-p)}&=\int d^d x e^{i p\cdot x} 
	\braket{T^{\m\n}(x)T^{\alpha\beta}(0)}=C_T \frac{ \pi^{d/2} \Gamma(-d/2)}{2^d (d-2)(d+1)\Gamma(d-2) }p^d 
	\Pi^{\mu\nu\alpha\beta}(p).  
\end{align}
Using the expression of the scalar (Euclidean) 2-point function 
\begin{equation}
	{B}_0(p_1^2)=\sdfrac{1}{\p^\frac{d}{2}}\int\,d^d \ell\ \frac{l}{\ell^2(\ell-p_1)^2}=\frac{ \left[\Gamma\left(\frac{d}{2}-1\right)\right]^2\Gamma\left(2-\frac{d}{2}\right)}{\Gamma\left(d-2\right)(p_1^2)^{2-\frac{d}{2}}}\label{B0ex}
\end{equation}
which is divergent for $d=2 k $, $k=1,2,3...$, it can also be rewritten in the form 
\begin{equation}
	\braket{T^{\m\n}(p)T^{\alpha\beta}(-p)}=4 C_T \left(\frac{\pi}{4}\right)^{d/2} \frac{1}{(d-2)^2 d (d+1)\Gamma(d/2-1)^2}\Pi^{\mu\nu\alpha\beta}(p)\, p^4  B_0(p^2).
\end{equation}
The singular nature of \eqref{firstform} in even dimensions emerges in DR from the appearance of the
$\Gamma(-d/2)$ factor, which can be regulated by an ordinary shift $d\to d- \epsilon$. After a redefinition of the constant $C_T\to c_T$ which absorbs the $d$-dependent prefactors, it takes the form 
\begin{align}
	\label{general}
	\braket{T^{\m\n}(p)T^{\alpha\beta}(-p)}=c_T\,\Pi^{\m\n\alpha\beta}(p)\,\Gamma\left(-\frac{d}{2}+\frac{\epsilon}{2}\right)\,p^{d- \epsilon}
\end{align}
where the constant $c_T$ is regular and arbitrary. Using the Lagrangian realization of the $TT$ in terms of the two free field theory sectors available in odd dimensions, it can be written as
\begin{align}
	\label{fform}
	\braket{T^{\mu_1\nu_1}(p)T^{\mu_2\nu_2}(-p)}&=\frac{\p^\frac{d}{2}\big(n_S+2(d-1)n_F\big)}{4(d-1)(d+1)}\,\Pi^{\mu_1\nu_1\mu_2\nu_2}(p)\,{B}_0(p^2)\,p^d\notag\\
	&=\frac{\p^\frac{d}{2}\,\big(n_S+2(d-1)n_F\big)\,d(d-2)}{16(d+1)(d-1)\Gamma\left(d-2\right)}\,\left[\Gamma\left(\frac{d}{2}-1\right)\right]^2\,\Pi^{\mu_1\nu_1\mu_2\nu_2}(p)\,\Gamma\left(-\frac{d}{2}\right)\,p^d
\end{align}
with $c_T$ matched in odd dimensions ($d>1$) according to the expression
\begin{equation}
	c_T=\frac{\p^\frac{d}{2}\,\big(n_S+2(d-1)n_F\big)\,d(d-2)}{16(d+1)(d-1)\Gamma\left(d-2\right)}\,\left[\Gamma\left(\frac{d}{2}-1\right)\right]^2.
\end{equation}
In even dimension we have a third (gauge) sector available and therefore it will be necessary to extend \eqref{fform} in order to perform a complete matching. We will address its renormalization in Section \ref{renorm}.\\
In the case of $d=3$ and $d=5$ we get
\begin{equation}
	c_T\ \ \mathrel{\stackrel{\makebox[0pt]{\mbox{\normalfont\tiny $d=3$}}}{=}}\ \  \frac{3\,\p^\frac{3}{2}\,\big(n_S+4 n_F\big)}{32(3+1)}\,\left[\Gamma\left(\frac{1}{2}\right)\right]^2=\frac{3\p^\frac{5}{2}}{128}\,\big(n_S+4n_F\big)\,
\end{equation}

\begin{equation}
	c_T\ \ \mathrel{\stackrel{\makebox[0pt]{\mbox{\normalfont\tiny $d=5$}}}{=}}\ \  \frac{15\,\p^\frac{5}{2}\,\big(n_S+2(5-1)n_F\big)}{64(5+1)\Gamma\left(5-2\right)}\,\left[\Gamma\left(\frac{3}{2}\right)\right]^2=\frac{5\p^\frac{7}{2}}{1024}\,\big(n_S+8n_F\big)\,
\end{equation}
We will be using these two expressions of $c_T$ in order to perform a comparison with the result of the transverse traceless $A_i$ given in \cite{Bzowski:2013sza} for odd dimensions. In such specific cases there are simplifications both from the exact and the perturbative solutions. In particular, the exact solutions turn into rational functions of the momenta, and, as we are going to show, they can be matched with the perturbative ones that we present below. \\
In $d=5$, as for all odd dimensions larger than 3, the general solution involves 3 independent constant, as we have mentioned, and we are short of 1 sector in order to match the general result. Nevertheless it is still possible to perform a matching between the two solutions, even though not in the most general case. \\
Clearly, in this case the perturbative results given below for $d=5$ correspond to a specific choice of the 3 constants of the solution of the conformal constraints. This, obviously, leaves open the issue whether arbitrary choices of all the three constants which appear in odd spacetime dimensions in the general solution correspond to a unitary theory or not, or whether it is possible to formulate, for odd values of values of $d > 3$, CFT's which do not have a free field theory realization. These may correspond to interacting CFT's.   
\section{Explicit results}
\label{d35}
\subsection{$d=3$ case}
We have shown in \secref{explicitd3} how to express the scalar integrals ${B}_0$ and ${C}_0$ in $d=3$.  
The explicit expression of the form factors in $d=3$, using the perturbative approach to one loop order, can be obtained by taking the limit $d\to3$ of \eqref{B01} and \eqref{C01} derived from the general diagrammatic expansion. We obtain
\begin{align}
	A_{1}^{d=3}(p_1,p_2,p_3)&=\frac{\p^3(n_S-4n_F)}{60(p_1+p_2+p_3)^6}\Big[p_1^3+6p_1^2(p_3+p_2)+(6p_1+p_2+p_3)\big((p_2+p_3)^2+3 p_2 p_3\big)\Big]\\
	A_{2}^{d=3}(p_1,p_2,p_3)&=\frac{\p^3(n_S-4n_F)}{60(p_1+p_2+p_3)^6}\Big[4p_3^2\big(7(p_1+p_2)^2+6p_1 p_2\big)+20p_3^3(p_1+p_2)+4p_3^4\notag\\
	&\hspace{4.4cm}+3(5p_3+p_1+p_2)(p_1+p_2)\big((p_1+p_2)^2+p_1 p_2\big)\Big]\notag\\
	&+\frac{\p^3\,n_F}{3(p_1+p_2+p_3)^4}\Big[p_1^3+4p_1^2(p_2+p_3)+(4p_1+p_2+p_3)\big((p_2+p_3)^2+p_2 p_3\big)\Big]\\
	A_{3}^{d=3}(p_1,p_2,p_3)&=\frac{\p^3(n_S-4n_F)\,p_3^2}{240(p_1+p_2+p_3)^4}\Big[28p_3^2(p_1+p_2)+3p_3\big(11(p_1+p_2)^2+6 p_1\,p_2\big)+7p_3^3\notag\\
	&+12(p_1+p_2)\big((p_1+p_2)^2+p_1p_2\big)\Big]\notag\\
	&+\frac{\p^3n_F\,p_3^2}{6(p_1+p_2+p_3)^3}\Big[3p_2(p_1+p_2)+2\big((p_1+p_2)^2+p_1p_2\big)+p_3^2\Big]\notag\\
	&-\frac{\p^3(n_s+4n_F)}{16(p_1+p_2+p_3)^2}\Big[p_1^3+2p_1^2(p_2+p_3)+(2p_1+p_2+p_3)\big((p_2+p_3)^2-p_2 p_3\big)\Big]
\end{align}
\begin{align}
	A_{4}^{d=3}(p_1,p_2,p_3)&=\frac{\p^3(n_S-4n_F)}{120(p_1+p_2+p_3)^4}\Big[(4p_3+p_1+p_2)\big(3(p_1+p_2)^4-3(p_1+p_2)^2p_1p_2+4p_1^2p_2^2\big)\notag\\
	&+9p_3^2(p_1+p_2)\big((p_1+p_2)^2-3p_1 p_2\big)-3p_3^5-12p_3^4(p_1+p_2)-9p_3^3\big((p_1+p_2)^2+2p_1 p_2\big)\Big]\notag\\
	&+\frac{\p^3\,n_F}{6(p_1+p_2+p_3)^3}\Big[(p_1+p_2)\big((p_1+p_2)^2-p_1 p_2\big)(p_1+p_2+3p_3)-p_3^4-3p_3^3(p_1+p_2)\notag\\
	&-6p_1p_2p_3^2\Big]-\frac{\p^3(n_s+4n_F)}{8(p_1+p_2+p_3)^2}\Big[p_1^3+2p_1^2(p_2+p_3)+(2p_1+p_2+p_3)\big((p_2+p_3)^2-p_2 p_3\big)\Big]\\
	A_{5}^{d=3}(p_1,p_2,p_3)&=\frac{\p^3(n_S-4n_F)}{240(p_1+p_2+p_3)^3}\Big[-3(p_1+p_2+p_3)^6+9(p_1+p_2+p_3)^4(p_1 p_2+p_2 p_3+p_1 p_3)\notag\\
	&+12(p_1+p_2+p_3)^2(p_1 p_2+p_2p_3+p_3 p_1)^2-33(p_1+p_2+p_3)^2p_1p_2p_3\notag\\
	&+12(p_1+p_2+p_3)(p_1p_2+p_2p_3+p_1p_3)p_1p_2p_3+8p_1^2p_2^2p_3^2\Big]\notag\\
	&+\frac{\p^3n_F}{12(p_1+p_2+p_3)^2}\Big[-(p_1+p_2+p_3)^5+3(p_1+p_2+p_3)^3(p_1p_2+p_2p_3+p_1p_3)\notag\\
	&+4(p_1+p_2+p_3)(p_1p_2+p_2p_3+p_1p_3)^2-11(p_1+p_2+p_3)^2p_1p_2p_3\notag\\
	&+4(p_1p_2+p_2p_3+p_1p_3)p_1p_2p_3\Big]-\frac{\p^3(n_S+4n_F)}{16}\Big[p_1^3+p_2^3+p_3^3\Big]
\end{align}

This is in agreement with the expression given in \cite{Bzowski:2013sza} in terms of he constant $\a_1,\a_2$ and $c_T$ if we choose 
(see \cite{Bzowski:2013sza})
\begin{align}
	\a_1=\frac{\p^3(n_S-4n_F)}{480}, \qquad \a_2=\frac{\p^3\,n_F}{6}, \qquad c_T=\frac{3\p^{5/2}}{128}(n_S+4n_F), \qquad c_g=0 
\end{align}
Notice that $c_g$ is a constant appearing in \cite{Bzowski:2013sza} related to the possibility of having a nonzero functional variation of the stress energy tensor respect to the metric ($\sim \delta T^{\mu\nu}(x)/\delta g_{\alpha\beta}(y))$ which is an extra contact term not included in our discussion.

\subsection{$d=5$ case}
In this case we have
\begin{equation}
	{C}_0(p_1^2,p_2^2,p_3^2)=\frac{\p^{3/2}}{p_1+p_2+p_3}.
\end{equation}
From \eqref{B0ex} the ${B}_0$ is calculated in $d=5$ as
\begin{equation}
	{B}_0(p_1^2)=-\frac{\p^{3/2}}{4} p_1.
\end{equation}
In the $d\to 5$ limit the $A_1$ form factor becomes, for instance, 
\begin{align}
	A_{1}^{d=5}(p_1,p_2,p_3)&=\frac{\p^4(n_S-4n_F)}{560(p_1+p_2+p_3)^7}\Big[(p_1+p_2+p_3)^2\big((p_1+p_2+p_3)^4+(p_1+p_2+p_3)^2(p_1p_2+p_2p_3+p_1p_3)\notag\\
	&\hspace{-1.9cm}+(p_1p_2+p_2p_3+p_1p_3)^2\big)+(p_1+p_2+p_3)\big((p_1+p_2+p_3)^2+5(p_1p_2+p_2p_3+p_1p_3)\big)p_1p_2p_3+10p_1^2p_2^2p_3^2\Big].
\end{align}
The remaining form factors are given in appendix of \cite{Coriano:2018bsy}. 
Their expressions are in agreement with those given in \cite{Bzowski:2013sza} when the 
corresponding constants (denoted by $\a_1$ and $\a_2$) are matched by the relations
\begin{align}
	\a_1=\frac{\p^4(n_S-4n_F)}{560 \times 72}, \qquad \a_2=\frac{\p^4\,n_F}{240}, \qquad c_T=\frac{5\p^{7/2}}{1024}(n_S+8n_F).
\end{align}
The case that we have analysed and their correspondence shows that we can safely move to $d=4$. In this case we will not 
attempt a comparison with the results of \cite{Bzowski:2013sza} which are far more involved and 
require the implementation of some recursion relations on the renormalized 3K integrals. In our case we will have to specialize our computation to $d=4$ with the inclusion of a third free field theory sector and extract the $A_i$'s after addressing their renormalization.

\section{The correlator in $d=4$ and the trace anomaly}
\begin{figure}[t]
	\centering
	\vspace{-1cm}
	\subfigure{\includegraphics[scale=0.2]{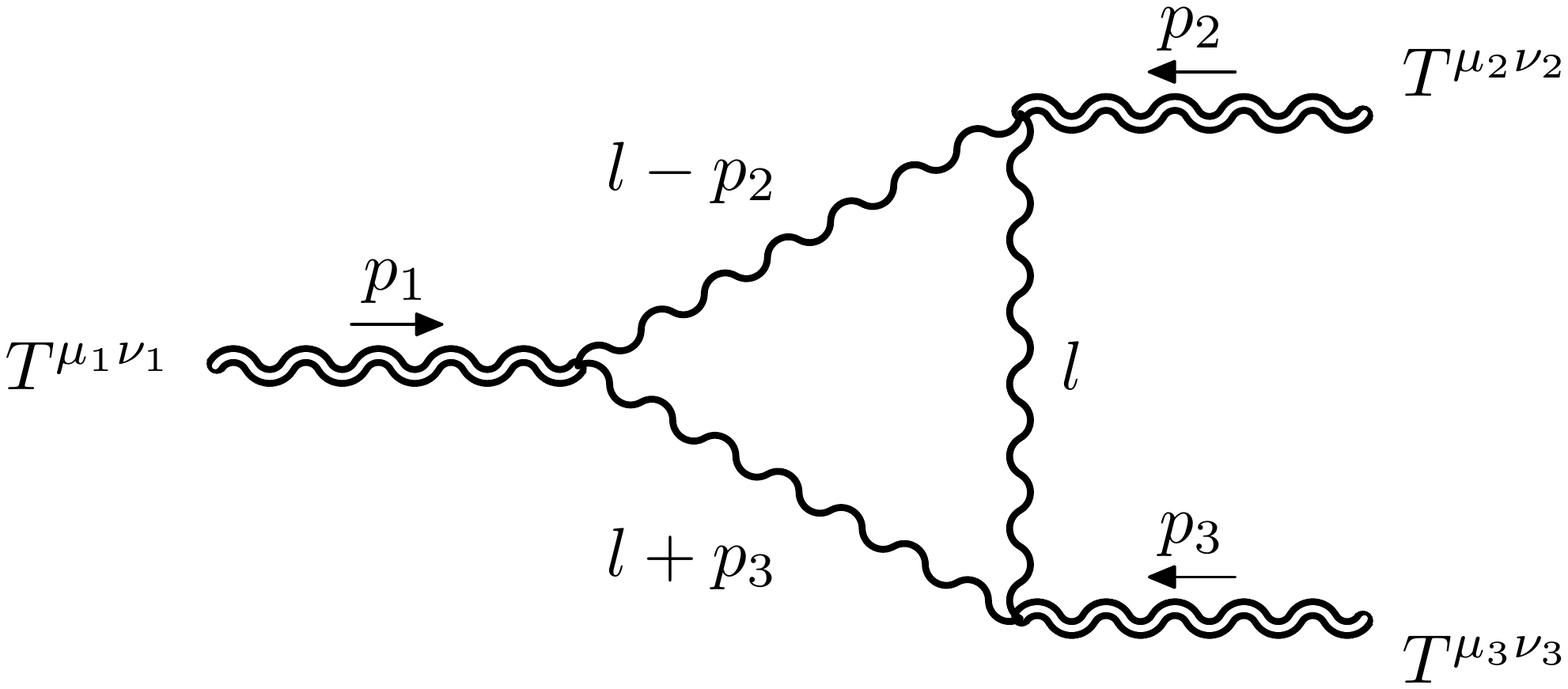}} \hspace{.3cm}
	\subfigure{\includegraphics[scale=0.2]{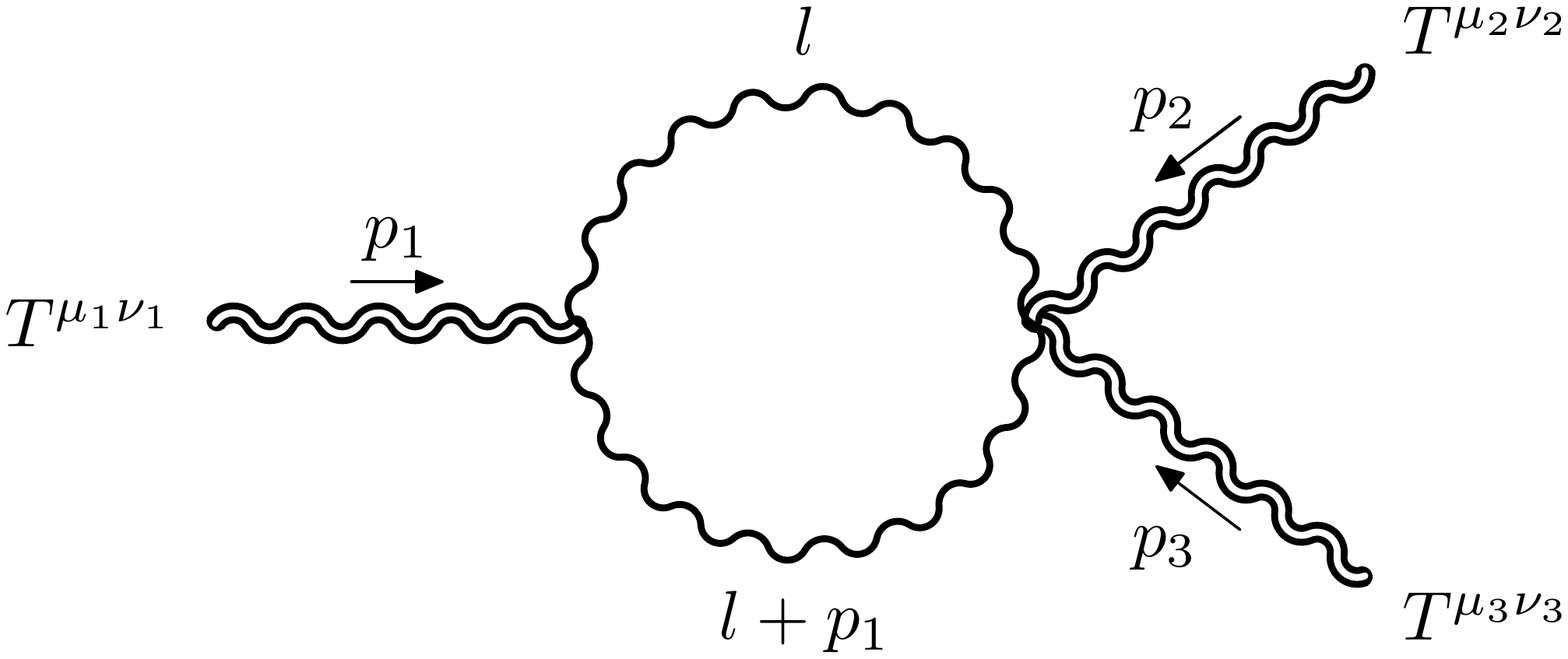}} \hspace{.3cm}
	\raisebox{.12\height}{\subfigure{\includegraphics[scale=0.16]{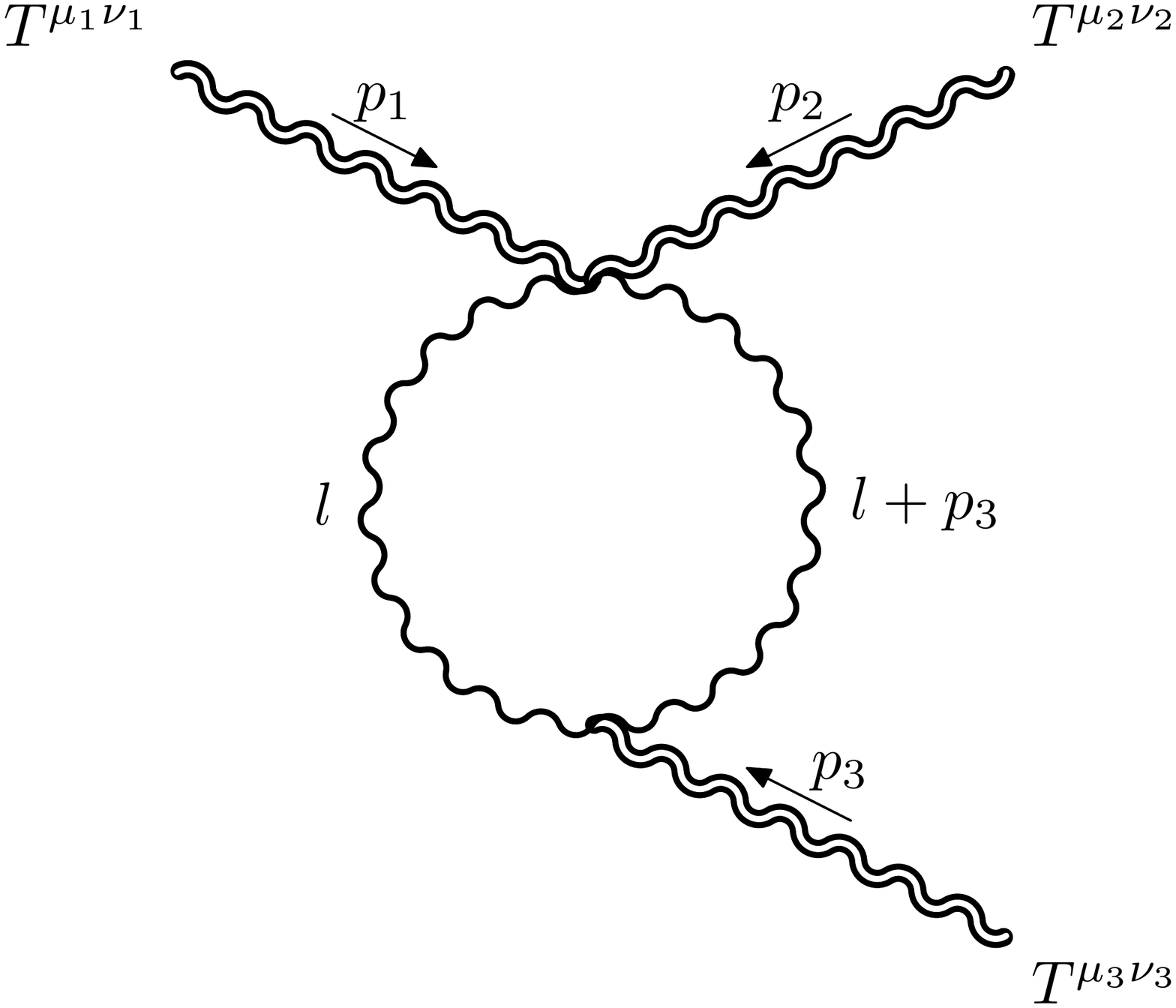}}\hspace{.3cm}}
	\raisebox{.12\height}{\subfigure{\includegraphics[scale=0.16]{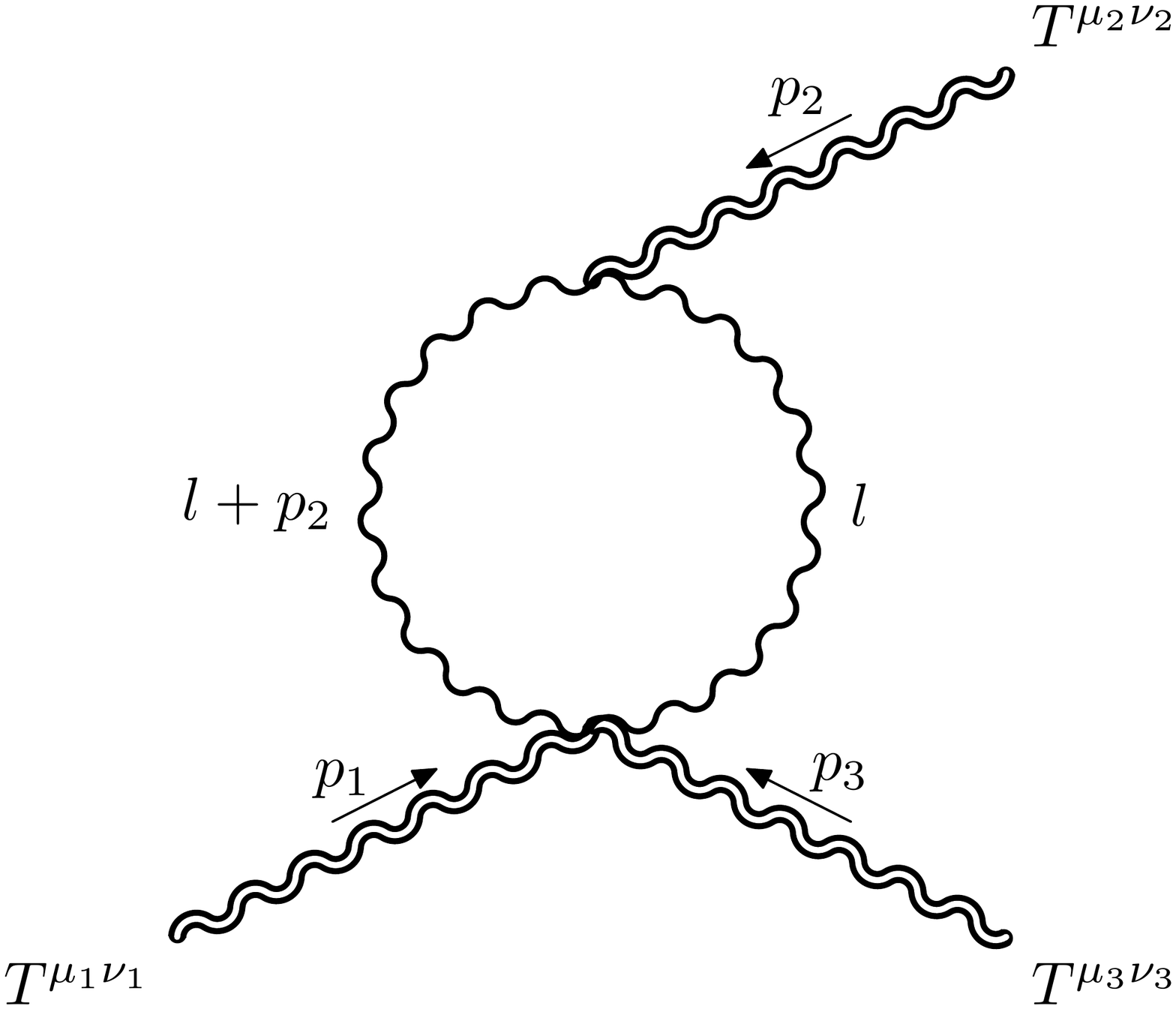}}}
	\vspace{-0.8cm}\caption{One-loop gauge diagrams for the three-graviton vertex.\label{Feynman3}}
\end{figure}
\subsection{Gauge and Ghost sectors}

We have to consider, as already mentioned, the contributions coming from the spin-1 sector, and in the one-loop approximation they correspond to the diagrams in \figref{Feynman3}. We have also to consider the contributions from the ghost, which can be calculated from the same type of diagrams given in \figref{Feynman3} but now with a ghost field running in the loop. A direct computation shows that the ghost and the gauge fixing contributions cancel.
Therefore we calculate in the one-loop approximation the contribution to the correlation function of the gauge sector, given by the diagrams in \figref{Feynman3}. These contributions can be written as 
\begin{align}
	\braket{T^{\m_1\n_1}(p_1)T^{\m_2\n_2}(p_2)T^{\m_3\n_3}(p_3)}_G=\, -V_{G}^{\m_1\n_1\m_2\n_2\m_3\n_3}(p_1,p_2,p_3)+\sum_{i=1}^3W_{G,i}^{\m_1\n_1\m_2\n_2\m_3\n_3}(p_1,p_2,p_3)\label{GaugeExpa}
\end{align}
for the triangle and the bubble topologies respectively. They are given by
\begin{align}
	&V_{G}^{\m_1\n_1\m_2\n_2\m_3\n_3}(p_1,p_2,p_3)=\notag\\
	&\qquad=\int \sdfrac{d^d\ell}{(2\pi)^d}\sdfrac{V^{\m_1\n_1\a_1\b_1}_{TAA}(\ell+p_3,\ell-p_2)\,\d_{\a_1\b_2}\,V^{\m_3\n_3\a_2\b_2}_{TAA}(\ell,\ell+p_3)\,\d_{\a_2\b_3}\,V^{\m_2\n_2\a_3\b_3}_{TAA}(\ell-p_2,\ell)\d_{\a_3\b_1}}{\ell^2(\ell-p_2)^2(\ell+p_3)^2}
\end{align}
\begin{align}
	W_{G,1}^{\m_1\n_1\m_2\n_2\m_3\n_3}(p_1,p_2,p_3)&=\frac{1}{2}\int \sdfrac{d^d\ell}{(2\pi)^d}\sdfrac{V^{\m_1\n_1\a_1\b_1}_{TAA}(\ell,\ell+p_1)\,\d_{\b_1\a_2}\,V^{\m_2\n_2\m_3\n_3\a_2\b_2}_{TTAA}(\ell+p_1,\ell)\,\d_{\a_1\b_2}}{\ell^2(\ell+p_1)^2}\\[2ex]
	W_{G,2}^{\m_1\n_1\m_2\n_2\m_3\n_3}(p_1,p_2,p_3)&=W_{G,1}^{\m_3\n_3\m_1\n_1\m_2\n_2}(p_3,p_1,p_2)\\[1ex]
	W_{G,3}^{\m_1\n_1\m_2\n_2\m_3\n_3}(p_1,p_2,p_3)&=W_{G,1}^{\m_2\n_2\m_1\n_1\m_3\n_3}(p_2,p_1,p_3).
\end{align}
One can show that the spin part of the two-graviton/two-fermion vertex does not contribute to the correlation function. \\
By acting with the transverse-traceless $\Pi$ projectors, we obtain the form factors $A_i$, $i=1,\dots,5$ in the fermion sector, in particular
\begin{align}
	\braket{t^{\m_1\n_1}(p_1)t^{\m_2\n_2}(p_2)t^{\m_3\n_3}(p_3)}_G&=\,\Pi^{\m_1\n_1}_{\a_1\b_1}(p_1)\Pi^{\m_2\n_2}_{\a_2\b_2}(p_2)\Pi^{\m_3\n_3}_{\a_3\b_3}(p_3)\notag\\
	&\times\bigg[ -V_{G}^{\a_1\b_1\a_2\b_2\a_3\b_3}(p_1,p_2,p_3)+\sum_{i=1}^3W_{G,i}^{\a_1\b_1\a_2\b_2\a_3\b_3}(p_1,p_2,p_3)\bigg]
\end{align}
Also in this case the number of gauge fields are kept arbitrary by the inclusion of an overall factor $n_G$.
\subsection{Divergences}

In $d=4$ the complete correlation function can be written as
\begin{equation}
	\braket{T^{\m_1\n_1}(p_1)T^{\m_2\n_2}(p_2)T^{\m_3\n_3}(p_3)}=\sum_{I=F,G,S}\,n_I\,\braket{T^{\m_1\n_1}(p_1)T^{\m_2\n_2}(p_2)T^{\m_3\n_3}(p_3)}_I
\end{equation}
also valid for the transverse traceless part of the correlator. In this case we encounter divergenes in the forms of single poles in $1/\epsilon$  ($\epsilon=(4-d)/2$). In this section we discuss the structures of such divergences and their elimination in DR using the two usual gravitational counterterms.\\
As a first remark, it is easy to realize for dimensional reasons and power counting that $A_{1}$ is UV finite. All other form factors have divergent parts explicitly given as
\begin{subequations}
	\begin{align}
		A_2^{Div}&=\frac{\p^2}{45\,\varepsilon}\,\big[26n_G-7n_F-2n_S\big]\\[1ex]
		A_3^{Div}&=\frac{\p^2}{90\,\varepsilon}\,\big[3(s+s_1)\big(6n_F+n_S+12n_G\big)+s_2(11n_F+62n_G+n_S)\big]\\[1ex]
		A_4^{Div}&=\frac{\p^2}{90\,\varepsilon}\,\big[(s+s_1)\big(29n_F+98n_G+4n_S\big)+s_2(43n_F+46n_G+8n_S)\big]\\[1ex]
		A_5^{Div}&=\frac{\p^2}{180\,\varepsilon}\bigg\{n_F \left(43 s^2-14 s (s_1+s_2)+43 s_1^2-14 s_1 s_2+43 s_2^2\right)\notag\\
		&\quad+2 \big[n_G \left(23 s^2+26 s (s_1+s_2)+23 s_1^2+26 s_1 s_2+23s_2^2\right)+2n_S\left(2 s^2-s (s_1+s_2)+2 s_1^2-s_1 s_2+2 s_2^2\right)\big]\bigg\}
	\end{align}\label{diverg}
\end{subequations}
and at this point we can proceed with their renormalization. 
\section{Renormalization of the $TTT$}
\label{renorm}
The renormalization of the 3-graviton vertex is obtained by the addition of 2 counterterms in the defining Lagrangian. In perturbation theory the one loop counterterm Lagrangian is
\begin{equation}
	S_{count}=-\sdfrac{1}{\varepsilon}\,\sum_{I=F,S,G}\,n_I\,\int d^dx\,\sqrt{-g}\bigg(\,\b_a(I)\,C^2+\b_b(I)\,E\bigg)
	\label{scount}
\end{equation}
corresponding to the Weyl tensor squared and the Euler density, omitting the extra 
$R^2$ operator which is responsible for the $\square R$ term in the trace anomaly \eqref{G}, having choosen the local part of anomaly 
$(\sim \beta_c\square R)$ vanishing ($\beta_c=0$). We refer to \cite{Coriano:2012wp} for a more detailed discussion of this point and of the finite renormalization needed to get from the general $\beta_c\neq 0$ to the $\beta_c=0$ case. By using the relation in \appref{Mvc}, the corresponding vertex counterterms are
\begin{align}
	&\braket{T^{\m_1\n_1}(p_1)T^{\m_2\n_2}(p_2)T^{\m_3\n_3}(p_3)}_{count}=\notag\\
	&\hspace{2cm}=-\sdfrac{1}{\varepsilon}\sum_{I=F,S,G}n_I\bigg(\b_a(I)\,V_{C^2}^{\m_1\n_1\m_2\n_2\m_3\n_3}(p_1,p_2,p_3)+\b_b(I)\,V_{E}^{\m_1\n_1\m_2\n_2\m_3\n_3}(p_1,p_2,p_3)\bigg)
\end{align} 
where
\begin{align}
	V_{C^2}^{\m_1\n_1\m_2\n_2\m_3\n_3}(p_1,p_2,p_3)&=8\int\,d^dx_1\,\,d^dx_2\,\,d^dx_3\,\,d^dx\,\bigg(\sdfrac{\d^3(\sqrt{-g}C^2)(x)}{\d g_{\m_1\n_1}(x_1)\d g_{\m_2\n_2}(x_2)\d g_{\m_3\n_3}(x_3)}\bigg)_{flat}\,e^{-i(p_1\,x_1+p_2\,x_2+p_3\,x_3)}\notag\\
	&\equiv 8\big[\sqrt{-g}\,C^2\big]^{\m_1\n_1\m_2\n_2\m_3\n_3}(p_1,p_2,p_3)\\[2ex]
	V_{E}^{\m_1\n_1\m_2\n_2\m_3\n_3}(p_1,p_2,p_3)&=8\int\,d^dx_1\,\,d^dx_2\,\,d^dx_3\,\,d^dx\,\bigg(\sdfrac{\d^3(\sqrt{-g}E)(x)}{\d g_{\m_1\n_1}(x_1)\d g_{\m_2\n_2}(x_2)\d g_{\m_3\n_3}(x_3)}\bigg)_{flat}\,e^{-i(p_1\,x_1+p_2\,x_2+p_3\,x_3)}\notag\\
	&\equiv 8\big[\sqrt{-g}\,E\big]^{\m_1\n_1\m_2\n_2\m_3\n_3}(p_1,p_2,p_3).\label{count}
\end{align}

These vertices satisfy the relations
\begin{align}
	\d_{\mu_1\nu_1}\,V_{C^2}^{\mu_1\nu_1\mu_2\nu_2\mu_3\nu_3}(p_1,p_2,p_3)&=4(d-4)\big[C^2\big]^{\mu_2\nu_2\mu_3\nu_3}(p_2,p_3)\notag\\
	&-8\,\bigg([C^2]^{\mu_2\nu_2\mu_3\nu_3}(p_1+p_2,p_3)+[C^2]^{\mu_2\nu_2\mu_3\nu_3}(p_2,p_1+p_3)\bigg)\\[2ex]
	\d_{\mu_1\nu_1}\,V_{E}^{\mu_1\nu_1\mu_2\nu_2\mu_3\nu_3}(p_1,p_2,p_3)&=4(d-4)\big[E\big]^{\mu_2\nu_2\mu_3\nu_3}(p_2,p_3)\\[2ex]
	p_{1\mu_1}\,V_{C^2}^{\mu_1\nu_1\mu_2\nu_2\mu_3\nu_3}(p_1,p_2,p_3)&=-4\,\bigg(p_2^{\nu_1}[C^2]^{\mu_2\nu_2\mu_3\nu_3}(p_1+p_2,p_3)+p_3^{\nu_1}[C^2]^{\mu_2\nu_2\mu_3\nu_3}(p_2,p_1+p_3)\bigg)\notag\\
	&\hspace{-0.7cm}+4\,p_{2\a}\bigg(\d^{\mu_2\nu_1}[C^2]^{\a\nu_2\mu_3\nu_3}(p_1+p_2,p_3)+\d^{\nu_2\nu_1}[C^2]^{\a\mu_2\mu_3\nu_3}(p_1+p_2,p_3)\bigg)\notag\\
	&\hspace{-0.7cm}+4\,p_{3\a}\bigg(\d^{\mu_3\nu_1}[C^2]^{\mu_2\nu_2\a\nu_3}(p_2,p_1+p_3)+\d^{\nu_3\nu_1}[C^2]^{\mu_2\mu_2\mu_3\a}(p_2,p_1+p_3)\bigg)\\[2ex]
	p_{1\mu_1}\,V_{E}^{\mu_1\nu_1\mu_2\nu_2\mu_3\nu_3}(p_1,p_2,p_3)&=0.
\end{align}

\section{Divergences of the two-point function: a worked out example}
Before coming to a discussion of the $TTT$, in this section we illustrate 
in some detail the way the generation of the extra tensor structures for such correlators takes place after renormalization. 
We will work out some of the intermediate steps first of the $TT$, for simplicity, presenting enough details which will be then applied to the $TTT$. 

We start by extending the analysis of Section \ref{compare} in the perturbative sector by including all the three sectors (scalar, fermion and gauge) in $d$ dimensions. This choice obviously violates conformal symmetry since the spin 1 contribution is not conformally invariant and it is responsible for an extra trace term proportional to $n_G$. One obtains
\begin{align}
	\braket{T^{\mu_1\nu_1}(p)T^{\mu_2\nu_2}(-p)}&=-\frac{\p^2\,p^4}{4(d-1)(d+1)}\,B_0(p^2)\,\P^{\mu_1\nu_1\mu_2\nu_2}(p)\Big[2(d-1)n_F+(2d^2-3d-8)n_G+n_S\Big]\notag\\
	&\hspace{1cm}+\frac{\p^2\,p^4\,n_G}{8(d-1)^2}(d-4)^2(d-2)\p^{\mu_1\nu_1}(p)\p^{\mu_2\nu_2}(p)\,B_0(p^2)\label{TTddim}
\end{align}
with a second contributon proportional to $n_G$. This term vanishes in $d=4$, as clear from the discussion below. \\
For this purpose, we recall that around $d=4$, the projectors are expanded according to the relation 
\begin{equation}
	\label{pexp}
	\P^{\,\mu_1\nu_1\mu_2\nu_2}(p)=\P^{(4)\,\mu_1\nu_1\mu_2\nu_2}(p)-\frac{2}{9}\varepsilon\,\pi^{\mu_1\nu_1}(p)\,\pi^{\mu_2\nu_2}(p)+O(\varepsilon^2). 
\end{equation}
This equations requires some clarification and we pause for a moment in order to illustrate its correct use. 

A consistent approach to the calculation is to perform all the tensor contractions in $d-$dimensions and only at the end move to $d=4$ in the limit of $\epsilon\to 0$. In this way one is reassured that the contraction of a metric tensor (in this case $\delta_{\mu}^{\mu}$, being us in the Euclidean case) gives $d$ and not $4$. The use of \eqref{pexp} is possible only if we are sure that there will not be any trace of the metric to perform. If these conditions are satisfied, then two methods of computation are equivalent and do not generate any ambiguity. \\
We illustrate this for the $TT$. Using \eqref{pexp} in \eqref{TTddim}, the latter takes the form

\begin{align}
	\label{result}
	\braket{T^{\mu_1\nu_1}(p)T^{\mu_2\nu_2}(-p)}&=-\frac{\p^2\,p^4}{4}\,\bigg(\frac{1}{\varepsilon}+\bar{B}_0(p^2)\bigg)\,\bigg(\P^{(4)\,\mu_1\nu_1\mu_2\nu_2}(p)-\frac{2}{9}\varepsilon\,\pi^{\mu_1\nu_1}(p)\,\pi^{\mu_2\nu_2}(p)+O(\varepsilon^2)\bigg)\notag\\
	&\hspace{-2cm}\times\Bigg[\bigg(\frac{2}{5}+\frac{4}{25}\varepsilon+O(\varepsilon^2)\bigg)n_F+\bigg(\frac{4}{5}-\frac{22}{25}\varepsilon+O(\varepsilon^2)\bigg)n_G+\bigg(\frac{1}{15}+\frac{16}{225}\varepsilon+O(\varepsilon^2)\bigg)n_S\Bigg]\notag\\
	&\hspace{-1cm}+\frac{\p^2\,p^4\,n_G}{8}\p^{\mu_1\nu_1}(p)\p^{\mu_2\nu_2}(p)\,\bigg(\frac{1}{\varepsilon}+\bar{B}_0(p^2)\bigg)\bigg[\frac{8}{9}\varepsilon^2+\frac{8}{27}\varepsilon^3+O(\varepsilon^4)\bigg]
\end{align}
where $\P^{ (4)\,\,\mu_1\nu_1\mu_2\nu_2}(p)$ is the transverse and traceless projector in $d=4$ and $\bar{B}_0(p^2)= 2 + \log(\mu^2/p^2)$ is the finite part in $d=4$ of the scalar integral in the $\overline{MS}$ scheme. As anticipated above, the last term of \eqref{result}, generated by the addition of a non-conformal sector ($\sim n_G$) vanishes separately as $\epsilon\to 0$. Finally, combining all the terms we obtain the regulated ($reg$) expression of the $TT$ around $d=4$ in the form

\begin{align}
	\label{regular}
	\braket{T^{\mu_1\nu_1}(p)T^{\mu_2\nu_2}(-p)}_{reg}&=-\frac{\p^2\,p^4}{60\,\varepsilon}\Pi^{(4)\,\mu_1\nu_1\mu_2\nu_2}(p)\left(6 n_F + 12 n_G + n_S\right)\notag\\
	&\hspace{-3cm}+\frac{\p^2\,p^4}{270}\p^{\mu_1\nu_1}(p)\p^{\mu_2\nu_2}(p)\left(6 n_F + 12 n_G + n_S\right)-\frac{\p^2\,p^4}{300}\bar{B}_0(p^2)\Pi^{\mu_1\nu_1\mu_2\nu_2}(p)\left(30n_F+60n_G+5n_S\right)\notag\\
	&\hspace{-2cm}-\frac{\p^2\,p^4}{900}\Pi^{\mu_1\nu_1\mu_2\nu_2}(p)\left(36n_F-198 n_G+16n_S\right)+O(\varepsilon)
\end{align}
The divergence in the previous expression can be removed through the one loop counterterm Lagrangian \eqref{scount}. In fact, the second functional derivative of $S_{count}$ with respect to the background metric gives 
\begin{align}
	\braket{T^{\mu_1\nu_1}(p)T^{\mu_2\nu_2}(-p)}_{count}&\equiv -\sdfrac{1}{\varepsilon}\sum_{I=F,S,G}\bigg(4\b_a(I)\,\big[\sqrt{-g}\,C^2\big]^{\m_1\n_1\m_2\n_2}(p,-p)\bigg)\notag\\
	&=-\frac{8}{\varepsilon}\frac{(d-3)\,}{(d-2)}p^4\P^{(d)\,\mu_1\nu_1\mu_2\nu_2}(p)\,\bigg(n_S\,\b_a(S)+n_F\,\b_a(F)+n_G\,\b_a(G)\,\bigg)
\end{align}
having used the relation $V_{E}^{\m_1\n_1\m_2\n_2}(p,-p)=0$. In particular, expanding around $d=4$ and using again \eqref{pexp} we obtain
\begin{align}
	\braket{T^{\mu_1\nu_1}(p)T^{\mu_2\nu_2}(-p)}_{count}&=-\frac{8\,p^4}{\varepsilon}\bigg(\P^{(4)\,\mu_1\nu_1\mu_2\nu_2}(p)-\frac{2}{9}\varepsilon\,\pi^{\mu_1\nu_1}(p)\,\pi^{\mu_2\nu_2}(p)+O(\varepsilon^2)\bigg)\,\bigg(\frac{1}{2}-\frac{\varepsilon}{2}+O(\varepsilon^2)\bigg)\notag\\
	&\hspace{1cm}\times\bigg(n_S\,\b_a(S)+n_F\,\b_a(F)+n_G\,\b_a(G)\,\bigg)\notag\\
	&\hspace{-3cm}=-\sdfrac{4}{\varepsilon}p^4\bigg(n_S\,\b_a(S)+n_F\,\b_a(F)+n_G\,\b_a(G)\,\bigg)\,\P^{(4)\,\mu_1\nu_1\mu_2\nu_2}(p)\notag\\
	&\hspace{-2cm}+4\, p^4\bigg(n_S\,\b_a(S)+n_F\,\b_a(F)+n_G\,\b_a(G)\,\bigg)\bigg[\P^{(4)\,\mu_1\nu_1\mu_2\nu_2}(p)+\frac{2}{9}\p^{\mu_1\nu_1}(p)\p^{\mu_2\nu_2}(p)\bigg]+O(\varepsilon)
\end{align}
which cancels the divergence arising in the two point function, if one chooses the parameters as in \eqref{choiceparm}.  
The renormalized 2-point function using  \eqref{choiceparm} then takes the form  \begin{align}
	\braket{T^{\mu_1\nu_1}(p)T^{\mu_2\nu_2}(-p)}_{Ren}&=\braket{T^{\mu_1\nu_1}(p)T^{\mu_2\nu_2}(-p)}+\braket{T^{\mu_1\nu_1}(p)T^{\mu_2\nu_2}(-p)}_{count}\notag\\
	&=-\frac{\p^2\,p^4}{60}\bar{B}_0(p^2)\Pi^{\mu_1\nu_1\mu_2\nu_2}(p)\left(6n_F+12n_G+n_S\right)\notag\\
	&\quad-\frac{\p^2\,p^4}{900}\P^{\mu_1\nu_1\mu_2\nu_2}(p)\big(126n_F-18n_G+31n_S\big)
	\label{tren}
\end{align}
Notice that the choice $\beta_c=0$ takes us to 
a final expression which is transverse and traceless. 
The same choice of parameters $\b_a,\,\b_b$ given in \eqref{choiceparm} removes the divergences in the three point function, as we are going to discuss below.

\section{Anomalous Conformal Ward Identities in $d=4$ and free field content}
The divergences arising in the form factors in $d=4$ and their renormalization induce a breaking of the conformal symmetry, thereby generating a set of anamalous CWI's.  In this section we will give the explicit form of the such identities in the presence of a trace anomaly.

\subsection{Primary anomalous CWI's and free field content}
The equations for the anomalous primary CWI's are generated after renormalization, starting from the $d$-dimensional expressions of the $A_i$'s  given in \cite{Coriano:2018bsy}. The renormalization procedure will involve only $B_0$.
The primary anomalous CWI's take the form
\begin{equation}
	\begin{split}
		& \textup{K}_{13}A^{Ren}_3=2A^{Ren}_2-\sdfrac{2\p^2}{45}\left(7n_F-26n_G+2n_S\right) \\
		&\textup{K}_{23}A^{Ren}_3=2A^{Ren}_2-\sdfrac{2\p^2}{45}\left(7n_F-26n_G+2n_S\right) \\[1.1ex]
		& \textup{K}_{13}A^{Ren}_4=-4A^{Ren}_2(p_2\leftrightarrow p_3)+\sdfrac{4\p^2}{45}\left(7n_F-26n_G+2n_S\right) \\
		&\textup{K}_{23}A^{Ren}_4=-4A^{Ren}_2(p_1\leftrightarrow p_3)+\sdfrac{4\p^2}{45}\left(7n_F-26n_G+2n_S\right) \\[1.1ex]
		& \textup{K}_{13}A^{Ren}_5=2\left[A^{Ren}_4-A^{Ren}_4(p_1\leftrightarrow p_3) \right]-\sdfrac{4\p^2}{9}(s-s_2)\left(5n_F+2n_G+n_s\right)\\
		&\textup{K}_{23}A^{Ren}_5=2\left[A^{Ren}_4-A^{Ren}_4(p_2\leftrightarrow p_3)\right] -\sdfrac{4\p^2}{9}(s_1-s_2)\left(5n_F+2n_G+n_s\right)
	\end{split}\label{PrimaryAnom}
\end{equation}
where now the differential operators $K_i$ take the form
\begin{equation}
	K_i=\frac{\partial^2}{\partial p_i^2}-\frac{3}{p_i}\frac{\partial}{\partial p_i}=4s_{i-1}\frac{\partial^2}{\partial s_{i-1}^2}-4\frac{\partial}{\partial s_{i-1}},\qquad i=1,2,3
\end{equation}
with the identification $s_0=s$. The $(p_1\leftrightarrow p_3)$ and $(p_2\leftrightarrow p_3)$ versions of the anomalous Ward identities can be obtained from \eqref{PrimaryAnom}. 
Using the expressions given in the appendix of \cite{Coriano:2018bsy}, we can identify the corresponding counterterms for the $A_i$, extracted from the 
transverse traceless parts of the vertices generated by the counterterm Lagrangian in \eqref{scount}, obtaining
\begin{align}
	A_2^{count}&=-\frac{16}{\varepsilon}\,\sum_{I=F,S,G}\,n_I\,\big[\b_a(I)+\b_b(I)\big]\notag\\
	A_3^{count}&=-\frac{8}{\varepsilon}\,\sum_{I=F,S,G}\,n_I\,\big[s_2\,\b_b(I)-(s+s_1)\b_a(I)\big]+o(\epsilon)\notag\\
	A_4^{count}&=-\frac{8}{\varepsilon}\,\sum_{I=F,S,G}\,n_I\,\big[(s+s_1-s_2)\,\b_b(I)-(s+s_1+3s_2)\b_a(I)\big]+o(\epsilon)\notag\\
	A_5^{count}&=-\frac{4}{\varepsilon}\,\sum_{I=F,S,G}\,n_I\,\big[-\big(s^2-2s(s_1+s_2)+(s_1-s_2)^2\big)\,\b_b(I)\notag\\
	&\hspace{3cm}-\big(3s^2-2s(s_1+s_2)+3s_1^2-2s_1s_2+3s_2^2\big)\b_a(I)\big]+o(\epsilon)
\end{align}
In order to cancel the divergences arising from the form factors, we need to choose the coefficient $\b_b(I)$ and $\b_a(I)$ as in \eqref{choiceparm}. The renormalized form factors can then be written as
\begin{align}
	A_2^{Ren}&=A_2^{Reg}\\
	A_3^{Ren}&=A_3^{Reg}-8\,(s+s_1+s_2)\sum_{I=F,S,G}n_I\,\b_a(I)
\end{align}
\begin{align}
	A_4^{Ren}&=A_4^{Reg}-16\,(s+s_1+s_2)\sum_{I=F,S,G}n_I\,\b_a(I)\\
	A_5^{Ren}&=A_5^{Reg}-8\,(s^2+s_1^2+s_2^2)\sum_{I=F,S,G}n_I\,\b_a(I)
\end{align}
where with ``reg'' we indicate those form factors which remain unmodified by the procedure, being finite. Such are $A_1$ and $A_2$.
\subsection{Secondary anomalous CWI's from free field theory }
The derivation of the secondary anomalous CWI's has been discussed within the general formalism in \cite{Bzowski:2013sza}  and in the perturbative approach in \cite{Coriano:2018bbe} in the case of the $TJJ$ correlator. The details of this analysis, which has been discussed at length in our previous work \cite{Coriano:2018bbe}, also in this case remain similar. We refer to \secref{secondary2} for a definition of the corresponding operators appearing in such equations and to \cite{Coriano:2018bbe}.  A lengthy computation gives
\begin{align}
	&L_6 A^{Ren}_1+R A^{Ren}_2-R A^{Ren}_2(p_2\leftrightarrow p_3)=0\\ 
	&L_4\,A^{Ren}_2+2p_1^2\,A^{Ren}_2+4RA_3-2RA^{Ren}_4(p_1\leftrightarrow p_3)=\frac{4\p^2\,p_1^2}{45}\left(7n_F-26n_G+2n_S\right)\\[1.5ex]
	&L_4\,A^{Ren}_2(p_1\leftrightarrow p_3)-R\,A^{Ren}_4+RA^{Ren}_4(p_2\leftrightarrow p_3)+2p_1^2(A^{Ren}_2(p_2\leftrightarrow p_3)-A^{Ren}_2)=\frac{2\p^2\,p_1^2}{45}\left(7n_F-26n_G+2n_S\right)\\
	&L_4\,A^{Ren}_2(p_2\leftrightarrow p_3)-4R\,A^{Ren}_3(p_2\leftrightarrow p_3)+2RA^{Ren}_4(p_1\leftrightarrow p_3)-2p_1^2A^{Ren}_2(p_2\leftrightarrow p_3)=0\\
	&L_2\,A^{Ren}_3(p_1\leftrightarrow p_3)+p_1^2(A^{Ren}_4-A^{Ren}_4(p_2\leftrightarrow p_3)=\frac{30\p^2}{225}(6n_F+12n_G+n_S)\big(s_2^2 \bar{B}_0(s_2)-s_1^2B_0^{Reg}(s_1)\big)\notag\\
	&\hspace{+1cm}-\frac{\p^2}{225}\bigg[n_F\big(55s^2+5s(29s_1+7s_2)+252(s_1^2-s_2^2)\big)+2n_G\big(155s^2+245s\,s_1-65s\,s_2-18(s_1^2-s_2^2)\big)\notag\notag\\
	&\hspace{+1.5cm}+n_S\big(5s^2+10s(2s_1+s_2)+62(s_1^2-s_2^2)\big)\bigg]\\
	&L_2\,A^{Ren}_4+2R\,A^{Ren}_5+8p_1^2A^{Ren}_3(p_2\leftrightarrow p_3)-2p_1^2(A^{Ren}_4+A^{Ren}_4(p_1\leftrightarrow p_3))=\notag\notag\\
	&\hspace{+1cm}-\frac{120\p^2\,s_1^2}{225}B_0(s_1)\big(6n_F+12n_G+n_sS\big)-\frac{4\p^2}{225}\Bigg[15s^2(6n_F+12n_G+n_S)+5s\,s_1(11n_F+62n_G)\notag\notag\\
	&\hspace{+1.5cm}+2s_1^2(126n_F-18n_G+31n_S)\bigg]\\
	&L_2\,A^{Ren}_4(p_2\leftrightarrow p_3)-2R\,A^{Ren}_5-8p_1^2A^{Ren}_3+2p_1^2(A^{Ren}_4(p_2\leftrightarrow p_3)+A^{Ren}_4(p_1\leftrightarrow p_3))=\notag\\
	&\hspace{+1cm}+\frac{2\p^2}{225} \bigg[60 s_2^2(6 n_F+12 n_G+n_S)B_0^{Reg}(s_2) +5 s \bigg(s (7 n_F-26 n_G+2 n_S)-s_1 (43 n_F+46 n_G+8 n_S)\bigg)\notag\\
	&\hspace{+1.5cm}-5 s s_2 (7 n_F-26 n_G+2 n_S)+4 s_2^2 (126 n_F-18 n_G+31 n_S)\bigg].
\end{align}
The most involved part of this analysis involves a rewriting of the differential action of the $L$ operators on $B_0$ and $C_0$. 
We have explicitly verified that the renormalized $A_i$ satisfy such equations confirming the consistency of the entire approach. 
\section{Reconstruction of the $\braket{TTT}$ in $d=4$}
In this section we will illustrate the reconstruction procedure for the $TTT$ using the perturbative realization of this correlator. In this case our goal will be to show how the separation of the vertex into a traceless part and an anomaly contribution takes place after renormalization. As already remarked in the introduction, the advantage of using a direct perturbative approach is to present for the 
transverse traceless sector of this vertex the simplest explicit form, in terms of the renormalized scalar 2- and 3-point functions. \\
The approach is obviously the standard one, where the renormalization is obtained by the addition to the bare vertex of the counterterms worked out in the previous two sections, but we will try to illustrate in some detail how the generation of the anomaly poles in the trace part takes place in these types of correlators. \\
We start from the bare local contributions in $d$ dimensions which take the form
\begin{align}
	\label{loc}
	\braket{t_{loc}^{\mu_1\nu_1}T^{\mu_2\nu_2}T^{\mu_3\nu_3}}&=\Big(\mathcal{I}^{\mu_1\nu_1}_{\a_1}(p_1)\,p_{1\b_1}+\frac{\p^{\mu_1\nu_1}(p_1)}{(d-1)}\d_{\a_1\b_1}\Big)\braket{T^{\a_1\b_1}T^{\mu_2\nu_2}T^{\mu_3\nu_3}}\notag\\
	&=-\frac{2\,\p^{\mu_1\nu_1}(p_1)}{(d-1)}\Big[\braket{T^{\mu_2\nu_2}(p_1+p_2)T^{\mu_3\nu_3}(p_3)}+\braket{T^{\mu_2\nu_2}(p_2)T^{\mu_3\nu_3}(p_1+p_3)}\Big]
	\notag\\
	&+\mathcal{I}^{\mu_1\nu_1}_{\a_1}(p_1)\Big\{-p_2^{\a_1}\braket{T^{\mu_2\nu_2}(p_1+p_2)T^{\mu_3\nu_3}(p_3)}-p_3^{\a_1}\braket{T^{\mu_2\nu_2}(p_2)T^{\mu_3\nu_3}(p_1+p_3)}\notag\\
	&+p_{2\b}\Big[\d^{\a_1\mu_2}\braket{T^{\b\mu_2}(p_1+p_2)T^{\mu_3\nu_3}(p_3)}+\d^{\a_1\nu_2}\braket{T^{\b\mu_2}(p_1+p_2)T^{\mu_3\nu_3}(p_3)}\Big]\notag\\
	&+p_{3\b}\Big[\d^{\a_1\mu_3}\braket{T^{\nu_2\mu_2}(p_2)T^{\b_3\nu_3}(p_1+p_3)}+\d^{\a_1\nu_3}\braket{T^{\mu_2\nu_2}(p_2)T^{\mu_3\b}(p_1+p_3)}\Big]\Big\}
\end{align}
which develop a singularity for $\epsilon\to 0$, with $\epsilon=(4-d)/2$, just like all the other contributions appearing in \eqref{tttdec}. We pause for a moment to describe the structure of this expression and comment on the general features of the regularization procedure.
\\ We perform all the tensor contractions in $d$ dimensions and in the final expression we set $d= 4 +\epsilon$. For example, if a projector such as $\Pi^{(d)}$ appears, we will be using Eq.~\eqref{pexp}, which relates $\Pi^{(d)}$ to $\Pi^{(4)}$, and so on. For instance, a projector such as $\pi^{\mu_1\nu_1}$ with open indices remains unmodified since it has no explicit $d$-dependence, unless it is contracted with a $\delta^{\mu\nu}$. It is then clear, from a cursory look at the right hand side of \eqref{loc} that the regulated expression of this expression involves a prefactor $1/(d-1)$, which is expanded around $d=4$ and the replacements of all the two point functions with the regulated expression given by Eq. \eqref{regular}, with the insertion of the appropriate momenta.\\
The corresponding counterterm is given by 
\begin{align}
	\label{locco}
	\braket{t_{loc}^{\mu_1\nu_1}T^{\mu_2\nu_2}T^{\mu_3\nu_3}}_{count}&=\Big(\mathcal{I}^{\mu_1\nu_1}_{\a_1}(p_1)\,p_{1\b_1}+\frac{\p^{\mu_1\nu_1}(p_1)}{(d-1)}\d_{\a_1\b_1}\Big)\braket{T^{\a_1\b_1}T^{\mu_2\nu_2}T^{\mu_3\nu_3}}_{(count)}\notag\\
	&\hspace{-3cm}=-\frac{1}{\varepsilon}\frac{(d-4)}{(d-1)}\p^{\mu_1\nu_1}(p_1)\bigg(4[E]^{\mu_2\nu_2\mu_3\nu_3}(p_2,p_3)+4[C^2]^{\mu_2\nu_2\mu_3\nu_3}(p_2,p_3)\bigg)\notag\\
	&\hspace{-3cm}+\frac{1}{\varepsilon}\frac{2}{(d-1)}\p^{\mu_1\nu_1}(p_1)\bigg(4[C^2]^{\mu_2\nu_2\mu_3\nu_3}(p_1+p_2,p_3)+4[C^2]^{\mu_2\nu_2\mu_3\nu_3}(p_2,p_1+p_3)\bigg)\notag\\
	&\hspace{-3cm}-\frac{1}{\varepsilon}\mathcal{I}^{\mu_1\nu_1}_{\a_1}(p_1)\bigg\{-4p_2^{\a_1}[C^2]^{\mu_2\nu_2\mu_3\nu_3}(p_1+p_2,p_3)-p_3^{\a_1}[C^2]^{\mu_2\nu_2\mu_3\nu_3}(p_2,p_1+p_3)\notag\\
	&\hspace{-3cm}+4p_{2\b}\Big[\d^{\a_1\mu_2}[C^2]^{\b\nu_2\mu_3\nu_3}(p_1+p_2,p_3)+\d^{\a_1\nu_2}[C^2]^{\mu_2\b\mu_3\nu_3}(p_1+p_2,p_3)\Big]\notag\\
\end{align}
\begin{align}
	&\hspace{0cm}+4p_{3\b}\Big[\d^{\a_1\mu_3}[C^2]^{\mu_2\nu_2\b\nu_3}(p_2,p_1+p_3)+\d^{\a_1\nu_3}[C^2]^{\mu_2\nu_2\mu_3\b}(p_2,p_1+p_3)\Big]\bigg\}.
\end{align}
where, for simplicity, we have absorbed the dependence on the total contributions to the beta functions $\beta_a$ and $\beta_b$ 
\begin{equation}
	\beta_{a,b}\equiv\sum_{I=f,s,G} \beta_{a,b} (I)
\end{equation}
into $[E]$ and $[C^2]$.\\
It is worth mentioning that all the divergent parts of the local term given in \eqref{loc} above are cancelled by the local parts of the counterterm \eqref{locco}. For its renormalized expression we obtain
\begin{align}
	\braket{t_{loc}^{\mu_1\nu_1}T^{\mu_2\nu_2}T^{\mu_3\nu_3}}_{Ren}&=\braket{t_{loc}^{\mu_1\nu_1}T^{\mu_2\nu_2}T^{\mu_3\nu_3}}+\braket{t_{loc}^{\mu_1\nu_1}T^{\mu_2\nu_2}T^{\mu_3\nu_3}}_{(count)}\notag\\
	&=\mathcal{V}_{loc\,0\,0 } +\braket{t_{loc}^{\mu_1\nu_1}T^{\mu_2\nu_2}T^{\mu_3\nu_3}}^{(4)}_{extra}
\end{align}
where
\begin{align}
	&\mathcal{V}_{loc\, 0 \, 0 }=-\frac{2\,\p^{\mu_1\nu_1}(p_1)}{3}\Big[\braket{T^{\mu_2\nu_2}(p_1+p_2)T^{\mu_3\nu_3}(-p_1-p_2)}_{Ren}+\braket{T^{\mu_2\nu_2}(p_2)T^{\mu_3\nu_3}(-p_2)}_{Ren}\Big]
	\notag\\
	&\quad+\mathcal{I}^{(4)\,\mu_1\nu_1}_{\a_1}(p_1)\Big\{-p_2^{\a_1}\braket{T^{\mu_2\nu_2}(p_1+p_2)T^{\mu_3\nu_3}(-p_1-p_2)}_{Ren}-p_3^{\a_1}\braket{T^{\mu_2\nu_2}(p_2)T^{\mu_3\nu_3}(-p_2)}_{Ren}\notag\\
	&\quad+p_{2\b}\Big[\d^{\a_1\mu_2}\braket{T^{\b\mu_2}(p_1+p_2)T^{\mu_3\nu_3}(-p_1-p_2)}_{Ren}+\d^{\a_1\nu_2}\braket{T^{\b\mu_2}(p_1+p_2)T^{\mu_3\nu_3}(-p_1-p_2)}_{Ren}\Big]\notag\\
	&\quad+p_{3\b}\Big[\d^{\a_1\mu_3}\braket{T^{\nu_2\mu_2}(p_2)T^{\b\nu_3}(-p_2)}_{Ren}+\d^{\a_1\nu_3}\braket{T^{\mu_2\nu_2}(p_2)T^{\mu_3\b}(-p_2)}_{Ren}\Big]\Big\}
\end{align}
with $\braket{TT}_{ren}$ given by \eqref{tren}. Notice the presence of an extra contribution coming from the local parts of counterterms that takes the explicit form
\begin{align}
	&\braket{t_{loc}^{\mu_1\nu_1}T^{\mu_2\nu_2}T^{\mu_3\nu_3}}^{(4)}_{extra}=\frac{\hat{\p}^{\mu_1\nu_1}(p_1)}{3 \, p_1^2}\bigg(4[E]^{\mu_2\nu_2\mu_3\nu_3}(p_2,p_3)+4[C^2]^{\mu_2\nu_2\mu_3\nu_3}(p_2,p_3)\bigg),
\end{align}
having defined 
\begin{align}
	\hat{\p}^{\mu\nu}(p)=(\delta^{\mu_1\nu_1}p^2 - p^\mu p^\nu)
\end{align}
which shows the emergence of an anomaly pole, similarly to the $TJJ$ cases  \cite{Giannotti:2008cv,Armillis:2009pq,Coriano:2018zdo}. \\
The renormalization of the other local contributions follows a similar pattern. In particular, the correlator with two $t_{loc}$ projections takes the form
\begin{align}
	&\braket{t_{loc}^{\mu_1\nu_1}t_{loc}^{\mu_2\nu_2}T^{\mu_3\nu_3}}_{Ren}= \mathcal{V}^{\mu_1\nu_1\mu_2\nu_2\mu_3\nu_3}_{loc\, loc \, 0} +\braket{t_{loc}^{\mu_1\nu_1}t_{loc}^{\mu_2\nu_2}T^{\mu_3\nu_3}}^{(4)}_{extra}
\end{align}
where
\begin{align}
	&\mathcal{V}^{\mu_1\nu_1\mu_2\nu_2\mu_3\nu_3}_{loc\, loc \, 0}=\Big(\mathcal{I}^{(4)\,\mu_2\nu_2}_{\a_2}(p_2)\,p_{2\b_2}+\frac{\p^{\mu_2\nu_2}(p_2)}{3}\d_{\a_2\b_2}\Big)\notag\\
	&\times\Bigg\{-\frac{2\,\p^{\mu_1\nu_1}(p_1)}{3}\Big[\braket{T^{\a_2\b_2}(p_1+p_2)T^{\mu_3\nu_3}(-p_1-p_2)}_{Ren}+\braket{T^{\a_2\b_2}(p_2)T^{\mu_3\nu_3}(-p_2)}_{Ren}\Big]
	\notag\\
	&\quad+\mathcal{I}^{(4)\,\mu_1\nu_1}_{\a_1}(p_1)\Big[-p_2^{\a_1}\braket{T^{\a_2\b_2}(p_1+p_2)T^{\mu_3\nu_3}(-p_1-p_2)}_{Ren}-p_3^{\a_1}\braket{T^{\a_2\b_2}(p_2)T^{\mu_3\nu_3}(-p_2)}_{Ren}\notag\\
	&\quad+p_{2\b}\Big(\d^{\a_1\a_2}\braket{T^{\b\b_2}(p_1+p_2)T^{\mu_3\nu_3}(-p_1-p_2)}_{Ren}+\d^{\a_1\b_2}\braket{T^{\b\a_2}(p_1+p_2)T^{\mu_3\nu_3}(-p_1-p_2)}_{Ren}\Big)\notag\\
	&\quad+p_{3\b}\Big(\d^{\a_1\mu_3}\braket{T^{\b_2\a_2}(p_2)T^{\b\nu_3}(-p_2)}_{Ren}+\d^{\a_1\nu_3}\braket{T^{\a_2\b_2}(p_2)T^{\mu_3\b}(-p_2)}_{Ren}\Big)\Big]\Bigg\}
\end{align}
in which we define 
\begin{align}
	\mathcal{I}^{(4)\,\mu\nu}_{\a}(p)\equiv \frac{1}{p^2}\left[2 p^{(\mu}\delta^{\nu)}_\alpha - 
	\frac{p_\alpha}{3}\left(\delta^{\mu\nu} +2\,\frac{p^\mu p^\nu}{p^2}\right)\right]
\end{align}
and with the presence of an extra term of the form
\begin{align}
	\braket{t_{loc}^{\mu_1\nu_1}t_{loc}^{\mu_2\nu_2}T^{\mu_3\nu_3}}^{(4)}_{extra}=\frac{\p^{\mu_1\nu_1}(p_1)}{3}\frac{\p^{\mu_2\nu_2}(p_2)}{3}\d_{\a_2\b_2}\bigg(4[E]^{\a_2\b_2\mu_3\nu_3}(p_2,p_3)+4[C^2]^{\a_2\b_2\mu_3\nu_3}(p_2,p_3)\bigg)
\end{align}
and finally the term with three insertions of $t_{loc}$
\begin{align}
	&\braket{t_{loc}^{\mu_1\nu_1}t_{loc}^{\mu_2\nu_2}t_{loc}^{\mu_3\nu_3}}_{Ren}= \mathcal{V}^{\mu_1\nu_1\mu_2\nu_2\mu_3\nu_3}_{loc\,loc\,loc} +\braket{t_{loc}^{\mu_1\nu_1}t_{loc}^{\mu_2\nu_2}t_{loc}^{\mu_3\nu_3}}^{(4)}_{extra}
\end{align}
with
\begin{align}
	&\mathcal{V}^{\mu_1\nu_1\mu_2\nu_2\mu_3\nu_3}_{loc\,loc\,loc}=\Big(\mathcal{I}^{(4)\,\mu_2\nu_2}_{\a_2}(p_2)\,p_{2\b_2}+\frac{\p^{\mu_2\nu_2}(p_2)}{3}\d_{\a_2\b_2}\Big)\Big(\mathcal{I}^{(4)\,\mu_3\nu_3}_{\a_3}(p_3)\,p_{3\b_3}+\frac{\p^{\mu_3\nu_3}(p_3)}{3}\d_{\a_3\b_3}\Big)\notag\\
	&\times\Bigg\{-\frac{2\,\p^{\mu_1\nu_1}(p_1)}{3}\Big[\braket{T^{\a_2\b_2}(p_1+p_2)T^{\a_3\b_3}(-p_1-p_2)}_{Ren}+\braket{T^{\a_2\b_2}(p_2)T^{\a_3\b_3}(-p_2)}_{Ren}\Big]
	\notag\\
	&\quad+\mathcal{I}^{(4)\,\mu_1\nu_1}_{\a_1}(p_1)\Big[-p_2^{\a_1}\braket{T^{\a_2\b_2}(p_1+p_2)T^{\a_3\b_3}(-p_1-p_2)}_{Ren}-p_3^{\a_1}\braket{T^{\a_2\b_2}(p_2)T^{\a_3\b_3}(-p_2)}_{Ren}\notag\\
	&\quad+p_{2\b}\Big(\d^{\a_1\a_2}\braket{T^{\b\b_2}(p_1+p_2)T^{\a_3\b_3}(-p_1-p_2)}_{Ren}+\d^{\a_1\b_2}\braket{T^{\b\a_2}(p_1+p_2)T^{\a_3\b_3}(-p_1-p_2)}_{Ren}\Big)\notag\\
	&\quad+p_{3\b}\Big(\d^{\a_1\a_3}\braket{T^{\b_2\a_2}(p_2)T^{\b\b_3}(-p_2)}_{Ren}+\d^{\a_1\b_3}\braket{T^{\a_2\b_2}(p_2)T^{\a_3\b}(-p_2)}_{Ren}\Big)\Big]\Bigg\}
\end{align}
where
\begin{align}
	\braket{t_{loc}^{\mu_1\nu_1}t_{loc}^{\mu_2\nu_2}t_{loc}^{\mu_3\nu_3}}^{(4)}_{extra}=\frac{\p^{\mu_1\nu_1}(p_1)\,\p^{\mu_2\nu_2}(p_2)\,\p^{\mu_3\nu_3}(\bar{p}_3)}{27}\d_{\a_2\b_2}\d_{\a_3\b_3}\bigg(4[E]^{\a_2\b_2\a_3\b_3}(p_2,\bar{p}_3)+4[C^2]^{\a_2\b_2\a_3\b_3}(p_2,\bar{p}_3)\bigg).
\end{align}
In summary, the counterterm cancels all the divergences arising in the 3-point function and from the local part of the counterterms there are extra contributions in the final renormalized $\braket{TTT}$ of the form
\begin{align}
	\braket{T^{\mu_1\nu_1}T^{\mu_2\nu_2}T^{\mu_3\nu_3}}^{(4)}_{extra}&=\left(\frac{\p^{\mu_1\nu_1}(p_1)}{3}\bigg(4[E]^{\mu_2\nu_2\mu_3\nu_3}(p_2,\bar{p}_3)+4[C^2]^{\mu_2\nu_2\mu_3\nu_3}(p_2,\bar{p}_3)\bigg)+(\text{perm.})\right)\notag\\
	&\hspace{-1cm}-\left(\frac{\p^{\mu_1\nu_1}(p_1)}{3}\frac{\p^{\mu_2\nu_2}(p_2)}{3}\d_{\a_2\b_2}\bigg(4[E]^{\a_2\b_2\mu_3\nu_3}(p_2,\bar{p}_3)+4[C^2]^{\a_2\b_2\mu_3\nu_3}(p_2,\bar{p}_3)\bigg) +(\text{perm.})\right)\notag\\
	&\hspace{-1cm}+\frac{\p^{\mu_1\nu_1}(p_1)}{3}\frac{\p^{\mu_2\nu_2}(p_2)}{3}\frac{\p^{\mu_3\nu_3}(\bar{p}_3)}{3}\d_{\a_2\b_2}\d_{\a_3\b_3}\Big(4[E]^{\a_2\b_2\a_3\b_3}(p_2,\bar{p}_3)+4[C^2]^{\a_2\b_2\a_3\b_3}(p_2,\bar{p}_3)\Big).
\end{align}
This extra contribution is exactly the anomalous part of the $TTT$, which in the flat limit becomes
\begin{align}
	\braket{T(p_1)T^{\mu_2\nu_2}(p_2)T^{\mu_3\nu_3}(\bar{p}_3)}_{anomaly}^{(4)}&=\big(4[E]^{\m_2\n_2\m_3\n_3}(p_2,p_3)+4[C^2]^{\m_2\n_2\m_3\n_3}(p_2,\bar{p}_3)\big)\\
	\braket{T(p_1)T(p_2)T^{\mu_3\nu_3}(\bar{p}_3)}_{anomaly}^{(4)}&=\d_{\a_2\b_2}\big(4[E]^{\a_2\b_2\m_3\n_3}(p_2,p_3)+4[C^2]^{\a_2\b_2\m_3\n_3}(p_2,\bar{p}_3)\big)\notag\\
	\braket{T(p_1)T(p_2)T(\bar{p}_3)}_{anomaly}^{(4)}&=\d_{\a_2\b_2}\d_{\a_3\b_3}\big(4[E]^{\a_2\b_2\a_3\b_3}(p_2,p_3)+4[C^2]^{\a_2\b_2\a_3\b_3}(p_2,\bar{p}_3)\big)
\end{align}
(with $T(p)\equiv \delta_{\m\n}T^{\m\nu}$).
The second order functional derivatives of the anomaly can be reconstructed using the expressions 
\begin{align}
	\big[E\big]^{\m_i\nu_i\m_j\nu_j}(p_i,p_j) &=\big[R_{\m\a\n\b}\,R^{\m\a \n\b}\big]^{\m_i\nu_i\m_j\nu_j}
	-4\,\big[R_{\m\n}R^{\m\n}\big]^{\m_i\nu_i\m_j\nu_j}
	+\big[ R^2\big]^{\m_i\nu_i\m_j\nu_j}\notag\\
	&\hspace{-2cm}=\bigg\{\big[R_{\m\a\n\b}\big]^{\m_i\nu_i}(p_i)\big[R^{\m\a \n\b}\big]^{\m_j\nu_j}(p_j)
	-4\,\big[R_{\m\n}\big]^{\m_i\nu_i}(p_i)\big[R^{\m\n}\big]^{\m_j\nu_j}(p_j)
	+\big[ R\big]^{\m_i\nu_i}(p_i)\big[R\big]^{\m_j\nu_j}(p_j)\bigg\}\notag\\[1.2ex]
	&\hspace{2cm}+\{(\mu_i,\nu_i,p_i)\leftrightarrow (\mu_j,\nu_j,p_j)\}\\[2ex]
	\big[C^2\big]^{\m_i\nu_i\m_j\nu_j}(p_i,p_j) &= \big[R_{\m\a\n\b}R^{\m\a \n\b}\big]^{\m_i\nu_i\m_j\nu_j}
	-2\,\big[R_{\m\n}R^{\m\n}\big]^{\m_i\nu_i\m_j\nu_j}
	+ \sdfrac{1}{3}\,\big[R^2\big]^{\m_i\nu_i\m_j\nu_j}\notag\\
	&\hspace{-2cm}=\bigg\{\big[R_{\m\a\n\b}\big]^{\m_i\nu_i}(p_i)\big[R^{\m\a \n\b}\big]^{\m_j\nu_j}(p_j)
	-2\,\big[R_{\m\n}\big]^{\m_i\nu_i}(p_i)\big[R^{\m\n}\big]^{\m_j\nu_j}(p_j)
	+\frac{1}{3}\big[ R\big]^{\m_i\nu_i}(p_i)\big[R\big]^{\m_j\nu_j}(p_j)\bigg\}\notag\\[1.2ex]
	&\hspace{2cm}+\{(\mu_i,\nu_i,p_i)\leftrightarrow (\mu_j,\nu_j,p_j)\},
\end{align}
for which we obtain
\begin{align}
	\braket{T^{\mu_1\nu_1}(p_1)T^{\mu_2\nu_2}(p_2)T^{\mu_3\nu_3}(\bar{p}_3)}^{(4)}_{extra}&=\left(\frac{\p^{\mu_1\nu_1}(p_1)}{3}\braket{T(p_1)T^{\mu_2\nu_2}(p_2)T^{\mu_3\nu_3}(\bar{p}_3)}^{(4)}_{anomaly}+(\text{perm.})\right)\notag\\
	&\hspace{-1cm}-\left(\frac{\p^{\mu_1\nu_1}(p_1)}{3}\frac{\p^{\mu_2\nu_2}(p_2)}{3}\braket{T(p_1)T(p_2)T^{\mu_3\nu_3}(\bar{p}_3)}^{(4)}_{anomaly}+(\text{perm.})\right)\notag\\
	&\hspace{-1cm}+\frac{\p^{\mu_1\nu_1}(p_1)}{3}\frac{\p^{\mu_2\nu_2}(p_2)}{3}\frac{\p^{\mu_3\nu_3}(\bar{p}_3)}{3}\braket{T(p_1)T(p_2)T(\bar{p}_3)}^{(4)}_{anomaly}.
\end{align}
\subsection{Summary}
To summarize, the full renormalized $\braket{TTT}$ in $d=4$ can be constructed using the renormalized transverse traceless and the local terms. In particular we find
\begin{align}
	&\braket{T^{\mu_1\nu_1}(p_1)T^{\mu_2\nu_2}(p_2)T^{\mu_3\nu_3}(\bar{p}_3)}^{(4)}=\braket{t^{\mu_1\nu_1}(p_1)t^{\mu_2\nu_2}(p_2)t^{\mu_3\nu_3}(\bar{p}_3)}^{(4)}_{Ren}\notag\\
	&\hspace{1cm}+\left(\braket{t_{loc}^{\mu_1\nu_1}(p_1)T^{\mu_2\nu_2}(p_2)T^{\mu_3\nu_3}(\bar{p}_3)}^{(4)}_{Ren}+(\text{cyclic perm.})\right)\notag\\
	&-\left(\braket{t_{loc}^{\mu_1\nu_1}(p_1)t_{loc}^{\mu_2\nu_2}(p_2)T^{\mu_3\nu_3}(\bar{p}_3)}^{(4)}_{Ren}+(\text{cyclic perm.})\right)+\braket{t_{loc}^{\mu_1\nu_1}(p_1)t_{loc}^{\mu_2\nu_2}(p_2)t_{loc}^{\mu_3\nu_3}(\bar{p}_3)}^{(4)}_{Ren}
\end{align} 
where the transverse and traceless parts are expressed as
\begin{align}
	&\braket{t^{\mu_1\nu_1}(p_1)t^{\mu_2\nu_2}(p_2)t^{\mu_3\nu_3}(\bar{p}_3)}^{(4)}_{Ren}=\Pi^{(4)\,\mu_1\nu_1}_{\a_1\b_1}(p_1)\Pi^{(4)\,\mu_2\nu_2}_{\a_2\b_2}(p_2)\Pi^{(4)\,\mu_3\nu_3}_{\a_3\b_3}(\bar{p}_3)\notag\\
	&\times\Big\{A_1^{Ren}\,p_2^{\a_1} p_2^{\b_1} \bar{p}_3^{\a_2} p_3^{\b_2} p_1^{\a_3} p_1^{\b_3}+ A_2^{Ren}\,\d^{\b_1\b_2} p_2^{\a_1} p_3^{\a_2} p_1^{\a_3} p_1^{\b_3} 
	+ A_2^{ren}\,(p_1 \leftrightarrow p_3)\, \d^{\b_2\b_3}  p_3^{\a_2} p_1^{\a_3} p_2^{\a_1} p_2^{\b_1} \notag\\
	&\hspace{0.8cm}+ A_2^{Ren}\,(p_2\leftrightarrow p_3)\, \d^{\b_3\b_1} p_1^{\a_3} p_2^{\a_1}  p_3^{\a_2} p_3^{\b_2}+ A_3^{Ren}\,\d^{\a_1\a_2} \d^{\b_1\b_2}  p_1^{\a_3} p_1^{\b_3} + A_3^{Ren}(p_1\leftrightarrow p_3)\,\d^{\a_2\a_3} \d^{\b_2\b_3}  p_2^{\a_1} p_2^{\b_1} \notag\\
	&\hspace{1.2cm}
	+ A_3^{Ren}(p_2\leftrightarrow p_3)\,\d^{\a_3\a_1} \d^{\b_3\b_1}  p_3^{\a_2} p_3^{\b_2} + A_4^{Ren}\,\d^{\a_1\a_3} \d^{\a_2\b_3}  p_2^{\b_1} p_3^{\b_2} + A_4^{Ren}(p_1\leftrightarrow p_3)\, \d^{\a_2\a_1} \d^{\a_3\b_1}  p_3^{\b_2} p_1^{\b_3} \notag\\
	&\hspace{3.5cm}+ A_4^{Ren}(p_2\leftrightarrow p_3)\, \d^{\a_3\a_2} \d^{\a_1\b_2}  p_1^{\b_3} p_2^{\b_1} + A_5 ^{Ren} \d^{\a_1\b_2}  \d^{\a_2\b_3}  \d^{\a_3\b_1}\Big\}
\end{align}
with the renormalized form factors given in Appendix of \cite{Coriano:2018bsy}. It can be further simplified in the form 
\begin{align}
	\braket{T^{\m_1\n_1}T^{\m_2\n_2}T^{\m_3\n_3}}_{Ren}= \braket{t^{\m_1\n_1} t^{\m_2\n_2} t^{\m_3\n_3}}_{Ren} + 
	\braket{T^{\m_1\n_1}T^{\m_2\n_2}T^{\m_3\n_3}}_{Ren\, l\,t} + \braket{T^{\m_1\n_1}T^{\m_2\n_2}T^{\m_3\n_3}}_{anomaly}\notag\\
\end{align}
and with the renormalized longitudinal traceless contribution ($l \, t$ ) given by 
\begin{align}
	\braket{ T^{\m_1\n_1}T^{\m_2\n_2}T^{\m_3\n_3}}_{Ren\,l \, t}&\equiv\left(\mathcal{V}^{\mu_1\nu_1\mu_2\nu_2\mu_3\nu_3}_{loc\,0\, 0} + \mathcal{V}^{\mu_1\nu_1\mu_2\nu_2\mu_3\nu_3}_{0\, loc\, 0}  + \mathcal{V}^{\mu_1\nu_1\mu_2\nu_2\mu_3\nu_3}_{0\, 0 \, loc} \right) \notag\\
	&-
	\left(\mathcal{V}^{\mu_1\nu_1\mu_2\nu_2\mu_3\nu_3}_{loc\,loc\, 0} + \mathcal{V}^{\mu_1\nu_1\mu_2\nu_2\mu_3\nu_3}_{0\, loc\, loc}  + \mathcal{V}^{\mu_1\nu_1\mu_2\nu_2\mu_3\nu_3}_{loc\, 0 \, loc} \right) + \mathcal{V}^{\mu_1\nu_1\mu_2\nu_2\mu_3\nu_3}_{loc\,loc\,loc}
\end{align}
and 
\begin{align}
	\label{expans2}
	& \braket{T^{\mu_1 \nu_1}({p}_1) T^{\mu_2 \nu_2}({p}_2) T^{\mu_3 \nu_3}({p}_3) }_{anomaly} = \frac{\hat\pi^{\mu_1 \nu_1}({p}_1)\hat\pi^{\mu_2 \nu_2}({p}_2) \hat\pi^{\mu_3 \nu_3}(\bar{p}_3)}{27 p_1^2 p_2^2 p_3^2}\braket{ T({p}_1)T({p}_2)T(\bar{p}_3)}_{anomaly} \notag\\
	& + \frac{\hat\pi^{\mu_2 \nu_2}({p}_2)}{3\, p_2^2} \braket{ T^{\mu_1 \nu_1}({p}_1) T({p}_2) T^{\mu_3 \nu_3}({p}_3)}_{anomaly} + \frac{\hat\pi^{\mu_3 \nu_3}({p}_3)}{3\, p_3^2} \braket{ T^{\mu_1 \nu_1}({p}_1) T^{\mu_2 \nu_2}({p}_2)  T({p}_3)}_{anomaly} \notag\\
	&+ \frac{\hat\pi^{\mu_1 \nu_1}({p}_1)}{3\, p_1^2} \braket{T({p}_1) T^{\mu_2 \nu_2}({p}_2) T^{\mu_3 \nu_3}({p}_3)}_{anomaly} - \: \frac{\hat\pi^{\mu_1 \nu_1}({p}_1) \hat\pi^{\mu_2 \nu_2}({p}_2)}{9\, p_1^2 p_2^2}\braket{ T({p}_1)T({p}_2)T^{\mu_3 \nu_3}({p}_3)}_{anomaly}  \notag\\
	&- \: \frac{\hat\pi^{\mu_2 \nu_2}({p}_2) \hat\pi^{\mu_3 \nu_3}({p}_2)}{9 p_2^2 p_3^2}\braket{T^{\mu_1 \nu_1}({p}_1)T(p_2)T({p}_3)}_{anomaly} - \frac{\hat\pi^{\mu_1 \nu_1}({p}_1)\hat \pi^{\mu_3 \nu_3}(\bar{p}_3)}{9 p_1^2 p_3^2}\braket{ T({p}_1) T^{\mu_2 \nu_2}({p}_2)T({p}_3)}_{anomaly}    .
\end{align}
As a final step, it is convenient to collect together the two renormalized contributions, the transverse and the longitudinal one, which are both traceless, into a single contribution 
\begin{align}
	\mathcal{V}^{\mu_1\nu_1\mu_2\nu_2\mu_3\nu_3}_{ traceless}\equiv \braket{t^{\m_1\n_1} t^{\m_2\n_2} t^{\m_3\n_3}}_{Ren} + \braket{ T^{\m_1\n_1}T^{\m_2\n_2}T^{\m_3\n_3}}_{Ren\,l \, t}
\end{align}
in order to cast the entire vertex in the form 
\begin{align}
	\braket{T^{\m_1\n_1}T^{\m_2\n_2}T^{\m_3\n_3}}_{Ren}=\mathcal{V}^{\mu_1\nu_1\mu_2\nu_2\mu_3\nu_3}_{traceless} + 
	\braket{T^{\m_1\n_1}T^{\m_2\n_2}T^{\m_3\n_3}}_{anomaly}.
\end{align}
We are going to comment briefly on the implications of these results at diagrammatic level.

\begin{figure}[t]
	\centering
	\vspace{2ex}
	\includegraphics[scale=0.57]{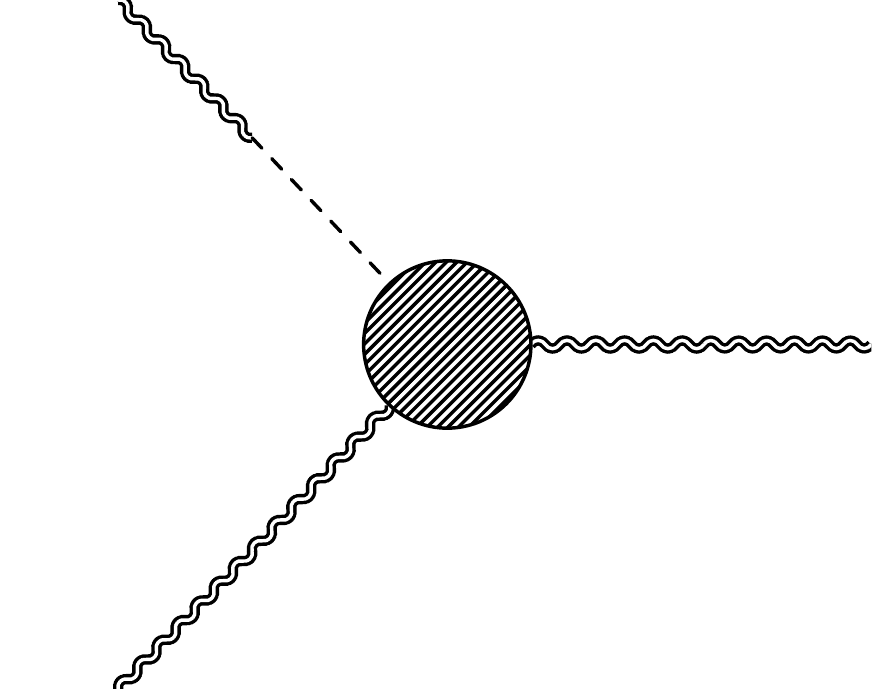}\hspace{1ex}
	\includegraphics[scale=0.57]{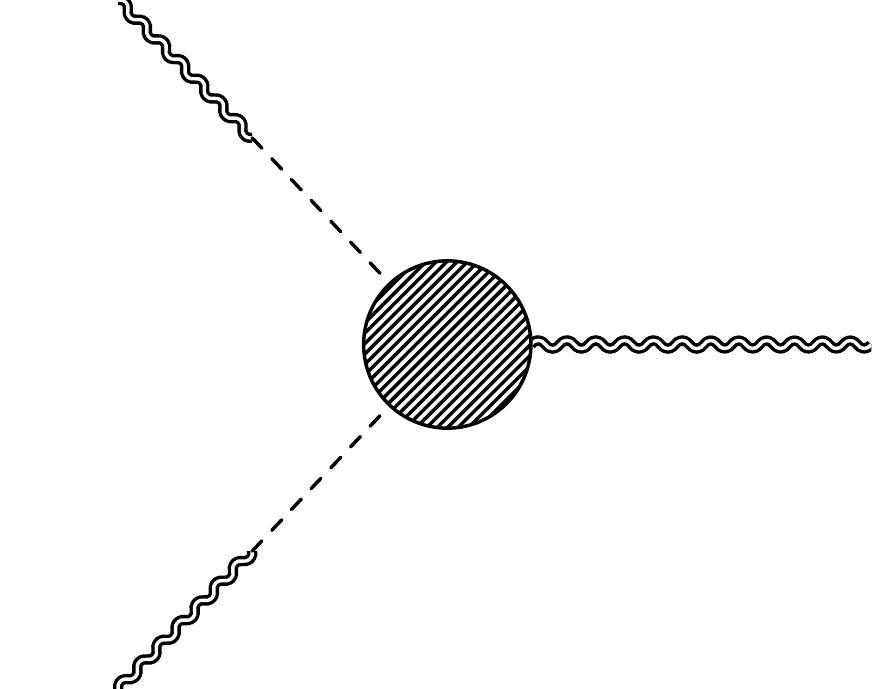}\hspace{1ex}
	\includegraphics[scale=0.57]{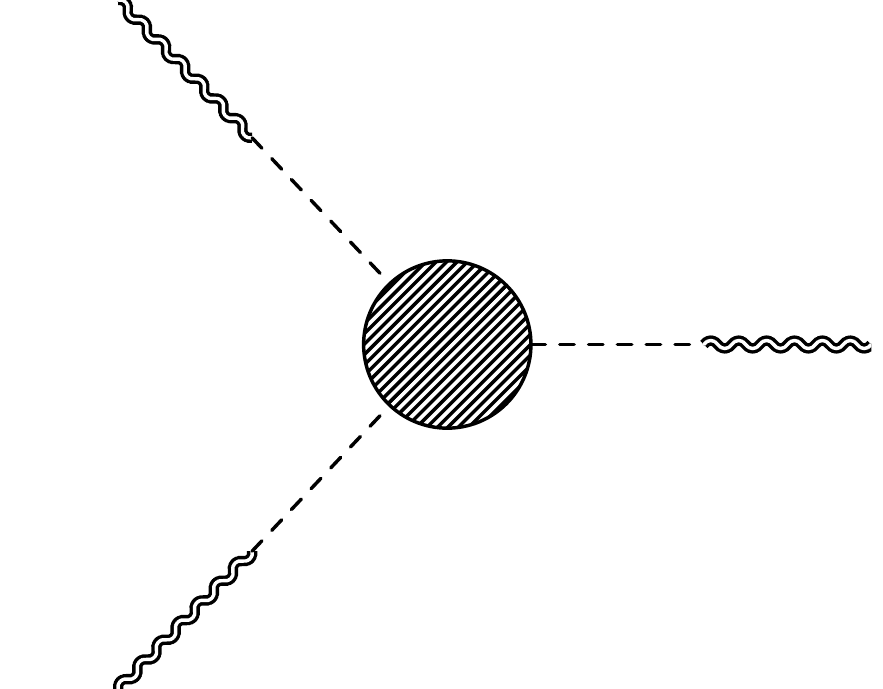}
	\vspace{2ex}
	\caption{Anomaly interactions mediated by the exchange of one, two or three poles. The poles are generated by the renormalization of the longitudinal sector of the $TTT$.}
	\label{dec}
\end{figure}

\subsection{The perturbative structure of the $TTT$ and the poles separation} 
The structure of the poles in the $TTT$ is summarized in Fig. \ref{dec} where we have denoted with a dashed line the exchange of one or more massless $(\sim 1/p_i^2)$ interactions. In configuration space such extra terms, related to the renormalization of the correlator, are the natural generalization of the typical anomaly pole interaction found, for instance, in the case of the $TJJ$, where the effect of the anomaly is in the generation of a nonlocal interaction of the form \cite{Giannotti:2008cv,Armillis:2009pq,Armillis:2010qk}
\begin{align}
	\mathcal{S}_{an}\sim \beta(e) \int d^4 x\, d^4 y R^{(1)}(x)  \left(\frac{1}{\square}\right)(x,y) F F(y)
\end{align}
with $F$ being the QED field strength and $\beta(e)$ the corresponding beta function of the gauge coupling. In the $TTT$ case, as one can immediately figure out from \eqref{expans2} such contributions can be rewritten as contribution to the anomaly action in the form
\begin{align}
	\mathcal{S}_{an}\sim \int d^4 x\, d^4 y R^{(1)}(x)  \left(\frac{1}{\square}\right)(x,y) \left( \beta_b E^{(2)}(y) + \beta_a (C^2)^{(2)}(y)\right)
\end{align}
and similar for the other contributions extracted from \eqref{expans2}. Notice that each $\hat\pi$ projector in \eqref{expans2} is accompanied by a corresponding anomaly (single) pole of the external invariants, generating contributions of the form $1/p_i^2$,  $1/(p_i^2 p_j^2) (i\neq j)$ and $1/(p_1^2 p_2^2 p_3^2)$, where multiple poles are connected to separate external graviton lines. Each momentum invariant appears as a single pole. One can use the correspondence 
\begin{equation}
	\frac{1}{p^2} \hat\pi^{\mu\nu} \leftrightarrow R^{(1)}\frac{1}{\square} 
\end{equation}
to include such nonzero trace contributions into the anomaly action. This involves a multiplication of the vertex by the external fields together with an integration over all the internal points. As shown in \cite{Coriano:2017mux} such nonzero trace contributions are automatically generated by the nonlocal conformal anomaly action, which accounts for the entire expression \eqref{expans2}. \\
The diagrammatic interpretation suggests a possible generalization of this result also to higher point functions, as one can easily guess, in a combination similar to that shown in Fig. \ref{dec}.\\
Notice that the numerators of such decompositions, which correspond to single, double and triple traces are, obviously, purely polynomial in the external invariants, being derived from the anomaly functional, which is local in momentum space.\\

\section{Conclusions}
We have presented a comparative study of the 3-graviton vertex $TTT$ in CFT's in momentum space. The conformal correlators' analysis is relatively new, beyond the standard Lagrangian approach, though the interest in this approach is growing \cite{Isono:2018rrb,Gillioz:2018mto}. Our analysis extends a previous work on the $TJJ$ correlator \cite{Coriano:2018bbe} and on the same $TTT$ vertex given in \cite{Coriano:2012wp}, based on similar approaches. Building on the analysis presented in \cite{Bzowski:2013sza, Bzowski:2017poo}, we have gone over the reconstruction program proposed in those works from a perturbative perspective. We have also presented an independent analysis of the solutions of the CWI's. This is based on a new approach which exploits some properties of the solutions of the hypergeometric systems of differential equations associated with the CWI's and equivalent to them. The method is alternative to the approach presented in \cite{Bzowski:2013sza}, which requires a rather complex analysis of the solutions' singularities, given in terms of 3-K integrals. 
The method has been extended by us also to 4-point functions, searching for special solutions of such systems, which are controlled by a larger class of hypergeometrics, respect to the simpler solutions found for 3-point functions discussed here we will show in future work.

The comparison with perturbation theory allows for drastic simplifications of the vertex results while keeping, for specific dimensions, the generality of the conformal 
(non-Lagrangian) solution. At the same time, having established a direct link between the two - i.e. the perturbative and the non-Lagrangian formulations - this opens the way to several independent analysis of this vertex - entirely based on the Feynman's expansion. \\
This would allow identifying the singularities - and henceforth the anomaly poles - present in such correlator, from a physical and straightforward perspective based on the analysis of the Landau conditions of the fundamental (1-loop) diagrams generated by the matching, as done in the simpler case of \cite{Coriano:2014gja} in a supersymmetric context.
The expression that we have presented of such vertex is the simplest one that can be written and down and in $d=4$ keeps its generality. It is possible to extend our analysis to higher (even) dimensions by the inclusion of antisymmetric forms, building on the analysis of \cite{Bastianelli:2000rs} as a third (beside scalar and fermions) sector, which would provide an extension of the approach presented in our work. 

Finally, we have also discussed the result fo the $TTT$ - for its renormalised expression - in terms of a homogenous (zero trace) contribution and an anomaly part. The anomaly (nonzero trace) part, is generated by the renormalisation of 
the local components of the $TTT$.  
Our detailed analysis shows that such contributions are not an artefact of the form factors' parameterisation or can be attributed to a 
specific decomposition but are a general feature of CFT's and is related to renormalisation. This is in agreement with the analysis \cite{Bzowski:2017poo} and, at the same time, with the predictions - limitedly to the anomaly part - coming from the nonlocal anomaly action \cite{Coriano:2017mux}, as we are going to show in the next chapter. 
\chapter{The conformal anomaly and the effective action}\label{chapter3}
\section{Introduction}
In a classical conformal invariant theory, such a massless Quantum Electrodynamics (QED), the dilatation operator generates a symmetry of the Lagrangian. Due to the renormalization procedure, this symmetry is broken. As a result, the stress energy tensor acquires a non-vanishing trace, a trace anomaly contribution generated by an effective action that is purely gravitational. In this approach, the matter is integrated out, and the action provides a semiclassical description of the interaction of the gravitational field with ordinary matter. 
The role of such effective action in quantum gravity has been debated for some time due to the various expressions that one can write down. \\
The nonlocal form of the action was initially introduced by Riegert \cite{1984PhLB..134...56R}. Discussions of such action in a cosmological context can be found in \cite{Antoniadis:2011ib,Antoniadis:2006wq}. Local and nonlocal formulations of such actions, with the possible inclusions of extra degrees of freedom in the form of a dilaton, have been recently reviewed by us in \cite{Coriano:2019dyc}.\\
In this chapter, we will study the origin of the trace anomaly as obtained in \cite{Birrell:1982ix}, with its explicit form first proposed in \cite{Deser:1976yx}. Then, using the approach of \cite{Riegert:1987kt}, we will construct a covariant, nonlocal action, which can be expanded around flat space so that the trace of its stress energy tensor takes the general form of the anomaly functional. Finally, using this anomaly effective action, we will show that the massless poles in the three-point correlation function $\braket{TTT}$ for a CFT in $d=4$ are precisely those obtained by the solution of the CWI's directly in flat space, by the methods of \cite{Bzowski:2013sza}. This equivalence is the content 
of \cite{Coriano:2017mux}. \\
In the next section, to make our discussion self-contained, we will review some standard features of the effective action using the Schwinger-DeWitt expansion and the heat-kernel regularization. More detailed discussions of these points can be found in \cite{Birrell:1982ix}. Then we will provide a full derivation of the nonlocal anomaly action, presenting the variational solution of the anomaly functional. \\
We remark that an anomaly action is not unique. It is an action which accounts for the trace/conformal anomaly, but it is defined modulo traceless contributions which remain, obviously, arbitrary, from the point of view of the solution of the variational problem. However, they may differ in the number of asymptotic degrees of freedom introduced in the action itself. With the therm "asymptotic" we refer to states which are part of the S-matrix derived from the effective action but are not part of the original theory or are directly coupled to the matter fields which have been integrated out in the process of deriving the anomaly action.\\
It is natural to introduce a dilaton as a Goldstone mode that couples to the divergence of a current which is non-conserved by the quantum corrections, and for the dilaton, this is given by the dilatation current. In the case of a conformal anomaly, this coupling is at most quartic in $d=4$ and can be worked out by the Noether method \cite{Coriano:2013nja,Coriano:2013xua}.
All the local forms of such actions introduce one extra degree of freedom, in the form of a dilaton field. 
In the context of supersymmetric theories, the dilaton field turns into a multiplet with a dilaton, an axion and an 
axino and can be described by a St\"uckelberg-like Lagrangian \cite{Coriano:2010ws} which has implications at a phenomenological level in the dark matter sector of supersymmetric extensions of the Standard Model. 
A dilaton could be the result of some strong dynamics and realizes a matching of the anomaly in the IR, the same anomaly that is identified, dynamically, by the emergence of anomaly poles in the UV expression of the correlator.  

\section{Path Integral formulation and the effective action}
The steps that take to the conformal anomaly may be quite diverse, 
for instance, one can rely on direct perturbative computations, which allow to identify some components of the anomaly functional on a diagrammatic basis. 
In this approach, one can consider correlators with several insertions of stress energy tensors and/or currents, sensitive to different components of the anomaly functional. For instance, the trace of the $TJJ$ accounts for the $F^2$ part of the anomaly, where $F$ is the field strength of the gauge field coupled to the $J$ current, but it is insensitive to other components, which come from gravity. For this reason, a complete picture of the trace anomaly requires other, more complex correlators containing multiple insertions of stress energy tensors.\\
One advantage of this approach, which can be performed in momentum space, is to bring us close to the core of an interacting theory, with the emergence of dynamical degrees of freedom. Ultimately, quantum field theory is a theory of massive and massless particles propagating in spacetime either as real or virtual 
(interpolating) states. Such effective interaction can be worked out directly in a traditional Feynman expansion. \\
A complementary/alternative approach starts from the vacuum persistence amplitude, which takes to the DeWitt-Schwinger or heat-kernel method. A critical feature of the method is to provide an expression for the anomaly action, which allows identifying the general structure of the anomaly functional. This is advantageous compared to the standard perturbative approach in momentum space, and for this reason, we will briefly summarize it in this section. \\  
We start from the functional integral in flat space
\begin{equation}
	Z[J]=\int\,\mathcal{D}\Phi\,\exp\left[iS_m[\Phi]+i\int d^dx\, J(x)\,\Phi(x)\right]\label{pathint}
\end{equation}
which is physically interpreted as the vacuum persistence amplitude $\braket{\text{out},0|0,\text{in}}$, with the normalization condition
\begin{equation}
	Z[0]\equiv\braket{\text{out},0|0,\text{in}}_{J=0}=\braket{0|0}=1.\label{normcond}
\end{equation}
for a generic action $S_m$.
{In a nontrivial background, considering $J=0$ in \eqref{pathint}, one is brought to consider the variation of $Z[0]$, which can be written as \cite{Schwinger452}
	\begin{equation}
		\d Z[0]=i\int\,\mathcal{D}\Phi\,\d S_m\,\exp\left(iS_m[\Phi]\right)=i\bra{\text{out},0}\d S_m\ket{0,\text{in}}\label{varia}.
	\end{equation}
	From the definition of the energy momentum tensor associated to the matter field and from \eqref{varia} one immediately obtains
	\begin{equation}
		\sdfrac{2}{\sqrt{-g}}\sdfrac{\d\,Z[0]}{\d\,g_{\m\n}}=i\bra{\text{out},0}T^{\m\n}\ket{0,\text{in}},\qquad T^{\mu\nu}\equiv\frac{2}{\sqrt{-g}}\sdfrac{\d\,S_m}{\d\,g_{\m\n}},\label{varia2}
	\end{equation}
	If we now express $Z[0]$ in terms of the effective action as
	\begin{equation}
		Z[0]=e^{i\mathcal{S}_{\scalebox{0.5}{$eff$}}},\qquad
		\mathcal{S}_{\scalebox{0.5}{$eff$}}=-i\ln\braket{\text{out},0|0,\text{in}}
	\end{equation}
	we can re-expressed \eqref{varia2} in the form
	\begin{equation}
		ie^{i\mathcal{S}_{\scalebox{0.5}{$eff$}}}\sdfrac{2}{\sqrt{-g}}\sdfrac{\d\,\mathcal{S}_{\scalebox{0.5}{$eff$}}}{\d\,g_{\m\n}}=i\bra{\text{out},0}T^{\m\n}\ket{0,\text{in}}\,.
	\end{equation}
	Recalling the definition of $Z[0]$ as the vacuum persistence amplitude, which is intimately related to the effective action via \eqref{varia2}, the expectation value of the energy momentum tensor can be expressed in terms of the effective action as
	\begin{equation}
		\sdfrac{2}{\sqrt{-g}}\sdfrac{\d\,\mathcal{S}_{\scalebox{0.5}{$eff$}}}{\d\,g_{\m\n}}=\frac{\bra{\text{out},0}T^{\m\n}\ket{0,\text{in}}}{\braket{\text{out},0|0,\text{in}}}=\braket{T^{\m\n}}\,.
	\end{equation}
	In a curved spacetime, the generating functional $Z[0]$ is evaluated similarly to the flat spacetime case, with the obvious replacements of the measure $d^dx$ by the covariant measure $\sqrt{g(x)}\,d^dx$ and of the delta function $\d^d(x-y)$ by $(\sqrt{g(y)})^{-1}\d^{d}(x-y)$. The latter property allows to write the equation
	\begin{equation}
		\int\,d^dx\,\sqrt{-g}\,\d^d(x-y)\,(\sqrt{g(y)})^{-1}=1.
	\end{equation}
	As an example, let's consider a scalar field coupled to gravity
	\begin{equation}
		\mathcal{L}(x)=\sdfrac{1}{2}\,\left[g^{\m\n}(x)\partial_\m\phi(x)\partial_\n\phi(x)-\left(m^2+\x\,R(x)\right) \phi^2(x)\right]
	\end{equation}
	where $\phi(x)$ is the scalar field, $m$ the mass of the field and $R$ is the Ricci scalar.  The constant $\x$ is in general assumed either to be zero, in the minimally coupled case, or $\x=1/4(d-2)/(d-1)$, for the conformally coupled case, and in $d=4$ gives $\xi=1/6$. 
	The equation of motion for the corresponding action $S=\int\,d^dx\,\sqrt{-g}\,\mathcal{L}(x)$ is
	\begin{equation}
		\sdfrac{1}{\sqrt{-g}}\sdfrac{\d\,S}{\d\,\phi}=(\square+m^2+\x\,R)\phi,
	\end{equation}
	where $\square=g^{\m\n}\nabla_\n\nabla_\m$ is the generalized d'Alembertian operator in a curved space, expressed in terms of covariant derivatives. The Feynman propagator $G_F$ for this equation has to satisfy the equation
	\begin{equation}
		\left[\square_x+m^2+\x\,R(x)\right] G_F(x,x')=-\frac{\d^d(x-x')}{\sqrt{g(x')}},\label{expand}
	\end{equation}
	where
	\begin{equation}
		iG_F(x,x')=\bra{0}T(\phi(x)\phi(x')\ket{0},
\end{equation}}
and \eqref{expand} has the formal solution 
\begin{equation}
	G_F(x,x')=-\left[\square_x+m^2+\xi\,R\right]^{-1}\frac{\d^d(x-x')}{\sqrt{g(x')}}.
\end{equation}
One can identify a relation between $Z[0]$ and the Feynman Green's function in the form
\begin{equation}
	Z[0]\propto [\det(-G_F)]^{\frac{1}{2}}=\exp\left[\sdfrac{1}{2}\Tr\ln(-G_F)\right],
\end{equation}
where the proportionally constant is metric-independent and $G_F$ has to be interpreted as an operator acting on a space of vectors $\ket{x}$, normalized by 
\begin{align}
	G_F(x,x')&=\bra{x}G_F\ket{x'}\\
	\braket{x|x'}&=(\sqrt{-g})^{-1}\d^d(x-x').
\end{align}
From the definition of the effective action then it follows that
\begin{equation}
	\mathcal{S}_{\scalebox{0.5}{$eff$}}=-i\ln Z[0]=-\sdfrac{i}{2}\Tr[\ln(-G_F)].\label{funct}
\end{equation} 
As ususal, in the operatorial formalism, the trace of an operator $M$ acting in this space is defined by
\begin{equation}
	\Tr\,M=\int\,d^dx\sqrt{-g}\,M_{xx}=\int\,d^dx\,\sqrt{-g}\bra{x}M\ket{x}.
\end{equation}
In order to make sense of the formal definition \eqref{funct}, we need to use an explicit representation of $G_F$. In the following section we shall use the DeWitt-Schwinger representation given by an integral over the proper time.

\section{Adiabatic expansion of Green's functions }
Details on the Schwinger-De Witt approach is summarized in \cite{Barvinsky:1984jd}. This approach is valid for arbitrary differential operators satisfying the condition of causality and gives at once the Green's function in an expansion around the light cone. Therefore, it is the most suitable tool for investigating the ultraviolet divergences for the calculation of specific counterterms, of $\b$-function and, in this case, of possible anomalies. \\
We are interested in the short distance behaviour of the Feynman propagator $G_F(x,x')$ in the limit $x\to x'$. Introducing Riemann normal coordinates $y^\m$ for the point $x$, with origin in $x'$, one may expand the metric tensor \cite{Kreyszig, Hatzinikitas:2000xe} around $y=0$ as
\begin{equation}
	g_{\m\n}(x)=\h_{\m\n}+\sdfrac{1}{3}R_{\m\a\n\b}y^\a y^\b-\sdfrac{1}{6}R_{\m\a\n\b;\g}y^\a y^\b y^\g+\left[\sdfrac{1}{20}R_{\m\a\n\b;\g\d}+\sdfrac{2}{45}R_{\a\m\b\l}R^{\l}_{\ \,\g\n\d}\right]y^\a y^\b y^\g y^\d+\dots
\end{equation}
where $\h_{\m\n}$ is the flat metric. We define $\mathcal{G}_F(x,x')$ and its Fourier transform as
\begin{align}
	\mathcal{G}_F(x,x')&=(g(x))^{\frac{1}{4}}\,G_F(x,x')\label{defG}\\
	\mathcal{G}_F(x,x')&=\sdfrac{1}{(2\pi)^d}\int\,d^dk\,e^{-iky}\,\mathcal{G}_F(k),\qquad\text{with}\quad ky=\h^{\a\b}k_\a\,y_\b.\label{Gfunc}
\end{align}
We now expand the equation \eqref{expand}, that defines the Feynman propagator, in normal coordinates, and after a Fourier transform we solve $\mathcal{G}_F(k)$ by iteration to any adiabatic order, i.e., to any order in derivatives of the metric. In this way, we construct the asymptotic expansion of Feynman propagator, which up to  order four is given by
\begin{align}
	\mathcal{G}_F(k)&\approx\,\sdfrac{1}{(k^2-m^2)}-\left(\sdfrac{1}{6}-\x\right)R\,\sdfrac{1}{k^2-m^2}+\sdfrac{i}{2}\left(\sdfrac{1}{6}-\x\right)R_{;\a}\partial^\a\left(\sdfrac{1}{(k^2-m^2)^2}\right)\notag\\
	&\qquad-\sdfrac{1}{3}\,a_{\a\b}\,\partial^\a\partial^\b\sdfrac{1}{(k^2-m^2)^2}+\left[\left(\sdfrac{1}{6}-\x\right)^2R^2+\sdfrac{2}{3}a^\l_{\ \l}\right]\sdfrac{1}{(k^2-m^2)^3}\label{matG}
\end{align}
where we have defined $\partial_\a=\partial/\partial k^\a$, and
\begin{align}
	a_{\a\b}=\sdfrac{1}{2}\left(\x-\sdfrac{1}{6}\right)R_{;\a\b}+\sdfrac{1}{120}R_{;\a\b}-\sdfrac{1}{40}R_{\a\b;\l}^{\quad\ \ \,\,\l}-\sdfrac{1}{30}R_\a^{\ \l}R_{\l\b}+\sdfrac{1}{60}R^{\k\ \l}_{\ \a\ \b}R_{\k\l}+\sdfrac{1}{60}R^{\l\m\k}_{\quad\ \a}R_{\l\m\k\b}\label{ageneric}
\end{align}
Substituting \eqref{matG} into \eqref{Gfunc}, we determine the corresponding asymptotic expansion of the Feynman propagator in coordinate space as
\begin{equation}
	\mathcal{G}_F(x,x')\approx\int\sdfrac{d^dk}{(2\pi)^d}\,e^{-iky}\left[a_0(x,x')+a_1(x,x')\left(-\sdfrac{\partial}{\partial m^2}\right)+a_2(x,x')\left(\sdfrac{\partial}{\partial m^2}\right)^2\right]\sdfrac{1}{k^2-m^2}\label{adiab}
\end{equation}
where 
\begin{equation}
	\begin{split}
		a_0(x,x')&=1 \\
		a_1(x,x')&=\left(\sdfrac{1}{6}-\x\right)R-\sdfrac{1}{2}\left(\sdfrac{1}{6}-\x\right)R_{;\a}y^\a-\sdfrac{1}{3}a_{\a\b}y^\a y^\b\\
		a_2(x,x')&=\sdfrac{1}{2}\left(\sdfrac{1}{6}-\x\right)^2R^2+\sdfrac{1}{3}a^\l_{\ \l}
	\end{split}\label{coefficientadiab}
\end{equation}
where all the geometric quantities on the right hand side of these last two equations evaluated at point $x'$. It is worth mentioning that in order to extract a time-ordered product in \eqref{Gfunc}, one has to perform the $k^0$ integral along the appropriate Feynman contour, a condition that translates in the substitution $m^2$ by $m^2-i\epsilon$.  
If one uses the integral representation
\begin{equation}
	\sdfrac{1}{k^2-m^2+i\epsilon}=-i\int_0^\infty\,ds\,e^{isk(k^2-m^2+i\epsilon)}
\end{equation}
in \eqref{adiab}, then the $d^dk$ integration may be performed explicitly leaving the $ds$ (proper time) integration to be performed as
\begin{equation}
	\mathcal{G}_F=-i(4\pi)^{-d/2}\int_0^\infty\,ids\,(is)^{-d/2}\exp\left[-im^2 s+\sdfrac{\s \,is}{2}\right]F(x,x';is),\label{Gnew}
\end{equation}
where $\s(x,x')=1/2\,y_\a y^\a$, while the function $F()$ has the following asymptotic adiabatic expansion
\begin{equation}
	F(x,x';is)\approx a_0(x,x')+a_1(x,x')is+a_2(x,x')(is)^2+\dots\label{aE}
\end{equation}

From the definition \eqref{defG}, then \eqref{Gnew} gives a representation of $G_F(x,x')$ originally derived by DeWitt as
\begin{equation}
	G_F^{DS}(x,x')=-i\D^{\frac{1}{2}}(x,x')(4\pi)^{-d/2}\int_0^\infty\,ids\,(is)^{-d/2}\exp\left[ -im^2s+\sdfrac{\s\,is}{2}\right] F(x,x';is)\label{DWSre}
\end{equation}
with 
\begin{equation}
	\D(x,x')=-\det[\partial_\m\partial_\n\s(x,x')][g(x)g(x')]^{-1/2},
\end{equation}
the Van Vleck determinant. It is simple to show that in the normal Riemann coordinates around $x'$, $\D$ reduces to $(\sqrt{g(x)})^{-1}$. {According to the treatment by DeWitt, one can extend the asymptotic expansion \eqref{aE} of $F$ to all the adiabatic orders writing
	\begin{equation}
		F(x,x';is)\approx\sum_{j=0}^\infty\,a_j(x,x')(is)^j
	\end{equation}
	where $a_0(x,x')=1$, and the remaining $a_j$ are determined recursively.  $G_F^{DS}(x,x')$ in \eqref{DWSre} is the DeWitt-Schwinger \emph{proper time representation} and it is a powerful tools to study the divergences of the effective action and a formidable tool to obtain the trace anomaly.
	
	\section{Effective Action and renormalization}
	We now apply the general construction of the generating function in terms of the Green's function to the case of curved spacetime. In analogy with the case of flat spacetime, we define the operator 
	\begin{equation}
		K_{xy}\equiv \left(\square_x+m^2-i\e+\x\,R\right)\,\sdfrac{\d^d(x-y)}{\sqrt{g(y)}}
	\end{equation}
	that can be treated  as a symmetric matrix $K$ with continuous indices with the properties
	\begin{align}
		\int\,d^dy\,\sqrt{g(y)}\,K_{xy}K_{yz}^{-1}&=\sdfrac{\d^d(x-z)}{\sqrt{g(z)}},\\
		\int\,d^dy\,\sqrt{g(y)}\,K_{xy}K_{yz}&=K_{xz}\\
		G_F(x,x')&=-K_{xx'}^{-1}\label{KGre},
	\end{align}
	where the last property has been derived by a comparison with \eqref{expand}. The relation between the operator $K$ and the Green's function $G_F$ is made manifest through \eqref{KGre}. We consider now the integral representation, previously in the analysis of $G_F$, for the operator $K^{-1}$ and in particular we get
	\begin{equation}
		K^{-1}=i\int_0^\infty \,e^{-iKs}ds.
	\end{equation}
	This result allows to make a comparison with \eqref{DWSre} when the $G_F(x,x')=\bra{x}G_F\ket{x'}$ is considered. In fact,
	\begin{align}
		\bra{x}G_F\ket{x'}=-K^{-1}_{xx'}=-i\int_0^\infty ds\,\bra{x}e^{-iKs}\ket{x'}
	\end{align}
	and by using the explicit result first given by DeWitt and Schwinger, we derive the representation of the exponential of the $K$ operator in the form 
	\begin{equation}
		\bra{x}e^{-iKs}\ket{x'}=i(4\pi)^{-d/2}\,\D^{\frac{1}{2}}(x,x')\,e^{-im^2s+is\,\s/2}\,F(x,x';is)\,(is)^{-d/2}.\label{DWSrep}
	\end{equation}
	In order to obtain the effective action $S_{eff}$ we consider 
	\begin{equation}
		\ln(-G_F)=-\ln(K)=\int_0^\infty\,e^{-iKs}\,(is)^{-1}\,i\,ds,\label{logG}
	\end{equation}
	which is valid up to the addition of a metric-independent infinite constant that can be neglected in our discussion. This relation is obtained by expanding the integral around $\Lambda=0$
	\begin{equation}
		\int_{\L}^{\infty}\,e^{-iKs}\,(is)^{-1}\,i\,ds=-Ei(-i\L\,K)\label{inte},
	\end{equation}
	with the expansion of the exponential integral function $Ei$ given by
	\begin{equation}
		Ei(x)=\g+\ln(-x)+O(x),\label{exp}
	\end{equation}
	and with $\g$ being the Euler's constant. In the same way we find that the $\ln(-G_F^{DS})$, in the DeWitt-Schwinger representation \eqref{DWSrep}, can be expressed as
	\begin{equation}
		\bra{x}\ln(-G_F^{DS})\ket{x'}=-\int_{m^2}^\infty\,G^{DS}_F(x,x')\,dm^2,
	\end{equation}
	where the integral with respect to $m^2$ is given by
	\begin{equation}
		\int_{m^2}^\infty\,\e^{-i\,m^2\,s}\,dm^2=e^{-i\,m^2\,s}\,(is)^{-1},
	\end{equation}
	followed by the use of \eqref{logG}.
	Recalling the definition of the effective action in terms of $\ln(-G_F^{DS})$ in \eqref{funct}, these expressions given so far yield
	\begin{equation}
		\mathcal{S}_{\scalebox{0.5}{$eff$}}=\sdfrac{i}{2}\int\,d^dx\,\sqrt{-g}\,\lim_{x'\to x}\int_{m^2}^\infty\,dm^2\,G_F^{DS}(x,x')=\sdfrac{i}{2}\int_{m^2}^\infty\,dm^2\,\int\,d^dx\,\sqrt{-g} G_F^{DS}(x,x),\label{Oneloop}
	\end{equation}
	where we have interchanged the order of integrations. This expression of $\mathcal{S}_{\scalebox{0.5}{$eff$}}$ is known as the \emph{one-loop effective action} because in \eqref{Oneloop} the $d^dx$ integration is seen to be precisely the expression for the one-loop contribution to the vacuum energy from matter fields. We may define an \emph{effective Lagrangian density} $L_{eff}$, as
	\begin{equation}
		L_{eff}(x)=\sdfrac{i}{2}\lim_{x'\to x}\int_{m^2}^\infty\,dm^2\,G_F^{DS}(x,x').
	\end{equation}
	which leads to the effective action once the integration with the measure $\sqrt{-g}\,d^dx$ is performed. \\
	In the following discussion we are going to extract and study the divergences of the effective Lagrangian. This analysis will be performed in $d=4$ and it will be helpful for the evaluation of the trace anomaly of the stress energy tensor. The explicit expression of $G_F^{DS}$ together with \eqref{aE} and \eqref{DWSre} manifest some divergences of $L_{eff}$ at the lower end of the $s$ integral when the limit $x'\to x$ is considered. The convergence in the upper limit of the integration, instead, is guaranteed by the $-i\epsilon$ that is implicitly added to $m^2$ in the DeWitt-Schwinger representation of $G_F$. The divergent part of $L_{eff}$, in the case $d=4$, is given by 
	\begin{equation}
		L_{eff}^{div}=-\lim_{x'\to x}\sdfrac{\D^{\frac{1}{2}}(x,x')}{32\pi^2}\int_0^\infty\,\sdfrac{ds}{s^3}\,e^{-i(m^2a-\s/2 s)}\left[a_0(x,x')+a_1(x,x')\,is+a_2(x,x')\,(is)^2\right]\label{DWexp}
	\end{equation}
	where the coefficients $a_0$, $a_1$ and $a_2$ are given by \eqref{coefficientadiab}, and the remaining terms in this asymptotic expansion, involving $a_3$ and higher, are finite in the limit $x\to x'$. The divergences of $L_{eff}$ reflect those of the stress energy tensor. Looking at the coefficients $a_i$ written in \eqref{DWexp}, one can immediately realize that they are entirely geometrical in the limit $x\to x'$ because of the vanishing of the normal coordinates in this limit. Then the coefficients will be given just in terms of the Riemann tensor $R_{\m\n\r\s}$ and of its contractions. The divergence arises from the action of the quantum matter fields, which are integrated out, and the result is an expression which depends only on the background field. The content of the quantum action is encoded in the finite contribution in $L_{eff}$. 
	
	When the limit $x\to x'$ is taken, the function  $\sigma=1/2y^\alpha\,y_\alpha$ goes to zero and generates divergences. Performing an asymptotic expansion of $L_{eff}$ and taking an analytic continuation of the dimension $d$ we find
	\begin{equation}
		L_{eff}\approx \sdfrac{(4\pi)^{-d/2}}{2}\,\sum_{j=0}^\infty\,a_j(x)\,\int_0^\infty i\,ds\,(is)^{j-1-d/2}\,e^{-im^2s}=\sdfrac{(4\pi)^{-d/2}}{2}\,\sum_{j=0}^\infty\,a_j(x)\,(m^2)^{d/2-j}\,\G(j-d/2)\label{expL}
	\end{equation}
	where $a_j(x)=a_j(x,x)$.We introduce an arbitrary regularization mass scale $\m$ \begin{equation}
		L_{eff}\approx \sdfrac{(4\pi)^{-d/2}}{2}\left(\sdfrac{m}{\m}\right)^{d-4}\,\sum_{j=0}^\infty\,a_j(x)\,m^{4-2j}\,\G(j-d/2).\label{exp2}
	\end{equation}
	When $d\to 4$ the first three terms of \eqref{exp2} generate divergences through the $\G$ functions that can be expanded around $d=4$ as 
	\begin{equation}
		\begin{split}
			\G\left(-\sdfrac{d}{2}\right)&=\sdfrac{4}{d(d-2)}\left(\sdfrac{2}{4-d}-\g\right)+O(d-4)\\
			\G\left(1-\sdfrac{d}{2}\right)&=\sdfrac{2}{2-d}\left(\sdfrac{2}{4-d}-\g\right)+O(d-4)\\
			\G\left(2-\sdfrac{d}{2}\right)&=\sdfrac{2}{4-d}-\g+O(d-4).
		\end{split}\label{gamma}
	\end{equation}
	In this way we have shown that the divergent structure of $L_{eff}$ is related just to the first three terms $a_0$, $a_1$ and $a_2$ and we obatin the expression
	\begin{equation}
		L_{div}=-(4\pi)^{-d/2}\left\{\sdfrac{1}{d-4}+\sdfrac{1}{2}\left[\g+\ln\left(\sdfrac{m^2}{\m^2}\right)\right]\right\}\left(\sdfrac{4m^2\,a_0}{d(d-2)}-\sdfrac{2m^2\,a_1}{d-2}+a_2\right)\label{exp3}+O((d-4))
	\end{equation}
	where we have used the expansion
	\begin{equation}
		\left(\sdfrac{m}{\m}\right)^{d-4}=1+\sdfrac{1}{2}(d-4)\ln\left(\sdfrac{m^2}{\m^2}\right)+O((d-4)^2).
	\end{equation}
	
	The functions $a_0$ ,$a_1$ and $a_2$ are given by taking the coincidence limits of \eqref{coefficientadiab} and \eqref{ageneric}
	\begin{align}
		a_0(x)&=1\\
		a_1(x)&=\left(\sdfrac{1}{6}-\x\right)\,R\\
		a_2(x)&=\sdfrac{1}{180}R_{\a\b\g\d}R^{\a\b\g\d}-\sdfrac{1}{180}R_{\a\b}R^{\a\b}-\sdfrac{1}{6}\left(\sdfrac{1}{5}-\x\right)\square R+\sdfrac{1}{2}\left(\sdfrac{1}{6}-\x\right)^2\,R^2,\label{a2}
	\end{align}
	
	{ showing that the divergent part of the effective Lagrangian $L_{eff}^{div}$, as given in \eqref{exp3}, is a purely geometrical expression. In order to absorb the divergences, we recall that the total action contains also the gravitational part, which is purely geometrical. In fact, the gravitational action is expressed in the Einstein-Hilbert form 
		\begin{equation}
			S_g=\int\,d^dx\,\sqrt{-g}\,L_g=\int\,d^dx\,\sqrt{-g}\sdfrac{1}{16\pi G_B}(R-2\L_B)
		\end{equation}
		in terms of the bare cosmological constant $\L_B$ and Newton's constant $G_B$. After incorporating the divergent part of the one loop effective action, the total gravitational Lagrangian density now becomes
		\begin{equation}
			L_g=\left\{\ -\left(A+\sdfrac{\L_B}{8\pi G_B}\right)+\left(B+\sdfrac{1}{16\pi G_B}\right)R-\sdfrac{a_2(x)}{(4\pi)^{d/2}}\left[\sdfrac{1}{d-4}+\sdfrac{1}{2}\left(\g+\ln\sdfrac{m^2}{\m^2}\right)\right]\right\}\label{Lg}
		\end{equation}
		where
		\begin{align}
			A&=\sdfrac{4m^4}{(4\pi)^{d/2}d(d-2)}\left\{\sdfrac{1}{d-4}+\sdfrac{1}{2}\left[\g+\ln\left(\sdfrac{m^2}{\m^2}\right)\right]\right\}\\
			B&=\sdfrac{2m^2}{(4\pi)^{d/2}(d-2)}\left(\sdfrac{1}{6}-\x\right)\left\{\sdfrac{1}{d-4}+\sdfrac{1}{2}\left[\g+\ln\left(\sdfrac{m^2}{\m^2}\right)\right]\right\}.
		\end{align}
		The effect of the new additions is to renormalise the cosmological constant as
		\begin{equation}
			\L\equiv \L_B+\sdfrac{32\pi m^4\,G_B}{(4\pi)^{d/2}d(d-2)}\left\{\sdfrac{1}{d-4}+\sdfrac{1}{2}\left[\g+\ln\left(\sdfrac{m^2}{\m^2}\right)\right]\right\},\label{reno}
		\end{equation}
		and this renormalized value is exactly the one we would consider as the physical one. Furthermore, by looking at the second term in \eqref{Lg}, we observe that also the Newton's gravitational constant is renormalized as
		\begin{equation}
			G=\sdfrac{G_B}{1+16\pi\,G_B\,B}.
		\end{equation}
		These procedure will be reflected in the gravitational field equation 
		\begin{equation}
			R_{\m\n}-\sdfrac{1}{2}R\,g_{\m\n}+\L\,g_{\m\n}=-8\pi G\braket{T_{\m\n}},
		\end{equation}
		with $\Lambda$ and $G$ renormalized constants. Finally, the last term in \eqref{Lg} is something new, absent in the usual Einstein-Hilbert Lagrangian. Indeed the factor $a_2(x)$ defines a higher order correction to the general theory of relativity that contains only second derivatives of the metric. The factor $a_2(x)$, as one can see from its definition in \eqref{a2}, is of adiabatic order four, i.e. it includes four derivatives of the metric. The renormalized gravitational action then becomes
		\begin{equation}
			S_g=\int\,d^dx\,\sqrt{-g}\,\left\{\sdfrac{1}{16\pi G}\Big(R-2\L\Big)-\sdfrac{a_2(x)}{(4\,\pi)^{d/2}}\left[\sdfrac{1}{d-4}+\sdfrac{1}{2}\left(\g+\ln\sdfrac{m^2}{\m^2}\right)\right]\right\}\label{Sg},
		\end{equation}
		giving for the left-hand side of the gravitational field equation the expression
		\begin{equation}
			R_{\m\n}-\sdfrac{1}{2}\,R\,g_{\m\n}+\L\,g_{\m\n}+\a\ \,^{\substack{\scalebox{0.5}{(1)}}}\!H_{\m\n}+\b\ \,^{\substack{\scalebox{0.5}{(2)}}}\!H_{\m\n}+\g\,H_{\m\n}\label{eqgrav}
		\end{equation}
		where 
		\begin{equation}
			\begin{split}
				\,^{\substack{\scalebox{0.5}{(1)}}}\!H_{\m\n}&\equiv\,\sdfrac{1}{\sqrt{-g}}\sdfrac{\d}{\d g^{\m\n}}\int\,d^dx\,\sqrt{-g}R^2=2R_{;\m\n}-2g_{\m\n}\square R-\sdfrac{1}{2}g_{\m\n}R^2+2R\,R_{\m\n}\\
				\,^{\substack{\scalebox{0.5}{(2)}}}\!H_{\m\n}&\equiv \,\sdfrac{1}{\sqrt{-g}}\sdfrac{\d}{\d g^{\m\n}}\int\,d^dx\,\sqrt{-g}\,R_{\a\b}R^{\a\b}=2R_{\m\ \ ;\n\a}^{\ \ \a}-\square R_{\m\n}-\sdfrac{1}{2}g_{\m\n}\square R+2R_\m^{\ \,\a}R_{\a\n}-\sdfrac{1}{2}g_{\m\n}R^{\a\b}R_{\a\b}\\
				H_{\m\n}&\equiv\,\sdfrac{1}{\sqrt{-g}}\sdfrac{\d}{\d g^{\m\n}}\int\,d^dx\,\sqrt{-g}R_{\a\b\g\d}R^{\a\b\g\d}\notag\\
				&=-\sdfrac{1}{2}g_{\m\n}R_{\a\b\g\d}R^{\a\b\g\d}+2R_{\m\a\b\g}R_\n^{\ \,\a\b\g}-4\square R_{\m\n}+2R_{;\m\n}-4R_{\m\a}R^\a_{\ \n}+4R^{\a\b}R_{\a\m\b\n},
			\end{split}\label{HH}
		\end{equation}
		as also pointed out in \cite{Barth:1983hb}. Notice that the term proportional to $\square\,R$ in the gravitational action \eqref{Sg} does not contribute to the gravitational field equation, because such a term is a covariant total divergence, and if we exclude special boundary effects, such as singularities etc., it will not contribute to the result. Furthermore, it is worth noting that in $d=4$ we have the relation
		\begin{equation}
			H_{\m\n}=-\,^{\substack{\scalebox{0.5}{(1)}}}\!H_{\m\n}+4\,^{\substack{\scalebox{0.5}{(2)}}}\!H_{\m\n},\label{ind}
		\end{equation}
		since the generalized Gauss-Bonnet theorem states that the integral of the Euler density, in $d=4$ is a topological invariant, and its metric variation vanishes identically
		\begin{equation}
			\frac{\delta}{\delta g^{\mu\nu}}\int\,d^dx\,\sqrt{-g}\,\left(R_{\a\b\d\g}R^{\a\b\d\g}-4R_{\a\b}R^{\a\b}+R^2\right)\,\overset{d=4}{=}\,0.
		\end{equation}
		The coefficients $\a$, $\b$ and $\g$ in \eqref{eqgrav} are explicitly given by
		\begin{align}
			\alpha&=\frac{1}{2(4\pi)^\frac{d}{2}}\left(\frac{1}{6}-\xi\right)^2\left[\sdfrac{1}{d-4}+\sdfrac{1}{2}\left(\g+\ln\sdfrac{m^2}{\m^2}\right)\right],\\
			\beta&=-\gamma=-\frac{1}{180(4\pi)^\frac{d}{2}}\left[\sdfrac{1}{d-4}+\sdfrac{1}{2}\left(\g+\ln\sdfrac{m^2}{\m^2}\right)\right],
		\end{align}
		and diverge as $d$ approaches the physical dimension $4$. In order to renormalise these divergences, one has to introduce in the gravitational action terms of adiabatic order four, multiplied by some bare coefficients $a_B$, $b_B$, $c_B$ that will be fixed by the renormalization procedure. However, because of \eqref{ind}, in $d=4$ only two of these coefficients are independent, and we may choose $c=0$. 
	}
	{
		\section{The conformal anomaly}
		We consider the case in which the classical action $S$ is invariant under conformal transformations. The requirement of conformal invariance at level of the action implies that the classical stress energy tensor is traceless. The conformal transformations, as pointed out in the previous sections, are essentially a rescaling of lengths at each spacetime point $x$ and the presence a fixed scale in the theory, like a mass, will break the symmetry. In order to apply the results of the section above, we have to consider the massless limit. However, all the higher order $(j>2)$ terms in DeWitt-Schwinger expansion of the effective Lagrangian are infrared divergent at $d=4$ as $m\to0$, but the expansion is still useful in the investigation of the ultraviolet divergent terms, like $a_j$ with $j=0,1,2$ in \eqref{expL}. 
		We may set $m=0$ immediately in the expansion, except for the terms related to $j=0,1$, because they are of positive power for $n\sim4$, and will vanish. The only non-vanishing term is for $j=2$ 
		\begin{equation}
			\sdfrac{(4\pi)^{-d/2}}{2}\left(\sdfrac{m}{\m}\right)^{d-4}\,a_2(x)\,\G(2-d/2),\label{div}
		\end{equation}
		which must be handled carefully. Taking the explicit expression of $a_2$ from \eqref{a2} with $\xi=\xi(d)$ given in the conformally coupled case, we may write the divergent term in the effective action arising from \eqref{div} as
		\begin{align}
			\mathcal{S}_{\scalebox{0.5}{$eff$}}^{\scalebox{0.5}{$(div)$}}&=\sdfrac{(4\pi)^{-d/2}}{2}\left(\sdfrac{m}{\m}\right)^{d-4}\,\G(2-d/2)\int\,d^dx\sqrt{-g}\,a_2(x)\notag\\
			&=\sdfrac{(4\pi)^{-d/2}}{2}\left(\sdfrac{m}{\m}\right)^{d-4}\,\G(2-d/2)\int\,d^dx\sqrt{-g}\,\left[b C^2(x)+b'\,E(x)\right]+O(d-4),\label{anoma1}
		\end{align}
		where
		\begin{equation}
			\begin{split}
				C^2&=R_{\a\b\g\d}R^{\a\b\g\d}-2R_{\a\b}R^{\a\b}+\sdfrac{1}{3}R^2\\
				E&=R_{\a\b\g\d}R^{\a\b\g\d}-4R_{\a\b}R^{\a\b}+R^2,
			\end{split}	\label{FeG}
		\end{equation}
		while the coefficients are
		\begin{equation}
			b=\sdfrac{1}{120},\qquad b'=-\sdfrac{1}{360}.
		\end{equation}
		
		In obtaining \eqref{anoma1} we have dropped the $\square R$ and $R^2$ terms from $a_2(x)$: the first because it is a total divergence, and so it will not contribute to the action, the second because its coefficient is proportional to $(d-4)^2$ when the conformal coupling $\x=\x(d)=\sdfrac{d-2}{4(d-1)}$ is inserted. In the limit $d\to4$ this coefficient removes the $(d-4)^{-1}$ singularity from the $\G$ function in \eqref{gamma}, causing the term to vanish. The appearance of the square Weyl tensor $C^2$ and the Euler density, that remain invariant under conformal transformation, leads to the property that, at $d=4$,  $\mathcal{S}^{\scalebox{0.5}{$(div)$}}_{\scalebox{0.5}{$eff$}}$ in the massless conformally coupled limit is invariant under conformal transformations.
		We compute the contribution of $\mathcal{S_{\scalebox{0.5}{$eff$}}}{}^{\hspace{-0.25cm}\scalebox{0.5}{(div)}}$ to the trace of the stress-tensor by considering one functional variation with respect to the metric and 
		using the identities 
		\begin{equation}
			\begin{split}
				\sdfrac{2}{\sqrt{-g}}g^{\m\n}\sdfrac{\d}{\d g^{\m\n}}\int\,d^dx\,\sqrt{-g}\,E&=-(d-4)\left(E-\sdfrac{2}{3}\square\,R\right),\\
				\sdfrac{2}{\sqrt{-g}}g^{\m\n}\sdfrac{\d}{\d g^{\m\n}}\int\,d^dx\,\sqrt{-g}\,C^2&=-(d-4)C^2,
			\end{split}\label{rel}
		\end{equation}
		we immediately obtain
		\begin{equation}
			\braket{T_\m^{\ \m}}_{div}=\sdfrac{2}{\sqrt{-g}}g^{\m\n}\sdfrac{\d\mathcal{S}^{\scalebox{0.5}{$(div)$}}_{\scalebox{0.5}{$eff$}}}{\d g^{\m\n}}=\sdfrac{(4\pi)^{-d/2}}{2}\left(\sdfrac{m}{\m}\right)^{d-4}(4-d)\G(2-d/2)\left[b'\left(E-\sdfrac{2}{3}\square R\right)+b\,C^2\right]+O(d-4).
		\end{equation}
		When the expansion around $d=4$ is taken we see that this expression becomes
		\begin{equation}
			\braket{T_\m^{\ \m}}_{div}=\frac{1}{(4\pi)^2}\left[b'\left(E-\sdfrac{2}{3}\square R\right)+b\,C^2\right].\label{anoma2}
		\end{equation}
		Since this result is independent of $m/\m$, which has been retained essentially as an infrared cut-off, we can finally set $m=0$, without changing the finite result \eqref{anoma2}. The feature of $\braket{T_{\m}^{\ \m}}_{div}$ to be local and a geometrical object comes directly from the properties of the $\mathcal{S}^{\scalebox{0.5}{$(div)$}}_{\scalebox{0.5}{$eff$}}$. 
		Now, because $\mathcal{S}_{\scalebox{0.5}{$eff$}}$ is conformally invariant in the massless and conformally coupled limit, the expectation value of the trace of the total stress tensor is zero
		\begin{equation}
			\left.\braket{T_\m^{\ \m}}\right|_{m=0,\x=1/6}=0.
		\end{equation}
		It follows that, if the divergent portion $\braket{T_{\m\n}}_{div}$ has acquired the trace \eqref{anoma2}, the finite, renormalized residue $\braket{T_{\m\n}}_{ren}$ must also have a trace, i.e. the negative of \eqref{anoma2}
		\begin{equation}
			\braket{T_{\m}^{\ \m}}_{ren}=-\sdfrac{1}{(4\pi)^2}\left[b'\left(E-\sdfrac{2}{3}\square R\right)+b\,C^2\right]=-\sdfrac{a_2}{(4\pi)^2}\label{anomalyincomp}
		\end{equation}
		This result is known as a conformal, or trace, anomaly. What it is written in \eqref{anomalyincomp} is the main part of the trace anomaly and it is present in any quantum theory that in $d=4$ incidentally breaks the conformal symmetry. Other terms are present in this relations and are related to the specific field content of the theory. For instance, if we consider a four-dimensional system of massless fields in interaction only with an external gravitational field and an external spin-1 gauge field, the divergences in the effective action will depend on the metric and its derivatives will involve terms like $R\,\square^{(d-4)/2}\,R$, $F^a_{\m\n}\,\square^{(d-4)/2}\,F^{\m\n}_a$, $\dots$ . In this case the only possible counterterm that can remove the infinities and will give a finite renormalized effective action is
		\begin{equation}
			\D\mathcal{S}_{\scalebox{0.5}{$eff$}}=\sdfrac{\m^{d-4}}{d-4}\int\,d^dx\,\sqrt{-g}\left[b\,C^2+b'\,E+c\,F^2\right],\label{countert}
		\end{equation}
		where $b$, $b'$ and $c$ are constants, while $E$ and $C^2$ are given by \eqref{FeG}. It is clear now that the trace anomaly will have a contribution coming from the field strength of the gauge field as $F^2=F^{a,\,\mu\nu}F_{\mu\nu}^a$. 
	}
	
	We have to clarify the role of $\m$ in \eqref{countert}, because it is an arbitrary parameter having the dimensions of a mass. It is analogous to the choice of renormalization point perturbation theory. $\mathcal{S}_{\scalebox{0.5}{$eff$}}$ will then involve terms like $R\,\ln(\square/\m^2)R$. The presence of such a dimensional parameter in the quantum theory is one way of understanding the appearance of conformal anomalies. Secondly, we have discarded total divergences in the integrand of $\D\mathcal{S}_{\scalebox{0.5}{$eff$}}$, but since we are at a generic dimension $d$, we must retain $G$, even though it is a topological invariant quantity and hence yields a zero metric variation when $d=4$. Then, no other contributions to $\D\mathcal{S}_{\scalebox{0.5}{$eff$}}$ are compatible with our requirement that $\D\mathcal{S}_{\scalebox{0.5}{$eff$}}$ has to be conformally invariant at $d=4$, i.e. to be proportional to 
	\begin{equation}
		\int\,d^4x\,\sqrt{-g}[b\,C^2+c\,F^2]=\int d^4x\,\sqrt{-g}\left[b\,C_{\m\n\r\s}C^{\m\n\r\s}+c\,e^2\,F^a_{\m\n}F^{\m\n}_a\right]_{d=4}
	\end{equation}
	where $C_{\m\n\r\s}$ is the Weyl tensor. Finally, the constants $b$, $b'$ and $c$ may be $d$-dependent. However, any such ambiguity will result only in finite terms which are conformal invariant at $d=4$ and hence do not contribute to the anomaly. The regularized tensor will be finite at $d=4$ and now, using also the identity
	\begin{equation}
		\sdfrac{2}{\sqrt{-g}}g_{\m\n}\sdfrac{\d}{\d g_{\m\n}}\int\,d^dx\,\sqrt{-g}\,F^2=(d-4)F^2,
	\end{equation}
	we find that 
	\begin{equation}
		T^\m_{\ \,\m}=b'\left(E-\sdfrac{2}{3}\square\,R\right)+b C^2+c F^2.\label{anom1}
	\end{equation}
	The $\square R$ term appearing in \eqref{anom1}, may be expressed as the metric variation of a local four-dimensional action, namely
	\begin{equation}
		\sqrt{-g}\,\square R=-\sdfrac{1}{6}g_{\m\n}\sdfrac{\d}{\d g_{\m\n}}\int\,d^4x\,\sqrt{-g}\, R^2.
	\end{equation}
	By contrast, there exists no local four-dimensional action such that, when functionally differentiated and traced, would yield the quadratic terms in the Riemann tensor, and thus no way of removing these anomalies by finite local additions to the Lagrangian. We have seen that a conformally invariant four dimensional theory
	\begin{equation}
		S_{CFT}=S_0+\sdfrac{1}{d-4}\int d^4x\,\sqrt{-g}\left[b C_{\m\n\r\s}C^{\m\n\r\s}+c\,e^2 F^a_{\m\n}F_a^{\m\n}\right]
	\end{equation}
	leads to a unique value of the $\square R$ anomaly given by \eqref{anom1}. One is free, of course, to consider some other theory obtained by ad hoc addition of a finite non conformal invariant term of the form $R^2$, i.e.
	\begin{equation}
		S=S_{CFT}+\sdfrac{\d'}{12}\int d^4x\,\sqrt{-g}R^2.
	\end{equation}
	The parameter $\d'$ could then be adjusted to yield extra $\square R$ contributions to $T^\m_{\ \m}$, or even to cancel out the $\square R$ term altogether by setting $\d'=\d$. In the present context of external gravitational fields, there is no great virtue in breaking the conformal invariance by hand in this fashion, but a completely satisfactory answer to this problem cannot be given as long as we confine our attention to a curved spacetime. Eventually, we must turning to the quantum dynamics of the gravitational field itself, and to the problem of constructing the correct Lagrangian for gravity plus matter. It is worth noting that any such theory is unlikely to exhibit conformal invariance anyway, but the anomalies will still survive to produce results which deviate from one's naive classical expectations. 
	We assume that no such conformal non-invariant additions have been made, and that the trace of the stress tensor is given by \eqref{anom1}. From the explicit calculations \cite{Duff:1993wm}, we may now fix the constants $b$ and $b'$ to be
	\begin{equation}
		b'=\sdfrac{1}{120(4\pi)^2}[N_S+6N_F+12N_V],\qquad b=-\sdfrac{1}{360(4\pi)^2}[N_S+11N_F+62N_V],
	\end{equation}
	where $N_S$, $N_F$ and $N_V$ are the numbers of scalars, spin-$1/2$ fermions and vectors in the theory, respectively. 
{\section{The anomaly effective action}
	In this section we are going to derive and discuss the nonlocal anomalous effective action in $d=4$ proposed in \cite{Riegert:1984kt}. We have shown in the previous section that, in a four-dimensional system of massless fields in interaction with external gravitational  and gauge fields, the quantum corrections induce on the trace of the energy-momentum tensor $T_{\mu\nu}$ an anomaly of the general form
	\begin{equation}
		T_A=\left(bC^2+b'E+b''\,\square\,R+dR^2+cF^2\right),\label{G}
	\end{equation}
	where the coefficients $b$, $b'$, $b''$, $c$, $d$ depend on the specific particle content (fermion, scalar, vector) and interactions of the theory, as described by the corresponding Lagrangian. The main idea in \cite{Riegert:1984kt} is to extract information about the exact effective action $\mathcal{S}_{\scalebox{0.5}{$eff$}}$ that is related to the trace of $T_{\mu\nu}$ as
	\begin{equation}
		\frac{2}{\sqrt{-g}}g_{\m\n}\frac{\d\mathcal{S}_{\scalebox{0.5}{$eff$}}}{\d g_{\m\n}}=T\label{trace1}
	\end{equation}
	using only the form of the trace anomaly equation \eqref{G}. This is a general procedure and it does not depend on the explicit form of the coefficients in \eqref{G}. We define the part of the effective action which generates the anomalous trace $T_A$ , $\mathcal{S}_{\scalebox{0.6}{anom}}$, the {anomaly effective action}. Looking at \eqref{G} we know that the term proportional to $\square R$ can be generated as a variation of 
	\begin{equation}
		-\sdfrac{2}{\sqrt{-g}}\,g_{\m\n}\sdfrac{\d}{\d g_{\m\n}}\int d^4x\,\sqrt{-g}\,R^2=-12\,\square \,R,\label{varRsq}
	\end{equation}
	for which the anomaly effective action will contain the contribution $c/(192\p^2)\int d^4x\sqrt{-g}R^2$. However, the other terms in \eqref{G} can't be obtained from the trace of the functional variation of an integral of simple scalar geometric quantities. For this reason, we proceed first by using a non-covariant approach in order to find such terms, and then we will try to identity the covariant nonlocal form of the anomaly effective action. We use the property of the Weyl group, and we consider a local conformal parametrization of the metric of the form $g_{\m\n}=e^{2\s}\bar g_{\m\n}$ for arbitrary $\sigma(x)$, where $\bar g_{\m\n}$ has a fixed determinant.  
	Inserting this expression into the trace anomaly equation we obtain
	\begin{equation}
		\frac{\d\mathcal{S}_{\scalebox{0.6}{anom}}}{\d\s}=\sqrt{-\bar{g}}\,e^{4\s}\,T_A.\label{an}
	\end{equation}
	The trace anomaly, by using the identity $E=C^2-2(R^{\a\b}R_{\a\b}-\frac{1}{3}R^2)$, can be rewritten in the form
	\begin{equation}
		T_A=\left[(b+b')C^2-2b\left(R^{\a\b}R_{\a\b}-\frac{1}{3}R^2\right)+b''\square\,R+dR^2+cF^2\right],\label{TA}
	\end{equation}
	and this simplifies the analysis. In fact, by conformal invariance, $C^\l_{\s\m\n}(g)=C^\l_{\s\m\n}(\bar g)\equiv\bar C^\l_{\s\m\n}$ and the gauge field  $F^i_{\m\n}$ are metric independent. Therefore we find that the $C^2$ and $F^2$ terms become
	\begin{equation}
		\begin{split}
			C^2&\equiv g^{\a\m}g^{\b\n}C^\l_{\g\a\b}C^\g_{\l\m\n}=e^{-4\s}\bar g^{\a\m}\bar g^{\b\n}\bar C^\l_{\g\a\b}\bar C^\g_{\l\m\n}\equiv e^{-4\s}\bar C^2\notag\\
			F^2&\equiv g^{\a\m}g^{\b\n}F^i_{\a\b}F_{i\m\n}=e^{-4\s}\bar g^{\a\m}\bar g^{\b\n}F^i_{\a\b}F_{i\m\n}\equiv e^{-4\s}\bar 
			F^2,\notag
		\end{split}
	\end{equation}
	and the equation to be solved to account for these terms is given by \eqref{an}
	\begin{equation}
		\frac{\d\mathcal{S}_{\scalebox{0.6}{anom}}^{C^2+F^2}}{\d\s}=\sqrt{-\bar{g}}\bigg[(b+b')\bar{C}^2+c\bar{F}^2\bigg].
	\end{equation}
	It admits the solution
	\begin{equation}
		\mathcal{S}_{\scalebox{0.6}{anom}}^{C^2+F^2}=\int d^4x\sqrt{-\bar{g}}\ \left[(b+b')\bar C^2+e\bar F^2\right]\s,\label{act}
	\end{equation}
	modulo an arbitrary functional, independent of $\s$, which in any 
	case can be added to the non-anomalous part of the effective action.
	The remaining piece of the anomaly to study is $R^{\a\b}R_{\a\b}-\frac{1}{3}R^2$ that, under a local conformal parametrization of the metric, behaves like
	\begin{align}
		\sqrt{-g}\left(R^{\a\b}R_{\a\b}-\frac{1}{3}R^2 \right)&=\sqrt{-\bar g}\left(\bar R^{\a\b}\bar R_{\a\b}-\frac{1}{3}\bar R^2 -4\bar R^{\a\b}\left(\bar\nabla_{\a}\bar\nabla_\b\s-\bar\nabla_{\a}\s\bar\nabla_\b\s\right)+2\bar R\bar\square\,\s \right.\notag\\
		&\qquad -4(\bar\square\,\s)^2-4\bar\square\,\s\bar\nabla_\a\s\bar\nabla^\a\s+4\bar\nabla_\a\bar\nabla_\b\s\bar\nabla^\a\bar\nabla^\b\s-8\bar\nabla^\a\nabla^\b\s\bar\nabla_\a\s\bar\nabla_\b\s\Big)\label{var1}
	\end{align}
	where the barred notation indicates that the dependence is on the metric $\bar{g}_{\mu\nu}$. In the next section we will illustrate a constructive method by which one can build an action, given its equation of motion, and we will apply it to \eqref{var1} in order to find the last missing piece of the anomaly effective action in its local form, before obtaining, at the last step, its nonlocal expression.  
}

{
	\section{Reconstruction of the local anomaly effective action}
	
	If we consider an action $A[\phi]$ which depends on a set of fields $\phi_\k(x)$, then its equations of motion are obtained through the stationary principle, with the variation of the action written as
	\begin{equation}
		\delta A=\int\,dx\,\sdfrac{\delta A}{\delta \phi_\k(x)}\delta\phi_\k(x).
	\end{equation}
}
Suppose now that $\delta A/\delta\phi_\k(x)$ is a given function of the fields and their derivatives. We can use the information about $\delta A/\delta\phi_k(x)$ to reconstruct the action $A[\phi]$. First of all we have to choose a reference configuration $\bar{\phi}_\k(x)$ of the fields. Then, in order to evaluate $A[\psi]$ for an arbitrary choice of fields $\phi_\k(x)=\psi_\k(x)$, we choose a ``path'' $\phi_\k(x,\l)$, $0\le\l\le1$, that begins at the reference point $\phi_\k(x,0)=\bar{\phi}_\k(x)$ and leads to the desired final point, $\phi_\k(x,1)=\psi_\k(x)$. For instance, a straight line path 
\[\phi_\k(x,\l)=\l\,\psi_\k(x)+(1-\l)\bar{\phi}_\k(x)\]
may be a convenient choice. Now consider the action $A[\phi(\l)]$ evaluated along the path; its derivative with respect to $\l$ is 
\begin{equation}
	\sdfrac{d}{d\l}A[\phi(\l)]=\int dx\left.\sdfrac{\delta A}{\delta\phi_\k(x)}\right|_{\phi=\phi(\l)}\sdfrac{\partial \phi_\k(x,\l)}{\partial\l},
\end{equation}
and the integration of this equation from $\l=0$ to $\l=1$ gives
\begin{equation}
	A[\psi]=\int dx\,\int_0^1\,d\l\,\left.\sdfrac{\delta A}{\delta\phi_\k(x)}\right|_{\phi=\phi(\l)}\sdfrac{\partial \phi_\k(x,\l)}{\partial\l}\,+\,A[\bar{\phi}].
\end{equation}
This equation determines $A[\psi]$, up to an arbitrary additive constant $A[\bar\phi]$. 
It is worth mentioning that it may be impossible to find an action that identically reproduces a given set of equations of motion as its Euler-Lagrange equations. Nevertheless, one can sometimes change the form of the equations without changing their content in such a way that the modified equations do follow from an action principle. 
{ We apply this method to find the corresponding term of \eqref{var1} in the anomaly effective action. The equation of motion that we consider is written in the form
	\begin{align}
		\frac{\d\mathcal{S}_{\scalebox{0.6}{anom}}^{C^2-E}}{\d\s}&=\sqrt{-\bar g}\left(\bar R^{\a\b}\bar R_{\a\b}-\frac{1}{3}\bar R^2 -4\bar R^{\a\b}\left(\bar\nabla_{\a}\bar\nabla_\b\s-\bar\nabla_{\a}\s\bar\nabla_\b\s\right)+2\bar R\bar\square\,\s \right.\notag\\
		&\qquad -4(\bar\square\,\s)^2-4\bar\square\,\s\bar\nabla_\a\s\bar\nabla^\a\s+4\bar\nabla_\a\bar\nabla_\b\s\bar\nabla^\a\bar\nabla^\b\s-8\bar\nabla^\a\nabla^\b\s\bar\nabla_\a\s\bar\nabla_\b\s\Big)\label{an1}.
	\end{align}
	Following the method above, in order to solve the equation \eqref{an1} and find $\mathcal{S}_{\scalebox{0.6}{anom}}^{C^2-E}[\s]$, we choose a path $\s(x,\l)$, $0\le\l\le1$ that begins from a reference point $\s(x,0)=0$ and goes to the final point $\s(x,1)=\tilde\s$. The simplest choice we are going to consider is a straight line, that in a parametric form is $\s(x,\l)=\l\tilde\s$. The variation of the action $\mathcal{S}_{\scalebox{0.5}{\text{anom}}}^{C^2-E}[\s]$ along this path is given by
	\begin{equation}
		\sdfrac{d}{d\l}\mathcal{S}_{\scalebox{0.6}{anom}}^{C^2-E}[\s(\l)]=\int d^4x\left.\frac{\d\mathcal{S}_{\scalebox{0.6}{anom}}}{\d \s}\right|_{\s=\s(\l)}\sdfrac{\partial \s(x,\l)}{\partial \l}.
	\end{equation}
	Integrating in the range of variations of the parameter $0\le\l\le1$ the action becomes
	\begin{align}
		\mathcal{S}_{\scalebox{0.6}{anom}}^{C^2-E}[\tilde{\s}]&=\int d^4x\int^1_0d\l\, \sqrt{-\bar g}\left(\bar R^{\a\b}\bar R_{\a\b}-\frac{1}{3}\bar R^2 +4\bar R^{\a\b}\left(\l\bar\nabla_{\a}\bar\nabla_\b\tilde\s-\l^2\bar\nabla_{\a}\tilde\s\bar\nabla_\b\tilde\s\right)-2\l\bar R\bar\square\,\tilde\s \right.\notag\\
		&\qquad -4\l^2(\bar\square\,\tilde\s)^2-4\l^3\bar\square\,\tilde\s\bar\nabla_\a\tilde\s\bar\nabla^\a\tilde\s+4\l^2\bar\nabla_\a\bar\nabla_\b\tilde\s\bar\nabla^\a\bar\nabla^\b\tilde\s-8\l^3\bar\nabla^\a\nabla^\b\tilde\s\bar\nabla_\a\tilde\s\bar\nabla_\b\tilde\s\Big)\tilde\s\label{CE}
	\end{align}
	and the integration over $\lambda$ is simple to achieve and generates constants. We integrate by parts, and by using the properties
	\begin{equation}
		\bar\nabla_\m\bar\square\,\tilde\s\equiv\bar\square\,\bar\nabla_\m\tilde\s+\bar R_{\m\n}\bar\nabla^\n\tilde\s,\qquad \bar\nabla_\m\bar R^{\m\n}\equiv\frac{1}{2}\bar\nabla^\n \bar R
	\end{equation} 
	we find the final form of the action for the term $C^2-E$ in the trace anomaly as
	\begin{align}
		\mathcal{S}_{\scalebox{0.6}{anom}}^{C^2-E}[\s]&=\int d^4x\sqrt{-\bar g}\left[\left(\bar R^{\a\b}\bar R_{\a\b}-\frac{1}{3}\bar R^2\right)\s -2\bigg(\bar R^{\a\b}-\frac{1}{2}g^{\a\b}\bar R\bigg)\bar\nabla_{\a}\s\bar\nabla_\b\s\right.+2\bar\nabla_\a\s\bar\nabla^\a\s\bar\square\,\s+(\bar\nabla_\a\s\bar\nabla^\a\s)^2\bigg]\label{CEF}.
	\end{align}
	It is straightforward to very that the variation with respect to $\s$ of $\mathcal{S}_{\scalebox{0.6}{anom}}^{C^2-E}[\s]$, obtained in \eqref{CEF}, produces exactly the trace relation \eqref{an1}.
	
	The remaining piece of the trace anomaly $T_A$ to consider is the $R^2$ term in \eqref{TA}. Following the same procedure discussed above, one can obtain the form of the action as
	\begin{equation}
		\mathcal{S}_{\scalebox{0.6}{anom}}^{R^2}=\int d^4 x\sqrt{-\bar g}\left\{\bar R^2\s+12\bar R\left(\frac{\s}{2}\bar\square\,\s+\frac{\s}{3}\bar\nabla^\a\s\bar\nabla_\a\s\right)+36\left[\frac{\s}{3}(\bar\square\,\s)^2+\frac{\s}{2}\bar\square\,\s\bar\nabla^\a\s\bar\nabla_\a\s+\frac{\s}{5}(\bar\nabla^\a\s\bar\nabla_\a\s)^2\right]\right\}.\label{SR2}
	\end{equation}
	However, this action, once it is functionally differentiated, does not reproduce the $R^2$ contribution in \eqref{TA}, and this issue is related to the fact that the functional derivative of \eqref{SR2} is not a symmetric kernel. Hence, we conclude that there exists no action, \emph{local} or \emph{nonlocal}, which has $R^2$ as its trace.\\
	Finally, we have found an action that reproduces all the parts of the trace anomaly, except for the $R^2$ piece, and it is written in a local non-covariant form. \\
	Now we will describe the procedure to obtain the nonlocal and covariant form of the anomaly action. 
	First of all, we consider the Weyl transformation $g_{\mu\nu}=e^{2\s}\bar g_{\mu\nu}$ of $\sqrt{-g} E$ and $\sqrt{-g} \square\, R$, commonsly used in the Weyl gauging 
	\cite{Coriano:2013nja,Codello:2012sn}\cite{Coriano:2013xua}, written explicitly as
	\begin{align}
		\sqrt{-g}\,E&=\sqrt{-\bar g}\left[\bar E+8\bar R^{\a\b}\left(\bar\nabla_\a\bar\nabla_\b\s-\bar\nabla_\a\s\bar\nabla_\b\right)-4\bar R\bar\square\,\s-8(\bar\square\,\s)^2\right.\notag\\
		&\qquad\qquad\qquad\qquad\qquad-\left.8\bar\square\,\s\bar\nabla^\a\s\bar\nabla_\a\s+8\bar\nabla^\a\bar\nabla^\b\s\bar\nabla_\a\bar\nabla_\b\s-16\bar\nabla^\a\bar\nabla^\b\s\bar\nabla_\a\s\bar\nabla_\b\s\right]\label{E1}\\
		\sqrt{-g}\,\square\, R&=\sqrt{-\bar g}\left[\bar\square\,\bar R-12R^{\a\b}\bar\nabla_\a\s\bar\nabla_\b\s-2
		\bar R\bar\square\,\s-2\bar\nabla_\a\bar R\bar\nabla^\a\s-6\bar\square^2\s\right.\notag\\
		&\left.\qquad\qquad +12(\bar\square\,\s)^2+12\bar\square\,\s\bar\nabla^\a\s\bar\nabla_\a\s+24\bar\nabla^\a\bar\nabla^\b\s\bar\nabla_\a\s\bar\nabla_\b\s-12\bar\nabla^\a\bar\nabla^\b\s\bar\nabla_\a\bar\nabla_\b\s\right]\label{R1},
	\end{align}
	and we observe that the combination 
	\begin{equation}
		\frac{1}{4}\sqrt{-g}\left(E-\frac{2}{3}\square\, R\right)=\sqrt{-\bar g}\left[\frac{1}{4}\left(\bar E-\frac{2}{3}\bar\square\, \bar R\right)+\bar\D_4\s\right]\label{point}
	\end{equation}
	is not anymore of order $\sigma^3$, but it is linear in $\sigma$ and $\D_4$ is a fourth order differential operator that behaves under Weyl transformations as 
	\begin{align}
		\D_4&\equiv \square^2+2 R^{\m\n}\nabla_\m\nabla_\n-\frac{2}{3}R\,\square\,+\frac{1}{3}(\nabla^\m\,R)\nabla_\m\\
		\sqrt{-g}\,\D_4&=\sqrt{-\bar g}\,\bar{\D}_4,\label{point2}
	\end{align}
	\cite{Riegert:1987kt,Antoniadis:1992xu,Antoniadis:1991fa,Mazur:2001aa}. Consequently, the operator $\Delta_4$, is the (unique) fourth-order conformally covariant differential operator acting on a field of zero scale dimension. Furthermore, $\Delta_4$ has the property to be self-adjoint 
	\begin{equation}
		\int d^4x\sqrt{-g}\,\y(\D_4\x)=\int d^4x\sqrt{-g}\,(\D_4\y)\x.\label{point3}
	\end{equation}
	where $\x$ and $\psi$ are scalar fields of zero scaling dimensions. The second property implies the existence of an action  
	\begin{equation}
		S[\xi]=\int\,d^4x\sqrt{-g}\left[\sdfrac{1}{2}(\square\,\xi)^2+R^{\mu\nu}\,\nabla_\mu\xi\,\nabla_\nu\xi-\sdfrac{1}{3}R\,\nabla^\mu\xi\,\nabla_\mu\xi\,\right].
	\end{equation}
	whose variation gives $\D_4\xi$. The cornerstone in the construction of the nonlocal and covariant form of the anomaly effective action is in this operator. We define the Green's function $D_4(x,y)$ inverse of $\D_4$ by
	\begin{equation}
		(\sqrt{-g}\,\D_4)_xD_4(x,y)=\d^4(x-y),\label{point4}
	\end{equation}
	that is conformally invariant, by virtue of the independence of $\d^4(x-y)$ with respect to the metric. We invert \eqref{point} using the properties of the operator $\Delta_4$ to find the explicit form of the function $\sigma(x)$, and we obtain
	\begin{equation}
		\s(x)=\int d^4y\,D_4(x,y)\left[\frac{1}{4}\sqrt{-g}\left(E-\frac{2}{3}\square\, R\right)\right]_y+\text{$\s$ independent terms}.
	\end{equation}
	Using the explicit expression of $\sigma$, we finally find the nonlocal but manifestly covariant anomaly effective action as
	\begin{equation}
		\mathcal{S}_{\rm anom}^{^{NL}}[g] =\frac {1}{4}\!\int \!d^4x\sqrt{-g_x}\, \left(E - \frac{2}{3}\sq R\right)_{\!x} 
		\int\! d^4x'\sqrt{-g_{x'}}\,D_4(x,x')\left[\frac{b'}{2}\, \left(E - \frac{2}{3}\sq R\right) +  b\,C^2\right]_{x'}
		\label{Snonl}
	\end{equation}
	with the condition $b''=-2/3b'$ in \eqref{TA}, and we have used the superscript $NL$ to indicate that this is the nonlocal form of the anomaly effective action. Finally, we can impose the condition on the coefficient $c=0$ in \eqref{TA} since a non-zero $R^2$ in this basis cannot be obtained from any effective action (local or not) \cite{Antoniadis:1992xu, Bonora:83, Mazur:2001aa}. 
}
\section{Anomalous Trace Ward Identities}
\label{Sec:TraceWI}
{
	In this and the next sections we will focus just on the gravitational part of the anomaly effective action. As we previously saw, the renormalization procedure gives rise to a non-vanishing trace of the energy momentum tensor. If one consider any correlation function involving at least two stress energy tensors in $d=4$, the trace Ward identity \eqref{tracettt} has to be modified in order to consider the anomalous trace contributions. As in the case of the conservation Ward Identities, the trace identities for the $n$-point functions may be derived by successive
	variation of the fundamental trace identity of the one-point function. In fact, considering only the $b$ and $b'$ terms, and rewriting \eqref{trace1} in the form
	\begin{equation}
		2\, g_{\m\n}(x)\,\frac{\d \mathcal{S}[g]}{\d g_{\m\n}(x)} =\mathcal{A} \equiv \sqrt{-g}\, \Big\{ b\, C^2 + b' \big(E - \tfrac{2}{3}\sq R\big) \Big\}
		\label{Adef}
	\end{equation}
	by varying again with respect to the metric, and finally evaluating with a flat space time Euclidean metric $\delta_{\m\n}$ gives
	\begin{equation}
		\delta_{\m_1\n_1}\braket{T^{\m_1\n_1}(x_1)T^{\m_2\n_2}(x_2)} = 2\, \frac{\d \mathcal{A}(x_1)}{\d g_{\m_2\n_2}(x_2)}\Bigg\vert_{flat}
		\label{twoptrace}
	\end{equation}
	for the two-point function, and 
	\begin{align}
		\delta_{\m_1\n_1}\braket{T^{\m_1\n_1}(x_1)T^{\m_2\n_2}(x_2)T^{\m_3\n_3}(x_3)}&=
		- 2\,\Big\{\d^4 (x_1-x_2) + \d^4(x_1-x_3)\Big\}\, \braket{T^{\m_2\n_2}(x_2)T^{\m_3\n_3}(x_3)} \notag\\
		&\hspace{2cm}+ \, 4 \, \frac{\d^2\mathcal{A}(x_1)}{\d g_{\m_2\n_2}(x_2)\d g_{\m_3\n_3}(x_3)}\Big|_{flat}
		\label{threeptrace}
	\end{align}
	for the three-point function. The corresponding results in momentum space are obtained by Fourier transforming the previous relations giving
	\begin{equation}
		\delta_{\a_1\b_1} \, \braket{T^{\a_1\b_1}(p)T^{\m_2\n_2}( -p)} =  2\, \tilde{\mathcal{A}}_1^{\m_2\n_2} (p)
		\label{twoptr}
	\end{equation}
	for the two point function and 
	\begin{align}
		&\d_{\a_1\b_1}\braket{T^{\a_1\b_1}(p_1)T^{\m_2\n_2}(p_2)T^{\m_3\n_3}(p_3)}\notag\\
		&\hspace{2cm}= - 2\,\braket{T^{\m_2\n_2}(p_1+p_2)T^{\m_3\n_3}(p_3)} -2\, \braket{T^{\m_2\n_2}(p_2)T^{\m_3\n_3}(p_1 + p_3) }
		+ \, 4 \, \tilde {\mathcal{A}}_2^{\m_2\n_2\m_3\n_3} (p_2,p_3)
		\label{threeptr}
	\end{align}
	for the three-point function, where 
	\begin{align}
		&(2\pi)^4 \,\d^4 (p_1+ \dots + p_{n+1})\, \tilde{ \mathcal{A}}_{n}^{\m_2\n_2\dots\m_{n+1}\n_{n+1}}(p_2,\dots ,p_{n+1}) \notag\\
		& \left.\qquad\qquad \equiv \int d^4x_1\dots d^4x_{n+1} \ e^{i p_1\cdot x_1 + \dots + i p_{n+1}\cdot x_{n+1}} \, 
		\frac{\d^{n}\mathcal{A}(x_1)}{\d g_{\m_2\n_2}(x_2)\dots \d g_{\m_{n+1}\n_{n+1}}(x_{n+1})}\right|_{flat}
		\label{Avardef}
	\end{align}
	is the Fourier transform of the $n^{th}$ variation of the anomaly in the flat space limit.} The locality of $\mathcal{A}(x_1)$ implies 
that $\tilde{\mathcal{A}}_{n}^{\m_2\n_2\dots\m_{n+1}\n_{n+1}}$ is a polynomial with only positive powers of the $p_j$, containing no 
$1/p_j^2$ pole terms or logarithms. We note also that if $b'' \neq 0$, the $b'' \sq R$ term may easily be included in $\mathcal{A}$, 
giving an additional local contribution to $\tilde {\mathcal{A}}_{n}$.

The trace identity (\ref{threeptr}) for the three-point function contains two terms involving the 2-point function, which would
usually be considered `non-anomalous,' since they are present even if $\tilde {\mathcal{A}}_2 =0$, notwithstanding the fact that the 2-point correlation function
itself carries an implicit dependence upon the first variation $\tilde{ \mathcal{A}}_1$ through (\ref{twoptr}). In addition, (\ref{threeptr}) contains the  
last term involving the second variation $\tilde {\mathcal{A}}_2$ which is anomalous. Clearly one may take additional variations of the fundamental trace identity 
(\ref{Adef}) with respect to the metric in order to obtain trace identities for higher $n+1$-point functions, and this pattern will continue 
with the hierarchy of trace identities, each implicitly dependent upon the $(n-1)-${th} and on the lower variations, as well as on the non-anomalous part of its trace Ward Identities. At each order we encounter an explicit new anomalous 
term involving the $n-${th} variation of the trace anomaly $\tilde {\mathcal{A}}_n$.

\section{The total effective action}
\label{Sec:AnomAct}
It is worth mentioning that the anomaly effective action found in \eqref{Snonl} satisfies the Wess-Zumino consistency condition, for which 
\begin{equation}
	\mathcal{S}_{\rm anom}[e^{2\s}\bar g] = \mathcal{S}_{\rm anom}[\bar g] + \G_{WZ}[\bar g;\s]
	\label{WZGamma}
\end{equation} 
for an arbitrary Weyl transformation of the metric $g_{\m\n}(x) = e^{2\s (x)} \,\bar g_{\m\n}(x)$, and whose variation is the anomaly, i.e. \eqref{Adef}. The latter equation can be written in the  equivalent form
\begin{equation}
	\frac{\d \G_{WZ}[\bar g;\s]}{\d \s (x)}=\sqrt{-g}\, \Big\{ b\, C^2 + b' \big(E - \tfrac{2}{3}\sq R\big) \Big\}\Big\vert_{g = e^{2\s}\bar g}.
	\label{traceiden}
\end{equation}
Moreover, the general form of the anomaly is the consequence of locality of the underlying QFT,  and the Wess-Zumino consistency condition exposed above. As shown in the previous secion, one constructs the Wess-Zumino functional in (\ref{WZGamma}), quartic in $\sigma$, as 
\begin{align}
	\G_{_{\!W\!Z}}[\bar g;\s] &= 2 b'\! \int\,d^4x\,\sqrt{-\bar g}\ \s\,\bar\D_4\,\s + \int\,dx\ \overline{\!\mathcal{A}} \, \s\notag\\
	& = b' \int\,d^4x\,\sqrt{-\bar g}\,\Big[2\,\s\,\bar\Delta_4\,\s + \left(\bar E - \tfrac{2}{3} \sqb \bar R\right)\s \Big]
	+ b \int\,d^4x\,\sqrt{-\bar g}\, \bar C^2\,\s.
	\label{WZfour}
\end{align}
We have seen that inverting the equation \eqref{point} in order to find $\s$ in terms of the Green's function of the fourth differential operator $\Delta_4$, we can write the Wess-Zumino functional in the form
\begin{equation}
	\G_{_{\!W\!Z}}[\bar g;\s] = \mathcal{S}_{\rm anom}^{^{NL}}[g=e^{2\s}\bar g] - \mathcal{S}_{\rm anom}^{^{NL}}[\bar g],
	\label{Weylshift}
\end{equation}
with $\mathcal{S}_{\rm anom}^{^{NL}}[g]$ the nonlocal form of the exact quantum 1PI effective action of the anomaly defined in \eqref{Snonl}. This part of the anomaly is the part which is not Weyl-invariant, and cannot be removed by any addition of local terms in the effective action, such as the $\int R^2$ associated with $b''$ term in \eqref{G} by (\ref{varRsq}). Moreover, there is the possibility of adding an arbitrary Weyl-invariant terms (local or not) in the effective action, producing no changes in the structure of \eqref{Adef}, but dropping out of the difference in (\ref{Weylshift}). In addition, these kind of terms cannot remove the
nonlocality in the Weyl non-invariant part of the anomaly action (\ref{Snonl}).
For instance, adding the nonlocal Weyl invariant term 
\begin{equation}
	\sdfrac {b^2}{8b'}\!\int \!d^4x\sqrt{-g_x}\, \big(C^2\big)_{x} \int\! d^4x'\sqrt{-g_{x'}}\,D_4(x,x') \big(C^2\big)_{x'}
	\label{SaddW}
\end{equation}
to \eqref{Snonl}, we will obtain an anomaly effective action written as
\begin{equation}
	\mathcal{S}_{\rm anom}^{^{NL}}[g]+\sdfrac {b^2}{8b'}\!\int \!d^4x\sqrt{-g_x}\, \big(C^2\big)_{x} \int\! d^4x'\sqrt{-g_{x'}}\,D_4(x,x') \big(C^2\big)_{x'}= \sdfrac {1}{8b'}\int d^4x \int d^4x'\, \mathcal{A}(x)\, D_4(x,x')\, \mathcal{A}(x')
	\label{Snonlsq}
\end{equation}
that still satisfy the trace relation and it is symmetrical in the invariants $E$ and $C^2$. 
As shown in \cite{Mottola:2006ew,Shapiro:1994ww}, one can derive a local form of the effective action by introducing a single new scalar field $\varphi$, to get
\begin{align}
	&&\hspace{-1.5cm} \mathcal{S}_{\rm anom}[g;\varphi] \equiv -\sdfrac{b'}{2} \int d^4x\,\sqrt{-g}\, \Big[ (\sq \varphi)^2 - 2 \big(R^{\m\n} - \tfrac{1}{3} R g^{\m\n}\big)
	(\nabla_\m\varphi)(\nabla_\n \varphi)\Big]\notag\\
	&& \hspace{1.5cm} +\, \sdfrac{1}{2}\,\int d^4x\,\sqrt{-g}\  \Big[b'\big(E - \tfrac{2}{3}\sq R\big) + b\,C^2 \Big]\,\varphi,
	\label{Sanom}
\end{align}
where $\varphi$ has canonical mass dimension zero. One can observe that, by using the form of $\mathcal{S}_{\rm anom}[g; \varphi]$, linear shifts in the spacetime scalar $\varphi$ are related to conformal transformations of the spacetime metric, and indeed
the Wess-Zumino consistency condition \eqref{Weylshift} implies the non-trivial relation
\begin{equation}
	\mathcal{S}_{\rm anom}[g; \varphi + 2 \s] = \mathcal{S}_{\rm anom}[e^{-2 \s} g; \varphi] + \mathcal{S}_{\rm anom}[g; 2 \s] .
	\label{SanomWZ}
\end{equation}
Note that although 
$\varphi$ is closely related to and couples to the conformal part of the metric tensor, $\varphi$ is an independent spacetime
scalar field and the local action (\ref{Sanom}) is fully coordinate invariant, unlike $\G_{_{\!W\!Z}}$ in (\ref{WZfour}), which 
depends separately upon $\bar g_{\m\n}$ and $\s$, and is therefore dependent on the conformal frame.
So far we have studied different part of the effective action and finally we can assert that the exact 1PI quantum effective action for a CFT is written as
\begin{equation}
	\mathcal{S} = \mathcal{S}_{\rm local}[g] + \mathcal{S}_{\rm inv}[g] + \mathcal{S}_{\rm anom}[g;\varphi]
	\label{genS}
\end{equation}
where $\mathcal{S}_{\rm local}[g]$ contains the local $\int R^2$ term, whose conformal variation (\ref{varRsq}) 
is associated with the $b'' \sq R$ term in \eqref{G},
while $\mathcal{S}_{\rm inv}[g]$ is an arbitrary Weyl-invariant term
\begin{equation}
	\mathcal{S}_{\rm inv}[e^{2\s}g] = \mathcal{S}_{\rm inv}[g]
\end{equation}
analogous to \eqref{SaddW}, previously used to obtain a symmetric form of the anomaly effective action. In general, this term is nonlocal and its expansion around flat space is responsible for the CWI's, instead of  $\mathcal{S}_{\rm anom}[g;\varphi]$ given by \eqref{Sanom} which is responsible for the anomalous trace \eqref{traceiden}. The form (\ref{genS}) of the decomposition 
of the quantum effective action was obtained in \cite{Mazur:2001aa} by considerations on the abelian group of local Weyl shifts, 
and its cohomology. The local and Weyl-invariant terms are elements of the trivial cohomology of the local Weyl group, while (\ref{Snonl}) or $\mathcal{S}_{\rm anom}[g;\varphi]$ is an element of the non-trivial cocycles of this cohomology, which is uniquely specified by the $b$ and $b'$ anomaly coefficients \cite{Antoniadis:1992xu,Mazur:2001aa,Bonora:83,Karakhanian:1994yd,Arakelian:1995ye}. 
A non-trivial test of the effective action \eqref{genS} and the correctness of the anomaly 
action \eqref{Sanom} is obtained by the reconstruction algorithm of \cite{Bzowski:2013sza} for $\braket{TTT}$ in CFTs in flat spacetime. One can show that the anomalous trace Ward Identities obeyed by $\braket{ TTT}$, must come entirely from the variation of $\mathcal{S}_{\rm anom}[g;\varphi]$, as pointed out in \cite{ Coriano:2018bsy,Mottola:2010}.
}
{
\section{Expansion of the effective action to the third order}
\label{Sec:VarAct}
In this section we discuss the anomaly part of the correlator $\braket{TTT}$ in CFTs in flat spacetime as computed directly from the effective action \eqref{genS}. In the next section we will show that this is exactly what is obtained from the explicit calculation in perturbation theory \cite{Coriano:2018bsy} and from the method proposed in \cite{Bzowski:2013sza}.
In order to obtain the contributions of the anomaly effective action to the three-point function we require the expansion of $\mathcal{S}_{\rm anom}[g;\varphi]$ to third order in deviations from flat space. The expansion can be performed by using the local form of the action \eqref{Sanom}, or equivalently the nonlocal version \eqref{Snonl}, as proposed in \cite{Coriano:2017mux}. The final covariant result does not depend on this choice, as expected. 
In this case we use the local form \eqref{Sanom}, and the consistent expansion  of $\mathcal{S}_{\rm anom}[g;\varphi]$ around flat space is 
expressed in terms both of the metric $g_{\m\n}$ and of $\varphi$
\begin{align}
	g_{\m\n} &= g_{\m\n}^{(0)} + g_{\m\n}^{(1)} + g_{\m\n}^{(2)} + \dots \equiv \eta_{\m\n} + h_{\m\n} + h_{\m\n}^{(2)} + \dots\\
	\varphi &= \varphi^{(0)} +  \varphi^{(1)} +  \varphi^{(2)}  + \dots
\end{align}
in \eqref{Sanom}. The expansion of the scalar field $\varphi$ can be performed order by order through its equation of motion 
\begin{equation}
	\sqrt{-g}\,\Delta_4\,\varphi =\sqrt{-g}\left[\frac{1}{2}\left(E-\frac{2}{3}\,\square R\right)+\frac{b}{2b'}\,C^2\right],
\end{equation}
for which we have the equations
\begin{align}
	&\hspace{4cm}\sqb^2 \varphi^{(0)} = 0 \label{eom0}\\
	&\hspace{-1.5cm}(\sqrt{-g} \D_4)^{(1)} \varphi^{(0)} + \sqb^2 \varphi^{(1)} = \left[\sqrt{-g}
	\left( \sdfrac{E}{2}- \sdfrac{\!\sq R\!}{3} + \sdfrac{b}{2b'}\, C^2 \right)\right]^{(1)}
	= - \sdfrac{\!1\!}{3}\, \sqb R^{(1)} \label{eom1}\\
	&\hspace{-2cm}(\sqrt{-g} \D_4)^{(2)} \varphi^{(0)} + (\sqrt{-g} \D_4)^{(1)} \varphi^{(1)} + \sqb^2 \varphi^{(2)} =
	\left[\sqrt{-g}\left(\sdfrac{E}{2}- \sdfrac{\!\sq R\!}{3} + \sdfrac{b}{2b'}\, C^2 \right)\right]^{(2)} \notag\\
	&\hspace{2cm}= \sdfrac{1}{2}E^{(2)} - \sdfrac{1}{3}\, [\sqrt{-g}\sq R]^{(2)} + \sdfrac{b}{2b'}\, [C^2]^{(2)} \label{eom2}
\end{align}
where $\sqb$ is the d'Alembertian wave operator in flat Minkowski spacetime, and we have used the fact that $E$ and $C^2$ are of second order in the curvature invariants while the Ricci scalar $R$ has an expansion that starts at first order. The equation of motion at zero order \eqref{eom0}, has the trivial solution $\varphi^{(0)}=0$ that corresponds to a choice of the boundary conditions appropriate for the standard Minkowski space vacuum. This condition is equivalent to taking $\braket{T^{\m\n}}_\eta = 0$ in flat Minkowski spacetime with no boundary effects. In this manner one solves the equations \eqref{eom1} and \eqref{eom2} recursively to get $\varphi^{(1)}$, $\varphi^{(2)}$ and $\varphi^{(n)}$ at all the orders in the expansion. 
We have now all the building blocks to express the expansion of the anomaly action at the third order, in order to proceed with a comparison against the anomaly part of the $\braket{TTT}$, given implicitly in the form
\begin{align}
	&\mathcal{S}_{\rm anom}^{(3)} =  - \sdfrac{b'}{2} \int d^4x \, \left\{2\,\varphi^{(1)} \sqb^2 \varphi^{(2)} +\varphi^{(1)} \big(\sqrt{-g} \D_4\big)^{(1)} \,\varphi^{\!(1)} \right\}\notag\\
	&\hspace{1.5cm} + \sdfrac{b'}{2} \int d^4x \left\{\left( - \sdfrac{2}{3} \sqb R^{(1)}\right) \varphi^{(2)} + \left(E^{(2)} - \sdfrac{2}{3}\, \sqrt{-g}\sq R\right)^{\!(2)} \varphi^{(1)} \right\}
	+  \sdfrac{b}{2} \int d^4x \, (C^2)^{(2)}\,\varphi^{(1)},
	\label{Sanom3a}
\end{align}
where 
\begin{equation}
	\big(\sqrt{-g} \D_4\big)^{\!(1)} =  \big(\sqrt{-g} \sq^2\big)^{\!(1)} + 2\, \partial_{\m} \left(R^{\m\n} - \sdfrac{1}{3} \eta^{\m\n} R\right)^{\!(1)}\partial_{\n}.
\end{equation}
After a lengthy but straightforward calculation we end up with the final expression
\begin{align}
	&\hspace{-5mm} \mathcal{S}_{\rm anom}^{(3)} =
	\sdfrac{b'}{9} \int\! d^4x \int\!d^4x'\!\int\!d^4x''\!\left\{\big(\partial_{\m} R^{(1)})_x\left(\sdfrac{1}{\sqb}\right)_{\!xx'}  
	\!\left(R^{(1)\m\n}\! - \!\sdfrac{1}{3} \eta^{\m\n} R^{(1)}\right)_{x'}\!
	\left(\sdfrac{1}{\sqb}\right)_{\!x'x''}\!\big(\partial_{\n} R^{(1)})_{x''}\right\}\notag\\
	&\hspace{-5mm}- \sdfrac{1}{6} \int\! d^4x\! \int\!d^4x'\! \left(b'\, E^{\!(2)} + b\,  [C^2]^{(2)}\right)_{\!x}\! \left(\sdfrac{1}{\sqb}\right)_{\!xx'} \!R^{(1)}_{x'}
	+ \sdfrac{b'}{18}  \int\! d^4x\, R^{(1)}\left(2\, R^{\!(2)} + (\sqrt{-g})^{(1)} R^{(1)}\right),
	\label{S3anom3}
\end{align}
where the last term is purely local. We observe that in spite of the presence of a double coincident pole $(\bar\square^2)^{-1}$ in $\varphi^{(2)}$ once \eqref{eom2} is inverted, the final expression of the anomaly contribution of the $\braket{TTT}$ has no double propagator
terms. The last term in \eqref{S3anom3} may be recognized as the expansion up to to third order of the covariant local action
\begin{equation}
	\sdfrac{b'}{18}  \int\! d^4x\, \sqrt{-g} \,R^2
	\label{bprimelocal}
\end{equation}
which, if subtracted from $S_{\rm anom}$ in (\ref{Sanom}), would cancel the $- \frac{2b'}{3} \sq R$ contribution to
the conformal anomaly resulting from $S_{\rm anom}$, upon using (\ref{varRsq}), and leaving just $b'\, E + b \,C^2$ for the trace. 
As previously mentioned, the result \eqref{S3anom3} may as well be derived from the nonlocal form of the anomaly action \eqref{Snonl}. We refer to \cite{Coriano:2017mux} for more details.
}

\section{The prediction of the anomaly action for the TTT}
\label{Sec:AnomTTT}
{
We have obtained the expansion at third order of the anomaly action in \eqref{S3anom3}, and we are going to calculate now the anomaly contribution to the $\braket{TTT}$ cdirectly from \eqref{S3anom3}. Since by \eqref{FeG}, both $E$ and $C^2$ are second order in curvature tensors, it suffices in (\ref{S3anom3}) to compute the Riemann tensor to first order
\begin{equation}
	R_{\m\a\n\beta}^{(1)} = \sdfrac{1}{2}\, \Big\{\!- \partial_\a\partial_\beta h_{\m\n}- \partial_\m\partial_\n h_{\a\beta} + \partial_\a\partial_\n h_{\b\m}
	+ \partial_\b\partial_\m h_{\a\n}\Big\}
\end{equation}
in the metric variation $h_{\m\n}$. All the contractions may be carried out with the use of the lowest order
flat space metric $\delta^{\m\n}$. In momentum space
\begin{equation}
	\int d^4x \, e^{ip\cdot x} \, R_{\m\a\n\b}^{(1)}(x) \equiv \big[R_{\m\a\n\b}^{(1)}\big]^{\m_1\n_1}(p)\, \tilde h_{\m_1\n_1}(p)
	\label{RieFour}
\end{equation}
which serves to defines the tensor polynomial
\begin{equation}
	\big[R_{\m\a\n\b}^{(1)}\big]^{\m_1\n_1}(p) = \sdfrac{1}{2}\, \Big\{\d^{(\m_1}_\a\, \d^{\n_1\!)\hspace{-4pt}}\,_\b\,p_\m\, p_\n
	+ \d^{(\m_1}_\m\, \d^{\n_1)}_\n\,p_\a\, p_\b  - \d_\b\,^{\hspace{-2pt}(\m_1}\, \d^{\n_1)}_\m\,p_\a \,p_\n - \d^{(\m_1}_\a\, \d^{\n_1)}_\n\,p_\b \,p_\m  \Big\}
	\label{Riemom}
\end{equation}
which has the contractions
\begin{equation}
	\big[R^{(1)}_{\m\n}\big]^{\m_1\n_1}(p) = \d^{\a\b}\, \big[R_{\m\a\n\b}^{(1)}\big]^{\m_1\n_1}(p)
	= \sdfrac{1}{2}\,\Big\{\d^{\m_1\n_1}\,p_\m\, p_\n  + \d^{(\m_1}_\m\, \d^{\n_1)}_\n\,p^2 
	- p^{(\m_1}\, \d^{\n_1)}_\m\,p_\n- p^{(\m_1} \d^{\n_1)}_\n\,p_\m \Big\} 
	\label{Riccip}
\end{equation}
and
\begin{equation}
	\big[R^{(1)}\big]^{\m_1\n_1}(p) = \d^{\m\n} \big[R^{(1)}_{\m\n}\big]^{\m_1\n_1}(p) = p^2 \d^{\m_1\n_1} - p^{\m_1}p^{\n_1}
	= p^2\,  \pi^{\m_1\n_1}(p)x
	\label{Ricscalp}
\end{equation}
defined in an analogous fashion to (\ref{RieFour}). 
}
We also require the squared contractions 
\begin{align}
&\big[R_{\m\a\n\b}^{(1)}R^{(1)\m\a \n\b}\big]^{\m_1\n_1\m_2\n_2} (p_1, p_2) \equiv
\big[R_{\m\a\n\b}^{(1)}\big]^{\m_1\n_1} (p_1) \big[R^{(1)\m\a \n\b}\big]^{\m_2\n_2}(p_2) 
\notag\\
&\hspace{1.5cm}= (p_1 \cdot p_2)^2\, \delta^{\m_1(\m_2}\delta^{\n_2)\n_1}
- 2\, (p_1\cdot p_2)\, p_1\,^{\hspace{-4pt}(\m_2}\delta^{\n_2)(\n_1}p_2\,^{{\hspace{-2pt}}\m_1)}
+ p_1^{\m_2}\,p_1^{\n_2}\,p_2^{\m_1}\,p_2^{\n_1}
\label{Riemsq}
\end{align}
and
\begin{align}
&\big[R_{\m\n}^{(1)}R^{(1)\m\n}\big]^{\m_1\n_1\m_2\n_2} (p_1, p_2) \equiv
\big[R_{\m\n}^{(1)}\big]^{\m_1\n_1} (p_1) \big[R^{(1)\m\n}\big]^{\m_2\n_2}(p_2) 
\notag\\
&\hspace{1.5cm}=\sdfrac{1}{4}\, p_1^2 \, \Big(p_2^{\m_1}\,  p_2^{\n_1}\, \delta^{\m_2\n_2}  
-  2\, p_2\,^{\hspace{-4pt}(\m_1}\delta^{\n_1)(\n_2}  p_2\,^{\hspace{-2.5pt}\m_2)}\Big)
+ \sdfrac{1}{4}\, p_2^2\, \Big(p_1^{\m_2}\,  p_1^{\n_2}\, \delta^{\m_1\n_1} 
-  2\,p_1\,^{\hspace{-4pt}(\m_1}\delta^{\n_1)(\n_2}  p_1\,^{\hspace{-2.5pt}\m_2)}\Big)\notag\\
&\hspace{1.6cm}+ \sdfrac{1}{4}\, p_1^2\ p_2^2\, \delta^{\m_1(\m_2}\delta^{\n_2)\n_1}
+ \sdfrac{1}{4}\, (p_1\cdot p_2)^2\, \delta^{\m_1\n_1}\delta^{\m_2\n_2} 
+ \sdfrac{1}{2} \, p_1^{(\m_1}\,p_2^{\n_1)}\,p_1^{(\m_2}\,p_2^{\n_2)}\notag\\
&\hspace{1.6cm}+  \sdfrac{1}{2} \, (p_1\cdot p_2)\,
\Big( p_1\,^{\hspace{-4pt}(\m_1}\, \delta^{\n_1)(\n_2}  p_2\,^{\hspace{-2.5pt}\m_2)}
-\delta^{\m_1\n_1}  \, p_1^{(\m_2}\,  p_2^{\n_2)} -\delta^{\m_2\n_2}  \, p_1^{(\m_1}\,  p_2^{\n_1)}\Big)\,.
\end{align}
With these expressions in hand, together with the simpler
\begin{equation}
\big[(R^{(1)})^2\big]^{\m_1\n_1\m_2\n_2} (p_1, p_2) \equiv
\big[R^{(1)}\big]^{\m_1\n_1} (p_1) \big[R^{(1)}\big]^{\m_2\n_2}(p_2)
= p_1^2\, p_2^2 \, \pi^{\m_1\n_1}(p_1)\, \pi^{\m_2\n_2}(p_2)
\label{Riccscalsq}
\end{equation}
we may express the third order anomaly action and its contribution to the 
three-point correlator in momentum space in the form
\begin{align}
&\braket{T^{\m_1\nu_1}(p_1)T^{\m_2\nu_2}(p_2)T^{\m_3\nu_3}(p_3)}_{anom}=\sdfrac{8}{3} \Big\{\pi^{\m_1\nu_1}(p_1)
\,\left[b'E^{(2)}+b(C^2)^{(2)}\right]^{\m_2\nu_2\m_3\nu_3}(p_2,p_3)+(\text{cyclic})\Big\}\notag\\
&\hspace{5cm} -\sdfrac{16b'}{9}\Big\{ \pi^{\m_1\nu_1}(p_1)\,Q^{\m_2\nu_2}(p_1,p_2,p_3)\, \pi^{\m_3\nu_3}(p_3)+(\text{cyclic})\Big\}\notag\\
&\hspace{4cm} +\sdfrac{16b'}{27\,}\, \pi^{\m_1\nu_1}(p_1)\,\pi^{\m_2\nu_2}(p_2)\,\pi^{\m_3\nu_3}(p_3)\,\Big\{p_3^2\, p_1\cdot p_2+(\text{cyclic})\Big\}
+ ({\rm local}) \label{AS3}
\end{align}
having taken into account a $2^3 = 8$ normalization factor in the definition of the correlator for $n=3$, and having summed over the $3$ cyclic permutations of the 
indices $(1,2,3)$. In (\ref{AS3})
\begin{align}
&Q^{\m_2\nu_2}(p_1,p_2,p_3) \equiv p_{1\m}\, [R^{\m\nu}]^{\m_2\nu_2}(p_2)\,p_{3\n} \notag\\
&= \sdfrac{1}{2} \,\Big\{(p_1\cdot p_2)(p_2\cdot p_3)\, \delta^{\m_2\n_2} 
+ p_2^2 \ p_1^{(\m_2}\, p_3^{\n_2)} - (p_2\cdot p_3) \, p_1^{(\m_2}\, p_2^{\n_2)} - (p_1\cdot p_2) \, p_2^{(\m_2}\, p_3^{\n_2)}\Big\}
\end{align}
by (\ref{Riccip}), and
\begin{subequations}
\begin{align}
	&\hspace{-1cm}\big[E^{(2)}\big]^{\m_i\nu_i\m_j\nu_j} =\big[R_{\m\a\n\b}^{(1)}R^{(1)\m\a \n\b}\big]^{\m_i\nu_i\m_j\nu_j}
	-4\,\big[R_{\m\n}^{(1)}R^{(1)\m\n}\big]^{\m_i\nu_i\m_j\nu_j}
	+\big[ \big(R^{(1)}\big)^2\big]^{\m_i\nu_i\m_j\nu_j}\\
	&\hspace{-1cm} \big[(C^2)^{(2)}\big]^{\m_i\nu_i\m_j\nu_j}= \big[R_{\m\a\n\b}^{(1)}R^{(1)\m\a \n\b}\big]^{\m_i\nu_i\m_j\nu_j}
	-2\,\big[R_{\m\n}^{(1)}R^{(1)\m\n}\big]^{\m_i\nu_i\m_j\nu_j}
	+ \sdfrac{1}{3}\,\big[(R^{(1)})^2\big]^{\m_i\nu_i\m_j\nu_j}
\end{align}
\end{subequations}
are given by (\ref{Riemsq})-(\ref{Riccscalsq}), with the corresponding momentum dependences $(p_i, p_j)$ suppressed, and (local)
refers to the third variation of (\ref{bprimelocal}), the purely local last term in (\ref{S3anom3}). 
{At this point, we show how to express \eqref{AS3} in a form that will be useful for its comparison with the perturbative one. It is a straightforward exercise 
in tensor algebra using (\ref{Riccip})-(\ref{Riccscalsq}) to verify that
\begin{align}
	&\delta_{\a_1\b_1}\,\braket{T^{\a_1\b_1}(p_1)T^{\m_2\n_2}(p_2)T^{\m_3\n_3}(p_3)}_{anom}\Big\vert_{p_3 = -(p_1 + p_2)} = \tilde{\mathcal{A}}^{\m_2\n_2\m_3\n_3}(p_2,\bar{p}_3)\notag\\
	&\hspace{2cm}=8 b\, \big[(C^2)^{(2)}\big]^{\m_2\n_2\m_3\n_3} (p_2, \bar{p}_3) + \,8b'\, \big[E^{(2)}\big]^{\m_2\n_2\m_3\n_3} (p_2, \bar{p}_3) 
	\label{traceS3}
\end{align}
where $\bar{p}_3=-p_1-p_2$,  giving on the right hand side the second variation of the trace anomaly. This is consistent with the explicitly anomalous contribution to the trace
identity presented in (\ref{threeptr}), provided again that the $\sq R$ contribution from the local term (\ref{bprimelocal}) 
is neglected. 
Taking an additional trace of (\ref{traceS3}), we find 
\begin{align}
	&\delta_{\a_1\b_1}\delta_{\a_3\b_3}\,\braket{T^{\a_1\b_1}(p_1)T^{\m_2\n_2}(p_2)T^{\a_3\b_3}(p_3)}\big\vert_{p_3 = -(p_1 + p_2)} 
	=  \, \delta_{\a_3\b_3}\tilde{\mathcal{A}}^{\m_2\n_2\a_3\b_3}(p_2,\bar{p}_3) \notag\\
	& \qquad = 16b'\, Q^{\m_2\n_2}(p_1,p_2,\bar{p}_3) \, + \ 8b'\,p_2^2\, \left( p_1^2  + p_1\cdot p_2\right)\pi^{\m_2\n_2}(p_2) .
	\label{dbltrace3}
\end{align}
Finally, we calculate the triple trace of \eqref{traceS3} to obtain
\begin{align}
	\delta_{\a_1\b_1}\delta_{\a_2\b_2}\delta_{\a_3\b_3}\,\braket{T^{\a_1\b_1}(p_1)T^{\m_2\n_2}(p_2)T^{\a_3\b_3}(p_3)}_{\scalebox{0.5}{$anom$}}\big\vert_{p_3 = -(p_1 + p_2)} &= \, \delta_{\a_2\b_2}\delta_{\a_3\b_3}\tilde{\mathcal{A}}^{\a_2\b_2\a_3\b_3}(p_2,\bar{p}_3)\notag\\
	&=
	16b' \left[ p_1^2\,p_2^2 - (p_1\cdot p_2)^2\right].
	\label{triptrace3}
\end{align}
Using these relations, we can write the anomaly contribution \eqref{AS3} to the $\braket{TTT}$ as 
\begin{align}
	&\braket{T^{\a_1\b_1}(p_1)T^{\m_2\n_2}(p_2)T^{\a_3\b_3}(p_3)}_{\scalebox{0.5}{$anom$}}=\sdfrac{1}{3}\, \pi^{\m_1\n_1}(p_1)\,\delta_{\a_1\b_1}\,\braket{T^{\a_1\b_1}(p_1)T^{\m_2\n_2}(p_2)T^{\a_3\b_3}(p_3)}_{\scalebox{0.5}{$anom$}}\notag\\
	&\hspace{2cm}+[(p_1,\m_1,\n_1)\leftrightarrow(p_2,\m_2,\n_2)]+[(p_1,\m_1,\n_1)\leftrightarrow(p_3,\m_3,\n_3)]\notag\\
	&\hspace{1.5cm}-\sdfrac{1}{9}\,\pi^{\m_1\n_1}(p_1)\,\pi^{\m_3\n_3}(p_3)\,\delta_{\a_1\b_1}\delta_{\a_3\b_3}\,\braket{T^{\a_1\b_1}(p_1)T^{\m_2\n_2}(p_2)T^{\a_3\b_3}(p_3)}_{\scalebox{0.5}{$anom$}} \notag\\
	&\hspace{2cm}+[(p1,\m_1,\n_1)\leftrightarrow(p_3,\m_3,\n_3)]+[(p_2,\m_2,\n_2)\leftrightarrow(p_3,\m_3,\n_3)]\notag\\
	&\quad+\sdfrac{1}{27}\,\pi^{\m_1\n_1}(p_1)\pi_2^{\m_2\n_2} (p_2)\pi^{\m_3\n_3}(p_3)\,\delta_{\a_1\b_1}\delta_{\a_2\b_2}\delta_{\a_3\b_3}\,\braket{T^{\a_1\b_1}(p_1)T^{\m_2\n_2}(p_2)T^{\a_3\b_3}(p_3)}_{\scalebox{0.5}{$anom$}}\,.
	\label{fin1}
\end{align}
and we will show that this is exactly what expected from the reconstruction method in \cite{Bzowski:2013sza} and from the explicit perturbative calculation in \cite{Coriano:2018bsy}. 

It is worth noting that the result in \eqref{fin1} generalizes the one obtained in several perturbative analysis in free-field 
theories for specific correlators such the $TJJ$.} The latter defines the most significant gravitational correction to a two-point function (in this case the photon propagator), at phenomenological level.   
The structure of the anomaly action, in this case, corresponds to the $F^2$ part -or gauge part - of the nonlocal anomaly functional and it is given by the expression
\begin{equation}
\label{pole}
\mathcal{S}_{pole}= - \frac{e^2}{ 36 \pi^2}\int d^4 x d^4 y \left(\square h(x) - \partial_\mu\partial_\nu h^{\mu\nu}(x)\right)  \square^{-1}_{x\, y} F_{\alpha\beta}(y)F^{\alpha\beta}(y).
\end{equation}
More details concerning this result can be found in the original works \cite{Giannotti:2008cv,Armillis:2009pq,Armillis:2009im,Armillis:2010qk} in the QED and QCD cases and in \cite{Coriano:2014gja} for supersymmetry, where the same pattern emerges from the analysis of the superconformal anomaly multiplet \cite{Coriano:2019dyc}.

\section{Conclusions}

The derivation of (\ref{Sanom}) and the decomposition (\ref{genS}) show that the anomaly action $\mathcal{S}_{\rm anom}$ cannot
and does not determine the Weyl invariant terms. This is a significant difference from the $d=2$ case, where {\it all} metrics are
locally conformally flat, and there are no undetermined Weyl invariant terms. In the special case of $d=2$ CFT, the non-local anomaly 
effective action is of the form $\int dx \int dy\, R(x)\,(\sq^{-1})_{xy} R(y)$ \cite{Polyakov:1981}, with $R$ the $d=2$ scalar curvature and $\sq^{-1}$ the Green's function inverse of the covariant wave operator, showing that the effect of the conformal anomaly involves an intermediate massless scalar exchange or isolated pole in momentum space. This pole may be interpreted as that of a propagating scalar field $\varphi$ introduced to cast the anomaly action in an equivalent local form. Its propagator gives rise to massless poles in all the higher point vertex functions 
$\braket{ T^{\m_1\n_1}(x_1) T^{\m_2\n_2}(x_2)T^{\m_3\n_3}(x_3)...}_g$  of multiple energy momentum tensors, obtained by varying the effective action 
with respect to the metric multiple times \cite{Blaschke:2014ioa}. This massless scalar exchange may be seen as an effective correlated two-particle 
$0^+$ state of the underlying quantum theory, similar to a Cooper pair of electrons in superconductivity, but appearing here in the Lorentz invariant vacuum state. The presence of a massless scalar pole in the three-point function (\ref{AS3}) shows that this pairing phenomenon occurs in $d=4$ as well.

What the anomaly effective action does determine are all the {\it anomalous} contributions to the higher point stress tensor
correlation functions, since the exact 1PI quantum action (\ref{genS}), is precisely the generating function for these connected
correlation functions in an arbitrary fixed background and $\mathcal{S}_{\rm anom}$ is the only term responsible for anomalous contributions in the trace Ward Identities. We have calculated the contribution of $\mathcal{S}_{\rm anom}$ to the three-point function, given explicitly by (\ref{AS3}), by three variations of the general curved space anomaly action.
The equality \eqref{fin1} shows that this is {\it precisely the same result} as that obtained for the anomalous trace parts
by the method of \cite{Bzowski:2013sza} of solving first the exact CWIs for the projected transverse, traceless parts of the
correlator directly in $d$-dimensional flat space, then reconstructing the full $\braket{TTT}$ by restoring the longitudinal and trace parts using the conservation and anomalous trace Ward Identities in the $d\rightarrow 4$ limit. This is an explicit verification of $\mathcal{S}_{\rm anom}$ for the anomalous trace parts of the full $\braket{TTT}$ in any $d=4$ CFT, including the presence of multiple pole terms in all the external invariants, predicted by the anomaly action. In the approach of \cite{Bzowski:2013sza} these multiple pole contributions are a simple consequence of the transverse projection operators in the reconstruction formula \eqref{fin1} for the trace parts of the three-point function. There is no doubt that these trace contributions, in complete 
agreement with the variation of the anomaly action have the correct analytic structure to satisfy the {\it anomalous} CWIs of a CFT
in the physical $d=4$ dimensions.

It is clear from this derivation that the massless pole contributions are unambiguously fixed and 
determined by the anomaly effective action, as a necessary consequence of the anomalous trace and conformal Ward Identities. 
That anomaly is generally associated with massless poles in $d=4$ has taken some time to recognize, although the prototype example was already provided by the $d=2$ Schwinger model and Polyakov action decades ago \cite{Blaschke:2014ioa}. Because of the coupling 
to $T^{\m\n}$, massless poles on the light cone lead to novel scalar gravitational effects on macroscopic scales, 
not described by Einstein's classical theory \cite{Mottola:2016mpl, Mottola:2006ew,AndMolMott:2003,Meissner:2016onk}. 
In particular, the appearance of a propagating effective massless scalar degree of freedom in $d=4$, explicitly represented by the local 
conformalon field $\varphi$ in (\ref{Sanom}) clearly has implications for gravity at low energies and at macroscopic scales \cite{Mottola:2006ew}, 
including the existence and propagation of scalar gravitational waves not present in classical General Relativity \cite{Mottola:2016mpl}.
Massless poles in $\braket{TTT}$ imply that there are long-range effects of stress tensor correlations on lightlike separated 
spacetime points \cite{Giannotti:2008cv,Armillis:2009pq}, similar to those already noted in the case of the chiral anomaly 
\cite{Giannotti:2008cv,Armillis:2009sm,Coriano:2008pg}. In the supersymmetric case, this behaviour is present in all the components of a superconformal anomaly multiplet \cite{Coriano:2014gja}. This light-cone behaviour is most clearly seen in a Lorentzian
momentum space representation.

It is essential to distinguish this generation of massless poles by anomalous Ward Identities, which represent an {\it explicit} 
breaking of conformal invariance by quantum effects, from the Goldstone massless poles generated by {\it spontaneous} 
symmetry breaking (SSB) of global symmetries by some field(s) acquiring a non-zero vacuum expectation value.  
Phenomenological dilaton effective actions are in the latter class. Note that the anomaly poles coming from $\mathcal{S}_{\rm anom}$ 
are not the result of SSB of conformal invariance in a broken phase, and not the result of introducing by hand a dilaton kinetic term, but rather a consequence of the quantum conformal anomaly itself explicitly breaking conformal invariance in
an otherwise conformally invariant CFT. 

Since the massless scalar poles are a necessary feature of the stress tensor correlation functions, determined by the anomaly effective action, 
which cannot be eliminated the addition of local or Weyl invariant terms, as shown explicitly in this chapter for $n=3$, and because of the possibly 
far-reaching physical implications as one of the very few accessible windows into gravity in the fully quantum regime, it is important to address 
such concerns for $n \ge 4$ in a Lorentzian momentum space representation directly in $d=4$. Clearly, the double coincident poles predicted by the anomaly effective action that are expected to appear in flat space amplitudes in four-point and higher-point point stress tensor correlation 
functions $\braket{ T^{\m_1\n_1}T^{\m_1\n_1}\dots T^{\m_n\n_n}}$ require additional investigation in order to settle this issue satisfactorily.

\chapter*{Conclusions}
\addcontentsline{toc}{chapter}{Conclusions}
\chaptermark{Conclusions}

We have studied the momentum space approach to the solution of the CWI's of CFT's in higher dimensions. Our work's goal has been to illustrate the essential steps needed to build tensor correlators starting from the scalar solutions, for 3-point functions. \\
In the case of 4-point functions, our attention has been centred around scalar correlators for which the CWI's are sufficient to isolate the unique solution if we enhance the symmetry with the addition of a dual conformal symmetry. Dual conformal symmetry in momentum space is obtained once the momentum variables are rewritten in a dual form, as the difference of coordinate-like variables and treated as ordinary correlators in such variables, mirroring the action of coordinate space. This enhancement of the symmetry is sufficient to fix the solutions also for such correlators.    
The solution of the conformal constraints are given in terms of triple-$K$ integrals and are expressed in terms of a set of constants, specific for each correlator and spacetime dimension.\\
We have presented a discussion of the intermediate steps in the description of two nontrivial correlators, the $TTO$ and the $TTT$, in a more pedagogical way, offering details that could help extend such methods to higher point function.\\ 
Several parallel studies have widened the goal of this activity, addressing issues such as the use of conformal blocks/CP symmetric blocks (Polyakov blocks) \cite{Isono:2019ihz,Isono:2018rrb,Isono:2019wex} \cite{Chen:2019gka}, the operator product expansion in momentum space \cite{Gillioz:2019iye}, as well as light-cone blocks \cite{Gillioz:2019lgs, Gillioz:2018mto,Gillioz:2018kwh}, analytic continuations to Lorentzian spacetimes \cite{Bautista:2019qxj} spinning correlators, and Yangian symmetry \cite{Loebbert:2020hxk,Loebbert:2016cdm} just to mention a few, motivated by CFT in momentum space. Related analysis have explored the link to Witten diagrams within the AdS/CFT correspondence \cite{Anand:2019lkt,Albayrak:2019yve}. At the same time, the extension of these investigations to de Sitter space has laid the foundations for new applications in cosmology \cite{Maldacena:2011nz, Arkani-Hamed:2018kmz,Baumann:2019oyu,Arkani-Hamed:2017fdk,Benincasa:2019vqr,Benincasa:2018ssx} \cite{Kundu:2014gxa,Almeida:2017lrq,Baumann:2020dch} and in gravitational waves \cite{Almeida:2019hhx}. Finally, investigations of such correlators in Mellin space \cite{Penedones:2010ue,Fitzpatrick:2011ia} \cite{Gopakumar:2016cpb,Gopakumar:2016wkt} offer a new perspective on the bootstrap program both in flat and in curved space \cite{Sleight:2019mgd,Sleight:2019hfp}, providing further insight into the operatorial structure of a given CFT, and connecting in a new way momentum space and Mellin variables.  \\
Undoubtedly, CWI's play a crucial role in this effort, with widespread applications both at zero and at finite temperature \cite{Ohya:2018qkr}.
Among all the possible correlators that one may investigate, those containing stress-energy tensors $(T)$  play a special role, due to the presence of the conformal anomaly \cite{Coriano:2012wp}. Analysis of 4-point functions have so far been limited to scalar correlators in flat \cite{Maglio:2019grh} \cite{Bzowski:2019kwd} and curved backgrounds \cite{Arkani-Hamed:2018kmz,Baumann:2019oyu}.
Interestingly, this analysis can be performed in parallel with the ordinary Lagrangian field theory approach, allowing to provide a particle interpretation of the breaking of the conformal symmetry. \\
Furthermore, we have concentrated on identifying the fundamental structure of the anomaly effective action, which has been widely debated in the former literature.\\
We have shown that the nonlocal version of such action correctly describes the anomaly structure of the 3-point function of stress energy tensors. \\
The specific analytic structure and massless poles predicted by the effective action are precisely what is obtained by reconstructing the trace parts from the CWI's. This indicates that the anomaly poles in momentum space are a necessary feature of the full $\braket{TTT}$ correlator in CFT. 
The addition of local or Weyl invariant terms to effective action cannot eliminate such contribution. \\
By matching general solutions of the CWI's to free field theories, one obtains the general expressions of correlators containing stress energy tensors, conserved currents and scalar operators of integer dimensions. The nonperturbative solutions of 3-point functions cannot be matched by such free field theories for correlators containing scalar operators of arbitrary scaling dimensions.

\appendix
\numberwithin{equation}{chapter}
\chapter{4-point function relations}
\section{Chain rules}\label{Appendix0}
In this appendix we summarize some important relations regarding the chain rules used in the derivation of the hypergeometric system of equations. They are given by
\begin{align}
	\frac{\partial^2}{\partial p_1^2} F(x,y)&=\frac{2x}{p_1^2}\ \partial_x F(x,y)+\frac{4x^2}{p_1^2}\ \partial_{xx}F,&& \frac{\partial^2}{\partial p_4^2} F(x,y)=\frac{2y}{p_4^2}\ \partial_y F(x,y)+\frac{4y^2}{p_4^2}\ \partial_{yy}F,\\[1.3ex]
	\frac{\partial^2}{\partial p_3^2} F(x,y)&=\frac{2x}{p_3^2}\ \partial_x F(x,y)+\frac{4x^2}{p_3^2}\ \partial_{xx}F,&& \frac{\partial^2}{\partial p_2^2} F(x,y)=\frac{2y}{p_2^2}\ \partial_y F(x,y)+\frac{4y^2}{p_2^2}\ \partial_{yy}F,\\[1.3ex]
	\frac{\partial}{\partial p_1} F(x,y)&=\frac{2x}{p_1} \ \partial_x F(x,y),&&\frac{\partial}{\partial p_4} F(x,y)=\frac{2y}{p_4} \ \partial_y F(x,y),\\[1.3ex]
	\frac{\partial}{\partial p_3} F(x,y)&=\frac{2x}{p_3} \ \partial_x F(x,y),&&\frac{\partial}{\partial p_2} F(x,y)=\frac{2y}{p_2} \ \partial_y F(x,y),\\[1.3ex]
	\frac{\partial}{\partial s}F(x,y)&=-\frac{2}{s}\big(x\,\partial_x F+y\,\partial_yF\big),&&\frac{\partial}{\partial t}F(x,y)=-\frac{2}{t}\big(x\,\partial_x F+y\,\partial_yF\big),
\end{align}
\begin{equation}
	\frac{\partial^2}{\partial s\partial t}F(x,y)=\frac{4}{st}\big[\big(
	x\,\partial_x +y\partial_y\big)F+\big(x^2\partial_{xx}+2xy\,\partial_{xy}+y^2\partial_{yy}\big)F\big],
\end{equation}
\begin{align}
	\left(p_1\frac{\partial}{\partial p_1}+p_2\frac{\partial}{\partial p_2}-p_3\frac{\partial}{\partial p_3}-p_4\frac{\partial}{\partial p_4}\right)F(x,y)&=\left(2x\,\partial_x+2y\,\partial_y -2x\,\partial_x-2y\,\partial_y\right)F(x,y)=0,\\[1.5ex]
	\left(p_1\frac{\partial}{\partial p_1}+p_4\frac{\partial}{\partial p_4}-p_3\frac{\partial}{\partial p_3}-p_2\frac{\partial}{\partial p_2}\right)F(x,y)&=\left(2x\,\partial_x+2y\,\partial_y -2x\,\partial_x-2y\,\partial_y\right)F(x,y)=0.
\end{align}

\section{3K integrals for 4-point functions}\label{AppendixA}
We summarize some relations concerning 3K integrals. We define 
\begin{equation}
	I_{\a\{\b_1,\b_2,\b_3\}}(p_1\,p_3; p_2\,p_4;s\,t)=\int_0^\infty\,dx\,x^\a\,(p_1p_3)^{\b_1}\,(p_2p_4)^{\b_2}\,(s\,t)^{\b_3}\,K_{\b_1}(p_1p_3\,x)\,K_{\b_2}(p_2p_4\,x)\,K_{\b_3}(st\,x)
\end{equation}
as in \eqref{3K}. The $K$ Bessel functions satisfy the relations
\begin{align}
	\frac{\partial}{\partial p}\big[p^\b\,K_\b(p\,x)\big]&=-x\,p^\b\,K_{\b-1}(p x)\\
	K_{\b+1}(x)&=K_{\b-1}(x)+\frac{2\b}{x}K_{\b}(x)
\end{align}
from which we obtain (omitting the argument in each integral as in \eqref{3K}) 
\begin{align}
	\frac{\partial}{\partial p_1}I_{\a\{\b_1,\b_2,\b_3\}}&=-p_1\,p_3^2\,I_{\a+1\{\b_1-1,\b_2,\b_3\}}\\
	\frac{\partial}{\partial p_3}I_{\a\{\b_1,\b_2,\b_3\}}&=-p_3\,p_1^2\,I_{\a+1\{\b_1-1,\b_2,\b_3\}}\\
	\frac{\partial}{\partial p_2}I_{\a\{\b_1,\b_2,\b_3\}}&=-p_2\,p_4^2\,I_{\a+1\{\b_1,\b_2-1,\b_3\}}\\
	\frac{\partial}{\partial p_4}I_{\a\{\b_1,\b_2,\b_3\}}&=-p_4\,p_2^2\,I_{\a+1\{\b_1,\b_2-1,\b_3\}}\\
	\frac{\partial}{\partial s}I_{\a\{\b_1,\b_2,\b_3\}}&=-s\,t^2\,I_{\a+1\{\b_1,\b_2,\b_3-1\}}\\
	\frac{\partial}{\partial t}I_{\a\{\b_1,\b_2,\b_3\}}&=-t\,s^2\,I_{\a+1\{\b_1,\b_2,\b_3-1\}}
\end{align}
and for the second derivative
\begin{align}
	\frac{\partial^2}{\partial p_1^2}I_{\a\{\b_1,\b_2,\b_3\}}&=-\,p_3^2\,I_{\a+1\{\b_1-1,\b_2,\b_3\}}+p_1^2\,p_3^4\,\,I_{\a+2\{\b_1-2,\b_2,\b_3\}}\\
	\frac{\partial^2}{\partial p_3^2}I_{\a\{\b_1,\b_2,\b_3\}}&=-\,p_1^2\,I_{\a+1\{\b_1-1,\b_2,\b_3\}}+p_3^2\,p_1^4\,\,I_{\a+2\{\b_1-2,\b_2,\b_3\}}\\
	\frac{\partial^2}{\partial p_2^2}I_{\a\{\b_1,\b_2,\b_3\}}&=-\,p_4^2\,I_{\a+1\{\b_1,\b_2-1,\b_3\}}+p_2^2\,p_4^4\,\,I_{\a+2\{\b_1,\b_2-2,\b_3\}}\\
	\frac{\partial^2}{\partial p_4^2}I_{\a\{\b_1,\b_2,\b_3\}}&=-\,p_2^2\,I_{\a+1\{\b_1,\b_2-1,\b_3\}}+p_4^2\,p_2^4\,\,I_{\a+2\{\b_1,\b_2-2,\b_3\}}\\
	\frac{\partial^2}{\partial s\partial t}I_{\a\{\b_1,\b_2,\b_3\}}&=-2\,s\,t\,I_{\a+1\{\b_1,\b_2,\b_3-1\}}+t^3\,s^3\,\,I_{\a+2\{\b_1,\b_2,\b_3-2\}}.
\end{align}
They can be rearranged using the relations
\begin{align}
	p_1^2\,p_3^2\,I_{\a+2\{\b_1-2,\b_2,\b_3\}}&=I_{\a+2\{\b_1,\b_2,\b_3\}}-2(\b_1-1)\,I_{\a+1\{\b_1-1,\b_2,\b_3\}}\\
	p_2^2\,p_4^2\,I_{\a+2\{\b_1,\b_2-2,\b_3\}}&=I_{\a+2\{\b_1,\b_2,\b_3\}}-2(\b_2-1)\,I_{\a+1\{\b_1,\b_2-1,\b_3\}}\\
	s^2\,t^2\,I_{\a+2\{\b_1,\b_2,\b_3-2\}}&=I_{\a+2\{\b_1,\b_2,\b_3\}}-2(\b_3-1)\,I_{\a+1\{\b_1,\b_2,\b_3-1\}}.
\end{align}

\section{4K integrals for Lauricella 4-point functions}\label{AppendixB}
We summarize some important relations about the 4K integrals. Defining the 4K integral as
\begin{equation}
	I_{\a\{\b_1,\b_2,\b_3,\b_4\}}(p_1,p_2,p_3,p_4)=\int_0^\infty\,dx\,x^\a\,\prod_{i=1}^4(p_i)^{\b_i}\,K_{\b_i}(p_i\,x)
\end{equation}
its first derivative with respect the mgnitudes of the momenta is given by
\begin{equation}
	p_i\frac{\partial}{\partial p_i}I_{\a\{\b_j\}}=-p_i^2\,I_{\a+1\{\b_j-\d_{ij}\}},\qquad i,j=1,\dots,4.
\end{equation}
One can show that the relation
\begin{equation}
	\int_0^\infty\,x^{\a+1}\frac{\partial}{\partial x}\left[\prod_{i=1}^4\,p_i^{\b_i}\,K_{\b_i}(p_i\,x)\right]=-\int_0^\infty\,\left[\frac{\partial x^{\a+1}}{\partial x}\right]\prod_{i=1}^4\,p_i^{\b_i}\,K_{\b_i}(p_i\,x)
\end{equation}
leads to the identity
\begin{equation}
	\sum_{i=1}^{4}p_i^2I_{\a+1\{\b_j-\d_{ij}\}}=(\a-\b_t+1)\,I_{\a\{\b_j\}},\qquad j=1,\dots,4
\end{equation}
where $\b_t=\b_1+\b_2+\b_3+\b_4$. 

\chapter{Transverse Ward Identities}\label{transWard}
In this appendix we illustrate the procedure for obtaining the canonical Ward identities related to the three local symmetry that one can have in the calculation of correlation functions of scalar, conserved current and conserved and traceless stress energy tensor operators. The first step is to couple the system to some background fields, and then require that the resulting generating functional is invariant  under {diffeomorphisms}, {gauge} and {Weyl} transformations. \\
In particular, we know that under a diffeomorphism $\x^\m$ the sources in the generating functional transform as
\begin{align}
	\d g^{\m\n}&=-(\nabla^\m\x^\n+\nabla^\n\x^\m),\\
	\d A^a_\m&=\x^\n\nabla_\n A_\m^a+\nabla_\m\x^\n\,A_\n^a,\\
	\d\phi_0^I&=\x^\n\partial_\n\phi_0^I,
\end{align}
where $\nabla$ is a Levi-Civita connection. Under a gauge symmetry transformation with parameter $\a^a$, the sources transform as
\begin{align}
	\d g^{\m\n}&=0,\\
	\d A^a_\m&=-D_\m^{ac}\a^c=-\partial_\m\a^a-f^{abc}A_\m^b\a^c,\\
	\d\phi_0^I&=-i\a^a(T^a_R)^{IJ}\phi^J_0,
\end{align}
where $T^a_R$ are matrices of a representation $R$ and $f^{abc}$ are structure constants of the group $G$. The gauge field transforms in the adjoint representation, while $\phi^I$ may transform in any representation $R$. The covariant derivative is $D_\m^{IJ}=\d^{IJ}\partial_\m-iA_\m^a(T_R^a)^{IJ}$.\\
The Ward Identities follow from the requirement that the generating functional $Z$ is invariant under these transformations and in particular under the variation
\begin{align}
	\d_\x&=\int\,d^dx\left[-(\nabla^\m\x^\n+\nabla^\n\x^\m)\sdfrac{\d}{\d g^{\m\n}}+(\x^\n\nabla_\n A_\m^a+\nabla_\m\x^\n\,A_\n^a)\sdfrac{\d}{\d A_\m^a}+\x^\n\partial_\n\phi_0^I\sdfrac{\d}{\d\phi_0^I}\right],\\
	\d_\a&=-\int\,d^dx\left[(\partial_\m\a^a-f^{abc}A_\m^b\a^c)\sdfrac{\d}{\d A_\m^a}+i\a^a(T^a_R)^{IJ}\phi^J_0\sdfrac{\d}{\d\phi_0^I}\right],
\end{align}
so that the canonical Ward Identities for the diffeomorphism and gauge transformations are respectively given by 
\begin{equation}
	\d_\x Z=0,\qquad \d_\a Z=0.
\end{equation}
Using the definitions of the $1$-point functions
\begin{align}
	\braket{T^{\mu\nu}(x)}=-\frac{2}{\sqrt{-g(x)}}\frac{\delta Z}{g_{\mu\nu}(x)}\Bigg|_{g=\d},\qquad\braket{J^{a\mu}(x)}=-\frac{1}{\sqrt{-g(x)}}\frac{\delta Z}{A^a_{\mu}(x)}\Bigg|_{A=0},\qquad \braket{\mO^I(x)}=-\frac{1}{\sqrt{-g(x)}}\frac{\delta Z}{\phi_0^I(x)}\Bigg|_{\phi=0},
\end{align}
the arbitrariness of the parameter $\a^a$, integrating by parts and also using the property $\sdfrac{1}{\sqrt{-g}}\partial_\m\sqrt{-g}=\G^\l_{\l\m}$ we find
\begin{align}
	\d_\a Z=0&=-\int\,d^dx\left[(\partial_\m\a^a-f^{abc}A_\m^b\a^c)\sdfrac{\d}{\d A_\m^a}+i\a^a(T^a_R)^{IJ}\phi^J_0\sdfrac{\d}{\d\phi_0^I}\right]Z\notag\\
	&=\int\,d^dx\sqrt{-g}\,\left[(\partial_\m\a^a-f^{abc}A_\m^b\a^c)\sdfrac{-1}{\sqrt{-g}}\sdfrac{\d}{\d A_\m^a}+i\a^a(T^a_R)^{IJ}\phi^J_0\,\,\sdfrac{-1}{\sqrt{-g}}\sdfrac{\d}{\d\phi_0^I}\right]Z\notag\\
	&=\int\,d^dx\sqrt{-g}\,\a^a\,\left[-\G^\l_{\m\l}\langle J^{\m a}(x)\rangle-(\partial_\m\d^{ab}+f^{acb}A_\m^c)\langle J^{\m b}(x)\rangle+i(T^a_R)^{IJ}\phi^J_0\,\,\langle \mathcal{O}_I(x)\rangle\right],
\end{align}
from which we get the first Ward Identity related to the gauge symmetry, expressed as
\begin{align}
	0&=D^{ab}_\m\langle J^{\m a}\rangle+\G^\l_{\m\l}\langle J^{\m a}(x)\rangle-i(T^a_R)^{IJ}\phi_0^J\langle\mathcal{O}_I\rangle\notag\\
	&=\nabla_\m\langle J^{\m a}\rangle+f^{abc}A_\m^b\langle J^{\m c}\rangle-i(T^a_R)^{IJ}\phi_0^J\langle\mathcal{O}_I\rangle\label{Ward1}.
\end{align}
The other Ward identities related to the diffeomorphism invariance will be
\begin{align}
	0=\d_\x Z&=\int\,d^dx\left[-(\nabla^\m\x^\n+\nabla^\n\x^\m)\sdfrac{\d}{\d g^{\m\n}}+(\x^\n\nabla_\n A_\m^a+\nabla_\m\x^\n\,A_\n^a)\sdfrac{\d}{\d A_\m^a}+\x^\n\partial_\n\phi_0^I\sdfrac{\d}{\d\phi_0^I}\right]Z\notag\\
	&=\int\,d^dx\,\sqrt{-g}\left[\sdfrac{1}{2}(\nabla^\m\x^\n+\nabla^\n\x^\m)\,\sdfrac{-2}{\sqrt{-g}}\sdfrac{\d}{\d g^{\m\n}}-(\x^\n\nabla_\n A_\m^a+\nabla_\m\x^\n\,A_\n^a)\sdfrac{-1}{\sqrt{-g}}\sdfrac{\d}{\d A_\m^a}-\x^\n\partial_\n\phi_0^I\sdfrac{-1}{\sqrt{-g}}\sdfrac{\d}{\d\phi_0^I}\right]Z\notag\\
	&=\int\,d^dx\,\sqrt{-g}\left[\sdfrac{1}{2}(\nabla^\m\x^\n+\nabla^\n\x^\m)\,\langle T_{\m\n}(x)\rangle-(\x^\n\nabla_\n A_\m^a+\nabla_\m\x^\n\,A_\n^a)\langle J^{\m a}(x)\rangle-\x^\n\partial_\n\phi_0^I\langle \mathcal{O}_I(x)\rangle\right]\notag\\
	&=\int\,d^dx\,\sqrt{-g}\x^\n\,\left[-\nabla^\m\,\langle T_{\m\n}(x)\rangle-\nabla_\n A_\m^a\langle J^{\m a}(x)\rangle+\nabla_\m\,(A_\n^a\langle J^{\m a}(x)\rangle)-\partial_\n\phi_0^I\,\langle \mathcal{O}_I(x)\rangle\right]
\end{align}
that leads to 
\begin{align}
	0&=\nabla^\m\,\langle T_{\m\n}(x)\rangle+\nabla_\n A_\m^a\,\langle J^{\m a}(x)\rangle-\nabla_\m\,A_\n^a\,\langle J^{\m a}(x)\rangle-A_\n^a\,\nabla_\m\,\langle J^{\m a}(x)\rangle+\partial_\n\phi_0^I\,\langle \mathcal{O}_I(x)\rangle\notag\\
	&=\nabla^\m\,\langle T_{\m\n}(x)\rangle-F_{\m\n}^a\,\langle J^{\m a}(x)\rangle-f^{abc}A_\n^a A_\m^b\langle J^{\m c}\rangle-A_\n^a\,\nabla_\m\,\langle J^{\m a}(x)\rangle+\partial_\n\phi_0^I\,\langle \mathcal{O}_I(x)\rangle.\label{Ward2}
\end{align}

Using the Ward Identity related to the gauge symmetry we obtain the final result 
\begin{align}
	0&=\nabla^\m\,\langle T_{\m\n}(x)\rangle-F_{\m\n}^a\,\langle J^{\m a}(x)\rangle-f^{abc}A_\n^a A_\m^b\langle J^{\m c}\rangle-A_\n^a\,\nabla_\m\,\langle J^{\m a}(x)\rangle+\partial_\n\phi_0^I\,\langle \mathcal{O}_I(x)\rangle\notag\\
	&=\nabla^\m\,\langle T_{\m\n}(x)\rangle-F_{\m\n}^a\,\langle J^{\m a}(x)\rangle-f^{abc}A_\n^a A_\m^b\langle J^{\m c}\rangle+A_\n^a\,\left[f^{abc}A_\m^b\langle J^{\m c}\rangle-i(T^a_R)^{IJ}\phi_0^J\langle\mathcal{O}_I\rangle\right]+\partial_\n\phi_0^I\,\langle \mathcal{O}_I(x)\rangle\notag\\
	&=\nabla^\m\,\langle T_{\m\n}(x)\rangle-F_{\m\n}^a\,\langle J^{\m a}(x)\rangle+D_\n^{IJ}\,\phi_0^I\,\langle \mathcal{O}_J(x)\rangle\label{transWardF}
\end{align}

The equations \eqref{Ward1} and \eqref{Ward2} are the Ward Identities for 1-point functions with sources turned on. These equations may then be differentiated with respect to the sources, with the aim of obtaining the corresponding Ward Identities for higher point functions. 

\section{Trace Ward Identities}

We have shown that the Lagrangian of a conformally invariant theory cannot contain dimensionful coupling constants. This constraint can be circumvented if we allow position dependent couplings, i.e., background fields and if we prescribe the correct transformation properties under Weyl transformation. Assume the operator $\mathcal{O}$ has a conformal dimension $\D$. For the term
\begin{equation}
	\int\,d^dx\,\,\mathcal{O}^{\m_1\dots\m_m}_{\n_1\dots\n_n}\ \phi^{\n_1\dots\n_n}_{\m_1\dots\m_m}
\end{equation}
to be invariant under the scalings, we must have
\begin{equation}
	\phi^{\n_1\dots\n_n}_{\m_1\dots\m_m}=(d-\D+m-n)\phi^{\n_1\dots\n_n}_{\m_1\dots\m_m}\,\s.
\end{equation}
For the metric, the gauge field and the scalar source we have the ordinary transformation rules,
\begin{align}
	\d_\s g_{\m\n}&=2g_{\m\n}\s,\\
	\d_\s A_\m^a&=0,\\
	\d_\s\phi_0&=(d-\D)\phi_0\,\s.
\end{align}
Let's then consider the case when the generating functional is free of the Weyl anomaly, for which
\begin{equation}
	\d_\s Z=0.\label{Weyl}
\end{equation}
The variation of the generating functional is realised by the following operator,
\begin{equation}
	\d_\s=\int\,d^dx\ \,\s\left[2g^{\m\n}\sdfrac{\d}{\d g^{\m\n}}+(d-\D)\phi^I\,\sdfrac{\d}{\d\phi^I_0}\right],
\end{equation}
and to be more specific, we can expand \eqref{Weyl} as
\begin{align}
	0=\d_\s Z&=\int\,d^dx\ \,\s\left[2g^{\m\n}\sdfrac{\d}{\d g^{\m\n}}+(d-\D)\phi^I\,\sdfrac{\d}{\d\phi^I_0}\right]Z\notag\\
	&=\int\,d^dx\ \sqrt{-g}\,\s\left[-g_{\m\n}\langle T^{\m\n}(x)\rangle-(d-\D)\phi^I\,\langle\mathcal{O}^I(x)\rangle\right].
\end{align}
In this case we find the following trace, or Weyl, Ward identity in the presence of sources 
\begin{equation}
	\langle T^\m_{\ \m}(x)\rangle=(\D-d)\phi_0^I\langle\mathcal{O}^I(x)\rangle.
\end{equation}
Also in this case, we can differentiate with respect to the sources in order to obtain the trace Ward identities for $n$-point functions.

\section{Identities with projectors \label{appendixB}}

The projectors are defined as
\begin{align}
	\pi^\m_\a(p)&=\d^\m_\a-\sdfrac{p^\m p_\a}{p^2},\\
	\Pi^{\m\n}_{\a\b}(p)&=\pi^{(\m}_\a\pi^{\n)}_\b-\sdfrac{1}{d-1}\pi^{\m\n}\pi_{\a\b},\label{Proj}\\
	\Pi^{\m\n\r\s}(p)&=\d^{\r\a}\d^{\s\b}\,\Pi^{\m\n}_{\a\b}(p)
\end{align}
where we define $\pi^{(\m}_\a\pi^{\n)}_\b=1/2(\pi^{\m}_\a\pi^{\n}_\b+\pi^{\n}_\a\pi^{\m}_\b)$. One derives the following identities 
\begin{align}
	p_\m\pi^{\m\n}(p)=p_\m\Pi^{\m\n}_{\a\b}(p)&=0,\\
	\d_{\m\n}\pi^{\m\n}(p)=\pi^\m_\m(p)&=d-1,\\
	\Pi^{\m\n\r}_{\ \ \ \ \ \r}(p)=\d_{\r\s}\Pi^{\m\n\r\s}(p)&=\pi_{\r\s}(p)\Pi^{\m\n\r\s}(p)=0,\\
	\Pi^{\m\r\n}_{\ \ \ \ \ \r}(p)=\d_{\r\s}\Pi^{\m\r\n\s}(p)=\pi_{\r\s}(p)\Pi^{\m\r\n\s}(p)&=\sdfrac{(d+1)(d-2)}{2(d-1)}\pi^{\m\n}(p),\\
	\Pi^{\m\n}_{\r\s}(p)\Pi^{\r\s}_{\m\n}(p)&=\sdfrac{1}{2}(d+1)(d-2),\\
	\pi^\m_\a(p)\pi^\a_\m(p)&\pi^\m_\n(p),\\
	\Pi^{\m\n}_{\a\b}(p)\Pi^{\a\b}_{\r\s}(p)&=\Pi^{\m\n}_{\r\s}(p),\\
	\Pi^{\m\n\r}_{\ \ \ \ \ \a}(p)\pi^{\a\s}(p)&=\Pi^{\m\n\r\s}(p),\\
	\Pi^{\m\n}_{\a\b}(p)\Pi^{\a\r\b\s}(p)&=\sdfrac{d-3}{2(d-1)}\Pi^{\m\n\r\s}(p).\\
\end{align}
Denoting with $\partial_\m$ the derivative respect to the $p$ momentum we find
\begin{align}
	\partial_\k\,\pi_{\m\n}(p)&=-\sdfrac{p_\m}{p^2}\pi_{\n\k}(p)-\sdfrac{p_\n}{p^2}\pi_{\m\k}(p),\\
	\partial_\k\Pi_{\m\n\r\s}&=-\sdfrac{p_\m}{p^2}\Pi_{\k\n\r\s}-\sdfrac{p_n}{p^2}\Pi_{\m\k\r\s}-\sdfrac{p_\r}{p^2}\Pi_{\m\n\k\s}-\sdfrac{p_\s}{p^2}\Pi_{\m\n\r\k},\\
	\pi^\m_\a\partial_\k\pi^\a_\n&=-\sdfrac{p_\n}{p^2}\pi^\m_\k,\\
	\pi^{\m\k}\partial_\a\pi^\a_\n-\pi^{\m\a}\partial_\a\pi^\k_\n&=-(d-2)\sdfrac{p_\n}{p^2}\pi^{\m\k}+\sdfrac{p^\k}{p^2}\pi^\m_\n,\\
	\Pi^{\m\n}_{\a\b}\partial_\k\Pi^{\a\b}_{\r\s}&=-\sdfrac{p_\r}{p^2}\Pi^{\m\n}_{\k\s}-\sdfrac{p_\s}{p^2}\Pi^{\m\n}_{\r\k},\\
	\Pi^{\m\n}_{\k\b}\partial_\a\Pi^{\a\b}_{\r\s}-\Pi^{\m\n\a}_{\quad\  \b}\partial_\a\Pi^{\b}_{\ \ \k\r\s}&=-\sdfrac{1}{2}\sdfrac{d-1}{p^2}[p_\r\Pi^{\m\n}_{\k\s}+p_\s\Pi^{\m\n}_{\r\k}]+\sdfrac{p_\k}{p^2}\Pi^{\m\n}_{\r\s},\\
	\Pi^{\m\n}_{\a\b}\,p^\l\partial_\l\Pi^{\a\b}_{\r\s}&=0
\end{align}
Analogous expressions with two derivatives are
\begin{align}
	\pi^\m_\a\partial^2\pi^\a_\n&=-\sdfrac{2}{p^2}\pi^\m_\n,\\
	p^\a\pi^\m_\b\partial_\a\partial_\k\pi^\b_\n&=\sdfrac{p_\n}{p^2}\pi^\m_\k,\\
	\Pi^{\m\n}_{\a\b}\partial^2\Pi^{\a\b}_{\r\s}&=-\sdfrac{4}{p^2}\Pi^{\m\n}_{\r\s},\\
	p^\g\Pi^{\m\n}_{\a\b}\partial_\g\partial_\k\Pi^{\a\b}_{\r\s}&=\sdfrac{p_\r}{p^2}\Pi^{\m\n}_{\k\s}+\sdfrac{p_\s}{p^2}\Pi^{\m\n}_{\r\k}.
\end{align}
For the semi-local operators we find
\begin{align}
	\Pi^{\m\n}_{\a\b}t^{\a\b}_{loc}&=0,\\
	\Pi^{\m\n}_{\a\b}\partial_\k t^{\a\b}_{loc}&=\sdfrac{2}{p^2}\Pi^{\m\n}_{\a\k}\,p_\b T^{\a\b},\\
	\Pi^{\m\n\k}_{\quad\ \k}\partial_\a t^{\a\b}_{loc}-\Pi^{\m\n\a}_{\quad\ \b}\partial_\a\,t^{\k\b}_{loc}&=\sdfrac{d-2}{p^2}\Pi^{\m\n\k}_{\quad\ \a} p_\b\,T^{\a\b}+\sdfrac{p^\b}{p^2}\Pi^{\m\n\k}_{\quad\ \a}\partial_\b (p_\r T^{\a\r})-\sdfrac{p^\k}{p^2}\Pi^{\m\n\a}_{\quad\ \b}\partial_\a(p_\r T^{\b\r}),\\
	\Pi^{\m\n}_{\a\b}\partial^2 t^{\a\b}_{loc}&=\sdfrac{4}{p^2}\Pi^{\m\n\a}_{\quad\ \b}\partial_\a(p_\r\,T^{\r\b}),\\
	p^\g\Pi^{\m\n}_{\a\b}\partial_\g\partial_\k\,t^{\a\b}_{loc}&=-\sdfrac{4}{p^2}\Pi^{\m\n}_{\a\k}p_\b\,T^{\a\b}+\sdfrac{2}{p^2}\Pi^{\m\n}_{\a\k}\,p^\b\partial_\b(p_\r T^{\a\r}),\\
\end{align}
\section{Properties of triple-K integrals \label{AppendixJ}}

The modified Bessel function of the second kind is defined by
\begin{equation}
	K_\n(x)=\sdfrac{\pi}{2\sin(\n x)}[I_{-\n}(x)-I_{\n}(x)],\ \ \n\in\mathbb Z.
\end{equation}
If $\n=\frac{1}{2}+n$, for $n$ integer, the Bessel function reduced to elementary functions
\begin{equation}
	K_\n(x)=\sqrt{\sdfrac{\pi}{2}}\,\sdfrac{e^{-x}}{\sqrt{x}}\,\sum_{j=0}^{\lfloor\,|\n|-1/2\rfloor}\ \frac{(|\n|-1/2+j)!}{j!(|\n|-1/2-j)!}\sdfrac{1}{(2x)^j},\ \ \n+1/2\in\mathbb{Z},
\end{equation}
where we have used the floor function. In particular
\begin{eqnarray}
	&\hspace{-2cm}K_{\frac{1}{2}}(x)=\sqrt{\sdfrac{\pi}{2}}\sdfrac{e^{-x}}{\sqrt{x}},\quad &K_{\frac{3}{2}}(x)=\sqrt{\sdfrac{\pi}{2}}\sdfrac{e^{-x}}{\sqrt{x^3}}(1+x),\notag\\
	&K_{\frac{5}{2}}(x)=\sqrt{\sdfrac{\pi}{2}}\sdfrac{e^{-x}}{\sqrt{x^5}},(x^2+3x+3),\quad
	&K_{\frac{7}{2}}(x)=\sqrt{\sdfrac{\pi}{2}}\sdfrac{e^{-x}}{\sqrt{x^5}}(x^3+6x^3+15x+5),
\end{eqnarray}
Using this expressions the triple-K integrals can be calculate in a very simple way. We can derive a useful expression in order regularise the triple-K integral in the region of non convergence. Considering $\b_i$ half-integers the triple-K integral read
\begin{align}
	I_{\a\{\b_1\,\b_2,\b_3\}}&=\int_0^\infty\,dx\,x^\a\,p_1^{\b_1}\,p_2^{\b_2}\,p_3^{\b_3}\,K_{\b_1}(p_1x)\,K_{\b_2}(p_2x)\,K_{\b_3}(p_3x)\notag\\[1.5ex]
	&\hspace{-1cm}=\sum_{k_1=0}^{|\b_1|-\frac{1}{2}}\ \,\sum_{k_2=0}^{|\b_2|-\frac{1}{2}}\ \,\sum_{k_3=0}^{|\b_3|-\frac{1}{2}}\ p_1^{\b_1-\frac{1}{2}-k_1}\,p_2^{\b_2-\frac{1}{2}-k_2}\,p_3^{\b_3-\frac{1}{2}-k_3}\,C_{k_1}(\b_1)\,C_{k_2}(\b_2)\,C_{k_3}(\b_3)\,\int_0^\infty\,dx\,x^{\a-k_t-\frac{3}{2}}\,e^{-p_t\,x}\notag\\
	&\hspace{-1cm}=\sum_{k_1=0}^{|\b_1|-\frac{1}{2}}\ \,\sum_{k_2=0}^{|\b_2|-\frac{1}{2}}\ \,\sum_{k_3=0}^{|\b_3|-\frac{1}{2}}\ p_1^{\b_1-\frac{1}{2}-k_1}\,p_2^{\b_2-\frac{1}{2}-k_2}\,p_3^{\b_3-\frac{1}{2}-k_3}\,p_t^{k_t-\a-\frac{1}{2}}\,C_{k_1}(\b_1)\,C_{k_2}(\b_2)\,C_{k_3}(\b_3)\,\G\left(\a-k_t-\sdfrac{1}{2}\right)\label{halfinteg}
\end{align}
where $k_t=k_1+k_2+k_3$ and $p_t=p_1+p_2+p_3$ and we have define $C_{k_i}(\b_i)$ as
\begin{equation}
	C_{k_i}(\b_i)\equiv \sqrt{\frac{\pi}{2^{2k_i+1}}}\,\frac{\left(|\b_i|-1/2+k_i\right)\,!}{k_i\,!\,\left(|\b_i|-1/2-k_i\right)\,!},
\end{equation}
and we have used the definition of the gamma function in order to write the integral
\begin{equation}
	\int_0^\infty\,dx\,x^{\a-k_t-\frac{3}{2}}\,e^{-p_t\,x}=p_t^{k_t-\a+\frac{1}{2}}\int_0^\infty\,dy\,y^{\a-k_t-\frac{3}{2}}\,e^{-\,y}=p_t^{k_t-\a+\frac{1}{2}}\,\G\left(\a-k_t-\frac{1}{2}\right)
\end{equation}
Using \eqref{halfinteg} we can calculate for instance the integrals
\begin{align}
	I_{\frac{9}{2}\left\{\frac{3}{2},\frac{3}{2},-\frac{1}{2}\right\}}&=\left(\sdfrac{\pi}{2}\right)^{3/2}\sdfrac{3(p_1^2+p_2^2)+p_3^2+12p_1\,p_2+4p_3(p_1+p_2)}{p_3(p_1+p_2+p_3)^4}\\[1.2ex]
	I_{\frac{7}{2}\left\{\frac{3}{2},\frac{3}{2},\frac{1}{2}\right\}}&=\left(\sdfrac{\pi}{2}\right)^{3/2}\sdfrac{2(p_1^2+p_2^2)+p_3^2+6 p_1\,p_2+3p_3(p_1+p_2)}{p_3(p_1+p_2+p_3)^3}
\end{align}
and any integrals with half-integer $\b_j$, $j=1,2,3$.

We now analyse the basic properties of the triple-K integrals. We have
\begin{equation}
	\begin{split}
		\sdfrac{\partial}{\partial p_n}I_{\a\{\b_j\}}&=-p_n\,I_{\a+1\{\b_j-\d_{jn}\}},\\
		I_{\a\{\b_j+\d_{jn}\}}&=p_n^2\,I_{\a\{\b_j-\d_{jn}\}}+2\b_n\,I_{\a-1\{\b_j\}},\\
		I_{\a\{\b_1\b_2,-\b_3\}}&=p_3^{-2\b_3}I_{\a\{\b_1\b_2,\b_3\}}
	\end{split}\label{identityBess}
\end{equation}
for any $n=1,2,3$, as follows from the basic Bessel function relations
\begin{align}
	\sdfrac{\partial}{\partial a}[a^\n K_\n(ax)]&=-x\,a^\n K_{\n-1}(ax),\\
	K_{\n-1}(x)+\sdfrac{2\n}{x}K_\n(x)&=K_{\n+1}(x),\\
	K_{\n-1}(p\,x)+\sdfrac{2\n}{p\,x}K_\n(p\,x)&=K_{\n+1}(p\,x),\\
	K_{-\n}(x)&=K_{\n}(x),\\
	\sdfrac{\partial}{\partial x}[a^\n K_\n(ax)]&=-\sdfrac{1}{2}a^{\n+1} (K_{\n-1}(ax)+K_{\n+1}(ax))\notag\\[0.8ex]
	&=-a^{\n+1}K_{\n+1}(ax)+\sdfrac{\n\,a^\n}{x}\,K_{\n}(ax)
\end{align}

The properties of the reduced triple-K integral is directly obtained from \eqref{identityBess}, in fact noticing that its definition given in \appref{solution} 
\begin{equation}
	J_{N\{k_j\}}=I_{\frac{d}{2}-1+N\{\D_j-\frac{d}{2}+k_j\}},
\end{equation}
is just a redefinition of the indices, we can obtain similar equations to \eqref{identityBess}. It is not difficult to show that
\begin{align}
	\sdfrac{\partial}{\partial p_n}\,J_{N\{k_j\}}&=-p_n\,J_{N+1\{k_j-\d_{jn}\}}\label{1dJ}\\
	\,J_{N\{k_j+\d_{jn}\}}&=p^2_n\ J_{N\{k_j-\d_{jn}\}}+2\left(\D_n-\sdfrac{d}{2}+k_n\right)\ J_{N-1\{k_j\}}\\
	\,J_{N+2\{k_j\}}&=p^2_n\ J_{N+2\{k_j-2\d_{jn}\}}+2\left(\D_n-\sdfrac{d}{2}+k_n-1\right)\ J_{N+1\{k_j-\d_{jn}\}}\\
	\sdfrac{\partial^2}{\partial p_n^2}\,J_{N\{k_j\}}&=J_{N+2\{k_j\}}-2\left(\D_n-\sdfrac{d}{2}+k_n-\sdfrac{1}{2}\right)\ J_{N+1\{k_j-\d_{jn}\}},\\
	K_n\,J_{N\{k_j\}}&\equiv\left(\sdfrac{\partial^2}{\partial p_n^2}+\sdfrac{(d+1-2\D_n)}{p_n}\sdfrac{\partial}{\partial p_n}\right)J_{N\{k_j\}}=J_{N+2\{k_j\}}-2\,k_n\,J_{N+1\{k_j-\d_{jn}\}}\label{Fund}.
\end{align}
\section{Fuchsian solutions of the primary CWI's for TTT correlator}
\label{fuchsTTT}
\subsection{$A_4$ solution}
Under the exchange of two momenta, $A_2(p_2\leftrightarrow p_3)$ becomes
\begin{align}
	A_2(p_2\leftrightarrow p_3)&= p_3^{d - 4}\sum_{a b} x^a y^{\frac{d}{2} - 2-a-b}\bigg[c^{(2)}(a,b)\,F_4\left(\alpha +2, \beta+2; \gamma, \gamma'; \sdfrac{x}{y}, \sdfrac{1}{y}\right)\notag\\
	&\hspace{-0.2cm}+ \frac{2\,c^{(1)}(a,b)}{\big(\b+2\big)}F_4(\alpha +3, \beta+2; \gamma, \gamma'; \sdfrac{x}{y}, \sdfrac{1}{y}\bigg)\bigg]. 
\end{align}
In order to solve the equation \eqref{eqA4} we will be needing the transformation property of $F_4$ given by \eqref{transfF4}.
Once plugged into the explicit expression on $A_4(p_2\leftrightarrow p_3)$, and separating its 4 indicial components $(a_i, b_i)$ we obtain
\begin{align}
	A_2(p_2\leftrightarrow p_3)&=p_3^{d-4}\Bigg\{\ c^{(2)}\left(0,\frac{d}{2}\right)\frac{(d-2)}{(d+2)}F_4\left(2-\frac{d}{2},2,1-\frac{d}{2},1-\frac{d}{2},x,y\right)\notag\\
	&\hspace{4cm}+\ c^{(1)}\left(0,\frac{d}{2}\right)\,\frac{d(d-2)}{(d+2)(d+4)}\,F_4\left(2-\frac{d}{2},2,1-\frac{d}{2},-\frac{d}{2},x,y\right)\Bigg\}\notag\\
	&\hspace{-2cm}+p_3^{d-4}x^{d/2}\Bigg\{\ c^{(2)}\left(\frac{d}{2},0\right)\,F_4\left(\frac{d}{2}+2,2,\frac{d}{2}+1,1-\frac{d}{2},x,y\right)+\ c^{(1)}\left(\frac{d}{2},0\right)\,\frac{d}{(d+4)}\,F_4\left(\frac{d}{2}+2,2,\frac{d}{2}+1,-\frac{d}{2},x,y\right)\Bigg\}\notag\\
	&\hspace{-2cm}+p_3^{d-4}y^{d/2}\Bigg\{\ c^{(2)}\left(0,0\right)\frac{(d+2)}{(d-2)}F_4\left(2,\frac{d}{2}+2,1-\frac{d}{2},\frac{d}{2}+1,x,y\right)\notag\\
	&\hspace{4cm}+ c^{(1)}\left(0,0\right)\,\frac{4(d+4)y}{(d-4)(d-2)}F_4\left(3,\frac{d}{2}+3,1-\frac{d}{2},\frac{d}{2}+2,x,y\right)\Bigg\}\notag\\
	&\hspace{-2cm}+p_3^{d-4}y^{d/2}x^{d/2}(-1)^d\Bigg\{-\ c^{(2)}\left(\frac{d}{2},\frac{d}{2}\right)F_4\left(d+2,\frac{d}{2}+2,\frac{d}{2}+1,\frac{d}{2}+1,x,y\right)\,\notag\\
	&\hspace{4cm}+\ 4c^{(1)}\left(\frac{d}{2},\frac{d}{2}\right)\frac{y}{d+2}F_4\left(d+3,\frac{d}{2}+3,\frac{d}{2}+1,\frac{d}{2}+2,x,y\right)\Bigg\}
\end{align}
and with a similar one for $A_4(p_1\leftrightarrow p_3)$, that we omit. At this stage, using the property of $F_4$ given in \eqref{exs} we can rearrange the expression of $A_2(p_2\leftrightarrow p_3)$ as
\begin{align}
	A_2(p_2\leftrightarrow p_3)&=p_3^{d-4}\Bigg\{\ c^{(2)}\left(0,\frac{d}{2}\right)\frac{(d-2)}{(d+2)}F_4\left(2-\frac{d}{2},2,1-\frac{d}{2},1-\frac{d}{2},x,y\right)\notag\\
	&\hspace{-1.5cm}+\ c^{(1)}\left(0,\frac{d}{2}\right)\,\frac{d(d-2)}{(d+2)(d+4)}\,F_4\left(2-\frac{d}{2},2,1-\frac{d}{2},-\frac{d}{2},x,y\right)\Bigg\}+p_3^{d-4}x^{d/2}\Bigg\{c^{(2)}\left(\frac{d}{2},0\right)\notag\\
	&\hspace{-1.5cm}\times\,F_4\left(\frac{d}{2}+2,2,\frac{d}{2}+1,1-\frac{d}{2},x,y\right)+\ c^{(1)}\left(\frac{d}{2},0\right)\,\frac{d}{(d+4)}\,F_4\left(\frac{d}{2}+2,2,\frac{d}{2}+1,-\frac{d}{2},x,y\right)\Bigg\}\notag\\
	&\hspace{1.5cm}+p_3^{d-4}y^{d/2}\Bigg\{\ c^{(2)}\left(0,0\right)\frac{(d+2)}{(d-2)}F_4\left(2,\frac{d}{2}+2,1-\frac{d}{2},\frac{d}{2}+1,x,y\right)\notag
\end{align}
\begin{align}
	&\hspace{1.5cm}+c^{(1)}\left(0,0\right)\,\frac{d(d+2)}{(d-4)(d-2)} \Bigg[F_4\left(2,\frac{d}{2}+2,1-\frac{d}{2},\frac{d}{2},x,y\right)-F_4\left(2,\frac{d}{2}+2,1-\frac{d}{2},\frac{d}{2}+1,x,y\right)\Bigg]\Bigg\}\notag\\
	&\hspace{2cm}+p_3^{d-4}y^{d/2}x^{d/2}(-1)^d\Bigg\{-\ c^{(2)}\left(\frac{d}{2},\frac{d}{2}\right)F_4\left(d+2,\frac{d}{2}+2,\frac{d}{2}+1,\frac{d}{2}+1,x,y\right)\,\notag\\
	&\hspace{1cm}+\ c^{(1)}\left(\frac{d}{2},\frac{d}{2}\right)\frac{2d}{(d+4)(d+2)}\Bigg[F_4\left(d+2,\frac{d}{2}+2,\frac{d}{2}+1,\frac{d}{2},x,y\right)-F_4\left(d+2,\frac{d}{2}+2,\frac{d}{2}+1,\frac{d}{2}+1,x,y\right)\Bigg]\Bigg\}.
\end{align}
A similar expression can be derived for $A_2(p_1\leftrightarrow p_3)$.
The  constants appearing in the solution for the form factor $A_4$ can be related as
\begin{align}
	&\begin{aligned}
		c_1^{(4)}\left(0,0\right)&=\frac{2}{(d+2)}c^{(2)}\left(0,\frac{d}{2}\right)+\frac{2}{(d+2)}c_3^{(4)}\left(0,0\right)\\
		c_1^{(4)}\left(0,\frac{d}{2}\right)&=-c^{(2)}\left(0,\frac{d}{2}\right)-\frac{(d+2)}{d}c_3^{(4)}\left(\frac{d}{2},0\right)\\
		c_1^{(4)}\left(\frac{d}{2},0\right)&=c_1^{(4)}\left(0,\frac{d}{2}\right)\\
		c_1^{(4)}\left(\frac{d}{2},\frac{d}{2}\right)&=\frac{2\sec\left(\frac{\p\,d}{2}\right)\left[\Gamma\left(1-\frac{d}{2}\right)\right]^2}{(d^2+6d+8)\Gamma(-d-1)}c^{(1)}\left(0,\frac{d}{2}\right)+\frac{2(-1)^d}{d+2}c^{(2)}\left(\frac{d}{2},\frac{d}{2}\right)\notag\\
		&-\frac{2(d+1)}{d}c_3^{(4)}\left(\frac{d}{2},\frac{d}{2}\right)-\frac{2(-1)^{d/2}\Gamma\left(1-\frac{d}{2}\right)\Gamma\left(-\frac{d}{2}\right)}{(d+2)\Gamma(-d-1)}c^{(2)}\left(0,\frac{d}{2}\right)\\[2ex]
	\end{aligned}\notag\\
\end{align}
\subsection{Relating the constants in the $A_5$ solution}
\label{conditionTTT}
The constants in the expression of $A_5$ are fixed as follows
\begin{align}
	\begin{aligned}[c]
		c^{(2)}\left(\frac{d}{2},0\right)&=c^{(2)}\left(0,\frac{d}{2}\right)\equiv C_2\\
		c^{(2)}\left(0,0\right)&=\frac{(d-2)}{(d+2)}\,C_2\\
		c^{(2)}\left(\frac{d}{2},\frac{d}{2}\right)&=\frac{\Gamma\left(-\frac{d}{2}\right)\,\Gamma\left(d+2\right)}{\Gamma\left(\frac{d}{2}\right)}C_2\\
	\end{aligned}
	\hspace{2cm}
	\begin{aligned}[c]
		c^{(3)}\left(\frac{d}{2},0\right)&=c^{(3)}\left(0,\frac{d}{2}\right)\equiv C_3\\[0.8ex]
		c^{(3)}\left(0,0\right)&=-C_3\\[1ex]
		c^{(3)}\left(\frac{d}{2},\frac{d}{2}\right)&=\frac{\Gamma\left(-\frac{d}{2}\right)\,\Gamma\left(d+1\right)}{\Gamma\left(\frac{d}{2}\right)}C_3\\
	\end{aligned}
\end{align}
\begin{align}
	\begin{aligned}[c]
		c^{(5)}\left(\frac{d}{2},0\right)&=c^{(5)}\left(0,\frac{d}{2}\right)=C_5=c^{(5)}\left(0,0\right)\\
		c^{(5)}\left(\frac{d}{2},\frac{d}{2}\right)&=-\frac{d^2\Gamma\left(-\frac{d}{2}-1\right)\Gamma(d+1)}{8\Gamma\left(\frac{d}{2}\right)}C_2+\frac{\Gamma\left(-\frac{d}{2}\right)\,\Gamma(d+1)}{2\Gamma\left(\frac{d}{2}+1\right)}C_5
	\end{aligned}
\end{align}
\begin{align}
	\begin{aligned}[c]
		c_1^{(5)}\left(\frac{d}{2},0\right)&=c_1^{(5)}\left(0,\frac{d}{2}\right)=-\frac{d^2}{2(d+2)(d+4)}C_1=c_1^{(5)}\left(0,0\right)\\
		c_1^{(5)}\left(\frac{d}{2},\frac{d}{2}\right)&=-\frac{d\,\Gamma\left(1-\frac{d}{2}\right)\Gamma(d+1)}{(d+2)(d+4)\Gamma\left(\frac{d}{2}\right)}C_1\\
		c_2^{(5)}\left(0,0\right)&=c_4^{(5)}\left(0,0\right)=-\frac{d}{2(d+2)}C_2\\
		c_2^{(5)}\left(\frac{d}{2},0\right)&=c_4^{(5)}\left(0,\frac{d}{2}\right)=\frac{d^2}{2(d+2)(d+4)}C_1-\frac{d}{d+2}C_2\\
		c_2^{(5)}\left(0,\frac{d}{2}\right)&=c_4^{(5)}\left(\frac{d}{2},0\right)=\frac{d}{d+2}C_2\notag\\
		c_2^{(5)}\left(\frac{d}{2},\frac{d}{2}\right)&=c_4^{(5)}\left(\frac{d}{2},\frac{d}{2}\right)=\frac{d\,\Gamma\left(-\frac{d}{2}-1\right)\Gamma(d+2)}{4(d+1)\Gamma\left(\frac{d}{2}\right)}C_2-\frac{d^2\,\Gamma\left(d+2\right)\Gamma\left(-\frac{d}{2}-1\right)}{4(d+1)(d+4)\,\Gamma\left(\frac{d}{2}\right)}C_1\\
		c^{(5)}_3\left(0,0\right)&=0\\
		c^{(5)}_3\left(0,\frac{d}{2}\right)&=c^{(5)}_3\left(\frac{d}{2},0\right)=-C_2\frac{d}{d+2}\\
		c^{(5)}_3\left(\frac{d}{2},\frac{d}{2}\right)&=-\frac{\pi\,d^2\csc\left(\frac{\pi\,d}{2}\right)\Gamma(d+1)}{8\Gamma\left(\frac{d}{2}+3\right)\Gamma\left(\frac{d}{2}\right)}C_1-\frac{\pi\,d(d+4)\csc\left(\frac{\pi\,d}{2}\right)\Gamma(d+1)}{4\Gamma\left(\frac{d}{2}+3\right)\Gamma\left(\frac{d}{2}\right)}C_2\\
		c_5^{(5)}\left(0,0\right)&=c_6^{(5)}\left(0,0\right)=0\\
		c_5^{(5)}\left(\frac{d}{2},0\right)&=c_6^{(5)}\left(0,\frac{d}{2}\right)=0\notag\\
		c_5^{(5)}\left(0,\frac{d}{2}\right)&=c_6^{(5)}\left(\frac{d}{2},0\right)=-\frac{d^2}{2(d+2)}C_2\notag\\
		c_5^{(5)}\left(\frac{d}{2},\frac{d}{2}\right)&=c_6^{(5)}\left(\frac{d}{2},\frac{d}{2}\right)=\frac{d^2\,\Gamma\left(-\frac{d}{2}-1\right)\,\Gamma(d+1)}{8\,\Gamma\left(\frac{d}{2}\right)}C_2
		\\
		c_7^{(5)}\left(0,0\right)&=-\frac{d^2}{2(d+2)}C_2\notag\\
		c_7^{(5)}\left(\frac{d}{2},0\right)&=c_7^{(5)}\left(0,\frac{d}{2}\right)=\frac{d^2}{2(d+2)}C_2\notag\\
		c_7^{(5)}\left(\frac{d}{2},\frac{d}{2}\right)&=-\frac{d^2\,\Gamma\left(-\frac{d}{2}-1\right)\,\Gamma(d+1)}{8\,\Gamma\left(\frac{d}{2}\right)}C_2
	\end{aligned}
\end{align}
\section{Metric variations of the counterterms}\label{Mvc}
In this appendix we list the metric variations of the counterterms. In particular we give them directly in the momentum using the definition of the Fourier transform in \eqref{count}. The metric variation are consider in the flat space-time limit and the first variation of the square of the metric, Riemann, Ricci and the scalar curvature are given as
\begin{align}
	\big[\sqrt{-g}\big]^{\m_i\n_i}&=\sdfrac{1}{2}\d^{\m_i\n_i}\\
	\big[R_{\m\a\n\b}\big]^{\m_i\n_i}(p_i)&=\frac{1}{2}\,\bigg(\d_\a^{(\m_i}\d^{\n_i)}_\b\,p_{i\m}\,p_{i\n}
	+\d_\m^{(\m_i}\d^{\n_i)}_\n\,p_{i\a}\,p_{i\b}-\d_\m^{(\m_i}\d^{\n_i)}_\b\,p_{i\a}\,p_{i\n}-\d_\a^{(\m_i}\d^{\n_i)}_\n\,p_{i\m}\,p_{i\b}\bigg)\\
	\big[R^{\m\a\n\b}\big]^{\m_i\n_i}(p_i)&=\frac{1}{2}\,\bigg(\d^{\a(\m_i}\d^{\n_i)\b}\,p_i^\m\,p_i^\n
	+\d^{\m(\m_i}\d^{\n_i)\n}\,p_i^\a\,p_i^\b-\d^{\m(\m_i}\d^{\n_i)\b}\,p_i^\a\,p_i^\n-\d^{\a(\m_i}\d^{\n_i)\n}\,p_i^\m\,p_i^\b\bigg)\\
	\big[R_{\m\n}\big]^{\m_i\n_i}(p_i)&=\frac{1}{2}\,\bigg(\d_\m^{(\m_i}\d^{\n_i)}_\n\,p_{i}^2
	+\d^{\m_i\n_i}\,p_{i\m}\,p_{i\n}-p_i^{(\m_i}\d^{\n_i)}_\m\,p_{i\n}-p_i^{(\m_i}\d^{\n_i)}_\n\,p_{i\m}\bigg)\\
	\big[R^{\m\n}\big]^{\m_i\n_i}(p_i)&=\frac{1}{2}\,\bigg(\d^{\m(\m_i}\d^{\n_i)\n}\,p_{i}^2
	+\d^{\m_i\n_i}\,p_i^\m\,p_i^\n-p_i^{(\m_i}\d^{\n_i)\m}\,p_i^\n-p_i^{(\m_i}\d^{\n_i)\n}\,p_i^\m\bigg)\\
	\big[R\big]^{\m_i\n_i}(p_i)&=\bigg(\d^{\m_i\n_i}\,p_{i}^2-p_i^{(\m_i}p_i^{\n_i)}\bigg)\\
	\big[\square R\big]^{\m_i\n_i}(p_i)&=p_i^2\,\bigg(p_i^{(\m_i}p_i^{\n_i)}-\d^{\m_i\n_i}\,p_{i}^2\bigg)
\end{align}
and the second variations of these object can be calculated in order to obtain
\begin{align}
	\big[R^{\,\b}_{\ \ \n\r\s}\big]^{\m_1\n_1\m_2\n_2}(p_1,p_2)&=\Big[-\frac{1}{2}\,\tilde F^{\m_1\n_1\b\epsilon}p_{1\s}\big(\tilde F^{\m_2\n_2}_{\e\n}p_{2\r}+\tilde F^{\m_2\n_2}_{\e\r}p_{2\n}-\tilde F^{\m_2\n_2}_{\n\r}p_{2\e}\big)\notag\\
	&\hspace{-1.5cm}-\frac{1}{2}\big(\tilde C^{\m_1\n_1\m_2\n_2\b}_{\hspace{1.4cm}\r}\ p_{2\n}-\tilde F^{\m_1\n_1\b\e}\,\tilde F^{\m_2\n_2}_{\n\r}\,p_{2\e}\big)p_{2\s}\notag\\
	&\hspace{-1.5cm}-\frac{1}{4}\big(\tilde F^{\m_1\n_1}_{\a\n}p_{1\s}+\tilde F^{\m_1\n_1}_{\a\s}p_{1\n}-\tilde F^{\m_1\n_1}_{\s\n}p_{1\a}\big)\big(\tilde F^{\m_2\n_2\b\a}\,p_{2\r}+\tilde F^{\m_2\n_2\b}_{\qquad\r}\,p_{2}^\a-\tilde F^{\m_2\n_2\a}_{\qquad\r}\,p_{2}^{\b}\big)\Big]-(\s\leftrightarrow \r)\\
	\big[R_{\m\n\r\s}\big]^{\m_1\n_1\m_2\n_2}(p_1,p_2)&=\d^{(\m_1}_\m\d^{\n_1)}_\b\big[R^{\,\b}_{\ \ \n\r\s}\big]^{\m_2\n_2}(p_2)+\d_{\m\b}\big[R^{\,\b}_{\ \ \n\r\s}\big]^{\m_1\n_1\m_2\n_2}(p_1,p_2)
\end{align}
\begin{align}
	\big[R^{\m\n\r\s}\big]^{\m_1\n_1\m_2\n_2}(p_1,p_2)&=\d^{\a\n}\d^{\b\r}\d^{\s\g}\big[R^{\,\m}_{\ \ \a\b\g}\big]^{\m_1\n_1\m_2\n_2}(p_1,p_2)\notag\\
	&\hspace{-1.5cm}-\big(\d^{\a(\m_1}\d^{\n_1)\n}\d^{\b\r}\d^{\s\g}+\d^{\a\n}\d^{\b(\m_1}\d^{\n_1)\r}\d^{\s\g}+\d^{\a\n}\d^{\b\r}\d^{\s(\m_1}\d^{\n_1)\g}\big)\big[R^{\,\b}_{\ \ \n\r\s}\big]^{\m_2\n_2}(p_2)
\end{align}
\begin{align}
	\big[R_{\n\s}\big]^{\m_1\n_1\m_2\n_2}(p_1,p_2)&=-\frac{1}{2}\tilde F^{\m_1\n_1\m_2\n_2}\left(p_{1\s}p_{2\n}-\sdfrac{1}{2}p_{1\n}p_{2\s}+p_{2\n}p_{2\s}\right)-\frac{1}{4}\d^{\m_2\n_2}\left(\tilde F^{\m_1\n_1}_{\a\n}\,p_{1\s}+\tilde F^{\m_1\n_1}_{\a\s}\,p_{1\n}\right)\,p_2^\a\notag\\
	&\hspace{-1.5cm}+\frac{1}{2}\big(\tilde C^{\m_1\n_1\m_2\n_2\e}_{\hspace{1.4cm}\n}\,p_{2\s}+\tilde C^{\m_1\n_1\m_2\n_2\e}_{\hspace{1.4cm}\s}\,p_{2\n}\big)\,(p_1+p_2)_{\e}+\frac{1}{2}F^{\m_2\n_2}_{\a\s}\tilde F^{\m_1\n_1}_{\b\n}\,p_1^\a\,p_2^\b\notag\\
	&\hspace{-1.5cm}-\frac{1}{2}\tilde F^{\m_2\n_2}_{\n\s}\,\tilde F^{\m_1\n_1\a\b}(p_1+p_2)_\a\,p_{2\b}-\frac{1}{2}\left(\tilde C^{\m_1\n_1\m_2\n_2}_{\hspace{1.25cm}\n\s}-\frac{1}{2}\d^{\m_2\n_2}\,\tilde F^{\m_1\n_1}_{\n\s}\right)\,p_1\cdot p_2
\end{align}
\begin{align}
	\big[R^{\n\s}\big]^{\m_1\n_1\m_2\n_2}(p_1,p_2)&=\d^{\n\a}\d^{\s\b}\big[R_{\a\b}\big]^{\m_1\n_1\m_2\n_2}(p_1,p_2)-\big(\d^{\n(\m_1}\d^{\n_1)\a}\d^{\s\b}+\d^{\n\a}\d^{\s(\m_1}\d^{\n_1)\b}\big)\big[R_{\a\b}\big]^{\m_2\n_2}(p_2)\\
	\big[R\big]^{\m_1\n_1\m_2\n_2}(p_1,p_2)&=-\left(p_2^2+\sdfrac{1}{4}p_1\cdot p_2\right)\,\tilde F^{\m_1\n_1\m_2\n_2}+\frac{1}{4}\,A^{\m_1\n_1\m_2\n_2}\,p_1\cdot p_2\notag\\
	&\hspace{-1.5cm}+\tilde C^{\m_1\n_1\m_2\n_2\a\b}\,(p_{1\a}+2p_{2\a})p_{2\b}-\d^{\m_2\n_2}\tilde F^{\m_1\n_1\a\b}\,(p_{1\a}+p_{2\a})p_{2\b}+\sdfrac{1}{2}\tilde C^{\m_2\n_2\m_1\n_1\a\b}\,p_{1\a}p_{2\b}
\end{align}
\begin{align}
	\big[\square R\big]^{\m_1\n_1\m_2\n_2}(p_1,p_2)&=\tilde F^{\m_1\n_1\m_2\n_2}\,\bigg[p_2^2(p_1+p_2)^2+\sdfrac{3}{2}(p_2^2+p_1\cdot p_2)\bigg]+\sdfrac{1}{2}\d^{\m_1\n_1}\tilde F^{\m_2\n_2\a\b}(p_1\cdot p_2)\,p_{2\a}p_{2\b}\notag\\
	&\hspace{-1.5cm}-\sdfrac{1}{2}\d^{\m_1\n_1}\d^{\m_2\n_2}(p_1\cdot p_2)\bigg[(p_1+p_2)^2-p_1\cdot p_2\bigg]+\d^{\m_2\n_2}F^{\m_1\n_1\a\b}p_{2\a}(p_1+p_2)_\b\bigg[(p_1+p_2)^2+p_2^2\bigg]\notag\\
	&\hspace{-1.5cm}-\tilde F^{\m_2\n_2\a\b}p_{2\a}p_{2\b}\,\tilde F^{\m_1\n_1\Gamma\d}p_{2\g}(p_1+p_2)_\d-(p_1+p_2)^2\,\tilde C^{\m_1\n_1\m_2\n_2\a\b}\bigg[2p_{2\a}p_{2\b}+p_{1\a}p_{2\b}+\sdfrac{1}{2}p_{2\a}p_{1\b}\bigg]
\end{align}
remembering that the order of indices of variation is important, because these second variation are not symmetrized. 

Using these relations it is possible to find the third variation of the counterterms. For instance the Weyl tensor counterterm is expressed as
\begin{align}
	\big[\sqrt{-g}\,C^2\big]^{\m_1\n_1\m_2\n_2\m_3\n_3}(p_1,p_2,p_3)&=\Bigg\{[\sqrt{-g}]^{\m_1\n_1}\bigg([R_{abcd}]^{\m_2\n_2}(p_2)[R^{abcd}]^{\m_3\n_3}(p_3)-\sdfrac{4}{d-2}[R_{ab}]^{\m_2\n_2}(p_2)[R^{ab}]^{\m_3\n_3}(p_3)\notag\\
	&\hspace{-3cm}+\sdfrac{2}{(d-2)(d-1)}[R]^{\m_2\n_2}(p_2)[R]^{\m_3\n_3}(p_3)\bigg)+\bigg([R_{abcd}]^{\m_1\n_1\m_2\n_2}(p_1,p_2)[R^{abcd}]^{\m_3\n_3}(p_3)\notag\\
	&\hspace{-3cm}+[R_{abcd}]^{\m_2\n_2}(p_2)[R^{abcd}]^{\m_1\n_1\m_3\n_3}(p_1,p_3)-\sdfrac{4}{d-2}[R_{ab}]^{\m_1\n_1\m_2\n_2}(p_1,p_2)[R^{ab}]^{\m_3\n_3}(p_3)\notag\\
	&\hspace{-3cm}-\sdfrac{4}{d-2}[R_{ab}]^{\m_2\n_2}(p_2)[R^{ab}]^{\m_1\n_1\m_3\n_3}(p_1,p_3)+\sdfrac{2}{(d-2)(d-1)}[R]^{\m_1\n_1\m_2\n_2}(p_1,p_2)[R]^{\m_3\n_3}(p_3)\notag\\
	&\hspace{-3cm}+\sdfrac{2}{(d-2)(d-1)}[R]^{\m_2\n_2}(p_2)[R]^{\m_1\n_1\m_3\n_3}(p_1,p_3)\bigg)\Bigg\}+\text{permutations}
\end{align}
where ``permutation'' indicate all the possible permutations of the indices $(\m_i,\n_i)$. In a same way the Euler density counterterm is given as
\begin{align}
	\big[\sqrt{-g}\,E\big]^{\m_1\n_1\m_2\n_2\m_3\n_3}(p_1,p_2,p_3)&=\Bigg\{[\sqrt{-g}]^{\m_1\n_1}\bigg([R_{abcd}]^{\m_2\n_2}(p_2)[R^{abcd}]^{\m_3\n_3}(p_3)-4[R_{ab}]^{\m_2\n_2}(p_2)[R^{ab}]^{\m_3\n_3}(p_3)\notag\\
	&\hspace{-4.7cm}+[R]^{\m_2\n_2}(p_2)[R]^{\m_3\n_3}(p_3)\bigg)+\bigg([R_{abcd}]^{\m_1\n_1\m_2\n_2}(p_1,p_2)[R^{abcd}]^{\m_3\n_3}(p_3)+[R_{abcd}]^{\m_2\n_2}(p_2)[R^{abcd}]^{\m_1\n_1\m_3\n_3}(p_1,p_3)\notag\\
	&\hspace{-4.7cm}-4[R_{ab}]^{\m_1\n_1\m_2\n_2}(p_1,p_2)[R^{ab}]^{\m_3\n_3}(p_3)-4[R_{ab}]^{\m_2\n_2}(p_2)[R^{ab}]^{\m_1\n_1\m_3\n_3}(p_1,p_3)+[R]^{\m_1\n_1\m_2\n_2}(p_1,p_2)[R]^{\m_3\n_3}(p_3)\notag\\
	&\hspace{-3cm}+[R]^{\m_2\n_2}(p_2)[R]^{\m_1\n_1\m_3\n_3}(p_1,p_3)\bigg)\Bigg\}+\text{permutations}.
\end{align}
\chapter{Vertices}\label{Appendix1}
We show in \figref{Figura2} and \figref{vertices} a list of all the vertices which are needed for the  perturbative expansion of the $TJJ$ and $TTT$ correlators. Here we list their explicit expressions. We use the letter $V$ to indicate the vertex, the subscript denotes the fields involved and the Greek indices are linked to the Lorentz structure of the space-time. Referring to \figref{vertices} and \figref{Figura2}, we take all the graviton momenta incoming as well as the momentum indicated as $k_1$, while $k_2$ is outgoing. In order to simplify the notation, we introduce the tensor components
\begin{align}
	A^{\m_1\n_1\m\n}&\equiv\d^{\mu_1\nu_1}\d^{\m\n}-2\d^{\m(\m_1}\d^{\n_1)\n}\\
	B^{\m_1\n_1\m\n}&\equiv\d^{\mu_1\nu_1}\d^{\m\n}-\d^{\m(\m_1}\d^{\n_1)\n}\\
	C^{\m_1\n_1\m_2\n_2\m\n}&\equiv\d^{\m(\m_1}\d^{\n_1)(\m_2}\d^{\n_2)\n}+\d^{\m(\m_2}\d^{\n_2)(\m_1}\d^{\n_1)\n}\\
	\tilde{C}^{\m_1\n_1\m_2\n_2\m\n}&\equiv\d^{\m(\m_1}\d^{\n_1)(\m_2}\d^{\n_2)\n}\\
	D^{\m_1\n_1\m_2\n_2\m\n}&\equiv\d^{\m_1\n_1}\d^{\m(\m_2}\d^{\n_2)\n}+\d^{\m_2\n_2}\d^{\m(\m_1}\d^{\n_1)\n}\\
	E^{\m_1\n_1\m_2\n_2\m\n}&\equiv\d^{\m_1\n_1}B^{\m_2\n_2\m\n}+C^{\m_1\n_1\m_2\n_2\m\n},\\
	F^{\a_1\a_2\m\n}&\equiv\d^{\a_1[\m}\d^{\n]\a_2}\\
	\tilde{F}^{\a_1\a_2\m\n}&\equiv\d^{\a_1(\n}\d^{\m)\a_2}\\
	\tilde{F}^{\a_1\a_2}_{\m\n}&\equiv\d^{(\a_1}_{\n}\d_{\m}^{\a_2)}\\
	G^{\m_1\n_1\a_1\a_2\m\n}&\equiv\d^{\m[\n}\d^{\a_2](\m_1}\d^{\n_1)\a_1}+\d^{\a_1[\a_2}\d^{\n](\m_1}\d^{\n_1)\m}\\
	H^{\m_1\n_1\m_2\n_1\a_1\a_2\m\n}&\equiv A^{\m_1\n_1\m\a_1}\tilde{F}^{\m_2\n_2\n\a_2}-A^{\m_2\n_2\m\a_1}\tilde{F}^{\m_1\n_1\n\a_2}\\
	I^{\m_1\n_1\m_2\n_2\a_1\a_1\m\n}&\equiv\d^{\m_1\n_1}D^{\m\a_1\n\a_2\m_2\n_2}-\sdfrac{1}{2}\d^{\a_1\m}\d^{\a_2\n}A^{\m_1\n_1\m_2\n_2}
\end{align}
where the round brackets denote symmetrization and the square brackets anti-symmetrization of the corresponding indices
\begin{align}
	\d^{\m(\m_1}\d^{\n_1)\n}&\equiv\sdfrac{1}{2}\bigg(\d^{\m\m_1}\d^{\n_1\n}+\d^{\m\n_1}\d^{\m_1\n}\bigg)\\
	\d^{\m[\m_1}\d^{\n_1]\n}&\equiv\sdfrac{1}{2}\bigg(\d^{\m\m_1}\d^{\n_1\n}-\d^{\m\n_1}\d^{\m_1\n}\bigg).
\end{align}
\begin{figure}[t]
	\centering
	\vspace{-.8cm}
	\raisebox{.14\height}{\subfigure{\includegraphics[scale=0.12]{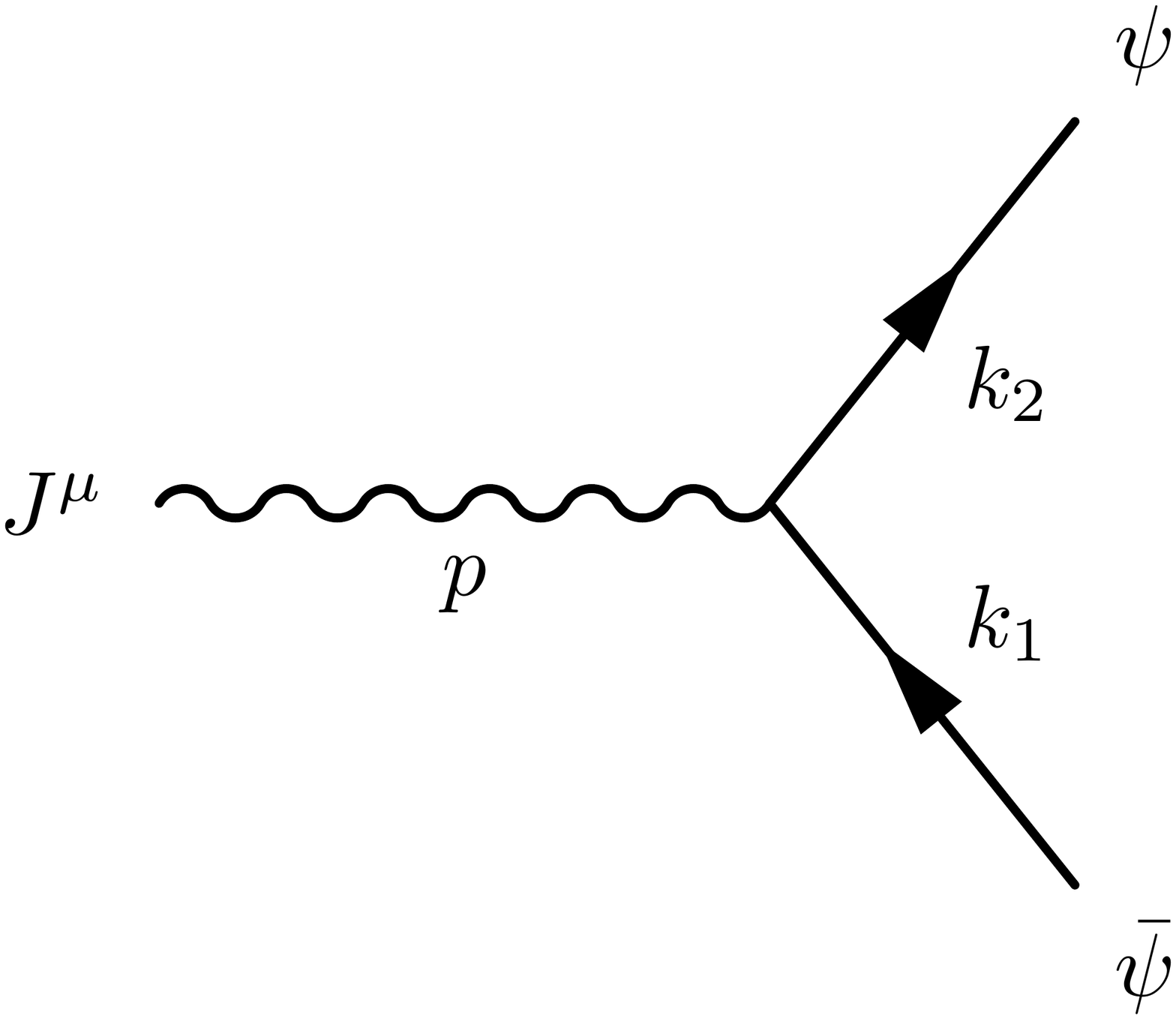}}} \hspace{.3cm}
	\raisebox{.05\height}{\subfigure{\includegraphics[scale=0.12]{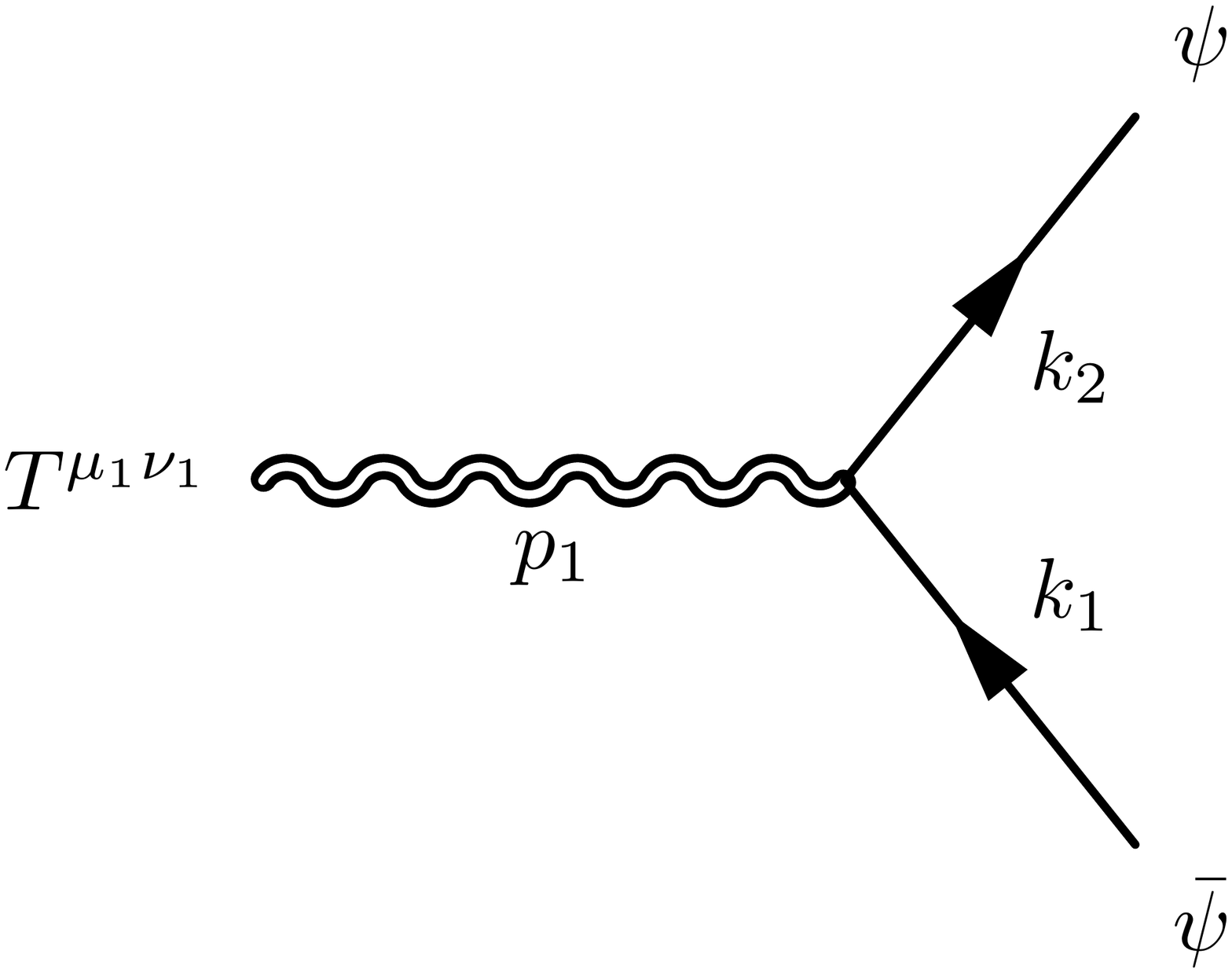}}} \hspace{.3cm}
	\raisebox{.12\height}{\subfigure{\includegraphics[scale=0.12]{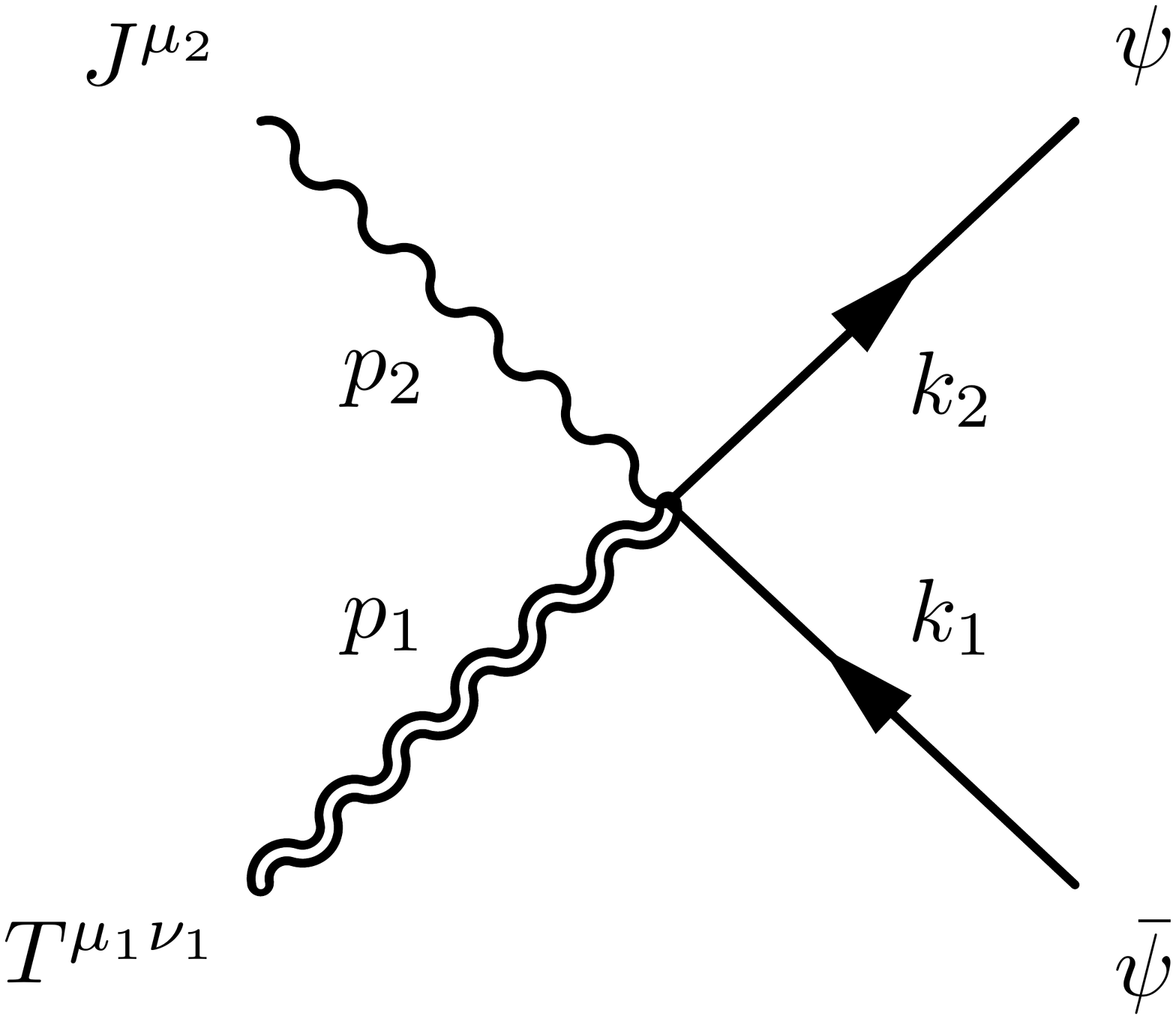}}}\hspace{.3cm}\vspace{-1.6cm}
	\vspace{-.6cm}
	\raisebox{.14\height}{\subfigure{\includegraphics[scale=0.12]{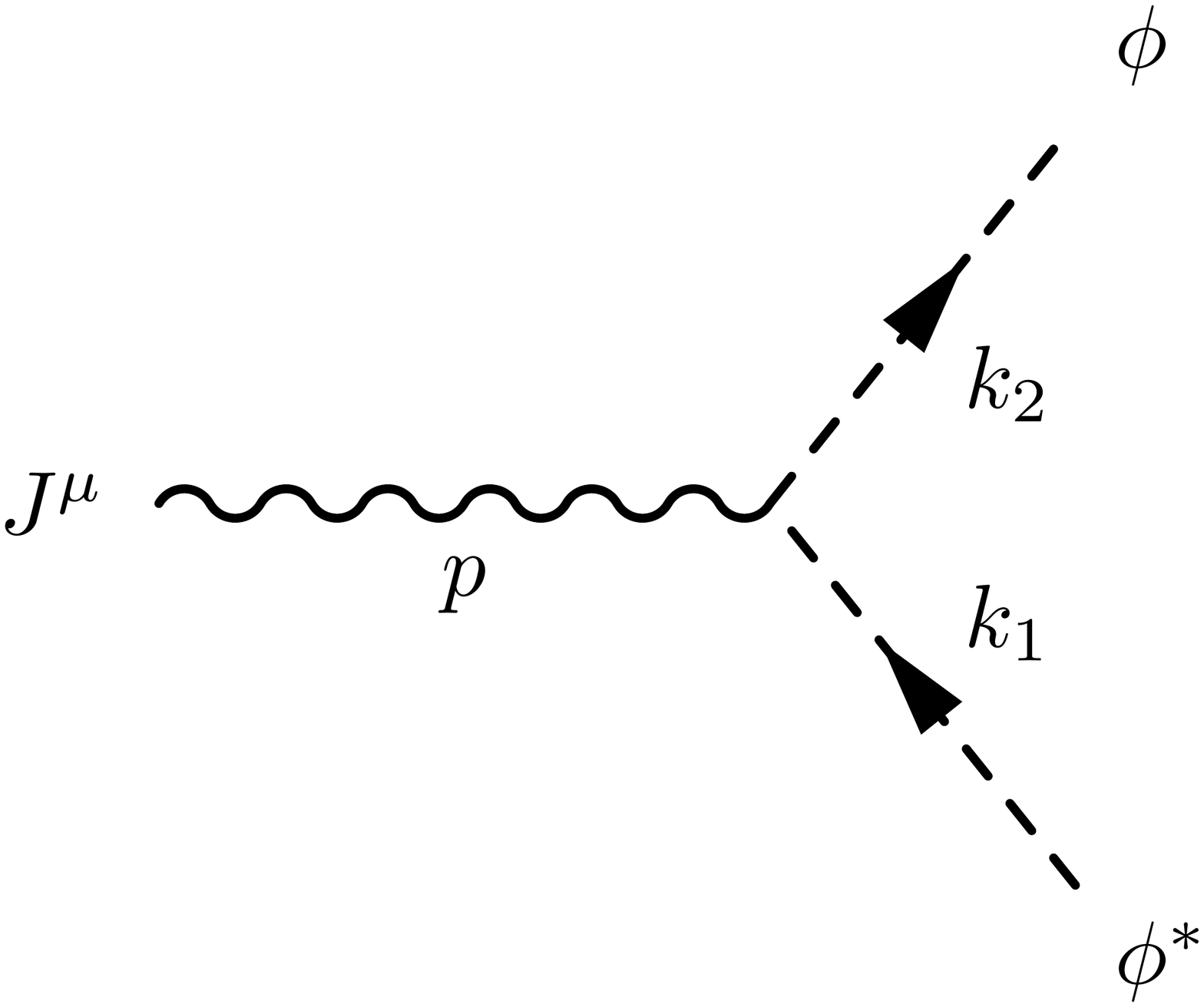}}} \hspace{.3cm}
	\raisebox{.05\height}{\subfigure{\includegraphics[scale=0.12]{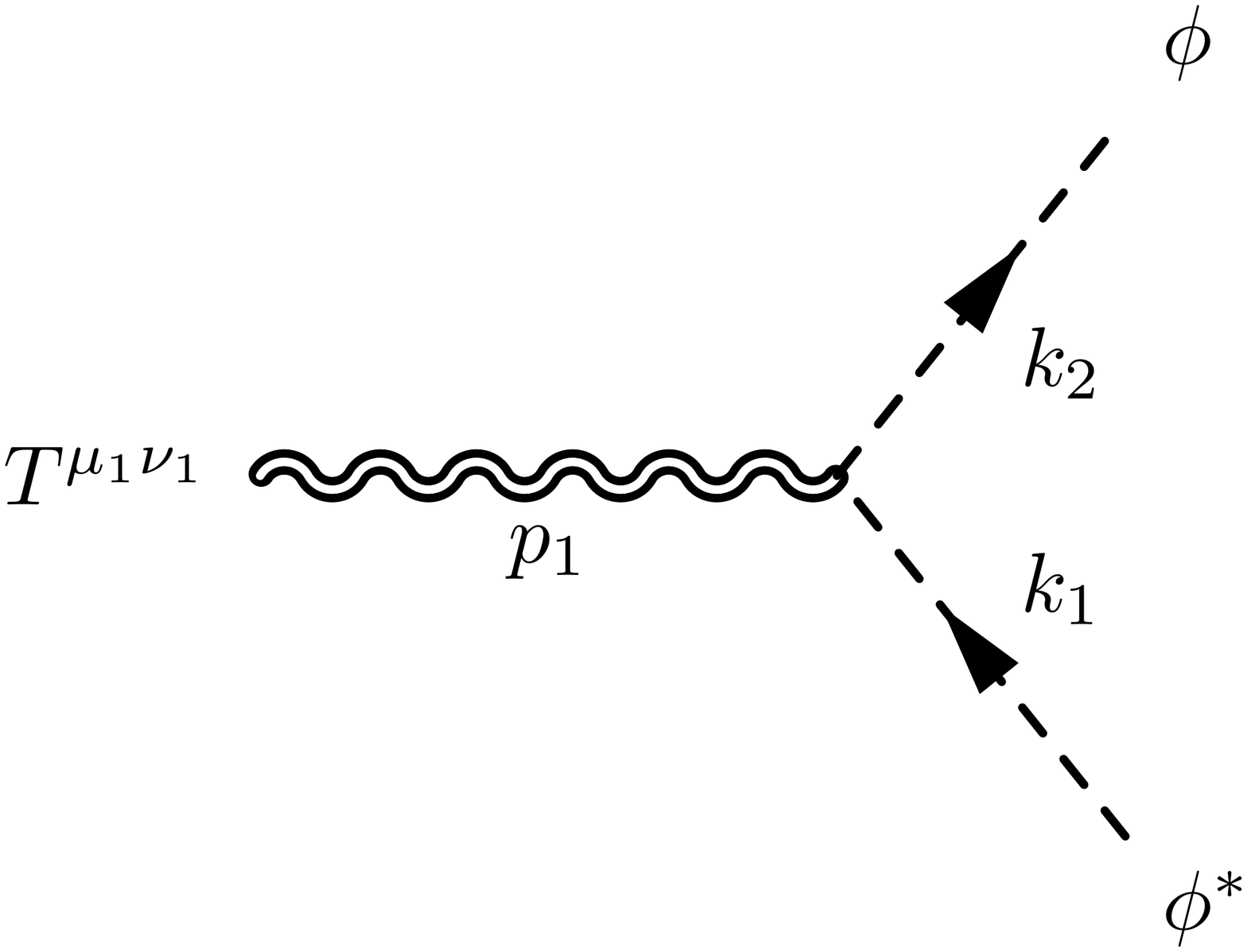}}} \hspace{.3cm}
	\raisebox{.12\height}{\subfigure{\includegraphics[scale=0.12]{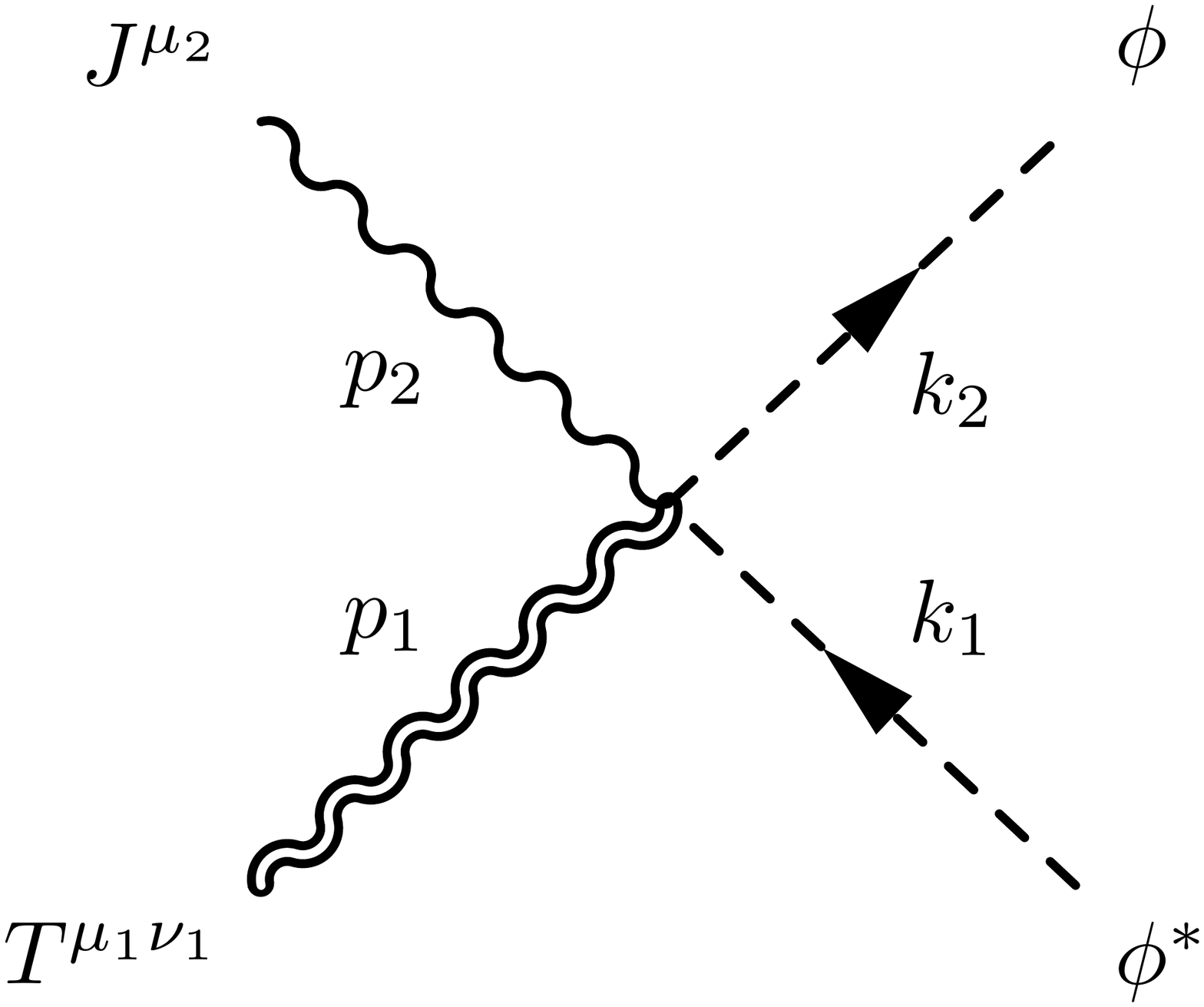}}}\hspace{.3cm}
	\raisebox{.12\height}{\subfigure{\includegraphics[scale=0.12]{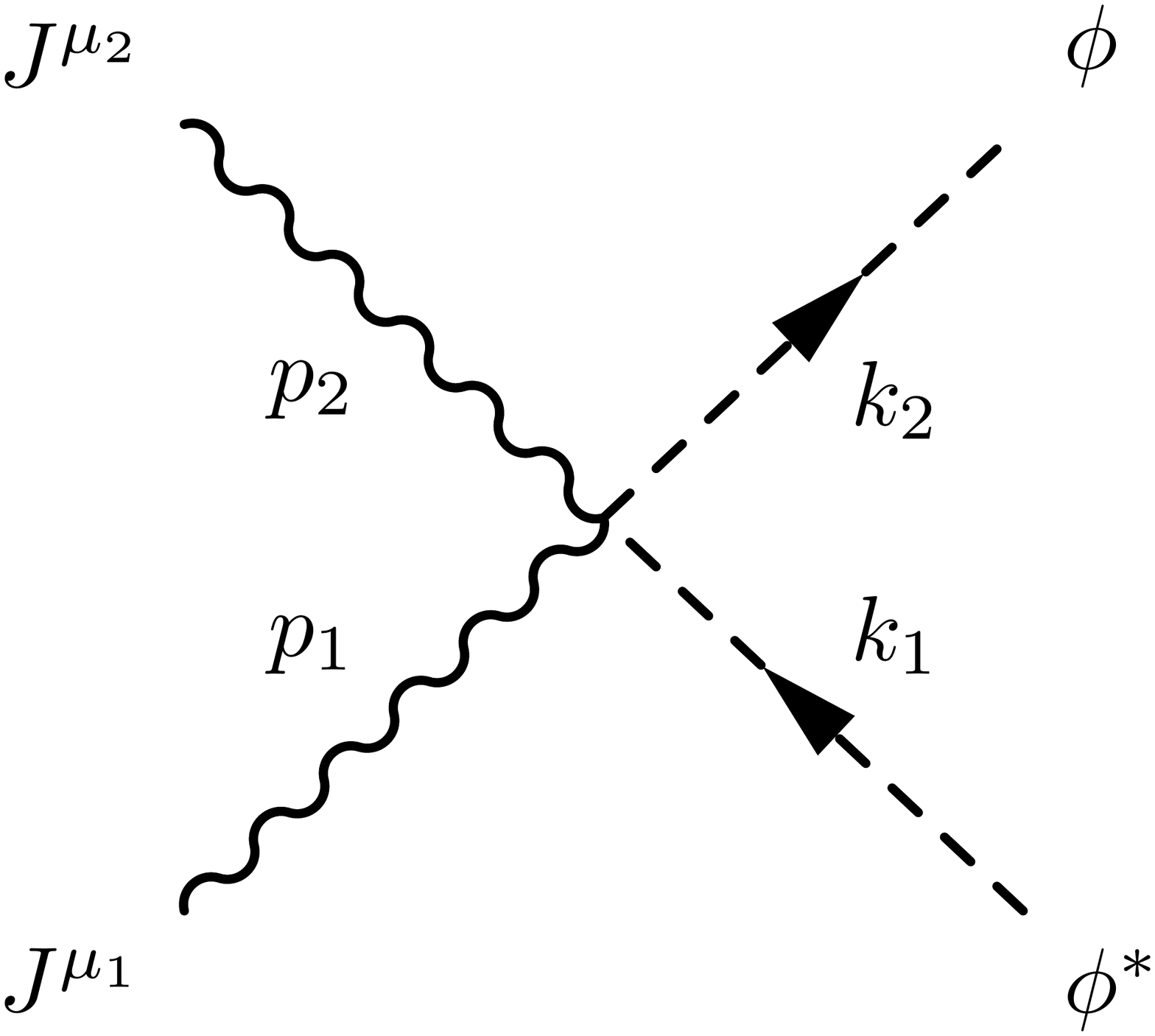}}}
	\vspace{-1cm}
	\caption{The vertices in QED and scalar QED. }\label{Figura2}
\end{figure}
In the scalar sectors we obtain
\begin{align}
	V^{\mu_1\nu_1}_{T\phi\phi}(k_1,k_2)&=\sdfrac{1}{2}A^{\m_1\n_1\m\n}\,k_{1\nu}\,k_{2\m}+\c\,B^{\m_1\n_1\m\n}\,(k_1-k_2)_\mu\,(k_1-k_2)_\nu\\[2ex]
	V^{\mu_1\nu_1\mu_2\nu_2}_{TT\phi\phi}(p_2,k_1,k_2)&=\left(\sdfrac{1}{4}A^{\m_1\n_1\m_2\n_2}\d^{\m\n}+C^{\m_1\n_1\m_2\m_2\m\n}-\sdfrac{1}{2}D^{\m_1\n_1\m_2\n_2\m\n}\right)\,k_{1\nu}\,k_{2\m}\notag\\
	&\hspace{-2cm}+\sdfrac{\c}{2}\bigg[\sdfrac{1}{2}\,\Big(E^{\m_1\n_1\m_2\n_2\m\n}-D^{\m_2\n_2\m\n\m_1\n_1}\Big)\,p_{2\m}p_{2\n}+\sdfrac{1}{2}\,\Big(E^{\m_1\n_1\m_2\n_2\m\n}-D^{\m_1\n_1\m\n\m_2\n_2}\Big)\,p_{2\m}(k_2-k_1)_\n\notag\\
	&\hspace{-1cm}+\Big(C^{\m_1\n_1\m_2\n_2\m\n}-D^{\m_1\n_1\m\n\m_2\n_2}\Big)(k_2-k_1)_\m(k_2-k_1)_\n\bigg]\\
	V^\m_{J\phi\phi^*}(k_1,k_2)&=ie\,(k_1^\m+k_2^\m)\\[2ex]
	V^{\m_1\n_1}_{T\phi\phi^*}(p_1k_1,k_2)& =\frac{1}{2}A^{\m\n\m_1\n_1}\,k_{1\m}k_{2\n}-\,\c\,\left(p_1^{\m_1}p_1^{\n_1}-\delta^{\m_1\n_1}p_1^2\right)\\[2ex]
	V^{\m_1\n_1\m_2}_{TJ\phi\phi^*}(k_1,k_2) &=-\frac{e}{2}\,A^{\m_1\n_1\m\m_2}\,(k_{1\m}-k_{2\m})\\[2ex]
	V^{\m_1\m_2}_{JJ\phi\phi^*}(k_1,k_2) &=-2\,e^2\,\delta^{\m_1\m_2}
\end{align}
where $\c=(d-2)/[8(d-1)]$ is the coefficient for the term of improvement.
In the fermion sector the relevant vertices are
\begin{align}
	V^\m_{J\psi\bar\psi}(k_1,k_2) &=e\,\g^\m\\[2ex]
	V^{\m_1\n_1\m_2}_{TJ\psi\bar\psi}(k_1,k_2) &=-\frac{e}{2}A^{\m_1\n_1\m_2\n}\,\gamma_\mu,\\[2ex]
	V^{\mu_1\nu_1}_{T\bar\psi\psi}(k_1,k_2)&=\sdfrac{1}{4}\,B^{\mu_1\nu_1\mu\nu}\,\gamma_\n\,(k_1-k_2)_\m\\[2ex]
	V^{\mu_1\nu_1\m_2\n_2}_{TT\bar\psi\psi}(p_2,k_1,k_2)&=\sdfrac{1}{8}\bigg[\d^{\m\n}A^{\m_1\n_1\m_2\n_2}-D^{\m_1\n_1\m_2\n_2\m\n}+C^{\m_1\n_1\m_2\n_2\m\n}+\tilde{C}^{\m_2\n_2\m_1\n_1\m\n}\bigg]\,\gamma_\n\,(k_1-k_2)_\m\notag\\
	&\hspace{1cm}+\sdfrac{1}{32}\tilde{C}^{\m_1\n_1\m_2\n_2\n\m}\,p_2^\sigma\,\left\{\gamma_\n,\left[\gamma_\m,\gamma_\sigma\right]\,\right\}
\end{align}
where the last term in the expression above is related to the spin connection. However, one can prove that this term does not contribute to the pinched 2-graviton diagrams. 

In the gauge sector we separate the gauge fixing contributions (GF) from the others, denoted with a subscript $M$, obtaining 
\begin{align}
	V^{\m_1\n_1\a_1\a_2}_{TAA,\,M}(k_1,k_2)&=\bigg(\d^{\m_1\n_1}F^{\a_1\m\n\a_2}+2\,G^{\m_1\n_1\a_1\a_2\m\n}\bigg)\,k_{1\m}\,k_{2\n}
\end{align}
\begin{align}
	V^{\m_1\n_1\m_2\n_2\a_1\a_2}_{TTAA,\,M}(k_1,k_2)&=\bigg[-\sdfrac{1}{2}A^{\m_1\n_1\m_2\n_2}F^{\m\a_1\n\a_2}+\d^{\m_2\n_2}\,G^{\m_1\n_1\a_1\a_2\m\n}+\d^{\m_1\n_1}\,G^{\m_2\n_2\a_1\a_2\m\n}\notag\\
	&\hspace{-2cm}-\big(\d^{\a_1\a_2}C^{\m_1\n_1\m_2\n_2\m\n}+\d^{\m\n}C^{\m_1\n_1\m_2\n_2\a_1\a_2}-\d^{\a_1\n}C^{\m_1\n_1\m_2\n_2\a_2\m}-\d^{\a_2\m}C^{\m_1\n_1\m_2\n_2\a_1\n}\big)\notag\\
	&\hspace{-2cm}-\big(\tilde{F}^{\m\n\m_1\n_1}\tilde{F}^{\m_2\n_2\a_1\a_2}+\tilde{F}^{\m\n\m_2\n_2}\tilde{F}^{\m_1\n_1\a_1\a_2}-\tilde{F}^{\m\a_2\m_1\n_1}\tilde{F}^{\m_2\n_2\a_1\n}-\tilde{F}^{\m\a_2\m_2\n_2}\tilde{F}^{\m_1\n_1\n\a_1}\big)\,\bigg]\,k_{1\m}\,k_{2\n}
\end{align}
\begin{figure}[t]
	\centering
	\vspace{-0.5cm}
	\subfigure{\includegraphics[scale=0.14]{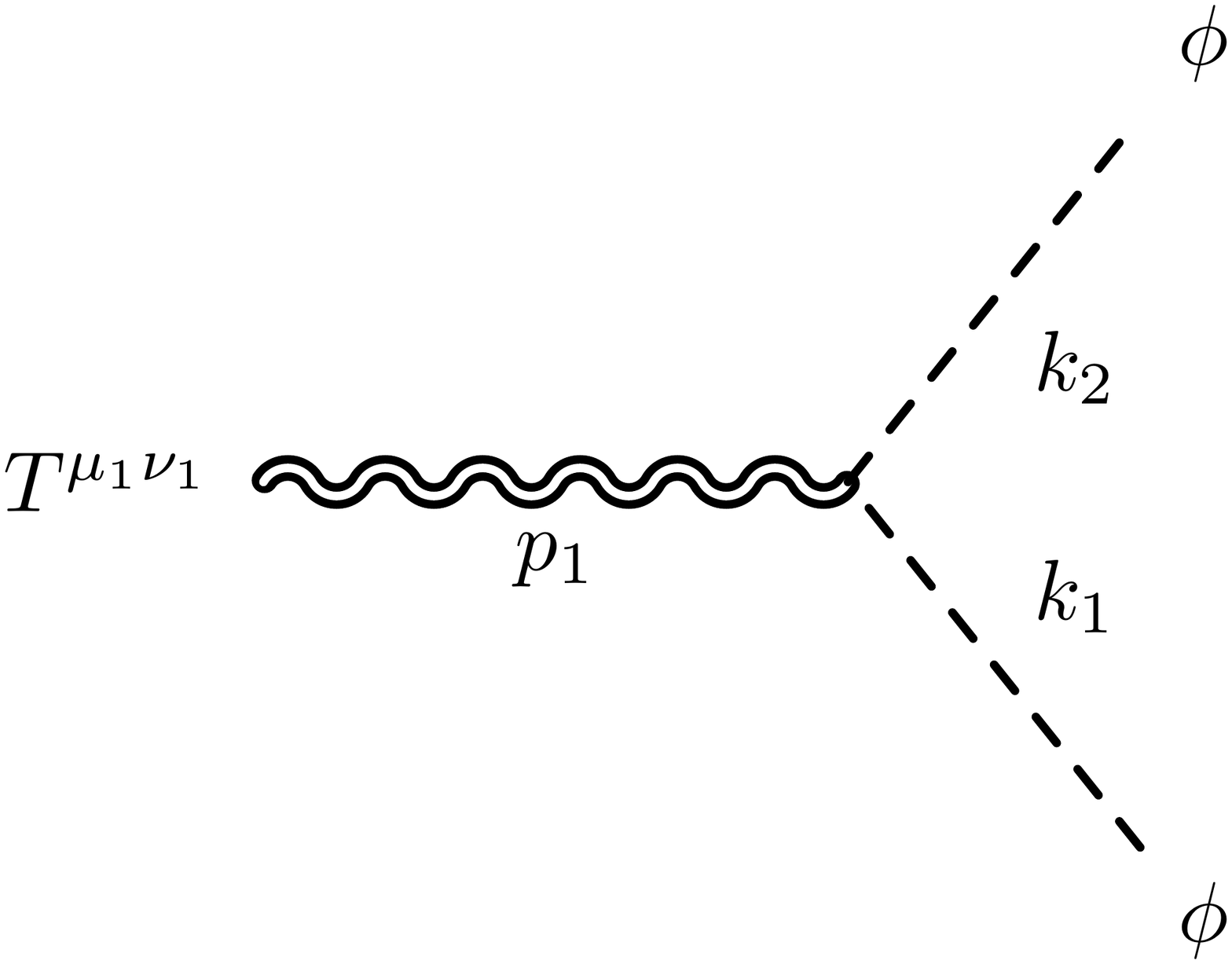}} \hspace{.3cm}
	\subfigure{\includegraphics[scale=0.14]{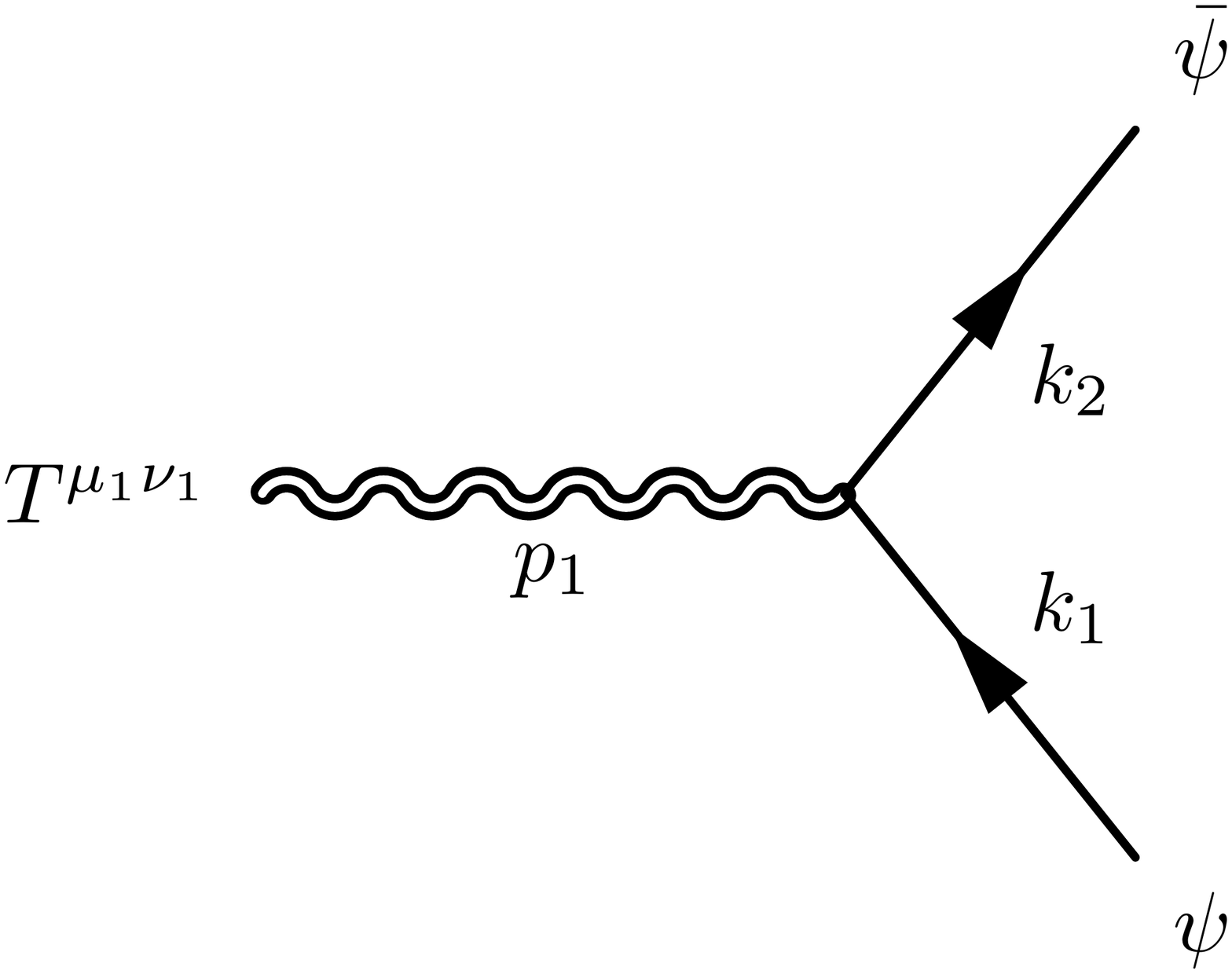}} \hspace{.3cm}
	\subfigure{\includegraphics[scale=0.14]{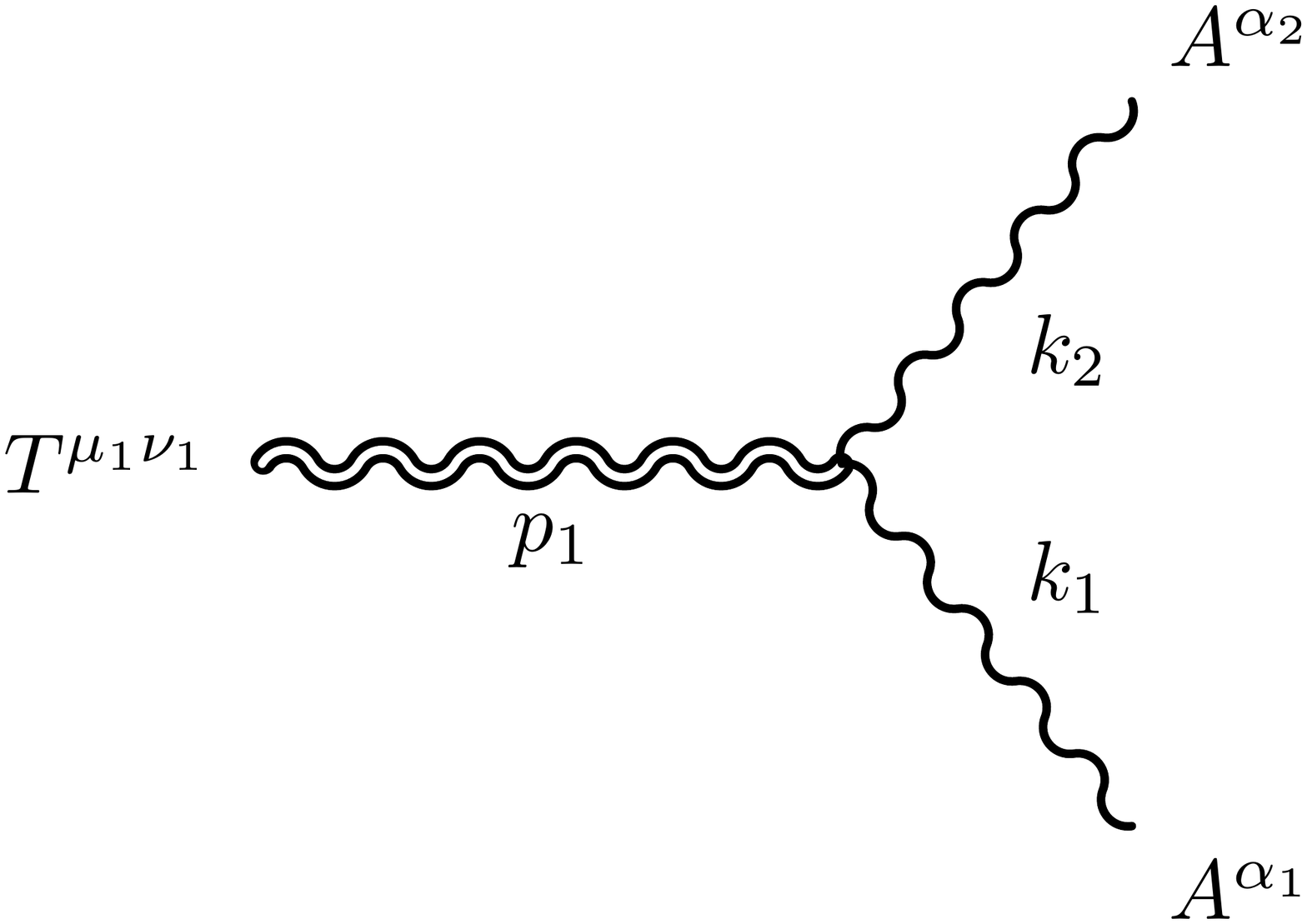}} \hspace{.3cm}
	\subfigure{\includegraphics[scale=0.14]{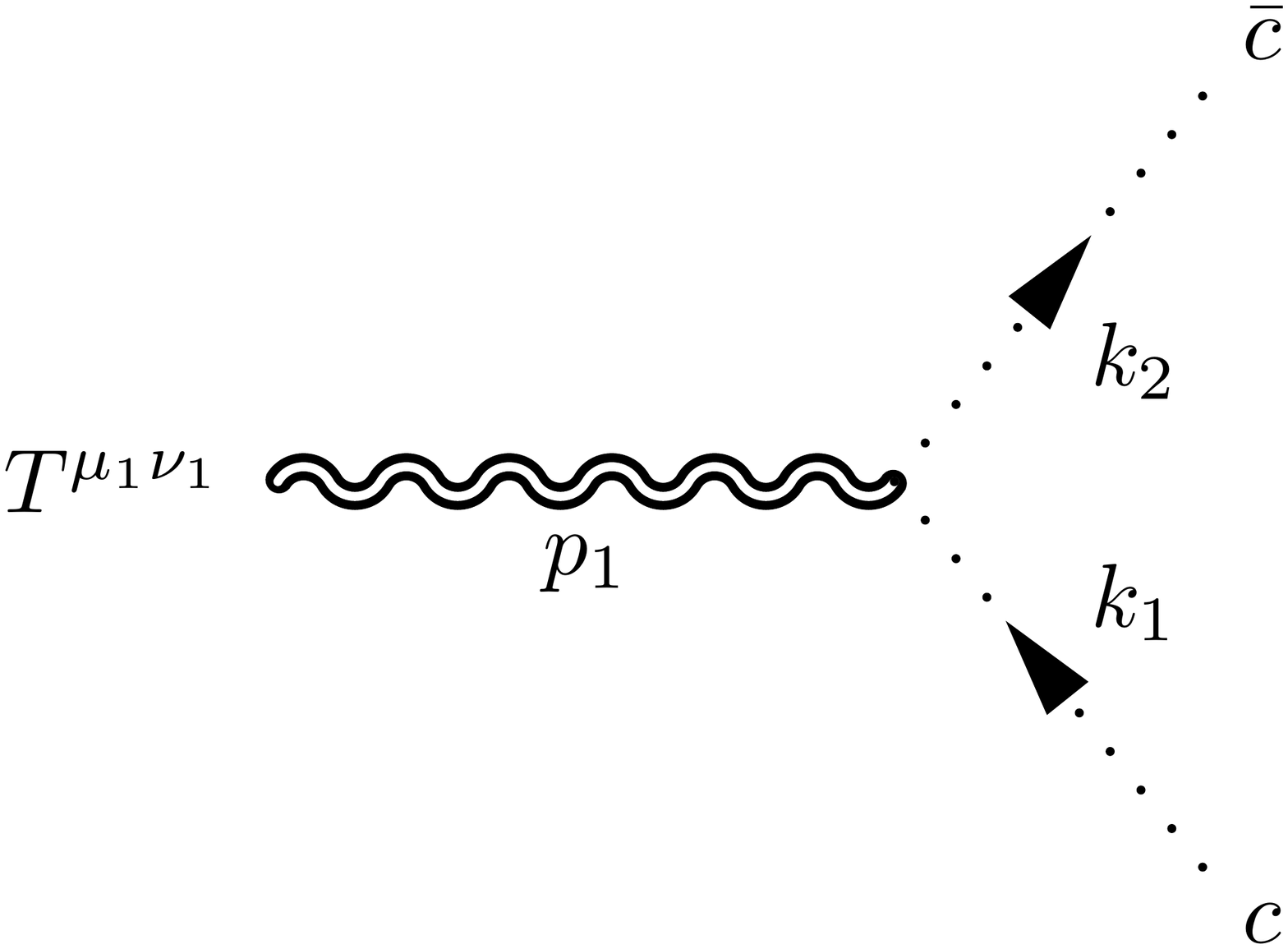}}
	\\
	\vspace{-1.5cm}
	\subfigure{\includegraphics[scale=0.14]{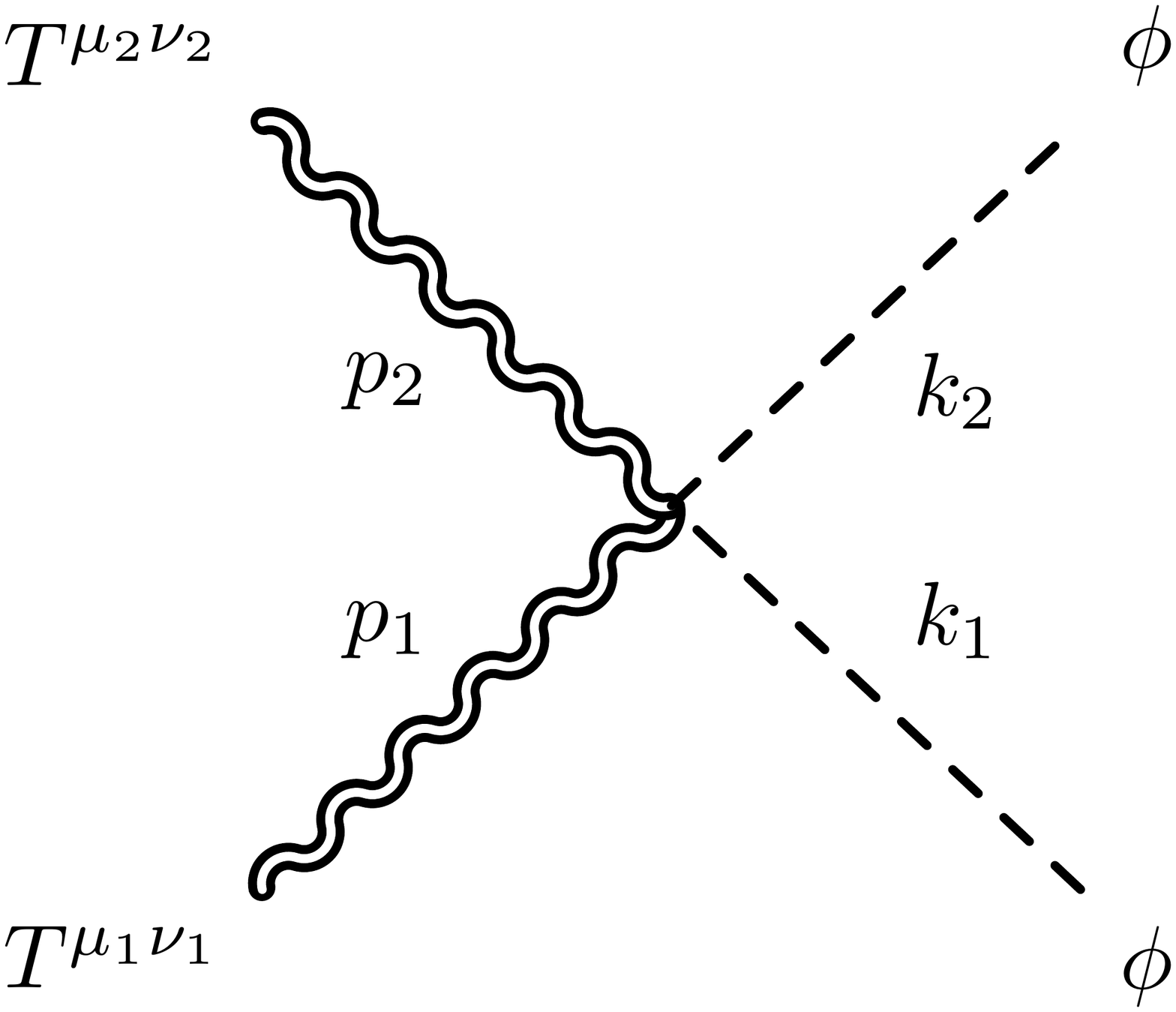}} \hspace{.3cm}
	\subfigure{\includegraphics[scale=0.14]{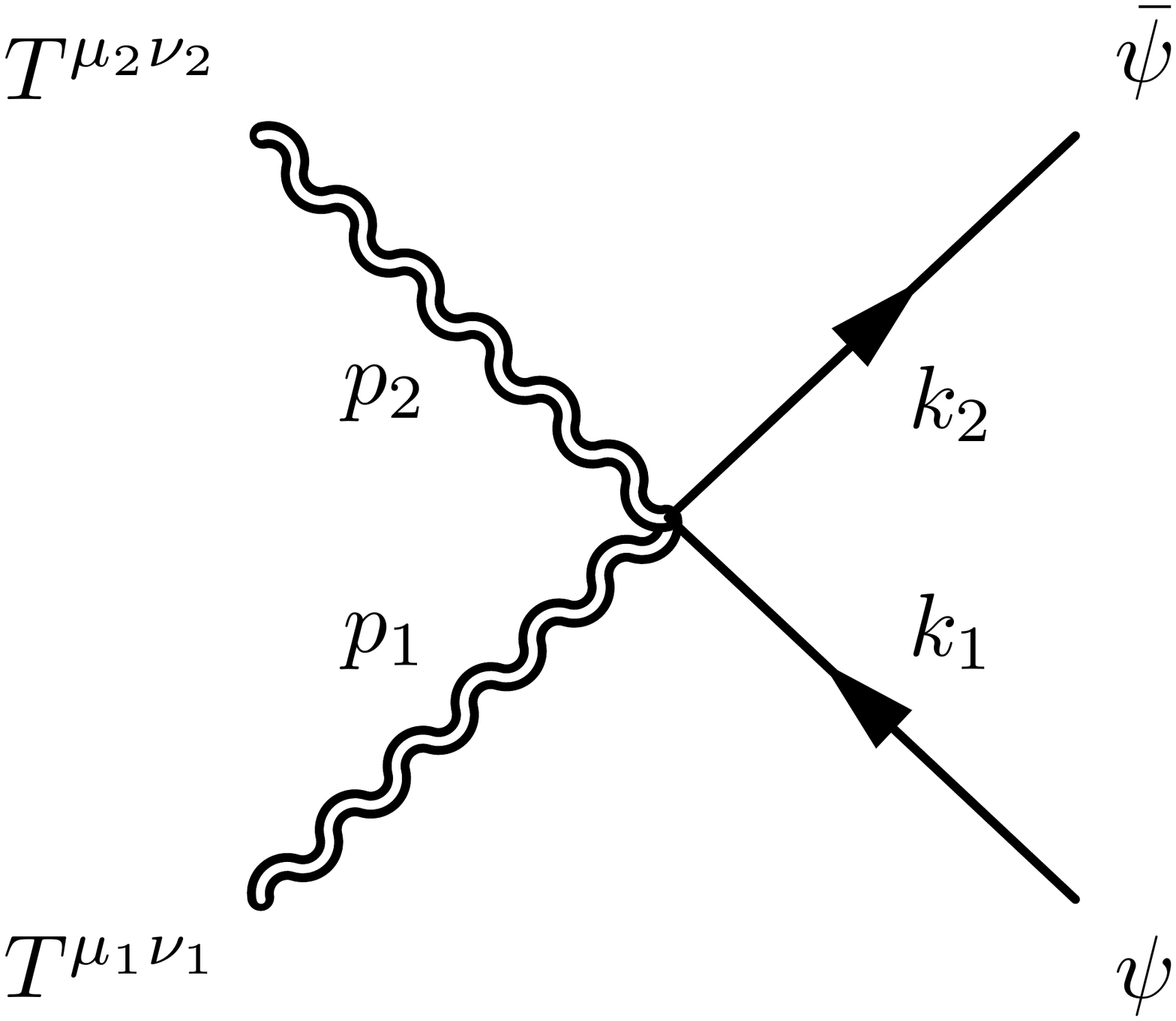}} \hspace{.3cm}
	\subfigure{\includegraphics[scale=0.14]{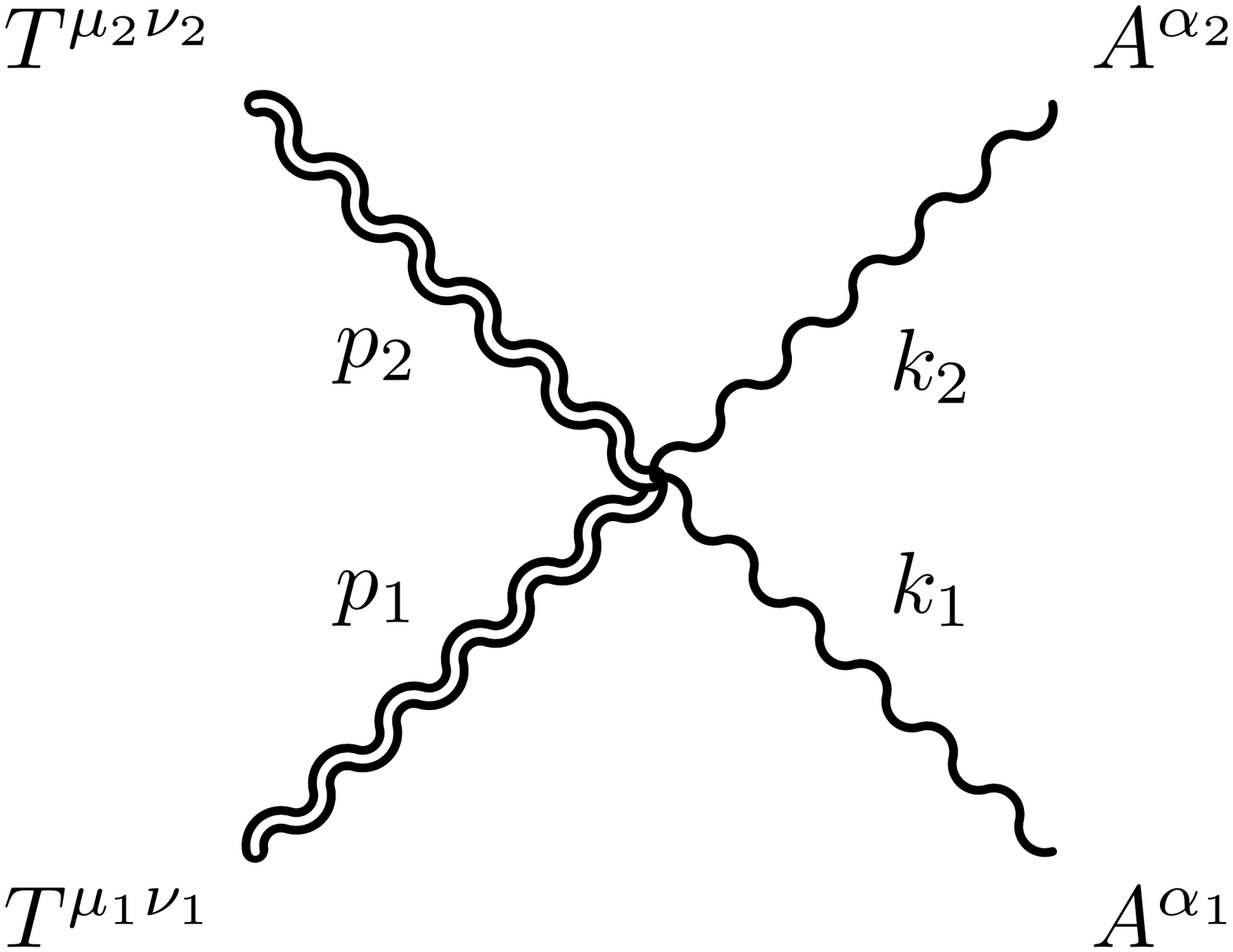}} \hspace{.3cm}
	\subfigure{\includegraphics[scale=0.14]{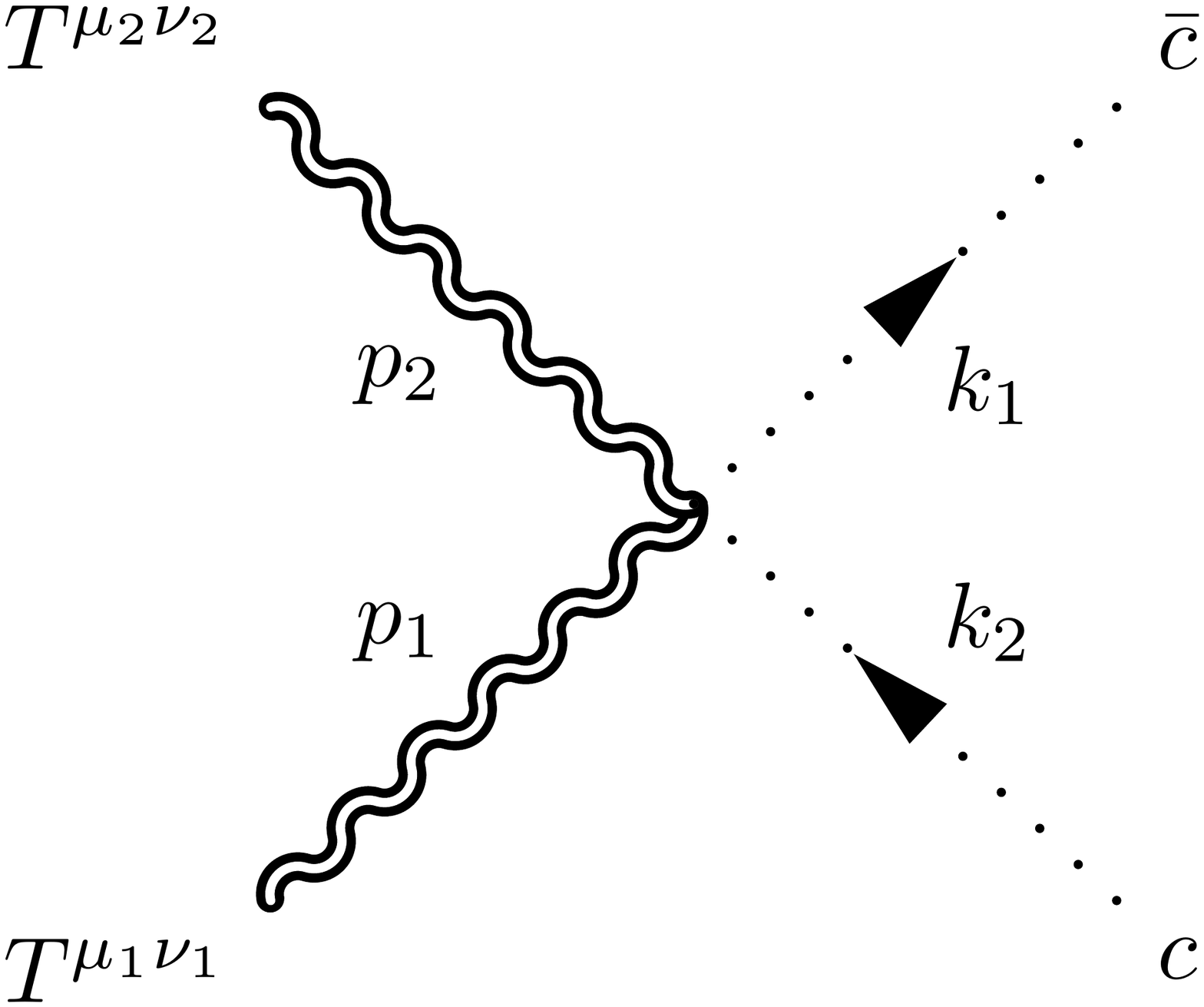}}
	\caption{Vertices used in the Lagrangian realization of the $TTT$ correlator.\label{vertices}}
\end{figure}
\begin{align}
	V^{\m_1\n_1\a_1\a_2}_{TAA,\,GF}(k_1,k_2)&=-\sdfrac{1}{2\xi}\bigg[-\d^{\m_1\n_1}\d^{\a_1\m}\d^{\a_2\n}\,k_{1\m}\,k_{2\n}+\big(\d^{\m_1\n_1}\tilde{F}^{\a_1\a_2\m\n}-2\tilde{C}^{\m_1\n_1\m\n\a_1\a_2}\big)\,k_{2\m}\,k_{2\n}\notag\\
	&\hspace{2cm}+\big(\d^{\m_1\n_1}\tilde{F}^{\a_1\a_2\m\n}-2\tilde{C}^{\m_1\n_1\m\n\a_2\a_1}\big)\,k_{1\m}\,k_{1\n}\,\bigg].
\end{align}

\begin{align}
	V^{\m_1\n_1\m_2\n_2\a_1\a_2}_{TTAA,\,GF}(p_2,k_1,k_2)&=-\sdfrac{1}{2\xi}\bigg[I^{\m_1\n_1\m_2\n_2\a_1\a_2\m\n}\,k_{1\m}\,k_{2\n}+H^{\m_2\n_2\m_1\n_1\a_2\a_1\m\n}\,p_{2\m}k_{1\n}\notag\\
	&\hspace{-3cm}+H^{\m_1\n_1\m_2\n_2\a_1\a_2\m\n}\,p_{2\m}k_{2\n}-\big(I^{\m_1\n_1\m_2\n_2\a_1\a_2\m\n}+A^{\m_2\n_2\a_2\n}\tilde{F}^{\m_1\n_1\m\a_1}-2\d^{\a_2\n}C^{\m_1\n_1\m_2\n_2\m\a_1}\big)\,k_{2\m}\,k_{2\n}\notag\\
	&\hspace{-3cm}-\big(I^{\m_1\n_1\m_2\n_2\a_2\a_1\m\n}+A^{\m_2\n_2\a_1\n}\tilde{F}^{\m_1\n_1\m\a_2}-2\d^{\a_1\n}C^{\m_1\n_1\m_2\n_2\m\a_2}\big)\,k_{1\m}\,k_{1\n}+\big(4\,\tilde{F}^{\m_1\n_1\n(\a_1}\tilde{F}^{\a_2)\m\m_2\n_2}\notag\\
	&\hspace{-2cm}-2\d^{\m_1\n_1}\tilde{C}^{\a_1\a_2\m_2\n_2\n\m}-2\d^{\m_2\n_2}\tilde{C}^{\a_1\a_2\m_1\n_1\m\n}+\d^{\m_1\n_1}\d^{\m_2\n_2}\tilde{F}^{\m\n\a_1\a_2}\big)\,p_{2\m}\,(p_2-k_2+k_1)_\n
	\bigg].
\end{align}
In the ghost sector we obtain
\begin{align}
	V^{\m_1\n_1}_{T\bar cc}(k_1,k_2)&=\sdfrac{1}{2}\,A^{\m_1\n_1\m\n}\,k_{1\m}\,k_{2\n}\\
	V^{\m_1\n_1\m_2\n_2}_{TT\bar cc}(k_1,k_2)&=\bigg(\sdfrac{1}{4}\d^{\m_2\n_2}\,A^{\m_1\n_1\m\n}-\sdfrac{1}{2}\,D^{\m_1\n_1\m\n\m_2\n_2}+C^{\m_1\n_1\m_2\n_2\m\n}\bigg)\,k_{1\m}\,k_{2\n}
\end{align}

\bibliographystyle{h-physrev-mod1}

\begin{thebibliography}{100}
	
	\bibitem{Kastrup:2008jn}
	H.~A. Kastrup,
	\newblock {\em {On the Advancements of Conformal Transformations and their
			Associated Symmetries in Geometry and Theoretical Physics}},
	\newblock Annalen Phys. {\bf 17}, 631 (2008), arXiv:0808.2730.
	
	\bibitem{DiFrancesco:1997nk}
	P.~Di~Francesco, P.~Mathieu, and D.~Senechal,
	\newblock {\em {Conformal Field Theory}} (Springer-Verlag, New York, 1997).
	
	\bibitem{Maldacena:2003nj}
	J.~M. Maldacena,
	\newblock {TASI 2003 lectures on AdS / CFT},
	\newblock in {\em {Progress in string theory. Proceedings, Summer School, TASI
			2003, Boulder, USA, June 2-27, 2003}}, pp. 155--203, 2003,
	arXiv:hep-th/0309246.
	
	\bibitem{Chernodub:2013kya}
	M.~N. Chernodub, A.~Cortijo, A.~G. Grushin, K.~Landsteiner, and M.~A.~H.
	Vozmediano,
	\newblock {\em {Condensed matter realization of the axial magnetic effect}},
	\newblock Phys. Rev. {\bf B89}, 081407 (2014), arXiv:1311.0878.
	
	\bibitem{Ambrus:2019khr}
	V.~E. Ambrus and M.~N. Chernodub,
	\newblock {\em {Helical vortical effects, helical waves, and anomalies of Dirac
			fermions}},
	\newblock (2019), arXiv:1912.11034.
	
	\bibitem{Chernodub:2019tsx}
	M.~N. Chernodub, C.~Corian\`o, and M.~M. Maglio,
	\newblock {\em {Anomalous Gravitational TTT Vertex, Temperature Inhomogeneity,
			and Pressure Anisotropy}},
	\newblock Phys. Lett. {\bf B802}, 135236 (2020), arXiv:1910.13727.
	
	\bibitem{Arjona:2019lxz}
	V.~Arjona, M.~N. Chernodub, and M.~A.~H. Vozmediano,
	\newblock {\em {Fingerprints of the conformal anomaly on the thermoelectric
			transport in Dirac and Weyl semimetals: Result from a Kubo formula}},
	\newblock Phys. Rev. {\bf B99}, 235123 (2019), arXiv:1902.02358.
	
	\bibitem{Gooth:2017mbd}
	J.~Gooth {\em et~al.},
	\newblock {\em {Experimental signatures of the mixed axial-gravitational
			anomaly in the Weyl semimetal NbP}},
	\newblock Nature {\bf 547}, 324 (2017), arXiv:1703.10682.
	
	\bibitem{Osborn:1993cr}
	H.~Osborn and A.~C. Petkou,
	\newblock {\em {Implications of Conformal Invariance in Field Theories for
			General Dimensions}},
	\newblock Ann. Phys. {\bf 231}, 311 (1994), arXiv:hep-th/9307010.
	
	\bibitem{Erdmenger:1996yc}
	J.~Erdmenger and H.~Osborn,
	\newblock {\em {Conserved currents and the energy momentum tensor in
			conformally invariant theories for general dimensions}},
	\newblock Nucl.Phys. {\bf B483}, 431 (1997), arXiv:hep-th/0103237.
	
	\bibitem{Coriano:2013jba}
	C.~Corian\`o, L.~Delle~Rose, E.~Mottola, and M.~Serino,
	\newblock {\em {Solving the Conformal Constraints for Scalar Operators in
			Momentum Space and the Evaluation of Feynman's Master Integrals}},
	\newblock JHEP {\bf 1307}, 011 (2013), arXiv:1304.6944.
	
	\bibitem{Bzowski:2013sza}
	A.~Bzowski, P.~McFadden, and K.~Skenderis,
	\newblock {\em {Implications of conformal invariance in momentum space}},
	\newblock JHEP {\bf 03}, 111 (2014), arXiv:1304.7760.
	
	\bibitem{Coriano:2018bsy}
	C.~Corian\`o and M.~M. Maglio,
	\newblock {\em {The general 3-graviton vertex ($TTT$) of conformal field
			theories in momentum space in $d=4$}},
	\newblock Nucl. Phys. {\bf B937}, 56 (2018), arXiv:1808.10221.
	
	\bibitem{Coriano:2018bbe}
	C.~Corian\`o and M.~M. Maglio,
	\newblock {\em {Exact Correlators from Conformal Ward Identities in Momentum
			Space and the Perturbative $TJJ$ Vertex}},
	\newblock Nucl. Phys. {\bf B938}, 440 (2019), arXiv:1802.07675.
	
	\bibitem{Coriano:2018zdo}
	C.~Corian\`o and M.~M. Maglio,
	\newblock {\em {Renormalization, Conformal Ward Identities and the Origin of a
			Conformal Anomaly Pole}},
	\newblock Phys. Lett. {\bf B781}, 283 (2018), arXiv:1802.01501.
	
	\bibitem{Giannotti:2008cv}
	M.~Giannotti and E.~Mottola,
	\newblock {\em {The Trace Anomaly and Massless Scalar Degrees of Freedom in
			Gravity}},
	\newblock Phys. Rev. {\bf D79}, 045014 (2009), arXiv:0812.0351.
	
	\bibitem{Armillis:2009pq}
	R.~Armillis, C.~Corian\`{o}, and L.~Delle~Rose,
	\newblock {\em {Conformal Anomalies and the Gravitational Effective Action: The
			$TJJ$ Correlator for a Dirac Fermion}},
	\newblock Phys. Rev. {\bf D81}, 085001 (2010), arXiv:0910.3381.
	
	\bibitem{Armillis:2009im}
	R.~Armillis, C.~Corian\`o, and L.~Delle~Rose,
	\newblock {\em {Anomaly Poles as Common Signatures of Chiral and Conformal
			Anomalies}},
	\newblock Phys. Lett. {\bf B682}, 322 (2009), arXiv:0909.4522.
	
	\bibitem{Armillis:2010qk}
	R.~Armillis, C.~Corian\`o, and L.~Delle~Rose,
	\newblock {\em {Trace Anomaly, Massless Scalars and the Gravitational Coupling
			of QCD}},
	\newblock Phys. Rev. {\bf D82}, 064023 (2010), arXiv:1005.4173.
	
	\bibitem{Coriano:2014gja}
	C.~Corian\`o, A.~Costantini, L.~Delle~Rose, and M.~Serino,
	\newblock {\em {Superconformal sum rules and the spectral density flow of the
			composite dilaton (ADD) multiplet in $\mathcal{N}=1$ theories}},
	\newblock JHEP {\bf 06}, 136 (2014), arXiv:1402.6369.
	
	\bibitem{Bastianelli:2004zp}
	F.~Bastianelli and C.~Schubert,
	\newblock {\em {One loop photon graviton mixing in an electromagnetic field.
			I}},
	\newblock JHEP {\bf 02}, 069 (2005), arXiv:gr-qc/0412095.
	
	\bibitem{Bastianelli:2007jv}
	F.~Bastianelli, U.~Nucamendi, C.~Schubert, and V.~M. Villanueva,
	\newblock {\em {One loop photon-graviton mixing in an electromagnetic field:
			Part 2}},
	\newblock JHEP {\bf 11}, 099 (2007), arXiv:0710.5572.
	
	\bibitem{Bastianelli:2012bz}
	F.~Bastianelli, O.~Corradini, J.~M. D\'avila, and C.~Schubert,
	\newblock {\em {On the low-energy limit of one-loop photon-graviton
			amplitudes}},
	\newblock Phys. Lett. {\bf B716}, 345 (2012), arXiv:1202.4502.
	
	\bibitem{Coriano:2012wp}
	C.~Coriano, L.~Delle~Rose, E.~Mottola, and M.~Serino,
	\newblock {\em {Graviton Vertices and the Mapping of Anomalous Correlators to
			Momentum Space for a General Conformal Field Theory}},
	\newblock JHEP {\bf 08}, 147 (2012), arXiv:1203.1339.
	
	\bibitem{Bastianelli:2019zrq}
	F.~Bastianelli and M.~Broccoli,
	\newblock {\em {Axial gravity and anomalies of fermions}},
	\newblock Eur. Phys. J. C {\bf 80}, 276 (2020), arXiv:1911.02271.
	
	\bibitem{Bonora:2014qla}
	L.~Bonora, S.~Giaccari, and B.~Lima~de Souza,
	\newblock {\em {Trace anomalies in chiral theories revisited}},
	\newblock JHEP {\bf 07}, 117 (2014), arXiv:1403.2606.
	
	\bibitem{Bonora:2017gzz}
	L.~Bonora {\em et~al.},
	\newblock {\em {Axial gravity, massless fermions and trace anomalies}},
	\newblock Eur. Phys. J. {\bf C77}, 511 (2017), arXiv:1703.10473.
	
	\bibitem{Mottola:2019nui}
	E.~Mottola and A.~V. Sadofyev,
	\newblock {\em {Chiral Waves on the Fermi-Dirac Sea: Quantum Superfluidity and
			the Axial Anomaly}},
	\newblock (2019), arXiv:1909.01974.
	
	\bibitem{Chernodub:2017jcp}
	M.~N. Chernodub, A.~Cortijo, and M.~A.~H. Vozmediano,
	\newblock {\em {A Nernst current from the conformal anomaly in Dirac and Weyl
			semimetals}},
	\newblock (2017), arXiv:1712.05386.
	
	\bibitem{Rinkel:2019kpo}
	P.~Rinkel, P.~Lopes, and I.~Garate,
	\newblock {\em {Influence of Landau levels on the phonon dispersion of Weyl
			semimetals}},
	\newblock Phys. Rev. B {\bf 99}, 144301 (2019).
	
	\bibitem{Mottola:2016mpl}
	E.~Mottola,
	\newblock {\em {Scalar Gravitational Waves in the Effective Theory of
			Gravity}},
	\newblock JHEP {\bf 07}, 043 (2017), arXiv:1606.09220,
	\newblock [Erratum: JHEP 09, 107 (2017)].
	
	\bibitem{Coriano:2019dyc}
	C.~Corian\`o, M.~M. Maglio, A.~Tatullo, and D.~Theofilopoulos,
	\newblock {Exact Correlators from Conformal Ward Identities in Momentum Space
		and Perturbative Realizations},
	\newblock in {\em {18th Hellenic School and Workshops on Elementary Particle
			Physics and Gravity (CORFU2018) Corfu, Corfu, Greece, August 31-September 28,
			2018}}, 2019, arXiv:1904.13174.
	
	\bibitem{Fradkin:1996is}
	E.~Fradkin and M.~Palchik,
	\newblock {\em {Conformal quantum field theory in D-dimensions}},
	\newblock (1996).
	
	\bibitem{Nakayama:2013is}
	Y.~Nakayama,
	\newblock {\em {Scale invariance vs conformal invariance}},
	\newblock Phys. Rept. {\bf 569}, 1 (2015), arXiv:1302.0884.
	
	\bibitem{Simmons_Duffin_2014}
	D.~Simmons-Duffin,
	\newblock {\em Projectors, shadows, and conformal blocks},
	\newblock Journal of High Energy Physics {\bf 2014} (2014).
	
	\bibitem{Simmons-Duffin:2016gjk}
	D.~Simmons-Duffin,
	\newblock {The Conformal Bootstrap},
	\newblock in {\em {Proceedings, Theoretical Advanced Study Institute in
			Elementary Particle Physics: New Frontiers in Fields and Strings (TASI 2015):
			Boulder, CO, USA, June 1-26, 2015}}, pp. 1--74, 2017, arXiv:1602.07982.
	
	\bibitem{Ferrara:1973yt}
	S.~Ferrara, A.~F. Grillo, and R.~Gatto,
	\newblock {\em {Tensor representations of conformal algebra and conformally
			covariant operator product expansion}},
	\newblock Annals Phys. {\bf 76}, 161 (1973).
	
	\bibitem{Weinberg:2010fx}
	S.~Weinberg,
	\newblock {\em {Six-dimensional Methods for Four-dimensional Conformal Field
			Theories}},
	\newblock Phys. Rev. {\bf D82}, 045031 (2010), arXiv:1006.3480.
	
	\bibitem{Costa:2011mg}
	M.~S. Costa, J.~Penedones, D.~Poland, and S.~Rychkov,
	\newblock {\em {Spinning Conformal Correlators}},
	\newblock JHEP {\bf 11}, 071 (2011), arXiv:1107.3554.
	
	\bibitem{SimmonsDuffin:2012uy}
	D.~Simmons-Duffin,
	\newblock {\em {Projectors, Shadows, and Conformal Blocks}},
	\newblock JHEP {\bf 04}, 146 (2014), arXiv:1204.3894.
	
	\bibitem{Dolan:2000ut}
	F.~A. Dolan and H.~Osborn,
	\newblock {\em {Conformal four point functions and the operator product
			expansion}},
	\newblock Nucl. Phys. {\bf B599}, 459 (2001), arXiv:hep-th/0011040.
	
	\bibitem{Poland:2018epd}
	D.~Poland, S.~Rychkov, and A.~Vichi,
	\newblock {\em {The Conformal Bootstrap: Theory, Numerical Techniques, and
			Applications}},
	\newblock (2018), arXiv:1805.04405.
	
	\bibitem{Poland:2016chs}
	D.~Poland and D.~Simmons-Duffin,
	\newblock {\em {The conformal bootstrap}},
	\newblock Nature Phys. {\bf 12}, 535 (2016).
	
	\bibitem{Capper:1975ig}
	D.~Capper and M.~Duff,
	\newblock {\em {Conformal Anomalies and the Renormalizability Problem in
			Quantum Gravity}},
	\newblock Phys.Lett. {\bf A53}, 361 (1975).
	
	\bibitem{Deser:1976yx}
	S.~Deser, M.~J. Duff, and C.~J. Isham,
	\newblock {\em {Nonlocal Conformal Anomalies}},
	\newblock Nucl. Phys. {\bf B111}, 45 (1976).
	
	\bibitem{Riegert:1984kt}
	R.~J. Riegert,
	\newblock {\em {A Nonlocal Action for the Trace Anomaly}},
	\newblock Phys. Lett. {\bf 134B}, 56 (1984).
	
	\bibitem{Coriano:2017mux}
	C.~Corian\`o, M.~M. Maglio, and E.~Mottola,
	\newblock {\em {TTT in CFT: Trace Identities and the Conformal Anomaly
			Effective Action}},
	\newblock Nucl. Phys. {\bf B942}, 303 (2019), arXiv:1703.08860.
	
	\bibitem{Maldacena:2011nz}
	J.~M. Maldacena and G.~L. Pimentel,
	\newblock {\em {On graviton non-Gaussianities during inflation}},
	\newblock JHEP {\bf 09}, 045 (2011), arXiv:1104.2846.
	
	\bibitem{Bzowski:2011ab}
	A.~Bzowski, P.~McFadden, and K.~Skenderis,
	\newblock {\em {Holographic predictions for cosmological 3-point functions}},
	\newblock JHEP {\bf 03}, 091 (2012), arXiv:1112.1967.
	
	\bibitem{Coriano:2012hd}
	C.~Corian\`o, L.~Delle~Rose, and M.~Serino,
	\newblock {\em {Three and Four Point Functions of Stress Energy Tensors in D=3
			for the Analysis of Cosmological Non-Gaussianities}},
	\newblock JHEP {\bf 12}, 090 (2012), arXiv:1210.0136.
	
	\bibitem{Arkani-Hamed:2018kmz}
	N.~Arkani-Hamed, D.~Baumann, H.~Lee, and G.~L. Pimentel,
	\newblock {\em {The Cosmological Bootstrap: Inflationary Correlators from
			Symmetries and Singularities}},
	\newblock (2018), arXiv:1811.00024.
	
	\bibitem{Benincasa:2018ssx}
	P.~Benincasa,
	\newblock {\em {From the flat-space S-matrix to the Wavefunction of the
			Universe}},
	\newblock (2018), arXiv:1811.02515.
	
	\bibitem{Arkani-Hamed:2018bjr}
	N.~Arkani-Hamed and P.~Benincasa,
	\newblock {\em {On the Emergence of Lorentz Invariance and Unitarity from the
			Scattering Facet of Cosmological Polytopes}},
	\newblock (2018), arXiv:181f1.01125.
	
	\bibitem{Arkani-Hamed:2017fdk}
	N.~Arkani-Hamed, P.~Benincasa, and A.~Postnikov,
	\newblock {\em {Cosmological Polytopes and the Wavefunction of the Universe}},
	\newblock (2017), arXiv:1709.02813.
	
	\bibitem{Henn:2014yza}
	J.~M. Henn and J.~C. Plefka,
	\newblock {\em {Scattering Amplitudes in Gauge Theories}},
	\newblock Lect. Notes Phys. {\bf 883}, pp.1 (2014).
	
	\bibitem{Benincasa:2013faa}
	P.~Benincasa,
	\newblock {\em {New structures in scattering amplitudes: a review}},
	\newblock Int. J. Mod. Phys. {\bf A29}, 1430005 (2014), arXiv:1312.5583.
	
	\bibitem{Maglio:2019grh}
	C.~Corian\`o and M.~M. Maglio,
	\newblock {\em {On Some Hypergeometric Solutions of the Conformal Ward
			Identities of Scalar 4-point Functions in Momentum Space}},
	\newblock JHEP {\bf 09}, 107 (2019), arXiv:1903.05047.
	
	\bibitem{Coriano:2020ccb}
	C.~Corian\`o and M.~M. Maglio,
	\newblock {\em {The Generalized Hypergeometric Structure of the Ward Identities
			of CFT's in Momentum Space in $d > 2$}},
	\newblock (2020), arXiv:2001.09622.
	
	\bibitem{Vidunas1}
	R.~Vidunas,
	\newblock {\em {Specialization of Appell's functions to univariate
			hypergeometric functions}},
	\newblock J. Math. Anal. Appl. {\bf 355}, 145 (2009).
	
	\bibitem{Appell}
	P.~Appell and K.~Kamp\`e~de Feri\`et,
	\newblock {\em { Fonctions hypergeometriques et hyperspheriques: polynomes
			d'Hermite}},
	\newblock Paris : Gauthier-Villars , 434 p. (1926).
	
	\bibitem{Coriano:2019nkw}
	C.~Corian\`o, M.~M. Maglio, and D.~Theofilopoulos,
	\newblock {\em {Four-Point Functions in Momentum Space: Conformal Ward
			Identities in the Scalar/Tensor case}},
	\newblock Eur. Phys. J. C {\bf 80}, 540 (2020), arXiv:1912.01907.
	
	\bibitem{Bzowski:2014qja}
	A.~Bzowski and K.~Skenderis,
	\newblock {\em {Comments on scale and conformal invariance}},
	\newblock JHEP {\bf 08}, 027 (2014), arXiv:1402.3208.
	
	\bibitem{Davydychev:1992xr}
	A.~I. Davydychev,
	\newblock {\em {Recursive algorithm of evaluating vertex type Feynman
			integrals}},
	\newblock J.Phys.A {\bf A25}, 5587 (1992).
	
	\bibitem{Bzowski:2015yxv}
	A.~Bzowski, P.~McFadden, and K.~Skenderis,
	\newblock {\em {Evaluation of conformal integrals}},
	\newblock JHEP {\bf 02}, 068 (2016), arXiv:1511.02357.
	
	\bibitem{Bzowski:2018fql}
	A.~Bzowski, P.~McFadden, and K.~Skenderis,
	\newblock {\em {Renormalised CFT 3-point functions of scalars, currents and
			stress tensors}},
	\newblock (2018), arXiv:1805.12100.
	
	\bibitem{Drummond:2006rz}
	J.~M. Drummond, J.~Henn, V.~A. Smirnov, and E.~Sokatchev,
	\newblock {\em {Magic identities for conformal four-point integrals}},
	\newblock JHEP {\bf 01}, 064 (2007), arXiv:hep-th/0607160.
	
	\bibitem{Drummond:2007aua}
	J.~M. Drummond, G.~P. Korchemsky, and E.~Sokatchev,
	\newblock {\em {Conformal properties of four-gluon planar amplitudes and Wilson
			loops}},
	\newblock Nucl. Phys. {\bf B795}, 385 (2008), arXiv:0707.0243.
	
	\bibitem{Drummond:2008vq}
	J.~M. Drummond, J.~Henn, G.~P. Korchemsky, and E.~Sokatchev,
	\newblock {\em {Dual superconformal symmetry of scattering amplitudes in N=4
			super-Yang-Mills theory}},
	\newblock Nucl. Phys. {\bf B828}, 317 (2010), arXiv:0807.1095.
	
	\bibitem{Usyukina:1992jd}
	N.~I. Usyukina and A.~I. Davydychev,
	\newblock {\em {An Approach to the evaluation of three and four point ladder
			diagrams}},
	\newblock Phys. Lett. {\bf B298}, 363 (1993).
	
	\bibitem{Usyukina:1993ch}
	N.~I. Usyukina and A.~I. Davydychev,
	\newblock {\em {Exact results for three and four point ladder diagrams with an
			arbitrary number of rungs}},
	\newblock Phys. Lett. {\bf B305}, 136 (1993).
	
	\bibitem{Broadhurst:1993ru}
	D.~J. Broadhurst and A.~L. Kataev,
	\newblock {\em {Connections between deep inelastic and annihilation processes
			at next to next-to-leading order and beyond}},
	\newblock Phys. Lett. {\bf B315}, 179 (1993), arXiv:hep-ph/9308274.
	
	\bibitem{Eden:1998hh}
	B.~Eden, P.~S. Howe, C.~Schubert, E.~Sokatchev, and P.~C. West,
	\newblock {\em {Four point functions in N=4 supersymmetric Yang-Mills theory at
			two loops}},
	\newblock Nucl. Phys. B {\bf 557}, 355 (1999), arXiv:hep-th/9811172.
	
	\bibitem{Eden:2000mv}
	B.~Eden, C.~Schubert, and E.~Sokatchev,
	\newblock {\em {Three loop four point correlator in N=4 SYM}},
	\newblock Phys. Lett. B {\bf 482}, 309 (2000), arXiv:hep-th/0003096.
	
	\bibitem{Coriano:1995fj}
	C.~Corian\`o and A.~R. White,
	\newblock {\em {Gauge theory high-energy behavior from j plane unitarity}},
	\newblock Nucl. Phys. {\bf B468}, 175 (1996), arXiv:hep-ph/9510329.
	
	\bibitem{Coriano:1996rj}
	C.~Corian\`o, A.~R. White, and M.~Wusthoff,
	\newblock {\em {NLO conformal symmetry in the Regge limit of QCD}},
	\newblock Nucl. Phys. {\bf B493}, 397 (1997), arXiv:hep-ph/9609405.
	
	\bibitem{Coriano:1994wk}
	C.~Corian\`o and A.~R. White,
	\newblock {\em {The Spectrum of the O(g**4) scale invariant Lipatov kernel}},
	\newblock Phys. Rev. Lett. {\bf 74}, 4980 (1995), arXiv:hep-ph/9411379.
	
	\bibitem{Coriano:1995hx}
	C.~Corian\`o and A.~White,
	\newblock {\em {t channel unitarity construction of small x kernels}},
	\newblock Acta Phys. Polon. {\bf B26}, 2005 (1995), arXiv:hep-ph/9511229.
	
	\bibitem{Bzowski:2015pba}
	A.~Bzowski, P.~McFadden, and K.~Skenderis,
	\newblock {\em {Scalar 3-point functions in CFT: renormalisation, beta
			functions and anomalies}},
	\newblock JHEP {\bf 03}, 066 (2016), arXiv:1510.08442.
	
	\bibitem{Kidonakis:1998nf}
	N.~Kidonakis, G.~Oderda, and G.~F. Sterman,
	\newblock {\em {Evolution of color exchange in QCD hard scattering}},
	\newblock Nucl. Phys. B {\bf 531}, 365 (1998), arXiv:hep-ph/9803241.
	
	\bibitem{Sterman:2002qn}
	G.~F. Sterman and M.~E. Tejeda-Yeomans,
	\newblock {\em {Multiloop amplitudes and resummation}},
	\newblock Phys. Lett. B {\bf 552}, 48 (2003), arXiv:hep-ph/0210130.
	
	\bibitem{Aybat:2006mz}
	S.~Aybat, L.~J. Dixon, and G.~F. Sterman,
	\newblock {\em {The Two-loop soft anomalous dimension matrix and resummation at
			next-to-next-to leading pole}},
	\newblock Phys. Rev. D {\bf 74}, 074004 (2006), arXiv:hep-ph/0607309.
	
	\bibitem{Bzowski:2017poo}
	A.~Bzowski, P.~McFadden, and K.~Skenderis,
	\newblock {\em {Renormalised 3-point functions of stress tensors and conserved
			currents in CFT}},
	\newblock (2017), arXiv:1711.09105.
	
	\bibitem{Prudnikov}
	B.~Prudnikov and Marichev,
	\newblock {\em {Integrals and series, vol 2}} (Gordon and Breach, 1992).
	
	\bibitem{2009PhLB..682..322A}
	R.~{Armillis}, C.~{Corian{\`o}}, and L.~{Delle Rose},
	\newblock {\em {Anomaly poles as common signatures of chiral and conformal
			anomalies}},
	\newblock Physics Letters B {\bf 682}, 322 (2009), arXiv:0909.4522.
	
	\bibitem{Armillis:2009sm}
	R.~Armillis, C.~Corian\`o, L.~Delle~Rose, and M.~Guzzi,
	\newblock {\em {Anomalous U(1) Models in Four and Five Dimensions and their
			Anomaly Poles}},
	\newblock JHEP {\bf 12}, 029 (2009), arXiv:0905.0865.
	
	\bibitem{Rinkel:2016dxo}
	P.~Rinkel, P.~L.~S. Lopes, and I.~Garate,
	\newblock {\em {Signatures of the chiral anomaly in phonon dynamics}},
	\newblock Phys. Rev. Lett. {\bf 119}, 107401 (2017), arXiv:1610.03073.
	
	\bibitem{Bastianelli:2012es}
	F.~Bastianelli, O.~Corradini, J.~M. Davila, and C.~Schubert,
	\newblock {\em {Photon-Graviton Amplitudes from the Effective Action}},
	\newblock Phys. Part. Nucl. {\bf 43}, 630 (2012), arXiv:1203.1689.
	
	\bibitem{Bastianelli:2016nuf}
	F.~Bastianelli and R.~Martelli,
	\newblock {\em {On the trace anomaly of a Weyl fermion}},
	\newblock JHEP {\bf 11}, 178 (2016), arXiv:1610.02304.
	
	\bibitem{Kataev:1996ce}
	A.~L. Kataev,
	\newblock {The Generalized Crewther relation: The Peculiar aspects of the
		analytical perturbative QCD calculations},
	\newblock in {\em {Continuous advances in QCD 1996. Proceedings, Conference,
			Minneapolis, USA, March 28-31, 1996}}, pp. 107--132, 1996,
	arXiv:hep-ph/9607426.
	
	\bibitem{Isono:2018rrb}
	H.~Isono, T.~Noumi, and G.~Shiu,
	\newblock {\em {Momentum space approach to crossing symmetric CFT
			correlators}},
	\newblock JHEP {\bf 07}, 136 (2018), arXiv:1805.11107.
	
	\bibitem{Gillioz:2018mto}
	M.~Gillioz,
	\newblock {\em {Momentum-space conformal blocks on the light cone}},
	\newblock (2018), arXiv:1807.07003.
	
	\bibitem{Bastianelli:2000rs}
	F.~Bastianelli, G.~Cuoghi, and L.~Nocetti,
	\newblock {\em {Consistency conditions and trace anomalies in six-dimensions}},
	\newblock Class. Quant. Grav. {\bf 18}, 793 (2001), arXiv:hep-th/0007222.
	
	\bibitem{1984PhLB..134...56R}
	R.~J. {Riegert},
	\newblock {\em {A non-local action for the trace anomaly}},
	\newblock Physics Letters B {\bf 134}, 56 (1984).
	
	\bibitem{Antoniadis:2011ib}
	I.~Antoniadis, P.~O. Mazur, and E.~Mottola,
	\newblock {\em {Conformal Invariance, Dark Energy, and CMB Non-Gaussianity}},
	\newblock JCAP {\bf 1209}, 024 (2012), arXiv:1103.4164.
	
	\bibitem{Antoniadis:2006wq}
	I.~Antoniadis, P.~O. Mazur, and E.~Mottola,
	\newblock {\em {Cosmological dark energy: Prospects for a dynamical theory}},
	\newblock New J. Phys. {\bf 9}, 11 (2007), arXiv:gr-qc/0612068.
	
	\bibitem{Birrell:1982ix}
	N.~D. Birrell and P.~C.~W. Davies,
	\newblock {\em {Quantum Fields in Curved Space}}Cambridge Monographs on
	Mathematical Physics (Cambridge Univ. Press, Cambridge, UK, 1984).
	
	\bibitem{Riegert:1987kt}
	R.~J. Riegert,
	\newblock {\em {A non-local action for the trace anomaly}},
	\newblock Phys.Lett. {\bf B134}, 56 (1984).
	
	\bibitem{Coriano:2013nja}
	C.~Corian\`o, L.~Delle~Rose, C.~Marzo, and M.~Serino,
	\newblock {\em {The dilaton Wess-Zumino action in six dimensions from Weyl
			gauging: local anomalies and trace relations}},
	\newblock Class. Quant. Grav. {\bf 31}, 105009 (2014), arXiv:1311.1804.
	
	\bibitem{Coriano:2013xua}
	C.~Corian\`o, L.~Delle~Rose, C.~Marzo, and M.~Serino,
	\newblock {\em {Conformal Trace Relations from the Dilaton Wess-Zumino
			Action}},
	\newblock (2013), arXiv:1306.4248.
	
	\bibitem{Coriano:2010ws}
	C.~Corian\`o, M.~Guzzi, and A.~Mariano,
	\newblock {\em {Relic Densities of Dark Matter in the U(1)-Extended NMSSM and
			the Gauged Axion Supermultiplet}},
	\newblock Phys. Rev. {\bf D85}, 095008 (2012), arXiv:1010.2010.
	
	\bibitem{Schwinger452}
	J.~Schwinger,
	\newblock {\em On the green{\textquoteright}s functions of quantized fields.
		i},
	\newblock Proceedings of the National Academy of Sciences {\bf 37}, 452 (1951),
	https://www.pnas.org/content/37/7/452.full.pdf.
	
	\bibitem{Barvinsky:1984jd}
	A.~O. Barvinsky and G.~A. Vilkovisky,
	\newblock {\em {THE GENERALIZED SCHWINGER-DE WITT TECHNIQUE AND THE UNIQUE
			EFFECTIVE ACTION IN QUANTUM GRAVITY}},
	\newblock Phys. Lett. {\bf 131B}, 313 (1983),
	\newblock [,141(1984)].
	
	\bibitem{Kreyszig}
	E.~Kreyszig,
	\newblock {\em Introduction to Differential Geometry and Riemannian Geometry}
	(University of Toronto Press, 1968).
	
	\bibitem{Hatzinikitas:2000xe}
	A.~Hatzinikitas,
	\newblock {\em {A Note on Riemann normal coordinates}},
	\newblock (2000), arXiv:hep-th/0001078.
	
	\bibitem{Barth:1983hb}
	N.~Barth and S.~Christensen,
	\newblock {\em {Quantizing Fourth Order Gravity Theories. 1. The Functional
			Integral}},
	\newblock Phys. Rev. D {\bf 28}, 1876 (1983).
	
	\bibitem{Duff:1993wm}
	M.~J. Duff,
	\newblock {\em {Twenty years of the Weyl anomaly}},
	\newblock Class. Quant. Grav. {\bf 11}, 1387 (1994), arXiv:hep-th/9308075.
	
	\bibitem{Codello:2012sn}
	A.~Codello, G.~D'Odorico, C.~Pagani, and R.~Percacci,
	\newblock {\em {The Renormalization Group and Weyl-invariance}},
	\newblock Class.Quant.Grav. {\bf 30}, 115015 (2013), arXiv:1210.3284.
	
	\bibitem{Antoniadis:1992xu}
	I.~Antoniadis, P.~O. Mazur, and E.~Mottola,
	\newblock {\em {Conformal symmetry and central charges in four-dimensions}},
	\newblock Nucl. Phys. {\bf B388}, 627 (1992), arXiv:hep-th/9205015.
	
	\bibitem{Antoniadis:1991fa}
	I.~Antoniadis and E.~Mottola,
	\newblock {\em {4-D quantum gravity in the conformal sector}},
	\newblock Phys. Rev. {\bf D45}, 2013 (1992).
	
	\bibitem{Mazur:2001aa}
	P.~O. Mazur and E.~Mottola,
	\newblock {\em {Weyl cohomology and the effective action for conformal
			anomalies}},
	\newblock Phys.Rev. {\bf D64}, 104022 (2001), arXiv:hep-th/0106151.
	
	\bibitem{Bonora:83}
	L.~Bonora, P.~Cotta-Ramusino, and C.~Reina,
	\newblock {\em {Conformal Anomaly and Cohomology}},
	\newblock Phys. Lett. {\bf B126}, 305 (1983).
	
	\bibitem{Mottola:2006ew}
	E.~Mottola and R.~Vaulin,
	\newblock {\em {Macroscopic effects of the quantum trace anomaly}},
	\newblock Phys. Rev. {\bf D74}, 064004 (2006), arXiv:gr-qc/0604051.
	
	\bibitem{Shapiro:1994ww}
	I.~L. Shapiro and A.~G. Zheksenaev,
	\newblock {\em {Gauge dependence in higher derivative quantum gravity and the
			conformal anomaly problem}},
	\newblock Phys. Lett. {\bf B324}, 286 (1994).
	
	\bibitem{Karakhanian:1994yd}
	D.~R. Karakhanian, R.~P. Manvelyan, and R.~L. Mkrtchian,
	\newblock {\em {Trace anomalies and cocycles of Weyl and diffeomorphism
			groups}},
	\newblock Mod. Phys. Lett. {\bf A11}, 409 (1996), arXiv:hep-th/9411068.
	
	\bibitem{Arakelian:1995ye}
	T.~Arakelian, D.~R. Karakhanian, R.~P. Manvelyan, and R.~L. Mkrtchian,
	\newblock {\em {Trace anomalies and cocycles of the Weyl group}},
	\newblock Phys. Lett. {\bf B353}, 52 (1995).
	
	\bibitem{Mottola:2010}
	E.~Mottola,
	\newblock {\em {New Horizons in Gravity: The Trace Anomaly, Dark Energy and
			Condensate Stars}},
	\newblock Acta Physica Polonica {\bf B41}, 2031 (2010), arXiv:1008.5006.
	
	\bibitem{Polyakov:1981}
	A.~M. Polyakov,
	\newblock {\em {Quantum Geometry of Bosonic Strings}},
	\newblock Phys. Lett. {\bf B103}, 207 (1981).
	
	\bibitem{Blaschke:2014ioa}
	D.~N. Blaschke, R.~Carballo-Rubio, and E.~Mottola,
	\newblock {\em {Fermion Pairing and the Scalar Boson of the 2D Conformal
			Anomaly}},
	\newblock JHEP {\bf 12}, 153 (2014), arXiv:1407.8523.
	
	\bibitem{AndMolMott:2003}
	P.~R. Anderson, C.~Molina-Paris, and E.~Mottola,
	\newblock {\em {Linear response, validity of semiclassical gravity, and the
			stability of flat space}},
	\newblock Phys. Rev. {\bf D67}, 024026 (2003), 0209075.
	
	\bibitem{Meissner:2016onk}
	K.~A. Meissner and H.~Nicolai,
	\newblock {\em {Conformal Anomalies and Gravitational Waves}},
	\newblock Phys. Lett. B {\bf 772}, 169 (2017), arXiv:1607.07312.
	
	\bibitem{Coriano:2008pg}
	C.~Corian\`o, M.~Guzzi, and S.~Morelli,
	\newblock {\em {Unitarity Bounds for Gauged Axionic Interactions and the
			Green-Schwarz Mechanism}},
	\newblock Eur. Phys. J. {\bf C55}, 629 (2008), arXiv:0801.2949.
	
	\bibitem{Isono:2019ihz}
	H.~Isono, T.~Noumi, and T.~Takeuchi,
	\newblock {\em {Momentum space conformal three-point functions of conserved
			currents and a general spinning operator}},
	\newblock JHEP {\bf 05}, 057 (2019), arXiv:1903.01110.
	
	\bibitem{Isono:2019wex}
	H.~Isono, T.~Noumi, and G.~Shiu,
	\newblock {\em {Momentum space approach to crossing symmetric CFT correlators.
			Part II. General spacetime dimension}},
	\newblock JHEP {\bf 10}, 183 (2019), arXiv:1908.04572.
	
	\bibitem{Chen:2019gka}
	H.-Y. Chen and H.~Kyono,
	\newblock {\em {On conformal blocks, crossing kernels and multi-variable
			hypergeometric functions}},
	\newblock JHEP {\bf 10}, 149 (2019), arXiv:1906.03135.
	
	\bibitem{Gillioz:2019iye}
	M.~Gillioz, X.~Lu, M.~A. Luty, and G.~Mikaberidze,
	\newblock {\em {Convergent Momentum-Space OPE and Bootstrap Equations in
			Conformal Field Theory}},
	\newblock (2019), arXiv:1912.05550.
	
	\bibitem{Gillioz:2019lgs}
	M.~Gillioz,
	\newblock {\em {Conformal 3-point functions and the Lorentzian OPE in momentum
			space}},
	\newblock (2019), arXiv:1909.00878.
	
	\bibitem{Gillioz:2018kwh}
	M.~Gillioz, X.~Lu, and M.~A. Luty,
	\newblock {\em {Graviton Scattering and a Sum Rule for the c Anomaly in 4D
			CFT}},
	\newblock JHEP {\bf 09}, 025 (2018), arXiv:1801.05807.
	
	\bibitem{Bautista:2019qxj}
	T.~Bautista and H.~Godazgar,
	\newblock {\em {Lorentzian CFT 3-point functions in momentum space}},
	\newblock (2019), arXiv:1908.04733.
	
	\bibitem{Loebbert:2020hxk}
	F.~Loebbert, J.~Miczajka, D.~M\"uller, and H.~M\"unkler,
	\newblock {\em {Massive Conformal Symmetry and Integrability for Feynman
			Integrals}},
	\newblock Phys. Rev. Lett. {\bf 125}, 091602 (2020), arXiv:2005.01735.
	
	\bibitem{Loebbert:2016cdm}
	F.~Loebbert,
	\newblock {\em {Lectures on Yangian Symmetry}},
	\newblock J. Phys. {\bf A49}, 323002 (2016), arXiv:1606.02947.
	
	\bibitem{Anand:2019lkt}
	N.~Anand, Z.~U. Khandker, and M.~T. Walters,
	\newblock {\em {Momentum space CFT correlators for Hamiltonian truncation}},
	\newblock (2019), arXiv:1911.02573.
	
	\bibitem{Albayrak:2019yve}
	S.~Albayrak and S.~Kharel,
	\newblock {\em {Towards the higher point holographic momentum space amplitudes
			II: Gravitons}},
	\newblock (2019), arXiv:1908.01835.
	
	\bibitem{Baumann:2019oyu}
	D.~Baumann, C.~Duaso~Pueyo, A.~Joyce, H.~Lee, and G.~L. Pimentel,
	\newblock {\em {The Cosmological Bootstrap: Weight-Shifting Operators and
			Scalar Seeds}},
	\newblock (2019), arXiv:1910.14051.
	
	\bibitem{Benincasa:2019vqr}
	P.~Benincasa,
	\newblock {\em {Cosmological Polytopes and the Wavefuncton of the Universe for
			Light States}},
	\newblock (2019), arXiv:1909.02517.
	
	\bibitem{Kundu:2014gxa}
	N.~Kundu, A.~Shukla, and S.~P. Trivedi,
	\newblock {\em {Constraints from Conformal Symmetry on the Three Point Scalar
			Correlator in Inflation}},
	\newblock JHEP {\bf 04}, 061 (2015), arXiv:1410.2606.
	
	\bibitem{Almeida:2017lrq}
	J.~P.~B. Almeida, J.~Motoa-Manzano, and C.~A. Valenzuela-Toledo,
	\newblock {\em {de Sitter symmetries and inflationary correlators in parity
			violating scalar-vector models}},
	\newblock JCAP {\bf 1711}, 015 (2017), arXiv:1706.05099.
	
	\bibitem{Baumann:2020dch}
	D.~Baumann, C.~Duaso~Pueyo, A.~Joyce, H.~Lee, and G.~L. Pimentel,
	\newblock {\em {The Cosmological Bootstrap: Spinning Correlators from
			Symmetries and Factorization}},
	\newblock (2020), arXiv:2005.04234.
	
	\bibitem{Almeida:2019hhx}
	J.~P. Beltr\'an~Almeida, J.~Motoa-Manzano, and C.~A. Valenzuela-Toledo,
	\newblock {\em {Correlation functions of sourced gravitational waves in
			inflationary scalar vector models. A symmetry based approach}},
	\newblock JHEP {\bf 09}, 118 (2019), arXiv:1905.00900.
	
	\bibitem{Penedones:2010ue}
	J.~Penedones,
	\newblock {\em {Writing CFT correlation functions as AdS scattering
			amplitudes}},
	\newblock JHEP {\bf 03}, 025 (2011), arXiv:1011.1485.
	
	\bibitem{Fitzpatrick:2011ia}
	A.~L. Fitzpatrick, J.~Kaplan, J.~Penedones, S.~Raju, and B.~C. van Rees,
	\newblock {\em {A Natural Language for AdS/CFT Correlators}},
	\newblock JHEP {\bf 11}, 095 (2011), arXiv:1107.1499.
	
	\bibitem{Gopakumar:2016cpb}
	R.~Gopakumar, A.~Kaviraj, K.~Sen, and A.~Sinha,
	\newblock {\em {A Mellin space approach to the conformal bootstrap}},
	\newblock JHEP {\bf 05}, 027 (2017), arXiv:1611.08407.
	
	\bibitem{Gopakumar:2016wkt}
	R.~Gopakumar, A.~Kaviraj, K.~Sen, and A.~Sinha,
	\newblock {\em {Conformal Bootstrap in Mellin Space}},
	\newblock Phys. Rev. Lett. {\bf 118}, 081601 (2017), arXiv:1609.00572.
	
	\bibitem{Sleight:2019mgd}
	C.~Sleight,
	\newblock {\em {A Mellin Space Approach to Cosmological Correlators}},
	\newblock (2019), arXiv:1906.12302.
	
	\bibitem{Sleight:2019hfp}
	C.~Sleight and M.~Taronna,
	\newblock {\em {Bootstrapping Inflationary Correlators in Mellin Space}},
	\newblock (2019), arXiv:1907.01143.
	
	\bibitem{Ohya:2018qkr}
	S.~Ohya,
	\newblock {\em {Conformal Ward-Takahashi Identity at Finite Temperature}},
	\newblock Springer Proc. Math. Stat. {\bf 255}, 271 (2017), arXiv:1801.02902.
	
	\bibitem{Bzowski:2019kwd}
	A.~Bzowski, P.~McFadden, and K.~Skenderis,
	\newblock {\em {Conformal 4-point functions in momentum space}},
	\newblock (2019), arXiv:1910.10162.
	
\end{thebibliography}

\newpage
\chapter*{}
\section*{Acknowledgements}
\addcontentsline{toc}{chapter}{Acknowledgements}
\chaptermark{Acknowledgements}

\noindent First of all, I would like to thank my advisor Claudio Corian\`o for his teachings over all these years. Thank you, Claudio, for teaching me how to develop and pursue ideas in the research world, for allowing me to meet and collaborate with scientists from different parts of the world.  \\
I want to thank Maxim Chernodub for his support and tireless enthusiasm for doing research. \\
I want to thank Paolo Benincasa for his fundamental support in recent years and his enlightening discussions and advice.\\
I want to thank Emil Mottola, Paul McFadden and Kostas Skenderis for the moments of discussion over the years. \\
Thanks to those who shared music, thoughts and moments with me.  
\end{document}